\documentclass[11pt,twoside,a4paper]{book}
\usepackage[utf8]{inputenc}
\usepackage[english]{babel}
\usepackage[dvips]{graphicx}
\usepackage{amsmath,amsthm,amssymb,esint,slashed,extarrows}
\usepackage[margin=3cm]{geometry}
\usepackage{float}
\usepackage{mathtools}
\usepackage{bm}
\usepackage{mathrsfs}
\usepackage{esvect}
\usepackage{hyperref}
\usepackage{feynmf}
\usepackage{simplewick}
\usepackage{anyfontsize}
\usepackage{setspace}
\usepackage{sectsty}
\usepackage{isotope}
\usepackage{afterpage}
\usepackage{xcolor}
\usepackage{multirow}
\usepackage{multicol}
\usepackage{array}
\usepackage{float}
\usepackage{cite}
\usepackage{simplewick}
\usepackage{chngcntr}
\usepackage{rotating}
\usepackage[printonlyused]{acronym}

\hypersetup{
    colorlinks=true,
    linkcolor=blue,
    filecolor=magenta,      
    urlcolor=cyan,
}

\title{PhD Thesis: Chiral Symmetries in the Standard Model of Particle Physics and its Extensions}
\author{Sergio Ferrando Solera}
\date{}

\begin{document}

\begin{titlepage}
\centering
{\includegraphics[width=\textwidth]{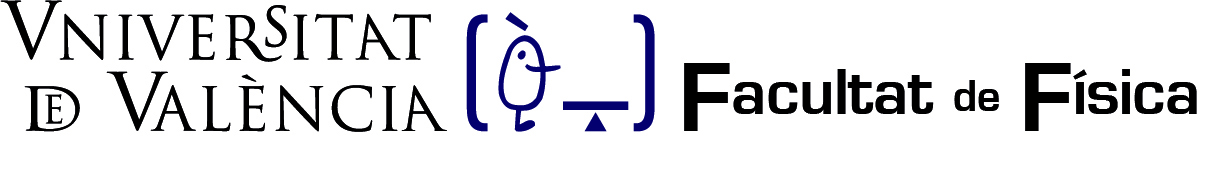}\par}
\vspace{0.3cm}
{\scshape DEPARTAMENT DE FÍSICA TEÒRICA \par}
\vspace{1.5cm}
{\includegraphics[width=0.5\textwidth]{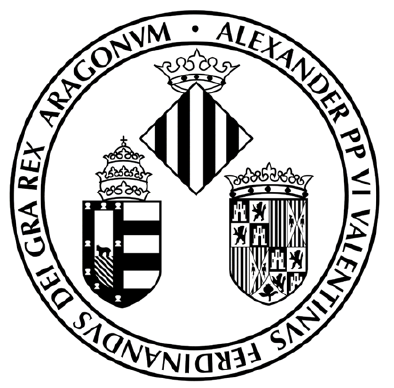}\par}
\vspace{1.5cm}
{\scshape\LARGE Chiral Symmetries in the Standard Model of Particle Physics and its Extensions \par}
\vspace{2cm}
{\large TESIS DOCTORAL \par}
{\large PROGRAMA DE DOCTORAT EN FÍSICA - 3126 \par}
\vspace{0.5cm}
{\large SERGIO FERRANDO SOLERA \par}
\vspace{0.5cm}
{\large Dirigida por: \par}
{\large Dr. Antonio Pich Zardoya \par}
{\large Dr. Luiz Henrique Vale Silva \par}
\vspace{0.5cm}
{\large Julio 2026 \par}
\end{titlepage}

\newpage
\thispagestyle{empty}
\mbox{}
\newpage




\frontmatter
\setcounter{page}{1}

\chapter*{Declaration of Authorship}

The author declares that the use of artificial intelligence tools in the preparation of this thesis was strictly limited to translation, language editing, and grammatical proofreading. No generative artificial intelligence tools were used to produce, draft, or synthesize the research content, arguments, or conclusions presented herein. All intellectual work, analysis, and writing remain the sole responsibility of the author. The research presented in this thesis is based on the series of articles and proceedings published by the author during his doctoral studies \cite{Solera:2023kwt,Solera:2024wjz,Solera:2024jrm,ValeSilva:2025,Solera:2026xnz}. Furthermore, the argumentation in the first chapter builds upon the framework established by Weinberg in \cite{Weinberg2}, focusing on the essential elements required for the development of this text.

\newpage
\thispagestyle{empty}
\mbox{}
\newpage

\cleardoublepage
\thispagestyle{empty}
\vspace*{5cm}
\begin{flushright}
   \large \textit{A Olivia, mi teoría del todo.}
\end{flushright}

\newpage
\thispagestyle{empty}
\mbox{}
\newpage

\chapter*{Agradecimientos}
\markboth{AGRADECIMIENTOS}{AGRADECIMIENTOS}

\hspace{15pt} En primer lugar, debo agradecer al Ministerio de Universidades (Gobierno de España), número de contrato FPU20/04279, por contribuir con la financiación de esta tesis doctoral.

Me gustaría agradecer, por sus aportaciones y conversaciones enriquecedoras, a todos los investigadores y colaboradores que han contribuido al desarrollo de los artículos escritos durante mi etapa doctoral: Chitrangada Devi, Prasanna Kumar Dhani, Andreas Hinzmann, Greg Landsberg, Jeongeun Lee, Emanuela Musumeci, José Zurita, Herm\`{e}s B\'{e}lusca-Ma\"{i}to, Marco Ardu, Shankha Banerjee, Joachim Brod, Leandro Cieri, Martín González Alonso, Kirsten Leslie, Zachary Polonsky, Jorge Portolés, Marcos N. Rabelo, Marc Riembau, Germán Rodrigo, Javier Fuentes Martín, Adrián Moreno Sánchez y Sebastian Jäger.

Quisiera dar mi más sincero agradecimiento a mis dos directores de tesis por su inestimable ayuda en todo este tiempo: Toni y Luiz.

Toni ha sido un pilar fundamental en mi desarrollo académico desde que, por la mayor de las fortunas, aceptó ser mi director de TFG hace ya 7 años. Él es la prueba de que una persona brillante (que siempre me sorprendía por su velocidad para comprender y aportar a los razonamientos que a mí tanto me había costado elaborar), puede ser a su vez el mayor ejemplo de humildad y de saber transmitir, de la mejor de las maneras, su pasión por la física a todas las personas que han tenido la suerte de ser sus alumnos.

Luiz, por su parte, ha sido la persona que ha hecho posible que este proyecto llegara a buen puerto. Siempre supo guiarme por el mejor de los caminos, y mantenerse firme en aquellos arduos momentos en los que parecía imposible alcanzar los resultados correctos (los operadores octete nos han restado algún que otro año de vida). Siempre guardaré el recuerdo de esas divertidas reuniones semanales plagadas de física y, por qué no decirlo, de MMA, culturismo, y otros muchos temas apasionantes.

A partir de este punto, me gustaría dar las gracias a ciertas personas que no han contribuido de forma directa a la confección de este texto, pero que sí han sido muy importantes para mí en este periodo.

En primer lugar, quisiera hacer una mención especial a las personas que han estado a mi lado en el mundo de la música, que tanta felicidad me han aportado en este tiempo. 

No puedo olvidar a los que considero mis tres grandes maestros musicales, aquellos que con su buen hacer y su amabilidad han contribuido en mayor medida a que yo pueda desarrollarme como músico, abrirme camino y disfrutar de la mejor manera de esta disciplina: mis queridos Cremades, Pascual y Juanjo. 

Una mención especial se merece también mi compañero de fagot y amigo Sisco. Siempre me sentiré orgulloso y le agradeceré que, con tanto cariño, me llame \textit{sensei}. Cuando él está a mi lado en un ensayo, por aburrido que este sea, logra que se convierta en un momento lleno de chistes y de buena música; haciendo que dejara de lado, por un momento, los problemas del día a día.

No puedo dejar de agradecer a todas las personas del CMP que me han hecho pasar ratos inolvidables: Amparo, Iván, Elena, Emma, Nuria, María, Marta y un muy largo etcétera.

Fuera del mundo de la música (aunque fanática de Hans Zimmer), mi estimada Susanna se merece un inmenso agradecimiento por todo lo que ha hecho y todo lo que ha aguantado conmigo. Ella, al igual que hizo Maxwell con la humanidad, me ayudó a encontrar la luz en el caos.

Me gustaría dedicar un momento a recordar con ternura (como siempre hago) a mi amigo Andrés, a quien le agradezco haberme acompañado en mis inicios en la aventura de la física, yendo juntos a las clases para las Olimpiadas. Un abrazo allí donde estés.

Quisiera centrarme en este punto en agradecer a todos mis queridos amigos que me han acompañado en este viaje. Todos ellos tienen ganado el cielo por aguantar mis interminables charlas sobre cine, música, literatura, anime, filosofía, economía, curiosidades de todo tipo, matemáticas y, en especial, física. Ciertamente, han aguantado muchas horas de comentarios friquis de física\dots ¡Así que podrán aguantar unos pocos más que vienen a continuación!

No puedo dejar de nombrar a Jorge, una persona más fuerte que la interacción que mantiene los núcleos unidos y con un corazón más grande que el alcance de la interacción gravitatoria. Le agradezco que sea una persona que, siempre con una amable sonrisa, sepa estar a tu lado en los momentos buenos y malos.

Me gustaría agradecer a mi amigo Xavi, cuyos intereses se mueven por caminos similares a los que sigue una partícula con movimiento browniano, todas las fascinantes conversaciones filosóficas, económicas y científicas que me ha permitido tener. Aunque sus gustos y aficiones puedan ser variables, aquello en lo que es verdaderamente constante es, precisamente, lo más importante: ser un buen amigo y una bellísima persona.

Agradezco a mi amigo Guille, con quien compartiré el corte de pelo de un agujero negro, todos esos grandes momentos que he compartido con él desde primaria hasta ahora. Aunque se sienta mal porque cree no saber ayudar, lo cierto es que siempre es una persona en quien confiar cuando tienes un problema.

Quisiera recordar también a mi amigo Pablo, que antes destacaba por su corte de pelo gluónico. Todo un atleta de élite que necesita unos buenos pectorales porque si no el corazón se le sale de tan grande que lo tiene. Es un amigo dulce, amable, sensible y trabajador, que va a lograr todo lo que se proponga, y a quien siempre agradeceré todas las risas que hemos compartido.

Jamás podría olvidar a mi inestimable amigo Álex que, al igual que la energía oscura en el universo, es una constante en mi vida. Tengo argumentos teóricos para saber desde cuándo somos amigos, pero ni me acuerdo del momento en el que te conocí de tanto tiempo que hace; supongo que sería durante la época inflacionaria (la física, no la mala). Jamás podré agradecerte lo suficiente todos los momentos que hemos compartido y que hayas estado siempre a mi lado.

Si debo dedicar las palabras que realmente merecen los demás amigos a quienes debo tanto, creo que se me pasaría la fecha de depósito de la tesis. Además, ya no se me ocurren más chistes físicos y no quiero usar la IA para ello, que pierde el toque personal, así que me tendré que contentar con nombrar y agradecer por los buenos momentos a otros muchos geniales amigos que han estado aquí este tiempo: Iván, Rubén, Víctor, Teresa, María Amparo y un largo etcétera.

No obstante, hay dos personas que han sido algo más que amigos durante estos cinco años; han sido unos hermanos, mis compañeros de doctorado, los \textit{Physics Yonkous}: Ezequiel y Pablo.

A Ezequiel, mi estimado \textit{nakama}, que en estos momentos está viviendo la vida de un gran navegante contemporáneo, viajando por el mundo mirando a las galaxias más lejanas creando los mejores mapas jamás vistos, le quiero agradecer todas y cada una de las palabras que hemos compartido durante estos años. Nuestras charlas han estado llenas de los temas más variopintos, de las reflexiones más profundas y, sobre todo, llenas de momentos de humor que hacían que se nos saltaran las lágrimas. 

Y qué decir de Pablo, mi Roger Federer particular, un amigo que siempre fue un ejemplo a seguir. Brillante, talentoso y trabajador a partes iguales, siempre le diré lo mismo: ``Te mereces todas y cada una de las cosas buenas que te pasen''. No sé si en el cántaro tendrá espacio para guardar todos los galardones que seguro acumulará con el tiempo, pero siempre ha tenido un hueco en su apretada agenda para dedicármelo cuando lo necesitaba, en especial, para enfrentarme a mi archienemiga: la burocracia. Quiero agradecerle lo buen amigo que siempre ha sido conmigo y todos los geniales momentos que hemos pasado juntos, desde el comienzo de la carrera, cuando hacíamos integrales en las mesas, hasta las comidas con Ezequiel, que tanta felicidad nos trajeron estos años del doctorado. 

Paso ahora a recordar a los familiares que han estado conmigo durante este periodo. Comenzaré con mis queridos abuelos Ángel y Teresa, a quienes agradezco el inmenso cariño que me han dado, especialmente cuando me cuidaban de pequeño. A mi abuela quiero decirle que, al menos por un tiempo, cumplí mi sueño de la infancia de convertirme en un científico. A mi abuelo, allí donde esté, que intento cumplir con su última enseñanza: hacer siempre lo que pueda, no más.

No puedo olvidar a mi tía Belvis y a Armando, a quienes agradezco que siempre tuvieran las puertas de su casa abiertas cuando yo lo necesitaba. Ellos trajeron al mundo a dos de las personas más especiales de mi vida, mis primos Pablo y Lucía, a quienes agradezco enormemente esos buenísimos momentos que me regalan todos los jueves en nuestras cenitas de \textit{anime}; no son conscientes ni jamás podrán imaginar cuánta alegría me traen. Son personas excepcionales con las que siempre espero con ansias reunirme.

Por supuesto, quisiera nombrar a Vicente y a Carmen, y agradecerles enormemente la acogida tan cálida y cariñosa que me han brindado durante estos años. Para mí ellos son la combinación perfecta entre la serenidad y la tormenta, el modelo a seguir de lo que es el amor: enfrentarse juntos a las dificultades de la vida, apoyando siempre a su compañero en sus proyectos vitales. Adoro todas las comidas de los domingos, llenas de buenos consejos y aprendizajes que me ayudan a tener los pies en el suelo, aunque siempre mirando a las estrellas. 

Obviamente, debo hablar de cuatro de los pilares de mi familia: mi hermano, Mari y mis dos queridos sobrinos: Aitana y Diego.

A Mari le agradezco la inmensa paciencia que ha tenido conmigo este tiempo; no es fácil atender a todas mis consultas, pero no sé qué sería de mí sin todos sus buenos consejos. Siempre busca cualquier momento libre que tiene para ayudar, en todos los sentidos, con la mayor ilusión y amabilidad a sus seres queridos. Pese a que de pequeño me enfadaba contigo porque creía que me quitabas el amor de mi hermano, lo cierto es que estuve muy equivocado: al final salí ganando, pues ese amor se sumó al tuyo. 

De pequeño no paraba de decir que quería un sobrinito\dots y finalmente ese regalo llegó multiplicado por dos, pues aparecieron Aitana y Diego. Les agradezco enormemente que ambos sean unas personas tan increíbles. Mi sobrina Aitana es dulce, amable, trabajadora y, sobre todo, ¡intensa! Mi sobrino Diego es un chico alegre y divertido, con un corazón inmenso y lleno de bondad. Aitano y Diega\dots ¡disculpad! Diego y Aitana (mi coco solo anda así así), no os imagináis cómo adoro veros crecer y presenciar en qué increíbles personas os estáis convirtiendo; vuestro tío siempre estará ahí para ayudaros.

Finalmente, tengo que hablar de mi tete. A él le agradezco que haya sido mi segundo padre, que de pequeño supiera enseñarme lo importante que es el esfuerzo y que fuera siempre un ejemplo a seguir. No sé si lo sabes, tete, pero cuando veo a mis profesoras de primaria me recuerdan que de pequeño yo solo sabía hablar de ti y de cuánto te admiraba (y lo sigo haciendo). No dejes nunca de tener la alegría y la curiosidad que tienes, ni de cuidar, a esas poquitas personas que de verdad te importan, tanto como las cuidas ahora.

Evidentemente, debo hablar de las dos personas que siempre han estado conmigo y a quienes les debo todo: mis padres. 

A mi madre, en especial le agradezco la alegría que siempre tiene y que se haya desvivido por su familia. Pese a que en la adolescencia me sacaba de mis casillas lo protectora que eras conmigo, ahora entiendo que solo lo hacías por lo importante que yo era para ti. También sé que cuando te cuento mis cosas de física no me comprendes, pero es que, parafraseando lo que se cuenta que le dijo Chaplin a Einstein, me quieres aunque no siempre me entiendes. Por más títulos que yo consiga sé que siempre me verás como a tu hijito pequeño y que dirás que muy listo para unas cosas y muy tonto para otras; estoy conforme y profundamente agradecido por ello.

Por supuesto, tengo que hablar de mi padre. Él es la persona que ha hecho posible que yo haya llegado hasta este punto de mi vida, algo por lo que le estaré eternamente agradecido. Las conversaciones que teníamos en la cama cuando yo era tan solo un niño fueron las que me hicieron convertirme en físico; tú fuiste capaz de saciar y fomentar mi gran curiosidad. No solo eso, tú creíste en mí y me llevaste en coche (te hice trabajar más que cuando eras taxista) al conservatorio, a las clases, a los premios\dots ¡a cualquier sitio que yo necesitara! Sin ello habría sido totalmente imposible que yo hubiera podido hacer todo lo que hice. Por más duro que te quieras hacer, sé que eres la persona con el corazón más bondadoso que existe y que dedicas hasta el último de tus esfuerzos a cuidar de tu familia. 

Ha sido para mí un orgullo y un regalo teneros a ambos como padres y os querré eternamente.

Por último, tengo que hablar de mi amor, de Olivia, la persona más importante de mi vida durante estos años. Yo, por más que lo intente, no sé cómo decirte lo que quiero decirte. Creo que ni una combinación de Cervantes y Beethoven podría expresar lo que mereces que exprese. Siempre has estado ahí, siempre, en cada instante, ya fuera de luz o de oscuridad. Siempre has sabido entenderme y apoyarme cuando yo más lo necesitaba, además de que me has hecho pasar los instantes más felices de mi vida. Me haces querer ser mejor persona. Repito, no sé cómo expresar todo lo que debo; simplemente\dots te lo agradezco todo.

\chapter*{Abstract}
\markboth{ABSTRACT}{ABSTRACT}

 Chiral gauge theories have become a central focus of research in theoretical particle physics, largely due to the explicit breaking of parity symmetry observed in the electroweak sector of the Standard Model. Several models have been developed to address the question of why this symmetry is broken at the energy scales of the current experiments; a prominent example being the Left-Right Symmetric Models (LRM), where this symmetry is restored at high energies, and its observed breaking is explained through a Spontaneous Symmetry Breaking (SSB) mechanism. These models introduce a new interaction governed by an \({\rm SU}(2)_{\rm R}\) group acting on the right-handed fermions, thereby predicting the existence of new \(W_R^{\pm}\) and \(Z_R\) gauge bosons. However, given the lack of direct detection of new physics at the LHC, current research has focused on setting lower bounds on these new particle masses. This absence of direct evidence has motivated two primary research avenues: first, the calculation of precision observables, which requires computing high-order radiative corrections in Quantum Field Theory; and second, the Effective Field Theory (EFT) approach. The latter, specifically the Standard Model Effective Field Theory (SMEFT) framework, enables a model-independent analysis (considering no further light-degrees of freedom) of the implications of potential new physics in current experimental data.

Radiative calculations are usually performed using dimensional regularization. However, in chiral gauge theories, this method requires a prescription for the \(D\)-dimensional definition of the \(\gamma_5\) matrix. To date, the only mathematically consistent definition known is the one given in the Breitenlohner-Maison-’t~Hooft-Veltman scheme. While robust, this prescription has the inconvenience of introducing gauge symmetry-breaking terms into the \(D\)-dimensional Lagrangian, that lead to obstructions of the Slavnov-Taylor identities in loop calculations. These obstructions can be remedied in anomaly-free theories by adding a proper set of gauge-variant divergent and finite (not necessarily evanescent) counterterms. One of the main objectives of this PhD thesis is to provide a pedagogical introduction to the method of cancellation of these spurious anomalies within the algebraic renormalization framework, and implement these techniques to derive the one-loop finite symmetry-restoring counterterms required for the renormalization of the four-fermion operators of the SMEFT.

Instead of effective models, one can also consider UV completions, like the aforementioned Left-Right Symmetric Models. They can be embedded into other new physics models, such as Grand Unified Theories based on larger gauge groups. This embedding has two important consequences: first, since parity restoration can occur at much higher energy scales, the renormalization group evolution of the coupling constants suggests that no discrete symmetries remain present in the LRMs at the TeV scale; second, they can be treated as EFTs, allowing for the inclusion of only the minimal scalar content to produce the SSB pattern, while generating fermion masses through dimension-5 effective operators. This effective LRM framework is developed in this work. A primary objective of the present PhD thesis is to derive lower bounds on the masses of the new gauge bosons and on the LR scale using collider observables. This analysis is performed for both the effective and traditional LR models, ensuring that the results remain independent of the other parameters of the theory.

\chapter*{Resumen}
\markboth{RESUMEN}{RESUMEN}

\hspace{15pt} Las teorías \textit{gauge} quirales se han convertido en un foco central de investigación en la física teórica de partículas, en gran medida debido a la ruptura explícita de la simetría de paridad observada en el sector electrodébil del Modelo Estándar. Se han desarrollado diversos modelos para abordar la cuestión de por qué esta simetría se rompe en las escalas de energía de los experimentos actuales; un ejemplo destacado son los Modelos de Simetría \textit{Left-Right} (LRM por sus siglas en inglés), donde esta simetría se restaura a altas energías y la ruptura observada se explica mediante un mecanismo de Ruptura Espontánea de la Simetría (SSB por sus siglas en inglés). Estos modelos introducen una nueva interacción gobernada por el grupo ${\rm SU}(2)_{\rm R}$ que actúa sobre los fermiones con quiralidad derecha, prediciendo así la existencia de nuevos bosones de gauge $W_R^{\pm}$ y $Z_R$. Sin embargo, dada la falta de detección directa de nueva física en el LHC, la investigación actual se ha centrado en establecer límites inferiores para las masas de estas nuevas partículas. Esta ausencia de evidencia directa ha motivado dos vías de investigación principales: primero, el cálculo de observables de precisión, que requiere obtener correcciones radiativas de alto orden en Teoría Cuántica de Campos; segundo, el enfoque de Teorías de Campos Efectivas (EFT por sus siglas en inglés). Este último, específicamente el marco de la Teoría de Campos Efectiva del Modelo Estándar (SMEFT por sus siglas en inglés), permite un análisis independiente del modelo (sin considerar nuevos grados de libertad ligeros) sobre las implicaciones de potenciales efectos de nueva física en los datos experimentales actuales.

Los cálculos radiativos se realizan habitualmente utilizando regularización dimensional. Sin embargo, en las teorías gauge quirales, este método requiere una prescripción para la definición de la matriz $\gamma_5$ en $D$ dimensiones. Actualmente, la única definición matemáticamente consistente conocida es la dada en el esquema de Breitenlohner-Maison-’t~Hooft-Veltman. Aunque es robusta, esta prescripción tiene el inconveniente de introducir términos que rompen la simetría gauge en el Lagrangiano de $D$ dimensiones, lo que provoca obstrucciones en las identidades de Slavnov-Taylor durante los cálculos de lazos (\textit{loops}). Estas obstrucciones pueden subsanarse en teorías libres de anomalías mediante la adición de un conjunto adecuado de contratérminos variantes \textit{gauge}, tanto divergentes como finitos (no necesariamente evanescentes). Uno de los objetivos principales de esta tesis doctoral es proporcionar una introducción pedagógica al método de cancelación de estas anomalías espurias dentro del marco de la renormalización algebraica, e implementar estas técnicas para derivar los contratérminos finitos necesarios para restaurar la simetría \textit{gauge} a un lazo, asociados a la renormalización de los operadores de cuatro fermiones de la SMEFT.

En lugar de modelos efectivos, uno puede considerar nuevas teorías de altas energías, como los Modelos de Simetría \textit{Left-Right} mencionados anteriormente. Estos pueden integrarse en otros modelos de nueva física, como las Teorías de Gran Unificación basadas en grupos de \textit{gauge} más amplios. Esta integración tiene dos consecuencias importantes: primero, dado que la restauración de la paridad puede ocurrir a escalas de energía mucho más altas, la evolución de las constantes de acoplamiento de acuerdo con las ecuaciones del grupo de renormalización, sugiere que las simetrías discretas no están presentes en los LRM a la escala de los TeV; en segundo lugar, estos pueden tratarse como EFT, lo que permite incluir únicamente el contenido escalar mínimo para producir el patrón de SSB, mientras que las masas de los fermiones se generan a través de operadores efectivos de dimensión 5. En este trabajo se desarrolla este marco de LRM efectivo. Un objetivo primordial de la presente tesis doctoral es derivar cotas inferiores para las masas de los nuevos bosones de \textit{gauge} y para la escala LR utilizando observables de colisionadores. Este análisis se realiza tanto para el modelo efectivo como para los tradicionales, asegurando que los resultados sean independientes de los demás parámetros de la teoría.

\chapter*{Resum}
\markboth{RESUM}{RESUM}

\hspace{15pt} Les teories \textit{gauge} quirals s'han convertit en un focus central de recerca en la física teòrica de partícules, en gran mesura a causa de la ruptura explícita de la simetria de paritat observada en el sector electrofeble del Model Estàndard. S'han desenvolupat diversos models per abordar la qüestió de per què aquesta simetria es trenca a les escales d'energia dels experiments actuals; un exemple destacat són els Models de Simetria \textit{Left-Right} (LRM per les seves sigles en anglès), on aquesta simetria es restaura a altes energies, i la seva ruptura observada s'explica mitjançant un mecanisme de Ruptura Espontània de la Simetria (SSB per les seves sigles en anglès). Aquests models introdueixen una nova interacció governada per un grup ${\rm SU}(2)_{\rm R}$ que actua sobre els fermions dretans, predient així l'existència de nous bosons de \textit{gauge} $W_R^{\pm}$ i $Z_R$. No obstant això, donada la manca de detecció directa de nova física a l'LHC, la recerca actual s'ha centrat a establir límits inferiors per a les masses d'aquestes noves partícules. Aquesta absència d'evidència directa ha motivat dues vies de recerca primàries: primer, el càlcul d'observables de precisió, que requereix computar correccions radiatives d'ordre superior en Teoria Quàntica de Camps; i segon, l'enfocament de la Teoria de Camps Efectiva (EFT  per les seves sigles en anglès). Aquest últim, específicament el marc de la Teoria de Camps Efectiva del Model Estàndard (SMEFT per les seves sigles en anglès), permet una anàlisi independent del model (sense considerar nous graus de llibertat lleugers) sobre les implicacions de la potencial nova física en les dades experimentals actuals.

Els càlculs radiatius es realitzen habitualment mitjançant la regularització dimensional. No obstant això, en les teories de \textit{gauge} quirals, aquest mètode requereix una prescripció per a la definició en $D$ dimensions de la matriu $\gamma_5$. Fins a la data, l'única definició matemàticament consistent que es coneix és la donada en l'esquema de Breitenlohner-Maison-’t~Hooft-Veltman. Tot i que la prescripció és robusta,  té l'inconvenient d'introduir termes de ruptura de la simetria de \textit{gauge} en el Lagrangià de $D$ dimensions, fet que condueix a obstruccions de les identitats de Slavnov-Taylor en els càlculs de llaç (\textit{loop}). Aquestes obstruccions es poden esmenar en teories lliures d'anomalies afegint un conjunt adequat de contra-termes variants \textit{gauge}, tant divergents com finits (no necessàriament evanescents). Un dels objectius principals d'aquesta tesi doctoral és proporcionar una introducció pedagògica al mètode de cancel·lació d'aquestes anomalies espúries dins del marc de la renormalització algebraica, i implementar aquestes tècniques per derivar els contra-termes finits necessaris per la restauració de la simetria \textit{gauge} a un llaç, associats a la renormalització dels operadors de quatre fermions de la SMEFT. 

En lloc de models efectius, es pot considerar noves teories d'altes energies, com els Models de Simetria \textit{Left-Right} esmentats anteriorment. Aquests es poden integrar en altres models de nova física, com les Teories de Gran Unificació basades en grups \textit{gauge} més amplis. Aquesta integració té dues conseqüències importants: en primer lloc, atès que la restauració de la paritat pot ocórrer a escales d'energia molt més elevades, l'evolució del grup de renormalització de les constants d'acoblament suggereix que no roman cap simetria discreta en els LRM a l'escala del TeV; en segon lloc, es poden tractar com a EFT, fet que permet incloure només el contingut escalar mínim per produir el patró de SSB, mentre que les masses dels fermions es generen mitjançant operadors efectius de dimensió 5. Aquest marc de treball del LRM efectiu es desenvolupa en aquest text. Un dels objectius primordials d'aquesta tesi doctoral és derivar límits inferiors per a les masses dels nous bosons de \textit{gauge} i per a l'escala LR utilitzant observables de col·lisionadors. Aquesta anàlisi es realitza tant per als models LR efectius com per als tradicionals, assegurant que els resultats romanguin independents de la resta de paràmetres de la teoria.

\chapter*{Resumen General}
\markboth{RESUMEN GENERAL}{RESUMEN GENERAL}

\hspace{15pt} La Teoría Cuántica de Campos (QFT por sus siglas en inglés) es el único marco conocido que unifica los principios de la mecánica cuántica y la relatividad especial. La primera es necesaria para comprender el comportamiento de los objetos microscópicos, mientras que la segunda explica la dinámica de los cuerpos que se mueven a velocidades cercanas a la de la luz, y la conversión masa-energía $E=mc^2$. Por consiguiente, las QFT son indispensables para dar sentido a los datos obtenidos en experimentos que involucran, por ejemplo, partículas subatómicas moviéndose a altas velocidades, o la desintegración de ciertos estados. El ejemplo más famoso de este tipo de experimentos son los aceleradores de partículas, como el Gran Colisionador de Hadrones (LHC por sus siglas en inglés). Además, las partículas con velocidades relativistas eran abundantes en los primeros instantes del universo, caracterizados por densidades y temperaturas extremadamente elevadas. Así, las QFT también desempeñan un papel crucial en la comprensión de las condiciones iniciales tras el \textit{Big Bang}. Son una herramienta fundamental en la exploración de las fronteras de la física y en la descripción de las interacciones de los constituyentes fundamentales de la materia.

Desde el punto de vista del autor de este texto, el descubrimiento más importante de la física en el siglo pasado, fue que las leyes de la naturaleza pueden derivarse, usando el formalismo de la teoría de campos, imponiendo que una determinada simetría se res\-peta. Ejemplos célebres son la simetría de Poincaré (simetría de Lorentz más traslaciones espaciotemporales), que es la base de la relatividad especial; o la invariancia bajo difeomorfismos, de la cual se obtienen las leyes de la gravitación en la relatividad general. En el contexto de la física de partículas, la importancia de las simetrías quedó clara tras el descubrimiento de simetrías globales como el isospín (extendido posteriormente a la simetría de sabor basada en el grupo ${\rm SU}(3)$), que resultó muy útil para explicar las propiedades e interacciones de los hadrones. Sin embargo, en este ámbito, las transformaciones más importantes no son las globales, sino las locales, las llamadas simetrías \textit{gauge}. Estas son transformaciones, generalmente basadas en grupos de Lie, que varían en los distintos puntos del espaciotiempo. A partir de la estructura geométrica de las transformaciones \textit{gauge}, las interacciones posibles entre los diferentes estados quedan completamente determinadas.  Las simetrías \textit{gauge} en QFT permitieron el desarrollo del Modelo Estándar (SM por sus siglas en inglés) de la física de partículas en el siglo pasado, la teoría que se utiliza actualmente para describir las interacciones (excepto la gravedad) entre las partículas elementales. La construcción de este modelo requirió el esfuerzo conjunto de físicos teóricos y experimentales, y culminó en 2012 con el descubrimiento del bosón de Higgs en el LHC por las colaboraciones ATLAS y CMS.

El Modelo Estándar se basa en el grupo \textit{gauge} ${\rm SU}(3)_{\rm QCD} \otimes {\rm SU}(2)_{\rm L} \otimes {\rm U}(1)_{\rm Y}$. Las interacciones fuertes se describen mediante la parte simple ${\rm SU}(3)_{\rm QCD}$ del grupo, mientras que el sector electrodébil queda determinado por la estructura de ${\rm SU}(2)_{\rm L} \otimes {\rm U}(1)_{\rm Y}$. Para los propósitos de este texto, la característica más notable del SM es que la interacción débil solo actúa sobre la componente levógira (${\rm SU}(2)_{\rm L}$) de los fermiones de la teoría. Esto tiene una consecuencia inmediata: la naturaleza es quiral, lo que significa que las interacciones distinguen entre las quiralidades izquierda y derecha de las partículas, por lo que la simetría de paridad se rompe; es decir, las leyes de la física cambian cuando se observan a través de un espejo. Actualmente se espera que el SM sea una teoría válida solo hasta cierta escala de energía, como consecuencia, es una Teoría de Campos Efectiva (EFT por sus siglas en inglés). No obstante, la naturaleza quiral de los fermiones y sus interacciones es una propiedad que debe ser considerada (o incluso explicada) por cualquier posible extensión de nueva física más allá del SM. Por tanto, esta tesis doctoral está dedicada al estudio de las simetrías \textit{gauge} quirales en la teoría cuántica de campos; particularmente, su papel en los cálculos de precisión (correcciones radiativas) en la extensión predilecta del SM, la Teoría de Campos Efectiva del Modelo Estándar (SMEFT por sus siglas en inglés), y cómo propuestas para la explicación de la estructura quiral de la naturaleza conducen a nuevas teorías físicas, en concreto los modelos \textit{Left-Right} (LRM por sus siglas en inglés).

El cálculo matemáticamente riguroso de las correcciones radiativas en teorías de \textit{gauge} quirales resulta altamente complejo debido a la presencia de la matriz $\gamma_5$. Hoy en día, la prescripción de regularización preferida para tener en cuenta las divergencias de las amplitudes en tales cálculos es la regularización dimensional. No obstante, aunque en este esquema se preserva la versión en cuatro dimensiones de la simetría de Lorentz, las simetrías de \textit{gauge} quirales se ven violadas incluso a nivel clásico cuando se emplea la prescripción de  Breitenlohner-Maison-'t~Hooft-Veltman (BMHV) para $\gamma_5$, que es el único método matemáticamente consistente que se conoce. Utilizando las herramientas de la renormalización algebraica, se puede demostrar que en esta situación existen obstrucciones a las identidades de Slavnov-Taylor (indispensables para la consistencia de la teoría), generadas por una inserción de un operador de ruptura de simetría (que es, simplemente, la transformación BRST del Lagrangiano en $D$-dimensiones) en diagramas 1PI. Dependiendo de su naturaleza, estas obstrucciones pueden clasificarse como: espurias (términos de orden $\varepsilon$), divergentes (términos de orden $1/\varepsilon$) y finitas (términos de orden $\varepsilon/\varepsilon$). Estas últimas pueden calcularse a partir de la parte divergente de los diagramas de Feynman utilizando el método de Bonneau. En particular, en una teoría que esté libre de anomalías físicas, estas obstrucciones finitas a las identidades de Slavnov-Taylor pueden eliminarse añadiendo el conjunto apropiado de contratérminos finitos.

Por otro lado, los modelos LR explican la ruptura de paridad a través de un mecanismo de Ruptura Espontánea de Simetría (SSB por sus siglas en inglés). En ellos se añade una interacción análoga a la débil para las partículas dextrógiras basada en el grupo \({\rm SU}(2)_{\rm R}\), por lo que aparecen nuevos bosones \textit{gauge} \(W_R^{\pm}\) y \(Z_R\). La simetría de paridad se restaura imponiendo que las constantes de acoplamiento de ambas interacciones sean iguales (\(g_L=g_R\)) en la nueva escala de energías LR \(v_R\), que debe ser muy superior, por razones fenomenológicas, a la escala electrodébil. No obstante, la situación de los modelos LR ha cambiado desde su introducción en la década de 1970. Se sabe que estos pueden integrarse en teorías de gran unificación (GUT por sus siglas en inglés), que pueden abordar el problema de la ruptura de la paridad. Debido a la evolución de las constantes de acoplamiento (o \textit{running}), es posible encontrar que $g_L \neq g_R$ en la escala $v_R$, por lo que resulta interesante estudiar estos modelos sin simetrías discretas adicionales. El problema es que esto aumentaría el número de parámetros libres, especialmente en el sector del sabor, debido a la existencia de una nueva matriz $V_R^{\text{CKM}}$ que no guarda relación con su homóloga levógira. En este escenario, los límites impuestos sobre las masas de los nuevos bosones de \textit{gauge} $W_R$ y $Z_R$, así como sobre la escala LR, se ven relajados.

\section*{Metodología y Objetivos}
\hspace{15pt}La metodología empleada en la presente tesis combina tres perspectivas diferentes: en primer lugar, una revisión de la literatura sobre las simetrías en las QFT, y el uso de los métodos de renormalización algebraica para reparar la rotura de la simetría \textit{gauge} quiral que aparece al utilizar regularización dimensional en el esquema BMHV; en segundo lugar, un cálculo puramente teórico mediante el \textit{software} adecuado para derivar los contratérminos finitos de restauración de simetría necesarios para los operadores de cuatro fermiones de la SMEFT con neutrinos dextrógiros a un \textit{loop}; por último, un estudio fenomenológico de los LRM con la propuesta y el análisis de uno nuevo basado en EFT, y la comparación entre las predicciones teóricas y los datos experimentales actuales para obtener cotas inferiores sobre las masas de las nuevas partículas y sobre la escala LR. Los objetivos principales de esta tesis doctoral son:

\begin{itemize}
    \item Comprender las implicaciones de las simetrías \textit{gauge} quirales en la consistencia matemática de una QFT, prestando especial atención a las condiciones de renormalizabilidad (específicamente en el contexto de las EFT) y a la cancelación de anomalías.
    \item Adquirir las capacidades necesarias para calcular los contratérminos finitos de restauración de simetría que son precisos en el contexto del esquema BMHV, dentro de la regularización dimensional de una teoría \textit{gauge} quiral, utilizando las herramientas de la renormalización algebraica.
    \item Calcular  dichos contratérminos finitos de restauración de simetría para los operadores de cuatro fermiones del SMEFT añadiendo neutrinos dextrógiros a un \textit{loop}.
    \item Proponer y desarrollar un nuevo modelo LR con solo dos dobletes escalares para romper espontáneamente la simetría, lo que requiere operadores efectivos de dimensión cinco para generar las masas de los fermiones.
    \item Obtener cotas inferiores para las masas de los nuevos bosones \textit{gauge} de los modelos LR y para la escala  LR que no dependan del resto de parámetros de la teoría, utilizando los datos experimentales actuales disponibles procedentes de experimentos de física de sabor y de colisionadores.
\end{itemize}

Para la consecución de estos objetivos el presente texto se ha organizado en un total de cinco capítulos cuyo contenido se resume a continuación.

\section*{Capítulo 1: Teorías Gauge. Renormalizabilidad y Consistencia}

\hspace{15pt}El primer capítulo contiene una revisión de la bibliografía acerca de cómo en las teorías \textit{gauge} la estructura de las divergencias está constreñida por las simetrías, y cómo las correcciones radiativas pueden romperlas dando lugar a las famosas anomalías cuánticas. A su vez, se muestra el papel fundamental que juegan los argumentos cohomológicos en este análisis.

Tal y como se ha mencionado, una teoría \textit{gauge} basada en un grupo de Lie \(G\) es aquella que es invariante bajo transformaciones locales asociadas a dicho grupo. Habitualmente se trabaja únicamente con transformaciones infinitesimales, de tal manera que si el modelo contiene un conjunto de campos de materia \(\psi_a\), la teoría es invariante bajo la transformación:

\begin{equation}
    \label{Resumen_Transformation_matter_fields}
    \delta\psi_{a}\left(x\right)=i\varepsilon^{\alpha}\left(x\right)\left(t_{\alpha}\right)^b_a\psi_{b}\left(x\right),
\end{equation}

\noindent donde los \(t_{\alpha}\) son generadores del grupo y los \(\varepsilon^{\alpha}(x)\) son parámetros infinitesimales que dependen del punto del espaciotiempo, lo que dota a la transformación de un carácter local. Para que una teoría sea invariante bajo transformaciones locales, es necesario añadir unos campos de espín uno conocidos como campos \textit{gauge} \(A_{\mu}^{\alpha}\left(x\right)\), que se transforman según la representación adjunta del grupo:

\begin{equation}
    \label{Resumen_Gauge_Transformation_Gauge_Field}
    \delta A_{\mu}^{\alpha}\left(x\right)=\partial_{\mu}\varepsilon^{\alpha}\left(x\right)+C^{\alpha}_{\,\,\,\gamma\beta}\varepsilon^{\beta}\left(x\right)A_{\mu}^{\gamma}\left(x\right),
\end{equation}

\noindent donde las \(C^{\alpha}_{\,\,\,\gamma\beta}\) son las constantes de estructura del álgebra de Lie. La construcción de un término cinético para el campo \textit{gauge} (y por tanto la construcción de un candidato a una teoría consistente) solo es posible si el grupo de Lie es un producto de grupos \(\rm{U}(1)\) y grupos simples. Estos últimos fueron clasificados en el catálogo de Cartan y corresponden con: \({\rm SO}(n)\), que son transformaciones que preservan la norma en un espacio vectorial real; \({\rm SU}(n)\), que preservan la norma en un espacio complejo; \({\rm USp}(n)\) que lo hacen en un espacio cuaterniónico; y los cinco grupos excepcionales \(G_2\), \(F_4\), \(E_6\), \(E_7\) y \(E_8\), que preservan estructuras octoniónicas.

La cuantización de una teoría \textit{gauge} es altamente no trivial dada la presencia de grados de libertad espurios en los \(A^{\alpha}_{\mu}(x)\); que por ser un vector Lorentz deberían tener cuatro, pero por la simetría \textit{gauge} tan solo tienen dos. Una posibilidad es trabajar en el \textit{gauge} axial, donde se impone la condición \(A_3^{\alpha}=0\) y se toma que el \(A_0^{\alpha}\) no es independiente de las otras componentes. El problema de esta condición es que no es invariante Lorentz, ya que no es invariante bajo rotaciones. Si se quiere preservar la simetría Lorentz para realizar cálculos radiativos, es más conveniente utilizar el formalismo de De Witt-Faddeev-Popov, donde se introducen unos nuevos campos fermiónicos de espín cero conocidos como \textit{ghosts} \(\omega_{\alpha}\) y \textit{antighosts} \(\omega_{\alpha}^{\ast}\), a los que se les asocia un número cuántico llamado número \textit{ghost}: 1 para \textit{ghosts} y -1 para \textit{antighosts}. Este formalismo cuenta con dos inconvenientes: el primero es que, aunque la simetría Lorentz se respeta, la simetría \textit{gauge} en el Lagrangiano se pierde al introducir un funcional \(f_{\alpha}\) para fijar el \textit{gauge} escogido; el segundo es que si se tomara un funcional de la forma \(f_{\alpha}=\partial_{\mu}A_{\alpha}^{\mu}+S_{\alpha\beta\gamma}A_{\beta\mu}A_{\gamma}^{\mu}\), con \(S_{\alpha\beta\gamma}\) un tensor totalmente simétrico, esto daría lugar a una teoría no renormalizable, ya que, en general, se requiere la introducción de contratérminos no simétricos en \(\alpha\), \(\beta\) y \(\gamma\).

Para solventar los inconvenientes antes mencionados, se utiliza el formalismo BRST. Aquí se observa que el Lagrangiano del método de De Witt-Faddeev-Popov es invariante bajo la transformación infinitesimal

\begin{equation}
    \label{Resumen_BRST_psi}
    \delta_{\theta}\psi_a=i\theta\omega_{\alpha}\left(t_{\alpha}\right)_{a}^b\psi_b,
\end{equation}
\vspace{-0.5cm}
\begin{equation}
    \label{Resumen_BRST_A}
    \delta_{\theta}A_{\alpha}^{\mu}=\theta\left[\partial^{\mu}\omega_{\alpha}+C_{\alpha\beta\gamma}A_{\beta}^{\mu}\omega_{\gamma}\right],
\end{equation}
\vspace{-0.5cm}
\begin{equation}
    \label{Resumen_BRST_Omega}
    \delta_{\theta}\omega_{\alpha}=-\frac{1}{2}\theta C_{\alpha\beta\gamma}\omega_{\beta}\omega_{\gamma},
\end{equation}
\vspace{-0.5cm}
\begin{equation}
    \label{Resumen_BRST_Omega_Ast}
    \delta_{\theta}\omega_{\alpha}^{\ast}=-\theta h_{\alpha},
\end{equation}
\vspace{-0.5cm}
\begin{equation}
    \label{Resumen_BRST_h}
    \delta_{\theta}h_{\alpha}=0,
\end{equation}

\noindent donde \(\theta\) es un parámetro infinitesimal anticonmutativo y los \(h_{\alpha}\) son unos campos escalares auxiliares llamados campos de Nakanishi-Lautrup. Una de las principales características de esta transformación es que es no lineal en los campos. Asociado a la transformación BRST está el operador \(s\), definido por la relación \(\delta_{\theta}(\cdot)\eqqcolon\theta s(\cdot)\), que tiene como particularidades que aumenta en una unidad el número \textit{ghost} del campo al que se le aplica y que es nilpotente, es decir, \(s^2=0\). Estas dos propiedades permiten definir una cohomología asociada a este operador, pues dados operadores \(\mathcal{O}\) de un cierto número \textit{ghost} que sean BRST cerrados \(s(\mathcal{O})=0\), se puede distinguir entre aquellos exactos \(\mathcal{O}=s(\mathcal{A})\) y aquellos que no lo son. El estudio de la cohomología del operador BRST es la clave en el estudio de la consistencia de las teorías \textit{gauge}. En concreto, en el formalismo BRST se considera que dos Lagrangianos son válidos para describir una misma teoría física si tan solo difieren en un operador BRST exacto de número \textit{ghost} cero.

Aunque el formalismo BRST es consistente y perfectamente válido, si se quiere estudiar teorías más generales (no basadas necesariamente en un álgebra de Lie cerrada) o que incluyan operadores efectivos de dimensión superior a cuatro, es conveniente trabajar con el método de Batalin-Vilkovisky, que también es útil en la discusión de las ecuaciones de movimiento. En este caso, a cada campo \(\chi_n\) de la teoría (físico o no), se le asocia un anticampo \(\chi_n^{\ddagger}\) que tiene la estadística opuesta y cuyo número \textit{ghost} es \({\rm gh}(\chi_n^{\ddagger})=-{\rm gh}\left(\chi_n\right)-1\). A continuación, se define el antiparéntesis de dos operadores como

\begin{equation}
    \label{Resumen_Definition_Antibracket}
    \left(F,G\right)\coloneqq\frac{\delta_RF}{\delta\chi_n}\frac{\delta_LG}{\delta\chi_n^{\ddagger}}-\frac{\delta_RF}{\delta\chi_n^{\ddagger}}\frac{\delta_LG}{\delta\chi_n},
\end{equation}

\noindent donde la suma sobre índices \(n\) incluye a los espaciotemporales, que permite introducir la versión generalizada del operador BRST dada por \(\left(S,\cdot\right)\), donde \(S\) es la acción. Se puede demostrar que se satisface \((S,S)=0\), lo que se conoce como la ecuación maestra; y puesto que el antiparéntesis tiene propiedades análogas al paréntesis de Poisson de la mecánica clásica, es fácil ver que el operador BRST generalizado es nilpotente \((S,(S,\cdot))=0\). Todos los razonamientos cohomológicos antes mencionados se trasladan a este caso, y se dice que dos acciones son equivalentes a la hora de describir una teoría si tan solo difieren en un operador bosónico de número \textit{ghost} cero que sea exacto bajo transformaciones BRST.

Las implicaciones a nivel cuántico de una teoría \textit{gauge} se estudian usando la Acción Efectiva Cuántica (QEA por sus siglas en inglés) \(\Gamma\). Esta no es más que un objeto análogo a una acción clásica, pero que contiene toda la información acerca de las correcciones cuánticas, y que se calcula a partir de los diagramas de Feynman que sean 1PI. Dado que la simetría BRST es no lineal, la QEA no necesariamente es invariante BRST, en su lugar satisface las llamadas identidades de Slavnov-Taylor, que en términos del antiparéntesis definido anteriormente se escriben como \(\left(\Gamma,\Gamma\right)=0\). A partir de ellas se puede demostrar con argumentos cohomológicos utilizando el formalismo de Batalin-Vilkovisky, que toda divergencia puede ser absorbida redefiniendo los parámetros del Lagrangiano, lo que significa que las teorías \textit{gauge} son renormalizables.

El argumento anterior supone que la integral de caminos es invariante \textit{gauge}, y dado que todas las acciones consideradas lo son, eso implica que también lo debe ser la medida. En teorías \textit{gauge} con fermiones quirales se puede dar que la medida no sea invariante aunque la acción sí lo sea, destruyendo así las implicaciones de la simetría a nivel cuántico, lo que se conoce como anomalía. El estudio de las anomalías se realiza analizando la cohomología de número \textit{ghost} uno del operador BRST generalizado, dado que se puede demostrar a partir de las condiciones de consistencia de Wess-Zumino, que si todos los operadores BRST cerrados son exactos, entonces las anomalías son espurias, pues se pueden eliminar añadiendo contratérminos al Lagrangiano. Esto es precisamente lo que sucede en teorías \textit{gauge} que no contienen factores \({\rm U}(1)\) o \({\rm SU}(n)\) con \(n\geq3\). Dado que el SM y las teorías LR se pueden entender como modelos a bajas energías de teorías basadas en \({\rm SO}(10)\), estas deben ser libres de anomalías. También el SMEFT, pues la medida de la integral de caminos de Feynman es la misma que en el SM.

\section*{Capítulo 2: El Esquema de Regularización Dimensional de Breitenlohner-Maison-'t~Hooft-Veltman}

\hspace{15pt}Los resultados sobre renormalizabilidad y ausencia de anomalías mostrados en el primer capítulo descansan sobre dos suposiciones previas: la primera, que la prescripción que se utiliza para definir las cantidades divergentes preserva las simetrías de la teoría; y la segunda, que las divergencias y las anomalías son locales. La primera cuestión se atiende con la definición de un esquema de regularización apropiado. La opción preferida actualmente es el esquema de regularización dimensional, cuya idea intuitiva es la siguiente: para manejar integrales divergentes de forma rigurosa, se calculan en un espaciotiempo \(D\)-dimensional donde estas converjan, dado que su resultado dependerá de \(D\), se puede hacer una prolongación analítica hasta \(D\to4\) para estudiar su comportamiento. Las divergencias aparecerán como polos del tipo \(1/(D-4)^n\), de manera que la forma más básica de eliminarlas es usando la prescripción de sustracción mínima, donde se asocia la parte infinita de las integrales a esos polos exclusivamente. Para teorías no quirales, el esquema de regularización dimensional preserva la simetría Lorentz y la simetría \textit{gauge}, aunque la definición precisa de las integrales en \(D\) dimensiones es delicada y es el objeto de estudio de la primera parte del segundo capítulo.

Demostrar que las divergencias son locales en cálculos a un \textit{loop} es trivial. La situación se complica en diagramas de mayor orden debido a la presencia de subdivergencias, es decir, divergencias que aparecen cuando algunos de los momentos internos de las amplitudes son llevados a infinito, y que potencialmente generan funciones de Green no locales. No obstante, si se utiliza el algoritmo BPH para la sustracción de subdivergencias, es posible demostrar el teorema de Weinberg con la ayuda de la \textit{forest formula}, que dice que una vez que las subdivergencias se eliminan, las funciones de Green son locales. Dicho de otra manera, cuando se consideran todos los diagramas de Feynman, incluyendo aquellos que involucran \textit{loops} con contratérminos obtenidos en cálculos de menor orden, las divergencias resultan ser locales, completando así la demostración de la renormalizabilidad de las teorías \textit{gauge}. La demostración del teorema de Weinberg es una parte fundamental del segundo capítulo.

En el siguiente apartado se presentan propiedades del esquema de regularización dimensional, como que preserva la unitariedad, la simetría de Poincaré en cuatro dimensiones y la causalidad. No obstante, el resultado más importante para la consecución del presente trabajo es que satisface el Principio de la Acción Cuántica (QAP por sus siglas en inglés). Este constituye la base de las herramientas de la renormalización algebraica, y su mayor implicación para la presente tesis es que muestra cómo calcular las posibles obstrucciones a las identidades de Slavnov-Taylor. En particular,  se tiene que 

\begin{equation}
    \label{Resumen_QAP}
    \left(\Gamma,\Gamma\right)=-2\Delta\cdot\Gamma,
\end{equation}

\noindent donde \(\Delta\cdot\Gamma\) representa inserciones únicas del operador de ruptura de simetría \(\Delta\), que a su vez está dado por la variación BRST de la acción \(D\)-dimensional:

\begin{equation}
    \label{Resumen_Delta_operator}
    \Delta\coloneqq s\left(\int{\rm d}^Dx\mathcal{L}^{(D)}\right).
\end{equation}

El punto clave es que, si se quiere realizar cálculos matemáticamente consistentes, el Lagrangiano \(D\)-dimensional de una teoría \textit{gauge} con fermiones quirales no es invariante BRST. Esto es debido a la presencia de la matriz \(\gamma_5\), cuya definición en \(D\)-dimensiones no es evidente. Aquella que permite cálculos matemáticamente bien definidos es la dada por el esquema BMHV, donde esta queda como un objeto puramente en cuatro dimensiones \(\gamma_5\coloneqq i\gamma^0\gamma^1\gamma^2\gamma^3\). Esta prescripción da lugar a propagadores fermiónicos puramente tetradimensionales,  que producen integrales no definidas en regularización dimensional. La solución a este problema es introducir campos ficticios \(\psi^f\) de quiralidad opuesta a los físicos que se utilizan únicamente para escribir un término cinético en \(D\) dimensiones para los fermiones, de manera que \(s\psi^f=0\). Este procedimiento tiene el inconveniente de que dicho término cinético no es invariante \textit{gauge}, por el contrario, su variación BRST viene dada por el término evanescente

\begin{equation}
    \label{Resumen_Kinetic_BRST_Breaking_1}
    s\left(i\bar{\psi}_i\slashed{\partial}\psi_i\right)=\omega_{\alpha}\left[\bar{\psi}_i\left(t^{\alpha}_{L,R}\right)_{ij}\left(\overrightarrow{\hat{\slashed{\partial}}}P_{R,L}+\overleftarrow{\hat{\slashed{\partial}}}P_{L,R}\right)\psi_j\right],
\end{equation}

\noindent donde se toma el primer subíndice (segundo) si el fermión físico es levógiro (dextrógiro), y \(\hat{\partial}\) representa la parte de la derivada en \(D-4\) dimensiones.

Este propagador variante \textit{gauge} introduce obstrucciones a las identidades de Salvnov-Taylor que se pueden calcular usando la Ec.~\eqref{Resumen_QAP}, a partir de inserciones únicas del operador 

\begin{equation}
    \label{Resumen_Symmetry_Breaking_Delta}
    \Delta=\int{\rm d}^Dx\omega_{\alpha}(x)\left[\bar{\psi}_i(x)\left(t^{\alpha}_{L,R}\right)_{ij}\left(\overrightarrow{\hat{\slashed{\partial}}}P_{R,L}+\overleftarrow{\hat{\slashed{\partial}}}P_{L,R}\right)\psi_j(x)\right].
\end{equation}

\noindent en diagramas 1PI. En teorías libres de anomalías físicas, estas obstrucciones podrán ser eliminadas añadiendo los contratérminos apropiados, tal y como se demuestra en el primer capítulo. Estos contratérminos pueden ser divergentes (\(\propto1/(D-4)\)), evanescentes (\(\propto(D-4)\)) debido a que el operador de ruptura de simetría lo es, y finitos (\(\propto(D-4)/(D-4)\)), dado que el carácter evanescente de dicho operador se puede compensar por una divergencia en los cálculos radiativos. Los contratérminos finitos son uno de los principales objetos de estudio de esta tesis, y estos se pueden obtener de forma simple a partir de las divergencias de diagramas 1PI con inserciones únicas de un operador de ruptura de simetría \textit{gauge} modificado, conocido como operador de Dvo\v{r}\'ak \(\check{\Delta}\). Este es el método de Bonneau, y dado que permite ver que las obstrucciones finitas a las identidades de Slavnov-Taylor también se pueden calcular a partir de divergencias (que son locales por el teorema de Weinberg), demuestra que las anomalías son locales, completando los razonamientos sobre anomalías del primer capítulo.

\section*{Capítulo 3: El Modelo Estándar y la Teoría Efectiva de Campos del Modelo Estándar}

\hspace{15pt}Como ya se ha mencionado, el SM es la teoría que se utiliza actualmente para comprender las propiedades e interacciones de las partículas elementales. Sin embargo, no ofrece respuestas a algunas de las preguntas más importantes sobre la naturaleza; a saber: el origen de las masas de los neutrinos, de qué está hecha la materia oscura o cuáles son las propiedades cuánticas de la gravedad. Como consecuencia, el SM no representa el paso final en la comprensión del universo; es necesario encontrar una extensión que explique todas las características mencionadas. No obstante, existen numerosos modelos que podrían cumplir estos requisitos, pero con los datos experimentales actuales no es posible determinar cuál es el correcto. Ante esta situación, resulta conveniente utilizar el enfoque \textit{bottom-up} de las EFT, e introducir la SMEFT. Además, hoy en día la presencia de divergencias en las QFT se entiende bajo este paradigma. Se cree que la existencia de infinitos en los cálculos radiativos se debe a que los momentos de integración en los \textit{loops} exploran energías donde la teoría ya no es válida, lo que implica que las teorías actuales solo funcionan hasta una determinada escala de energía.

La idea principal de la SMEFT es la siguiente: es sabido que, hasta unos pocos TeV, los datos experimentales pueden describirse perfectamente utilizando el contenido de partículas del SM con su simetría \textit{gauge} quiral ${\rm SU}(3)_{\rm QCD} \otimes {\rm SU}(2)_{\rm L} \otimes {\rm U}(1)_{\rm Y}$. Por lo tanto, independientemente de cuál sea la nueva extensión física del SM, a bajas energías esta se comporta como una teoría con el contenido de partículas y las simetrías del SM. Así, los efectos de nueva física aparecerán en la presencia de operadores de dimensión mayor que cuatro, los cuales están ausentes en el SM. Siguiendo el paradigma de Weinberg, para tener en cuenta todos los posibles efectos de nueva física, es necesario construir el Lagrangiano más general con operadores de dimensión arbitraria (dependiendo de la precisión deseada) que respete las simetrías del SM y solo contenga los estados físicos que aparecen en él. 

El tercer capítulo es una revisión de la literatura dedicada a la presentación de los ingredientes principales del SM y de la SMEFT hasta operadores de dimensión seis. En concreto, cómo la cancelación de anomalías en el SM es muy delicada, pues las hipercargas toman uno de los dos únicos valores posibles que hacen la teoría consistente; cómo el hecho de que las partículas elementales tengan masa requiere de la existencia del bosón de Higgs, si esta proviene de un mecanismo de SSB; el hecho de que las masas de los neutrinos se pueden generar en la SMEFT a partir del único operador de dimensión cinco posible, el operador de Weinberg; la existencia de una base completa y mínima, la base de Warsaw, para operadores de dimensión seis; y que en una teoría efectiva las simetrías accidentales que se encuentran con operadores de dimensión inferior a cuatro, tienden a romperse, ejemplos de esto son el número bariónico \(B\), el número leptónico \(L\) y la diferencia \(B-L\).

\section*{Capítulo 4: Contratérminos Finitos para la Restauración de la Simetría con una Inserción de los Operadores de Cuatro Fermiones de SMEFT a un Loop}

\hspace{15pt}El estudio de la SMEFT es uno de los campos más prolíficos para los fenomenólogos. Esto se debe a que puede emplearse para analizar todas las posibles teorías nuevas de manera simultánea y compararlas con los datos experimentales actuales. No obstante, las predicciones del SM coinciden perfectamente con lo observado, por lo que solo es posible hallar nuevos efectos mediante cálculos de alta precisión. En consecuencia, las correcciones radiativas de los observables de la SMEFT han cobrado una mayor importancia en los últimos años.  Dado que la SMEFT es una teoría quiral, si se utiliza la regularización dimensional de forma matemáticamente consistente, es necesario introducir contratérminos que restauren la simetría. Sin embargo, hasta este momento, los cálculos se realizaban siempre utilizando esquemas que no son matemáticamente consistentes, aunque pueden proporcionar respuestas correctas para observables particulares a un determinado orden. La validez de estos métodos se vuelve incierta al realizar cálculos de orden superior, mientras que la regularización dimensional en el esquema BMHV, añadiendo el conjunto adecuado de contratérminos de restauración de simetría, es perfectamente consistente.

El cuarto capítulo presenta uno de los resultados originales obtenidos por el autor de esta tesis: el cálculo de los contratérminos finitos necesarios para la restauración de la simetría \textit{gauge} asociados a los operadores de cuatro fermiones de la SMEFT, incluyendo neutrinos dextrógiros y utilizando los métodos de la renormalización algebraica. Dado que el cálculo se realiza en la fase no rota, es conveniente introducir campos ficticios con quiralidades opuestas a los físicos incluso cuando ambas están presentes en la teoría, ya que así simetrías globales como la hipercarga se respetan. Estos resultados se comparan con los obtenidos por otro grupo de manera paralela, usando un método innovador basado en la introducción de espuriones para restaurar la simetría. Aunque la inmensa mayoría de operadores y coeficientes de Wilson concuerdan, en el momento presente todavía quedan una pequeñas diferencias.

\section*{\boldmath Capítulo 5: Modelos Left-Right. Cotas Generales en las Masas de los Bosones Gauge \(W_R\) y \(Z_R\)}


\hspace{15pt}Aunque el uso de la SMEFT es una buena estrategia para buscar efectos de nueva física en los datos experimentales, lo cierto es que no proporciona una descripción de las posibles nuevas interacciones, solo puede emplearse para constreñir sus efectos a bajas energías. Por lo tanto, para obtener una comprensión más amplia de la naturaleza, es necesario proponer una nueva teoría con nuevas interacciones y/o nuevas partículas; este es el enfoque \textit{up-bottom}, la construcción de modelos. Una de las extensiones más relevantes del SM se basa en intentar explicar su naturaleza quiral, proponiendo que la ruptura de la paridad proviene de una SSB a la escala \(v_R\); estos son los modelos LR. Como ya se ha dicho, en estas teorías se añade una nueva interacción de tipo débil basada en el grupo ${\rm SU}(2)_{\rm R}$, lo que conduce al grupo de Lie \({\rm SU}(3)_{\rm QCD}\otimes{\rm SU}(2)_{\rm L}\otimes{\rm SU}(2)_{\rm R}\otimes{\rm U}(1)_{\rm X}\). Se puede restaurar una especie de simetría de paridad imponiendo que los acoplamientos de las dos interacciones ${\rm SU}(2)$ sean iguales, es decir, $g_L = g_R$. Una característica destacable de las teorías basadas en este grupo de Lie es que las cargas de las partículas están totalmente determinadas (salvo un factor multiplicativo global) por la condición de cancelación de anomalías.

Tradicionalmente se han estudiado dos realizaciones específicas de la SSB en los modelos LR. En ambas se añade un bidoblete \(\Phi\) responsable de la masa de los quarks y de los leptones cargados,  la diferencia es que en un caso hay dos dobletes adicionales \(\chi_L\) y \(\chi_R\), y en el otro dos tripletes \(\Delta_L\) y \(\Delta_R\). La mayor diferencia entre ambos modelos es que en el primero los neutrinos son partículas de Dirac cuya masa proviene únicamente de su interacción con \(\Phi\), y en el segundo son partículas de Majorana cuyo término de masa tiene contribuciones de la interacción con los tripletes. No obstante, como se ha dicho, hoy en día también se sabe que los modelos LR pueden integrarse en grupos más grandes de gran unificación, como ${\rm SO}(10)$, por lo que la restauración de la paridad puede ocurrir a escalas de energía aún más elevadas. Esto implica que tiene cabida considerar modelos efectivos donde la masa de los fermiones se genera a través de operadores de dimensión mayor a cuatro, y se considera únicamente el sector escalar mínimo necesario para romper la simetría. De esta forma, uno de los resultados originales del autor de esta tesis es proponer un modelo LR efectivo donde los términos de masa provienen de operadores de dimensión cinco análogos al operador de Weinberg, y el sector escalar contiene únicamente a los dobletes \(\chi_L\) y \(\chi_R\).

En este contexto donde los modelos LR son teorías a bajas energías de otros modelos mayores, las simetrías discretas adicionales que normalmente se imponen pueden romperse debido a la evolución de las constantes de acoplo, de manera que se tiene \(g_L\neq g_R\). Como se ha dicho, esto aumenta en gran medida el número de parámetros libres de la teoría, de forma que las cotas inferiores sobre la escala \(v_R\) y las masas de los nuevos bosones \textit{gauge} \(W_R^{\pm}\) y \(Z_R\) se ven reducidas. Si uno intenta obtener cotas directas sobre estos parámetros de forma independiente a los demás usando observables de sabor, se va a encontrar con serias dificultades debido a la presencia de la matriz \(V_{R}^{\rm CKM}\) en el cálculo de los observables. La situación se agrava en el modelo efectivo, pues las amplitudes de los diagramas de caja son divergentes. Además, el autor de este texto demostró que las posibles cotas que uno obtendría no son muy superiores a aquellas que provienen de observables de colisionadores, por lo que estos van a ser el centro de este análisis.

En el quinto capítulo se presenta otro de los resultados originales del autor de esta tesis: la obtención de cotas inferiores sobre la escala \(v_R\)  y las masas de los nuevos bosones \textit{gauge} \(W_R^{\pm}\) y \(Z_R\) independientes del resto de los parámetros de la teoría, a partir de los datos del Run 2 del LHC. Para ello, se busca (con la ayuda del teorema de equivalencia) el valor de los parámetros del modelo que hace que la sección eficaz de producción y decaimiento de los nuevos bosones \textit{gauge} en ciertos procesos, sea mínima, de forma que cualquier otro valor conducirá a una cota inferior mayor sobre sus masas. Esto se hace dentro de los límites de la teoría de perturbaciones para los dos modelos tradicionales con dobletes y tripletes, y para el modelo efectivo. Las cotas directas para la masa del \(Z_R\) se derivan del proceso \(pp\rightarrow Z_RX\rightarrow \ell^+\ell^-X\), con \(X\) un estado arbitrario. En el caso del \(W_R\) se utiliza \(pp\rightarrow W_R\rightarrow j_1j_2\) si los neutrinos son de Majorana (modelo efectivo y modelo con tripletes) y \(pp\rightarrow W_R\rightarrow \ell \bar{\nu}_R\) si son de Dirac (modelo con dobletes). Dado que las masas de ambos bosones están relacionadas mediante el ángulo de mezcla \(\gamma\) análogo al ángulo de Weinberg \(\theta_W\), es posible obtener cotas indirectas sobre las masas que conducen a límites inferiores complementarios, de los que también se obtiene una cota mínima para \(v_R\).

\section*{Conclusiones y Futuras Líneas de Investigación}

\hspace{15pt}Las conclusiones que se derivan de los resultados estudiados y obtenidos en la presente tesis son las siguientes:

\begin{itemize}
    \item La simetría \textit{gauge} puede emplearse, utilizando el formalismo BRST y el de Batalin-Vilkovisky, para restringir las divergencias de las funciones de Green en cálculos perturbativos. Se puede demostrar que los infinitos pueden eliminarse mediante redefiniciones de los parámetros originales y campos del Lagrangiano. Esta afirmación también es válida para las EFT.
    \item En una teoría \textit{gauge} quiral, es posible encontrar obstrucciones a las identidades de Slavnov-Taylor, conocidas como anomalías \textit{gauge}, que pueden clasificarse en físicas y espurias. En el SM, el SMEFT y los modelos LR, todas las anomalías \textit{gauge} son espurias, por lo que pueden eliminarse añadiendo el conjunto adecuado de contratérminos al Lagrangiano.
    \item La única prescripción conocida para la definición de $\gamma_5$ en regularización dimensional que es matemáticamente consistente  es el esquema BMHV, el cual introduce la ruptura de la simetría \textit{gauge} incluso a nivel clásico en teorías quirales, debido a la necesidad de regularizar los propagadores fermiónicos.
    \item Utilizando el QAP y el método de Bonneau, se puede demostrar que las obstrucciones finitas a las identidades de Slavnov-Taylor pueden calcularse a partir de las divergencias de los diagramas 1PI con inserciones simples de un operador de ruptura de simetría, el operador de  Dvo\v{r}\'ak.
    \item A partir del cálculo de los contratérminos finitos de restauración de simetría necesarios para la renormalización a un \textit{loop} de los operadores de cuatro fermiones del SMEFT$+\nu_R$, se ha demostrado explícitamente que la teoría es libre de anomalías a ese orden en la expansión perturbativa. 
    \item La mayor dificultad al utilizar las técnicas de renormalización algebraica en este cálculo radica en encontrar la preimagen de la transformación BRST de los contratérminos finitos.
    \item Puesto que los modelos LR pueden integrarse en otras teorías más allá del SM, resulta natural estudiarlos en el caso en que no se asuman simetrías discretas adicionales, y proponer un modelo donde las masas de los fermiones se generen mediante operadores efectivos. 
    \item Obtener cotas independientes del modelo para las masas de los bosones gauge $W_R$ y $Z_R$, y para la escala $v_R$, utilizando observables de sabor, requerirá un ajuste global (\textit{global fit}) para tener en cuenta los demás parámetros libres.
    \item Es posible hallar tales cotas globales, suponiendo únicamente que la teoría es perturbativa, utilizando los datos de física de colisionadores procedentes del Run 2 del experimento LHC.
\end{itemize}

A su vez, se derivan las posibles futuras líneas de investigación:

\begin{itemize}
    \item El cálculo  usando los métodos de la renormalización algebraica del conjunto completo de contratérminos finitos de restauración de simetría necesarios para llevar a cabo la renormalización a un \textit{loop} del SMEFT$+\nu_R$, más allá de los operadores de cuatro fermiones estudiados aquí.
    \item La aplicación de estos resultados para el cálculo de observables físicos a un \textit{loop}, y realizar comparaciones con otros esquemas de renormalización.
    \item El diseño de un \textit{software} adecuado para automatizar el cálculo de los contratérminos finitos dentro del formalismo de la renormalización algebraica.
    \item El cálculo (probablemente utilizando el software mencionado anteriormente) de los contratérminos de restauración de simetría en el SMEFT (o en otras teorías físicas de interés) a órdenes superiores y/o considerando operadores de dimensión superior.
    \item Una actualización de las cotas de colisionadores obtenidas para las masas de los bosones \textit{gauge} $W_R$ y $Z_R$, y para la escala LR $v_R$, utilizando los datos del Run 3 del LHC cuando estén disponibles.
    \item Un cálculo de las cotas para las masas de los nuevos bosones gauge en otras teorías más allá del SM, utilizando las técnicas desarrolladas aquí, para obtener resultados con independencia de los demás parámetros físicos.
    \item La obtención de una base mínima de operadores de dimensión seis en el modelo LR efectivo, y sus aplicaciones a resultados fenomenológicos.
    \item Un \textit{global fit} que incluya observables de precisión electrodébiles y de sabor para restringir los parámetros de los modelos LR en el caso en que no se impongan simetrías discretas.
\end{itemize}

\thispagestyle{empty}

\tableofcontents

\hypersetup{linkcolor=black}
\listoftables \addcontentsline{toc}{chapter}{List of Tables}
\newpage
\thispagestyle{empty}
\mbox{}
\newpage

\listoffigures \addcontentsline{toc}{chapter}{List of Figures}

\newpage
\thispagestyle{empty}
\mbox{}
\newpage

\hypersetup{linkcolor=blue}
\chapter*{List of Abbreviations} 
\addcontentsline{toc}{chapter}{List of Abbreviations}

The abbreviations employed in this thesis are:

\begin{table}[h]
\begin{tabular}{ll}
    \textbf{1PI} & 1 Particle Irreducible \\
    \textbf{BMHV} & Breitenlohner-Maison-'t Hooft-Veltman \\
    \textbf{BPH} & Bogoliubov-Parasiuk-Hepp \\
    \textbf{BPHZ} & Bogoliubov-Parasiuk-Hepp-Zimmermann \\
    \textbf{BRST} &  Becchi-Rouet-Stora-Tyutin  \\
    \textbf{CKM} & Cabibbo-Kobayashi-Maskawa \\
    \textbf{CP} &  Charge-Parity \\
    \textbf{EFT} &  Effective Field Theory \\
    \textbf{EW} &  Electroweak \\
    \textbf{LEFT} & Low Energy Effective Field Theory \\
    \textbf{LHC} & Large Hadron Collider \\
    \textbf{LR} & Left-Right \\
    \textbf{LRM} & Left-Right Model \\
    \textbf{MS} & Minimal Subtraction \\
    \textbf{NLO} & Next to Leading Order \\
    \textbf{NP} & New Physics \\
    \textbf{PDF} & Parton Distribution Function \\
    \textbf{PMNS} &  Pontecorvo-Maki-Nakagawa-Sakata \\
    \textbf{QAP} & Quantum Action Principle \\
    \textbf{QCD} & Quantum Chromodynamics \\
    \textbf{QEA} & Quantum Effective Action \\
    \textbf{QED} & Quantum Electrodynamics \\
    \textbf{QFT} &  Quantum Field Theory  \\
    \textbf{RGE} &  Renormalization Group Equation  \\
    \textbf{SM} &  Standard Model \\
    \textbf{SMEFT} & Standard Model Effective Field Theory \\
    \textbf{SSB} & Spontaneous Symmetry Breaking \\
    \textbf{UV} & Ultraviolet\\
    \textbf{VEV} & Vacuum Expectation Value \\
\end{tabular}
\end{table}

\numberwithin{equation}{section}
\newpage
\thispagestyle{empty}
\mbox{}
\newpage
\mainmatter
\thispagestyle{empty}\cleardoublepage\thispagestyle{empty}
\setcounter{page}{1}
\chapter*{Introduction}
\addcontentsline{toc}{chapter}{Introduction}
\markboth{INTRODUCTION}{INTRODUCTION}

Quantum Field Theory (QFT) is a framework that unifies the principles of quantum mechanics and special relativity. The former is needed to understand the behavior of microscopic objects, while the latter explains the dynamics of bodies that move at velocities near the speed of light, and the mass-energy conversion \(E=mc^2\). Consequently, QFTs are completely necessary to explain the data obtained from experiments that involve, for instance, subatomic particles moving at high speeds or the decay of a certain state. The most famous example where experiments of this kind take place are particle accelerators, such as the Large Hadron Collider (LHC). Furthermore, microscopic states with relativistic velocities were abundant at the origin of the universe, characterized by extremely high densities and temperatures. Thus, QFTs also play a crucial role in understanding the initial conditions after the Big Bang. They are a fundamental tool in the exploration of the frontiers of physics and in the description of the interactions of the fundamental constituents of matter. 

In the view of the author of this text, the most important discovery in physics in the previous century was that the laws of nature can be derived, using the formalism of field theory, from the assumption that a certain symmetry is satisfied. Famous examples are the Poincaré symmetry (Lorentz symmetry plus spacetime translations), which is the basis of special relativity; or the invariance under diffeomorphisms, from which the laws of gravitation in general relativity are obtained. In the context of particle physics, the importance of symmetries was clear from the discovery of global symmetries like isospin (later extended to the \({\rm SU}(3)\) flavor symmetry), which was very useful in the explanation of the properties and interactions of the hadrons. However, in this context, the most important transformations are not the global ones but the local ones, the so-called \textit{gauge symmetries}. These are transformations, usually based on Lie groups, that are different in different points of the spacetime. From the geometric structure of gauge transformations, the possible interactions between the different states are completely fixed. The major achievement of gauge-invariant QFTs was the development of the Standard Model (SM) of particle physics in the last century, the theory that is used nowadays to describe the interactions (except gravity) between elementary particles. The construction of this model required the combination of the efforts of theoretical and experimental physicists, and culminated in 2012 when the Higgs boson was discovered in the LHC by the ATLAS and CMS groups.

The SM is based on the gauge Lie group  \({\rm SU}(3)_{\rm QCD}\otimes{\rm SU}(2)_{\rm L}\otimes{\rm U}(1)_{\rm Y}\). Strong interactions are described by the \({\rm SU}(3)_{\rm QCD}\) simple part of the group, while the electroweak sector is fixed from the structure of \({\rm SU}(2)_{\rm L}\otimes{\rm U}(1)_{\rm Y}\). For the purpose of this text, the most remarkable feature of the SM is that the weak interaction only acts on the left-handed part (\({\rm SU}(2)_{\rm L}\)) of the fermions of the theory. This has an immediate consequence, \textit{nature is chiral}. This means that the interactions distinguish between the left- and right-handed chiralities of the particles, so the parity symmetry is broken, such that physics laws change when observed through a mirror. Furthermore, although the SM is expected to be a theory that is valid only up to a certain energy scale, the chiral nature of the fermions of the theory and their interactions is a feature that needs to be taken into account (or even explained) by any possible new physics extension. This PhD thesis is thus devoted to the study of chiral gauge symmetries in quantum field theory; particularly, their role in the precision calculations in the extensions of the SM, and how the explanation of the chiral structure of nature leads to the proposal of new physical theories. This text is structured in five chapters that study the following topics:

\section*{Chapter 1: Gauge Theories. Renormalizability and Consistency}

From a mathematical perspective, one of the major difficulties in the calculations in QFTs is the appearance of divergences in radiative calculations beyond the tree-level approximation. However, for theories with a gauge symmetry the structure of those infinities is completely fixed. At the quantum level, gauge transformations are promoted to the so-called \textit{BRST transformations}, which are non-linear, nilpotent and involve anticommuting parameters and a new set of fields: the ghosts, antighosts and Nakanishi-Lautrup fields. Using functional methods, it is shown that BRST invariance leads to the \textit{Zinn-Justin equation}, which represents the implications of BRST symmetry in the \textit{Quantum Effective Action} (QEA), a classical-like object that contains all the information about the quantum effects. From this equation, it is possible to prove that, in a theory containing only operators with mass scale up to four (in four spacetime dimensions), the divergences generated from quantum corrections must be BRST invariant. This result can be generalized using the Batalin-Vilkovisky formalism to general Effective Field Theories (EFTs), which contain operators of arbitrary mass scale, and are one of the main subjects of study of phenomenological particle physicists nowadays. As a consequence, \textit{the divergences in QFTs with a gauge symmetry can be reabsorbed by redefining the fields and the original parameters of the Lagrangian, such as masses or couplings}. From the modern perspective, this property is known as \textit{renormalizability}.

However, the previous result relies on several subtle considerations. For instance, in a gauge theory with chiral fermions (the main subject of this text), the classical gauge symmetry might not be a symmetry of the associated quantum theory. This is a so-called \noindent{anomaly}, and its presence produces obstructions to the Zinn-Justin equation that might make the theory non-renormalizable, in the sense that some divergence will not respect the symmetry requirements. However, using the \textit{Wess-Zumino consistency conditions}, it is possible to see that a certain type of anomalies might be eliminated by adding counterterms to the Lagrangian. It will be shown that all of the anomalies present in the theories studied in this text are of this kind, they are said to be spurious anomalies. 

In the first chapter of the thesis, all the concepts mentioned above with the proofs of the main results will be presented; namely: classical gauge theories studying which are the Lie groups compatible with the structure of the Lagrangian, the necessity to introduce the ghosts and antighosts to quantize the theory in a Lorentz invariant gauge, the BRST and Batalin-Vilkovisky formalisms, the conditions that the non-linear symmetries impose on the QEA with a proof of the Zinn-Justin equation, the proof of the renormalizability of gauge theories, a study of anomalies in chiral theories with a derivation of the Wess-Zumino consistency conditions and a proof of the fact that in the theories analyzed here all the anomalies are spurious. This is done following the excellent presentation in \cite{Weinberg2}, with special attention to the main results needed for this thesis.

\section*{Chapter 2: The Breitenlohner-Maison-'t~Hooft-Veltman Dimensional Reg\-u\-lar\-i\-za\-tion Scheme}

The previous renormalizability result also relies on a more fundamental one, which is the fact that the divergences generated by quantum corrections are \textit{local}, in the sense that the Green's functions can be written as polynomials on the fields and their derivatives at the same spacetime point. However, to prove this statement, it is convenient to give a precise mathematical definition for the divergences in the amplitudes. This is known as a \textit{regularization scheme}. It is better to chose one where most part of the underlying symmetries of the theory (namely Lorentz or gauge symmetries) are respected; and for this purpose, \textit{dimensional regularization} has been the preferred option over the last decades. The main idea of dimensional regularization is to evaluate the integrals in a \(D\)-dimensional spacetime where it converges, and then take the analytical continuation to \(D=4\). Divergences will appear as poles on \(D-4\), so it is possible to identify the divergent part of the amplitudes precisely with those terms. This is only one of the infinitely many possible \textit{renormalization schemes}, known as \textit{Minimal Subtraction} (MS), and it is the one that will be used in this text. 

With this precise definition of divergences, it is straightforward to show that in one-loop calculations they are local. However, this is not so obvious when multi-loop calculations are involved. To study this situation, it is necessary to use an algorithm to cancel the subdivergences (the ones that appear when some particular momenta go to infinity). This is provided by the BPH prescription and the forest formula. Using them, it is possible to prove Weinberg's theorem, which is precisely the statement of the fact that \textit{all divergences are local}. This is enough to complete the proof of the renormalizability of gauge theories in the absence of anomalies. Nonetheless, anomalies can also come from finite contributions, so this result is not enough to see that they are associated with local operators.

To deal with anomalies, it is useful to work with a regularization scheme that satisfies the \textit{Quantum Action Principle} (QAP), which relates any transformation of the Green's functions to the way it acts upon the interaction part of the Lagrangian, and it is one of the main ingredients of \textit{algebraic renormalization}. Apart from preserving Lorentz invariance (at least the four-dimensional part in chiral theories) and unitarity, dimensional regularization also satisfies the QAP. In this context, it is possible to show that the possible anomalies of the theory can be computed from 1 Particle Irreducible (1PI) diagrams with single insertions of a symmetry breaking operator, generated by the BRST transformation of the \(D\)-dimensional Lagrangian. In non-chiral theories, it is always possible to find a \(D\)-dimensional extension of the Lagrangian that is BRST invariant. However, there is no unique way to define the \(\gamma_5\) matrix in spaces with \(D\neq4\). The only known mathematically consistent way to do this is by using the \textit{Breitenlohner-Maison-'t~Hooft-Veltman} (BMHV) scheme, where the spacetime is separated into its four- and \((D-4)\)-dimensional parts, and \(\gamma_5\) is kept as a purely four-dimensional object. With this definition, to get well defined amplitudes in chiral gauge theories, one can introduce for each fermion an associated fictitious field with opposite chirality in the fermion kinetic terms, that leads to evanescent operators that break the gauge symmetry. This procedure generates spurious anomalies.

When calculating the obstructions to the Zinn-Justin equation, one finds three types of contributions: evanescent terms, that vanish in the limit \(D\to4\); divergent contributions, that are eliminated in the process of cancellation of divergences; and \textit{finite terms}, that appear when the \(D-4\) factor due to the evanescent symmetry-breaking operator is canceled by the pole in \(D-4\) of a certain loop-amplitude. The identification of those finite contributions is one of the main objectives of this text. There is an extremely powerful technique to isolate those contributions which is called the \textit{Bonneau method}. It shows that all the finite symmetry-breaking terms can be computed from the divergent part of diagrams with a modified symmetry breaking operator.  This also completes the proof of the fact that anomalies are local and, consequently, of the fact that the theories studied in this thesis are anomaly-free.

In the second chapter, precise definitions of the \(D\)-dimensional integrals and of the objects appearing there will be given. The Weinberg's theorem will be proved introducing the BPH prescription and the forest formula, and there is also a proof of the fact that dimensional regularization satisfies the QAP. The main features of the BMHV scheme are introduced, including the fact that it breaks the gauge symmetry. Then, the Bonneau identities will be proved, and it will be shown how to employ them to eliminate the spurious anomalies generated by finite parts of the amplitudes. This second chapter is a pedagogical compilation of the results presented in the references therein. 

\section*{Chapter 3: The Standard Model and the Standard Model Effective Field Theory of Particle Physics}

As already mentioned, the SM is the theory that is currently used to understand the properties and interactions of the elementary particles. However, it does not offer answers to some of the most important questions about nature, namely, the origin of the neutrino masses, what dark matter is made off or what the quantum properties of gravity are. As a consequence, the SM is not the final step in the understanding of the universe, as it is necessary to find an extension that explains all the mentioned features. Nevertheless, there are plenty of models that can fulfill these requirements, but from the current experimental data it is not possible to know which one is the correct one. In this situation, it is convenient to use the bottom-up approach of EFTs and introduce the \textit{Standard Model Effective Field Theory} (SMEFT). Furthermore, nowadays the presence of divergences in QFTs is understood using this paradigm. It is believed that the presence of infinities in radiative calculations is due to the fact that the integration momenta in the loops explore energies where the theory is not valid, meaning that the current theories only work up to a certain energy scale. Nevertheless, the appearance of divergences does not automatically invalidate the predictive power of the theory. 

The main idea of the SMEFT is the following: it is known that, up to a few TeV, the experimental data can be described using the particle content of the SM with the associated \({\rm SU}(3)_{\rm QCD}\otimes{\rm SU}(2)_{\rm L}\otimes{\rm U}(1)_{\rm Y}\) chiral gauge symmetry. Therefore, with independence of what the new physical extension of the SM is (as long as no new light degrees of freedom are introduced), at low energies it behaves as a theory with the particle content and the symmetries of the SM. Thus, the new physics effects will appear in the presence of operators of dimension greater than four, which are absent in the SM. Following Weinberg's paradigm, in order to take into account all the possible new physics effects, one needs to construct the most general Lagrangian with operators of arbitrary dimensionality (depending on the desired precision) that respects the symmetries of the SM and only contains the physical states appearing therein. The third chapter is devoted to the presentation of the main ingredients of the SM and of the SMEFT up to dimension-6 operators.

\section*{Chapter \!4: \!\!\! Symmetry-restoring \!  finite \! counterterms \! of \! SMEFT four-fermion operator insertions at one loop}

The study of the SMEFT has been one of the most prolific fields  for phenomenologists over the last years. This is because it can be used to analyze all the possible new theories at the same time (within the scope already mentioned), and compare with the results of the current experiments. Nevertheless, the remarkable agreement between the SM and experimental data motivates the pursuit of new effects by performing high-precision calculations. Consequently, radiative corrections to the SMEFT observables had become more important in the last years. Since the SMEFT is a chiral theory, if one uses dimensional regularization in a mathematically consistent way, it is necessary to introduce symmetry-restoring counterterms. However, up to this moment, the calculations were done (with some exceptions) using schemes which are not mathematically consistent, but can provide correct answers for particular observables at a certain loop order. The validity of these methods becomes unclear when doing higher order loop calculations, while dimensional regularization in the BMHV scheme adding the proper set of symmetry-restoring counterterms is perfectly consistent.

The purpose of the fourth chapter is to present the one-loop symmetry-restoring finite counterterms associated with the four-fermion operators. To extend the scope of the analysis, also operators involving right-handed neutrinos are considered. The calculation was  done by the author of this text in \cite{Solera:2026xnz}. The tools of algebraic renormalization were used, and since the computation is performed in the unbroken phase, fictitious fields with opposite chiralities were added for all of the fermions of the chiral theory. The complete set of finite counterterms needed for one-loop calculations in the MS scheme, and as a first step for higher order corrections, is shown. These results are compared with the ones obtained by another group using an elegant new method that involves the addition of a spurion field to restore the gauge symmetry in \(D\) dimensions.

\section*{\boldmath Chapter 5: \!\! Left-Right Models. \!\! General Bounds on the Masses of the \(W_R\) and \(Z_R\) Gauge Bosons}

Although the use of the SMEFT is a good strategy to look for new physics effects in the experimental data, the fact is that it does not consist of a description of the new high-energy interactions, it can only be used to constrain the possible new theories once their low-energy effects are known. Thus, to have a broader understanding of nature it is necessary to propose a new theory, with new interactions and/or new particles; this is the top-down approach, the model-building. One of the most relevant extensions of the SM is based on trying to explain its chiral nature, proposing that the breaking of parity comes from \textit{Spontaneous Symmetry Breaking} (SSB): these are the so-called \textit{Left-Right} (LR) models. In these theories, a new weak-like interaction based on the \({\rm SU(2)}_{\rm R}\) chiral gauge group is added, leading (in the simplest case) to the \({\rm SU}(3)_{\rm QCD}\otimes{\rm SU}(2)_{\rm L}\otimes{\rm SU}(2)_{\rm R}\otimes{\rm U}(1)_{\rm X}\) Lie group. One can restore a kind of a parity-symmetry imposing that the couplings to the two \({\rm SU}(2)\) interactions are equal \(g_L=g_R\). The fact that parity is not a good symmetry at the energies currently explored and that no new physical states have been found, can be explained using an SSB mechanism where this new gauge symmetry is broken at a high-energy scale \(v_R\).

Nowadays, it is also known that LR models can be embedded into larger groups of grand unification, like \({\rm SO}(10)\), so the restoration of parity can occur at even higher energy scales. If one does not impose this additional discrete symmetry to the LR models, the number of free parameters increases, so it is highly non-trivial to put constraints on the new scale \(v_R\) that are independent of them. The purpose of the fifth chapter is precisely to find lower bounds on the masses of the new \(W_R\) and \(Z_R\) gauge bosons, and on the LR scale \(v_R\), making no further reference to the other free parameters. This is based on the works done by the author of this text in \cite{Solera:2023kwt,Solera:2024wjz,Solera:2024jrm}. The chapter starts with a presentation of the LR models including the three possible SSB mechanisms studied, one of them based on an EFT with two scalar doublets was introduced for the first time in \cite{Solera:2023kwt}. Then, there is a quantitative discussion about the use of flavor observables for this purpose, and the bounds obtained from collider experiments are presented.

\section*{Methodology and Objectives}

The methodology employed in the present thesis combines three different perspectives: first of all, a review of the literature about symmetries in QFTs, and the use of the methods of algebraic renormalization to restore the chiral gauge symmetry when using dimensional regularization with the BMHV scheme; second, a purely theoretical calculation using the appropriate software to derive the finite symmetry-restoring counterterms needed for the four-fermion operators of SMEFT extended to include right-handed neutrinos at one loop; finally, a phenomenological study of the LR models with the proposal and the analysis of a new one based on EFTs, and the comparison between theoretical predictions and the current experimental data, to obtain lower bounds on the masses of the new particles and on the LR scale. The main objectives of this PhD thesis are:

\begin{itemize}
    \item To develop an understanding about the implications of chiral gauge symmetries in the mathematical consistency of a QFT, paying special attention to the renormalizability conditions (specifically in the context of EFTs) and the cancellation of anomalies.
    \item To acquire the abilities needed to calculate the finite symmetry-restoring counterterms that are necessary in the context of the BMHV scheme in the dimensional regularization of a chiral gauge theory, paying specific attention to the tools of algebraic renormalization.
    \item To calculate those finite symmetry-restoring counterterms for the four-fermion operators of the SMEFT plus right-handed neutrinos at one loop.
    \item To propose and develop a new LR model with only two doublets to spontaneously break the symmetry, which requires effective dimension-5 operators to generate the masses of the fermions.
    \item To find lower bounds on the masses of the new gauge bosons of the LR models and on the LR scale which do not depend on the rest of the parameters of the theory using the available current experimental data from flavor and collider experiments.
\end{itemize}

\thispagestyle{empty}

\chapter{Gauge Theories. Renormalizability and Consistency}
\markboth{1. GAUGE THEORIES. RENORMALIZABILITY AND CONSISTENCY}{}

\section*{Introduction}

The ideas underlying gauge theories stem from the unification of the principles of quantum mechanics with the requirement that the mathematical description of nature must be \textit{local}. The first postulate of quantum physics states that physical systems are represented by an element of a projective Hilbert space. In the context of Quantum Field Theory (QFT) this principle leads to the fact that the Lagrangian that describes the interactions of a fermion represented by a Dirac field \(\psi\left(x\right)\) is invariant under a global phase redefinition or, in other words, under a global \({\rm U}(1)\) transformation. Indeed, it is easy to show that the Lagrangian of a free Dirac field \(\mathcal{L}=\bar{\psi}\left(x\right)\left(i\slashed{\partial}-m\right)\psi\left(x\right)\) is invariant under \(\psi\left(x\right)\rightarrow e^{iQ\theta}\psi\left(x\right)\), as long as \(\theta\) does not depend on the spacetime coordinates \(\theta\neq\theta\left(x\right)\). It is at this point where the principle of locality becomes relevant. The phase convention that we use in this PhD thesis written in Valencia should not be affected by the phase convention that Steven Weinberg took when he wrote his books on QFT thirty years ago in Austin \cite{Weinberg1,Weinberg2}. As a consequence, one is compelled to introduce a vector field \(A_{\mu}\left(x\right)\) that gets transformed as \(A_{\mu}\left(x\right)\rightarrow A_{\mu}\left(x\right)+Q\partial_{\mu}\theta\left(x\right)\) in order to have a Lagrangian invariant under local \({\rm U(1)}\) transformations. The most general one including operators up to dimension four, apart from arbitrary normalization constants and charge-parity (CP) breaking effects, that satisfies this requirement is \(\mathcal{L}=\bar{\psi}\left(x\right)\left(i\slashed{D}-m\right)\psi\left(x\right)-\frac{1}{4}F_{\mu\nu}F^{\mu\nu}\), with \(D_{\mu}\coloneqq\partial_{\mu}-ieQA_{\mu}\left(x\right)\) and \(F_{\mu\nu}\coloneqq\partial_{\mu}A_{\nu}\left(x\right)-\partial_{\nu}A_{\mu}\left(x\right)\), but note that this Lagrangian is precisely the one that describes the interactions between a fermion with charge \(Q\) and an electromagnetic field. Furthermore, the conservation of the electric charge is obtained applying Noether's theorem to the global \({\rm U}(1)\) transformation.

A local internal transformation is called a \textit{gauge} transformation and the quantities that remain unchanged under it are said to be gauge invariant. It has been shown before that the unification of the ideas of quantum mechanics and locality (a requirement from special relativity) results in the concept of gauge symmetry, a concept which is powerful enough to predict Maxwell's equations. The extension of the ideas from an abelian  \({\rm U}(1)\) group into more general non-abelian groups of transformations \cite{Yang_Mills,Utiyama} will be presented throughout this chapter. Ultimately, this will lead to the construction of realistic physical theories like the Standard Model (SM) of particle physics, the Standard Model Effective Field Theory (SMEFT) and the Left-Right (LR) models, all of them being the main subject of study of this thesis. Furthermore, the ideas of De Witt-Faddeev-Popov, Becchi-Rouet-Stora-Tyutin (BRST) and Batalin-Vilkovisky will also be presented. They allow for the quantization of these theories and the study of the possible inconsistencies coming from the breaking of the underlying gauge symmetry at the quantum level, the so-called \textit{anomalies}.

\section{Classical Gauge Theories}

In this section the main features of gauge theories at the classical level are presented, focusing the attention in objects that are invariant under infinitesimal gauge transformations associated with a \textit{Lie group},\footnote{The procedures described here can be extended to a more general class of symmetries.} the only kind of gauge transformation considered in this text. Therefore, imagine that a physical theory contains a set of matter fields \(\psi_{a}\left(x\right)\) labeled by the subindex \(a\) that get transformed as

\begin{equation}
    \label{Transformation_matter_fields}
    \delta\psi_{a}\left(x\right)=i\varepsilon^{\alpha}\left(x\right)\left(t_{\alpha}\right)^b_a\psi_{b}\left(x\right),
\end{equation}

\noindent where \(\varepsilon^{\alpha}\left(x\right)\) is a set of infinitesimal gauge parameters and \(t_{\alpha}\) is a set of matrices which correspond to the generators of a Lie group and satisfy the commutation relation

\begin{equation}
    \label{Commutation_rule}
    \left[t_{\alpha},t_{\beta}\right]=iC^{\gamma}_{\,\,\,\alpha\beta}t_{\gamma},
\end{equation}

\noindent with \(C^{\gamma}_{\,\,\,\alpha\beta}\) a set of real numbers known as the structure constants. Obviously, they have to be antisymmetric under the exchange \(\alpha\leftrightarrow\beta\). The space of generators with the commutator forms a so-called Lie algebra. From the Jacobi identity

\begin{equation}
    \label{Jacobi_Identity}
    \left[t_{\alpha},\left[t_{\beta},t_{\gamma}\right]\right]+\left[t_{\gamma},\left[t_{\alpha},t_{\beta}\right]\right]+\left[t_{\beta},\left[t_{\gamma},t_{\alpha}\right]\right]=0
\end{equation}

\noindent it is straightforward to derive the constraint

\begin{equation}
    \label{Jacobi_Identity_Structure_Constants}
    C^{\epsilon}_{\,\,\,\alpha\delta}C^{\delta}_{\,\,\,\beta\gamma}+C^{\epsilon}_{\,\,\,\gamma\delta}C^{\delta}_{\,\,\,\alpha\beta}+C^{\epsilon}_{\,\,\,\beta\delta}C^{\delta}_{\,\,\,\gamma\alpha}=0.
\end{equation}

\noindent It is important to remark that any set of matrices satisfying~\eqref{Commutation_rule} furnish a representation of the \textit{Lie algebra}.\footnote{In realistic theories like the SM the different matter fields may transform under different representations of the gauge group. In order to take this into account, the introduction of another index \(m\) for the matter fields \(\psi^{m}_{a}\left(x\right)\) and for the generators \(t^m\) is needed. The present discussion can be easily generalized to that case so, for the moment, this consideration is omitted with the purpose of alleviating the notation.} A case of great importance is the one given by the matrices \(\left(t^{\rm ad}_{\alpha}\right)^{\beta}_{\gamma}\coloneqq iC^{\beta}_{\,\,\,\alpha\gamma}\), which is known as the adjoint representation.

In order to build objects that are invariant under the transformation~\eqref{Transformation_matter_fields}, the so-called \textit{covariant derivative} (a derivative that transforms in the same way as the operator on which it acts) is introduced. In the case of the matter fields it is given by

\begin{equation}
    \label{Covariant_Derivative_Matter_Fields}
    \left(D_{\mu}\psi\left(x\right)\right)_a\coloneqq \partial_{\mu}\psi_a\left(x\right)-iA_{\mu}^{\alpha}\left(x\right)\left(t_{\alpha}\right)_a^b\psi_b\left(x\right),
\end{equation}

\noindent where the \(A_{\mu}^{\alpha}\left(x\right)\) are the \textit{gauge fields}. They can be interpreted from the differential-geometric point of view as connection one-forms and they must transform under infinitesimal gauge transformations as

\begin{equation}
    \label{Gauge_Transformation_Gauge_Field}
    \delta A_{\mu}^{\alpha}\left(x\right)=\partial_{\mu}\varepsilon^{\alpha}\left(x\right)+C^{\alpha}_{\,\,\,\gamma\beta}\varepsilon^{\beta}\left(x\right)A_{\mu}^{\gamma}\left(x\right)=\partial_{\mu}\varepsilon^{\alpha}\left(x\right)+i\varepsilon^{\beta}\left(x\right)(t_{\beta}^{\rm ad})^{\alpha}_{\gamma}A_{\mu}^{\gamma}\left(x\right),
\end{equation}

\noindent so they belong to the adjoint representation. As previously mentioned, a straightforward calculation shows that the covariant derivative transforms in the same way as \(\psi_a\left(x\right)\):

\begin{equation}
    \label{Transformation_Covariant_Derivative}
    \delta\left(D_{\mu}\psi\left(x\right)\right)_a=i\varepsilon^{\alpha}\left(x\right)\left(t_{\alpha}\right)_a^bD_{\mu}\psi_b\left(x\right).
\end{equation}

The \textit{field strength tensor} \(F_{\mu\nu}^{\alpha}\left(x\right)\) (equivalent to the curvature tensor) can also be considered. It is defined as

\begin{equation}
    \label{Curvature_Tensor}
    \left(\left[D_{\mu},D_{\nu}\right]\psi\left(x\right)\right)_a\coloneqq-i\left(t_{\alpha}\right)_a^bF_{\mu\nu}^{\alpha}\left(x\right)\psi_b\left(x\right).
\end{equation}

\noindent A direct calculation shows that \(F_{\mu\nu}^{\alpha}\left(x\right)\) is given by

\begin{equation}
    \label{Definition_F}
    F_{\mu\nu}^{\alpha}\left(x\right)=\partial_{\mu}A_{\nu}^{\alpha}\left(x\right)-\partial_{\nu}A_{\mu}^{\alpha}\left(x\right)+C^{\alpha}_{\,\,\,\beta\gamma}A_{\mu}^{\beta}\left(x\right)A_{\nu}^{\gamma}\left(x\right)
\end{equation}

\noindent and it gets transformed under the adjoint representation:

\begin{equation}
    \label{Transformation_F}
    \delta F_{\mu\nu}^{\alpha}\left(x\right)=i\varepsilon^{\beta}\left(x\right)(t^{\rm ad}_{\beta})^{\alpha}_{\gamma}F^{\gamma}_{\mu\nu}\left(x\right)=\varepsilon^{\beta}\left(x\right)C^{\alpha}_{\gamma\beta}F^{\gamma}_{\mu\nu}\left(x\right).
\end{equation}

The gauge transformations of \(\psi_a\), \(F_{\mu\nu}^a\) and their covariant derivatives do not include any derivative of \(\varepsilon^{\alpha}\), which means that any object invariant under global transformations that only depends on these operators will also remain unchanged under local transformations. Consequently, the Lagrangian of a gauge theory will be constructed from Lorentz and gauge invariant combinations of these objects. It is appropriate to begin with the kinetic term for the gauge fields since \textit{it will constrain the allowed Lie groups from which we can get a dynamical theory}. The only Lorentz and possibly gauge invariant operator\footnote{Another possibility is \(\mathcal{L}'_{{\rm Kin},A}=-\frac{1}{2}g'_{\alpha\beta}\varepsilon^{\mu\nu\rho\sigma}F^{\alpha}_{\mu\nu}F^{\beta}_{\rho\sigma}\), but it is just a total derivative so it will produce no contribution in perturbative calculations, although it has important P- and CP-breaking effects which come from non-perturbative considerations.} of dimension four (up to arbitrary normalization constants) that contains a quadratic term in \(A_{\mu}^{\alpha}\) is

\begin{equation}
    \label{General_Kinetic_Term}
    \mathcal{L}_{{\rm Kin},A}=-\frac{1}{4}g_{\alpha\beta}F_{\mu\nu}^{\alpha}F^{\beta\mu\nu},
\end{equation}

\noindent where \(g_{\alpha\beta}\) is a constant matrix that must satisfy the following properties: it has to be real in order to have a hermitian Lagrangian; due to the fact that \(F_{\mu\nu}^{\alpha}F^{\beta\mu\nu}\) is symmetric under the exchange \(\alpha\leftrightarrow\beta\), the matrix \(g\) can always be taken to be symmetric; and if we want to have a well-defined vacuum for the associated QFT it has to be positive-definite. Furthermore, the requirement of~\eqref{General_Kinetic_Term} being gauge invariant leads to the condition:

\begin{equation}
    \label{Requirement_g}
    g_{\alpha\beta}C^{\beta}_{\,\,\,\gamma\delta}+g_{\gamma\beta}C^{\beta}_{\,\,\,\alpha\delta}=0.
\end{equation}

There is a fundamental theorem \cite{Weinberg2,GLASHOW1961437} which proves that \textit{there exists such a real symmetric positive-definite matrix \(g_{\alpha\beta}\)  if and only if the Lie algebra can be decomposed as a direct sum of \({\rm U}(1)\)s and commuting compact simple algebras.}\footnote{A compact Lie algebra is one for which the associated Lie group is compact in the topological sense. This condition is equivalent to the matrix \(M_{\alpha\beta}\coloneqq{\rm tr}\left\{t^{\rm ad}_{\alpha}t^{\rm ad}_{\beta}\right\}=-C^{\gamma}_{\,\,\,\alpha\delta}C^{\delta}_{\,\,\,\beta\gamma}\) being positive-definite. A simple Lie algebra \(\mathcal{G}\) is one for which there is no subalgebra \(\mathcal{H}\) such that the commutator of any element of \(\mathcal{H}\) with any element of \(\mathcal{G}\) belongs to \(\mathcal{H}\).} Moreover, there exists a complete classification of these algebras done by Élie Cartan \cite{cartan1894structure,Georgi:2000vve}. They consist of the algebras associated with the special unitary groups \({\rm SU}(n)\) for \(n\geq2\), the special orthogonal groups \({\rm SO}(n)\) except for \(n=1,2\) and \(4\), the unitary symplectic groups \({\rm USp}(2n)\) for \(n\geq1\) and five exceptional cases: \(G_2\), \(F_4\), \(E_6\), \(E_7\) and \(E_8\).\footnote{It is interesting to say that the \({\rm SO}(n)\), \({\rm SU}(n)\) and \({\rm USp}(n)\) groups can be interpreted as the groups of rotation matrices acting over \(n\)-dimensional real, complex or quaternion vectors, respectively. The five exceptional Lie groups preserve algebraic structures that involve the octonions. This is highly remarkable since these four number systems are known to be the only four real normed division algebras according to the Hurwitz's theorem.} It is important to remark that not all of these groups are totally independent: \({\rm USp}(2)\) is isomorphic to \({\rm SU}(2)\) and \({\rm SU}(2)\), \({\rm USp}(4)\) and \({\rm SU}(4)\) are the simply connected covering groups of \({\rm SO}(3)\), \({\rm SO}(5)\) and \({\rm SO}(6)\), respectively. Consequently, if a gauge theory is based on a Lie group it must be a direct product of the ones mentioned. For instance, the Lie group of the SM and SMEFT is \({\rm SU}(3)_{\rm QCD}\otimes{\rm SU}(2)_{\rm L}\otimes{\rm U}(1)_{\rm Y}\), while for LR models it is \({\rm SU}(3)_{\rm QCD}\otimes{\rm SU}(2)_{\rm L}\otimes{\rm SU}(2)_{\rm R}\otimes{\rm U}(1)_{\rm B-L}\).\footnote{In the cases where the group can be decomposed as a product of different subgroups it will be convenient to multiply all of the generators associated with different parts of the algebra by a constant \(g\), which is known as the coupling constant. For simple groups this constant is the only number needed to describe how strong the interactions are, because all of the particles are coupled to the gauge fields in the same way if they are not singlets. Only for the \({\rm U}(1)\)s the particles can have different charges.}

It can be shown \cite{Weinberg2,GLASHOW1961437} that the algebra is a direct sum of \({\rm U}(1)\)s and compact simple algebras if and only if it is possible to take a basis of generators for which the structure constants are totally antisymmetric. In this case, they will be written as \(C^{\alpha}_{\,\,\,\beta\gamma}\equiv C_{\alpha\beta\gamma}\). Moreover, in this situation it is also always possible \cite{Weinberg2} to take a normalization of the gauge fields, generators and structure constants in such a way that \(g_{\alpha\beta}=\delta_{\alpha\beta}\), leaving the kinetic term as

\begin{equation}
    \label{Kinetic_Term}
    \mathcal{L}_{{\rm Kin},A}=-\frac{1}{4}F_{\mu\nu}^{\alpha}F^{\mu\nu}_{\alpha}.
\end{equation}

\noindent At this point, it is important to remark that \textit{it is not possible to construct a gauge theory for a Lie group with} \(C_{\alpha\beta\gamma}\neq0\) \textit{in which the gauge fields are not self-interacting} since the kinetic term always contains cubic and quartic couplings, so it is also interesting to study the theory described just by the Lagrangian in Eq.~\eqref{Kinetic_Term} with no additional operators, which is known as the pure gauge theory.

As mentioned in \cite{Weinberg2}, it can be proved that all of the finite representations of a compact Lie group are unitary, so the corresponding Lie algebra only has hermitian generators \(t_{\alpha}^{\dagger}=t_{\alpha}\). Consequently, \(i\bar{\psi}_a\slashed{D}\psi_a\) can be used as a kinetic term for the spin-\(1/2\) matter fields and \(\left(D_{\mu}\phi\right)^{\dagger}D^{\mu}\phi\) for the spin-0 case. Then, only considering operators up to dimension four, the most general Lagrangian has the form

\begin{equation}
    \label{Gauge_Lagrangian}
    \mathcal{L}=-\frac{1}{4}F_{\mu\nu}^{\alpha}F^{\mu\nu}_{\alpha}+\mathcal{L}_M\left(\psi,D\psi\right),
\end{equation}

\noindent where now \(\psi\) is accounting for fermions and scalars. It is important to remark that mass terms for the \(A_{\alpha}^{\mu}\) are forbidden by gauge invariance.\footnote{ However, in two dimensions Quantum Electrodynamics (QED) exhibits massive photons. This is known as the Schwinger model \cite{Schwinger:1962tp} and there is no inconsistency. The mass term is generated at one-loop and it is associated with a non-local gauge invariant operator in the Lagrangian. It only appears in two dimensions because in that situation the photons only have one physical degree of freedom, as they can be expressed as a total derivative of a scalar field \cite{Weinberg1}.}  The associated equations of motion read

\begin{equation}
    \label{Equations_Motion}
    D_{\mu}F^{\mu\nu}_{\alpha}=J^{\nu}_{\alpha},
\end{equation}

\noindent with

\begin{equation}
    \label{Matter_Current}
    J_{\alpha}^{\mu}\coloneqq i\frac{\partial \mathcal{L}_M}{\partial\left(D_{\mu}\psi\right)}t_{\alpha}\psi.
\end{equation}

\noindent Since \(F_{\alpha}^{\mu\nu}\) gets transformed under the adjoint representation of the gauge group, then:

\begin{equation}
    \label{Covariant_Derivative_F}
    D^{\rho}F^{\mu\nu}_{\alpha}\coloneqq\partial^{\rho}F_{\alpha}^{\mu\nu}-i(t^{\rm ad}_{\beta})_{\alpha\gamma}A^{\rho}_{\beta}F_{\gamma}^{\mu\nu}=\partial^{\rho}F_{\alpha}^{\mu\nu}+C_{\alpha\beta\gamma}A^{\rho}_{\beta}F_{\gamma}^{\mu\nu}.
\end{equation}

\noindent Using Noether's theorem applied to the global symmetries associated with the gauge symmetry, one can get the following gauge covariant conservation laws:

\begin{equation}
    \label{Gauge_Covariant_Conservation_Law}
    D_{\mu}J^{\mu}_a=0,
\end{equation}

\noindent where \(J^{\mu}_a\) gets transformed under the adjoint representation. Finally, from Eqs.~\eqref{Curvature_Tensor} and~\eqref{Covariant_Derivative_F} and using the Jacobi identity for the commutators containing three covariant derivatives, the inhomogeneous Maxwell's equations are obtained:

\begin{equation}
    \label{Inhomogeneous_Maxwell_Non_Abelian}
    D^{\mu}F^{\nu\rho}_{\alpha}+D^{\rho}F^{\mu\nu}_{\alpha}+D^{\nu}F^{\rho\mu}_{\alpha}=0.
\end{equation}

\section{Quantization in the Axial Gauge}

In this section, the attention is focused on the quantization of the theory associated with the Lagrangian in Eq.~\eqref{Gauge_Lagrangian} so, for the moment, effective operators will not be studied. The quantization of gauge theories is highly non-trivial due to the fact that \textit{ not all of the degrees of freedom of the four-potential} \(A_{\mu}^{\alpha}\) \textit{are physical}. For example, consider the canonical momenta \(\Pi_{\alpha}^{\mu}\) associated with it:

\begin{equation}
    \label{Canonical_Momentum}
    \Pi_{\alpha}^{\mu}\coloneqq\frac{\partial\mathcal{L}}{\partial\left(\partial_0A_{\alpha\mu}\right)}=F_{\alpha}^{\mu0}.
\end{equation}

\noindent The antisymmetry of the tensor \(F_{\alpha}^{\mu\nu}\) yields  a vanishing momentum \(\Pi_{\alpha}^{0}=0\), so the condition \(\{A_{\alpha0}\left(t,\textbf{x}\right),\Pi_{\beta}^{0}\left(t,\textbf{y}\right)\}=\delta_{\alpha\beta}\delta^{3}\left(\textbf{x}-\textbf{y}\right)\) for the Poisson bracket cannot be satisfied. As a consequence, the canonical commutation relations cannot be imposed. Therefore, the field \(A_{\alpha0}\) will not be considered as an independent degree of freedom, its value will be determined by the other fields and momenta. Furthermore, there is an additional\footnote{Having two constraints is related to the fact that, even though each of the gauge fields \(A_{\alpha}\) has four independent components, there are only two physical polarizations.} constraint coming from the equations of motion of \(A_{\alpha0}\):

\begin{equation}
    \label{Second_Constraint}
    \begin{split}
           0=\frac{\partial\mathcal{L}}{\partial A_{\alpha0}}-\partial_{\mu}\frac{\partial\mathcal{L}}{\partial\left(\partial_{\mu}A_{\alpha0}\right)}&=\partial_{\mu}F^{\mu0}_{\alpha}+C_{\alpha\beta\gamma}A_{\beta\mu}F_{\gamma}^{\mu0}-J_{\alpha}^0\\&=\partial_{k}\Pi_{\alpha}^{k}+C_{\alpha\beta\gamma}A_{\beta k}\Pi_{\gamma}^k-J_{\alpha}^0,
    \end{split}
\end{equation}

\noindent where \(k=1,2,3\). Eq.~\eqref{Second_Constraint} shows that the value of one of the gauge fields can be determined from the values of the rest of them and the associated momenta.

In order to quantize the theory,  the gauge fixing condition\footnote{Note that this condition breaks Lorentz invariance, because it cannot be imposed for different observers at the same time.} \(A_{\alpha3}=0\) will be imposed. This particular choice is known as the \textit{axial gauge}, in which the only degrees of freedom are \(A_{\alpha i}\) with  \(i=1,2\), \(\psi_a\) and the associated momenta. From Eq.~\eqref{Canonical_Momentum} the following expression for the  conjugated momenta of \(A_{\alpha i}\) is derived:

\begin{equation}
    \label{Momenta_Axial_Gauge}
    \Pi_{\alpha i}=F_{\alpha i 0}=\partial_{i}A_{\alpha0}-\partial_{0}A_{\alpha i}+C_{\alpha\beta\gamma}A_{\beta i}A_{\gamma 0}.
\end{equation}

\noindent Now, the constraint in Eq.~\eqref{Second_Constraint} will be used to write \(A_{\alpha 0}\) in terms of the other degrees of freedom. For this purpose, note that the \(F^{k0}\)  satisfy the equalities \(F^{i0}_{\alpha}=-\Pi_{\alpha i}\) and \(F^{30}_{\alpha}=-\partial_3A_{\alpha}^0\) in the axial gauge. Thus, the following differential equation for \(A_{\alpha}^0\) is obtained:

\begin{equation}
    \label{Constraint_A0}
    \left(\partial_3\right)^2A_{\alpha}^0=-\partial_i\Pi_{\alpha i}-C_{\alpha\beta\gamma}A_{\beta i}\Pi_{\gamma i}-J_{\alpha}^0.
\end{equation}

\noindent This gives \(A_{\alpha}^0\) as a function of the dynamical degrees of freedom by solving Eq.~\eqref{Constraint_A0} with the appropriate boundary conditions. In order to prove this statement, the matter part of the Lagrangian needs to be studied. It is easy to see that the canonical momenta associated with \(\psi_a\) are simply

\begin{equation}
    \label{Canonical_Momenta_Matter}
    \pi_a\coloneqq\frac{\partial\mathcal{L}}{\partial\left(\partial_0\psi_a\right)}=\frac{\partial\mathcal{L}_M}{\partial\left(D_0\psi_a\right)},
\end{equation}

\noindent given the definition in Eq.~\eqref{Covariant_Derivative_Matter_Fields}. A straightforward calculation using Eq.~\eqref{Matter_Current} shows that 

\begin{equation}
    \label{Matter_Current_Canonical_Momenta}
    J_{\alpha}^0=i\pi_a\left(t_{\alpha}\right)^a_{b}\psi_b,
\end{equation}

\noindent confirming that \(A_{\alpha0}\) can be written as a function of the other fields and momenta integrating the constraint~\eqref{Constraint_A0}. Finally, with these ingredients it is possible to construct the Hamiltonian of the theory in the axial gauge:

\begin{equation}
    \label{Hamiltonian_Axial_Gauge}
    \begin{split}
        \mathcal{H}&\coloneqq \Pi_{\alpha}^i  \partial_0A_{\alpha i} +\pi_a\partial_0\psi_a-\mathcal{L} \\& =\Pi_{\alpha i}\left(\Pi_{\alpha i} -\partial_iA_{\alpha0}-C_{\alpha\beta\gamma}A_{\beta i}A_{\gamma 0}\right) +\pi_a\partial_0\psi_a\\&-\frac{1}{2}F_{\alpha0i}F_{\alpha0i}+\frac{1}{4}F_{\alpha ij}F_{\alpha ij}+\frac{1}{2}F_{\alpha i3}F_{\alpha i3}-\frac{1}{2}F_{\alpha 03}F_{\alpha 03}-\mathcal{L}_M\\&=\mathcal{H}_M-\Pi_{\alpha i}\left(\partial_iA_{\alpha0}+C_{\alpha\beta\gamma}A_{\beta i}A_{\gamma 0}\right)+\frac{1}{2}\Pi_{\alpha i}\Pi_{\alpha i}\\&+\frac{1}{4}F_{\alpha ij}F_{\alpha ij}+\frac{1}{2}\partial_3A_{\alpha i}\partial_3A_{\alpha i}-\frac{1}{2}\partial_3A_{\alpha0}\partial_3A_{\alpha0},
    \end{split}
\end{equation}

\noindent with \(\mathcal{H}_M\coloneqq \pi_a\partial_0\psi_a-\mathcal{L}_M\).

In order to quantize the theory using the path integral formalism, the integration over the dynamical degrees of freedom (\(A_{\alpha i}\), \(\psi_a\) and the associated momenta) of the factor \(\exp\{iI\}\) with

\begin{equation}
    \label{Action_Path_Integral}
    I\coloneqq\int{\rm d}^4x\left(\Pi_{\alpha}^i  \partial_0A_{\alpha i} +\pi_a\partial_0\psi_a-\mathcal{H} \right)
\end{equation}

\noindent has to be performed. Imagine that \(A_{\alpha 0}\) is also a degree of freedom over which one can integrate. In that case, since the Hamiltonian in Eq.~\eqref{Hamiltonian_Axial_Gauge} is at most quadratic in that field, the prescription of Gaussian integration can be used. Indeed, taking the appropriate boundary conditions, the following relation can be obtained: \(-\frac{1}{2}\int{\rm d}^4x\partial_3A_{\alpha0}\partial_3A_{\alpha0}=\frac{1}{2}\int{\rm d}^4x{\rm d}^4yA_{\alpha0}\left(x\right)\delta_{\alpha\beta}\delta^4\left(x-y\right)\left(\partial_3\right)^2A_{\beta0}\left(y\right)\). In these cases, it can be shown \cite{Weinberg1} that the integration over the field \(A_{\alpha 0}\) in the Gaussian prescription results in the integrand evaluated at the stationary point of the action up to a constant multiplicative factor which contains the functional determinant of \(\delta_{\alpha\beta}\delta^4\left(x-y\right)\left(\partial_3\right)^2\). The stationary point is determined by 

\begin{equation}
    \label{Stationary_A0}
    \frac{\delta I}{\delta A_{\alpha0}}=-\left(\partial_3\right)^2A_{\alpha0}-\partial_i\Pi_{\alpha i}-C_{\alpha\beta\gamma}A_{\beta i}\Pi_{\gamma i}-J_{\alpha}^0=0.
\end{equation}

\noindent Note that this condition is precisely the constraint showed in Eq.~\eqref{Constraint_A0}. Therefore, integrating also over \(A_{\alpha 0}\) the restriction will be automatically imposed. Consequently, the integration variables are \(A_{\alpha i}\), \(\psi_a\), the associated canonical momenta and \(A_{\alpha0}\). Now, since \(I\) is a quadratic form in \(\Pi_{\alpha i}\left(x\right)\) with field independent kernel \(\frac{1}{2}\delta_{\alpha\beta}\delta_{ij}\delta^4\left(x-y\right)\), the integration over \(\Pi_{\alpha i}\) using again the Gaussian prescription can easily be done. Assuming the same situation for the matter part, it is only necessary to evaluate the integrand at the stationary points:

\begin{equation}
     \label{Stationary_Pi}
     \frac{\delta I}{\delta \Pi_{\alpha i}}=-\Pi_{\alpha i}-\partial_0A_{\alpha i}+\partial_iA_{\alpha 0}+C_{\alpha\beta\gamma}A_{\beta i}A_{\gamma 0}=-\Pi_{\alpha i}+F_{\alpha i0}=0
\end{equation}

\noindent and

\begin{equation}
    \label{Stationary_pi}
    \frac{\delta I}{\delta \pi_{a}}=\partial_0\psi_a-\frac{\partial \mathcal{H}_M}{\partial \pi_{a}}=0.
\end{equation}

\noindent Then, the factor \(I\) in the exponent is simply

\begin{equation}
    \label{FactorI}
    \begin{split}
        I\coloneqq\int{\rm d}^4x&\left(\pi_a\frac{\partial \mathcal{H}_M}{\partial \pi_{a}}-\mathcal{H}_M+F_{\alpha i0}F_{\alpha i0}-\frac{1}{2}F_{\alpha i0}F_{\alpha i0}\right.\\&\left.-\frac{1}{4}F_{\alpha ij}F_{\alpha ij}-\frac{1}{2}\partial_3A_{\alpha i}\partial_3A_{\alpha i}+\frac{1}{2}\partial_3A_{\alpha0}\partial_3A_{\alpha0}\right)=\int{\rm d}^4x\left.\mathcal{L}\right|_{A_{\alpha3}=0}.
    \end{split}
\end{equation}

\noindent Thus, in the axial gauge the path integral formalism leads to the integration of the exponential of \(i\) times the original action evaluated at \(A_{\alpha3}=0\) with the variables \(A_{\alpha 0}\), \(A_{\alpha 1}\), \(A_{\alpha 2}\) and the matter fields \(\psi_a\). Finally, one can integrate over the four components of \(A_{\alpha\mu}\) including the appropriate Dirac delta:

\begin{equation}
    \label{T_Ordered_Axiual_Gauge}
    \left<T\left\{\mathcal{O}_1\mathcal{O}_2\dots\right\}\right>_{\rm Vac}\propto\int\mathcal{D}A_{\alpha\mu}\mathcal{D}\psi_a\left(\mathcal{O}_1\mathcal{O}_2\dots\right)\delta\left(A_{\alpha3}\right)\exp\left\{i\int{\rm d}^4x\mathcal{L}\right\},
\end{equation}

\noindent where \(\mathcal{D}\phi_i\) represents \(\prod_{i,x}d\phi_{i}\left(x\right)\).

\section{The De Witt-Faddeev-Popov Method} \label{sec_Faddeev_Popov}

The problem with the quantization in the axial gauge is that \textit{ the condition} \(A_{\alpha3}=0\) \textit{explicitly breaks Lorentz invariance} so, in order to make full use of the power of the covariant formalism, it is necessary to change the gauge. Thus, it is necessary to make a gauge transformation in Eq.~\eqref{T_Ordered_Axiual_Gauge} (which is equivalent to a change of variables in the path integral) assuming the gauge invariance condition

\begin{equation}
    \label{Path_Integral_Gauge_Transformation}
    \mathcal{D}A_{\Lambda\alpha\mu}\mathcal{D}\psi_{\Lambda a}\left(\mathcal{O}_{\Lambda1}\mathcal{O}_{\Lambda2}\dots\right)\exp\left\{i\int{\rm d}^4x\mathcal{L}_{\Lambda}\right\}=\mathcal{D}A_{\alpha\mu}\mathcal{D}\psi_a\left(\mathcal{O}_1\mathcal{O}_2\dots\right)\exp\left\{i\int{\rm d}^4x\mathcal{L}\right\},
\end{equation}

\noindent where \(\Lambda\) represents the set of parameters of a gauge transformation. Generally speaking, only the T-ordered products of gauge invariant operators are considered, so the condition \(\mathcal{O}_{\Lambda i}=\mathcal{O}_i\) is automatically satisfied. In the same way, the majority of Lagrangians of this text are gauge invariant,\footnote{The exception is the \(D\)-dimensional extension of the chiral Lagrangians, one of the main objects of study of this thesis. The generalization of the tools presented here for those cases is considered in Sec.~\ref{sec:BMHV}.} for those cases we have \(\mathcal{L}_{\Lambda}=\mathcal{L}\). Furthermore, it is also easy to prove that \textit{the measure element} \(\mathcal{D}A_{\mu\alpha}\) \textit{is invariant under infinitesimal gauge transformations} \(A_{\lambda\alpha}^{\mu}=A_{\alpha}^{\mu}+\partial^{\mu}\lambda_{\alpha}+C_{\alpha\beta\gamma}A_{\beta}^{\mu}\lambda_{\gamma}\) with infinitesimal parameters \(\lambda_{\alpha}\left(x\right)\). The associated change in the measure is given by \(\mathcal{D}A_{\lambda\alpha\mu}={\rm det}\left(\mathcal{J}\right)\mathcal{D}A_{\alpha\mu}\) with the Jacobian matrix \(\mathcal{J}\) defined by

\begin{equation}
    \label{Jacobian_Matrix}
    \mathcal{J}_{\alpha\mu x,\beta\nu y}\coloneqq\frac{\delta A_{\lambda\alpha\mu}\left(x\right)}{\delta A_{\beta\nu}\left(y\right)}=\delta^{4}\left(x-y\right)\delta^{\nu}_{\mu}\left(\delta_{\alpha\beta}+C_{\alpha\beta\gamma}\lambda_{\gamma}\left(x\right)\right).
\end{equation}

\noindent At first order in \(\lambda\) the Jacobian is the identity because the trace of the antisymmetric structure constants is zero (\({\rm det}(id+\varepsilon A)\approx1+\varepsilon{\rm tr}\left(A\right)\) for small \(\varepsilon\)). As a consequence, the measure \(\mathcal{D}A_{\mu\alpha}\) is gauge invariant. The question of whether the same situation for the measure of the matter fields \(\mathcal{D}\psi_a\) is the same is crucial and it will be studied in Sec. \ref{sec:Anomalies}, but for the moment it is assumed that this is the case.

In order to change the gauge, one needs to show that the functional integral in Eq.~\eqref{T_Ordered_Axiual_Gauge} can be expressed, up to a multiplicative constant factor with no physical meaning, as

\begin{equation}
    \label{Functional_Integral_FP}
    \mathcal{I}\coloneqq\int\mathcal{D}\phi_nG\left[\phi\right]B\left[f\left[\phi\right]\right]{\rm det}\left(F\left[\phi\right]\right),
\end{equation}

\noindent with \(\phi_n\) representing the matter and gauge fields, \(G\left[\phi\right]\) a functional that satisfies the condition of \(\mathcal{D}\phi_nG\left[\phi\right]\) being invariant under gauge transformations, \(B\left[f\right]\) an arbitrary functional of the so-called gauge-fixing functional \(f\left[\phi\right]\) and 

\begin{equation}
    \label{F_Determinant}
    F_{\alpha x,\beta y}\left[\phi\right]\coloneqq\left.\frac{\delta f_{\alpha}\left[\phi_{\lambda};x\right]}{\delta\lambda_{\beta}\left(y\right)}\right|_{\lambda=0}.
\end{equation}

Given the assumption of~\eqref{Path_Integral_Gauge_Transformation}, it is clear that \(G\left[\phi\right]\) can be identified with the functional \(\left(\mathcal{O}_1\mathcal{O}_2\dots\right)\exp\left\{i\int{\rm d}^4x\mathcal{L}\right\}\). Furthermore, one can simply take \(f_{\alpha}\left[\phi;x\right]=A_{\alpha3}\left(x\right)\) and \(B\left[f\right]=\delta\left(f\right)=\delta\left(A_{\alpha3}\left(x\right)\right)\). Finally, from the condition \(A_{\alpha 3}=0\) of the axial gauge, it is clear that the infinitesimal gauge transformation results in \(f_{\alpha}\left[\phi_{\lambda};x\right]=A_{\lambda\alpha3}=\partial_3\lambda_{\alpha}\left(x\right)\) so \(F_{\alpha x,\beta y}\left[\phi\right]=\delta_{\alpha\beta}\partial_3\delta^{4}\left(x-y\right)\). Therefore, the determinant \({\rm det}\left(F\left[\phi\right]\right)\) is field independent so its contribution is just a non-physical multiplicative constant factor, completing the proof of the statement.

Now, the main theorem of the De Witt-Faddeev-Popov method \cite{PhysRev.162.1195,FADDEEV196729} is presented: \textit{ the integral in Eq.~\eqref{Functional_Integral_FP} does not depend on the gauge-fixing functional and only depends on} \(B\left[f\left[\phi\right]\right]\) \textit{trough a non-physical multiplicative factor.} In order to prove this theorem, it is necessary to consider the change of variables associated with a gauge transformation with a set of parameters \(\Lambda\):

\begin{equation}
    \label{Change_Variables_Functional_Integral_FP}
    \begin{split}
        \mathcal{I}&\coloneqq\int\mathcal{D}\phi_nG\left[\phi\right]B\left[f\left[\phi\right]\right]{\rm det}\left(F\left[\phi\right]\right)=\int\mathcal{D}\phi_{\Lambda n}G\left[\phi_{\Lambda}\right]B\left[f\left[\phi_{\Lambda}\right]\right]{\rm det}\left(F\left[\phi_{\Lambda}\right]\right)\\&=\int\mathcal{D}\phi_{n}G\left[\phi\right]B\left[f\left[\phi_{\Lambda}\right]\right]{\rm det}\left(F\left[\phi_{\Lambda}\right]\right),
    \end{split}
\end{equation}

\noindent where the invariance of  \(\mathcal{D}\phi_nG\left[\phi\right]\) has been used. First of all, since \(\mathcal{I}\) does not depend on \(\Lambda\) one can integrate at both sides over it with a weighting functional \(W\left[\Lambda\right]\)

\begin{equation}
    \label{Integration_Weighting_Functional}
    \mathcal{I}\int\mathcal{D}\Lambda_{\alpha}W\left[\Lambda\right]=\int\mathcal{D}\phi_nG\left[\phi\right]\underbrace{\int\mathcal{D}\Lambda_{\alpha}W\left[\Lambda\right]B\left[f\left[\phi_{\Lambda}\right]\right]{\rm det}\left(F\left[\phi_{\Lambda}\right]\right)}_{\eqqcolon C\left[\phi\right]}.
\end{equation}

\noindent Now, consider the functional \(F\left[\phi_{\Lambda}\right]\) defined by

\begin{equation}
    \label{F_Determinant_Lambda}
    F_{\alpha x,\beta y}\left[\phi_{\Lambda}\right]\coloneqq\left.\frac{\delta f_{\alpha}\left[\left(\phi_{\Lambda}\right)_{\lambda};x\right]}{\delta\lambda_{\beta}\left(y\right)}\right|_{\lambda=0}.
\end{equation}

\noindent Due to the fact that only gauge transformations associated with a Lie group are considered, the composition of \(\Lambda\) with \(\lambda\) is again a gauge transformation that can be denoted as \(\bar{\Lambda}\left(x;\Lambda,\lambda\right)\), so it is possible to write \(\left(\phi_{\Lambda}\right)_{\lambda}\equiv\phi_{\bar{\Lambda}\left(\Lambda,\lambda\right)}\). Then, using the chain rule, the following relation is obtained:

\begin{equation}
    \label{Chain_Rule}
    \begin{split}
        F_{\alpha x,\beta y}\left[\phi_{\Lambda}\right]&=\int{\rm d}^4z\left.\frac{\delta f_{\alpha}[\phi_{\bar{\Lambda}\left(\Lambda,\lambda\right)};x]}{\delta \bar{\Lambda}^{\gamma}\left(z\right)}\right|_{\bar{\Lambda}(\Lambda,0)=\Lambda}\left.\frac{\delta \bar{\Lambda}^{\gamma}\left(z;\Lambda,\lambda\right)}{\delta \lambda^{\beta}\left(y\right)}\right|_{\lambda=0}\\&=\int{\rm d}^4z\underbrace{\left.\frac{\delta \bar{\Lambda}^{\gamma}\left(z;\Lambda,\lambda\right)}{\delta \lambda^{\beta}\left(y\right)}\right|_{\lambda=0}}_{\eqqcolon\mathcal{R}_{\beta y }^{\gamma z}\left[\Lambda\right]}\underbrace{\frac{\delta f_{\alpha}[\phi_{\Lambda};x]}{\delta \Lambda^{\gamma}\left(z\right)}}_{\eqqcolon\mathcal{F}_{\alpha x,\gamma z}\left[\phi,\Lambda\right]}.
    \end{split}
\end{equation}

From the properties of the determinant of the product of matrices, it is possible to write the functional determinant as \({\rm det}\left(F\left[\phi_\Lambda\right]\right)={\rm det}\left(\mathcal{F}\left[\phi,\Lambda\right]\right){\rm det}\left(\mathcal{R}\left[\Lambda\right]\right)\). In this case, the arbitrary weighting functional \(W\left[\Lambda\right]\) could be taken to be equal to \(1/{\rm det}\left(\mathcal{R}\left[\Lambda\right]\right)\) canceling the second factor. Consequently, the functional \(C\left[\phi\right]\) is simply

\begin{equation}
    \label{Functional_C}
    C\left[\phi\right]=\int\mathcal{D}\Lambda_{\alpha}{\rm det}\left(\mathcal{F}\left[\phi,\Lambda\right]\right)B\left[f\left[\phi_{\Lambda}\right]\right].
\end{equation}

\noindent Note from its definition that the determinant of \(\mathcal{F}\) is the Jacobian of the transformation between \(f_{\alpha}[\phi_{\Lambda};x]\) and \(\Lambda^{\gamma}\left(z\right)\). Therefore, Eq.~\eqref{Functional_C} reads

\begin{equation}
    \label{Functional_C2}
    C\left[\phi\right]=\int\mathcal{D}\Lambda_{\alpha}{\rm det}\left(\mathcal{F}\left[\phi,\Lambda\right]\right)B\left[f\left[\phi_{\Lambda}\right]\right]=\int\mathcal{D}fB\left[f\right]=C,
\end{equation}

\noindent where \(C\) is a constant. Finally, from Eq.~\eqref{Integration_Weighting_Functional} one can find that

\begin{equation}
    \label{FP_Theorem_Result}
    \mathcal{I}=\frac{C\int\mathcal{D}\phi_nG\left[\phi\right]}{\int\mathcal{D}\Lambda_{\alpha}W\left[\Lambda\right]},
\end{equation}
\noindent so \(\mathcal{I}\) does not depend on \(f\) and only depends on \(B\) through the non-physical constant \(C\), completing the proof of the theorem.

Then, instead of \(B\left[f\left[\phi\right]\right]=\delta\left(f\left[\phi\right]\right)=\delta\left(A_{\alpha3}\right)\), a functional that can be easily managed using the Gaussian prescription for integration will be taken. For example:

\begin{equation}
    \label{Functional_B}
    B\left[f\right]=\exp\left\{-\frac{i}{2\xi}\int{\rm d}^4xf_{\alpha}\left(x\right)f_{\alpha}\left(x\right)\right\},
\end{equation}

\noindent with \(\xi\) an arbitrary real number known as the gauge fixing parameter. Furthermore, in order to have Lorentz invariance, the gauge fixing functional can be taken to be 

\begin{equation}
    \label{Gauge_Fixing_Function}
    f_{\alpha}\left(x\right)=\partial_{\mu}A^{\mu}_{\alpha}.
\end{equation}

\noindent This gauge choice is known as the generalized \(\xi\)-gauge and it has a well-defined propagator for the gauge fields.

It is still necessary to know the effect of \({\rm det}\left(F\left[\phi\right]\right)\) with this choice of the gauge fixing functional. Again, for this purpose, it is useful to write the determinant as a Gaussian integral. Then, one can use that, up to a non-physical multiplicative factor, the integrals over Grassmann variables satisfy the relation

\begin{equation}
    \label{Determinant_Grassmann_Variable}
    {\rm det}\left(F\left[\phi\right]\right)\propto\int\mathcal{D}\omega^{\ast}_{\alpha}\mathcal{D}\omega_{\alpha}\exp\left\{i\int{\rm d}^4x{\rm d}^4y\omega^{\ast}_{\alpha}\left(x\right)\omega_{\beta}\left(y\right)F_{\alpha x,\beta y}\left[\phi\right]\right\},
\end{equation}

\noindent where \(\omega_{\alpha}\left(x\right)\) and \(\omega^{\ast}_{\alpha}\left(x\right)\) are independent\footnote{It is important to remark that they are not related by any kind of charge conjugation transformation.} anti-commuting variables known as \textit{ghosts} and \textit{antighosts}, respectively. They are needed to compensate the non-physical degrees of freedom of the vector potential. It is clear that, even though they satisfy Fermi statistics, these new fields are Lorentz scalars. It may seem that this is a violation of the spin-statistics theorem, but in the next section it will be shown that there is no contradiction because these are not physical degrees of freedom, they are just auxiliary variables. However, \textit{their contribution is crucial in the calculation of loop diagrams.}

It is still necessary to compute the matrix \(F_{\alpha x,\beta y}\left[\phi\right]\). For this purpose, consider the definition of \(f_{\alpha}\left[\phi_{\lambda};x\right]\):

\begin{equation}
    \label{Gauge_Fixing_Xi_Gauge}
    f_{\alpha}\left[\phi_{\lambda};x\right]=\partial_{\mu}A^{\mu}_{\lambda\alpha}=\partial_{\mu}A^{\mu}_{\alpha}+\square\lambda_{\alpha}+C_{\alpha\beta\gamma}\partial_{\mu}(A_{\beta}^{\mu}\lambda_{\gamma}).
\end{equation}

\noindent Thus, it is straightforward to prove that

\begin{equation}
    \label{F_XI_Gauge}
    F_{\alpha x,\beta y}\left[\phi\right]=\delta_{\alpha\beta}\square\delta^4\left(x-y\right)-C_{\alpha\beta\gamma}\partial_{\mu}\left(\delta^{4}\left(x-y\right)A_{\gamma}^{\mu}\right),
\end{equation}

\noindent so the determinant can be written as

\begin{equation}
    \label{Determinant_Ghost_Interaction}
    {\rm det}\left(F\left[\phi\right]\right)\propto\int\mathcal{D}\omega^{\ast}_{\alpha}\mathcal{D}\omega_{\alpha}\exp\left\{i\int{\rm d}^4x\left[-\partial_{\mu}\omega_{\alpha}^{\ast}\partial^{\mu}\omega_{\alpha}+C_{\alpha\beta\gamma}\left(\partial_{\mu}\omega^{\ast}_{\alpha}\right)\omega_{\beta}A^{\mu}_{\gamma}\right]\right\}.
\end{equation}

\noindent Therefore, using the De Witt-Faddeev-Popov formalism, the T-ordered products of gauge invariant operators in a Lorentz invariant general \(\xi\)-gauge are calculated by

\begin{equation}
    \label{T_Ordered_Faddeev_Popov}
    \begin{split}
        \left<T\left\{\mathcal{O}_1\mathcal{O}_2\dots\right\}\right>_{\rm Vac}&\propto\int\mathcal{D}A_{\alpha\mu}\mathcal{D}\psi_a\mathcal{D}\omega_{\alpha}^{\ast}\mathcal{D}\omega_{\alpha}\left(\mathcal{O}_1\mathcal{O}_2\dots\right)\exp\left\{i\int{\rm d}^4x\left[\mathcal{L}_M-\frac{1}{4}F_{\alpha\mu\nu}F^{\mu\nu}_{\alpha}\right.\right.\\&\left.\left.-\frac{1}{2\xi}\left(\partial_{\mu}A_{\alpha}^{\mu}\right)\left(\partial_{\nu}A_{\alpha}^{\nu}\right)-\partial_{\mu}\omega_{\alpha}^{\ast}\partial^{\mu}\omega_{\alpha}+C_{\alpha\beta\gamma}\left(\partial_{\mu}\omega^{\ast}_{\alpha}\right)\omega_{\beta}A^{\mu}_{\gamma}\right]\right\}.
    \end{split}
\end{equation}

This can be interpreted as if a new non-gauge invariant Lagrangian with new degrees of freedom (the ghosts and antighosts) is introduced. It is important to remark that the presence of these fields in a theory with \(C_{\alpha\beta\gamma}=0\) is, in general, also necessary if we take another gauge fixing functional \(f_{\alpha}\left[\phi;x\right]\).

\section{BRST Quantization}\label{sec:BRST_Quantization}

\begin{figure}
    \centering
    \includegraphics[scale=0.20]{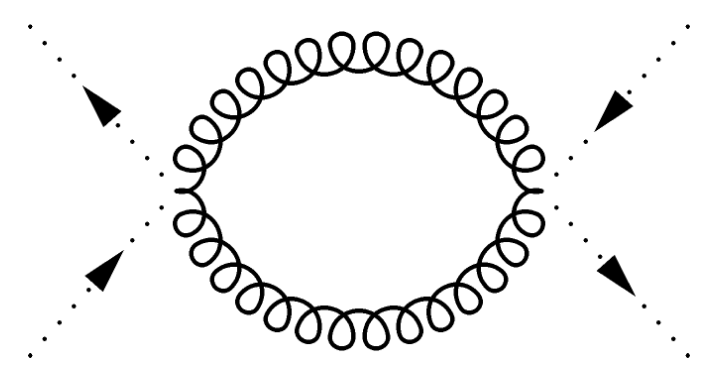} 
    \caption[One loop divergent Feynman diagram associated with \(\omega\omega^{\ast}\rightarrow\omega\omega^{\ast}\) that appears when using the gauge fixing functional \(f_{\alpha}=\partial_{\mu}A_{\alpha}^{\mu}+S_{\alpha\beta\gamma}A_{\beta\mu}A_{\gamma}^{\mu}\).]{One loop divergent Feynman diagram associated with \(\omega\omega^{\ast}\rightarrow\omega\omega^{\ast}\) that appears when using the gauge fixing functional \(f_{\alpha}\left[\phi;x\right]=\partial_{\mu}A_{\alpha}^{\mu}+S_{\alpha\beta\gamma}A_{\beta\mu}A_{\gamma}^{\mu}\).}
    \label{fig:4_Ghost_Amplitude}
\end{figure}

Even though the De Witt-Faddeev-Popov formalism has the advantage of being manifestly Lorentz invariant it still requires the explicit breaking of gauge symmetries. Furthermore, this method \textit{does not lead in general to a renormalizable theory}. It can be shown that the Lagrangian in Eq.~\eqref{T_Ordered_Faddeev_Popov} is renormalizable \cite{PhysRevD.5.3121,PhysRevD.5.3137,HOOFT1972318,PhysRevD.7.1049,PhysRevD.9.933}, which means that any divergence can be reabsorbed by a shifting of its free parameters and field redefinitions. Nonetheless, as shown before, Green's functions have no dependence on \(f_{\alpha}\left[\phi;x\right]\), so the theory should be renormalizable for any choice of the gauge fixing functional. But if instead of~\eqref{Gauge_Fixing_Function} one considers \(f_{\alpha}\left[\phi;x\right]=\partial_{\mu}A_{\alpha}^{\mu}+S_{\alpha\beta\gamma}A_{\beta\mu}A_{\gamma}^{\mu}\), with \(S_{\alpha\beta\gamma}\) a totally symmetric tensor in the group indices,\footnote{Its existence is guaranteed, for instance, in any \({\rm SU}(N)\) theory with \(N\geq3\).} the matrix \(F_{\alpha x,\beta y}\left[\phi\right]\) will contain terms of the form \(\delta^4\left(x-y\right)S_{\alpha\gamma\delta}C_{\gamma\epsilon\beta}A_{\epsilon\mu}A_{\delta}^{\mu}\), which lead to interaction vertices in the Lagrangian with two gauge bosons, one ghost and one antighost: \(S_{\alpha\gamma\delta}C_{\beta\gamma\epsilon}\omega_{\alpha}^{\ast}\omega_{\beta}A_{\delta\mu}A_{\epsilon}^{\mu}\). Then, divergent contributions can be found in one-loop diagrams associated with the amplitude of \(\omega^{\ast}\omega\rightarrow\omega^{\ast}\omega\) \cite{Weinberg2} (see Fig. \ref{fig:4_Ghost_Amplitude}). Such divergences should be canceled by the addition of a vertex with four ghosts in the Lagrangian, something that cannot appear in the De Witt-Faddeev-Popov formalism, which only contains quadratic terms in these fields (Eq.~\eqref{Determinant_Grassmann_Variable}). Therefore, \textit{at the quantum level the classical gauge symmetry must be promoted to something else that leads to a consistent renormalizable theory.} In order to find the quantum analogue of gauge invariance, it will be shown that there is an underlying symmetry in the De Witt-Faddeev-Popov formalism, the \textit{BRST symmetry} \cite{Becchi:1974md,BECCHI1976287,Tyutin:1975qk}.

First of all, consider the calculation of the functional integral

\begin{equation}
    \label{Integral_Nakanishi_Lautrup}
    B\left[f\right]\coloneqq\int\mathcal{D}h_{\alpha}\exp\left\{i\int{\rm d}^4x\left[\frac{\xi}{2}h_{\alpha}h_{\alpha}+f_{\alpha}h_{\alpha}\right]\right\},
\end{equation}

\noindent with \(h_{\alpha}\left(x\right)\) auxiliary commutative variables known as \textit{Nakanishi-Lautrup fields} \cite{10.1143/PTP.35.1111,Lautrup:1967zz}. Using the prescription of Gaussian integration, it is only necessary to evaluate the integrand at its stationary point with respect to \(h_{\alpha}\), given by the condition \(h_{\alpha}=-f_{\alpha}/\xi\):

\begin{equation}
    \label{Calculation_Integral_Nakanishi_Lautrup}
    B\left[f\right]\coloneqq\int\mathcal{D}h_{\alpha}\exp\left\{i\int{\rm d}^4x\left[\frac{\xi}{2}h_{\alpha}h_{\alpha}+f_{\alpha}h_{\alpha}\right]\right\}\propto\exp\left\{-\frac{i}{2\xi}\int{\rm d}^4xf_{\alpha}f_{\alpha}\right\},
\end{equation}

\noindent which is just the \(B\left[f\left[\phi\right]\right]\) functional chosen in the previous section. Thus, defining \(\Delta_{\alpha}\left(x\right)\coloneqq\int{\rm d}^4yF_{\alpha x, \beta y}\left[\phi\right]\omega_{\beta}\left(y\right)\), the T-ordered product of Eq.~\eqref{T_Ordered_Faddeev_Popov} can be written integrating over the Nakanishi-Lautrup fields as

\begin{equation}
    \label{T-Ordered_Nakanishi_Lautrup}
    \begin{split}
        \left<T\left\{\mathcal{O}_1\mathcal{O}_2\dots\right\}\right>_{\rm Vac}&\propto\int\mathcal{D}A_{\alpha\mu}\mathcal{D}\psi_a\mathcal{D}\omega_{\alpha}^{\ast}\mathcal{D}\omega_{\alpha}\mathcal{D}h_{\alpha}\left(\mathcal{O}_1\mathcal{O}_2\dots\right)\\&\times\exp\left\{iI_{\rm New}\left[\psi_a,A_{\alpha}^{\mu},\omega^{\ast}_{\alpha},\omega_{\alpha},h_{\alpha}\right]\right\},
    \end{split}
\end{equation}

\noindent with 

\begin{equation}
    \label{New_Action}
    I_{\rm New}\left[\psi_a,A_{\alpha}^{\mu},\omega^{\ast}_{\alpha},\omega_{\alpha},h_{\alpha}\right]\coloneqq\int{\rm d}^4 x\left\{\mathcal{L}\left[\psi_a,A_{\alpha}^{\mu}\right]+\omega^{\ast}_{\alpha}\Delta_{\alpha}+h_{\alpha}f_{\alpha}+\frac{\xi}{2}h_{\alpha}h_{\alpha}\right\}.
\end{equation}

The crucial point is that the action in Eq.~\eqref{New_Action} is invariant under the so-called infinitesimal BRST fermionic transformation

\begin{equation}
    \label{BRST_psi}
    \delta_{\theta}\psi_a=i\theta\omega_{\alpha}\left(t_{\alpha}\right)_{a}^b\psi_b,
\end{equation}
\vspace{-0.5cm}
\begin{equation}
    \label{BRST_A}
    \delta_{\theta}A_{\alpha}^{\mu}=\theta\left[\partial^{\mu}\omega_{\alpha}+C_{\alpha\beta\gamma}A_{\beta}^{\mu}\omega_{\gamma}\right],
\end{equation}
\vspace{-0.5cm}
\begin{equation}
    \label{BRST_Omega}
    \delta_{\theta}\omega_{\alpha}=-\frac{1}{2}\theta C_{\alpha\beta\gamma}\omega_{\beta}\omega_{\gamma},
\end{equation}
\vspace{-0.5cm}
\begin{equation}
    \label{BRST_Omega_Ast}
    \delta_{\theta}\omega_{\alpha}^{\ast}=-\theta h_{\alpha},
\end{equation}
\vspace{-0.5cm}
\begin{equation}
    \label{BRST_h}
    \delta_{\theta}h_{\alpha}=0,
\end{equation}

\noindent with \(\theta\) being an anticommuting constant infinitesimal parameter. This means that, defining \(\delta_{\theta}\chi_n\eqqcolon\theta s\chi_n\), where \(\chi_n\) represents any of the fields of the new action, then \(s\chi_n\) has the opposite statistics to \(\chi_n\) except for \(h_{\alpha}\), which is invariant. It is also important to remark that this transformation acts non-linearly on the fields. This fact will play a crucial role in the study of the symmetry at the quantum level. Furthermore, defining a new charge called \textit{ghost number} \({\rm gh}(\cdot)\) in such a way that \({\rm gh}\left(\omega_{\alpha}\right)=1\), \({\rm gh}\left(\omega^{\ast}_{\alpha}\right)=-1\) and \({\rm gh}\left(\psi_a\right)={\rm gh}\left(A_{\alpha}^{\mu}\right)={\rm gh}\left(h_{\alpha}\right)=0\), it is clear that \({\rm gh}\left(s\chi\right)={\rm gh}\left(\chi\right)+1\) except for the Nakanishi-Lautrup fields.

One of the most important features of the BRST transformation is that it is \textit{nilpotent}, which means that \(\delta_{\theta}\left(\delta_{\theta}F\right)=0\) for any functional \(F\left[\chi_n\right]\). The proof relies on a direct calculation using Eqs.~\eqref{BRST_psi} to~\eqref{BRST_h} that shows that \(\delta_{\theta}\left(s\chi_n\right)=0\) for any \(\chi_n\) \cite{Weinberg2}. Then, one can compute the BRST transformation of the product of two fields \(\chi_1\) and \(\chi_2\):

\begin{equation}
    \label{BRST_Two_Fields}
    \begin{split}
        \delta_{\theta}\left(\chi_1\chi_2\right)&=\left(\delta_{\theta}\chi_1\right)\chi_2+\chi_1\left(\delta_{\theta}\chi_2\right)=\theta\left(s\chi_1\right)\chi_2+\chi_1\theta\left(s\chi_2\right)\\&=\theta\left[\left(s\chi_1\right)\chi_2\pm\chi_1\left(s\chi_2\right)\right]\eqqcolon \theta s\left(\chi_1\chi_2\right),
    \end{split}
\end{equation}

\noindent where the plus sign is for the case where \(\chi_1\) is bosonic and the minus sign for the case where it is fermionic. Using the fact that \(\delta_{\theta}\left(s\chi_{1,2}\right)=0\), the following relation is obtained:

\begin{equation}
    \label{2BRST_Two_Fields}
    \delta_{\theta}\left(s\chi_1\chi_2\right)=\left(s\chi_1\right)\theta\left(s\chi_2\right)\pm\theta\left(s\chi_1\right)\left(s\chi_2\right)=\theta\left[\mp\left(s\chi_1\right)\left(s\chi_2\right)\pm\left(s\chi_1\right)\left(s\chi_2\right)\right]=0,
\end{equation}

\noindent because the statistics of \(s\chi_n\) is opposite to the statistics of \(\chi_n\). Thus, by the principle of mathematical induction it is clear that \(s^2\) acting on a product of any number of \(\chi_n\) vanishes. As a consequence, \(s^2F\left[\chi_n\right]=0\) for any functional, so one can simply write the nilpotency condition \(s^2=0\).

Now, it will be shown that the action in Eq.~\eqref{New_Action} is BRST invariant. For this purpose, note that, when acting on the physical fields \(\psi_a\) and \(A_{\alpha}^{\mu}\), the BRST transformation is just a gauge transformation with parameter \(\varepsilon_{\alpha}\left(x\right)=\theta\omega_{\alpha}\left(x\right)\). Thus, from the fact that \(\mathcal{L}\left[\psi_a,A_{\alpha}^{\mu}\right]\) is gauge invariant it follows that it is also BRST invariant. The BRST transformation of the gauge fixing functional is just

\begin{equation}
    \label{BRST_Gauge_Fixing}
    \delta_{\theta}f_{\alpha}\left[\phi;x\right]=\int{\rm d}^4 y \left.\frac{\delta f_{\alpha}\left[\phi_{\lambda};x\right]}{\delta \lambda_{\beta}\left(y\right)}\right|_{\lambda=0}\theta\omega_{\beta}\left(y\right)=\theta\int{\rm d}^4 y F_{\alpha x,\beta y}\left[\phi\right]\omega_{\beta}\left(y\right)=\theta\Delta_{\alpha}\left[\phi;x\right].
\end{equation}

\noindent Therefore, using Eqs.~\eqref{BRST_Omega_Ast} and~\eqref{BRST_h} it is straightforward to see that 

\begin{equation}
    \label{BRST_New_Lagrangian}
    -s\left(\omega_{\alpha}^{\ast}f_{\alpha}+\frac{\xi}{2}\omega_{\alpha}^{\ast}h_{\alpha}\right)=\omega^{\ast}_{\alpha}\Delta_{\alpha}+h_{\alpha}f_{\alpha}+\frac{\xi}{2}h_{\alpha}h_{\alpha},
\end{equation}

\noindent which is precisely the second part of the new action, so it can be written in terms of a new operator \(\Psi\) as  \(I_{\rm New}\left[\chi_n\right]=\int{\rm d}^4 x\mathcal{L}\left[\phi_m\right]+s\Psi\left[\chi_n\right]\) and, as a consequence of the nilpotency of \(s\), \(sI_{\rm New}=0\). Thus, an underlying symmetry of the theory associated with gauge invariance has been found. In the De Witt-Faddeev-Popov formalism gauge invariance is related to the fact that Green's functions should not depend on the choice of the gauge fixing functional. Furthermore, a change in the gauge fixing functional only leads to a change in the functional \(\Psi\). Therefore, requiring gauge invariance implies that the theory should be invariant under any change \(\tilde{\delta}\Psi\) restricted to the condition of \(s\Psi\) being quadratic in the ghost fields. But this condition is precisely what makes the theory inconsistent at the quantum level since, in general, it is not renormalizable. Consequently, the requirement of gauge invariance will be generalized to BRST invariance in order to avoid this problem.

In the BRST formalism the tree-level action is given by

\begin{equation}
    \label{Action_BRST}
    I_{\rm BRST}\left[\phi_n,\omega_{\alpha},\omega_{\alpha}^{\ast},h_{\alpha}\right]=I_0\left[\phi_n\right]+s\Psi\left[\phi_n,\omega_{\alpha},\omega_{\alpha}^{\ast},h_{\alpha}\right],
\end{equation}

\noindent where \(I_0\left[\phi_n\right]\) is a BRST invariant bosonic functional with ghost number zero of the matter and gauge fields and \(\Psi\) \textit{is any fermionic functional}\footnote{Now, it is no longer assumed that \(\Psi\) is linear in the ghost fields, terms of the form \(\omega^{\ast}\omega^{\ast}\omega\) can be considered. They lead to the desired quartic ghost interactions in the action.} \textit{of ghost number} \(-1\). Furthermore, it is assumed that two theories that only differ in \(\Psi\) are completely equivalent. Since the actions are BRST invariant then \(I_1-I_2\) is a closed form \(s\left(I_1-I_2\right)=0\), so it is only necessary to know if it is also exact \(I_1-I_2=s\Psi\). Therefore, the study of whether two actions represent the same physical situation is the study of the cohomology of the nilpotent BRST operator \(s\). \textit{In order to prove that a theory is renormalizable one only has to show that any divergence is associated with a BRST-exact operator.}


Now, the implications of the invariance under any change \(\tilde{\delta}\Psi\) at the quantum level will be shown. The matrix elements of physical states should be unaffected under this transformation:

\begin{equation}
    \label{BRST_Matrix_Elements}
    \tilde{\delta}\left<a|b\right>=i\left<a\right|\tilde{\delta}I_{\rm New}\left|b\right>=i\left<a\right|s\tilde{\delta}\Psi\left|b\right>=0.
\end{equation}

\noindent Associated with the BRST transformation one can define a fermionic charge operator \(Q\) such that for any operator \(\Phi\)

\begin{equation}
    \label{Fermionic_Charge}
    \delta_{\theta}\Phi=i\left[\theta Q,\Phi\right]=i\theta\left[Q,\Phi\right]_{\mp},
\end{equation}

\noindent so \(s\Phi=i\left[Q,\Phi\right]_{\mp}\) where \(\left[\cdot,\cdot\right]_{-}\) represents the commutator, \(\left[\cdot,\cdot\right]_{+}\) represents the anticommutator, the minus sign is taken when \(\Phi\) is bosonic and the plus sign when it is fermionic. The nilpotency condition can be written as \(0=-ss\Phi=\left[Q,\left[Q,\Phi\right]_{\mp}\right]_{\pm}=\left[Q^2,\Phi\right]_{-}\). This requirement can only be satisfied if \(Q^2\) vanishes or it is proportional to the identity, but since \(Q^2\) raises by two units the ghost number it cannot be equal to the unit operator so \(Q^2=0\), i.e., \(Q\) is nilpotent. Furthermore, Eq.~\eqref{BRST_Matrix_Elements} can also be written as

\begin{equation}
    \label{BRST_Matrix_Elements2}
    \left<a\right|[Q,\tilde{\delta}\Psi]_{+}\left|b\right>=0.
\end{equation}

In order for this condition to hold for any change of \(\Psi\), the action of the operator \(Q\) on physical states must satisfy

\begin{equation}
    \label{BRST_Physical_States}
    \left<a\right|Q=Q\left|b\right>=0.
\end{equation}

\noindent It has been shown that for any gauge theory \textit{the space of physical states contains no ghosts or antighosts} and that all of the physical states have a positive definite norm and lead to well-defined probabilities \cite{Curci:1976yb,10.1143/PTP.60.1869,10.1143/PTPS.66.1}. Therefore, as previously announced, \textit{there is no problem with the fact that these auxiliary fields do not satisfy the spin-statistics theorem.} Finally, consider the state \(\left|\tilde{a}\right>=\left|a\right>+Q\left|\alpha\right>\) with arbitrary \(\left|\alpha\right>\) (physical or not). From \(Q^2=0\) it is clear that \(\left|\tilde{a}\right>\) satisfies the condition in Eq.~\eqref{BRST_Physical_States}, so it is a physical state. Moreover, computing the product with any other physical state \(\left|b\right>\) one obtains \(\left<b|\tilde{a}\right>=\left<b|a\right>+\left<b\right|Q\left|\alpha\right>=\left<b|a\right>\), so all of the matrix elements remain invariant. Consequently, both \(\left|a\right>\) and \(\left|\tilde{a}\right>\) represent the same physical state. Thus, the question of whether two physical quantum states are equivalent or not is solved studying the cohomology of the operator \(Q\).

\section{The Batalin-Vilkovisky Formalism}\label{sec:Batalin_Vilkovisky}

The BRST formalism presented in the previous section gives a consistent way to quantize gauge theories \cite{Becchi:1974md}. Nonetheless, another procedure can be introduced which is even more general because it allows for the quantization of a broader class of theories including the ones that contain symmetries with an associated open algebra like, for instance, supergravity. It is known as the Batalin-Vilkovisky formalism \cite{Fradkin:1975cq,Fradkin:1977hw,BATALIN1977309,Fradkin:1977xi,BATALIN198127,Batalin:1984ss,Batalin:1985qj} and, even in the context of this thesis where only Yang-Mills theories will be considered, it is particularly useful since \textit{it makes the study of renormalizability in Effective Field Theories (EFTs) and the derivation of the consistency conditions of quantum anomalies straightforward}. Furthermore, the tools required for the description of this method are the same that are needed in the study of the BRST symmetry at the quantum level.

It is appropriate to begin with the introduction of a new set of fields known as \textit{antifields}. They are denoted by \(\chi_n^{\ddagger}\) and there is one of them for each field \(\chi_{n}\) of the theory (note that we are including the ghosts and the Nakanishi-Lautrup fields) is needed. The classical action is defined as:

\begin{equation}
    \label{Classical_Action_Antifields}
    S[\chi_n,\chi^{\ddagger}_n]\coloneqq I_0\left[\phi_n\right]+\left(s\chi_n\right)\chi_n^{\ddagger},
\end{equation}

\noindent with the \(I_0\left[\phi_n\right]\) presented in Eq.~\eqref{Action_BRST} and \(n\) representing also the spacetime indices over which one integrates to define the action. Therefore, since the action must be a bosonic functional of ghost number zero, then statistics of the antifields has to be the same as the one of the BRST transformation of the fields (so \(\chi_n^{\ddagger}\) has the opposite statistics to \(\chi_n\)) and \({\rm gh}(\chi_n^{\ddagger})=-{\rm gh}\left(\chi_n\right)-1\). A simple calculation gives:

\begin{equation}
    \label{Calculation_Master_Equation}
    \begin{split}
        \frac{\delta_R S}{\delta\chi_n^{\ddagger}}\frac{\delta_LS}{\delta\chi_n} &=\left(s\chi_n\right)\left[\frac{\delta_L I_0}{\delta\chi_n}+\frac{\delta_L\left(s\chi_m\right)}{\delta\chi_n}\chi_m^{\ddagger}\right]=\left(s\phi_r\right)\frac{\delta_L I_0}{\delta\phi_r}+\left(s\chi_n\right)\frac{\delta_L\left(s\chi_m\right)}{\delta\chi_n}\chi_m^{\ddagger}\\&=sI_0+\left(s^2\chi_m\right)\chi_m^{\ddagger},
    \end{split}
\end{equation}

\noindent where \(\delta_{L,R}\) denotes the differentiation from left- and right-hand side, respectively. Using that the action \(I_0\) is BRST invariant and the BRST operator \(s\) is nilpotent the following expression is obtained:

\begin{equation}
    \label{Master_Equation}
     \frac{\delta_R S}{\delta\chi_n^{\ddagger}}\frac{\delta_LS}{\delta\chi_n}=0.
\end{equation}

This is the so-called \textit{master equation} and it is the fundamental formula of the Batalin-Vilkovisky formalism. In this procedure, Eq.~\eqref{Master_Equation} is read in the opposite way, assuming that the master equation holds, the \(I_0\) has to be BRST invariant and the BRST transformation must be nilpotent, because both the terms with no antifields and with one antifield must be canceled independently. In order to find more connections between the two formalisms, one can consider the following particular value for the antifields:

\begin{equation}
    \label{Antifields_Functional_Psi}
    \chi_n^{\ddagger}=\frac{\delta\Psi\left[\chi_n\right]}{\delta \chi_n},
\end{equation}

\noindent where \(\Psi\left[\chi_n\right]\) is a functional of ghost number \(-1\), which leads to the  needed condition \({\rm gh}(\chi_n^{\ddagger})=-{\rm gh}\left(\chi_n\right)-1\). Moreover, the assumption on the statistics of the antifields implies that it has to be fermionic, because its derivative with respect to a boson will be a fermion and its derivative with respect to a fermion will be a boson. For this reason, it is not necessary to distinguish between left- and right-differentiation. Thus, Eq.~\eqref{Classical_Action_Antifields} reads

\begin{equation}
    \label{Antifields_Acion_BRST_Action}
    S[\chi_n]\coloneqq I_0\left[\phi_n\right]+\left(s\chi_n\right)\frac{\delta\Psi\left[\chi_n\right]}{\delta \chi_n}=I_0\left[\phi_n\right]+s\Psi\left[\chi_n\right]=I_{\rm BRST}\left[\chi_n\right].
\end{equation}

Nevertheless, allowing the classical action to be a general non-linear functional of the antifields, one can derive from the master equation that \(I_0\left[\phi\right]\) (the part of the action expansion with only matter and gauge fields) will be invariant under a transformation with an associated commutation relation that might be closed\footnote{This means that the commutator of two transformations produces a transformation of the same class.} only if the equations of motion \(\delta I_0/\delta\phi=0\) are satisfied. Indeed, the associated generalized Jacobi identity can also be found. This is a more general class of symmetries than the ones that can be described by the BRST formalism and it is precisely the situation in supergravity models \cite{PhysRevD.13.3214,Kallosh:1978de,DEWIT1978389,VANNIEUWENHUIZEN1981189}. Nonetheless, only theories associated with a gauge group with a closed algebra will be studied here, so the actions will always have the form of Eq.~\eqref{Classical_Action_Antifields}, although most part of the following statements also apply to the general case.

Due to the fact that \(S\) is a bosonic functional and \(\chi_n\) and \(\chi_n^{\ddagger}\)
 have the opposite statistics, either \(\delta_{R}S/\delta\chi_n^{\ddagger}\) or \(\delta_{L}S/\delta\chi_n\) is also bosonic and left- and right-differentiation can be exchanged up to a minus sign:

 \begin{equation}
     \label{Master_Equation_2}
     \frac{\delta_{R}S}{\delta\chi_n^{\ddagger}}\frac{\delta_{L}S}{\delta\chi_n}=\frac{\delta_{L}S}{\delta\chi_n}\frac{\delta_{R}S}{\delta\chi_n^{\ddagger}}=-\frac{\delta_{R}S}{\delta\chi_n}\frac{\delta_{L}S}{\delta\chi_n^{\ddagger}}=-\frac{1}{2}\left(\frac{\delta_{R}S}{\delta\chi_n}\frac{\delta_{L}S}{\delta\chi_n^{\ddagger}}-\frac{\delta_{R}S}{\delta\chi_n^{\ddagger}}\frac{\delta_{L}S}{\delta\chi_n}\right).
 \end{equation}

\noindent Thus, defining the so-called \textit{antibracket} \(\left(\cdot,\cdot\right)\) of two functionals \(F\) and \(G\) as

\begin{equation}
    \label{Definition_Antibracket}
    \left(F,G\right)\coloneqq\frac{\delta_RF}{\delta\chi_n}\frac{\delta_LG}{\delta\chi_n^{\ddagger}}-\frac{\delta_RF}{\delta\chi_n^{\ddagger}}\frac{\delta_LG}{\delta\chi_n},
\end{equation}

\noindent then the master equation can be expressed as

\begin{equation}
    \label{Master_Equation_Antibracket}
    \left(S,S\right)=0.
\end{equation}

It is clear that this antibracket is some sort of generalization of the Poisson bracket of classical field theory. From its definition in Eq.~\eqref{Definition_Antibracket}, the following three properties can be derived \cite{Weinberg2}:

\begin{itemize}
    \item[1.] \(\left(F,G\right)=\pm\left(G,F\right)\) where the plus sign is taken if both \(F\) and \(G\) are bosonic and the minus sign otherwise.
    \item[2.] \(\left(F,GH\right)=\left(F,G\right)H\pm G\left(F,H\right)\) where the minus sign is taken if \(F\) is bosonic and \(G\) is fermionic and the plus sign otherwise.
    \item[3.] \(s_{123}\left(F_1,\left(F_2,F_3\right)\right)+s_{312}\left(F_3,\left(F_1,F_2\right)\right)+s_{231}\left(F_2,\left(F_3,F_1\right)\right)=0\) where \(s_{ijk}=-1\) if \(F_i\) and \(F_{k}\) are bosonic and \(s_{ijk}=1\) otherwise.
\end{itemize}

 Using these properties it will be shown that the transformation defined by

\begin{equation}
    \label{Generalized_BRST_Fields}
    \hat{\delta}_{\theta}\chi_n\coloneqq\theta\frac{\delta_R S}{\delta \chi_n^{\ddagger}}=-\theta\left(S,\chi_n\right),
\end{equation}

\begin{equation}
    \label{Generalized_BRST_AntiFields}
    \hat{\delta}_{\theta}\chi_n^{\ddagger}\coloneqq-\theta\frac{\delta_R S}{\delta \chi_n}=-\theta(S,\chi_n^{\ddagger}),
\end{equation}

\noindent with \(\theta\) an infinitesimal fermionic constant parameter, acts as a \textit{generalized BRST transformation}. First of all, it is necessary to prove how a general functional gets transformed. This will be done by mathematical induction. Assuming that functionals that only contain products up to a certain number of fields and antifields satisfy \(\hat{\delta}_{\theta}F=-\theta\left(S,F\right)\) (the base case is the transformation of the fields and antifields themselves in Eqs.~\eqref{Generalized_BRST_Fields} and~\eqref{Generalized_BRST_AntiFields}), the following relation is obtained: 

\begin{equation}
    \label{Generalized_BRST_Product}
    \hat{\delta}_{\theta}\left(FG\right)=-\theta\left(S,F\right)G-F\theta\left(S,G\right)=-\theta\left[\left(S,F\right)G\pm F\left(S,G\right)\right],
\end{equation}

\noindent where the minus sign corresponds to the case where \(F\) is fermionic and the plus sign to the case where it is bosonic. Therefore, using the relation for the antibracket of a product of functionals one gets:

 \begin{equation}
    \label{Generalized_BRST_Product_Result}
    \hat{\delta}_{\theta}\left(FG\right)=-\theta\left(S,FG\right),
\end{equation}

\noindent so for any functional that can be written as the sum of products of fields and antifields \(\hat{\delta}_{\theta}F=-\theta\left(S,F\right)\). Consequently, using Eq.~\eqref{Master_Equation_Antibracket} it is clear that the action is invariant under the generalized BRST transformation \(\hat{\delta}_{\theta}S=-\theta\left(S,S\right)=0\). From the generalized Jacobi identity of the antibracket, the nilpotency condition can be derived:

\begin{equation}
    \label{Generalized_Jacobi_SS}
    0=\mp\left(S,\left(S,F\right)\right)\mp\left(F,\left(S,S\right)\right)-\left(S,\left(F,S\right)\right),
\end{equation}

\noindent where the upper sign corresponds to the case where \(F\) is bosonic and the lower sign to the case where it is fermionic. Under the same circumstances \(\left(F,S\right)=\pm\left(S,F\right)\) so

\begin{equation}
    \label{Generalized_Jacobi_SS2}
    0=\mp2\left(S,\left(S,F\right)\right)\mp\left(F,\left(S,S\right)\right).
\end{equation}

\noindent Using the master equation \(\left(S,S\right)=0\) one finally obtains that \textit{the generalized BRST transformation is nilpotent}:

\begin{equation}
    \label{Nilpotent_Generalized_BRST}
    \left(S,\left(S,F\right)\right)=0.
\end{equation}

The discussion here is so-far completely general, but now assume that \(S\) takes the form of Eq.~\eqref{Classical_Action_Antifields} and \(F\) does not contain antifields. In this case

\begin{equation}
    \label{Generalized_BRST_Original_BRST}
     \hat{\delta}_{\theta}F=-\theta\left(S,F\right)=\theta\frac{\delta_RS}{\delta\chi_n^{\ddagger}}\frac{\delta_LF}{\delta\chi_n}=\theta\left(s\chi_n\right)\frac{\delta_LF}{\delta\chi_n}=\theta sF=\delta_{\theta}F,
\end{equation}

\noindent so the generalized BRST transformation is reduced to the original one. Then, from the properties of the antibracket, the proof of the fact that the original BRST transformation is nilpotent is quite simple. This will be a general situation, using the Batalin-Vilkovisky procedure some of the otherwise cumbersome derivations will become straightforward.

Moving again to the general scenario, the connection between the cohomological arguments of the previous section and this quantization procedure becomes clear once one realizes that, if \(S\) is a solution of the master equation, so it is \(\tilde{S}\coloneqq S+\left(\delta F,S\right)\), with \(\delta F\) an infinitesimal functional:

\begin{equation}
    \label{General_Solution_Master_Equation}
    \begin{split}
        (\tilde{S},\tilde{S})&=\left(S,S\right)+\left(S,\left(\delta F,S\right)\right)+\left(\left(\delta F,S\right),S\right)=0,
    \end{split}
\end{equation}

\noindent where the nilpotency condition has been used. This \(\delta F\) must be fermionic and have ghost number \(-1\) in order for \(\tilde{S}\) to be a bosonic functional of ghost number zero, so it is the generalization of the \(\tilde{\delta}\Psi\) in the BRST formalism. Therefore, it is assumed that \textit{if two actions only differ in a term of the form} \(\left(\delta F,S\right)\) \textit{they are equivalent.}  

Consider the particular value of Eq.~\eqref{Antifields_Functional_Psi} for the antifields. In that case, it is possible to define

\begin{equation}
    \label{General_Action_Psi}
    I_{\Psi}\left[\chi_n\right]\coloneqq S\left[\chi_n,\frac{\delta\Psi\left[\chi_n\right]}{\delta\chi_n}\right].
\end{equation}

\noindent In order to see how the ambiguity \(\tilde{S}=S+\left(\delta F,S\right)\) in the solutions of the master equation is realized in this scenario, the particular case \(\delta F[\chi_n,\chi_n^{\ddagger}]=\varepsilon \tilde{\Psi}\left[\chi_n\right]\) (with \(\varepsilon\) a commuting infinitesimal parameter) is studied:

\begin{equation}
    \label{Ambiguity_Action}
    \tilde{I}_{\Psi}\left[\chi_n\right]=S\left[\chi_n,\frac{\delta\Psi\left[\chi_n\right]}{\delta\chi_n}\right]+\varepsilon\frac{\delta_R\tilde{\Psi}}{\delta\chi_n}\frac{\delta_L S}{\delta\chi_n^{\ddagger}}=S\left[\chi_n,\frac{\delta}{\delta\chi_n}\left(\Psi+\varepsilon\tilde{\Psi}\right)\right].
\end{equation}

\noindent This means that physical quantities are invariant under \(\Psi\) redefinitions,\footnote{For the moment, only infinitesimal transformations have been considered. The generalization to finite \(\Psi\) redefinitions is presented in Sec. \ref{sec:Renormalization_Gauge_Theories}.} which is precisely the situation in the BRST formalism.

Finally, the realization of this ambiguity at the quantum level is shown. The assumption is that  there is no meaning in the particular choice of \(\Psi\) for the functional 

\begin{equation}
    \label{Partition_Function_Psi}
    Z_{\Psi}\coloneqq\int\mathcal{D}\chi_n\exp\left\{iI_{\Psi}\right\}.
\end{equation}

\noindent Thus, considering a shift \(\delta \Psi\left[\chi_n\right]\) in \(\Psi\left[\chi_n\right]\), the associated shift in \(Z_{\Psi}\) is given by

\begin{equation}
    \label{Shift_Z}
    \delta Z_{\Psi}=i\int\mathcal{D}\chi_n\exp\left\{iI_{\Psi}\right\}\left.\frac{\delta_RS\left[\chi,\chi^{\ddagger}\right]}{\delta\chi_n^{\ddagger}}\right|_{\chi^{\ddagger}=\delta\Psi/\delta\chi}\frac{\delta\left(\delta\Psi\right)}{\delta\chi_n}.
\end{equation}

\noindent After integration by parts this becomes

\begin{equation}
    \label{Shift_Z2}
    \begin{split}
        \delta Z_{\Psi}&=-i\int\mathcal{D}\chi_n\frac{\delta_L}{\delta\chi_n}\left\{\exp\left\{iI_{\Psi}\right\}\left.\frac{\delta_RS\left[\chi,\chi^{\ddagger}\right]}{\delta\chi_n^{\ddagger}}\right|_{\chi^{\ddagger}=\delta\Psi/\delta\chi}\right\}\delta\Psi\\&=\int\mathcal{D}\chi_n\exp\left\{iI_{\Psi}\right\}\left\{\frac{\delta_LI_{\Psi}}{\delta\chi_n}\frac{\delta_RS}{\delta\chi_n^{\ddagger}}-i\frac{\delta_L}{\delta\chi_n}\frac{\delta_R}{\delta\chi_n^{\ddagger}}S\right\}_{\chi^{\ddagger}=\delta\Psi/\delta\chi}\delta\Psi.
    \end{split}
\end{equation}

\noindent Defining the operator \(\Delta\coloneqq\dfrac{\delta_R}{\delta\chi_n}\dfrac{\delta_{L}}{\delta\chi_n^{\ddagger}}\) and using Eq.~\eqref{Master_Equation_2} the previous equation reads:

\begin{equation}
    \label{Shift_Z3}
    \delta Z_{\Psi}=-\frac{1}{2}\int\mathcal{D}\chi_n\exp\left\{iI_{\Psi}\right\}\left\{\left(S,S\right)-2i\Delta S\right\}_{\chi^{\ddagger}=\delta\Psi/\delta\chi}\delta\Psi,
\end{equation}

\noindent where the substitution \(\chi^{\ddagger}_n=\delta\Psi/\delta\chi_n\) is made before the calculation of \(\delta_{L,R}/\delta \chi_n\). Therefore, since the ambiguity  \(\tilde{S}\coloneqq S+\left(\delta F,S\right)\) at the quantum level requires that for all \(\Psi\) the quantity \(\delta Z_{\Psi}\) must vanish, then

\begin{equation}
    \label{Quantum_Master_Equation}
    \left(S,S\right)-2i\hbar\Delta S=0
\end{equation}

\noindent at \(\chi^{\ddagger}_n=\delta \Psi/\delta\chi_n\). The \(\hbar\) has been introduced at the end of the calculation just to show that the second term corresponds to loop contributions. \textit{This is the so-called quantum master equation and it is the consistency condition for the symmetry to hold at the quantum level.} At the classical level the action is simply assumed to satisfy \(\left(S,S\right)=0\), but in order to preserve the associated symmetry at the quantum level also \(\Delta S=0\) is needed, which is a non-trivial requirement. \textit{The presence of a non-vanishing} \(\Delta S\) \textit{means that the symmetry could be anomalous.}

\markboth{1. GAUGE THEORIES. RENORMALIZABILITY AND CONSISTENCY}{1.6. THE QEA AND THE ZINN-JUSTIN EQUATION}
\section{Symmetries at the Quantum Level. The Quantum Effective Action and the Zinn-Justin Equation}
\markboth{1. GAUGE THEORIES. RENORMALIZABILITY AND CONSISTENCY}{1.6. THE QEA AND THE ZINN-JUSTIN EQUATION}
In this section the implications of the presence of classical symmetries at the quantum level will be studied, with particular focus on the BRST symmetry. For this purpose, it is useful to introduce the \textit{Quantum Effective Action} (QEA), an object that can be interpreted as \textit{an action that contains all of the information about quantum corrections}. In order to give its definition, it is necessary to begin with the \textit{partition function}

\begin{equation}
    \label{Partition_Function}
    Z\left[J\right]\coloneqq\int\mathcal{D}\chi_n\exp\left\{iI\left[\chi\right]+i\int{\rm d}^4x \chi_nJ_n\right\},
\end{equation}

\noindent where the \(J_n\) are a set of classical external sources and \(I\left[\chi_n\right]\) will be identified with the BRST invariant action in Eq.~\eqref{Action_BRST}. It is well known \cite{Weinberg2} that the partition function allows for the calculation of the vacuum expectation values from the relation

\begin{equation}
    \label{VEV_Partition_Function}
    \left[\frac{-i\delta_R}{\delta J_{r_1}\left(y_1\right)}\cdots\frac{-i\delta_R}{\delta J_{r_s}\left(y_s\right)}Z\left[J\right]\right]_{J=0}=\int\mathcal{D}\chi_n\left[\chi_{r_1}\left(y_1\right)\cdots\chi_{r_s}\left(y_{r_s}\right)\right]\exp\left\{iI\left[\chi_n\right]\right\}.
\end{equation}

From the point of view of perturbation theory, \(Z\left[J\right]\) is computed as the sum of all vacuum-vacuum amplitudes with the external classical sources \(J_n\). It is also appropriate to define \(iW\left[J\right]\) as the addition of all connected contributions, so \(\left(iW\left[J\right]\right)^N\) contains all the amplitudes with \(N\) disconnected components. It can be shown \cite{Weinberg2} that, in order to get the correct symmetry factors for the calculation of \(Z\left[J\right]\), one needs to divide over \(N!\):

\begin{equation}
    \label{Connected_Disconnected}
    Z\left[J\right]=\sum_{N=0}^{\infty}\frac{1}{N!}\left(iW\left[J\right]\right)^N=\exp\left\{iW\left[J\right]\right\}.
\end{equation}

It is useful to consider the \textit{Legendre transformation} of the functional \(W\left[J\right]\), for which the so-called classical fields are necessary:

\begin{equation}
    \label{Classical_Field}
    X^{J}_n\left(x\right)\coloneqq\frac{\delta_R W\left[J\right]}{\delta J_n\left(x\right)}=-i\frac{1}{Z\left[J\right]}\frac{\delta_R Z\left[J\right]}{\delta J_n\left(x\right)}.
\end{equation}

\noindent They are proportional to the vacuum expectation value of the associated quantum fields in the presence of the external currents. Assuming that the previous expression can be inverted, one can define \(X_n\) as the particular value of \(X_n^J\) that is obtained if one takes the classical current \(J_n\left(x\right)\) to be equal to a certain function \(J_{n}^{X}\left(x\right)\). Therefore, the Legendre transformation of \(W\left[J\right]\) is given by

\begin{equation}
    \label{Quantum_Effective_Action}
    \Gamma\left[X\right]\coloneqq-\int{\rm d}^4xX_n\left(x\right)J^{X}_{n}\left(x\right)+W\left[J_{X}\right].
\end{equation}

\noindent This \(\Gamma\left[X\right]\) is also known as the Quantum Effective Action \cite{DeWitt:1964oba,Jona-Lasinio:1964zvf}. It satisfies the properties of the usual Legendre transformations:

\begin{equation}
    \label{Inverse_Legendre}
    \begin{split}
        \frac{\delta_L \Gamma\left[X\right]}{\delta X_m\left(y\right)}&=-       J_m^X\left(y\right)\mp\int{\rm d}^4xX_n\left(x\right)\frac{\delta_L J^{X}_{n}\left(x\right)}{\delta X_m\left(y\right)}\pm\int{\rm d}^4x \left.\frac{\delta_R W\left[J\right]}{\delta J_n\left(x\right)}\right|_{J=J_X}\frac{\delta_L J^X_n\left(x\right)}{\delta X_m\left(y\right)}\\&=-J_m^X\left(y\right),
    \end{split}
\end{equation}

\noindent where the upper sign corresponds to the case where \(X_m\left(y\right)\) is bosonic and the lower sign to the case where it is fermionic. In the absence of external currents Eq.~\eqref{Inverse_Legendre} becomes \(\delta_L\Gamma\left[X\right]/\delta X=0\), which means that the possible values of the classical fields lead to a stationary point in the QEA.

The main property of the QEA is the fact that it is an action that encodes all of the quantum effects of the theory. The proof of this statement begins with the definition of the quantity

\begin{equation}
    \label{W_Constant_g}
    \exp\left\{iW_{\Gamma}\left[J,g\right]\right\}\coloneqq \int\mathcal{D}\chi_n\exp\left\{ig^{-1}\left[\Gamma\left[\chi\right]+\int{\rm d}^4x\chi_n\left(x\right)J_n\left(x\right)\right]\right\},
\end{equation}

\noindent where \(g\) is an arbitrary constant. The associated Feynman rules have a propagator which is proportional to \(g\) and vertices whose couplings are proportional to \(g^{-1}\). Consequently, a Feynman diagram with \(V\) vertices and \(I\) internal lines has an associated amplitude proportional to \(g^{I-V}\). Using Euler's formula, \(I-V\) can be related to the number of loops \(L\) of the graph: \(L=I-V+1\). Thus, a diagram with \(L\) loops is proportional to \(g^{L-1}\), so denoting the \(L\)-loop contributions to \(W_{\Gamma}\left[J,g\right]\) as \(W_{\Gamma}^{\left(L\right)}\left[J\right]\), one can write

\begin{equation}
    \label{Loop_Decomposition_W_Gamma}
    W_{\Gamma}\left[J,g\right]=\sum_{L=0}^{\infty}g^{L-1}W_{\Gamma}^{\left(L\right)}\left[J\right].
\end{equation}

\noindent Now, it is necessary to pay attention on the tree-level contribution of \(W_{\Gamma}\left[J,g\right]\). For this purpose, the limit \(g\to0\) can be taken, since it will make the \(L=0\) term in Eq.~\eqref{Loop_Decomposition_W_Gamma} the dominant one. In this case, Eq.~\eqref{W_Constant_g} becomes

\begin{equation}
    \label{W_Constant_G_Infinity}
    \lim_{g\to0}\exp\left\{iW_{\Gamma}\left[J,g\right]\right\}\propto\exp\left\{ig^{-1}\left[\Gamma\left[X_J\right]+\int{\rm d}^4xX^J_n\left(x\right)J_n\left(x\right)\right]\right\},
\end{equation}

\noindent where the \(X^J_n\) are the values of \(\chi_n\) for which the stationary phase is obtained, given by the condition: 

\begin{equation}
    \label{Stationary_Phase}
    \left.\frac{\delta_L\Gamma\left[\chi\right]}{\delta\chi_n\left(x\right)}\right|_{\chi=X_J}=-J_n\left(x\right).
\end{equation}

\noindent Therefore, the \(g^{-1}\) term in the expansion of \(iW_{\Gamma}\left[J,0\right]\) satisfies the relation 

\begin{equation}
    \label{0_Loop_Term_W}
    W_{\Gamma}^{(0)}\left[J\right]=\Gamma\left[X_J\right]+\int{\rm d}^4xX^J_n\left(x\right)J_n\left(x\right),
\end{equation}

\noindent which is precisely \(W\left[J\right]\). Taking the limit \(g=1\) in Eq.~\eqref{Loop_Decomposition_W_Gamma} it is easy to see that the tree-level part of \(W_{\Gamma}\left[J,1\right]\) is equal to \(W\left[J\right]\) and, consequently, the tree-level part of Eq.~\eqref{W_Constant_g} at \(g=1\) is just the partition function in Eq.~\eqref{Partition_Function}:

\begin{equation}
    \label{Partition_Function_QEA}
    Z\left[J\right]=\int\limits_{\substack{\textrm{Tree}}}\mathcal{D}\chi_n\exp\left\{i\Gamma\left[\chi\right]+i\int{\rm d}^4x\chi_n\left(x\right)J_n\left(x\right)\right\}.
\end{equation}

\noindent Equivalently, one can also write

\begin{equation}
    \label{W_Function_QEA}
    iW\left[J\right]=\int\limits_{\substack{\textrm{Connected}\\ \textrm{Tree}}}\mathcal{D}\chi_n\exp\left\{i\Gamma\left[\chi\right]+i\int{\rm d}^4x\chi_n\left(x\right)J_n\left(x\right)\right\}.
\end{equation}

The tree-level diagrams are the contributions of the partition function that can be understood from the classical Lagrangian. This means that the QEA is equivalent to a classical action but with interactions that contain all of the effects of the quantum corrections. The QEA can be computed by adding all of the \textit{1 Particle Irreducible} (1PI)\footnote{These are diagrams that remain connected when any of their internal lines is cut.} Feynman diagrams \cite{Goldstone:1962es}. This is due to the fact that a general amplitude can be decomposed as a tree-level graph whose vertices are 1PI subdiagrams. 



Since the QEA encodes all of the information about quantum effects, it is useful to know the restrictions that BRST invariance imposes on it in order to understand the behavior of symmetries at the quantum level. Consider a general set of transformations of the form

\begin{equation}
    \label{Field_Transformation}
    \chi_n\left(x\right)\rightarrow\chi_n\left(x\right)+\varepsilon F_n\left[x;\chi\right],
\end{equation}

\noindent where \(\varepsilon\) is an infinitesimal (commuting or anticommuting) parameter and \(F_n\left[x;\chi\right]\) a general functional of the fields and the spacetime coordinates. The generalized version of Eq.~\eqref{Path_Integral_Gauge_Transformation}, where the quantity \(\mathcal{D}\chi_n\exp\left\{iI\left[\chi\right]\right\}\) remains invariant under Eq.~\eqref{Field_Transformation}, is assumed. Thus, one can make the change of variables

\begin{equation}
    \label{Change_Variables_Partition_Function}
    \begin{split}
        Z\left[J\right]&=\int\mathcal{D}\left(\chi_n\left(x\right)+\varepsilon F_n\left[x;\chi\right]\right)\exp\left\{iI\left[\chi+\varepsilon F\right]+i\int{\rm d}^4x\left(\chi_n\left(x\right)+\varepsilon F_n\left[x;\chi\right]\right)J_n\right\}\\&=\int\mathcal{D}\chi_n\exp\left\{iI\left[\chi\right]+i\int{\rm d}^4x\left(\chi_n\left(x\right)+\varepsilon F_n\left[x;\chi\right]\right)J_n\right\}\\&\approx\int\mathcal{D}\chi_n\exp\left\{iI\left[\chi\right]+i\int{\rm d}^4x\chi_n\left(x\right)J_n\left(x\right)\right\}\left[1+i\varepsilon\int{\rm d}^4xF_n\left[x;\chi\right]J_n\left(x\right)\right]\\&=Z\left[J\right]\left[1+\frac{i}{Z\left[J\right]}\varepsilon\int{\rm d}^4x\int\mathcal{D}\chi_nF_n\left[x;\chi\right]J_n\left(x\right)\right.\\& \hspace{3cm}\left.\times\exp\left\{iI\left[\chi\right]+i\int{\rm d}^4x\chi_n\left(x\right)J_n\left(x\right)\right\}\right]
    \end{split}
\end{equation}

\noindent so the functional \(F_n\left[x;\chi\right]\) must satisfy the restriction

\begin{equation}
    \label{VEV_F_Restriction}
    \int{\rm d}^4x\left<F_n\left(x\right)\right>_{J}J_n\left(x\right)=0
\end{equation}

\noindent with

\begin{equation}
    \label{VEV_F_Definition}
    \left<F_n\left(x\right)\right>_{J}\coloneqq\frac{1}{Z\left[J\right]}\int\mathcal{D}\chi_nF_n\left[x;\chi\right]\exp\left\{iI\left[\chi\right]+i\int{\rm d}^4x\chi_n\left(x\right)J_n\left(x\right)\right\}.
\end{equation}

 \noindent Using Eq.~\eqref{Inverse_Legendre}, the following constraint on the QEA is imposed:

\begin{equation}
    \label{Slavnov_Taylor}
    \int{\rm d}^4x\left<F_n\left(x\right)\right>_{J_{\chi}}\frac{\delta_L\Gamma\left[\chi\right]}{\delta \chi_n\left(x\right)}=0.
\end{equation}

\noindent These are known as the \textit{Slavnov-Taylor identities} \cite{Slavnov:1972fg,Taylor:1971ff} and they represent the restrictions that the presence of symmetries imposes on the possible operators that may appear at the QEA from interactions at the quantum level. It is easy to show \cite{Weinberg2} that, if \(F_n\left[x;\chi\right]\) is a linear functional of the fields \(\chi_n\left(x\right)\), then \(\left<F_n\left(x\right)\right>_{J_{\chi}}=F_n\left[x;\chi\right]\) and the QEA is invariant under the transformation~\eqref{Field_Transformation}. Nevertheless, \textit{the BRST transformation is not linear, which means that quantum corrections can produce effective interactions that are not BRST invariant.}

There is a method which makes the study of the constraints on the QEA of non-linear but nilpotent symmetry transformations manageable. It was developed by Zinn-Justin \cite{Zinn-Justin:1974ggz} and it relies on taking (instead of Eq.~\eqref{Partition_Function}) the partition function 

\begin{equation}
    \label{Partition_Function_Antifields}
    Z\left[J,K\right]\coloneqq\int\mathcal{D}\chi_n\exp\left\{iI\left[\chi\right]+i\int{\rm d}^4x \chi_n\left(x\right)J_n\left(x\right)+i\int{\rm d}^4x\left(s\chi_n\left(x\right)\right)K_n\left(x\right)\right\},
\end{equation}

\noindent with \(s\chi_n\left(x\right)\) the BRST transformation of \(\chi_n\) and \(K_n\left(x\right)\) an external source analogous to the antifields of Sec. \ref{sec:Batalin_Vilkovisky} with the same bosonic or fermionic behavior. The vacuum expectation values of the different operators can still be obtained using the relation

\begin{equation}
    \label{VEV_Partition_Function_Antifield}
    \left[\frac{-i\delta_R}{\delta J_{r_1}\left(y_1\right)}\cdots\frac{-i\delta_R}{\delta J_{r_s}\left(y_s\right)}Z\left[J,K\right]\right]_{J,K=0}=\int\mathcal{D}\chi_n\left[\chi_{r_1}\left(y_1\right)\cdots\chi_{r_s}\left(y_{r_s}\right)\right]\exp\left\{iI\left[\chi_n\right]\right\}.
\end{equation}

\noindent One can also define the functional \(W\left[J,K\right]\coloneqq-i\ln\left(Z\left[J,K\right]\right)\), the classical fields 

\begin{equation}
    \label{Classical_Fields_Antifield}
    X_{J,K}^n\left(x\right)\coloneqq\frac{\delta_R}{\delta J_n\left(x\right)}W\left[J,K\right]
\end{equation}

\noindent and (inverting this relation considering the definition of the \(J_{X,K}\) for which \(X_{J,K}\) takes a particular value \(X\)) the QEA as the Legendre transformation:

\begin{equation}
    \label{Quantum_Effective_Action_Antifields}
    \Gamma\left[X,K\right]\coloneqq-\int{\rm d}^4xX_n\left(x\right)J^{n}_{X,K}\left(x\right)+W\left[J_{X,K},K\right],
\end{equation}

\noindent which satisfies the equation

\begin{equation}
    \label{Inverse_Legendre_Antifield}
        \frac{\delta_L \Gamma\left[X,K\right]}{\delta X_n\left(x\right)}=-J^n_{X,K}\left(x\right).
\end{equation}

It is clear that the addition of the BRST source \(K_n\left(x\right)\) does not change the interpretation of the quantum effective action as a classical action that contains all of the quantum effects. Furthermore, due to the fact that the BRST transformation is nilpotent, \(s\chi_n\) (which takes the place of the \(F\) previously introduced) is invariant under the change of variables \(\chi_n\to\chi_n+\delta_{\theta}\chi_n\) in the partition function. Therefore, the previous derivation of the Slavnov-Taylor identities still holds and the condition

\begin{equation}
    \label{Slavnov_Taylor_Antifields}
    \int{\rm d}^4x\left<s\chi_n\left(x\right)\right>_{J_{\chi,K},K}\frac{\delta_L\Gamma\left[\chi,K\right]}{\delta \chi_n\left(x\right)}=0
\end{equation}

\noindent with

\begin{equation}
    \label{VEV_Antifields}
    \begin{split}
    \left<s\chi_n\left(x\right)\right>_{J_{\chi,K},K}\coloneqq\frac{1}{Z\left[J_{\chi,K},K\right]}&\int\mathcal{D}\chi_n\left(s\chi_n\left(x\right)\right)\exp\biggl\{iI\left[\chi\right]\\&\left.+i\int{\rm d}^4x\left[\chi_nJ^n_{\chi,K}+\left(s\chi_n\right)K_n\right]\right\}
    \end{split}
\end{equation}

\noindent is satisfied. But in this case, from Eq.~\eqref{Quantum_Effective_Action_Antifields} one gets

\begin{equation}
    \label{QEA_K_Derivative}
    \begin{split}
         \frac{\delta_R\Gamma\left[\chi,K\right]}{\delta K_n\left(x\right)}&=\left.\frac{\delta_RW\left[J,K\right]}{\delta K_n\left(x\right)}\right|_{J=J_{\chi,K}}+\int{\rm d}^4y\left.\frac{\delta_RW\left[J,K\right]}{\delta J_m\left(y\right)}\right|_{J=J_{\chi,K}}\frac{\delta_R J_{\chi,K}^m\left(y\right)}{\delta K_n\left(x\right)}\\&-\int{\rm d}^4y\chi_m\left(y\right)\frac{\delta_R J^m_{\chi,K}\left(y\right)}{\delta K_n\left(x\right)}
    \end{split}
\end{equation}

 \noindent and using Eqs.~\eqref{Partition_Function_Antifields},~\eqref{Classical_Fields_Antifield} and~\eqref{VEV_Antifields} it is straightforward to prove that

\begin{equation}
    \label{QEA_K_Derivative1}
    \begin{split}
        \frac{\delta_R\Gamma\left[\chi,K\right]}{\delta K_n\left(x\right)}&=\left.\frac{\delta_RW\left[J,K\right]}{\delta K_n\left(x\right)}\right|_{J=J_{\chi,K}}\\&=\frac{1}{Z\left[J_{\chi,K},K\right]}\int\mathcal{D}\chi_n\left(s\chi_n\left(x\right)\right)\exp\left\{iI\left[\chi\right]+i\int{\rm d}^4x\left[\chi_nJ^n_{\chi,K}+\left(s\chi_n\right)K_n\right]\right\}\\&=\left<s\chi_n\left(x\right)\right>_{J_{\chi,K},K}.
    \end{split}
\end{equation}

\noindent Thus, in this formalism the Slavnov-Taylor identities can be written as

\begin{equation}
    \label{Zinn-Justin}
    \int{\rm d}^4x\frac{\delta_R\Gamma\left[\chi,K\right]}{\delta K_n\left(x\right)}\frac{\delta_L\Gamma\left[\chi,K\right]}{\delta \chi_n\left(x\right)}=0.
\end{equation}

This is the so-called \textit{Zinn-Justin equation} and it will become crucial in the study of the renormalizability of gauge theories in the following section. As one can see, it is analogous to the master equation of the Batalin-Vilkovisky formalism so, defining the antibracket 

\begin{equation}
    \label{Definition_Antibracket_Zinn_Justin}
    \left(F,G\right)\coloneqq\int{\rm d}^4x\left[\frac{\delta_RF}{\delta\chi_n\left(x\right)}\frac{\delta_LG}{\delta K_n \left(x\right)}-\frac{\delta_RF}{\delta K_n\left(x\right)}\frac{\delta_LG}{\delta\chi_n\left(x\right)}\right],
\end{equation}

\noindent which satisfies the same properties as the antibracket of Eq.~\eqref{Definition_Antibracket}, it is easy to show that the Zinn-Justin equation can also be expressed as

\begin{equation}
    \label{Zinn_Justin_Antibracket}
    \left(\Gamma,\Gamma\right)=0.
\end{equation}

\section{Renormalizability of Gauge Theories} \label{sec:Renormalization_Gauge_Theories}

In this section the study of the renormalizability of gauge theories is presented. As previously pointed out in Sec. \ref{sec:BRST_Quantization}, \textit{a QFT is said to be renormalizable in the modern sense if all divergences that appear in calculations with a regularization procedure that preserves the symmetries of the model}\footnote{In Sec.~\ref{sec:BMHV} it will be shown that this is not the case for dimensional regularization in chiral theories.} \textit{can be reabsorbed by shiftings of the parameters of the original Lagrangian and field redefinitions} \cite{Weinberg:1978kz}. It can be shown that this is the case for general \(\xi\)-gauges if the Lagrangian only contains operators up to dimension four. Here the general discussion \cite{Gomis:1995jp} is presented using the Batalin-Vilkovisky formalism since, as already mentioned, it is appropriate for theories that also include effective interactions.

In order to see the relation between the Zinn-Justin equation and the Batalin-Vilkovisky procedure, the condition \(\chi_n^{\ddagger}=K_n+\delta\Psi/\delta\chi_n\) is imposed in Eq.~\eqref{Classical_Action_Antifields}, leading to the partition function in Eq.~\eqref{Partition_Function_Antifields}. This connection between \(\chi_n^{\ddagger}\) and  \(K_n\) is a particular case of the so-called anticanonical transformations, the ones that preserve the antibrackets \((\chi_n,\chi_m^{\ddagger})=\delta_{nm}\) and \((\chi_n,\chi_m)=(\chi^{\ddagger}_n,\chi_m^{\ddagger})=0\). By analogy with the situation in the context of classical field theory, it can be shown that any antibracket can be computed with respect to any set of coordinates related by an anticanonical transformation and it will give the same result \cite{Weinberg2}. Thus, it is convenient to calculate \(\left(\Gamma,\Gamma\right)\) using \(\chi_n^{\ddagger}\) instead of \(K_n\).

The action is assumed to satisfy the master equation \(\left(S,S\right)=0\) and some linear structural constraints on its operators,\footnote{One example of a structural constraint is to impose that the action can only contain operators with zero or one antifields.} meaning that if \(\mathcal{S}_1\) and \(\mathcal{S}_2\) respect the restrictions, then so does \(\alpha_1\mathcal{S}_1+\alpha_2\mathcal{S}_2\), with \(\alpha_1\) and \(\alpha_2\) two complex numbers. It can be expanded in powers of \(\hbar\) as

\begin{equation}
    \label{Loop_Expansion_S}
    S=S_R+\sum_{n=1}^{\infty}\hbar^{n}S_n,
\end{equation}

\noindent where \(S_R\) is the renormalized action (a finite object) and \(S_n\) are the different infinite counterterms needed for \(n\)-loop calculations. Due to the fact that the master equation must be satisfied for all of the different terms in the \(\hbar\) expansion, the following constraints are derived:

\begin{equation}
    \label{Renormalized_Master_Equation}
    \left(S_R,S_R\right)=0
\end{equation}

\noindent and 

\begin{equation}
    \label{Master_Equation_Counterterms}
    2\left(S_R,S_N\right)+\sum_{n=1}^{N-1}\left(S_n,S_{N-n}\right)=0.
\end{equation}

In Sec. \ref{sec:Batalin_Vilkovisky} it has been mentioned that there is an ambiguity in the definition of the classical action, since \(\tilde{S}=S+\left(\delta F,S\right)\) also satisfies the master equation. Therefore, it may happen that only some infinities can be canceled by the counterterms \(S_n\) but the rest are reabsorbed by redefinitions of the action. However, in order to study this possibility, it is necessary to extend the so-called infinitesimal anticanonical transformation \(S\rightarrow S+\left(\delta F,S\right)\) with an infinitesimal generating functional \(\delta F\) to the finite scenario. For this purpose, it is appropriate to define the smooth curve \(S\left(t\right)\) with parameter \(t\in\left[0,1\right]\) in such a way that \(S\left(0\right)=S\) and \(S\left(1\right)=\tilde{S}\), where \(\tilde{S}\) is the finite anticanonical transformation of \(S\). Two infinitesimally close points of the curve are related by an infinitesimal anticanonical transformation \(S\left(t+{\rm d}t\right)=S\left(t\right)+\left(F\left(t\right){\rm d}t,S\left(t\right)\right)\), with \(F\left(t\right)\) being a fermionic generating functional. Therefore, \(S\left(1\right)\) can, indeed, be interpreted as a finite anticanonical transformation because it is the smooth composition of infinitesimal ones. From the relation \(S\left(t+{\rm d}t\right)=S\left(t\right)+\left(F\left(t\right){\rm d}t,S\left(t\right)\right)\) it is straightforward to show that 

\begin{equation}
    \label{Finite_Anticanonical}
    \frac{{\rm d}}{{\rm d}t}S\left(t\right)=\left(F\left(t\right),S\left(t\right)\right),
\end{equation}

\noindent which, given a generating functional \(F\left(t\right)\), is the only condition needed to obtain \(\tilde{S}\). One can take the arbitrary generating functional to be expressed as the loop-expansion

\begin{equation}
    \label{Generating_Functional}
    F\left(t\right)\coloneqq \sum_{n=1}^{\infty}n\hbar^nt^{n-1}F_n,
\end{equation}

\noindent with \(F_n\) a set of arbitrary functionals with no \(t\) or \(\hbar\) dependence. Consequently, the associated finite anticanonical transformation is

\begin{equation}
    \label{Finite_Anticanonical_Result}
    \begin{split}
        \tilde{S}&=S\left(1\right)=\sum_{n=0}^{\infty}\frac{1}{n!}\left.\frac{{\rm d}^n}{{\rm d}t^n}S\left(t\right)\right|_{t=0}=S(0)+\left(F\left(0\right),S\left(0\right)\right)+\frac{1}{2}\left.\frac{{\rm d}}{{\rm d}t}\left(F\left(t\right),S\left(t\right)\right)\right|_{t=0}+\cdots\\&=S+\hbar\left(F_1,S\right)+\hbar^2\left(F_2,S\right)+\frac{1}{2}\hbar^2\left(F_1,\left(F_1,S\right)\right)+\cdots\\&=S_R+\hbar\left\{S_1+\left(F_1,S_R\right)\right\}+\hbar^2\left\{S_2+\left(F_2,S_R\right)+\left(F_1,S_1 \right)+\frac{1}{2}\left(F_1,\left(F_1,S_R\right)\right)\right\}+\cdots
    \end{split}
\end{equation}

This finite anticanonical transformation is just the exponential map of the infinitesimal one and it can be interpreted as a field redefinition into a new set of renormalized fields and antifields that are related to the original ones by an anticanonical transformation. It is important to remark that, for a particular order in the loop expansion of the exponential map, there is only one term that contains the functional \(F_N\), which comes from \(\frac{1}{N!}\left.\frac{{\rm d}^N}{{\rm d}t^N}S\left(t\right)\right|_{t=0}\rightarrow\frac{1}{N!}\left(\left.\frac{{\rm d}^{N-1}}{{\rm d}t^{N-1}}F\left(t\right)\right|_{t=0},S\right)=\hbar^N\left(F_N,S\right)\rightarrow \hbar^N\left(F_N,S_R\right)\). Therefore, \textit{in order to cancel the divergences that appear in \(N\)-loop calculations, one can use both the counterterms} \(S_N\) \textit{that satisfy the structural constraints in Eq.~\eqref{Master_Equation_Counterterms} and terms of the form} \(\left(F_N,S_R\right)\).

If the regularization procedure preserves the symmetries of the theory, the form of the possible divergences is constrained by the Zinn-Justin equation. The QEA can be decomposed as

\begin{equation}
    \label{Loop_Expansion_QEA}
    \Gamma=\sum_{N=0}^{\infty}\Gamma_N,
\end{equation}

\noindent with \(\Gamma_N\) containing the \(N\)-th order loop contributions. Due to the fact that the tree-level effects can be described only by the renormalized action, then the condition \(\Gamma_0=S_R\) must be satisfied. Furthermore, the Zinn-Justin equation also needs to hold order by order in the loop expansion:

\begin{equation}
    \label{Loop_Expansion_Zinn_Justin}
    \sum_{n=0}^N\left(\Gamma_n,\Gamma_{N-n}\right)=0.
\end{equation}

Imagine that all of the divergences associated with \(n\)-loop diagrams have been removed for every \(n<N\). Therefore, the only remaining infinities that may appear in \(N\)-th order calculations are contained in \(\Gamma_N\) and can be denoted as \(\Gamma_{N,\infty}\). Due to the fact that the Zinn-Justin equation needs to be satisfied for both the finite and divergent parts of the QEA independently, it is easy to show that:

\begin{equation}
    \label{Constraint_Infinities}
    \left(S_R,\Gamma_{N,\infty}\right)=0,
\end{equation}

\noindent where the symmetry of the antibracket of two bosonic operators has been used. Thus, Eq.~\eqref{Constraint_Infinities} implies that, if the theory is renormalized order by order in the loop expansion, then the divergent parts are constrained by being closed forms of ghost number zero under the operator \(\left(S_R,\cdot\right)\). The constraint satisfied by the divergent part of the QEA is not the same as the constraint satisfied by the counterterms \(S_N\). However, if one finds a particular solution \(S_N^0\) of Eq.~\eqref{Master_Equation_Counterterms}, then it is always possible to define the counterterm  \(S_N=S_N^0+\bar{S}_N\) as long as both \(S_N^0\) and \(\bar{S}_N\) satisfy the same structural constraints as \(S_R\) and \(\left(S_R,\bar{S}_N\right)=0\).\footnote{This can easily be done if \(S_R\) is of the form of Eq.~\eqref{Classical_Action_Antifields} and \(\bar{S}_N\) is taken to be a gauge invariant functional of the matter and gauge fields, because in that case \(\left(S_R,\cdot\right)=-s\left(\cdot\right)\). This is precisely what is needed for the renormalization of gauge theories.}

It has been shown \cite{Barnich:1994ve,Barnich:1994db,Barnich:1994mt} using only cohomological arguments that the most general solution to the equation \(\left(S_R,\mathcal{O}\right)=0\) (with \(\mathcal{O}\) an operator written in terms of local functionals\footnote{The fact that the divergences are local will be proved in Sec.~\ref{sec:Weinberg_Theorem}.} with ghost number zero) for a gauge theory associated with a semisimple group in addition to \({\rm U}(1)\) factors is given by

\begin{equation}
    \label{Cohomological_Solution}
    \mathcal{O}=\int{\rm d}^4x\mathcal{I}+\left(\mathcal{F},S_R\right)+c_{\alpha}^{\Delta}\int{\rm d}^4x\left\{J_{\Delta}^{\mu}B_{\mu}^{\alpha}+t_{\Delta}^n\phi_n^{\ddagger}\omega^{\alpha}_{B}\right\},
\end{equation}

\noindent where \(\mathcal{I}\) is a gauge invariant polynomial on the matter and gauge fields, \(\mathcal{F}\) is an arbitrary fermionic functional that can be written in terms of local operators of ghost number \(-1\), \(c_{\alpha}^{\Delta}\)  are a set of coupling constants,\footnote{Note that the couplings are the same for both \(J_{\Delta}^{\mu}B_{\mu}^{\alpha}\) and \(t_{\Delta}^n\phi_n^{\ddagger}\omega^{\alpha}_{B}\), so only the antifield independent part is needed in order to know whether these contributions are present \cite{Barnich:1994ve,Henneaux:1992ig}.} \(B_{\mu}^{\alpha}\) are the gauge fields associated with abelian factors of the group, \(\omega^{\alpha}_{B}\) are the corresponding abelian ghosts, \(J_{\Delta}^{\mu}\) are a set of gauge invariant conserved currents coming from a symmetry (not necessarily the one associated with \(B_{\mu}^{\alpha}\)) denoted by \(\Delta\) and \(t_{\Delta}^n\) are the generators of that symmetry for the matter and gauge fields \(\phi_n\).

The operator \(\int{\rm d}^4x\mathcal{I}\) can always be absorbed by the counterterms \(\bar{S}_N\)\footnote{Sometimes this requires enlarging the structural constraints of \(S\). For instance, in bosonic QED it is necessary to add the gauge invariant quartic coupling \(\lambda\left(\phi^{\dagger}\phi\right)^2\) in order to obtain a renormalizable theory, so it is not possible to impose the structural constraint \(\lambda=0\).} and the exact form \(\left(\mathcal{F},S_R\right)\) can be absorbed, as previously mentioned, using an anticanonical transformation into a renormalized set of fields and antifields. Therefore, if the gauge group is semisimple this cohomological argument shows that the theory is renormalizable, but if additional abelian \(\mathrm{U}\left(1\right)\) factors are present (like in the SM or in LR theories) some additional considerations have to be taken. First of all, looking at Eq.~\eqref{Classical_Action_Antifields}, it is clear that for gauge theories the structural constraint imposed on the action can be written in the general form

\begin{equation}
    \label{Structural_Constraint_Action}
    S=I\left[\phi\right]+\omega^{A}t_A^B\left[\phi\right]\phi_B^{\ddagger}+\frac{1}{2}C_{\,\,\,AB}^C\omega^{A}\omega^{B}\omega_C^{\ddagger}-h^A\omega_{A}^{\ast\ddagger},
\end{equation}

\noindent where the sum (integration) over all the indices including spacetime coordinates is taken, \(t^B_A\left[\phi\right]\) represents a generalized version of the generators of the gauge symmetry \(\delta_{\varepsilon}\phi^B=\varepsilon^At_A^B\left[\phi\right]\) and \(C_{ABC}\) are the structure constants. Thus, it is clear that all the terms in Eq.~\eqref{Cohomological_Solution} can be reabsorbed by the classical action if the way in which the \({\rm U}\left(1\right)\) transformations act on the fields is changed in order for the \(J_{\Delta}^{\mu}\) to be the associated conserved currents. However, if there are no possible divergences in these types of terms, they cannot appear in \(\Gamma_{N,\infty}\). Furthermore, it can be shown, generalizing for EFTs the theorems that state that the divergences in \({\rm U}\left(1\right)\) theories appearing in the vertices of interactions with gauge bosons can be understood as a part of the renormalization of the photon wave-function \cite{Weinberg1}, that the possible \(c_{\Delta}J_{\Delta}^{\mu}B_{\mu}\) factors have to be finite \cite{Weinberg2,Gomis:1995jp}. Consequently, \textit{gauge theories are renormalizable.}\footnote{ It has been shown that this is the case to all orders in perturbation theory. However, due to the fact that the perturbation series in QFT are generally asymptotic series that do not converge, they do not reproduce the whole theory. Apart from the possible presence of non-trivial fixed points of the \(\beta\) function, lattice calculations point to the fact that theories like QED or \(\phi^4\) that contain Landau poles in the UV are not renormalizable beyond perturbation theory. The situation for theories with asymptotic freedom is still unknown and it is a part of one of the Millennium Prize Problems.}

This reasoning can be extended to other theories like quantum gravity in the presence of Yang-Mills fields \cite{Barnich:1995ap}, showing that all of the divergences can be reabsorbed if the action is allowed to be the most general functional invariant under diffeomorphisms. Nevertheless, in other cases one should enlarge the structural constraints in order to reabsorb all the infinities and it is not known if this procedure can generally be done \cite{Weinberg2}.

\section{Anomalies. Wess-Zumino Consistency Conditions} \label{sec:Anomalies}

Up to this point, the fact that the measure \(\mathcal{D}\phi_n\) is invariant under gauge transformations has always been assumed. In Sec. \ref{sec_Faddeev_Popov} it was shown that this is the case for the gauge bosons, but the situation with the matter fields still needs to be studied. Consider a set of fermions \(\psi_n\) and their Dirac adjoints \(\bar{\psi}_n\coloneqq\psi^{\dagger}_n\gamma^0\). Assuming that the \(\psi_n\) transform as \(\psi\to U\psi\) with \(U\) a unitary matrix, it is easy to show that the transformation of the Dirac adjoints is \(\bar{\psi}\to\bar{\psi}\bar{U}\), where \(\bar{U}\coloneqq\gamma^0U^{\dagger}\gamma^{0}\). Since \(\psi_n\) and \(\bar{\psi}_n\) are a set of Grassmann variables, the transformation of the measure is \cite{Weinberg1}:

\begin{equation}
    \label{Transoformation_Measure_Fermions}
    \mathcal{D}\psi_n\mathcal{D}\bar{\psi}_n\to\left({\rm det}\,\mathcal{U}{\rm det}\,\bar{\mathcal{U}}\right)^{-1}\mathcal{D}\psi_n\mathcal{D}\bar{\psi}_n,
\end{equation}

\noindent with \(\mathcal{U}_{nx,my}\coloneqq U\left(x\right)_{nm}\delta^4\left(x-y\right)\) and \(\bar{\mathcal{U}}_{nx,my}\coloneqq\bar{U}\left(x\right)_{nm}\delta^4\left(x-y\right)\). If \(U\) does not contain any Dirac index, which is the situation with non-chiral transformations, then \(\bar{U}=U^{\dagger}\) and the measure is gauge invariant. However, for chiral transformations like \(U\left(x\right)=\exp\left\{i\gamma_5\alpha\left(x\right)t\right\}\) with \(t\) the group generator, it is easy to see from the relations \((\gamma^0)^2=1\) and \(\left\{\gamma_5,\gamma^0\right\}=0\) that \(\bar{U}=U\), which leads to

\begin{equation}
    \label{Transoformation_Measure_Chiral_Fermions}
    \mathcal{D}\psi_n\mathcal{D}\bar{\psi}_n\to\left({\rm det}\,\mathcal{U}\right)^{-2}\mathcal{D}\psi_n\mathcal{D}\bar{\psi}_n.
\end{equation}

\textit{As a consequence of the non-invariance of the measure, breakings of the gauge symmetry at the quantum level can appear in theories with chiral-fermions} \cite{Fujikawa:1979ay,Alvarez-Gaume:1984zlq}. These are the previously announced \textit{anomalies}. It is important to remark that \textit{their appearance may depend on the regularization procedure},  implying that (as it will be shown) their presence may have no physical meaning since they could be canceled by the addition of local counterterms to the action. For instance, the \(\gamma^0\) factors could have been eliminated in the calculation of \(\det\,\bar{\mathcal{U}}\), because they correspond to a unitary transformation. But the definition of divergent determinants depends on the regularization scheme and the ones without those \(\gamma^0\) factors may lead to the breaking of Lorentz symmetry \cite{Weinberg2}.

The form of any possible symmetry breaking (physical or not) is constrained by the so-called \textit{Wess-Zumino consistency conditions} \cite{Wess:1971yu}, whose derivation in the Batalin-Vilkovisky formalism is trivial. At the quantum level, the presence of a nilpotent symmetry transformation leads to the Zinn-Justin equation,\footnote{This is the case for a gauge symmetry, so it may seem that the following reasoning only applies to that scenario. Nevertheless, global transformations can also be studied gauging them and then taking the limit of the gauge couplings equal to zero.} which is written in terms of its loop expansion in Eq.~\eqref{Loop_Expansion_Zinn_Justin}. At one-loop, the condition \(\left(S_R,\Gamma_1\right)=0\) is obtained, but in the case where anomalies are present this antibracket does not vanish:

\begin{equation}
    \label{One_Loop_Anomaly}
    \left(S_R,\Gamma_1\right)=G_1,
\end{equation}

\noindent where \(G_1\) must be a fermionic functional of ghost number one.\footnote{The presence of this function is related to having a non-vanishing \(\Delta S\) in the quantum master equation.} Using the fact that the operator \(\left(S_R,\cdot\right)\) is nilpotent, the one-loop Wess-Zumino consistency condition in the antifield formalism is obtained:

\begin{equation}
    \label{One_Loop_Wess_Zumino}
    \left(S_R,G_1\right)=0.
\end{equation}

This means that the possible anomalies are functionals of ghost number one that are closed under the operator \(\left(S_R,\cdot\right)\). A particular solution is the one where \(G_1\) can be expressed as \(G_1=\left(S_R,F_1\right)\) with \(F_1\) a local bosonic functional of ghost number zero. In this scenario, the condition \(\left(S_R,\Gamma_1-F_1\right)=0\) can be derived and since \(F_1\) is a local functional it can be understood as a correction to the action. \textit{In this case, the anomaly is not physical because its effect can be eliminated by the addition of the appropriate counterterms.} This is the situation for a symmetry-breaking that comes from a regularization procedure that does not respect the symmetries of a consistent theory.

The Wess-Zumino consistency conditions can also be studied in the BRST formalism. In that case, one can simply assume that the BRST transformation of the one-loop QEA does not vanish \(s\Gamma_1=G_1\). Then, from the nilpotency of \(s\), the consistency condition \(sG_1=0\) is obtained and, again, if \(G_1=sF_1\) with \(F_1\) a local functional, the anomaly is not physical since it can be eliminated by the addition of a set of gauge-variant counterterms to the action. Nevertheless, this formalism is not suited for the study of the consistency of a symmetric theory. The cancellation of anomalies using \(\left(S_R,F_1\right)\) is more general than using \(sF_1\), because it includes the addition of counterterms containing antifields. Furthermore, since the BRST transformation is not linear, there might be no anomaly in a theory having a non-vanishing \(s\Gamma\), the anomaly is associated with having a \(\left(\Gamma,\Gamma\right)=G\neq0\). For these reasons, the Wess-Zumino conditions will be derived for an arbitrary order in the loop expansion using the Batalin-Vilkovisky formalism.

Assume that the theory is well-defined so the anomalies can be canceled by the addition of the appropriate set of counterterms. Then, imagine that the symmetry breakings in \(M\)-loop calculations with \(M<N\) have been eliminated so:

\begin{equation}
    \label{Cancelation_M_Loop_Anomaly}
    \sum_{L=0}^M \left(\Gamma_L,\Gamma_{M-L}\right)=2\left(S_R,\Gamma_M\right)+\sum_{L=1}^{M-1}\left(\Gamma_L,\Gamma_{M-L}\right)=0.
\end{equation}

\noindent The \(N\)-loop anomaly function is 

\begin{equation}
    \label{N_Loop_Anomaly}
    G_N\coloneqq2\left(S_R,\Gamma_N\right)+\sum_{M=1}^{N-1}\left(\Gamma_M,\Gamma_{N-M}\right),
\end{equation}

\noindent so using the nilpotency of \(\left(S_R,\cdot\right)\) the relation

\begin{equation}
    \label{S_R_N_Loop_Anomaly}
    \left(S_R,G_N\right)=\sum_{M=1}^{N-1}\left(S_R,\left(\Gamma_M,\Gamma_{N-M}\right)\right)
\end{equation}

\noindent is obtained. From the Jacobi identity of the antibracket for bosonic operators and Eq.~\eqref{Cancelation_M_Loop_Anomaly} this can be written as

\begin{equation}
    \label{S_R_N_Loop_Anomaly_2}
    \left(S_R,G_N\right)=-2\sum_{M=1}^{N-1}\left(\left(S_R,\Gamma_M\right),\Gamma_{N-M}\right)=\sum_{M=1}^{N-1}\sum_{L=1}^{M-1}\left(\left(\Gamma_L,\Gamma_{M-L}\right),\Gamma_{N-M}\right).
\end{equation}

\noindent Finally, using again the Jacobi identity \cite{Weinberg2} it can be shown that the double sum vanishes:

\begin{equation}
    \left(S_R,G_N\right)=0.
\end{equation}

To summarize, for any anomaly (physical or not) the Wess-Zumino consistency condition \(\left(S_R,G_1\right)=0\) must be satisfied, because it follows directly from the nilpotency of  \(\left(S_R,\cdot\right)\). But \textit{if a theory is free of physical anomalies, then} \(\left(S_R,G_N\right)=0\) \textit{must hold at any order in perturbation theory}. Therefore, since any anomaly function of the form \(G_N=\left(S_R,F_N\right)\) with \(F_N\) a local functional can be canceled by the addition of counterterms, then physical anomalies can only be present if there are functionals of ghost number one closed under \(\left(S_R,\cdot\right)\) but not exact. Using again only cohomological arguments, it can be shown \cite{Barnich:1994ve,Barnich:1994db,Barnich:1994mt} that the possible candidate anomalies\footnote{ The fact that these anomalies are local in the context of dimensional regularization will be studied in Sec.~\ref{sec:Bonneau}. It is important to remark that here only anomalies that appear in perturbative calculations are considered.} for a gauge theory associated with a semisimple group in addition to \({\rm U}(1)\) factors  are 

\begin{equation}
    \label{Possible_Anomalies}
    \begin{split}
        G&=c_1\int{\rm d}^4x\varepsilon^{\mu\nu\rho\sigma}{\rm tr}\left\{\omega\left[\partial_{\mu}A_{\nu}\partial_{\rho}A_{\sigma}+\frac{3}{2}\partial_{\mu}A_{\nu}A_{\rho}A_{\sigma}\right]\right\}\\&+c_2^{\alpha}\int{\rm d}^4x\varepsilon^{\mu\nu\rho\sigma}\partial_{\mu}B^{\alpha}_{\nu}{\rm tr}\left\{\omega Y^{\alpha}\partial_{\rho}A_{\sigma}\right\}+c_3\int{\rm d}^4x\mathcal{I}_{\alpha}\omega^{\alpha}_B\\&+c_{4\alpha\beta}^{\Delta}\int{\rm d}^4x\left(J_{\Delta}^{\mu}B_{\mu}^{\alpha}\omega_B^{\beta}+\frac{1}{2}t^n_{\Delta}\phi_n^{\ddagger}\omega_B^{\alpha}\omega_B^{\beta}\right),
    \end{split}
\end{equation}

\noindent where the objects \(\omega\coloneqq\omega^{\alpha}t_{\alpha}\) and \(A_{\mu}\coloneqq A^{\alpha}_{\mu}t_{\alpha}\) (with \(t_{\alpha}\) the generators of the group) are defined; \(Y^{\alpha}\) represents the hypercharge operator associated with the abelian gauge boson \(B_{\mu}^{\alpha}\); \(\mathcal{I}_{\alpha}\) is a BRST-invariant polynomial in the matter and gauge fields and \(c_{4\alpha\beta}^{\Delta}=-c^{\Delta}_{4\beta\alpha}\).\footnote{Like in Eq.~\eqref{Cohomological_Solution}, the antifield dependent solution is completely determined from the independent part \cite{Barnich:1994ve,Henneaux:1992ig}.}

\begin{figure}
    \centering
    \includegraphics[scale=0.20]{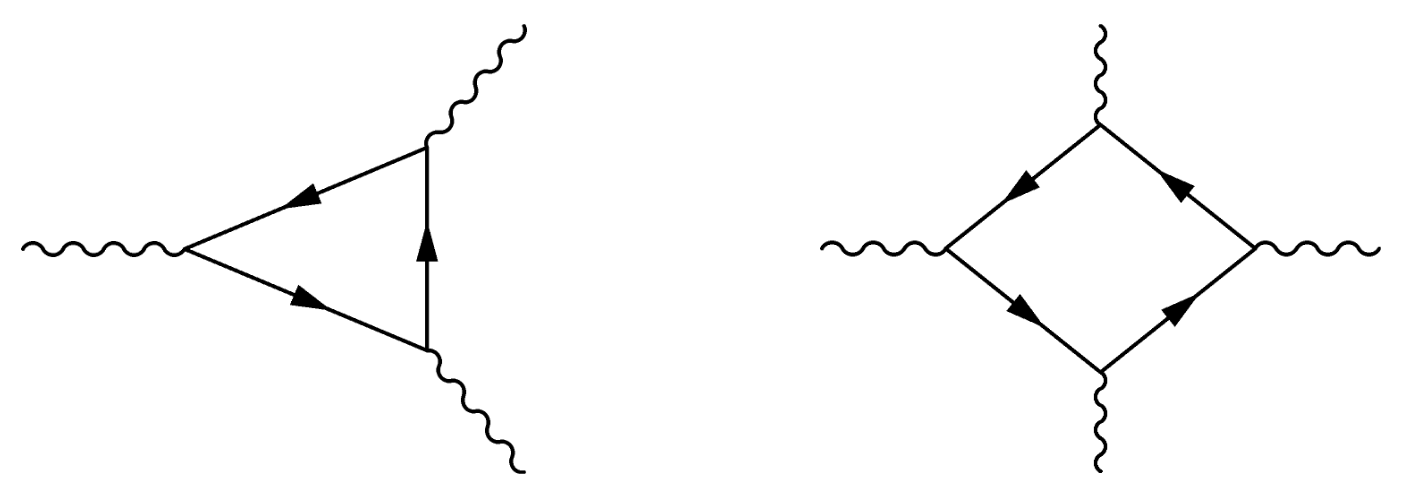} 
    \caption[Triangle (left) and square (right) Feynman diagrams that contribute to the ABBJ anomaly at one loop.]{Triangle (left) and square (right) Feynman diagrams that contribute to the ABBJ anomaly at one loop. Given the Wess-Zumino consistency conditions the amplitude of one of them can be fully determined from the other.}
    \label{fig:Anomalies}
\end{figure}

The first term in Eq.~\eqref{Possible_Anomalies} is the only one that appears in gauge theories associated with a semisimple gauge group and it is known as the \textit{Adler-Bardeen-Bell-Jackiw (ABBJ) anomaly} \cite{Bell:1969ts,Adler:1969gk,Bardeen:1969md,Adler:1969er}. The second factor is analogous to the first one but in the case where \({\rm U}(1)\) subgroups are present. Indeed, the first part of the ABBJ anomaly coincides with this term if the correspondence of the hypercharge operator \(Y^{\alpha}\) with the generator of the \({\rm U}(1)\) group is made. These symmetry breaking contributions appear in the calculation of the triangle diagrams (Fig. \ref{fig:Anomalies}) and its cancellation is crucial for the consistency of the theory. Given that \(\varepsilon^{\mu\nu\rho\sigma}{\rm tr}\left\{t_{\alpha}t_{\beta}t_{\gamma}\right\}\omega^{\alpha}\partial_{\mu}A_{\nu}^{\beta}\partial_{\rho}A_{\sigma}^{\gamma}=\varepsilon^{\mu\nu\rho\sigma}{\rm tr}\left\{t_{\alpha}t_{\gamma}t_{\beta}\right\}\omega^{\alpha}\partial_{\mu}A_{\nu}^{\beta}\partial_{\rho}A_{\sigma}^{\gamma}\) and \({\rm tr}\left\{Yt_{\alpha}t_{\beta}\right\}\propto Y\delta_{\alpha\beta}\), only the totally symmetric part \(D_{\alpha\beta\gamma}\coloneqq\frac{1}{2}{\rm tr}\left\{t_{\alpha}\left(t_{\beta}t_{\gamma}+t_{\gamma}t_{\beta}\right)\right\}\) of the traces needs to be considered. Furthermore, in chiral theories it is common to include all fermions of the same chirality into one object:

\begin{equation}
    \label{Same_Chirality_Fermions}
    \Psi_{L,R}\coloneqq\left(
    \begin{array}{c}
    P_{L,R}\psi \\
    \gamma^{0}P_{L,R}\mathcal{C}\psi^{\ast}
    \end{array}\right),
\end{equation}

\noindent where \(P_{L,R}\) represent the left- and right-handed projectors, respectively; and \(\mathcal{C}\) the charge conjugation operator. Then, \(\Psi_{L,R}\) gets transformed under the generator

\begin{equation}
    \label{Chiral_Generator}
    t_{\alpha}\coloneqq\left(
    \begin{array}{cc}
       t_{\alpha}^{L,R}  & 0 \\
         0 & -(t_{\alpha}^{R,L})^{T} 
    \end{array}\right).
\end{equation}

As a consequence, in order for the triangle anomalies not to appear, the needed condition is

\begin{equation}
    \label{Vanishing-Symmetric}
    D_{\alpha\beta\gamma}=\frac{1}{2}\left[{\rm tr}\left\{t_{\alpha}^L\left\{t_{\beta}^L,t_{\gamma}^L\right\}\right\}-{\rm tr}\left\{t_{\alpha}^R\left\{t_{\beta}^{R},t_{\gamma}^R\right\}\right\}\right]=0.
\end{equation}

Using group theory arguments it can be shown that, for any possible fermion content, this condition is automatically satisfied for gauge theories that do not contain \({\rm SU}(n)\) with \(n\geq3\) or \({\rm U}(1)\) subgroups \cite{Georgi:1972bb,10.1063/1.1704831,10.1063/1.1704832}. As a consequence, any semisimple gauge theory with no \({\rm SU}(n)\) (for \(n\geq3\)) factors is free of anomalies at any order in perturbation theory. \textit{The SM and LR models are consistent QFTs with no symmetry breaking} because they can be understood as low energy limits of \({\rm SO}(10)\) grand unified models \cite{Fritzsch:1974nn,PhysRevLett.52.1072,PhysRevD.30.1052,PhysRevD.31.1718}. \textit{The SMEFT should also be free of anomalies}\footnote{This has been explicitly checked in some cases \cite{Bonnefoy:2020tyv,Feruglio:2020kfq,Cohen:2023hmq,Fuentes-Martin:2025meq}; in fact, in this text it is shown that four-fermion operators do not introduce anomalies at one-loop order.} given the fact that these inconsistencies come, as mentioned before, from a gauge-variant measure in the path integral, but the measure of the SMEFT is the same as the SM one, so it is gauge invariant.

\thispagestyle{empty}

\chapter{The Breitenlohner-Maison-'t~Hooft-Veltman Dimensional Regularization Scheme}
\markboth{2. THE BMHV DIMENSIONAL REGULARIZATION SCHEME}{}

\section*{Introduction}

In the previous chapter it has been shown that gauge theories are anomaly-free and renormalizable if the symmetry group satisfies certain conditions. Nonetheless, to make practical calculations it is necessary to define a prescription that associates a certain value to the divergent integrals that appear in the computation of the probability amplitudes of different processes. Apart from the presence of infrared divergences, the infinities usually appear when integrating up to large momenta in the loop diagrams. Given the fact that nowadays it is strongly believed that the theories that physicist use to describe nature are just effective theories, a natural regularization prescription is to integrate up to an ultraviolet cutoff \(\Lambda_{\rm NP}\). Nevertheless, this scheme breaks Lorentz invariance because a maximum value of the integration momentum depends on the particular observer. Both Poincaré and gauge invariance are powerful tools in actual calculations so it is important that \textit{the choice of the regularization scheme respects as many symmetries as possible}. Nowadays, an often preferred option following this prescription is to use \textit{dimensional regularization}.

However, an argument presented in \cite{Bruque:2018bmy} shows that one needs to be very careful in formalizing the precise definitions of the regularization prescription. Their argument goes as follows: assume that the regularization scheme in an arbitrary \(D\)-dimensional space respects the shift invariance of the integrals so

\begin{equation}
    \label{Shift_Invariance}
    \int_R{\rm d}^Dk\left[f\left(k+p\right)-f\left(k\right)\right]=0,
\end{equation}

\noindent where \(\int_{R}\) represents the regularized value. Then, taking the Taylor expansion up to order \(p\) in the integrand it is easy to show that, for a general \(p\):

\begin{equation}
    \label{Vanishing_Derivative_Integral}
    \int_R{\rm d}^Dk\frac{\partial}{\partial k_{\mu}}f\left(k\right)=0.
\end{equation}

\noindent Now consider the following divergent integral in a 2-dimensional Euclidean space:

\begin{equation}
   \label{Euclidean_Integral} 
   I_{ij}\coloneqq\int{\rm d}^2\mathbf{k}\frac{k_ik_j}{\left(\mathbf{k}^2+m^2\right)^2}
\end{equation}

\noindent The easiest way to calculate its regularized value is to use the fact that

\begin{equation}
    \label{Auxiliar_Integral}
    0=\int_R{\rm d}^2\mathbf{k}\frac{\partial}{\partial k_{i}}\left(\frac{k_j}{\mathbf{k}^2+m^2}\right)=\int_R{\rm d}^2\mathbf{k}\left[\frac{\delta_{ij}}{\mathbf{k}^2+m^2}-2\frac{k_ik_j}{\left(\mathbf{k}^2+m^2\right)^2}\right],
\end{equation}

\noindent so assuming that the regularization scheme respects the linearity of the integrals \(I_{ij}^R\) can be written as

\begin{equation}
    \label{Euclidean_Integral_2}
    I_{ij}^R=\frac{1}{2}\delta_{ij}\int_R{\rm d}^2\mathbf{k}\frac{1}{\mathbf{k}^2+m^2}=\frac{1}{2}\delta_{ij}\int_R{\rm d}^2\mathbf{k}\frac{\mathbf{k}^2+m^2}{\left(\mathbf{k}^2+m^2\right)^2}=\frac{1}{2}\delta_{ij}\left[I_{kk}^R+\int_R{\rm d}^2\mathbf{k}\frac{m^2}{\left(\mathbf{k}^2+m^2\right)^2}\right].
\end{equation}

\noindent Finally, the second integral is convergent and it is easy to evaluate. Assuming that for convergent integrals the regularization scheme gives the proper finite result one gets

\begin{equation}
    \label{Euclidean_Integral_3}
    I_{ij}^R=\frac{1}{2}\delta_{ij}\left[I_{kk}^R+\pi\right].
\end{equation}

\noindent This means that this prescription is not compatible with the commutation of the \(\delta_{ij}\) operator and the integration since \(I_{ij}^R\neq\delta_{ij}I_{kk}^R/2\), showing that some of the basic intuitive properties might not hold at the same time.

One can go even further with this argument, consider the following determinant:

\begin{equation}
    \label{determinant_genuine}
    {\rm Det}\left(\mu_1\dots\mu_m;\nu_1\dots\nu_m\right)\coloneqq\left|
    \begin{array}{ccc}
    \delta_{\mu_1\nu_1} & \cdots & \delta_{\mu_1\nu_m} \\
    \vdots & \ddots & \vdots \\
    \delta_{\mu_m\nu_1} & \cdots & \delta_{\mu_m\nu_m}
    \end{array}\right|.
\end{equation}

\noindent It is clear that if \(m>D\) the object vanishes because some indices are repeated and consequently there are at least two identical columns and rows. Now take the 2-dimensional quantity in the Euclidean space:

\begin{equation}
    \label{Determinant_Quantity}
    {\rm Det}\left(ikl;jmn\right)p_1^kp_2^lp_3^mp_4^nI^R_{ij}
\end{equation}

\noindent If one uses the fact that in a two dimensional space that determinant is zero from the beginning then the object vanishes, but if one uses its definition and makes the tensor contractions the result is

\begin{equation}
    \label{Determinant_Result}
    {\rm Det}\left(ikl;jmn\right)p_1^kp_2^lp_3^mp_4^nI^R_{ij}= \pi\left[\left(p_1p_4\right)\left(p_2p_3\right)-\left(p_1p_3\right)\left(p_2p_4\right)\right],
\end{equation}

\noindent which is non-vanishing for general momenta in two dimensions, just consider the case with \(p_1\parallel p_4\), \(p_2\parallel p_3\), \(p_1\perp p_3\) and \(p_2\perp p_4\). This shows that \textit{there might be inconsistencies when applying relations that only hold in a particular dimensional space to objects appearing in regularized integrals}.\footnote{Notice that this is also the case even if one always stays in a particular dimension during the whole calculation.} In order to avoid these problems, in the first section of this chapter the precise definitions of the dimensional regularization scheme will be given following the presentation done in \cite{Collins_1984}.

Furthermore, apart from a regularization scheme it is also necessary to define an algorithm to reabsorb the divergences appearing in the amplitudes, this is known as a renormalization scheme and the one that will be used in this text is minimal subtraction. In the second section, a sketch of the proof that dimensional regularization plus minimal subtraction define a procedure that makes any amplitude with an arbitrary number of loops finite is given. The proof relies on showing that the needed counterterms can be expressed as a set of local operators, which also completes the proof of the first chapter of the fact that gauge theories are renormalizable.

In the third section the proof that dimensional regularization satisfies the \textit{Quantum Action Principle} (QAP) is presented. \textit{This principle is the fundamental property needed to use the techniques of algebraic renormalization}, a procedure to deal with theories that contain anomalies that may (or may not) be generated by the regularization procedure. Indeed, in the fourth section the only known mathematically consistent version of dimensional regularization, the Breitenlohner-Maison-'t Hooft-Veltman (BMHV) scheme, is presented and it is shown that it breaks the gauge symmetries of a theory containing chiral fermions. Consequently, in the fifth section the way to identify the counterterms needed to restore the gauge symmetry using the techniques of algebraic renormalization and the Bonneau method is described. This will also complete the study of the anomalies shown in the first chapter since it will be shown that the possible obstructions to the Slavnov-Taylor identities are local.

\section{Dimensional Regularization} \label{sec:Dimensional_Regularization}

The dimensional regularization scheme was introduced in \cite{tHooft:1972tcz, Bollini:1972ui, Ashmore:1972uj} and it relies on the idea that \textit{the UV divergent integrals are usually convergent if the dimensionality of the spacetime is low enough}. For instance, consider

\begin{equation}
    \label{Div_int}
    \int{\rm d}^Dp\frac{1}{\left(p^2-m^2\right)^2},
\end{equation}

\noindent by simple power counting it is clear that the integral converges for \(D\leq3\) and diverges otherwise. Therefore, it is possible to evaluate the expression for small \(D\), take the analytic continuation to \(D=4\) and isolate the divergent part. Since only the dimension of the spacetime has been changed, all of the symmetry properties of the theory are preserved with the exception that will be studied in Sec.~\ref{sec:BMHV}. Nonetheless, as pointed out in the introduction of this chapter, it is necessary to define the properties of \(D\)-dimensional integration in a mathematically rigorous way. This was first done in \cite{Wilson:1972cf}. For example, even though the associated values to the integrals behave as \(D\)-dimensional objects, the fact is that \textit{the Lorentz algebra is indeed infinite dimensional}. The reason for this is that one could use this method to make calculations in theories that live in a spacetime with arbitrarily large physical dimension \(D_0\), but with amplitudes that only converge if \(D\) is low enough. In such a theory one can find \(D_0\) linearly-independent external momenta, so in order to keep this property (and other ones related to the \(\gamma^{\mu}\) matrices) while doing calculations considering that \(D_0\) can be arbitrary, the dimensionality of the spacetime must be infinite.\footnote{This property is related to the problem pointed out in the introduction, where it was shown that using results that are only valid for a particular dimension of the spacetime might lead to inconsistencies. If the Lorentz algebra is considered to be infinite dimensional during the calculations, in the previous example one cannot have a vanishing determinant.}

The four basic axioms that the \(D\)-dimensional integration operation (in an infinite dimensional spacetime) must satisfy are the following:

\begin{itemize}
    \item[1.] \textit{Linearity}:  
    \begin{equation}
         \label{linearity}
         \int{\rm d}^Dp\left[\alpha f\left(p\right)+\beta g\left(p\right)\right]=\alpha\int{\rm d}^Dpf\left(p\right)+\beta\int{\rm d}^Dpg\left(p\right).
    \end{equation}
    \item[2.] \textit{Scaling}:
    \begin{equation}
        \label{scaling}
        \int{\rm d}^Dpf\left(sp\right)=s^{-D}\int{\rm d}^Dpf\left(p\right).
    \end{equation}
    \item[3.] \textit{Translation invariance}:
    \begin{equation}
        \label{Translation_invariance}
        \int{\rm d}^Dpf\left(p+q\right)=\int{\rm d}^Dpf\left(p\right).
    \end{equation}
    \item[4.] \textit{Lorentz invariance of the results}.
\end{itemize}

\noindent Indeed, it can be shown \cite{Wilson:1972cf} that any integration satisfying these axioms is uniquely defined apart from a normalization that is usually fixed by the Gaussian prescription of the integral \(\int{\rm d}^Dp\exp\left\{-p^2\right\}=\pi^{D/2}\).

Now it is the moment to show the precise definition of the prescription for the \(D\)-dimensional integration operation. This is going to be done in the Euclidean space since one can always go from the Minkowski spacetime to it using a Wick rotation. Consider the object

\begin{equation}
    \label{D_Dim_Integral}
    I\coloneqq\int{\rm d}^D\mathbf{p}f\left(\mathbf{p},\mathbf{q}_1,\dots,\mathbf{q}_N\right),
\end{equation}

\noindent where the \(\mathbf{q}_i\) represent a set of Euclidean vectors of external momenta. One needs to give a meaning to \({\rm d}^D\mathbf{p}\) for the case where \(D\) is not an integer in order to perform the analytic continuation mentioned before. The first step of the prescription is to split the integration momentum \(\mathbf{p}\) into its part parallel to the \(\mathbf{q}_i\) and the rest: \(\mathbf{p}=\mathbf{p}_{\parallel}+\mathbf{p}_{\perp}\) so

\begin{equation}
    \label{D_Dim_Integral_2}
    I=\int{\rm d}^J\mathbf{p}_{\parallel}\int{\rm d}^{D-J}\mathbf{p}_{\perp}f\left(\mathbf{p}_{\perp},\mathbf{p}_{\parallel},\mathbf{q}_1,\dots,\mathbf{q}_N\right),
\end{equation}

\noindent where \(J\) is the dimension of the parallel subspace and it has been used that \(\mathbf{p}_{\perp}\cdot\mathbf{p}_{\parallel}=\mathbf{p}_{\perp}\cdot\mathbf{q}_i=0\) for every \(i\). Note that the parallel and perpendicular subspaces are not uniquely defined. Since the previous integral does not depend on the direction of \(\mathbf{p}_{\perp}\), it is possible to evaluate it using polar coordinates

\begin{equation}
    \label{D_Dim_Integral_3}
    I=\frac{2\pi^{\left(D-J\right)/2}}{\Gamma\left(\left(D-J\right)/2\right)}\int{\rm d}^J\mathbf{p}_{\parallel}\int_0^{\infty}{\rm d}p_{\perp}p_{\perp}^{D-J-1}f\left(p_{\perp},\mathbf{p}_{\parallel},\mathbf{q}_1,\dots,\mathbf{q}_N\right),
\end{equation}

\noindent where the factor \(2\pi^{(D-J)/2}/\Gamma((D-J)/2)\) results from the integration of the angular part in a \(D-J\) dimensional space. But now, the expression of \(I\) is well-defined even for complex values of \(D\) in terms of ordinary integrals. 

As already mentioned, the integrals should be convergent in the UV region for small enough \(D\) and clearly this is the case for Eq.~\eqref{D_Dim_Integral_3} as long as the function goes to zero fast enough for \(p_{\perp}\to\infty\). Nevertheless, if \(D\) is too small one might face problems in the region \(p_{\perp}\to0\) due to the presence of the factor \(p_{\perp}^{D-J-1}\) and find a situation where the integral does not converge for any value of the dimension, so it is necessary to give a meaning to the expressions for those cases. This is done by taking the analytic continuation to small \(D\) of the convergent integrals. For simplicity, consider the expression

\begin{equation}
    \label{Integral_Continuation}
    \int{\rm d}^D\mathbf{p}f\left(\mathbf{p}^2\right)=\frac{2\pi^{D/2}}{\Gamma\left(D/2\right)}\int_0^{\infty}{\rm d}p p^{D-1}f\left(p^2\right).
\end{equation}

\noindent Assume that it does not diverge at \(p\to\infty\) for some positive value of \(D\) and that \(f\left(p^2\right)\) is analytic at \(p\to 0\). Therefore, the integral is convergent in the region \(0<\mathfrak{Re}\left(D\right)\leq D_{\rm max}\). Now, it is possible to define an analytic continuation for negative values of the real part of \(D\) from the expression

\begin{equation}
    \label{Integral_Continuation_2}
    \begin{split}
    \int{\rm d}^D\mathbf{p}f\left(\mathbf{p}^2\right)=\frac{2\pi^{D/2}}{\Gamma\left(D/2\right)}&\left\{\int_C^{\infty}{\rm d}p p^{D-1}f\left(p^2\right)\right.\\&\left.+\int_0^C{\rm d}pp^{D-1}\left[f\left(p^2\right)-f\left(0\right)\right]+f\left(0\right)\frac{C^D}{D}\right\}.
    \end{split}
\end{equation}

\noindent This is well-defined in the range \(-2<\mathfrak{Re}\left(D\right)\leq D_{\rm max}\) and it is equal to Eq.~\eqref{Integral_Continuation} in \(0<\mathfrak{Re}\left(D\right)\leq D_{\rm max}\) for any positive value of the constant \(C\).\footnote{It will be shown that the pole at \(D=0\) is canceled by the pole of \(\Gamma\left(D/2\right)\).} In order to prove this, it is clear that the first integral converges for \(\mathfrak{Re}\left(D\right)\leq D_{\rm max}\) since the region \(p\to 0\) has been eliminated introducing the auxiliary cutoff \(C\). Now, in the second integral one can consider the Taylor expansion of the function \(f\left(p^2\right)\) around \(p^2=0\) since by assumption it is analytical at that point so: \(p^{D-1}\left[f\left(p^2\right)-f\left(0\right)\right]\sim p^{D+1}f'\left(0\right)+\mathcal{O}\left(p^{D+3}\right)\). The integral of \(p^{1+D}\) converges at \(p\to 0\) for \(-2<\mathfrak{Re}\left(D\right)\), hence the range of validity of Eq.~\eqref{Integral_Continuation_2}. Finally, in the region \(-2<\mathfrak{Re}\left(D\right)<0\) it is possible to take the limit \(C\to\infty\) and get

\begin{equation}
    \label{Integral_Continuation_3}
    \int{\rm d}^D\mathbf{p}f\left(\mathbf{p}^2\right)=\frac{2\pi^{D/2}}{\Gamma\left(D/2\right)}\int_0^{\infty}{\rm d}pp^{D-1}\left[f\left(p^2\right)-f\left(0\right)\right].
\end{equation}

It is possible to generalize this procedure for arbitrary small values of the dimension just by subtracting a bigger part of the Taylor expansion of \(f\left(p^2\right)\)

\begin{equation}
    \label{Integral_Continuation_4}
    \begin{split}
    \int{\rm d}^D\mathbf{p}f\left(\mathbf{p}^2\right)=\frac{2\pi^{D/2}}{\Gamma\left(D/2\right)}&\left\{\int_C^{\infty}{\rm d}p p^{D-1}f\left(p^2\right)\right.\\&+\int_0^C{\rm d}pp^{D-1}\left[f\left(p^2\right)-\sum_{n=0}^l\frac{f^{(n)}\left(0\right)}{n!}\left(p^2\right)^n\right]\\&\left.+\sum_{n=0}^l\frac{f^{(n)}\left(0\right)}{n!}\frac{C^{2n+D}}{2n+D}\right\}.
    \end{split}
\end{equation}

\noindent Considering that the second integrand behaves as \(p^{2l+D+1}\) as \(p\to0\) then the range of validity of the expression is \(-2l-2<\mathfrak{Re}\left(D\right)<D_{\rm max}\). Again, in the region \(-2l-2<\mathfrak{Re}\left(D\right)<-2l\) it is possible to take the limit \(C\to\infty\) to obtain

\begin{equation}
    \label{Integral_Continuation_5}
    \int{\rm d}^D\mathbf{p}f\left(\mathbf{p}^2\right)=\frac{2\pi^{D/2}}{\Gamma\left(D/2\right)}\int_0^{\infty}{\rm d}pp^{D-1}\left[f\left(p^2\right)-\sum_{n=0}^l\frac{f^{(n)}\left(0\right)}{n!}\left(p^2\right)^n\right].
\end{equation}

\noindent Just by taking a big enough \(l\) one gets an expression for the integral that is convergent for any negative real part of \(D\). Furthermore, if \(f\left(p^2\right)\) has a power law singularity at \(p^2=0\) that leads to an infrared divergence, it can also be eliminated subtracting the Laurent expansion instead of the Taylor series.

Finally, it is interesting to study what happens for even negative values of the real part of the dimension. Consider, for instance, \(D/2=-l+\varepsilon\) where \(\varepsilon\) is a small quantity. In that case, it is possible to study the divergent behavior of the integral in the range \(-2l<\mathfrak{Re}\left(D\right)<-2l+2\) at \(p\to0\):

\begin{equation}
    \label{Negative_Integer_Values}
    \begin{split}
        \int_0^{\infty}{\rm d}pp^{D-1}&\left[f\left(p^2\right)-\sum_{n=0}^{l-1}\frac{f^{(n)}\left(0\right)}{n!}\left(p^2\right)^n\right]\sim\lim_{C\to0}\int_C^{\infty}{\rm d}pp^{D-1}\frac{f^{(l)}\left(0\right)}{l!}p^{2l}\\&\sim-\frac{f^{(l)}\left(0\right)}{2l!}\lim_{C\to0}\frac{\left(C^2\right)^{D/2+l}}{\left(D/2+l\right)}\sim-\frac{f^{(l)}\left(0\right)}{2l!}\frac{1}{\varepsilon}.
    \end{split}
\end{equation}

\noindent But the factor \(2\pi^{D/2}/\Gamma\left(D/2\right)\) behaves as \(2l!\left(-\pi\right)^l\varepsilon\) around \(D/2=-l\) due to the pole of the \(\Gamma\) function. Then, the result of the integral for those cases is

\begin{equation}
     \label{Negative_Integer_Values_2}
     \int{\rm d}^{-2l}\mathbf{p}f\left(\mathbf{p}^2\right)=-\left(-\pi\right)^lf^{(l)}\left(0\right).
\end{equation}

A summary of the prescription for the \(D\)-dimensional integration operation is now given. First of all, one splits the space into the parallel and the perpendicular parts to the external momenta according to Eq.~\eqref{D_Dim_Integral_3}. Then, if there is some positive value of \(D\) for which the resulting integral converges at both \(p_{\perp}\to\infty\) and \(p_{\perp}\to0\) one takes that as the result of the operation. Otherwise, one uses Eq.~\eqref{Integral_Continuation_5} as the definition in order to eliminate the bad behavior at \(p_{\perp}\to0\) for small enough \(D\). This result is going to be a complex function of \(D\) so it is possible to make the analytical continuation into the physical dimension of the space at which a pole in the integral usually appears.

Now, there are several properties that one needs to prove in order to show that this definition of the \(D\)-dimensional integration is indeed well-defined. Only a brief comment is presented here, the rigorous demonstrations can be found at \cite{Collins_1984}. First of all, since a particular splitting of the momentum space has been chosen in order to define the integrals, it is necessary to check whether the prescription is independent of how one does it as long as the parallel space includes all of the external momenta vectors. The proof consists in showing that the combination of Eqs.~\eqref{D_Dim_Integral_3} and~\eqref{Integral_Continuation_5} remains unchanged if the dimension of the parallel space is raised. Therefore, to see the equivalence between two different splittings one only needs to embed the two different parallel spaces into a larger common one where the result of the integral is exactly the same. Furthermore, it is important to prove that the four axioms mentioned above are indeed satisfied by this prescription. The proof relies on the fact that the \(D\)-dimensional integration has been defined in terms of ordinary integrals respecting the rotational covariance (Lorentz covariance before considering the Wick rotation). Finally, it can be shown that if an integral is convergent at \(D=4\) (or at the general physical dimension of the space) this prescription gives the correct result after the analytic continuation.

For practical purposes, the only integral that is usually required at one-loop calculations\footnote{The definition of multiple-loop integrals (with several integration variables) is completely analogous to the one presented for one-loop expressions, but considering that the parallel subspace for one variable also includes the rest of the integration momenta, apart from the external ones.} is

\begin{equation}
    \label{1_Loop_Integral}
    I\left(\alpha,\beta,M^2;D\right)\coloneqq\int{\rm d}^Dp\frac{\left(p^2\right)^\alpha}{\left(p^2-M^2\right)^\beta}
\end{equation}

\noindent in the Minkowski spacetime. After performing the Wick rotation of the 0-th component of the momentum one obtains \(I\left(\alpha,\beta,M^2;D\right)=i\left(-1\right)^{\alpha-\beta}I^E\left(\alpha,\beta,M^2;D\right)\) with

\begin{equation}
    \label{1_Loop_Integral_Euclidean}
    I^E\left(\alpha,\beta,M^2;D\right)\coloneqq\int{\rm d}^D\mathbf{p}\frac{\left(\mathbf{p}^2\right)^\alpha}{\left(\mathbf{p}^2+M^2\right)^\beta}=\frac{2\pi^{D/2}}{\Gamma\left(D/2\right)}\int_0^{\infty}{\rm d}p \frac{\left(p^2\right)^{\alpha+(D-1)/2}}{\left(p^2+M^2\right)^\beta}.
\end{equation}

\noindent Convergence at \(p=0\) requires that \(-2\alpha<D\) while convergence at \(p\to\infty\) puts the constraint \(D<2\beta-2\alpha\). The evaluation of the integral is rather simple if \(\alpha>0\) and \(\beta>\alpha\), because in that case there is a region of positive \(D\) for which there are no divergences. In that situation one can implement the usual integration techniques:

\begin{equation}
    \label{1_Loop_Integral_Euclidean_2}
    \begin{split}
    I^E\left(\alpha,\beta,M^2;D\right)&=\frac{2\pi^{D/2}}{\Gamma\left(D/2\right)}\int_0^{\infty}{\rm d}p \frac{\left(p^2\right)^{\alpha+(D-1)/2}}{\left(p^2+M^2\right)^\beta}\hspace{0.3cm}\left|\begin{array}{l}
    u=p^2/M^2\\{\rm d}u=2p{\rm d}p/M^2 \end{array}\right|\\ \\&=\frac{\pi^{D/2}}{\Gamma\left(D/2\right)}\left(M^2\right)^{\alpha-\beta+D/2} \int_0^{\infty}{\rm d}u\frac{u^{\alpha-1+D/2}}{\left(1+u\right)^{\beta}} \hspace{0.3cm}\left|\begin{array}{l}
    x=1/(1+u)\\{\rm d}x=-x^2{\rm d}u \end{array}\right| \\ \\&=\frac{\pi^{D/2}}{\Gamma\left(D/2\right)}\left(M^2\right)^{\alpha-\beta+D/2}\int_0^1{\rm d}xx^{\beta-\alpha-D/2-1}\left(1-x\right)^{\alpha+D/2-1} \\ \\&=\frac{\pi^{D/2}}{\Gamma\left(D/2\right)}\left(M^2\right)^{\alpha-\beta+D/2}\frac{\Gamma\left(\beta-\alpha-D/2\right)\Gamma\left(\alpha+D/2\right)}{\Gamma\left(\beta\right)}.
    \end{split}
\end{equation}

\noindent Therefore, the complete result of the integral in the Minkowski spacetime for that range of parameters is

\begin{equation}
    \label{1_Loop_Integral_Result}
    I\left(\alpha,\beta,M^2;D\right)=i\left(-1\right)^{\alpha-\beta}\pi^{D/2}\left(M^2\right)^{\alpha-\beta+D/2}\frac{\Gamma\left(\beta-\alpha-D/2\right)\Gamma\left(\alpha+D/2\right)}{\Gamma\left(\beta\right)\Gamma\left(D/2\right)}.
\end{equation}

\noindent But this is a definition of \(I\left(\alpha,\beta,M^2;D\right)\) as an analytic function for all values of its parameters, so this can be taken as the general result coming from the analytic continuation. An important particular case is the one with \(\beta=0\) and arbitrary \(\alpha\). In that situation, since \(\Gamma\left(\beta\right)\sim1/\beta\) near that pole, the result vanishes. Thus, in the dimensional regularization prescription

\begin{equation}
    \label{Integral_Power_of_p}
    \int{\rm d}^Dp \left(p^2\right)^{\alpha}=0.
\end{equation}

\noindent An important particular case of this is \(\int{\rm d}^Dp=0\).

Usually, any other one-loop integral can be reduced to the one in Eq.~\eqref{1_Loop_Integral} using tensor contractions, Lorentz invariance or techniques like Feynman parametrization. In order to use these properties it is necessary to define the \(D\)-dimensional versions of the objects of the theory. First of all, consider the Euclidean metric tensor \(\delta_{ij}\) (the definition is completely analogous in the Minkowski space). The two main properties that one might require for the definition are that the contraction with itself gives the dimension \(\delta_{ij}\delta^{ij}=D\) and that it commutes with integration in such a way that

\begin{equation}
    \label{Commutation_Integration}
    \delta_{ij}\int{\rm d}^D\mathbf{p}p^ip^jf\left(\mathbf{p}^2\right)=\int{\rm d}^D\mathbf{p}\mathbf{p}^2f\left(\mathbf{p}^2\right).
\end{equation}

\noindent But having for instance \(\delta_{ij}\delta^{ij}=D\) is highly non-trivial considering the fact that, indeed, the dimension of the space is infinite. One can define the action of the covariant 2-tensor \(\delta\) on a contravariant tensor \(T\) in terms of its components as \(\delta\left(T\right)=\sum_iT^{ii}\), but since the dimension of the space is not finite there is no guarantee that this sum converges. The definition needs to give that result if the sum converges and satisfy the other properties in the general case.

The usual definition\footnote{In the Minkowski space an additional \(-i\) factor is needed in order to cancel the one produced by the Wick rotation.} of the action of the metric tensor \(\delta\) in dimensional regularization is given in terms of the following integral with the Dirac delta:

\begin{equation}
    \label{Definition_Metric_Tensor}
    \delta\left(T\right)\coloneqq\frac{D\Gamma\left(D/2\right)}{\pi^{D/2}}\int{\rm d}^D\mathbf{p}\sum_{ij}T^{ij}p_ip_j\delta^{(1)}\left(\mathbf{p}^2-1\right).
\end{equation}

\noindent If the sum \(\sum_{ij}T^{ij}\) is well-defined it commutes with ordinary integration and one gets

\begin{equation}
    \label{Definition_Metric_Tensor_1}
    \delta\left(T\right)\coloneqq\frac{D\Gamma\left(D/2\right)}{\pi^{D/2}}\sum_{ij}T^{ij}\int{\rm d}^D\mathbf{p}p_ip_j\delta^{(1)}\left(\mathbf{p}^2-1\right).
\end{equation}

\noindent In order to evaluate that integral one considers \(p_i\) and \(p_j\) as a part of the parallel space. For instance, if \(i\neq j\):

\begin{equation}
    \label{Integral_Evaluation_metric_tensor}
    \int{\rm d}^D\mathbf{p}p_ip_j\delta\left(\mathbf{p}^2-1\right)=\int_{-\infty}^{\infty}{\rm d}p_ip_i\int_{-\infty}^{\infty}{\rm d}p_jp_j\int{\rm d}^{D-2}\mathbf{p}_{\perp}\delta^{(1)}\left(p_i^2+p_j^2+\mathbf{p}^2_{\perp}-1\right)=0,
\end{equation}

\noindent which vanishes due to the parity of the integrals. If \(i=j\) then

\begin{equation}
    \label{Integral_Evaluation_metric_tensor_1}
    \begin{split}
    \int{\rm d}^D\mathbf{p}p_ip_j\delta^{(1)}\left(\mathbf{p}^2-1\right)&=\int_{-\infty}^{\infty}{\rm d}p_i\int{\rm d}^{D-1}\mathbf{p}_{\perp}p_i^2\delta^{(1)}\left(p_i^2+\mathbf{p}^2_{\perp}-1\right)\\&=\int_{-\infty}^{\infty}{\rm d}p_i\int{\rm d}^{D-1}\mathbf{p}_{\perp}\sum_{j=1}^D\frac{p_j^2}{D}\delta^{(1)}\left(p_i^2+\mathbf{p}^2_{\perp}-1\right) \\&=\frac{1}{D}\int{\rm d}^D\mathbf{p}\mathbf{p}^2\delta^{(1)}\left(\mathbf{p}^2-1\right)=\frac{2\pi^{D/2}}{D\Gamma\left(D/2\right)}\int_0^{\infty}{\rm d}pp^{D+1}\delta^{(1)}\left(p^2-1\right) \\&=\frac{\pi^{D/2}}{D\Gamma\left(D/2\right)}\int_0^{\infty}{\rm d}p^2\left(p^2\right)^{D/2}\delta^{(1)}\left(p^2-1\right)=\frac{\pi^{D/2}}{D\Gamma\left(D/2\right)}.
    \end{split}
\end{equation}

\noindent Therefore, the result of Eq.~\eqref{Definition_Metric_Tensor_1} is just \(\sum_iT^{ii}\) if the trace converges. Nonetheless, this definition can be used even for situations where this is not the case. For instance, one can consider

\begin{equation}
    \label{Metric_Tensor_Contraction}
    \begin{split}
    \delta\left(\delta\right)&=\frac{D\Gamma\left(D/2\right)}{\pi^{D/2}}\int{\rm d}^D\mathbf{p}\sum_{ij}\delta^{ij}p_ip_j\delta^{(1)}\left(\mathbf{p}^2-1\right)= \frac{D\Gamma\left(D/2\right)}{\pi^{D/2}}\int{\rm d}^D\mathbf{p}\mathbf{p}^2\delta^{(1)}\left(\mathbf{p}^2-1\right)\\&=2D\int_0^{\infty}{\rm d}pp^{D+1}\delta^{(1)}\left(p^2-1\right)=D\int_{0}^{\infty}{\rm d}p^2\left(p^2\right)^{D/2}\delta^{(1)}\left(p^2-1\right)=D.
    \end{split}
\end{equation}

\noindent This is one of the required properties for the metric tensor, the contraction with itself must give the dimension \(D\). The commutation with the integration operation is just a consequence of the fact that it is possible to exchange the order of integration \cite{Collins_1984}.

Finally, it is also necessary to define the \(\gamma^{\mu}\) matrices in \(D\) dimensions in order to do calculations in theories with fermions. Their basic properties are:

\begin{equation}
    \label{Anticommutator_Gamma_Matrices}
    \left\{\gamma_{\mu},\gamma_{\nu}\right\}=2\eta_{\mu\nu}\mathbb{I}
\end{equation}

\noindent and

\begin{equation}
    \label{Conjugate_Gamma_Matrix}
    \gamma_{\mu}^{\dagger}=\gamma_{0}\gamma_{\mu}\gamma_{0}.
\end{equation}

\noindent However, since the dimension of the Lorentz algebra is not finite the index \(\mu\) runs from zero to infinity and a definition of this infinite-dimensional matrices is required. This is done by induction being the base case, for instance, the well-known \(\gamma^{\mu}\) matrices in four dimensions. Consider a representation of dimension \(2^{\omega}\) of the Dirac algebra \(\gamma^{\mu}_{(\omega)}\) with index \(0\leq\mu\leq2\omega-1\). Then, the associated infinite dimensional \(\gamma^{\mu}\) matrix is defined as

\begin{equation}
    \label{Infinite_Dimensional_Gamma}
    \gamma^{\mu}\coloneqq\left(\begin{array}{ccc}
    \gamma_{(\omega)}^{\mu} & 0 & \cdots \\
    0 &  \gamma^{\mu}_{(\omega)} & \cdots \\
    \vdots & \vdots & \ddots
    \end{array}\right),
\end{equation}

\noindent for \(0\leq\mu\leq2\omega-1\).  To extend Eq.~\eqref{Infinite_Dimensional_Gamma} to higher values of the \(\mu\) index one defines by induction a representation of dimension \(2^{\omega+1}\) of the Dirac algebra \(\gamma^{\mu}_{(\omega+1)}\). For example, in the case \(0\leq\mu\leq2\omega-1\) this is done as

\begin{equation}
    \label{Next_Representation}
    \gamma^{\mu}_{(\omega+1)}\coloneqq\left(\begin{array}{cc}
    \gamma_{(\omega)}^{\mu} & 0  \\
    0 &  \gamma^{\mu}_{(\omega)} 
    \end{array}\right),
\end{equation}

\noindent an expression that is consistent with Eq.~\eqref{Infinite_Dimensional_Gamma}. It is possible to construct the object

\begin{equation}
    \label{Gamma_5_like}
    \tilde{\gamma}_{(\omega)}\coloneqq i^{\omega-1}\gamma_{(\omega)}^0\cdots\gamma_{(\omega)}^{2\omega-1},
\end{equation}

 \noindent which for \(\omega=2\) is the \(\gamma_5\) matrix and generally speaking it satisfies \(\tilde{\gamma}_{(\omega)}^{\dagger}=\tilde{\gamma}_{(\omega)}\), \(\tilde{\gamma}_{(\omega)}^2=\mathbb{I}\) and \(\{\tilde{\gamma}_{(\omega)},\gamma_{(\omega)}^{\mu}\}=0\). Using \(\tilde{\gamma}_{(\omega)}\) the \(\gamma_{(\omega+1)}^{2\omega}\) and \(\gamma_{(\omega+1)}^{2\omega+1}\) matrices can be chosen to be

\begin{equation}
    \label{Next_Representation_1}
    \gamma^{2\omega}_{(\omega+1)}\coloneqq\left(\begin{array}{cc}
    0 &  \tilde{\gamma}_{(\omega)}  \\
     -\tilde{\gamma}_{(\omega)} & 0 
    \end{array}\right)
\end{equation}

\noindent and 

\begin{equation}
    \label{Next_Representation_2}
    \gamma^{2\omega+1}_{(\omega+1)}\coloneqq\left(\begin{array}{cc}
    0 &  i\tilde{\gamma}_{(\omega)}  \\
     i\tilde{\gamma}_{(\omega)} & 0 
    \end{array}\right).
\end{equation}

\noindent It is straightforward to prove that these two matrices satisfy the properties of Eqs.~\eqref{Anticommutator_Gamma_Matrices} and~\eqref{Conjugate_Gamma_Matrix}. Using induction and Eq.~\eqref{Infinite_Dimensional_Gamma} this completes the definition of the infinite dimensional \(\gamma^{\mu}\) matrices for all \(\mu\) index values.

It is well-known that using Eqs.~\eqref{Anticommutator_Gamma_Matrices} and~\eqref{Conjugate_Gamma_Matrix} it is enough to simplify any product of \(\gamma^{\mu}\) matrices. However, it is still necessary to define the trace of products of these objects given the fact that since they are infinite dimensional the usual sum of the diagonal elements does not usually converge. Consequently, \textit{it is possible to define a trace that does not satisfy the usual linear or cyclic properties} \cite{Kreimer:1989ke,Korner:1991sx}. Nonetheless, in order to have a mathematically consistent dimensional regularization scheme it is better to preserve them both. Therefore, one can introduce the following definition

\begin{equation}
    \label{Definition_Trace}
    {\rm tr}\{M\}\coloneqq\left({\rm tr}\{\mathbb{I}\}\right)\lim_{N\to\infty}\frac{1}{N}\sum_{i}^NM^{ii}
\end{equation}

\noindent from which the usual properties can be easily derived. Given the fact that there is no \(D\)-dependence in the definition of the \(\gamma^{\mu}\) matrices, the only dependence on the dimension in the trace is in \({\rm tr}\{\mathbb{I}\}=f\left(D\right)\). Different choices of the function \(f\) lead to a renormalization group transformation \cite{Collins_1984}. The usual convention is to take \(f\left(D\right)=4\), i.e., the physical dimension.

The \(D\)-dimensional definition of the objects \(\gamma_5\) and \(\varepsilon_{\mu\nu\rho\sigma}\) that appear in chiral theories requires special considerations that will be presented in Sec.~\ref{sec:BMHV}.

\section{Renormalization Theory: BPH Prescription, Forest Formula and Weinberg's Theorem} \label{sec:Weinberg_Theorem}
\markboth{2. THE BMHV DIMENSIONAL REGULARIZATION SCHEME}{2.2. BPH PRESCRIPTION, FOREST FORMULA AND WEINBERG'S THEOREM}
As shown in the first chapter, if the renormalization procedure preserves the symmetries of a gauge theory then it is possible to absorb all the divergences appearing in perturbative calculations by adding the proper set of counterterms to the Lagrangian. But this result relies on the assumption that \textit{all the infinities in the integrals can be written as polynomials in the external momenta and the masses of the particles involved}, otherwise it is not possible to find a local polynomial in the fields that generates the needed counterterms. In this section a sketch of the proof that divergences are local is presented, following the work done in \cite{Collins_1984}. This result is known as \textit{Weinberg's theorem} \cite{Weinberg:1959nj}.

First of all, it is necessary to define a renormalization scheme, i.e., a prescription to isolate and subtract the divergent part of the amplitudes. Given the fact that in dimensional regularization the infinities appear as poles of the \(\Gamma\) function around \(D=4\) \cite{Speer:1974cz,Breitenlohner:1977hr,Breitenlohner:1975hg,Breitenlohner:1976te}, it is natural to use the \textit{Minimal Subtraction} (MS) scheme \cite{tHooft:1973mfk}, where the divergent parts of the integrals are associated just with the terms of the Laurent expansion of the \(\Gamma\) function that contain negative powers of \((D-4)\). This is the prescription that will be used in this section, but there are infinitely many other possibilities. For instance, since the presence of the coefficient \(\ln(4\pi)-\gamma_E\approx1.954\) coming from the 0-th order term of the Laurent series of the \(\Gamma\) function leads to a slower convergence of the perturbative expansion, it is also natural to use the \(\overline{\rm MS}\) scheme \cite{Bardeen:1978yd}, where those factors are also considered as a part of the divergence. Other possibilities include the on-shell schemes, which are based on the idea that certain amplitudes might lead directly to the measurable physical quantities of the theory, like masses, charges, etc.

\begin{figure}
    \centering
    \includegraphics[scale=0.18]{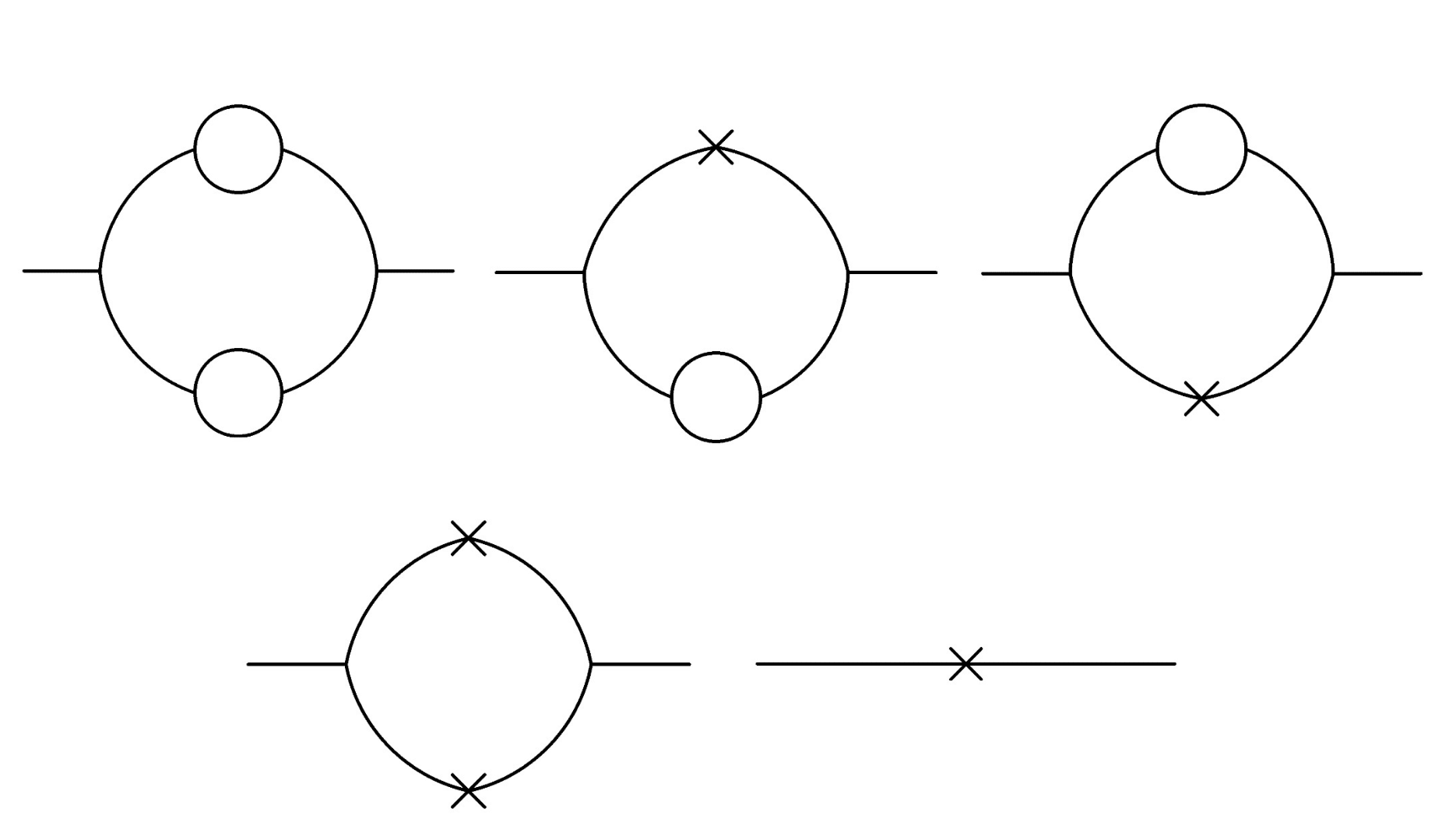} 
    \caption[Example of a three-loop contribution (top-left panel) to the corrections of the propagator in a scalar theory with cubic interactions.]{Example of a three-loop contribution (top-left panel) to the corrections of the propagator in a scalar theory with cubic interactions. The other Feynman diagrams are the ones with the counterterm-insertions needed to eliminate both the overall divergence and subdivergences.}
    \label{fig:3_Loop_Diagrams}
\end{figure}

It is not difficult to show that the Weinberg's theorem is valid for one-loop amplitudes, the problem arises when considering integrals that involve several integration momenta. Looking at Eq.~\eqref{1_Loop_Integral_Result} it is clear that \(D\)-dimensional one-loop integrals lead to a result which is a polynomial in some mass scale \(M^2\), that can include the external momenta. When expanding that factor around \(D=4\) one usually obtains something like \((M^2)^{a}(1+(D-4)\ln(M^2/\mu^2)+\cdots)\), where \(\mu\) is a mass scale required by the dimensional regularization prescription as it will be shown in Sec.~\ref{Sec:QAP}. Considering that one-loop integrals can have at most a simple pole at \(D=4\) due to the presence of one \(\Gamma\) function, the full result has the form \((M^2)^{a}(1/(D-4)+\ln(M^2/\mu^2)+\cdots)\), so in the MS scheme the divergences are just polynomials in \(M^2\) that can be eliminated adding local operators of the fields to the Lagrangian. But consider a two-loop amplitude that can be computed just by applying the result in Eq.~\eqref{1_Loop_Integral_Result} two times, one for each integration variable. In that situation one still has the \((M^2)^{a}(1+(D-4)\ln(M^2/\mu^2)+\cdots)\) dependence, but now there might be two divergent \(\Gamma\) functions, one of which comes from the case where one integration variable goes to infinity (performing the first integral) while the other one remains finite. This is a so-called \textit{subdivergence} and repeating the reasoning done before it leads to a term of the form \(\ln(M^2/\mu^2)/(D-4)\) in the result. This divergence is not a local polynomial in the external momenta if \(M^2\) depends on some \(p^2\). The fact is that, apart from the cancellation of the overall divergence (the case where all the integration momenta go to infinity at the same time), cancellations of the subdivergences are also needed. This is naturally done when considering divergent diagrams that contain some counterterms as one of their interaction vertices (see the example in Fig.~\ref{fig:3_Loop_Diagrams}). Thus, \textit{a way to systematically cancel the subdivergences is needed in order to prove Weinberg's theorem}.

The first algorithm to cancel all divergences that it is going to be presented here was introduced by Bogoliubov and Parasiuk \cite{Bogoliubov:1957gp} with corrections from Hepp \cite{Hepp:1966eg}. It is called the \textit{BPH prescription} and it relies on a recursive definition of a renormalized amplitude. First of all, it is necessary to define a subtraction operator \(T\) in such a way that \(T\circ A\) gives the divergent part of a certain amplitude \(A\). The definition of this operator depends on the renormalization scheme that one is using. In the MS scheme with dimensional regularization \(T\) isolates the poles in \(D-4\) of the integral. Now consider a Feynman diagram \(G\), in the BPH prescription the value of its associated renormalized amplitude \(R\left(G\right)\) is

\begin{equation}
    \label{BPH_Renormalized_Amplitude}
    R\left(G\right)=\bar{R}\left(G\right)-T\circ\bar{R}\left(G\right),
\end{equation}

\noindent where \(\bar{R}\left(G\right)\) is the amplitude with all of its subdivergences subtracted. Therefore, \(T\circ\bar{R}\left(G\right)\) isolates the overall divergence, the one where all of the internal momenta get large at the same rate. Furthermore, \(\bar{R}\left(G\right)\) is defined as

\begin{equation}
    \label{BPH_Renormalized_Amplitude_1}
    \bar{R}\left(G\right)=U\left(G\right)+\sum_{\gamma\varsubsetneq G}C_{\gamma}\left(G\right),
\end{equation}

\noindent where \(U\left(G\right)\) is the original unrenormalized amplitude and \(\gamma\) represents any subgraph of \(G\) (any subset of internal momenta) except \(G\) itself. The counterterm \(C_{\gamma}\left(G\right)\) eliminates the overall divergence of the subgraph \(\gamma\) when all of its subdivergences have already been subtracted. This is the reason why this definition is recursive and indeed this process ends when there are no further divergences in any subgraph and then the counterterm is just zero. 

The second algorithm was introduced by Zimmermann \cite{Zimmermann:1969jj} and it solves the BPH recursion leading to the so-called \textit{forest formula}. The idea is to consider all the sets (including the empty set) whose elements are connected 1PI subgraphs of a certain graph \(G\), in such a way that if two subgraphs belong to the same set they are either nested or disjoint, i.e., non-overlapping. Such sets are called \textit{forests}, and the point is that, by construction, there is a one-to-one correspondence between them and sets of all the lines of the graph in which some collection of internal momenta \(k_1\), \(k_2\), \(\dots\), \(k_n\) appear. Consequently, \textit{different forests are associated with different possible subdivergences}. The set of all the forests of a graph \(G\) is denoted by \(\mathcal{F}\left(G\right)\) and if \(G\) itself does not belong to a particular forest then it is called a \textit{normal forest}, otherwise it is said to be a \textit{full forest}.

Given these definitions, the forest formula for the renormalized value of a certain amplitude associated with a particular graph \(G\) is simply

\begin{equation}
    \label{forest_formula}
    R\left(G\right)=\sum_{F\in\mathcal{F}\left(G\right)}\prod_{\gamma\in F}\left(-T_{\gamma}\right)U\left(G\right),
\end{equation}

\noindent where the operator \(T_{\gamma}\) eliminates the overall divergence associated with the subgraph \(\gamma\). It is important to remark that the product of subtraction operators goes from smaller to bigger graphs if they are nested. Furthermore, notice that the unrenormalized amplitude \(U\left(G\right)\) appears when considering the forest \(F=\emptyset\). It is still necessary to prove that the values of \(R\left(G\right)\) in Eqs.~\eqref{BPH_Renormalized_Amplitude} and~\eqref{forest_formula} coincide given the fact that the BPH algorithm gives, by definition, the value of a renormalized amplitude.

The proof of the forest formula goes as follows: assume that

\begin{equation}
    \label{forest_formula_proof}
    \bar{R}\left(G\right)=\sum_{F\in\bar{\mathcal{F}}\left(G\right)}\prod_{\gamma\in F}\left(-T_{\gamma}\right)U\left(G\right),
\end{equation}

\noindent where  \(\bar{\mathcal{F}}\left(G\right)\) is the set of all of the normal forests of \(G\). Then, if this is true, it is clear that

\begin{equation}
    \label{forest_formula_proof_1}
    \begin{split}
    R\left(G\right)&=\sum_{F\in\mathcal{F}\left(G\right)}\prod_{\gamma\in F}\left(-T_{\gamma}\right)U\left(G\right)\\&=\sum_{F\in\bar{\mathcal{F}}\left(G\right)}\prod_{\gamma\in F}\left(-T_{\gamma}\right)U\left(G\right)+\sum_{F\in\mathcal{F}(G)/\bar{\mathcal{F}}\left(G\right)}\prod_{\gamma\in F}\left(-T_{\gamma}\right)U\left(G\right)\\&=\bar{R}\left(G\right)-T_{G}\circ\sum_{F\in\bar{\mathcal{F}}\left(G\right)}\prod_{\gamma\in F}\left(-T_{\gamma}\right)U\left(G\right)=\bar{R}\left(G\right)-T_{G}\circ \bar{R}\left(G\right),
    \end{split}
\end{equation}

\noindent where it has been used that the set of full forest can be decomposed as unions of \(\left\{G\right\}\) with normal forests and the fact that, by construction, \(T_G\) is always the last operator to act. But this is precisely the expression in Eq.~\eqref{BPH_Renormalized_Amplitude}, so indeed the only result that still needs to be shown is that Eq.~\eqref{forest_formula_proof} coincides with Eq.~\eqref{BPH_Renormalized_Amplitude_1}. For this purpose, it is clear that for every forest one can always find a unique set of biggest graphs \(\left\{M_1,\dots,M_n\right\}\) in the sense that they are disjoint \(M_i\cap M_j=\emptyset\) and any other subgraph belongs to some of them. Therefore, it is possible to write Eq.~\eqref{forest_formula_proof} as 

\begin{equation}
    \label{forest_formula_proof_2}
    \begin{split}
    \bar{R}\left(G\right)=U\left(G\right)&+\sum_{M_1,\dots,M_n}\prod_{i=1}^{n}\left(-T_{M_i}\right)\left\{\sum_{F_1\in\bar{\mathcal{F}}\left(M_1\right)}\cdots\sum_{F_n\in\bar{\mathcal{F}}\left(M_n\right)}\prod_{\gamma_1\in F_1}\left(-T_{\gamma_1}\right)\times\right.\\&\left.\cdots\times\prod_{\gamma_n\in F_n}\left(-T_{\gamma_n}\right)U\left(G\right)\right\},
    \end{split}
\end{equation}

\noindent where the term \(U\left(G\right)\) again comes from the forest associated with the empty set. Notice that the order of application of the subtraction operators for disjoint graphs is irrelevant. To complete the proof it is necessary to use the induction hypothesis in the size of the graphs. Eq.~\eqref{forest_formula_proof} is clearly true for the base case where there is only one divergent subgraph in \(G\), because that means that \(\bar{\mathcal{F}}\left(G\right)=\left\{\emptyset,\left\{\gamma\right\}\right\}\) and \(\bar{R}\left(G\right)=U\left(G\right)-T_{\gamma}U\left(G\right)\). Now assume that the expression holds for bigger graphs up to a certain number on internal divergent subgraphs. Then, the terms \(\sum_{F_i\in\bar{\mathcal{F}}\left(M_i\right)}\prod_{\gamma_i\in F_i}\left(-T_{\gamma_i}\right)U\left(G\right)\) in Eq.~\eqref{forest_formula_proof_2} give the amplitude \(U\left(G\right)\) but subtracting all the subdivergences of the subgraphs of \(M_{i}\). Finally, the product \(\prod_{i=1}^n\left(-T_{M_i}\right)\) eliminates the overall divergences of the subgraphs \(M_i\) as required by the BPH algorithm. This completes the proof.

The BPH prescription can be used to prove Weinberg's theorem, which consists of two statements:
\begin{itemize}
    \item[1.]\textit{Consider a 1PI graph \(G\) with negative degree of divergence \(\delta\left(G\right)\).\footnote{The overall degree of divergence \(\delta\left(G\right)\) is determined just by power counting of the integration momenta in the numerator and denominator of the integrands.} Then the graph is finite once all its subdivergences have been removed, i.e., \(\bar{R}\left(G\right)\) is finite.}
    \item[2.] \textit{The overall divergence of a 1PI graph \(G\) is a polynomial of degree \(\delta\left(G\right)\) in the external momenta and masses of the involved particles.}
\end{itemize}

In order to deal with the masses of the inner particles in the propagators it is possible to use an argument shown and developed in more detail in \cite{Collins:1974bg}. The idea is that one can expand the denominators in powers of the masses as

\begin{equation}
    \label{propagator_expansion}
    \frac{1}{p^2-m^2}=\frac{1}{p^2}+\frac{m^2}{p^4}+\frac{m^4}{p^6}+\mathcal{O}\left(m^8\right)
\end{equation}

\noindent up to a term where the integral is finite just by power counting. As a consequence, it is easy to see that proving Weinberg's theorem for massless theories (the counterterms are polynomials in the external momenta) is enough to show that the result holds in the massive scenario (the counterterms are polynomials in the external momenta and masses).

To prove the theorem it is convenient to start with the first statement. Consider a 1PI graph \(G\) with negative degree of divergence. Given the BPH prescription it is clear that all the subdivergences coming from regions where some momenta go to infinity at the same rate, while others remain finite, are eliminated leading to a finite amplitude for \(G\). However, it is still necessary to show that no further divergences coming from the case where several momenta go to infinity at different rates can appear. This scenario is studied using the induction hypothesis. Assume that Weinberg's theorem holds for the subgraphs of \(G\), being the base case the one-loop diagrams. By this hypothesis, the regions of integration where some momenta remain finite while the others go to infinity not necessarily at the same rate do not contribute to the divergence once the counterterms are added. Furthermore, since \(G\) has a negative degree of divergence, it remains finite when all the momenta go to infinity at the same rate. The only case that still needs to be studied is the one where all the momenta go to infinity but at different rates. 

To study this possibility it is convenient to denote all the large momenta of the integral as \(K\)  and use \(k\) for all the smaller ones. Now, one performs the integration over the \(K\)s expanding the integrand for \(K\gg k\). The integration \(\int_{K>k}{\rm d}K\) in this region can be decomposed as \(\int_{k}^{\infty}{\rm d}K=\int_{0}^{\infty}{\rm d}K-\int_{0}^k{\rm d}K\) and the possible infinity produced by the first integral is canceled when adding the appropriate counterterm. Moreover, assuming Weinberg's theorem for smaller graphs, the second term produces a polynomial in \(k\) up to the degree of divergence of the subgraphs. As a consequence, this first integration leads to a polynomial in \(k\) whose degree can be determined by power counting of the number of momenta \(K\). Given the fact that the superficial degree of divergence of the graph \(G\) is negative, the final integration over \(k\) does not produce a divergence. This completes the proof of the first statement of Weinberg's theorem.

To finish the proof by induction, it is also necessary to show that the second statement of the theorem is true, but now it is possible to use the first one. Consider again a 1PI graph \(G\) of the same size, now with superficial degree of divergence \(\delta\left(G\right)>0\). The renormalized amplitude can be written using the BPH algorithm as

\begin{equation}
    \label{Renormalized_Weinberg_Amplitude}
    R\left(G\right)=U\left(G\right)+\sum_{\gamma\subseteq G}C_{\gamma}\left(G\right)=\bar{R}\left(G\right)+C\left(G\right),
\end{equation}

\noindent where \(C\left(G\right)\) is the counterterm needed to eliminate the overall divergence of \(G\). Now, it is necessary to differentiate \(\delta\left(G\right)+1\) times with respect to the external momenta, because each differentiation lowers the degree of divergence of an amplitude by one unit. To see this one can simply take the particular example of \(\partial_p\int{\rm d}^Dk\frac{1}{(k+p)^2}=-2\int{\rm d}^Dk\frac{1}{(k+p)^3}\). The differentiation is applied both in the amplitude and in the counterterms, so the crucial point here is whether the derivatives with respect to the external momenta of the \(C_{\gamma}(G)\) are the needed counterterms for the derivatives of the amplitude. For the MS scheme with dimensional regularization this is clearly the case, because integration and differentiation commute and the counterterms needed only have to eliminate the derivatives with respect to \(p\) of the poles of the function, which are the derivatives of the original counterterms. As a consequence, \(\partial^{\alpha}_p\bar{R}\left(G\right)=\bar{R}\left(\partial_{p}^{\alpha}G\right)\) and differentiating \(\delta\left(G\right)+1\) times one gets an amplitude with negative degree of divergence and all the subdivergences subtracted. The first result of Weinberg's theorem states that this amplitude is finite. Therefore, the \(\left(\delta\left(G\right)+1\right)\)-th derivative with respect to \(p\) of the overall counterterm has to vanish, which means that it is a polynomial of degree \(\delta\left(G\right)\) in the external momenta. \textit{This completes the proof of Weinberg's theorem and hence the proof of the fact that gauge theories are renormalizable}.

\section{Properties of Dimensional Regularization: the Quantum Action Principle and Algebraic Renormalization} \label{Sec:QAP}

\markboth{2. THE BMHV DIMENSIONAL REGULARIZATION SCHEME}{2.3. QUANTUM ACTION PRINCIPLE AND ALGEBRAIC RENORMALIZATION}

A renormalization prescription not only needs to be able to eliminate all the divergences that appear in the calculations of the amplitudes, but also has to preserve fundamental properties of QFTs. Some of the most important ones are summarized in the review \cite{Belusca-Maito:2023wah} and they are:

\begin{itemize}
    \item[1.] \textit{Unitarity}. The \(S\) matrix must satisfy the relation
    \begin{equation}
        \label{unitarity}
        SS^{\dagger}=S^{\dagger}S=\mathbb{I}.
    \end{equation}

    \item[2.] \textit{Poincaré Invariance}. If \(U(\Lambda,a)\) is a representation in the Hilbert space of a Poincaré transformation with translation parameter \(a\) and associated Lorentz transformation \(\Lambda\), then
    \begin{equation}
        \label{Poincaré_Invariance}
        U\left(\Lambda,a\right)SU\left(\Lambda,a\right)^{-1}=S.
    \end{equation}

    \item[3.] \textit{Causality}. The \(S\) matrix can be written in terms of the operators of the Lagrangian \(\mathcal{L}\left(x\right)=\sum_i\mathcal{O}_i\left(x\right)\) using its Dyson series \cite{Dyson:1949ha} as:
    \begin{equation}
        \label{Dyson_Series}
        S=\mathbb{I}+\sum_{n=1}^{\infty}\frac{i^n}{n!}\int{\rm d}^4x_1\cdots{\rm d}^4x_n\sum_{i_1,\dots, i_n}T\left[\mathcal{O}_{i_1}\left(x_1\right)\cdots\mathcal{O}_{i_n}\left(x_n\right)\right],
    \end{equation}
    \noindent where \(T\left[\cdot\right]\) represents the time-ordered product. The requirement of causality can be expressed in terms of it as
    \begin{equation}
        \label{Causality}
        \begin{split}
        T\left[\mathcal{O}_{i_1}\left(x_1\right)\cdots\mathcal{O}_{i_n}\left(x_n\right)\right]&=T\left[\mathcal{O}_{i_{j_1}}\left(x_{j_1}\right)\cdots\mathcal{O}_{i_{j_m}}\left(x_{j_m}\right)\right]\\&\times T\left[\mathcal{O}_{i_{j_{m+1}}}\left(x_{j_{m+1}}\right)\cdots\mathcal{O}_{i_{j_n}}\left(x_{j_n}\right)\right],
        \end{split}
    \end{equation}
    \noindent if the set \(\left\{x_{j_1},\dots,x_{j_m}\right\}\) is out of the light-cone of the set \(\left\{x_{j_{m+1}},\dots,x_{j_n}\right\}\).
\end{itemize}

A proof that it is possible to construct a renormalization prescription that satisfies all those properties was presented in \cite{Bogoliubov:1957gp,Hepp:1966eg} and it is based on the BPH algorithm. One important case was developed by Zimmermann \cite{Zimmermann:1969jj}: it is called the BPHZ scheme and it relies on the forest formula. The idea of the BPH scheme is that, since divergences are polynomials of degree \(\delta(G)\) in the external momenta according to Weinberg's theorem, one can simply eliminate those parts of the amplitudes taking the Taylor expansion around \(p=0\) of the results of the integrals. The BPHZ scheme goes even further and does not even require to do the integration, because those terms in the Taylor expansion can already be identified at the integrand level and it is possible to eliminate them from the beginning. In order to deal with subdivergences one can simply use the forest formula. This BPHZ prescription is extremely interesting since the integrals that one needs to perform are already finite, which means that no regulator is needed, showing that the physical results of a renormalization procedure should not depend on the method that one uses. This crucial result is summarized in the main theorem of renormalization theory: \textit{any two renormalization constructions satisfying the requirements of unitarity, causality and Poincaré invariance can only differ by a finite redefinition of the parameters of the Lagrangian} \cite{Popineau:2016lhf}. Furthermore, as shown in \cite{Breitenlohner:1977hr}, dimensional regularization with minimal subtraction also satisfies these requirements and therefore it is equivalent to the BPHZ prescription.

The principal interest of this section is to study another fundamental property that may be satisfied by a renormalization procedure: the \textit{Quantum Action Principle}. This is a relation that plays a fundamental role in the study of symmetry-breaking, particularly in the context of algebraic renormalization where, along with the Wess-Zumino consistency conditions, is one of the two main ingredients. The validity of this principle in the context of dimensional regularization was first established in \cite{Breitenlohner:1977hr}. The derivation that is going to be shown here was presented in \cite{Stockinger:2005gx}.

As mentioned in \cite{Belusca-Maito:2023wah}, one of the main advantages of dimensional regularization is that this prescription can be studied from the Lagrangian point of view. In this sense, the basic relation of perturbation theory to calculate \(n\)-point Green's functions can be written as

\begin{equation}
    \label{Perturbation_Theory}
    \left<\Omega\right|T\left[\phi_1\left(x_1\right)\cdots\phi_n\left(x_n\right)\right]\left|\Omega\right>=\frac{\left<0\right|T\left[\phi_1\left(x_1\right)\cdots\phi_n\left(x_n\right)\exp\left\{i\mu^{D-4}\int{\rm d}^Dx\mathcal{L}_{\rm int}^{(D)}\right\}\right]\left|0\right>}{\left<0\right|T\left[\exp\left\{i\mu^{D-4}\int{\rm d}^Dx\mathcal{L}_{\rm int}^{(D)}\right\}\right]\left|0\right>},
\end{equation}

\noindent where \(\left|\Omega\right>\) represents the vacuum of the interactive theory, \(\left|0\right>\) the vacuum of the free field theory, \(\mathcal{L}_{\rm int}^{D}\) is the extension to \(D\) dimensions of the interaction part of the Lagrangian and \(\mu\) is a mass parameter that takes into account the change from four to \(D\) in the energy scale of the Lagrangian. As a consequence, the mass scale of \(\mathcal{L}^{(D)}_{\rm int}\) in Eq.~\eqref{Perturbation_Theory} is four. Now consider a transformation of the fields whose action on a general operator can be obtained applying 

\begin{equation}
    \label{transformation_operator}
    \delta\coloneqq\int{\rm d}^Dx\delta\phi_i(x)\frac{\delta}{\delta\phi_i(x)}.
\end{equation}

\noindent Acting on the \(n\)-point Green's function one gets

\begin{equation}
    \label{transformation_green_function}
    \begin{split}
    &\delta\left<\Omega\right|T\left[\phi_1\left(x_1\right)\cdots\phi_n\left(x_n\right)\right]\left|\Omega\right>=\sum_{k=1}^n\left<\Omega\right|T\left[\phi_1\left(x_1\right)\cdots\delta\phi_k(x_k)\cdots\phi_n\left(x_n\right)\right]\left|\Omega\right>\\&=\frac{\sum_{k=1}^n\left<0\right|T\left[\phi_1\left(x_1\right)\cdots\delta\phi_k(x_k)\cdots\phi_n\left(x_n\right)\exp\left\{i\mu^{D-4}\int{\rm d}^Dx\mathcal{L}_{\rm int}^{(D)}\right\}\right]\left|0\right>}{\left<0\right|T\left[\exp\left\{i\mu^{D-4}\int{\rm d}^Dx\mathcal{L}_{\rm int}^{(D)}\right\}\right]\left|0\right>}.
    \end{split}
\end{equation}

Now, it will be shown that the numerator of Eq.~\eqref{transformation_green_function} satisfies the relation

\begin{equation}
    \label{Greens_Function_Lagrangian}
    \begin{split}
        &i\sum_{k=1}^n\left<0\right|T\left[\phi_1\left(x_1\right)\cdots\delta\phi_k(x_k)\cdots\phi_n\left(x_n\right)\exp\left\{i\mu^{D-4}\int{\rm d}^Dx\mathcal{L}_{\rm int}^{(D)}\right\}\right]\left|0\right>\\&=\left<0\right|T\left[\phi_1\left(x_1\right)\cdots\phi_n\left(x_n\right)\left(\mu^{D-4}\int{\rm d}^Dy\delta\mathcal{L}^{(D)}\right)\exp\left\{i\mu^{D-4}\int{\rm d}^Dx\mathcal{L}_{\rm int}^{(D)}\right\}\right]\left|0\right>.
    \end{split}
\end{equation}

\noindent For this purpose, it is convenient to split \(\delta\mathcal{L}^{(D)}\) in terms of its kinetic and interaction parts \(\delta\mathcal{L}^{(D)}=\delta\mathcal{L}_0^{(D)}+\delta\mathcal{L}_{\rm int}^{(D)}\). In order to simplify the analysis, the \(\mu^{D-4}\) factors will be reabsorbed by the Lagrangians. Furthermore, it is always possible to write 

\begin{equation}
    \label{kinetic_part}
    \mathcal{L}_0^{(D)}=\frac{1}{2}\phi_iD_{ij}^{(D)}\phi_j,
\end{equation}

\noindent with \(D_{ij}^{(D)}\) being a \(D\)-dimensional symmetric differential operator. Consequently, it is straightforward to see that \(\delta\mathcal{L}_0^{(D)}=(\delta\phi_i)D_{ij}^{(D)}\phi_j\). Thus, the \(N\)-th order term of the perturbative expansion in the second part of the equality in Eq.~\eqref{Greens_Function_Lagrangian} is just

{\small\begin{equation}
    \label{N_perturbative_Expansion}
    \begin{split}
     &\frac{1}{N!}\left<0\right|T\biggl[\phi_1\left(x_1\right)\cdots\phi_n\left(x_n\right)\left(\int{\rm d}^Dy\delta\mathcal{L}^{(D)}_0\right)\overbrace{\left(i\int{\rm d}^Dx\mathcal{L}_{\rm int}^{(D)}\right)\cdots\left(i\int{\rm d}^Dx\mathcal{L}_{\rm int}^{(D)}\right)}^{N\,{\rm times}}\biggl]\left|0\right>\\&+\frac{1}{(N-1)!}\left<0\right|T\biggl[\phi_1\left(x_1\right)\cdots\phi_n\left(x_n\right)\left(\int{\rm d}^Dy\delta\mathcal{L}^{(D)}_{\rm int}\right)\underbrace{\left(i\int{\rm d}^Dx\mathcal{L}_{\rm int}^{(D)}\right)\cdots\left(i\int{\rm d}^Dx\mathcal{L}_{\rm int}^{(D)}\right)}_{(N-1)\,{\rm times}}\biggl]\left|0\right>.
    \end{split}
\end{equation}}

\noindent Now consider the first summand; there are three possible types of Wick contractions with the fields of the kinetic term:

\begin{equation}
    \label{N_perturbative_Expansion_Contraction}
     \frac{1}{N!}\left<0\right|T\biggl[\phi_1\cdots\contraction[3ex]{}{\phi_k}{\cdots\phi_n\!\biggl(\int{\rm d}^Dy\delta\phi_iD_{ij}^{(D)}}{\!\phi_j}\phi_k\cdots\phi_n\!\biggl(\int{\rm d}^Dy\delta\contraction{}{\phi}{\,D_{ij}^{(D)}}{\phi}\phi_iD_{ij}^{(D)}\contraction[2ex]{\,}{\phi_j}{\!\left(i\int{\rm d}^Dx\mathcal{L}_{\rm int}^{(D)}\right)}{\cdots}\phi_j\biggl)\!\left(i\int{\rm d}^Dx\mathcal{L}_{\rm int}^{(D)}\right)\cdots\left(i\int{\rm d}^Dx\mathcal{L}_{\rm int}^{(D)}\right)\!\biggl]\!\left|0\right>.
\end{equation}

\noindent The contraction of the two fields of the kinetic term leads to a loop integral, because they are two objects evaluated at the same point of the spacetime, with a propagator \(P_{jk}\) going from \(\phi_j\) to some \(\phi_k\) of \(\delta\phi_i\). It is important to keep in mind that \(\delta\phi_i\) is not necessarily a linear transformation. Assuming that the dimensional regularization scheme preserves the fundamental relation \(D^{(D)}_{ij}P^{(D)}_{j_k}=i\delta_{ik}\delta^{(D)}(0)\) (see Sec.~\ref{sec:BMHV}) it is clear that this loop integral cannot have an inner momentum as long as the field transformation does not introduce additional derivatives. It has been shown that in dimensional regularization (Sec.~\ref{sec:Dimensional_Regularization}) the integral \(\int{\rm d}^Dp\) vanishes, so there is no contribution from this contraction.

The contraction with the \(\phi_k\left(x_k\right)\) leads to the following contribution:

\begin{equation}
    \label{contraction_2}
    \begin{split}
        &\phi_1(x_1)\cdots\contraction[2ex]{}{\phi_k(x_k)}{\cdots\phi_n(x_n)\int{\rm d}^Dy\delta\phi_i(y)D_{ij}^{D}}{\phi_j(y)}\phi_k(x_k)\cdots\phi_n(x_n)\int{\rm d}^Dy\delta\phi_i(y)D_{ij}^{(D)}\phi_j(y)\\&\rightarrow\phi_1(x_1)\cdots\left(\int{\rm d}^Dy\delta\phi_i(y)D_{ij}^{(D)}(y)P^{(D)}_{jk}(y-x_k)\right)\cdots\phi_n(x_n)\\&=\phi_1(x_1)\cdots\left(\int{\rm d}^Dy\delta\phi_i(y)i\delta_{ik}\delta^{(D)}(y-x_k)\right)\cdots\phi_n(x_n)\\&=i\phi_1(x_1)\cdots\delta\phi_k(x_x)\cdots\phi_n(x_n).
    \end{split}
\end{equation}

\noindent Doing all these contractions one would get precisely the \(N\)-th order contribution of the perturbative expansion of the first term of the equality in Eq.~\eqref{Greens_Function_Lagrangian}. As a consequence, in order to complete the proof it is only necessary to show that the contraction with the interaction Lagrangian cancels the second term of the sum in Eq.~\eqref{N_perturbative_Expansion}. This can be easily done by direct calculation:

\begin{equation}
    \label{contraction_interaction_lagrangian}
    \begin{split}
        &\biggl(\int{\rm d}^Dy\delta\phi_i(y)D_{ij}^{(D)}(y)\contraction[2ex]{}{\phi}{\biggl)\!\biggl(i\int{\rm d}^Dx}{\mathcal{L}_{\rm int}^{(D)}(x)}\phi_j(y)\biggl)\!\biggl(i\int{\rm d}^Dx\mathcal{L}_{\rm int}^{(D)}(x)\biggl)\\&\rightarrow i\int{\rm d}^Dy{\rm d}^Dx\delta\phi_{i}(y)D_{ij}^{(D)}(y)P_{jk}^{(D)}(y-x)\frac{\delta\mathcal{L}^{(D)}_{\rm int}(x)}{\delta\phi_k(x)}\\&=i\int{\rm d}^Dy{\rm d}^Dx\delta\phi_{i}(y)i\delta_{ik}\delta^{(D)}(y-x)\frac{\delta\mathcal{L}^{(D)}_{\rm int}(x)}{\delta\phi_k(x)}=-\int{\rm d}^Dy\delta\phi_i(y)\frac{\delta\mathcal{L}^{(D)}_{\rm int}(y)}{\delta\phi_i(y)}\\&=-\int{\rm d}^Dy\delta\mathcal{L}^{(D)}_{\rm int}(y).
    \end{split}
\end{equation}

\noindent Therefore, taking into account that there are \(N\) possible contractions of this kind that lead to the combinatorial factor \(N/N!=1/(N-1)!\), one gets the desired cancellation. As a consequence, writing again explicitly the \(\mu^{D-4}\) factors:

\begin{equation}
    \label{Final_Result_Contractions}
    \begin{split}
        &i\delta\left<\Omega\right|T\left[\phi_1\left(x_1\right)\cdots\phi_n\left(x_n\right)\right]\left|\Omega\right>\\&=\frac{\left<0\right|T\left[\phi_1\left(x_1\right)\cdots\phi_n\left(x_n\right)\left(\mu^{D-4}\int{\rm d}^Dy\delta\mathcal{L}^{(D)}\right)\exp\left\{i\mu^{D-4}\int{\rm d}^Dx\mathcal{L}_{\rm int}^{(D)}\right\}\right]\left|0\right>}{\left<0\right|T\left[\exp\left\{i\mu^{D-4}\int{\rm d}^Dx\mathcal{L}_{\rm int}^{(D)}\right\}\right]\left|0\right>}\\&=\left<\Omega\right|T\left[\phi_1\left(x_1\right)\cdots\phi_n\left(x_n\right)\Delta\right]\left|\Omega\right>,
    \end{split}
\end{equation}

\noindent where the operator \(\Delta\) is just

\begin{equation}
    \label{Delta_operator}
    \Delta\coloneqq\mu^{D-4}\int{\rm d}^Dx\delta\mathcal{L}^{(D)}.
\end{equation}

The result in Eq.~\eqref{Final_Result_Contractions} is known as the quantum action principle. It means that \textit{the variation of \(n\)-point Green's functions can be computed from diagrams with single insertions of the operator obtained from the transformation of the \(D\)-dimensional action}. If the change of the fields corresponds to a BRST transformation, this means that if it is possible to construct a \(D\)-dimensional BRST-invariant extension of the classical action, then no anomalies can be present. This is always the case for non-chiral gauge theories, as mentioned in the previous chapter anomalies can only appear in theories with chiral fermions, where this extension is not always possible (see Sec.~\ref{sec:BMHV}).

In the context of the study of anomalies, the QAP can be used to derive the following fundamental relation \cite{Belusca-Maito:2023wah}:

\begin{equation}
    \label{QAP_QEA}
    \int{\rm d}^Dx\frac{\delta_R\Gamma\left[\chi,K\right]}{\delta K_n\left(x\right)}\frac{\delta_L\Gamma\left[\chi,K\right]}{\delta \chi_n\left(x\right)}=-\frac{1}{2}\left(\Gamma,\Gamma\right)=\Delta\cdot\Gamma,
\end{equation}

\noindent where the \textit{obstruction \(\Delta\cdot\Gamma\) to the Zinn-Justin equation is computed from 1PI Feynman diagrams with single insertions of the symmetry-breaking operator obtained from the BRST-variation of the \(D\)-dimensional action} \(\Delta=s(\mu^{D-4}\int{\rm d}^Dx\mathcal{L}^{(D)})\). As mentioned in the first chapter, if the result of the calculation of these diagrams can be expressed as the BRST variation of a local operator, then the anomaly is not physical and it can be interpreted as an artifact of the regularization procedure.

\section{Gauge Symmetry-Breaking in Chiral Theories} \label{sec:BMHV}

The dimensional regularization scheme faces some difficulties when dealing with theories that contain objects whose properties only hold in four dimensions. This is precisely what happens in the context of models with chiral fermions, where the \(\gamma_5\) matrix is present. A good review of the problem of defining a \(D\)-dimensional analogue of this object can be found in \cite{Jegerlehner:2000dz}. The definition of the \(\gamma_5\) matrix is purely four-dimensional \(\gamma_5\coloneqq i\gamma^0\gamma^1\gamma^2\gamma^3\) and since the dimension of the Dirac algebra in dimensional regularization is not finite, it is not possible to extend this definition. However, due to the fact that, in practical calculations in four dimensions, only properties like \(\left\{\gamma^{\mu},\gamma_5\right\}=0\), \(\gamma_5^2=\mathbb{I}\) and \(\gamma_5^{\dagger}=\gamma_5\) are used, one might only try to maintain them. The problem is that \textit{the anticommutation rule leads to mathematical inconsistencies in \(D\)-dimensions}. To see this, just consider the trace \({\rm tr}\left\{\gamma^{\mu}\gamma^{\nu}\gamma^{\rho}\gamma^{\sigma}\gamma_5\right\}\). In an infinite-dimensional Dirac algebra it is always possible to take a \(\gamma^{\alpha}\) with \(\alpha\neq0\) that is not present there, so it anticommutes with all the other matrices. As a consequence, one obtains:

\begin{equation}
    \label{gamma5_inconsistent_trace}
    \begin{split}
        {\rm tr}\left\{\gamma^{\mu}\gamma^{\nu}\gamma^{\rho}\gamma^{\sigma}\gamma_5\right\}&=-{\rm tr}\left\{\gamma^{\alpha}\gamma^{\alpha}\gamma^{\mu}\gamma^{\nu}\gamma^{\rho}\gamma^{\sigma}\gamma_5\right\}={\rm tr}\left\{\gamma^{\alpha}\gamma^{\mu}\gamma^{\nu}\gamma^{\rho}\gamma^{\sigma}\gamma_5\gamma^{\alpha}\right\}\\&={\rm tr}\left\{\gamma^{\alpha}\gamma^{\alpha}\gamma^{\mu}\gamma^{\nu}\gamma^{\rho}\gamma^{\sigma}\gamma_5\right\}=- {\rm tr}\left\{\gamma^{\mu}\gamma^{\nu}\gamma^{\rho}\gamma^{\sigma}\gamma_5\right\},
    \end{split}
\end{equation}

\noindent which means that \( {\rm tr}\left\{\gamma^{\mu}\gamma^{\nu}\gamma^{\rho}\gamma^{\sigma}\gamma_5\right\}=0\). Therefore, there is no possible way to make an analytic continuation of the result of a calculation where this trace appears to \(D=4\), where \( {\rm tr}\left\{\gamma^{\mu}\gamma^{\nu}\gamma^{\rho}\gamma^{\sigma}\gamma_5\right\}=-4i\varepsilon^{\mu\nu\rho\sigma}\) with the prescription \(\varepsilon^{0123}=1\). Another example of an object with properties that are purely four-dimensional is the Levi-Civita tensor \(\varepsilon^{\mu\nu\rho\sigma}\). One of them is the product

\begin{equation}
    \label{Levi_Civita_Dim4}
    \varepsilon_{\mu_1\dots\mu_4}\varepsilon_{\nu_1\dots\nu_4}=-\sum_{\pi\in S_4}\mathrm{sign}\left(\pi\right)\prod_{i=1}^4\eta_{\mu_i\nu_{\pi\left(i\right)}},
\end{equation}

\noindent where \(S_n\) is the group of permutations of \(n\) objects, \(\mathrm{sign}\left(\pi\right)\) is the signature of the permutation.

The existence of these difficulties in the dimensional regularization scheme were already noticed by 't Hooft and Veltman in \cite{tHooft:1972tcz}. They proposed a possible solution whose mathematical consistency was proven by Breitenlohner and Maison in \cite{Breitenlohner:1977hr,Breitenlohner:1975hg,Breitenlohner:1976te}. This prescription is called the BMHV scheme after their names and \textit{it is the only known algebraically consistent way to use dimensional regularization}. This means that the results that one obtains with manipulations of the Dirac algebra are unique and the limit \(D\to4\) is well-defined, because it reproduces the correct expressions. The idea is to split the \(D\)-dimensional space into four- and \((D-4)\)-dimensional parts. For this purpose, the metric tensors are used as projectors:

\begin{equation}
    \label{BMHV_Metrics}
    \textrm{dim}\,D:\hspace{0.2cm} \eta_{\mu\nu}, \hspace{1cm} \textrm{dim}\,4:\hspace{0.2cm} \bar{\eta}_{\mu\nu}, \hspace{1cm} \textrm{dim}\,\left(D-4\right):\hspace{0.2cm} \hat{\eta}_{\mu\nu}, 
\end{equation}

\noindent where

\begin{equation}
    \label{Metric_Projectors}
    \hat{\eta}_{\mu\nu}\coloneqq\left\{
    \begin{array}{ll}
       \eta_{\mu\nu}, &\textrm{if} \,\,\mu\geq4\,\, \textrm{or}\,\, \nu\geq 4  \\
        0,  &  \textrm{otherwise}
    \end{array}\right., \hspace{2cm} \bar{\eta}_{\mu\nu}\coloneqq\eta_{\mu\nu}-\hat{\eta}_{\mu\nu}.
\end{equation}

\noindent The properties that are derived from these definitions are:

\begin{equation}
    \label{Projectors}
    \begin{array}{lll}
        \eta^{\mu\alpha}\hat{\eta}_{\alpha\nu}=\hat{\eta}^{\mu\alpha}\hat{\eta}_{\alpha\nu}=\hat{\eta}^{\mu}_{\nu}, &  \hspace{1cm}  \eta^{\mu\alpha}\bar{\eta}_{\alpha\nu}=\bar{\eta}^{\mu\alpha}\bar{\eta}_{\alpha\nu}=\bar{\eta}^{\mu}_{\nu}, &  \hspace{1cm} \hat{\eta}^{\mu\alpha}\bar{\eta}_{\alpha\nu}=0,  \\ & & \\
         \eta^{\mu}_{\mu}=D, & \hspace{1cm} \hat{\eta}^{\mu}_{\mu}=D-4,  & \hspace{1cm} \bar{\eta}^{\mu}_{\mu}=4.
    \end{array} 
\end{equation}

\noindent Then, the projections of Lorentz vectors and Dirac matrices into these subspaces are defined as

\begin{equation}
    \label{Projections_Vectors_Matrices}
    \begin{array}{lll}
        p^{\mu}\coloneqq\eta^{\mu}_{\nu}\,p^{\nu}, &  \hspace{1cm}  \hat{p}^{\mu}\coloneqq\hat{\eta}^{\mu}_{\nu}\,p^{\nu}, &  \hspace{1cm} \bar{p}^{\mu}\coloneqq\bar{\eta}^{\mu}_{\nu}\,p^{\nu},  \\ & & \\
         \gamma^{\mu}\coloneqq\eta^{\mu}_{\nu}\,\gamma^{\nu}, &  \hspace{1cm}  \hat{\gamma}^{\mu}\coloneqq\hat{\eta}^{\mu}_{\nu}\,\gamma^{\nu}, &  \hspace{1cm} \bar{\gamma}^{\mu}\coloneqq\bar{\eta}^{\mu}_{\nu}\,\gamma^{\nu}.
    \end{array} 
\end{equation}

\noindent The properties that they satisfy are

\begin{equation}
    \label{Trace_Gamma}
    \textrm{tr}\left\{\gamma^{\mu}\right\}=0\,\,\textrm{if}\,\,D\neq1.
\end{equation}

\begin{equation}
    \label{Projectors-Commutators}
    \begin{array}{ll}
        \eta^{\mu}_{\nu}\,\hat{\gamma}^{\nu}=\hat{\gamma}^{\mu}, &  \hspace{1cm}  \eta^{\mu}_{\nu}\,\bar{\gamma}^{\nu}=\bar{\gamma}^{\mu}, \\ & \\  \hat{\eta}^{\mu}_{\nu}\,\hat{\gamma}^{\nu}=\hat{\gamma}^{\mu}, & \hspace{1cm}  \bar{\eta}^{\mu}_{\nu}\,\bar{\gamma}^{\nu}=\bar{\gamma}^{\mu}, \\ & \\
         \hat{\eta}^{\mu}_{\nu}\,\bar{\gamma}^{\nu}=0,  & \hspace{1cm} \bar{\eta}^{\mu}_{\nu}\,\hat{\gamma}^{\nu}=0, 
    \end{array} \hspace{2cm}
    \begin{array}{ll}
         \left\{\gamma_{\mu},\hat{\gamma}_{\nu}\right\}\coloneqq2\hat{\eta}_{\mu\nu}\mathbb{I}, &  \hspace{0.5cm}   \left\{\hat{\gamma}_{\mu},\hat{\gamma}_{\nu}\right\}\coloneqq2\hat{\eta}_{\mu\nu}\mathbb{I}, \\ & \\  \left\{\gamma_{\mu},\bar{\gamma}_{\nu}\right\}\coloneqq2\bar{\eta}_{\mu\nu}\mathbb{I}, &  \hspace{0.5cm}   \left\{\bar{\gamma}_{\mu},\bar{\gamma}_{\nu}\right\}\coloneqq2\bar{\eta}_{\mu\nu}\mathbb{I}, \\ & \\
          \left\{\hat{\gamma}_{\mu},\bar{\gamma}_{\nu}\right\}=0.  &  
    \end{array} 
\end{equation}

Finally, \textit{\(\varepsilon_{\mu\nu\rho\sigma}\) and \(\gamma_5\) are kept as purely four-dimensional objects}. For instance, the Levi-Civita tensor is defined as 

\begin{eqnarray}
    \label{Levi_Civita_Definition}
    \varepsilon^{\mu\nu\rho\sigma}\coloneqq\left\{
    \begin{array}{rl}
        1 & \textrm{if} \,\, \left(\mu\nu\rho\sigma\right) \,\,\textrm{is an even permutation of}\,\, \left(0123\right)  \\
        -1 & \textrm{if} \,\, \left(\mu\nu\rho\sigma\right) \,\,\textrm{is an odd permutation of}\,\, \left(0123\right)  \\
        0 & \textrm{otherwise}
    \end{array}\right.,
\end{eqnarray}

\noindent so it satisfies 

\begin{equation}
    \label{Levi_Civita_Properties}
    \begin{array}{ll}
        \varepsilon_{\mu\nu\rho\sigma}\,\hat{\eta}^{\sigma}_{\tau}=0, & \hspace{1cm} \varepsilon_{\mu_1\dots\mu_4}=\textrm{sign}\left(\pi\right)\varepsilon_{\mu_{\pi\left(1\right)}\dots\mu_{\pi\left(4\right)}},   \\ & \\
        \varepsilon_{\mu\nu\rho\sigma}\,\bar{\eta}^{\sigma}_{\tau}=\varepsilon_{\mu\nu\rho\tau}, & \hspace{1cm} \varepsilon_{\mu_1\dots\mu_4}\varepsilon_{\nu_1\dots\nu_4}=-\sum\limits_{\pi\in S_4}\mathrm{sign}\left(\pi\right)\prod\limits_{i=1}^4\bar{\eta}_{\mu_i\nu_{\pi\left(i\right)}}, \\ & \\
        \varepsilon_{\mu\nu\rho\sigma}\,\eta^{\sigma}_{\tau}=\varepsilon_{\mu\nu\rho\tau}. &
    \end{array}
\end{equation}

\noindent The \(\gamma_5\) matrix can be defined as 

\begin{equation}
    \label{gamma_5_definition}
    \gamma_5\coloneqq i\,\gamma^0\gamma^1\gamma^2\gamma^3=-\frac{i}{4!}\varepsilon_{\mu\nu\rho\sigma}\gamma^{\mu}\gamma^{\nu}\gamma^{\rho}\gamma^{\sigma},
\end{equation}

\noindent leading to the following relations:

\begin{equation}
    \label{Gamma_5_Properties}
    \begin{array}{ll}
        \textrm{tr}\left\{\gamma_5\right\}=0, & \hspace{1cm} \gamma_5^2=\mathbb{I},   \\ & \\
        \left\{\gamma_5,\bar{\gamma}^{\mu}\right\}=0, & \hspace{1cm} \left\{\gamma_5,\gamma^{\mu}\right\}=\left\{\gamma_5,\hat{\gamma}^{\mu}\right\}=2\gamma_5\hat{\gamma}^{\mu}, \\ & \\
        \left[\gamma_5,\hat{\gamma}^{\mu}\right]=0, & \hspace{1cm} \left[\gamma_5,\gamma^{\mu}\right]=\left[\gamma_5,\bar{\gamma}^{\mu}\right]=2\gamma_5\bar{\gamma}^{\mu}.
    \end{array}
\end{equation}

\noindent Notice that now it does not anticommute with all the  \(\gamma^{\mu}\), but only with its four-dimensional part.

At this point, it is necessary to define a \(D\)-dimensional extension of the Lagrangian. There are several possible choices and they will all give consistent results, but some of them are more convenient than others for practical calculations. For instance, currents like \(\bar{\psi}_L\gamma^{\mu}\xi_L\) have three equivalent expressions in four dimensions: \(\bar{\psi}\gamma^{\mu}P_L\xi\), \(\bar{\psi}P_R\gamma^{\mu}\xi\) and \(\bar{\psi}P_R\gamma^{\mu}P_L\xi\). The fact is that they are not the same in \(D\) dimensions in the BMHV scheme, because \(\gamma_5\) does not anticommute with \(\gamma^{\mu}\). As pointed out in \cite{Ferrari:1994ct}, the choices \(\bar{\psi}\gamma^{\mu}P_L\xi\) and \(\bar{\psi}P_R\gamma^{\mu}\xi\) are not compatible with hermiticity. For example, it is easy to see that \(\left(\bar{\psi}\gamma^{\mu}P_L\xi\right)^{\dagger}=\bar{\xi}\bar{\gamma}^{\mu}P_L\psi+\bar{\xi}\hat{\gamma}^{\mu}P_R\psi\neq\bar{\xi}\gamma^{\mu}P_L\psi\) and it can also be shown that \(\bar{\psi}\gamma^{\mu}P_L\xi=\bar{\psi}P_R\bar{\gamma}^{\mu}P_L\xi+\bar{\psi}P_L\hat{\gamma}^{\mu}P_L\xi\), so the addition of spurious right-handed components in \(D\)-dimensions in this case is mandatory. Nowadays, the most symmetric choice \(\bar{\psi}P_R\gamma^{\mu}P_L\xi\)  is the preferred one for practical calculations. However, in the BMHV scheme this is a purely four-dimensional operator  \(\bar{\psi}P_R\gamma^{\mu}P_L\xi=\bar{\psi}P_R\bar{\gamma}^{\mu}P_L\xi\). In dimensional regularization there is no problem with having interactions that are purely four-dimensional, so this convention will be kept in this text. Nonetheless, the extension of the kinetic terms requires a deeper study.

Having four-dimensional kinetic terms, like \(i\bar{\psi}_L\slashed{\partial}\psi_L\) or \(i\bar{\psi}_R\slashed{\partial}\psi_R\), that are associated with a four-dimensional propagator \(1/\bar{\slashed{p}}\) is problematic in the context of dimensional regularization. First of all, they might lead to ill-defined amplitudes when performing \(D\)-dimensional integration over a 4-dimensional integrand \(\int{\rm d}^Dp/\bar{\slashed{p}}=\int{\rm d}^4p/\bar{\slashed{p}}\int{\rm d}^{D-4}p=\infty\cdot0\). Furthermore, the derivation of the QAP done in the previous section requires a \(D\)-dimensional propagator, otherwise it might not be valid. To solve this problem in a theory where only one physical chirality of the fermions is present\footnote{Consider, for instance, the SM, where there are only left-handed neutrinos.} \textit{it is necessary to add non-interacting fictitious fermions with the opposite chirality}. For example, in a model with only right-handed fermions \(\psi_R\) it is possible to use the spurious left-handed fictitious fermion \(\psi_L^f\) to write a \(D\)-dimensional propagator

\begin{equation}
    \label{D_Dim_Propagator}
    i\bar{\psi}\slashed{\partial}\psi=i\bar{\psi}_R\bar{\slashed{\partial}}\psi_R+i\bar{\psi}^f_L\bar{\slashed{\partial}}\psi^f_L+i\bar{\psi}_R\hat{\slashed{\partial}}\psi_L^f+i\bar{\psi}_L^f\hat{\slashed{\partial}}\psi_R.
\end{equation}

\noindent The associated propagator is now the usual \(i/\slashed{p}\), which leads to well-defined amplitudes in dimensional regularization. However, \textit{it explicitly breaks the gauge symmetries of the theory} due to the presence of the \((D-4)\)-dimensional part, the so-called evanescent contribution. The fictitious fields have to be non-interacting in order not to change the physical model when going to \(D\)-dimensions, so they have to be BRST invariant \(s\psi^f_{L,R}=0\). This means that the evanescent part of the propagator transforms as:

\begin{equation}
    \label{Kinetic_BRST_Breaking_1}
    s\left(i\bar{\psi}_i\slashed{\partial}\psi_i\right)=\omega_{\alpha}\left[\bar{\psi}_i\left(t^{\alpha}_{L,R}\right)_{ij}\left(\overrightarrow{\hat{\slashed{\partial}}}P_{R,L}+\overleftarrow{\hat{\slashed{\partial}}}P_{L,R}\right)\psi_j\right],
\end{equation}

\noindent where the first subscript in the generators and the projectors is taken when the physical fields are left-handed, the second one when they are right-handed and  the different \(\psi_i\) correspond to different fields that transform under a certain representation of the gauge group.

The situation is different if fields with both chiralities are already present in the theory, like the fermions in the SM except neutrinos. There are several choices of the \(D\)-dimensional extension of the Lagrangian, although some of them might be more convenient than others depending on the model \cite{Ebert:2024xpy}. The two main possibilities are:

\begin{itemize}
    \item[1.] \textit{Propagators with physical fields.} The most obvious option is to use the already present physical fields \(\psi_L\) and \(\psi_R\) to construct the propagator:
    
    \begin{equation}
    \label{D_Dim_Propagator_1}
    i\bar{\psi}\slashed{\partial}\psi=i\bar{\psi}_R\bar{\slashed{\partial}}\psi_R+i\bar{\psi}_L\bar{\slashed{\partial}}\psi_L+i\bar{\psi}_R\hat{\slashed{\partial}}\psi_L+i\bar{\psi}_L\hat{\slashed{\partial}}\psi_R.
\end{equation}

    \noindent The problem is that this approach not only breaks gauge symmetries, but also global ones when the left- and right-handed fermions have different quantum numbers, like the hypercharge.

    \item[2.] \textit{Propagators with fictitious fields.} In this approach one still adds the fictitious fields \(\psi_L^f\) and \(\psi_R^f\) to the theory and constructs the propagators like in Eq.~\eqref{D_Dim_Propagator}. It is possible to add quantum numbers to these spurious fermions in order to preserve global symmetries, even though they do not transform under the action of the gauge group. However, this extension is not appropriate to deal with theories that contain massive fermions.\footnote{In chiral theories the presence of massive fermions is usually associated with spontaneous symmetry breaking and working in the broken phase.} The addition of a Dirac mass term leads to the Lagrangian \cite{Ebert:2024xpy}:

    \begin{equation}
    \label{lagrangian_Broken_Phase}
    \begin{split}
        \mathcal{L}&=i\left(\bar{\psi}_L+\bar{\psi}_R^{f}\right)\slashed{\partial}\left(\psi_L+\psi_R^{f}\right)+i\left(\bar{\psi}_R+\bar{\psi}_L^{f}\right)\slashed{\partial}\left(\psi_R+\psi_L^{f}\right)-m\left(\bar{\psi}_L\psi_R+\bar{\psi}_R\psi_L\right)\\&=\left(\bar{\psi}_L+\bar{\psi}_R^{f},\bar{\psi}_R+\bar{\psi}_L^{f}\right)\left(\begin{array}{cc}
            i\slashed{\partial} & -mP_R  \\
            -mP_L & i\slashed{\partial}
        \end{array}\right)\left(\begin{array}{c}
        \psi_L+\psi_R^{f} \\
        \psi_R+\psi_L^{f}
        \end{array}\right)
    \end{split}
    \end{equation}

    \noindent with associated propagator matrix

    \begin{equation}
    \label{Propagator_Broken_Phase}
    \left(\begin{array}{cc}
        \dfrac{p^2\slashed{p}-m^2\bar{\slashed{p}}P_L}{p^4-m^2\bar{p}^2} & \dfrac{m\left[\bar{p}^2P_L+\hat{p}^2P_R+\hat{\slashed{p}}\bar{\slashed{p}}\left(P_L-P_R\right)\right]}{p^4-m^2\bar{p}^2} \\
        \dfrac{m\left[\bar{p}^2P_R+\hat{p}^2P_L+\hat{\slashed{p}}\bar{\slashed{p}}\left(P_R-P_L\right)\right]}{p^4-m^2\bar{p}^2} & \dfrac{p^2\slashed{p}-m^2\bar{\slashed{p}}P_R}{p^4-m^2\bar{p}^2} 
    \end{array}\right),
    \end{equation}

    \noindent making the calculations cumbersome.
    
\end{itemize}

\noindent All of the computations that require a \(D\)-dimensional extension of the Lagrangian in this text will be done in the unbroken phase, so the preferred option is the second one.

As mentioned in the previous section, the presence of a term in the Lagrangian that breaks the gauge symmetry leads to an obstruction to the Slavnov-Taylor identity, that can be computed using Eq.~\eqref{QAP_QEA} due to the fact that all these extensions contain a \(D\)-dimensional propagator. Using Eq.~\eqref{Kinetic_BRST_Breaking_1}, it is clear that the symmetry breaking operators \(\Delta\) needed to calculate \(\Delta\cdot\Gamma\) will have the form

\begin{equation}
    \label{Symmetry_Breaking_Delta}
    \Delta=\int{\rm d}^Dx\omega_{\alpha}(x)\left[\bar{\psi}_i(x)\left(t^{\alpha}_{L,R}\right)_{ij}\left(\overrightarrow{\hat{\slashed{\partial}}}P_{R,L}+\overleftarrow{\hat{\slashed{\partial}}}P_{L,R}\right)\psi_j(x)\right].
\end{equation}

 \noindent Nonetheless, in the previous chapter it was shown that if the anomalies are not physical (like in the SM, SMEFT and LR models) they can be eliminated when adding a set of appropriate counterterms to the Lagrangian. In the context of dimensional regularization, the anomalies have to be canceled in the four-dimensional limit, so at one loop:

 \begin{equation}
     \label{Cancelation_Anomalies_Dim_Reg}
     \lim_{D\to4}\left(S_R^{(4)},\Gamma_{1}+I_{\rm sct}+I_{\rm fct}\right)=0,
 \end{equation}

\noindent where \(S_R^{(4)}\) is the four-dimensional regularized action,\footnote{It is necessary to use the four-dimensional action in order to define the generalized BRST operator because the \(D\)-dimensional action is not gauge invariant anymore so it does not satisfy the master equation. In the calculations done in this text the presence of antifields is not required, so it is also possible to write Eq.~\eqref{Cancelation_Anomalies_Dim_Reg} using the usual \(s\) operator.} \(\Gamma_1\) the one-loop QEA and \(I_{\rm sct}\) and \(I_{\rm fct}\) the counterterms needed to restore the symmetry. It is important to distinguish between the singular counterterms \(I_{\rm sct}\) and the finite ones \(I_{\rm fct}\) because the former appear naturally in the process of elimination of divergences and the latter have to be added by hand. These \textit{finite counterterms are there due to the presence of divergences in the loop calculations and the insertion of an evanescent symmetry-breaking operator}, that lead to \(\varepsilon/\varepsilon\) contributions in the computations of diagrams, with \(\varepsilon\coloneqq 4-D\). The way to obtain them is discussed in the next section. Finally, it is important to remark that the \(\Delta\) operators will be different in higher loop calculations, that are not going to be done in this text. This is because of the fact that the calculation of \(s\left(S\right)\) not only needs to take into account the presence of a gauge-variant propagator, but also the addition of gauge-variant counterterms found at lower loop orders.

Nowadays this approach based on algebraic renormalization has been used to calculate the full renormalization of an abelian chiral gauge theory with only right-handed fermions, no scalars and operators up to dimension four at four loops \cite{vonManteuffel:2025swv}. 
The calculations for the non-abelian scenario at one loop without scalars can be found at \cite{Martin:1999cc,Cornella:2022hkc} and with scalars (the SM being a particular case) in \cite{Belusca-Maito:2020ala}. There are also results for the gauge group \({\rm SU}(2)\) without scalars at two loops \cite{Kuhler:2025znv}. Furthermore, for EFTs like the Low-Energy Effective Field Theory (LEFT) there are results of the one-loop full renormalization \cite{Naterop:2023dek} and calculations of the renormalization group equations at two loops \cite{Naterop:2024cfx}. Finally, the one-loop renormalization of the SMEFT was presented at \cite{Fuentes-Martin:2025meq}, but using a recent new method based on the introduction of a spurion field \cite{OlgosoRuiz:2024dzq} instead of using the traditional algebraic renormalization.

\section{Bonneau Method and Finite Counterterms} \label{sec:Bonneau}

There are three types of contributions to the obstruction of the \(D\)-dimensional Slavnov-Taylor identities that can come from the calculation of 1PI diagrams with single insertions of the symmetry-breaking operator \(\Delta\). First of all, there are singular terms that are eliminated in the process of cancellation of divergences by the singular gauge-variant counterterms \(I_{\rm sct}\). Due to the fact that the \(\Delta\) operator is evanescent, one can also find contributions that vanish when the limit \(D\to4\) is taken. Finally, there are also finite \(\varepsilon/\varepsilon\) terms that appear as a consequence of the presence of an evanescent operator in a divergent graph. These contributions are problematic because they survive when the limit \(D\to4\) (\(\varepsilon\to0\)) is taken, leading to non-physical anomalies in the four-dimensional theory that have to be canceled by appropriate counterterms. Since these contributions are finite, one might think that the whole computation of the amplitudes is needed in order to identify them. However, \textit{there is a method developed by Guy Bonneau that permits to isolate those finite terms by looking at the divergent part of certain integrals} \cite{Bonneau:1980zp,Bonneau:1979jx}, which makes their calculation much more manageable. A sketch of the proof of the method will now be presented.

First of all, since the appearance of a symmetry-breaking operator \(\Delta\) in the BMHV scheme is due to the fact that the \(\gamma_5\) matrix does not anticommute with all the other \(\gamma^{\mu}\) matrices \(\left\{\gamma_5,\gamma_{\mu}\right\}=2\hat{\eta}_{\mu\nu}\gamma_5\gamma^{\nu}\), then it must have the form \(\Delta=-\hat{\eta}_{\mu\nu}\Delta^{\mu\nu}\), where \(\Delta^{\mu\nu}\) is not an evanescent operator. Now consider the unrenormalized amplitude of a 1PI graph \(G_{\Delta_{\mu\nu}}\) that contains a single insertion of the operator \(\Delta_{\mu\nu}\): \(U_{\mu\nu}(G_{\Delta_{\mu\nu}})\). This amplitude can be decomposed as:

\begin{equation}
    \label{Decomposition_Bonneau}
    U_{\mu\nu}(G_{\Delta_{\mu\nu}})=A_{\mu\nu}(G_{\Delta_{\mu\nu}})+B_{\mu\nu}(G_{\Delta_{\mu\nu}}),
\end{equation}

\noindent where \(A_{\mu\nu}(G_{\Delta_{\mu\nu}})\) is the part that does not produce an \(\varepsilon\) when contracted with \(\hat{\eta}^{\mu\nu}\) and \(B_{\mu\nu}(G_{\Delta_{\mu\nu}})\) the part that does. Therefore, by contracting with \(-\hat{\eta}^{\mu\nu}\) one obtains the amplitude associated with the same 1PI graph, but with \(\Delta\) as a special vertex \(G_{\Delta}\):

\begin{equation}
    \label{Decomposition_Bonneau_2}
    U(G_{\Delta})=-A(G_{\Delta})+\varepsilon B(G_{\Delta}), 
\end{equation}

\noindent where \(A(G_{\Delta})\coloneqq\hat{\eta}^{\mu\nu}A_{\mu\nu}(G_{\Delta_{\mu\nu}})\) and \(B(G_{\Delta})\coloneqq-\hat{\eta}^{\mu\nu}B_{\mu\nu}(G_{\Delta_{\mu\nu}})/\varepsilon\).

The second step is trying to relate the renormalized amplitude of the graph \(G_{\Delta}\) with the renormalized amplitude of the graph \(G_{\Delta_{\mu\nu}}\). This is not trivial due to the presence of the factor \(\varepsilon\) in Eq.~\eqref{Decomposition_Bonneau_2}, because \textit{the subtractions of divergences will not commute with the contraction with \(\hat{\eta}^{\mu\nu}\)}. The expressions for \(R(G_{\Delta})\) and \(R_{\mu\nu}(G_{\Delta_{\mu\nu}})\) will be obtained from the forest formula in Eq.~\eqref{forest_formula}. There are three sets of graphs to be considered: the set of graphs \(F_V\) that contain the special vertex \(V\), the set of graphs \(F_A\) that do not contain the special vertex but are totally contained in the graphs that do contain it, and the set of all the other graphs \(F_B\). The subtractions of the graphs of \(F_A\) and \(F_B\) will be the same for \(\Delta\) and \(\Delta_{\mu\nu}\), the problem arises when studying the diagrams in \(F_V\). Consider the smallest graph \(\gamma^{\Delta}_1\) of \(F_V\) with \(V=\Delta\). The subtraction operator in the forest formula acts on the overall divergence, so it is necessary to study the associated renormalized amplitude when all the subdivergences (coming from the set of diagrams \(F_A\)) have been subtracted. Using Eq.~\eqref{Decomposition_Bonneau_2} it can be decomposed as

\begin{equation}
    \label{Decompositon_Renormalized}
    \bar{R}(\gamma^{\Delta}_1)=-\bar{A}(\gamma^{\Delta}_1)+\varepsilon \bar{B}(\gamma^{\Delta}_1).
\end{equation}

\noindent In dimensional regularization with the MS scheme the subtraction operator eliminates the singular part of the amplitude \(-{\rm s.p.}(\bar{R}(\gamma^{\Delta}_1))\), which is given by

\begin{equation}
    \label{singular_part_decomposition_renormalized}
    \begin{split}
        -{\rm s.p.}(\bar{R}(\gamma^{\Delta}_1))&=-{\rm s.p.}(-\bar{A}(\gamma^{\Delta}_1))-{\rm s.p.}(\varepsilon \bar{B}(\gamma^{\Delta}_1))\\&=(-\hat{\eta}^{\mu\nu})[-{\rm s.p.}(\bar{A}_{\mu\nu}(\gamma^{\Delta_{\mu\nu}}_1))]-{\rm s.p.}(\varepsilon \bar{B}(\gamma^{\Delta}_1)),
    \end{split}
\end{equation}

\noindent where it has been used that the contraction of \(-\hat{\eta}^{\mu\nu}\) with \(\bar{A}_{\mu\nu}(\gamma^{\Delta_{\mu\nu}}_1)\) does not introduce an \(\varepsilon\) factor by definition.

Now it is necessary to work with the term \(-{\rm s.p.}(\varepsilon \bar{B}(\gamma^{\Delta}_1))\). For this purpose, a simple theorem of complex calculus is going to be proved. Consider a meromorphic function \(f(\varepsilon)\) with Laurent expansion

\begin{equation}
    \label{Laurent_Expansion}
    f\left(\varepsilon\right)=\sum_{n=1}^p\frac{f_{-n}}{\varepsilon^n}+\sum_{n=0}^{\infty}f_n\varepsilon^n,
\end{equation}

\noindent where \(p\) is the order of the pole at \(\varepsilon=0\). Then, it is easy to see that

\begin{equation}
    \label{singular_part_1}
    \begin{split}
         -\textrm{s.p.}\left(\varepsilon f\left(\varepsilon\right)\right)&=-\varepsilon\sum_{n=2}^{p}\frac{f_{-n}}{\varepsilon^{n}}=-\varepsilon\left(-\frac{f_{-1}}{\varepsilon}+\sum_{n=1}^{p}\frac{f_{-n}}{\varepsilon^{n}}\right)=-\varepsilon\,{\rm s.p.}(f(\varepsilon))+{\rm r.s.p.}(f(\varepsilon)),
    \end{split}
\end{equation}

\noindent where \({\rm r.s.p.}(f(\varepsilon))\) is the residue at the simple pole of \(f(\varepsilon)\). Taking this into account it is clear that 

\begin{equation}
    \label{singular_part_decomposition_renormalized_2}
    -{\rm s.p.}(\varepsilon \bar{B}(\gamma^{\Delta}_1))=-\varepsilon\,{\rm s.p.}(\bar{B}(\gamma^{\Delta}_1))+{\rm r.s.p.}(\bar{B}(\gamma^{\Delta}_1)).
\end{equation}

 At this point, it is possible to decompose \(\varepsilon\) in terms of contractions of the  metric tensors \(\varepsilon=-\hat{\eta}^{\mu\nu}\eta_{\mu\nu}\). Then, \(\eta_{\mu\nu}\) can be introduced into \({\rm s.p.}(\bar{B}(\gamma^{\Delta}_1))\) because it does not change the pole structure. Finally, the \(\eta_{\mu\nu}\) can be distributed in the appropriate places of the amplitude \(\bar{B}(\gamma^{\Delta}_1)\) using relations like \(\eta_{\mu\nu}=(\gamma_{\mu}\gamma_{\nu}+\gamma_{\nu}\gamma_{\mu})/2\) in order to reproduce \(\bar{B}_{\mu\nu}(\gamma^{\Delta_{\mu\nu}}_1)\). As a consequence, one gets

\begin{equation}
    \label{singular_part_decomposition_renormalized_3}
    \begin{split}
        -{\rm s.p.}(\bar{R}(\gamma^{\Delta}_1))&=(-\hat{\eta}^{\mu\nu})[-{\rm s.p.}(\bar{A}_{\mu\nu}(\gamma^{\Delta_{\mu\nu}}_1))]-{\rm s.p.}(\varepsilon \bar{B}(\gamma^{\Delta}_1))\\&=(-\hat{\eta}^{\mu\nu})[-{\rm s.p.}(\bar{A}_{\mu\nu}(\gamma^{\Delta_{\mu\nu}}_1))]+(-\hat{\eta}^{\mu\nu})[-{\rm s.p.}(\bar{B}_{\mu\nu}(\gamma^{\Delta_{\mu\nu}}_1))]+{\rm r.s.p.}(\bar{B}(\gamma^{\Delta}_1))\\&=(-\hat{\eta}^{\mu\nu})[-{\rm s.p.}(\bar{R}_{\mu\nu}(\gamma_1^{\Delta_{\mu\nu}}))]+{\rm r.s.p.}(\bar{B}(\gamma^{\Delta}_1)).
    \end{split}
\end{equation}

\noindent Therefore, the renormalized amplitude for the \(\gamma_1^{\Delta}\) graph can be written as

\begin{equation}
    \label{renormalized_amplitude_difference}
    R(\gamma^{\Delta}_1)=(-\hat{\eta}^{\mu\nu})R_{\mu\nu}(\gamma_1^{\Delta_{\mu\nu}})+{\rm r.s.p.}(\bar{B}(\gamma^{\Delta}_1)),
\end{equation}

\noindent which is the so-called \textit{Bonneau identity} at one loop.

For the general case it is necessary to define an operator \(\mathcal{O}_{\gamma}(M)\circ A\) that replaces the result of the graph \(\gamma\) by \(M\) into the amplitude \(A\). Considering that the difference between the renormalized value of \(G_{\Delta}\) and \(G_{\Delta_{\mu\nu}}\) is given only by the diagrams in \(F_V\), it is possible to generalize Eq.~\eqref{renormalized_amplitude_difference} into the \textit{Bonneau identity}

\begin{equation}
    \label{Bonneau_Identity}
    R\left(G_{\Delta}\right)-(-\hat{\eta}^{\mu\nu})R_{\mu\nu}\left(G_{\Delta_{\mu\nu}}\right)=\sum_{\gamma_i^{\Delta}}\mathcal{O}_{\gamma_i^{\Delta}}({\rm r.s.p.}(\bar{B}(\gamma_i^{\Delta})))\circ R(G/\gamma_i^{\Delta}).
\end{equation}

\noindent The renormalized value of the graph \(G_{\Delta}\) with \(\Delta\) being the symmetry-breaking operator gives the finite part of the obstruction to the \(D\)-dimensional Slavnov-Taylor identities. Taking the limit \(D\to4\) the evanescent contributions are canceled and one obtains the \(\varepsilon/\varepsilon\) parts. Since \(\hat{\eta}^{\mu\nu}\) is evanescent they are

\begin{equation}
    \label{Bonneau_Identity_D=4}
    \lim_{D\to4}R\left(G_{\Delta}\right)=\lim_{D\to4}\sum_{\gamma_i^{\Delta}}\mathcal{O}_{\gamma_i^{\Delta}}({\rm r.s.p.}(\bar{B}(\gamma_i^{\Delta})))\circ R(G/\gamma_i^{\Delta}).
\end{equation}

\noindent This means that \textit{those terms are obtained by calculating the residue at the simple pole of certain amplitudes}. It is important to remark that, with this method, the finite contribution to the anomalies is also obtained from the calculation of divergent graphs where all the subdivergences have been subtracted. From Weinberg's theorem this means that those terms are also local, so \textit{the anomalies that appear in perturbation theory are local}, completing the proof of the previous chapter.

Finally, there is a simple way to isolate \({\rm r.s.p.}(\bar{B}(\gamma_i^{\Delta}))\). Consider the tensor \(\check{\eta}^{\mu\nu}\), which is going to be referred as the Dvo\v{r}\'ak metric in this text and is defined as \(\check{\eta}^{\mu\nu}\coloneqq-\hat{\eta}^{\mu\nu}/\varepsilon\). Therefore, it has the same properties as the \((D-4)\)-dimensional projector but with unit trace \(\check{\eta}^{\mu}_{\mu}=1\). It is possible to define the operator \(\check{\Delta}\coloneqq\check{\eta}^{\mu\nu}\Delta_{\mu\nu}\) and calculate the amplitude of the diagram \(G_{\check{\Delta}}\) with a single insertion of it. The result is simply

\begin{equation}
    \label{Check_Insertion}
    U\left(G_{\check{\Delta}}\right)=\check{\eta}^{\mu\nu}A_{\mu\nu}(G_{\Delta_{\mu\nu}})+B(G_{\Delta}),
\end{equation}

\noindent  because the \(\varepsilon\) factor that appeared with the \(B(G_{\Delta})\) term is canceled by the \(1/\varepsilon\) of the definition of \(\check{\eta}^{\mu\nu}\). Then, repeating the analysis done before with the forest formula one gets:

\begin{equation}
    \label{Check_Insertion_1}
    \bar{R}(\gamma^{\check{\Delta}}_i)=\check{\eta}^{\mu\nu}\bar{A}_{\mu\nu}(\gamma^{\Delta_{\mu\nu}}_i)+\bar{B}(\gamma^{\Delta}_i).
\end{equation}

\noindent Thus, to isolate the term \({\rm r.s.p.}(\bar{B}(\gamma_i^{\Delta}))\) it is only necessary to calculate

\begin{equation}
    \label{Check_Insertion_2}
    {\rm r.s.p.}(\bar{B}(\gamma_i^{\Delta}))={\rm r.s.p.}\left(\left.\bar{R}(\gamma^{\check{\Delta}}_i)\right|_{\check{\eta}=0}\right),
\end{equation}

\noindent \textit{the residue at the simple pole of amplitudes with all the subdivergences subtracted and single insertions of \(\check{\Delta}\) taking the limit \(\check{\eta}^{\mu\nu}=0\)}. This leads to the following result for the finite obstruction of the 4-dimensional Slavnov-Taylor identities:

\begin{equation}
    \label{Bonneau_Identity_D=4-1}
    \lim_{D\to4}R\left(G_{\Delta}\right)=\lim_{D\to4}\sum_{\gamma_i^{\Delta}}\mathcal{O}_{\gamma_i^{\Delta}}\left(\left.{\rm r.s.p.}\left(\bar{R}(\gamma^{\check{\Delta}}_i)\right|_{\check{\eta}=0}\right)\right)\circ R(G/\gamma_i^{\Delta}).
\end{equation}

\thispagestyle{empty}

\chapter{The Standard Model and the Standard Model Effective Field Theory of Particle Physics}
\markboth{3. THE SM AND SMEFT OF PARTICLE PHYSICS}{}

\section*{Introduction}

The \textit{Standard Model} of particle physics \textit{is the theory that is currently used to understand the interactions between the fundamental constituents of matter}. Its development is an achievement that required the combined efforts of a huge number of theoretical and experimental physicist. In references \cite{Weinberg:2004kv,Quigg:2015cfa} historical reviews are presented. The first section of this chapter introduces the fermion content of the model, associated with representations of the group \({\rm SU}(3)_{\rm QCD}\otimes{\rm SU}(2)_{\rm L}\otimes{\rm U}(1)_{\rm Y}\), apart from the required gauge bosons. That is enough to study the consistency of the theory from the point of view of the cancellation of anomalies, as already mentioned in the first chapter. In the second section the \textit{Goldstone theorem} is presented, along with its implications in the \textit{Spontaneous Symmetry Breaking} (SSB) mechanism, needed to understand why at low energies the physical states seem to be representations of the gauge group \({\rm SU}(3)_{\rm QCD}\otimes{\rm U}(1)_{\rm QED}\). Moreover, it is also shown how to quantize the model in a general \(R_{\xi}\) gauge. The SSB requires the introduction of the famous \textit{Higgs boson}, \textit{responsible for the origin of the masses of the fundamental particles}. The implications of having \textit{three massive generations of fermions} in the SM is shown in the third section. This is the basis of the so-called \textit{flavor physics}, which is one of the most important sectors to test the SM and look for new physics.

In spite of the success of the SM in explaining the huge amount of experimental results obtained over the last decades \cite{ParticleDataGroup:2024cfk}, \textit{it is nowadays believed that it is not the final theory of nature}. For instance, it does not explain the origin of neutrino masses \cite{ParticleDataGroup:2024cfk,Strumia:2006db,Mohapatra:2005wg,Gonzalez-Garcia:2007dlo,deGouvea:2016qpx,SajjadAthar:2021prg} and it has no candidate for a dark matter particle \cite{ParticleDataGroup:2024cfk,Bertone:2004pz,Hooper:2009zm,Klasen:2015uma,Lisanti:2016jxe}. \textit{It is currently thought to be an effective field theory} that only works up to a certain scale of new physics \(\Lambda_{\rm NP}\). Then, in the fourth section of this chapter the basic concepts of effective field theories are introduced, e.g., the \textit{integration of the heavy degrees of freedom}, the \textit{matching conditions} and the \textit{decoupling theorem}, which states that, under certain circumstances, \textit{the short-distance effects are negligible at low energies}. Finally, in the last section, the Lagrangian of the Standard Model Effective Field Theory is presented, paying special attention to the operators of dimension five and dimension six in the \textit{Warsaw basis}, which is the one that will be used in the calculations of the next chapter. 

\section{The Strong and Electroweak Interactions} \label{sec:Strong_EW}

The SM describes \textit{three of the four fundamental forces of nature: the strong, weak and electromagnetic interactions}. The former was first understood as a gauge theory based on the \({\rm SU}(3)_{\rm QCD}\) group in \cite{Fritzsch:1973pi}. The idea is that the elementary matter particles that feel the strong interaction, the \textit{quarks}, have an \textit{inner unobservable degree of freedom that can take three values} which is named \textit{color}; this is the reason why this theory is called \textit{Quantum Chromodynamics} (QCD). The three possible colors are blue, green and red, and the physics should not depend on this arbitrary label. This means that the associated Lagrangian is invariant under \({\rm SU}(3)_{\rm QCD}\) complex rotations of the quark triplets \(q_i=(q^b_i,q^g_i,q_i^r)^T\), which transform under the fundamental representation of this group. The subindex \(i=1,\dots,6\) stands for the six \textit{flavors} (types) of quarks of the SM: up (\(u\)), down (\(d\)), strange (\(s\)) \cite{Gell-Mann:1964ewy,Zweig:1964ruk,Zweig:1964jf,Bloom:1969kc,Breidenbach:1969kd}, charm (\(c\)) \cite{Bjorken:1964gz,Glashow:1970gm,Rosner:1998vy}, bottom (\(b\)) \cite{Kobayashi:1973fv,E288:1977xhf} and top (\(t\)) \cite{Kobayashi:1973fv,D0:1995jca}. 

Weak interactions can change the flavors of quarks and are responsible for the \(\beta\) decays. Consequently, particles that are unaffected by the strong force, the so-called \textit{leptons}, do feel the weak interaction. These are the electron (\(e\)), the muon (\(\mu\)) \cite{Street:1937me}, the tau (\(\tau\)) \cite{Perl:1975bf} and the associated electronic (\(\nu_{e}\)) \cite{Cowan:1956rrn}, muonic (\(\nu_{\mu}\)) \cite{Danby:1962nd} and tauonic (\(\nu_{\tau}\)) \cite{DONUT:2000fbd} neutrinos. The \(\beta\) decay was perfectly described at low energies by an effective Hamiltonian introduced by Fermi: \(\mathcal{H}=G_FJ_{\mu}^{\dagger}J^{\mu}\), with \(J^{\mu}=\bar{n}\gamma^{\mu}p+\bar{e}\gamma^{\mu}\nu_e\) and \(G_F\approx1.17\cdot10^{-5}\,{\rm GeV}^{-2}\). This suggests that an associated gauge theory will need the up-type and down-type quarks and the neutral and charged leptons to be included in doublets of an \({\rm SU}(2)\) group. Moreover, the observed approximate conservation of the lepton and baryon number per families \cite{Glashow:1970gm} indicates that the doublets are precisely \((u,d)^{T}\), \((c,s)^{T}\), \((t,b)^{T}\), \((\nu_e,e)^{T}\), \((\nu_{\mu},\mu)^{T}\) and \((\nu_{\tau},\tau)^{T}\). Furthermore, evidence shows \cite{Lee:1956qn,Wu:1957my} that \textit{only the left-handed part of the fermions can feel the weak force}. A consistent theory that included all these considerations based on the gauge group \({\rm SU(2)_{\rm L}\otimes{\rm U}(1)_{\rm Y}}\) was developed in \cite{Glashow:1961tr,Weinberg:1967tq,Salam:1968rm}. Notice that an additional \({\rm U}(1)\) group has been added in order to take into account the electromagnetic interactions. This is the reason why this theory is called the \textit{electroweak theory}.

As a summary, the fermions of this theory furnish the following representations of the SM gauge group \({\rm SU}(3)_{\rm QCD}\otimes{\rm SU}(2)_{\rm L}\otimes{\rm U}(1)_{\rm Y}\):

\begin{equation}
    \label{Fermions_Representations}
    \begin{split}
        &q_L^{i}=\left(\begin{array}{c}
        u_L^{i}\\d_L^{i}
        \end{array}\right)\sim(\mathbf{3},\mathbf{2})_{y_q},\hspace{0.8cm} u_R^{i}\sim(\mathbf{3},\mathbf{1})_{y_u}, \hspace{0.8cm} d_R^{i}\sim(\mathbf{3},\mathbf{1})_{y_d},\\&\ell_L^i=\left(\begin{array}{c}
        \nu_L^{i}\\e_L^i\end{array}\right)\sim(\mathbf{1},\mathbf{2})_{y_{\ell}},\hspace{0.8cm} e_R^i\sim(\mathbf{1,\mathbf{1}})_{y_e},
    \end{split}
\end{equation}
 
\noindent where the superindex \(i=1,2,3\) stands for the \textit{family number}. The first family includes the up and down quarks and the electronic leptons, the second family the charm and strange quarks and the muonic leptons and the third family the top and bottom quarks and the tauonic leptons. Notice that \textit{right-handed neutrinos have not been introduced}. The reason for that is that, in the SM, they will be \textit{sterile}, meaning that they get transformed under the \((\mathbf{1},\mathbf{1})_0\) representation of the group, so they will not have gauge interactions. Furthermore, it was thought that neutrinos were massless, so the right-handed counterpart was not needed. It is important to remark that the existence of a fourth family of fermions is completely excluded since the discovery of the Higgs boson, as it will be explained in Sec.~\ref{sec:CKM}.

The generator of the group \({\rm U}(1)_{\rm Y}\) is called the \textit{hypercharge} and its possible values are highly constrained by the requirement of cancellation of anomalies. Consider the triangle diagrams presented in Sec.~\ref{sec:Anomalies} that contain the gauge boson associated with this generator as some external line. Due to the fact that the trace of the generators of \({\rm SU}(3)_{\rm QCD}\) and \({\rm SU(2)}_{\rm L}\) is zero, the only non-vanishing amplitudes are the ones that include two gauge fields from \({\rm SU(2)}_{\rm L}\) or \({\rm SU}(3)_{\rm QCD}\) and one from \({\rm U}(1)_{\rm Y}\), or three gauge fields from \({\rm U}(1)_{\rm Y}\). Using the fact that the cancellation of the addition of all those diagrams requires the factor \( D_{\alpha\beta\gamma}=[{\rm tr}\{t_{\alpha}^L\{t_{\beta}^L,t_{\gamma}^L\}\}-{\rm tr}\{t_{\alpha}^R\{t_{\beta}^{R},t_{\gamma}^R\}\}]/2\) (Eq.~\eqref{Vanishing-Symmetric}) to vanish, this leads to the following conditions on the gauge couplings \cite{Weinberg2}:

\begin{itemize}
    \item \({\rm SU}(3)_{\rm QCD}-{\rm SU}(3)_{\rm QCD}-{\rm U}(1)_{\rm Y}\):
    \begin{equation}
        \label{Anomaly331}
        2y_q-y_u-y_d=0.
    \end{equation}
    \item \({\rm SU}(2)_{\rm L}-{\rm SU}(2)_{\rm L}-{\rm U}(1)_{\rm Y}\):
       \begin{equation}
        \label{Anomaly221}
            3y_q+y_{\ell}=0.
    \end{equation}
       \item \({\rm U}(1)_{\rm Y}-{\rm U}(1)_{\rm Y}-{\rm U}(1)_{\rm Y}\):
       \begin{equation}
        \label{Anomaly111}
            6y_q^3-3y_u^3-3y_d^3+2y_{\ell}^3-y_e^3=0.
    \end{equation}
\end{itemize}

\noindent In the first case, it has been used that only quarks can interact with the \({\rm SU}(3)_{\rm QCD}\) gauge bosons and that there are two left-handed quarks in the doublet. For the second case, one uses the fact that only the left-handed doublets interact with the \({\rm SU}(2)_{\rm L}\) gauge bosons and that there are three colors of quarks. Finally, in the last scenario one only needs to count how many fermions transform under the same representation of the \({\rm U}(1)_{\rm Y}\) group.

There is another important kind of anomaly that must be considered to study the consistency of the theory: the gravitational anomaly. As it was shown in \cite{Delbourgo:1972xb}, fermion loops with one gauge boson and two gravitons produce an anomaly of the form

\begin{equation}
    \label{Gravitational_Anomaly}
    {\rm tr}\{t_L-t_R\}\varepsilon^{\mu\nu\rho\sigma}R_{\mu\nu\kappa\lambda}R_{\rho\sigma}^{\,\,\,\,\,\kappa\lambda},
\end{equation}

\noindent where \(R_{\mu\nu\rho\sigma}\) is the Riemann tensor. This obstruction to the Slavnov-Taylor identities is automatically canceled for the \({\rm SU}(N)\) groups since \({\rm tr}\{t\}=0\). However, it yields to the following restriction when considering the generator of the hypercharge:

\begin{itemize}
    \item \({\rm Graviton}-{\rm Graviton}-{\rm U}(1)_{\rm Y}\):
    \begin{equation}
        \label{Anomaly_Gravity}
        6y_q-3y_u-3y_d+2y_{\ell}-y_e=0.
    \end{equation}
\end{itemize}

\noindent Apart from the freedom of interchanging the hypercharges of \(u_R\) and \(d_R\) there are only two solutions to these equations \cite{Weinberg2}:

\begin{itemize}
    \item \textit{Solution 1}: \(y_u/y_q=4\), \(y_d/y_q=-2\), \(y_{\ell}/y_q=-3\) and \(y_e/y_q=-6\).
    \item \textit{Solution 2}: \(y_u=-y_d\) and \(y_q=y_{\ell}=y_e=0\).
\end{itemize}

\noindent The electric charges that will be derived from the second solution do not correspond to what is observed in nature, so the only possibility is the first one. In this text, the normalization that will be taken is \(y_q=1/6\), so \(y_u=2/3\), \(y_d=-1/3\), \(y_{\ell}=-1/2\) and \(y_e=-1\). As a consequence, the electromagnetic charge and the hypercharge generators are related by \(Q=T_L^3+Y\). The fermion content of the theory transforms under the following representations of the SM gauge group:

\begin{equation}
    \label{Fermions_Representations_2}
    \begin{split}
        &q_L^{i}=\left(\begin{array}{c}
        u_L^{i}\\d_L^{i}
        \end{array}\right)\sim(\mathbf{3},\mathbf{2})_{1/6},\hspace{0.8cm} u_R^{i}\sim(\mathbf{3},\mathbf{1})_{2/3}, \hspace{0.8cm} d_R^{i}\sim(\mathbf{3},\mathbf{1})_{-1/3},\\&\ell_L^i=\left(\begin{array}{c}
        \nu_L^{i}\\e_L^i\end{array}\right)\sim(\mathbf{1},\mathbf{2})_{-1/2},\hspace{0.8cm} e_R^i\sim(\mathbf{1,\mathbf{1}})_{-1}.
    \end{split}
\end{equation}

The Lagrangian that contains the kinetic terms and interactions between fermions and gauge bosons is simply

\begin{equation}
    \label{Lagrangian_SM_Fermion_Gauge}
    \mathcal{L}=i\sum_f\bar{\psi}_f\slashed{D}\psi_f-\frac{1}{4}\sum_{\alpha=1}^8G^{\alpha\mu\nu}G^{\alpha}_{\mu\nu}-\frac{1}{4}\sum_{i=1}^3W^{i\mu\nu}W_{\mu\nu}^i-\frac{1}{4}B^{\mu\nu}B_{\mu\nu},
\end{equation}

\noindent where the subindex \(f\) stands for all the fermions of the SM. The covariant derivatives are given by:

\begin{equation}
    \label{Covariant_Derivative_SM}
    D_{\mu}\psi=\left(\partial_{\mu}-ig'YB_{\mu}-ig_LT_L^iW^i_\mu-ig_ST_S^{\alpha}G^\alpha_{\mu}\right)\psi.
\end{equation}

\noindent As mentioned in the first section, it is convenient to separate the semisimple group into its different simple parts and the \({\rm U}(1)\) factor introducing the coupling constants \(g'\), \(g_L\) and \(g_S\). The gauge fields associated with the \({\rm U}(1)_{\rm Y}\), \({\rm SU(2)}_{\rm L}\) and \({\rm SU}(3)_{\rm QCD}\) components of the group are \(B_{\mu}\), \(W_{\mu}^i\) and \(G^{\alpha}_{\mu}\), respectively. \(B_{\mu\nu}\), \(W_{\mu\nu}^i\) and \(G_{\mu\nu}^{\alpha}\) are the corresponding field-strength tensors. The eight particles associated with the \(G_{\mu}^{\alpha}\) fields are called gluons. Furthermore, it is usual to define \(W^{\pm}_{\mu}\coloneqq(W^1_{\mu}\mp iW^2_{\mu})/\sqrt{2}\), because these states are in correspondence with the physical \(W^{\pm}\) bosons of the weak interaction. The \(B_{\mu}\) and \(W_{\mu}^3\) fields are two neutral gauge bosons that are eigenstates of the interaction basis, but not of the mass basis.

\section{Spontaneous Symmetry Breaking and the Higgs Boson} \label{sec:SSB}

As mentioned in the first chapter, mass terms for the gauge bosons are forbidden by the symmetries of the theory. Indeed, the photon and the gluons are known to be massless, but the \(W^{\pm}\) particles and a neutral mediator of the weak interaction, the \(Z^{0}\), have an extremely high mass: \(M_W=80.369\pm0.013\,{\rm GeV}\) and \(M_Z=91.1880\pm0.0020\,{\rm GeV}\) \cite{ParticleDataGroup:2024cfk}. Furthermore, since the fermions are chiral it is not possible to add a Dirac mass term for them without breaking the symmetry. Only a hypothetical right-handed neutrino (which is not included in the SM) can acquire a Majorana mass because it would transform as a singlet under the gauge group. This means that the physical states that are observed in nature do not seem to respect the underlying symmetries needed for the consistency of the theory. The solution for this relies on the concept of \textit{Spontaneous Symmetry Breaking} and its implementation in the electroweak theory was done in \cite{Weinberg:1967tq,Salam:1968rm}. The discussion of these ideas will closely follow the presentation done in \cite{Weinberg2}.

The main idea behind SSB is that, even though the full theory might be invariant under a symmetry transformation, the ground state (the vacuum in a QFT) may not explicitly manifest it. This is analogous to the situation in a ferromagnetic material. The interactions are invariant under rotations, but when the magnet is cooled down the spins are aligned toward a privileged direction that seems to break this symmetry. It is important to remark that the rotational symmetry is still there and, due to it, this particular direction could have been any one. In order to incorporate these ideas to QFT in the perturbative regime\footnote{In the non-perturbative scenario it is possible to have fermion-bilinears that take a non-vanishing vacuum expectation value \(\left<\bar{\psi}\psi\right>_{\rm VAC}\neq0\). This is analogous to the SSB of the chiral symmetry in QCD.} it is necessary to have a set of fields whose \textit{Vacuum Expectation Value} (VEV) is non-vanishing. This is not possible with spinors or fermions because it will lead to the SSB of the Lorentz symmetry, which is known to be satisfied. As a consequence, it is necessary to add a set of scalars to the Lagrangian of the SM.

Consider that the path integral is invariant under the following linear infinitesimal transformation of the scalar fields:

\begin{equation}
    \label{Transformation_Scalars}
    \phi_n\left(x\right)\to\phi_n\left(x\right)+i\varepsilon\sum_mt_{nm}\phi_m\left(x\right).
\end{equation}

\noindent For simplicity, it will be considered that the scalar fields are real since it is always possible to decompose a complex field into its real and imaginary parts. As mentioned in the first chapter, if there is a linear symmetry transformation that preserves the Lagrangian and the measure then the QEA is also invariant under it, so

\begin{equation}
    \label{Transformation_QEA_Scalar}
    \sum_{n,m}\int\frac{\delta\Gamma[\phi]}{\delta\phi_n(x)} t_{nm}\phi_m(x){\rm d}^4x=0.
\end{equation}

\noindent Consider that the \(\phi_n(x)\) take the same value for all the points of the spacetime. In
that case the QEA is usually written as \(\Gamma[\phi]=-\mathcal{V}_4V[\phi]\), where \(\mathcal{V}_4\) is the four-dimensional volume and \(V[\phi]\) the \textit{quantum effective potential}. This decomposition relies on the fact that the QEA is an action that contains all the information about the quantum properties of the theory, so the integration of the corresponding Lagrangian density with constant fields leads to the volume of the spacetime times the Lagrangian density and since the derivatives vanish only the potential survives. In terms of \(V[\phi]\) Eq.~\eqref{Transformation_QEA_Scalar} reads

\begin{equation}
    \label{Transformation_QEP}
    \sum_{nm}\frac{\delta V[\phi]}{\delta\phi_n}t_{nm}\phi_m=0.
\end{equation}

\noindent Taking the second variation with respect to \(\phi_p\) one gets the condition

\begin{equation}
    \label{Second_Variation_QEP}
    \sum_n\frac{\delta V[\phi]}{\delta\phi_n}t_{np}+\sum_{nm}\frac{\delta^2V[\phi]}{\delta\phi_n\delta\phi_p} t_{nm}\phi_m=0.
\end{equation}

Now suppose that the constant values of the \(\phi_n\) are precisely the ones that minimize the quantum effective potential, i.e., the vacuum expectation values \(\bar{\phi}_n\). Then, Eq.~\eqref{Second_Variation_QEP} leads to

\begin{equation}
    \label{Second_Variation_QEP_Minimum}
    \sum_{nm}\left.\frac{\delta^2V[\phi]}{\delta\phi_n\delta\phi_p}\right|_{\phi=\bar{\phi}} t_{nm}\bar{\phi}_m=0.
\end{equation}

\noindent The second derivative of the QEA can be computed from all 1PI graphs with \(\phi_n\) and \(\phi_p\) as external lines with zero momenta in the case where those fields are constant. But this is simply the inverse of the propagator at \(p=0\), \(\Delta^{-1}_{np}(p=0)\), so

\begin{equation}
    \label{Propagator_Condition}
    \sum_{n}\Delta^{-1}_{np}(p=0)\left(\sum_mt_{nm}\bar{\phi}_m\right)=0.
\end{equation}

\noindent As a consequence, if the vacuum is not invariant under a certain symmetry, then the propagator has a pole at \(p=0\) implying that there is an associated massless particle. These massless scalars are called Nambu-Goldstone bosons \cite{Nambu:1960xd,Goldstone:1961eq} who found them in specific situations associated with superconductivity. A general proof of their existence appears in \cite{Goldstone:1962es}.

It may seem that finding additional massless particles makes the situation even more difficult, but the presence of these bosons in gauge theories has important implications. Imagine that the physical model has a symmetry associated with a \textit{compact} group \(G\) which is spontaneously broken into a subgroup \(H\). This means that the vacuum is invariant under any transformation of \(H\). It is going to be shown that, in this situation, it is always possible to find a transformation \(\gamma\left(x\right)\) of \(G\) such that for any field configuration

\begin{equation}
    \label{Rotation_Goldstone}
    \phi_n\left(x\right)=\sum_m\gamma_{nm}\left(x\right)\tilde{\phi}_m\left(x\right),
\end{equation}

\noindent where the set \(\tilde{\phi}_n(x)\) is free of Goldstone bosons. First of all, from Eq.~\eqref{Propagator_Condition} it is clear that these Goldstone modes are linear combinations of the vectors \(\sum_mt^{\alpha}_{nm}\bar{\phi}_m\), where \(t^{\alpha}\) is a broken generator of \(G\). Since the group is compact and the scalars are real, it is possible to work with orthogonal representations of it. Therefore, the condition of \(\tilde{\phi}_n(x)\) being free of Goldstone modes can be expressed as

\begin{equation}
    \label{Free_of_Goldstones}
    \sum_{nm}\tilde{\phi}_n(x)t^{\alpha}_{nm}\bar{\phi}_m=0.
\end{equation}

Consider the quantity 

\begin{equation}
    \label{Quantity_Proof}
    C_{\phi}(g)\coloneqq\sum_{nm}\phi_{n}(x)g_{nm}\tilde{\phi}_m,
\end{equation}

\noindent where \(g\) is an element of \(G\). Since the group is compact, this function of \(g\) has to achieve a stationary value for some element \(g=\gamma(x)\). Taking an infinitesimal variation \(\delta g=i\sum_{\alpha}\varepsilon_{\alpha}gt^{\alpha}\) around that stationary point one gets:

\begin{equation}
    \label{Stationary_Variation}
    \begin{split}
        0=\delta C_{\phi}(\gamma(x))&=i\sum_{\alpha}\varepsilon_{\alpha}\sum_{nmp}\phi_n(x)\gamma_{np}(x)t^{\alpha}_{pm}\bar{\phi}_m\\&=i\sum_{\alpha}\varepsilon_{\alpha}\sum_{mp}\left(\sum_n\gamma^{-1}_{pn}(x)\phi_n(x)\right)t^{\alpha}_{pm}\bar{\phi}_m,
    \end{split}
\end{equation}

\noindent where it has been used that \(\gamma^{T}=\gamma^{-1}\). Since this is true for every variation around \(\gamma\) then

\begin{equation}
    \label{Stationary_Variation_2}
    \sum_{mp}\left(\sum_n\gamma^{-1}_{pn}(x)\phi_n(x)\right)t^{\alpha}_{pm}\bar{\phi}_m=0,
\end{equation}

\noindent proving that \(\sum_n\gamma^{-1}_{pn}(x)\phi_n(x)\) is free of Goldstone modes and therefore it is possible to find such \(\gamma(x)\) of \(G\) that makes Eq.~\eqref{Free_of_Goldstones} hold.

Up to this point the reasoning is completely general. However, if the group \(G\) is associated with a local symmetry of the theory then Eq.~\eqref{Free_of_Goldstones} implies that it is possible to find a gauge transformation such that the Goldstone modes are rotated away. Such gauge is called \textit{unitary gauge} and due to the gauge invariance of the theory its existence implies that these massless scalars are not physical. Nonetheless, the existence of those degrees of freedom has important implications. Consider the kinetic term in such a theory of the fields \(\tilde{\phi}_n(x)\):

\begin{equation}
    \label{Kinetic_Term_Free_Goldstones}
    \mathcal{L}_{K\phi}=\frac{1}{2}\sum_{n}\left(D_{\mu}\tilde{\phi}_n(x)\right)^2=\frac{1}{2}\sum_n\left(\partial_{\mu}\tilde{\phi}_n(x)-i\sum_{m\alpha}t^{\alpha}_{nm}A_{\alpha\mu}(x)\tilde{\phi}_m(x)\right)^2.
\end{equation}

\noindent It is convenient to write explicitly the VEV \(v_n\) of the fields \(\tilde{\phi}_n(x)=v_n+\tilde{\phi}_{n}'(x)\). Thus, the quadratic part of the previous Lagrangian is obtained from:

\begin{equation}
    \label{Kinetic_Term_Free_Goldstones_Quadratic}
    \mathcal{L}_{K\phi}^{\rm Quad}=\frac{1}{2}\sum_n\left(\partial_{\mu}\tilde{\phi}_n(x)-i\sum_{m\alpha}t^{\alpha}_{nm}A_{\alpha\mu}(x)v_m\right)^2. 
\end{equation}

\noindent From Eq.~\eqref{Free_of_Goldstones} it is clear that the cross term \(\sum_{nm\alpha}\partial_{\mu}\tilde{\phi}_n(x)t^{\alpha}_{nm}v_mA_{\alpha}^{\mu}\) vanishes so 

\begin{equation}
    \label{Kinetic_Term_Free_Goldstones_Quadratic_2}
    \mathcal{L}_{K\phi}^{\rm Quad}=\frac{1}{2}\sum_n\partial_{\mu}\tilde{\phi}_n(x)\partial^{\mu}\tilde{\phi}^n(x)+\frac{1}{2}\sum_{\alpha\beta}\mu_{\alpha\beta}^2A_{\alpha\mu}A^{\mu}_{\beta},
\end{equation}

\noindent with

\begin{equation}
    \label{Mass_Term_Gauge_Bosons}
    \mu_{\alpha\beta}^{2}\coloneqq-\sum_{nmp}t^{\alpha}_{nm}t^{\beta}_{np}v_mv_l.
\end{equation}

\noindent This means that gauge fields associated with broken generators get a mass while the others remain massless. In other words, the degrees of freedom of the Goldstone modes are absorbed to produce the longitudinal polarization of those vectors.

In the case of the SM it is necessary to spontaneously break the \({\rm SU}(2)_{\rm L}\otimes{\rm U}(1)_{\rm Y}\) group into \({\rm U}(1)_{\rm QED}\). This means that the only unbroken linear combination of generators will be \(Q=T_L^3+Y\). The interactions between the scalars and the fermions must give a mass to these particles through the so-called \textit{Yukawa couplings}. For instance, taking the case of the electron, the easiest way to generate a Dirac mass term \(-m_e\bar{e}_Le_R+{\rm h.c.}\) is to introduce a doublet \(\Phi=(\phi^+,\,\phi^0)^{T}\) and the following operator in the Lagrangian:

\begin{equation}
    \label{Yukawa_Electron}
    \mathcal{L}_{Ye}=-y_e\left(\bar{\nu}_{L},\,\bar{e}_L\right)\left(\begin{array}{c}
        \phi^+ \\
         \phi^0 
    \end{array}\right)e_R+{\rm h.c.}
\end{equation}

\noindent Therefore, the doublet gets transformed under the representation \(\Phi\sim(\mathbf{1},\mathbf{2})_{1/2}\) of the gauge group. As already mentioned, the electric charge must be an unbroken generator \(Q\left<\Phi\right>_{\rm Vac}=0\), leading to the condition \(\left<\phi^+\right>_{\rm Vac}=0\). Furthermore, it is always possible to make a gauge rotation (a rotation into the unitary gauge) that eliminates \(\phi^+\) and makes \(\phi^0\) a real scalar field. As a consequence, it is possible to decompose the doublet in that gauge as

\begin{equation}
    \label{Doublet_VEV}
    \Phi=\frac{1}{\sqrt{2}}\left(\begin{array}{c}
        0 \\
        v+H 
    \end{array}\right),
\end{equation}

\noindent where \(v\) represents the VEV of the doublet and \(H\) the scalar field obtained after the complex rotation.

Apart from the Yukawa couplings (that will be studied in more detail in the next section), the most general renormalizable Lagrangian that contains the doublet is

\begin{equation}
    \label{Lagrangian_Doublet}
    \mathcal{L}_{\Phi}=\left|\left(\partial_{\mu}-ig_LT_L^iW_{\mu}^i-\frac{i}{2}g'B_{\mu}\right)\Phi\right|^2-\frac{\mu^2}{2}\Phi^{\dagger}\Phi-\frac{\lambda}{4}\left(\Phi^{\dagger}\Phi\right)^2.
\end{equation}

\noindent Using Eq.~\eqref{Doublet_VEV} it is possible to show that the previous Lagrangian is equivalent to

\begin{equation}
    \label{Lagrangian_Doublet_Expansion}
    \begin{split}
        \mathcal{L}_{\Phi}&=\frac{1}{2}\partial_{\mu}H\partial^{\mu}H+\frac{1}{4}g_L^2v^2\left(1+\frac{H}{v}\right)^2W_{\mu}^+W^{\mu-}+\frac{1}{2}\frac{v^2}{4}\left(1+\frac{H}{v}\right)^2\left(g_LW_{\mu}^3-g'B_{\mu}\right)^2\\&-\frac{\mu^2}{4}(v+H)^2-\frac{\lambda}{16}\left(v+H\right)^4.
    \end{split}
\end{equation}

\noindent At tree level, the value of the VEV is obtained from the minimum of the potential \(V(v)\coloneqq\mu^2v^2/4+\lambda v^4/16\), which is placed at  \(v=\sqrt{-2\mu^2/\lambda}\). This implies that \(\mu^2<0\), since \(\lambda>0\) in order to have a minimum and not a maximum. Therefore Eq.~\eqref{Lagrangian_Doublet_Expansion} reads

\begin{equation}
    \label{Lagrangian_Doublet_Expansion_2}
    \begin{split}
        \mathcal{L}_{\Phi}&=-V_{\rm min}+\frac{1}{2}\partial_{\mu}H\partial^{\mu}H-\frac{1}{2}(-\mu^2)H^2-\frac{\lambda v}{4}H^3-\frac{\lambda}{16}H^4\\&+\frac{1}{4}g_L^2v^2\left(1+\frac{H}{v}\right)^2W_{\mu}^+W^{\mu-}+\frac{1}{2}\frac{v^2}{4}\left(1+\frac{H}{v}\right)^2\left(g_LW_{\mu}^3-g'B_{\mu}\right)^2.
    \end{split}
\end{equation}

\noindent As a consequence, associated with \(H\) there is a massive particle (with mass \(-\mu^2>0\) at tree-level) which is called the Higgs boson \cite{Englert:1964et,Higgs:1964pj,Guralnik:1964eu}. The announcement of its discovery in the year 2012 by the ATLAS \cite{ATLAS:2012yve} and CMS \cite{CMS:2012qbp} collaborations completed the SM. Recent measurements show that the mass of this particle is \(M_H=125.20\pm0.11\,{\rm GeV}\) \cite{ParticleDataGroup:2024cfk}.

 It has been shown that only one of the four degrees of freedom of the doublet is physical, which means that the remaining three are used to give masses to the vector bosons. Indeed, looking at Eq.~\eqref{Lagrangian_Doublet_Expansion_2} is clear that, at tree level, the \(W^{\pm}\) bosons have a mass \(M_W=\frac{1}{2}g_Lv\) and a coupling to the Higgs field that is proportional to its square. However, it is also clear that the interaction eigenstates \(W_{\mu}^3\) and \(B_{\mu}\) are not mass eigenstates. Indeed, its mass term can be written as

 \begin{equation}
     \label{Mass_Term_Neutral}
      \frac{1}{2}\frac{v^2}{4}\left(W_{\mu}^3,\, B_{\mu}\right)\left(\begin{array}{cc}
          g_L^2 & -g_Lg' \\
          -g_Lg' & g'^2
      \end{array}\right)\left(\begin{array}{c}
        W^{3\mu} \\
         B^{\mu}  
      \end{array}\right).
 \end{equation}

\noindent The mass eigenstates are obtained diagonalizing this matrix. Defining the Weinberg angle \(\theta_W\) as

\begin{equation}
    \label{Cos_Sin_Weinberg}
    \cos\theta_W\coloneqq \frac{g_L}{\sqrt{g_L^2+g'^2}}, \hspace{0.6cm} \sin\theta_W=\frac{g'}{\sqrt{g_L^2+g'^2}},
\end{equation}

\noindent it is easy to show that the massless eigenstate associated with the photon is \(A_{\mu}=\cos\theta_WB_{\mu}+\sin\theta_WW^3_{\mu}\) and the massive one associated with the \(Z^0\) is \(Z_{\mu}=\cos\theta_WW_{\mu}^3-\sin\theta_WB_{\mu}\). Its mass is given at tree level by \(M_Z=\frac{1}{2}v\sqrt{g_L^2+g'^2}\), so it is related with the one of the \(W^{\pm}\) bosons by

\begin{equation}
    \label{Mass_Bosons_Relation}
    M_W=M_Z\cos\theta_W.
\end{equation}

\noindent Furthermore, the coupling of the \(Z\) with the Higgs field is also proportional to the square of its mass, while the photon does not couple to this scalar:

\begin{equation}
    \label{Lagrangian_Doublet_Expansion_3}
    \begin{split}
        \mathcal{L}_{\Phi}&=-V_{\rm min}+\frac{1}{2}\partial_{\mu}H\partial^{\mu}H-\frac{1}{2}(-\mu^2)H^2-\frac{\lambda v}{4}H^3-\frac{\lambda}{16}H^4\\&+M_W^2\left(1+\frac{H}{v}\right)^2W_{\mu}^+W^{\mu-}+\frac{1}{2}M_Z^2\left(1+\frac{H}{v}\right)^2Z_{\mu}Z^{\mu}.
    \end{split}
\end{equation}

The problem with the unitary gauge is that the propagator of a massive vector field behaves as \(\Delta(k)\sim\mathcal{O}(k^0)\) in the limit \(k\to\infty\). As a consequence, the fact that the theory is renormalizable is not straightforward to prove. However, it is possible to work in a different gauge where those propagators are well-behaved, but the Goldstone bosons are present so the physical particle content of the theory is not manifest. These are the so-called Renormalizable \(\xi\)-gauges \cite{Fujikawa:1972fe}, usually denoted as \(R_{\xi}\) gauges. In the expansion of the Lagrangian in Eq.~\eqref{Kinetic_Term_Free_Goldstones} in a general gauge with \(\phi_n(x)=v_n+\phi_n'(x)\) there is a problematic quadratic term \(-i\sum_{nm\alpha}\partial_{\mu}\phi'_nt^{\alpha}_{nm}A_{\alpha}^{\mu}v_m\) which vanishes in the unitary gauge because \(\tilde{\phi}_n'\) is free of Goldstone bosons. Nevertheless, there is another possible way to eliminate that contribution. Following the construction in Sec.~\ref{sec_Faddeev_Popov} it is possible to add the following functional to the path integral:

\begin{equation}
    \label{Functional_Gauge_Fixing_SSB}
    B[f]=\exp\left\{\frac{-i}{2\xi}\int{\rm d}^4x\sum_{\alpha}f_{\alpha}f_{\alpha}\right\},
\end{equation}

\noindent with the function \(f_{\alpha}\) being:

\begin{equation}
    \label{Function_Gauge_Fixing_SSB}
    f_{\alpha}=\partial_{\mu}A_{\alpha}^{\mu}+i\xi\sum_{nm}t^{\alpha}_{nm}\phi'_nv_m.
\end{equation}

\noindent The introduction of this functional is equivalent to adding the gauge-fixing term \(\mathcal{L}_{\rm gf}=-\frac{1}{2\xi}\sum_{\alpha}f_{\alpha}f_{\alpha}\) to the Lagrangian. Therefore, the crossed term in the product of the functions \(f_{\alpha}\) cancels the term \(-i\sum_{nm\alpha}\partial_{\mu}\phi'_nt^{\alpha}_{nm}A_{\alpha}^{\mu}v_m\).

It is important to remark that the results of the unitary gauge are obtained when \(\xi\to\infty\) because only the stationary phase of \(B[f]\) can survive and it corresponds to the condition \(\sum_{nm\alpha}\partial_{\mu}\phi'_nt^{\alpha}_{nm}A_{\alpha}^{\mu}v_m=0\). In the limit \(\xi\to0\) the condition of the Landau gauge \(\partial_{\mu}A^{\mu}_{\alpha}=0\) is derived. As shown in Sec.~\ref{sec_Faddeev_Popov}, it is necessary to add the ghosts and antighosts to the Lagrangian. With this gauge fixing function the contribution is 

\begin{equation}
    \label{Ghosts_SSB}
    \mathcal{L}_{\rm ghost}=\omega_{\alpha}^{\ast}\left\{\square\omega_{\alpha}-\sum_{\beta\gamma}C_{\alpha\beta\gamma}\partial_{\mu}\left(\omega_{\beta}A_{\gamma}^{\mu}\right)-\xi\sum_{n\beta}\left(t_{\alpha}v\right)_n\omega_{\beta}\left[\left(t_{\beta}v\right)_n+\left(t_\beta\phi'\right)_n\right]\right\},
\end{equation}

\noindent so the ghosts obtain a mass proportional to the parameter \(\xi\). It is also possible to show \cite{Weinberg2} that the propagator of a gauge field in this case is

\begin{equation}
    \label{Propagator_Gauge_Rxi}
    \Delta_{\mu\nu}=\frac{-i}{k^2-M^2}\left(\eta_{\mu\nu}-\frac{k_{\mu}k_{\nu}(1-\xi)}{k^2-\xi M^2}\right),
\end{equation}

\noindent where \(M\) is the mass that it will acquire after the SSB. As it can be seen, now there 
is no problem with the behavior at \(k\to\infty\) for \(\xi\) fixed.

\section{Massive Fermions and Flavor Physics. The CKM Matrix} \label{sec:CKM}

Apart from giving masses to the gauge bosons, the Higgs mechanism can also produce Dirac mass terms for the fermions through the already mentioned Yukawa interactions. Consider the charge conjugated field \(\Phi^c\coloneqq i\sigma_2\Phi^{\ast}=\left(\phi^{0\dagger},-\phi^-\right)^{T}\), where \(\sigma_2\) is the second Pauli matrix and \(\phi^-=\phi^{+\dagger}\). It is easy to show that \(\Phi^c\sim(\mathbf{2},\mathbf{1})_{-1/2}\), so the most general Lagrangian with operators up to dimension four that combines scalars and fermions is

\begin{equation}
    \label{Yukawa_Lagrangian}
    \mathcal{L}_Y=-\sum_{ij}\left\{Y^d_{ij}\,\bar{q}^i_L\Phi d^j_{R}+Y_{ij}^u\,\bar{q}^i_L\Phi^c u_R^j+Y_{ij}^e\,\bar{\ell}_L^i\Phi e_R^j\right\}+{\rm h.c.},
\end{equation}

\noindent where the indices \(i\) and \(j\) run over the three families of the SM and, consequently, \(Y^d\), \(Y^u\) and \(Y^e\) are three \(3\times3\) matrices that contain the so-called Yukawa couplings. In the unitary gauge, the previous Lagrangian reads

\begin{equation}
    \label{Yukawa_Lagrangian_Unitary_Gauge}
    \mathcal{L}_Y=-\frac{1}{\sqrt{2}}\left(v+H\right)\sum_{ij}\left\{Y^d_{ij}\,\bar{d}^i_L d^j_{R}+Y_{ij}^u\,\bar{u}^i_L u_R^j+Y_{ij}^e\,\bar{e}_L^i e_R^j\right\}+{\rm h.c.}
\end{equation}

\noindent Therefore, the term with the VEV produces  the following three mass-matrices:

\begin{equation}
    \label{Mass_Matrices_Quarks}
    M^d\coloneqq\frac{1}{\sqrt{2}}vY^d,\hspace{0.8cm} M^u\coloneqq\frac{1}{\sqrt{2}}vY^u,\hspace{0.8cm} M^e\coloneqq\frac{1}{\sqrt{2}}vY^e.
\end{equation}

\noindent It is important to remark that \textit{there are no mass-terms for the neutrinos since the right-handed counterparts have not been introduced}.

Now, it is necessary to diagonalize these matrices and write the Lagrangian in terms of the mass eigenstates. For this purpose, consider the fact that the Lagrangian in Eq.~\eqref{Lagrangian_SM_Fermion_Gauge} is invariant under an \({\rm SU}(3)^5_f\) flavor transformation \(q_L^i\to U^q_{ij}q^j_L\), \(u_R^i\to U^u_{ij}u^j_R\), \(d_R^i\to U^d_{ij}d^j_R\), \(\ell_L^i\to U^{\ell}_{ij}\ell^j_L\) and \(e_R^i\to U^e_{ij}e^j_R\). These complex rotations lead to the following transformation in the mass matrices: 

\begin{equation}
    \label{Transformation_Mass_Matrices}
    M^d\to U^{q\dagger}M^dU^d, \hspace{0.8cm} M^u\to U^{q\dagger}M^uU^{u}, \hspace{0.8cm} M^e\to U^{\ell\dagger}M^eU^e.
\end{equation}

\noindent There is an important theorem that states that any complex matrix \(M\) can be decomposed as \(M=UDV^{\dagger}\), where \(U\) and \(V\) are unitary matrices and \(D\) is a real diagonal positive semi-definite (no negative values in the diagonal) matrix. This is known as \textit{singular value decomposition}. Therefore, it is possible to use the transformations in Eq.~\eqref{Transformation_Mass_Matrices} to diagonalize, for instance, \(M^e\) and \(M^u\). In the leptonic sector this leads to the fact that it is always possible to write the Lagrangian in terms of physical fields where there are no interactions between different families. This implies that the lepton number (the number of leptons minus the number of antileptons) per family is conserved in the SM.\footnote{This is not true when considering quantum effects. The only quantum numbers associated with a global symmetry that are not anomalous in the SM are the differences between the lepton numbers per family and \(B-L\), where \(B\) and \(L\) are the total baryon and lepton numbers, respectively.} It is important to remark that the singular value decomposition is not unique because \(M\) is also diagonalized by \(D=(UK)^{\dagger}M(VK)\), where \(K\) is a diagonal matrix of phases.

The mass matrix of the down-type quarks is still not in a diagonal form. Since \(U^q\) has been used to diagonalize \(M^u\) but \(U^d\) is still arbitrary, it is only necessary to introduce one additional unitary matrix. Therefore, it is possible to obtain the mass eigenstates from the following general transformations:

\begin{equation}
    \label{Diagonalization_Mass_Matrices}
    \begin{split}
        &u_L^i\to (U^qK^u)_{ij}u_L^j,\hspace{0.8cm}u_R^i\to(U^uK^u)_{ij}u_R^j, \hspace{0.8cm} d_L^i\to(U^qU^{d_L}K^d)_{ij}d_L^j, \\& d_R^i\to(U^dK^d)_{ij}d_R^j, \hspace{0.8cm} \ell_L^i\to U^{\ell}_{ij}\ell_L^j, \hspace{0.8cm} e_R^i\to U^e_{ij}e_R^j.
    \end{split}
\end{equation}

\noindent They diagonalize the mass matrices and leave the interaction  between the Higgs field and the fermions in the unitary gauge as

\begin{equation}
    \label{Yukawa_Lagrangian_Unitary_Gauge_2}
    \mathcal{L}_Y=-\left(1+\frac{H}{v}\right)\sum_{i}\left\{\mathcal{M}^d_{i}\,\bar{d}^i_L d^i_{R}+\mathcal{M}^u_i\,\bar{u}^i_L u_R^i+\mathcal{M}^e_i\,\bar{e}_L^i e_R^i\right\}+{\rm h.c.},
\end{equation}

\noindent where \(\mathcal{M}^d=(m_d,m_s,m_b)\), \(\mathcal{M}^u=(m_u,m_c,m_t)\) and \(\mathcal{M}^e=(m_e,m_{\mu},m_{\tau})\). It is important to remark that \textit{the couplings between the Higgs and the gauge bosons are quadratic in their masses and the couplings with the fermions are just proportional to them}. The latter property is the reason why it is nowadays known that there are only three families of fermions. The existence of new quarks with a mass high enough not to have been detected by the LHC will increase the probability of production of Higgs bosons to a value that is already excluded \cite{Eberhardt:2012gv,Eberhardt:2012sb}.

As already mentioned, the transformation in Eq.~\eqref{Diagonalization_Mass_Matrices} leaves the leptonic sector invariant. Nonetheless, since the transformations of the \(u_L\) and \(d_L\) fields are different, the situation is not the same with the quarks \cite{Glashow:1970gm}. Due to the fact that all the matrices are unitary, the neutral currents \(\bar{u}_{L,R}\gamma^{\mu}u_{L,R}\) and \(\bar{d}_{L,R}\gamma^{\mu}d_{L,R}\) remain invariant, so \textit{there are no Flavor Changing Neutral Currents (FCNC) in the SM at tree level}. However, it is easy to see that the interactions between the fermions and the \(W^{\pm}\) bosons can be written as

\begin{equation}
    \label{Charged_Current_Lagrangian}
    \mathcal{L}_{CC}=\frac{1}{\sqrt{2}}g_LW_{\mu}^{+}\sum_{ij}\bar{u}^i_L\gamma^{\mu}\left(V_{\rm CKM}\right)_{ij}d_L^j+{\rm h.c.},
\end{equation}

\noindent where \(V_{\rm CKM}\coloneqq K^{u\dagger}U^{d_L}K^d\) is known as the Cabibbo-Kobayashi-Maskawa matrix \cite{Cabibbo:1963yz,Kobayashi:1973fv} and it is the basis of the so-called quark flavor physics, because it induces flavor changing charged currents. \(V_{\rm CKM}\) is a completely general \(3\times3\) unitary matrix, so it contains nine free parameters, three of them are real and six of them are complex. However, there are six arbitrary phases in \(K^u\) and \(K^d\), although one of them has no physical meaning because it is associated with the conservation of the total baryon number. This means that it is possible to rotate away five phases in such a way that there is only one physical complex parameter in \(V_{\rm CKM}\). This is the only sizable complex parameter in the SM, so it is the only source of CP violation and thus the only possible explanation for the baryogenesis in the origin of the Universe. Nevertheless, it is known that starting with equal amounts of matter and antimatter its effect is not enough to explain the current observations. It is important to remark that these phases only appear if the number of generations is greater than or equal to three \cite{Kobayashi:1973fv}.

There are different parametrizations of the CKM matrix. The Chau-Keung parametrization is:

\begin{equation}
    \label{Parametrization_CKM}
     \begin{split}
     V_{\rm CKM}&=\left(
     \begin{array}{ccc}
         1 & 0 & 0 \\
         0 & c_{23} & s_{23} \\
         0 & -s_{23} & c_{23}
     \end{array}
     \right)\left(
     \begin{array}{ccc}
         c_{13} & 0 & s_{13}\,e^{-i\delta} \\
         0 & 1 & 0 \\
         -s_{13}\,e^{i\delta} & 0 & c_{13}
     \end{array}
     \right)\left(
     \begin{array}{ccc}
         c_{12} & s_{12} & 0 \\
         -s_{12} & c_{12} & 0\\
         0 & 0 & 1
     \end{array}
     \right)\\
     &=\left(
     \begin{array}{ccc}
        c_{12}c_{13}  & s_{12}c_{13} & s_{13}\,e^{-i\delta}  \\
        -s_{12}c_{23}-c_{12}s_{23}s_{13}\,e^{i\delta}  & c_{12}c_{23}-s_{12}s_{23}s_{13}\,e^{i\delta} & s_{23}c_{13} \\
         s_{12}s_{23}-c_{12}c_{23}s_{13}\,e^{i\delta} & -c_{13}s_{23}-s_{12}c_{23}s_{13}\,e^{i\delta} & c_{23}c_{13}
     \end{array}
     \right),
 \end{split}
\end{equation}

\noindent where \(s_{ij}\) and \(c_{ij}\) are short-hand notations for \(\sin\theta_{ij}\) and \(\cos\theta_{ij}\), respectively. Nowadays, from a global fit of the SM, the measured values of the matrix elements are \cite{ParticleDataGroup:2024cfk}:

\begin{equation}
    \label{CKM_Measurements}
    \left|V_{\rm CKM}\right|=\left(
    \begin{array}{ccc}
    0.97435\pm0.00016 & 0.22501\pm0.00068 & 0.003732^{+0.000090}_{-0.000085} \\
    0.22487\pm0.00068  & 0.97349\pm0.00016 & 0.04183^{+0.00079}_{-0.00069} \\
    0.00858^{+0.00019}_{-0.00017} & 0.04111^{+0.00077}_{-0.00068} & 0.999118^{+0.000029}_{-0.000034}
    \end{array}\right),
\end{equation}

\noindent and the CP-violation phase is \(\delta=1.147\pm0.026\,{\rm rad}\). This hierarchical structure of the CKM matrix can also be easily appreciated in the Wolfenstein parametrization \cite{Wolfenstein:1983yz}, where a small \(\lambda\) describes the data:

\begin{equation}
    \label{Wolfenstein}
    V_{\rm CKM}=\left(
    \begin{array}{ccc}
    1-\lambda^2/2 & \lambda & A\lambda^3\left(\rho-i\eta\right) \\
    -\lambda & 1-\lambda^2/2 & A\lambda^2 \\
    A\lambda^3(1-\rho-i\eta) & -A\lambda^2 & 1
    \end{array}\right)+\mathcal{O}\left(\lambda^4\right).
\end{equation}

\noindent The values of the Wolfenstein parameters obtained from the global fit of the SM are \(\lambda=0.22501\pm0.00068\), \(A= 0.826^{+0.016}_{-0.015}\), \(\bar{\rho}=0.1591\pm0.0094\) and \(\bar{\eta}=0.3523^{+0.0073}_{-0.0071}\). It is clear that the CKM matrix has an almost diagonal structure and in principle there is no reason for this. This is part of the so-called \textit{flavor puzzle}, which also includes other aspects like the hierarchy of the fermion masses.

\section{Effective Field Theories} \label{sec:EFTs}

It is nowadays believed that neither the SM of particle physics nor General Relativity (GR) are the final theories to describe the fundamental interactions of the universe. Both of them are thought to be EFTs, i.e., theories that only work up to a certain energy scale usually referred as \(\Lambda_{\rm NP}\). \textit{The Lagrangian of an EFT contains an infinite set of all the operators that respect an underlying symmetry} \cite{Weinberg:1978kz}. These operators are classified into three groups depending on their mass scale dimension \(d\). If the spacetime dimension is denoted as \(D_0\) one can find:

\begin{itemize}
    \item[1.] \textit{Relevant Operators}. Operators with \(d<D_0\), whose effects become important at low energies. In \(D_0=4\) these are only the kinetic and mass terms and a \(\phi^3\) interaction in a theory with scalars \cite{Weinberg1}.
    \item[2.] \textit{Irrelevant Operators}. Operators with \(d>D_0\), whose effects are suppressed at low energies \(E\ll\Lambda_{\rm NP}\). Indeed, a theory containing these terms in the Lagrangian usually violates unitarity for high enough energies. The scale at which unitarity is broken indicates the scale \(\Lambda_{\rm NP}\) for which a \textit{UV completion} is needed in order to explain the interactions.
    \item[3.] \textit{Marginal Operators}. Operators with \(d=D_0\), whose behavior at high and low energies depends on the quantum corrections. For instance, from the Renormalization Group Equations (RGEs), it is well-known that the electric charge of QED grows at high energies and it is small at low scales, while the coupling constant \(g_S\) of QCD has the opposite behavior (asymptotic freedom).
\end{itemize}

In this sense, the difference between the SM and GR is that in the former it is possible to find both relevant and marginal operators because the theory only contains fields with spin less or equal to one. However, the gravitational field in GR is described with a massless field of spin two (the hypothetical graviton) and all the interactions of such particles are produced by irrelevant operators \cite{Weinberg1}. As a consequence, it is possible to have a consistent description at low energies of the three forces of the SM using only operators that are renormalizable in the classical sense, while the current description of gravity is known to fail at an energy scale equal to the Planck mass (\(\sim 10^{19}\,{\rm GeV} \)). 

In this section the \textit{top-down} approach will be presented. The UV-complete model is taken to be known and one integrates out the heavy degrees of freedom producing a low-energy effective Lagrangian.\footnote{It is important to remark that EFTs are even more general, in the sense that the QFT that one finds at low energies might not be generated by another QFT. For instance, QFTs like supergravity are known to be the low energy limits of string theories, which are not QFTs in the usual sense.} Consider a theory containing a set of light fields \(\phi_i\) and a set of heavy fields \(\Phi_i\). The partition function is given by

\begin{equation}
    \label{Partition_Function_EFT}
    Z_{\rm UV}[J_{\phi},J_{\Phi}]=\int\mathcal{D}\phi_i\mathcal{D}\Phi_i\exp\left\{i\int{\rm d}^4x\left[\mathcal{L}_{\rm UV}[\phi_i,\Phi_i]+J^i_{\phi}\phi_i+J^i_{\Phi}\Phi_i\right]\right\}.
\end{equation}

\noindent The main goal in this approach is to construct a partition function \(Z_{\rm EFT}[J_{\phi}]\) that only contains the light fields \(\phi_i\) but reproduces the physical amplitudes of the UV theory that only include them as external lines. This condition is imposed by taking

\begin{equation}
    \label{Partition_Functions_Equivalence}
    Z_{\rm EFT}[J_{\phi}]=Z_{\rm UV}[J_{\phi},0].
\end{equation}

\noindent From Eq.~\eqref{Partition_Functions_Equivalence} an effective Lagrangian \(\mathcal{L}_{\rm EFT}[\phi_i]\) will be obtained. Therefore, it is necessary to perform the functional integration of the heavy fields in Eq.~\eqref{Partition_Function_EFT}. Although this approach has been used to do the \textit{matching} between both theories at one loop \cite{Henning:2014wua,delAguila:2016zcb,Henning:2016lyp,Fuentes-Martin:2016uol,Zhang:2016pja} and recently at two loops \cite{Fuentes-Martin:2023ljp,Fuentes-Martin:2024agf}, only the tree-level results, where the method is trivial, will be presented here.

The tree-level amplitudes correspond to the case where the quantum fields are treated as classical objects that satisfy the equations of motion. This means that the functional integration is simply done by substituting the heavy fields \(\Phi_i\) by the values that are obtained from the conditions \(\delta\mathcal{L}_{\rm UV}/\delta\Phi_i=0\). As an example, consider the following Lagrangian of a theory with a light and a heavy scalars \(h\) and \(H\), respectively:

\begin{equation}
    \label{Lagrangian_Toy_Model}
    \mathcal{L}=\frac{1}{2}\partial_{\mu}h\partial^{\mu}h-\frac{1}{2}m^2h^2+\frac{1}{2}\partial_{\mu}H\partial^{\mu}H-\frac{1}{2}M^2H^2-\frac{\lambda}{4!}h^4-\frac{c}{2}h^2H.
\end{equation}

\noindent Assuming that \(M\gg m\), the effective Lagrangian for energies \(E\ll M\) is obtained by substituting in Eq.~\eqref{Lagrangian_Toy_Model} the solution of:

\begin{equation}
    \label{EOM_Toy_Model}
    \frac{\delta\mathcal{L}}{\delta\Phi}=\left(\square+M^2\right)H-\frac{c}{2}h^2=0.
\end{equation}

\noindent From Eq.~\eqref{EOM_Toy_Model} one obtains \(H=\frac{c}{2}\left(\square+M^2\right)^{-1}h^2\), but this leads to a non-local Lagrangian. For this reason, considering that the term \(\square h\) is of the order of the square of the energy \(E^2\) of a certain process and that \(E^2/M^2\ll1\) in the low-energy limit, one expands the inverse of the propagator up to a certain power of the inverse of \(M^2\): \(H=\frac{c}{2M^2}(1-\frac{\square}{M^2}+\mathcal{O}(E^4/M^4))h^2\). This leads to the effective Lagrangian:

\begin{equation}
    \label{Effective_Lagrangian_Toy_Model}
    \mathcal{L}_{\rm Eff}=\frac{1}{2}\partial_{\mu}h\partial^{\mu}h-\frac{1}{2}m^2h^2-\left(\frac{\lambda}{4!}+\frac{3c^2}{8M^2}\right)h^4+\frac{3c^2}{8M^4}h^2\square h^2+\mathcal{O}(M^{-6}).
\end{equation}

\noindent The effects of integrating out the heavy field at tree level are corrections to the \(h^4\) interaction and the introduction of a tower of operators with \(d>4\). 

It is clear that, at tree level, the contributions of the heavy fields are suppressed by powers of \(E/M\) at low energies. This result is not obvious if one includes the loop corrections, because they involve integrals over momenta that also explore high energies. However, there is a famous theorem proved by Appelquist and Carazzone \cite{Appelquist:1974tg} called the \textit{decoupling theorem} which states that \textit{the effects of the heavy fields are either absorbed by redefinitions of the parameters of the Lagrangian in the renormalization procedure or are suppressed at low energies by powers of \(E/M\)}. There are important exceptions to this theorem, for instance in theories with a spontaneously broken gauge symmetry where the masses are generated through interaction terms \cite{Pich:1998xt}. One famous case has to do with the mass of the Higgs boson, where the quantum corrections make it proportional to the heavy scales of the theory. This is known as the \textit{hierarchy problem}.

Even in the cases where the decoupling theorem holds, it might not be manifest depending on the regularization procedure that one uses. For example, in the cutoff scheme it is natural to integrate momenta up to the scale of new physics \(\Lambda_{\rm NP}\), but then the suppression by inverse powers of it in the couplings of the non-renormalizable operators will be affected. For this reason, the use of dimensional regularization is preferred.  Nonetheless, even in those cases using the MS or \(\overline{\rm MS}\) schemes one might find apparent inconsistencies in the running of the \(\beta\) functions. For example, it is well known that the one-loop \(\beta\) function for the strong coupling of QCD in this scheme is \(\beta_{g_S}=-g_S^3(11/4-n_f/6)/4\pi^2\), so it depends on the number of flavors of quarks \(n_f\). This means that, if one integrates out the top quark (or any other heavy flavor compared to the energy scale of interest while staying in the perturbative regime of QCD), and calculates the \(\beta\) function in that case, it might seem that one obtains the wrong running, because the number of flavors would be lower than six. These heavy fermions need to be integrated out at low energies, since they produce large logarithms (\(\ln(\mu/m_t)\) with \(\mu\ll m_t\)) that break the perturbative expansion. The coupling constants in the original and in the effective theory are different, but the physical results are the same once the appropriate matching is performed.

Now, the procedure to make the matching at loop level will be presented. First of all, one calculates a certain amplitude in the UV-complete model and in the effective theory. For the latter case it is necessary to include in the effective Lagrangian all the allowed non-redundant (not equivalent by field redefinitions or the use of equations of motion) operators compatible with the symmetries of the theory that might contribute to that amplitude at a certain order in the mass expansion. Since the physical amplitudes should be the same, equating them one obtains a relation between the parameters of both Lagrangians, which now depend on the energy scale \(\mu\) since quantum corrections are considered.  Due to the fact that the light particles are the same in both cases their \(\ln(\mu/m)\) contributions are canceled when equating the amplitudes. One has to do the matching between the parameters at a certain energy scale, so it is convenient to use \(\mu=M\) in order to cancel the \(\ln(\mu/M)\) that produce large logarithms that might break the perturbative expansion at low energies. This is known as \textit{improved perturbation theory} and it is practical for calculations at low energies even when the full theory is known. Finally, one calculates the running of the parameters in the effective model from \(\mu=M\) to the desired low-energy scale.

\section{Standard Model Effective Field Theory} \label{sec:SMEFT}

In this section, the \textit{bottom-up} approach of the EFTs will be presented considering a particular example of great importance, the SMEFT. In the previous section it was assumed that the UV-complete model was known, but this may not always be the case. For instance, nowadays it is not known what is the final theory that describes nature. However, using EFTs one can still obtain information about the new-physics effects. The idea is that, no matter what  the interactions at high energies are or what the nature of the heavy particles is, evidence shows that, at low energies, the particle content is the one of the SM\footnote{Apart from the possible existence of light particles that are difficult to detect like right-handed neutrinos.} and the interactions must respect the \({\rm SU}(3)_{\rm QCD}\otimes{\rm SU}(2)_{\rm L}\otimes{\rm U}(1)_{\rm Y}\) symmetry. For this reason, the EFT generated by any possible UV-complete model at the scale of the SM should be described by a Lagrangian of the form:

\begin{equation}
    \label{SMEFT_Lagrangian}
    \mathcal{L}_{\rm SMEFT}=\mathcal{L}_{\rm SM}+\frac{1}{\Lambda_{\rm NP}}\sum_ic_i^{(5)}\mathcal{O}_i^{(d=5)}+\frac{1}{\Lambda^2_{\rm NP}}\sum_ic_i^{(6)}\mathcal{O}_i^{(d=6)}+\cdots,
\end{equation}

\noindent where the \(\mathcal{O}_i^{(d=d_0)}\) represent all the possible operators of a certain mass scale \(d_0\) which contain the particles of the SM and respect its associated gauge symmetry, and \(c_i^{(d_0)}\) are the corresponding couplings often called \textit{Wilson coefficients}.

As a particular example of the construction of one of these operators, it is not difficult to find \(\mathcal{O}_i^{(d=5)}\). The only objects that are available are fermions, scalars, field-strength tensors and their covariant derivatives. In the case without fermions, in order to respect \({\rm SU}(2)_{\rm L}\) invariance, one needs to have an even number of scalar doublets, but then it is necessary to add a structure with one, three or five Lorentz indices and there is no possible contraction of them that respects Lorentz invariance. Since the mass-scale of the fermions is \(3/2\), these operators of dimension five can only contain up to two of them and the case with one fermion is forbidden by Lorentz invariance. In order to respect the \({\rm SU}(3)_{\rm QCD}\) symmetry, the only possibilities are operators with a pair of a quark and an antiquark or operators with leptons. Starting with the quarks, if they have the same chirality it is necessary to add an odd number of \(\gamma^{\mu}\) matrices. If one uses three or more of them, it is necessary to make contractions with covariant derivatives or field-strength tensors, but that leads to operators with dimension six or more. Using only one \(\gamma^{\mu}\) matrix the contraction is done with a covariant derivative, but this leads to an operator of dimension four and the only possibility is to add a scalar doublet, which will break gauge invariance. As a consequence, the pairs of quarks and antiquarks need to have opposite chiralities. In order to get an operator of dimension five that respects Lorentz invariance in that case, one needs to add two covariant derivatives and one scalar doublet or three scalar doublets. Both cases break gauge invariance, so it is not possible to construct such operators with quark-antiquark pairs and this reasoning also holds for lepton-antilepton pairs. The only remaining possibility is to construct an operator with two leptons or two antileptons. There is only one kind of such operators, the so-called \textit{Weinberg operators} \cite{Weinberg:1979sa}: 

\begin{equation}
    \label{Weinberg Operators}
    \mathcal{L}_{\rm SMEFT}^{(d=5)}=\frac{1}{\Lambda_{\rm NP}}\sum_{ij}c^{(5)}_{ij}(\bar{\ell}^{ic}_L\Phi^{c\ast})(\Phi^{c\dagger}\ell_{L}^j)+{\rm h.c.}
\end{equation}

It is easy to see that these operators violate both lepton number and \(B-L\) by two units. This is a general result of the EFTs, the accidental symmetries that one finds for a renormalizable Lagrangian are broken when considering all the possible operators. This is one of the reasons why it is nowadays believed that only gauge symmetries are the true internal symmetries of nature. When the VEV of the doublet is taken, this term produces \textit{Majorana masses} for the left-handed neutrinos

\begin{equation}
    \label{Majorana_Masses_Weinberg}
    \mathcal{L}_{\rm Majorana}=\frac{1}{2}\frac{v^2}{\Lambda_{\rm NP}}\sum_{ij}c^{(5)}_{ij}\bar{\nu}^{ic}_L\nu^j_L+{\rm h.c.}
\end{equation}

\noindent The masses are proportional to \(v^2/\Lambda_{\rm NP}\), so this could in principle explain why their measured values are so small. However, in order to reproduce the current observations (assuming couplings of order one), the scale of new physics should be of the order of \(\Lambda_{\rm NP}\sim 10^{15}\,{\rm GeV}\) (something that might be in contradiction with the measured value of the Higgs boson mass). This is the reason why it is believed that the SMEFT indeed contains different high-energy scales. Some of the sectors of the new physics might preserve symmetries like \(B\), \(L\) and \(B-L\), so the bounds on the effects of the associated operators are not so strong. On the other hand, all the operators of odd dimension violate \(B-L\) by an odd number of units, while some operators of even dimension (as the dimension six \(B\)-violating ones in Tab.~\ref{tab:Four_Fermion_Warsaw}) violate either \(B\) or \(L\) but preserve \(B-L\) \cite{Kobach:2016ami,Helset:2019eyc}, so the associated Wilson coefficients are extremely constrained.

The first complete basis of independent dimension six operators (Tabs.~\ref{tab:non_Four_Fermion_Warsaw} and~\ref{tab:Four_Fermion_Warsaw}) was presented in \cite{Grzadkowski:2010es} and it is called the \textit{Warsaw basis}. It contains the 2499 operators that are needed to obtain the \textit{on-shell} matrix elements, which means that they are not related by the equations of motion. Furthermore, they use \textit{Fierz identities} to reduce the number of independent operators. These identities only hold in a four-dimensional spacetime, which means that one needs to be careful when doing loop calculations in dimensional regularization. In that case, it is usually necessary to take into account the presence of evanescent operators, which are just the difference between two operators that are related by a strictly four-dimensional expression such as a Fierz identity and, consequently, vanish in the limit \(D\to4\). The bases of operators of dimension seven \cite{Lehman:2014jma,Liao:2016hru}, eight \cite{Li:2020gnx,Lehman:2015coa,Murphy:2020rsh}, nine \cite{Li:2020xlh,Liao:2020jmn} and up to twelve \cite{Harlander:2023psl} are also known. Furthermore, Green's bases (the ones needed for off-shell calculations) have also been obtained for operators of dimension six \cite{Harlander:2023psl}, seven \cite{Zhang:2023kvw} and eight \cite{Ren:2022tvi}. When adding extra degrees of freedom on top of the SM (such as light RH neutrinos) the basis of effective operators has to be extended. A similar construction holds in presence of additional symmetries, as shown in Sec.~\ref{sec:Effective_LR} when discussing left-right models.

\begin{table}[t] 
\centering
\renewcommand{\arraystretch}{1.5}
{
\small
\begin{tabular}{||c|c||c|c||c|c||} 
\hline \hline
\multicolumn{2}{||c||}{$X^3$} & 
\multicolumn{2}{|c||}{$\Phi^6$~ and~ $\Phi^4 D^2$} &
\multicolumn{2}{|c||}{$\psi^2\Phi^3$}\\
\hline
$Q_G$                & $f^{ABC} G_\mu^{A\nu} G_\nu^{B\rho} G_\rho^{C\mu} $ &  
$Q_\Phi$       & $(\Phi^\dag\Phi)^3$ &
$Q_{e\Phi}$           & $(\Phi^\dag \Phi)(\bar \ell_p e_r \Phi)$\\
$Q_{\widetilde G}$          & $f^{ABC} \widetilde G_\mu^{A\nu} G_\nu^{B\rho} G_\rho^{C\mu} $ &   
$Q_{\Phi\Box}$ & $(\Phi^\dag \Phi)\raisebox{-.5mm}{$\Box$}(\Phi^\dag \Phi)$ &
$Q_{u\Phi}$           & $(\Phi^\dag \Phi)(\bar q_p u_r \widetilde{\Phi})$\\
$Q_W$                & $\varepsilon^{IJK} W_\mu^{I\nu} W_\nu^{J\rho} W_\rho^{K\mu}$ &    
$Q_{\Phi D}$   & $\left(\Phi^\dag D^\mu\Phi\right)^\star \left(\Phi^\dag D_\mu\Phi\right)$ &
$Q_{d\Phi}$           & $(\Phi^\dag \Phi)(\bar q_p d_r \Phi)$\\
$Q_{\widetilde W}$          & $\varepsilon^{IJK} \widetilde W_\mu^{I\nu} W_\nu^{J\rho} W_\rho^{K\mu}$ &&&&\\    
\hline \hline
\multicolumn{2}{||c||}{$X^2\Phi^2$} &
\multicolumn{2}{|c||}{$\psi^2 X\Phi$} &
\multicolumn{2}{|c||}{$\psi^2\Phi^2 D$}\\ 
\hline
$Q_{\Phi G}$     & $\Phi^\dag \Phi\, G^A_{\mu\nu} G^{A\mu\nu}$ & 
$Q_{eW}$               & $(\bar \ell_p \sigma^{\mu\nu} e_r) \tau^I \Phi W_{\mu\nu}^I$ &
$Q_{\Phi \ell}^{(1)}$      & $(\Phi^\dag i\,\raisebox{2mm}{\boldmath ${}^\leftrightarrow$}\hspace{-4mm} D_\mu\,\Phi)(\bar \ell_p \gamma^\mu \ell_r)$\\
$Q_{\Phi\widetilde G}$         & $\Phi^\dag \Phi\, \widetilde G^A_{\mu\nu} G^{A\mu\nu}$ &  
$Q_{eB}$        & $(\bar \ell_p \sigma^{\mu\nu} e_r) \Phi B_{\mu\nu}$ &
$Q_{\Phi \ell}^{(3)}$      & $(\Phi^\dag i\,\raisebox{2mm}{\boldmath ${}^\leftrightarrow$}\hspace{-4mm} D_\mu^{\,I}\,\Phi)(\bar \ell_p \tau^I \gamma^\mu \ell_r)$\\
$Q_{\Phi W}$     & $\Phi^\dag \Phi\, W^I_{\mu\nu} W^{I\mu\nu}$ & 
$Q_{uG}$        & $(\bar q_p \sigma^{\mu\nu} T^A u_r) \widetilde{\Phi}\, G_{\mu\nu}^A$ &
$Q_{\Phi e}$            & $(\Phi^\dag i\,\raisebox{2mm}{\boldmath ${}^\leftrightarrow$}\hspace{-4mm} D_\mu\,\Phi)(\bar e_p \gamma^\mu e_r)$\\
$Q_{\Phi\widetilde W}$         & $\Phi^\dag \Phi\, \widetilde W^I_{\mu\nu} W^{I\mu\nu}$ &
$Q_{uW}$               & $(\bar q_p \sigma^{\mu\nu} u_r) \tau^I \widetilde{\Phi}\, W_{\mu\nu}^I$ &
$Q_{\Phi q}^{(1)}$      & $(\Phi^\dag i\,\raisebox{2mm}{\boldmath ${}^\leftrightarrow$}\hspace{-4mm} D_\mu\,\Phi)(\bar q_p \gamma^\mu q_r)$\\
$Q_{\Phi B}$     & $ \Phi^\dag \Phi\, B_{\mu\nu} B^{\mu\nu}$ &
$Q_{uB}$        & $(\bar q_p \sigma^{\mu\nu} u_r) \widetilde{\Phi}\, B_{\mu\nu}$&
$Q_{\Phi q}^{(3)}$      & $(\Phi^\dag i\,\raisebox{2mm}{\boldmath ${}^\leftrightarrow$}\hspace{-4mm} D_\mu^{\,I}\,\Phi)(\bar q_p \tau^I \gamma^\mu q_r)$\\
$Q_{\Phi\widetilde B}$         & $\Phi^\dag \Phi\, \widetilde B_{\mu\nu} B^{\mu\nu}$ &
$Q_{dG}$        & $(\bar q_p \sigma^{\mu\nu} T^A d_r) \Phi\, G_{\mu\nu}^A$ & 
$Q_{\Phi u}$            & $(\Phi^\dag i\,\raisebox{2mm}{\boldmath ${}^\leftrightarrow$}\hspace{-4mm} D_\mu\,\Phi)(\bar u_p \gamma^\mu u_r)$\\
$Q_{\Phi WB}$     & $ \Phi^\dag \tau^I \Phi\, W^I_{\mu\nu} B^{\mu\nu}$ &
$Q_{dW}$               & $(\bar q_p \sigma^{\mu\nu} d_r) \tau^I \Phi\, W_{\mu\nu}^I$ &
$Q_{\Phi d}$            & $(\Phi^\dag i\,\raisebox{2mm}{\boldmath ${}^\leftrightarrow$}\hspace{-4mm} D_\mu\,\Phi)(\bar d_p \gamma^\mu d_r)$\\
$Q_{\Phi\widetilde WB}$ & $\Phi^\dag \tau^I \Phi\, \widetilde W^I_{\mu\nu} B^{\mu\nu}$ &
$Q_{dB}$        & $(\bar q_p \sigma^{\mu\nu} d_r) \Phi\, B_{\mu\nu}$ &
$Q_{\Phi u d}$   & $i(\widetilde{\Phi}^\dag D_\mu \Phi)(\bar u_p \gamma^\mu d_r)$\\
\hline \hline
\end{tabular}
}
\caption[Warsaw basis for the complete set of non-redundant dimension six operators of SMEFT other than the four-fermion ones.]{Warsaw basis for the complete set of non-redundant dimension six operators of SMEFT other than the four-fermion ones. The dual field-strength tensors are defined as \(\widetilde{F}_{\mu\nu}\coloneqq\frac{1}{2}\varepsilon_{\mu\nu\rho\sigma}F^{\rho\sigma}\). Table obtained from \cite{Grzadkowski:2010es}.}
\label{tab:non_Four_Fermion_Warsaw}
\end{table}

\begin{sidewaystable}
\centering
\renewcommand{\arraystretch}{1.5}
\begin{tabular}{||c|c||c|c||c|c||}
\hline\hline
\multicolumn{2}{||c||}{$(\bar LL)(\bar LL)$} & 
\multicolumn{2}{|c||}{$(\bar RR)(\bar RR)$} &
\multicolumn{2}{|c||}{$(\bar LL)(\bar RR)$}\\
\hline
$Q_{\ell\ell}$        & $(\bar \ell_p \gamma_\mu \ell_r)(\bar \ell_s \gamma^\mu \ell_t)$ &
$Q_{ee}$               & $(\bar e_p \gamma_\mu e_r)(\bar e_s \gamma^\mu e_t)$ &
$Q_{\ell e}$               & $(\bar \ell_p \gamma_\mu \ell_r)(\bar e_s \gamma^\mu e_t)$ \\
$Q_{qq}^{(1)}$  & $(\bar q_p \gamma_\mu q_r)(\bar q_s \gamma^\mu q_t)$ &
$Q_{uu}$        & $(\bar u_p \gamma_\mu u_r)(\bar u_s \gamma^\mu u_t)$ &
$Q_{\ell u}$               & $(\bar \ell_p \gamma_\mu \ell_r)(\bar u_s \gamma^\mu u_t)$ \\
$Q_{qq}^{(3)}$  & $(\bar q_p \gamma_\mu \tau^I q_r)(\bar q_s \gamma^\mu \tau^I q_t)$ &
$Q_{dd}$        & $(\bar d_p \gamma_\mu d_r)(\bar d_s \gamma^\mu d_t)$ &
$Q_{\ell d}$               & $(\bar \ell_p \gamma_\mu \ell_r)(\bar d_s \gamma^\mu d_t)$ \\
$Q_{\ell q}^{(1)}$                & $(\bar \ell_p \gamma_\mu \ell_r)(\bar q_s \gamma^\mu q_t)$ &
$Q_{eu}$                      & $(\bar e_p \gamma_\mu e_r)(\bar u_s \gamma^\mu u_t)$ &
$Q_{qe}$               & $(\bar q_p \gamma_\mu q_r)(\bar e_s \gamma^\mu e_t)$ \\
$Q_{\ell q}^{(3)}$                & $(\bar \ell_p \gamma_\mu \tau^I \ell_r)(\bar q_s \gamma^\mu \tau^I q_t)$ &
$Q_{ed}$                      & $(\bar e_p \gamma_\mu e_r)(\bar d_s\gamma^\mu d_t)$ &
$Q_{qu}^{(1)}$         & $(\bar q_p \gamma_\mu q_r)(\bar u_s \gamma^\mu u_t)$ \\ 
&& 
$Q_{ud}^{(1)}$                & $(\bar u_p \gamma_\mu u_r)(\bar d_s \gamma^\mu d_t)$ &
$Q_{qu}^{(8)}$         & $(\bar q_p \gamma_\mu T^A q_r)(\bar u_s \gamma^\mu T^A u_t)$ \\ 
&& 
$Q_{ud}^{(8)}$                & $(\bar u_p \gamma_\mu T^A u_r)(\bar d_s \gamma^\mu T^A d_t)$ &
$Q_{qd}^{(1)}$ & $(\bar q_p \gamma_\mu q_r)(\bar d_s \gamma^\mu d_t)$ \\
&&&&
$Q_{qd}^{(8)}$ & $(\bar q_p \gamma_\mu T^A q_r)(\bar d_s \gamma^\mu T^A d_t)$\\
\hline\hline
\multicolumn{2}{||c||}{$(\bar LR)(\bar RL)$ and $(\bar LR)(\bar LR)$} &
\multicolumn{4}{|c||}{$B$-violating}\\\hline
$Q_{\ell edq}$ & $(\bar \ell_p^j e_r)(\bar d_s q_t^j)$ &
$Q_{duq}$ & \multicolumn{3}{|c||}{$\epsilon^{\alpha\beta\gamma} \epsilon_{jk} 
 \left[ (d^\alpha_p)^T C u^\beta_r \right]\left[(q^{\gamma j}_s)^T C \ell^k_t\right]$}\\
$Q_{quqd}^{(1)}$ & $(\bar q_p^j u_r) \epsilon_{jk} (\bar q_s^k d_t)$ &
$Q_{qqu}$ & \multicolumn{3}{|c||}{$\epsilon^{\alpha\beta\gamma} \epsilon_{jk} 
  \left[ (q^{\alpha j}_p)^T C q^{\beta k}_r \right]\left[(u^\gamma_s)^T C e_t\right]$}\\
$Q_{quqd}^{(8)}$ & $(\bar q_p^j T^A u_r) \epsilon_{jk} (\bar q_s^k T^A d_t)$ &
$Q_{qqq}$ & \multicolumn{3}{|c||}{$\epsilon^{\alpha\beta\gamma} \epsilon_{jn} \epsilon_{km} 
  \left[ (q^{\alpha j}_p)^T C q^{\beta k}_r \right]\left[(q^{\gamma m}_s)^T C \ell^n_t\right]$}\\
$Q_{\ell equ}^{(1)}$ & $(\bar \ell_p^j e_r) \epsilon_{jk} (\bar q_s^k u_t)$ &
$Q_{duu}$ & \multicolumn{3}{|c||}{$\epsilon^{\alpha\beta\gamma} 
  \left[ (d^\alpha_p)^T C u^\beta_r \right]\left[(u^\gamma_s)^T C e_t\right]$}\\
$Q_{\ell equ}^{(3)}$ & $(\bar \ell_p^j \sigma_{\mu\nu} e_r) \epsilon_{jk} (\bar q_s^k \sigma^{\mu\nu} u_t)$ &
& \multicolumn{3}{|c||}{}\\
\hline\hline
\end{tabular}
\caption[Warsaw basis for the complete set of non-redundant dimension six four-fermion operators of SMEFT.]{Warsaw basis for the complete set of non-redundant dimension six four-fermion operators of SMEFT. Table obtained from \cite{Grzadkowski:2010es}.}
\label{tab:Four_Fermion_Warsaw}
\end{sidewaystable}

\thispagestyle{empty}

\chapter{Symmetry-restoring finite counterterms of SMEFT four-fermion operator insertions at one loop}
\markboth{4. FINITE RENORMALIZATION OF SMEFT FOUR-FERMION OPERATORS}{}

\section*{Introduction}

This chapter is devoted to the one-loop calculation of the symmetry-restoring finite counterterms associated with the SMEFT\(+\nu\) four-fermion operators. This is based on the work done by the author of this text in \cite{Solera:2026xnz}. The calculation was performed using the traditional techniques of algebraic renormalization and the one-loop Bonneau identity, presented in Sec.~\ref{sec:Bonneau}. The scheme used is dimensional regularization with the BMHV prescription for \(\gamma_5\) plus minimal subtraction adding a set of fictitious fields. The results will be compared with those obtained in \cite{Fuentes-Martin:2025meq}, where they used a novel method which will be referred here as the \textit{spurion method} \cite{OlgosoRuiz:2024dzq}. In this approach, a new spurion field \(\Omega\) is added in such a way that the fermion propagator becomes gauge invariant in \(D\) dimensions. Then, the one-loop QEA is computed and the limit \(\Omega\to\mathbb{I}\) is taken to obtain the original theory with the required symmetry-restoring counterterms. The main advantage of this method is that there is \textit{no need to calculate the preimage of the BRST transformation}, which, as it will be shown, is one of the most demanding parts. However, up to this date \textit{the agreement between the two results is not complete}.

The sections of this chapter are organized as follows: in the first one, the scheme and the techniques used for the calculation are presented; in the next part, the computation of the amplitudes associated with the finite contributions to the obstruction of the Slavnov-Taylor identities are shown, following the Bonneau method; finally, in the third section, the preimage of the BRST transformation is obtained, leading to the required finite counterterms, which will be compared with the results using the spurion approach. To simplify the reading, the different parts of the calculation are separated according to the number of Higgs fields involved. 

\section{\boldmath Extension of SMEFT\(+\nu\) to \(D\) Dimensions} \label{D_SMEFT}

The theory considered is the SM $ \mathcal{L}_{\rm SM} $, with right-handed neutrinos to enlarge the scope of the analysis, and dimension-6 operators: $ \mathcal{L} = \mathcal{L}_{{\rm SM} + \nu} + \sum_i C_i Q_i $, where

\begin{equation}
\label{eq:SM_nuR_Lagrangian}
\begin{split}
	\mathcal{L}_{{\rm SM} + \nu} &= - \frac{1}{4} G^{A}_{\mu \nu} G^{A \mu \nu}_{} - \frac{1}{4} W^{I}_{\mu \nu} W^{I \mu \nu}_{} - \frac{1}{4} B^{}_{\mu \nu} B^{\mu \nu}_{} \\
    &+(D_\mu \Phi^\dagger) \, (D^\mu \Phi) - \lambda \left( \Phi^\dagger \Phi - \frac{1}{2} v^2 \right)^2 + \sum_{\psi = q_L, u_R, d_R, \ell_L, e_R, \nu_R} i\bar{\psi} \slashed{D} \psi \\
	&- \left[ \frac{1}{2} (\bar{\nu}^{C}_R M_\nu \nu_R) + \Phi^{\dagger } \bar{d}_R Y_d q_{L} + \Phi^{c\dagger } \bar{u}_R Y_u q_{L} + \Phi^{\dagger } \bar{e}_R Y_e \ell_{L} + \Phi^{c\dagger } \bar{\nu}_R Y_\nu \ell_{L} + {\rm h.c.} \right].
\end{split}
\end{equation}

\noindent The object \(\bar{\nu}^{C}_R M_\nu \nu_R\) is a Majorana mass term for right-handed neutrinos, it will be studied in more detail in the next chapter. From now on, in this part of the text the chiralities \(L\) and \(R\) of the fermions will not be indicated to alleviate the notation. The gauge fixing functional used (not explicitly added in the previous expression) corresponds to the one of a general \(R_{\xi}\) gauge, as pointed in the calculation of the ``wine-glass'' diagram in the following section. There is no particular advantage in using the background field gauge as done in \cite{Fuentes-Martin:2025meq} if only four-fermion operators are considered.

\begin{table}[t!]
    \centering
    \renewcommand{\arraystretch}{1.2}
    \begin{tabular}{l}
        $ ( \bar{L} L ) ( \bar{L} L ) $ \\
        \hline
        $ Q^{}_{\ell \ell} = (\bar{\ell} \gamma_\mu \ell) (\bar{\ell} \gamma^\mu \ell) $ \\
        $ Q^{(1)}_{q q} = (\bar{q} \gamma_\mu q) (\bar{q} \gamma^\mu q) $ \\
        $ Q^{(3)}_{q q} = (\bar{q} \gamma_\mu \tau^I q) (\bar{q} \gamma^\mu \tau^I q) $ \\
        $ Q^{(1)}_{\ell q} = (\bar{\ell} \gamma_\mu \ell) (\bar{q} \gamma^\mu q) $ \\
        $ Q^{(3)}_{\ell q} = (\bar{\ell} \gamma_\mu \tau^I \ell) (\bar{q} \gamma^\mu \tau^I q) $ \\
    %
        \phantom{} \\
        $ ( \bar{R} R ) ( \bar{R} R ) $ \\
        \hline
        $ Q^{}_{y y} = (\bar{y} \gamma_\mu y) (\bar{y} \gamma^\mu y) $ \\
        $ Q^{}_{x x} = (\bar{x} \gamma_\mu x) (\bar{x} \gamma^\mu x) $ \\
        $ Q^{}_{y x} = (\bar{y} \gamma_\mu y) (\bar{x} \gamma^\mu x) $ \\
        $ Q_{\nu e} = (\bar{\nu} \gamma_\mu \nu) (\bar{e} \gamma^\mu e) $ \\
        $ Q^{(1)}_{u d} = (\bar{u} \gamma_\mu u) (\bar{d} \gamma^\mu d) $ \\
        $ Q^{(8)}_{u d} = (\bar{u} \gamma_\mu T^A u) (\bar{d} \gamma^\mu T^A d) $ \\
        $ Q_{d u \nu e} = (\bar{d} \gamma_\mu u) (\bar{\nu} \gamma^\mu e) $ \\
        $[ Q_{d u \nu e}^\dagger = (\bar{u} \gamma_\mu d) (\bar{e} \gamma^\mu \nu) ]$ \\
    \end{tabular}
    \hspace{10mm}
    \begin{tabular}{l}
        \phantom{} \\
        $ ( \bar{L} L ) ( \bar{R} R ) $ \\
        \hline
        $ Q^{}_{\ell y} = (\bar{\ell} \gamma_\mu \ell) (\bar{y} \gamma^\mu y) $ \\
        $ Q^{}_{\ell x} = (\bar{\ell} \gamma_\mu \ell) (\bar{x} \gamma^\mu x) $ \\
        $ Q^{}_{q y} = (\bar{q} \gamma_\mu q) (\bar{y} \gamma^\mu y) $ \\
        $ Q^{(1)}_{q x} = (\bar{q} \gamma_\mu q) (\bar{x} \gamma^\mu x) $ \\
        $ Q^{(8)}_{q x} = (\bar{q} \gamma_\mu T^A q) (\bar{x} \gamma^\mu T^A x) $ \\
    %
        \phantom{} \\
        $ ( \bar{L} R ) ( \bar{R} L ) $ + h.c. \\
        \hline
        $ Q^{}_{\ell e d q} = (\bar{\ell}^m e) (\bar{d} q_m) $ \\
        $ Q^{}_{\ell \nu u q} = (\bar{\ell}^m \nu) (\bar{u} q_m) $ \\
    %
        \phantom{} \\
        $ ( \bar{L} R ) ( \bar{L} R ) $ + h.c. \\
        \hline
        $ Q_{\ell \nu \ell e} = (\bar{\ell}^m \nu) \epsilon_{mn} (\bar{\ell}^n e) $ \\
        $ Q^{(1)}_{q u q d} = (\bar{q}^m u) \epsilon_{mn} (\bar{q}^n d) $ \\
        $ Q^{(8)}_{q u q d} = (\bar{q}^m T^A u) \epsilon_{mn} (\bar{q}^n T^A d) $ \\
        $ Q^{(1)}_{\ell e q u} = (\bar{\ell}^m e) \epsilon_{mn} (\bar{q}^n u) $ \\
        $ Q^{(3)}_{\ell e q u} = (\bar{\ell}^m \sigma_{\mu \nu} e) \epsilon_{mn} (\bar{q}^n \sigma^{\mu \nu} u) $ \\
        $ Q^{(1)}_{\ell \nu q d} = (\bar{\ell}^m \nu) \epsilon_{mn} (\bar{q}^n d) $ \\
        $ Q^{(3)}_{\ell \nu q d} = (\bar{\ell}^m \sigma_{\mu \nu} \nu) \epsilon_{mn} (\bar{q}^n \sigma^{\mu \nu} d) $ \\
    \end{tabular}
    \caption[Four-fermion baryon- and lepton-number-conserving operators of the so-called Warsaw basis extended to include right-handed neutrinos $\nu$.]{Four-fermion baryon- and lepton-number-conserving operators of the so-called Warsaw basis extended to include right-handed neutrinos $\nu$, where $ x = u, d $ and $ y = e, \nu $; $q$ ($\ell$) are weak-isospin doublet quarks (leptons), and $u, d$ ($e, \nu$) are weak-isospin singlet quarks (leptons). Flavor or generation indices are omitted; when indicated in the text (e.g., as in the Wilson coefficient $ C_{\ell e d q; f i j k} $) they correspond to the fields above in that same ordering (i.e., $ (\bar{\ell}^m_f e_i) (\bar{d}_j q_{m; k}) $).}
    \label{tab:dimension_6}
\end{table}

The set of non-renormalizable operators is introduced in Tab.~\ref{tab:dimension_6}. For completeness, and since the calculation is very similar, the basis of operators of Ref.~\cite{Grzadkowski:2010es} is extended to include right-handed neutrinos.\footnote{A complete off-shell basis for the \(\nu\)SMEFT and the matching to the \(\nu\)LEFT can be found in \cite{Chala:2020vqp}. Possible UV completions were studied in \cite{Beltran:2023ymm}, and the one-loop RGEs were derived in \cite{Ardu:2024tzb,Datta:2020ocb,Chala:2020pbn}.} It is important to remark that these right-handed neutrinos are not the fictitious fields presented in Sec.~\ref{sec:BMHV}, needed to have a well-regulated propagator for the left-handed counterparts. Operators that violate total baryon or lepton numbers are found in \cite{Alonso:2014zka,Liao:2016qyd}. Nonetheless,  with these operators it is not possible to contract together two fermions, and, as it will be shown in the following section, those contractions appear in the only possible non-vanishing topologies that contribute to the obstruction of the Slavnov-Taylor identities. Therefore, at one loop, this category of operators does not require the introduction of finite counterterms, see also Refs.~\cite{Naterop:2025lzc,Banik:2025wpi}.

 It is sometimes more convenient to consider operators in a different basis with respect to the Warsaw one \cite{Grzadkowski:2010es}, following the use of Fierz identities
\begin{eqnarray}\label{eq:Fierz_1}
    && (\gamma_\mu P_L)_{i j} (\gamma^\mu P_L)_{k l} = - (\gamma_\mu P_L)_{i l} (\gamma^\mu P_L)_{k j} \,, \\
    && (\gamma_\mu P_L)_{i j} (\gamma^\mu P_R)_{k l} = 2 (P_R)_{i l} (P_L)_{k j} \,, \\
    && (P_L)_{i j} (P_L)_{k l} = \frac{1}{2} (P_L)_{i l} (P_L)_{k j} + \frac{1}{8} (\sigma_{\mu \nu} P_L)_{i l} (\sigma^{\mu \nu} P_L)_{k j} \,, \\
    && (\sigma_{\mu \nu} P_L)_{i j} (\sigma^{\mu \nu} P_L)_{k l} = 6 (P_L)_{i l} (P_L)_{k j} - \frac{1}{2} (\sigma_{\mu \nu} P_L)_{i l} (\sigma^{\mu \nu} P_L)_{k j} \,, \label{eq:Fierz_4}
\end{eqnarray}

\noindent where $i, j, k, l$ are spinorial indices, and $ \sigma^{\mu \nu} = i [\gamma^\mu, \gamma^\nu] / 2 $; while other identities are found by interchanging the chiral projectors $ P_L \leftrightarrow P_R $. Of course, fermions anticommute, and this leads to an additional sign when presenting these identities along with fermionic operators. When employing these identities, it is useful to consider the following ones for the generators of $\rm{SU}(3)$
\begin{eqnarray}\label{eq:GellMann_identity_1}
    && \delta_{\alpha \tilde{\alpha}} \delta_{\beta \tilde{\beta}} = \frac{1}{N_c} \delta_{\alpha \tilde{\beta}} \delta_{\beta \tilde{\alpha}} + 2 T^A_{\alpha \tilde{\beta}} T^A_{\beta \tilde{\alpha}} \,, \\
    && T^A_{\alpha \tilde{\alpha}} T^A_{\beta \tilde{\beta}} = \frac{C_A}{N_c} \delta_{\alpha \tilde{\beta}} \delta_{\beta \tilde{\alpha}} - \frac{1}{N_c} T^A_{\alpha \tilde{\beta}} T^A_{\beta \tilde{\alpha}} \,, \label{eq:GellMann_identity_2}
\end{eqnarray}
where $ \alpha, \tilde{\alpha}, \beta, \tilde{\beta} $ are color indices, and $ C_A = \frac{1}{2} \left( N_c - \frac{1}{N_c} \right) $. Analogous expressions hold for the generators $T_L^I$ of ${\rm SU(2)}_{\rm L}$, with $N_c$ replaced by $2$. The two bases differ in general by evanescent operators in \(D\) dimensions. Since the symmetry-breaking operator is itself an evanescent operator, and since only the infinite parts of diagrams of ghost number one with single insertions of it are needed, the choice of the basis has no effect on the results that will be displayed (apart from overall constant factors resulting from the use of the Fierz identities). This has been explicitly checked doing the same calculation in two different bases related by a Fierz identity and obtaining the same result. Therefore, \textit{there is no contribution from evanescent operators to the finite symmetry restoring counterterms at one loop}.

The calculations in this text will be performed in the unbroken phase. Consequently, the most convenient scheme (as pointed in Sec.~\ref{sec:BMHV}) is the one where fictitious fields with opposite chiralities are added for each chiral fermion of the theory, regardless of whether their physical counterparts with opposite chiralities are already present in the Lagrangian.  Global symmetries, like the one associated with hypercharge conservation, will be preserved. The symmetry breaking operators \(\Delta\) coming from the BRST transformation of the propagators are the ones shown in Eq.~\eqref{Symmetry_Breaking_Delta}, where the fields can be \(\psi=q,u,d,\ell,e\) and \(\nu\); and the generators \(t_{L,R}^{\alpha}\) are the generators of \({\rm SU}(3)_{\rm QCD}\), \({\rm SU}(2)_{\rm L}\) and \({\rm U}(1)_{\rm Y}\). It is important to remark that the associated interaction vertices involve a ghost, a fictitious fermion and a physical one. Note that, as already mentioned in Sec.~\ref{sec:BMHV}, there are no other interaction vertices that involve the fictitious fields. Consequently, \textit{all the other interactions of the Lagrangian are kept as purely four-dimensional}.

\section{Calculation of the Obstruction to the Slavnov-Taylor Identities} \label{Calculation_Obstruction_ST}
\markboth{4. FINITE RENORMALIZATION OF SMEFT FOUR-FERMION OPERATORS}{4.2. OBSTRUCTION TO THE SLAVNOV-TAYLOR IDENTITIES}

As shown in Sec.~\ref{Sec:QAP}, from the methods of algebraic renormalization it is possible to write the obstructions to the Slavnov-Taylor identities as \(\left(\Gamma,\Gamma\right)=-2\Delta\cdot\Gamma\), where \(\Delta\cdot\Gamma\) represents the 1PI diagrams with single insertions of the symmetry breaking operator \(\Delta\). Using the Bonneau method at one loop, the finite contributions to \(\Delta\cdot\Gamma\) are obtained from the \textit{divergent part} of the amplitude \(I_{\varepsilon}^{G^{\check{\Delta}}}\) of 1PI diagrams \(G\) with single insertions of the symmetry breaking operator \(\check{\Delta}\coloneqq\check{\eta}^{\mu\nu}\Delta_{\mu\nu}\)  taking the limit

\begin{equation}
    \label{Calculation_Renormalized_Amplitudes_D=4}
    \left.\Delta\cdot\Gamma\right|_{1\,{\rm Loop}}^{\rm Finite}=\lim_{D\to4}\textrm{r.s.p.}\left(\left.I_{\varepsilon}^{G^{\check{\Delta}}}\right|_{\check{\eta}=0}\right).
\end{equation}

\noindent This is precisely what will be done in this section.

\textit{Only divergent Green's functions of ghost number one with single insertions of the symmetry-breaking Dvo\v{r}\'ak operator contribute}. They can be divided according to the matter fields involved: none, one, and two scalar fields, all having two external fermions. It will be shown that the Green's functions with Superficial Degree of Divergence zero (\({\rm SDD}=0\)) are convergent, the amplitudes with four fermion fields being a particular case. As it will be made explicit in the following, given the basis of operators already discussed, cases of interest involve closed (that require the computation of a trace of Dirac matrices) and/or open fermion lines,\footnote{In theories with renormalizable interactions as in the SM, the analogous case of having an open fermion line (with an internal vector or scalar being exchanged) and more than one external vector boson or scalar legs does not require any finite renormalization. This can be easily understood based on identifying the superficial degree of divergence, and using the same arguments as presented here for the properties of Green's functions. See also Ref.~\cite{ValeSilva:2025}.}
although this distinction is not important at this order due to the possible use of Fierz identities, as discussed in the previous section. The calculations are performed in the BMHV scheme using \texttt{FeynCalc} \cite{Mertig:1990an,Shtabovenko:2016sxi,Shtabovenko:2020gxv,Shtabovenko:2023idz}. When indicated, the momenta $ k_g, k_h, k_v $ are all incoming.

The Feynman rule associated with the symmetry breaking operator in Eq.~\eqref{Symmetry_Breaking_Delta} has several properties that simplify the calculation of the divergent part of the one-loop diagrams. For instance, consider that the ghost with incoming momentum \(k_g\) is attached to two internal fermion propagators: one of them with incoming momentum \(k\) and the other with outgoing momentum \(k+k_g\). The symmetry breaking vertex transforms physical fermions into fictitious ones and vice versa; but, given the chosen prescription for the \(D\)-dimensional Lagrangian, only physical particles can appear in any other interaction vertex of a Feynman diagram. Consequently, a chiral flip (i.e., a transformation between physical and fictitious fields) is needed in one and only one of the fermion propagators, see the last term in Eq.~\eqref{D_Dim_Propagator}. Therefore, the combination of the chiral flip with the interaction vertex will produce the Dirac structure 

\begin{equation}
    \label{Dirac_Structure_SB}
    (\bar{\slashed{k}}+\bar{\slashed{k}}_g)\left[\hat{\slashed{k}}P_{R,L}-(\hat{\slashed{k}}+\hat{\slashed{k}}_g)P_{L,R}\right]\hat{\slashed{k}}+(\hat{\slashed{k}}+\hat{\slashed{k}}_g)\left[\hat{\slashed{k}}P_{R,L}-(\hat{\slashed{k}}+\hat{\slashed{k}}_g)P_{L,R}\right]\bar{\slashed{k}}
\end{equation}

\noindent in the numerator of the integrands, where the first term represents the case with the outgoing fermion being always physical and the incoming one having experienced a chiral flip, and the second term represents the opposite situation.\footnote{Eq.~\eqref{Dirac_Structure_SB} means that the contributions are linear in the symmetry breaking source given the fact that only single insertions are required.}

\begin{figure}
    \centering
    \includegraphics[scale=0.2]{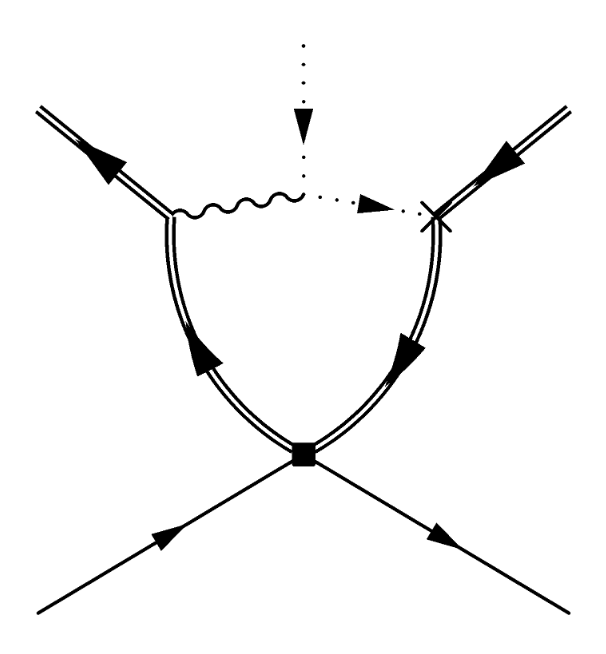} \hspace{15mm} \includegraphics[scale=0.182]{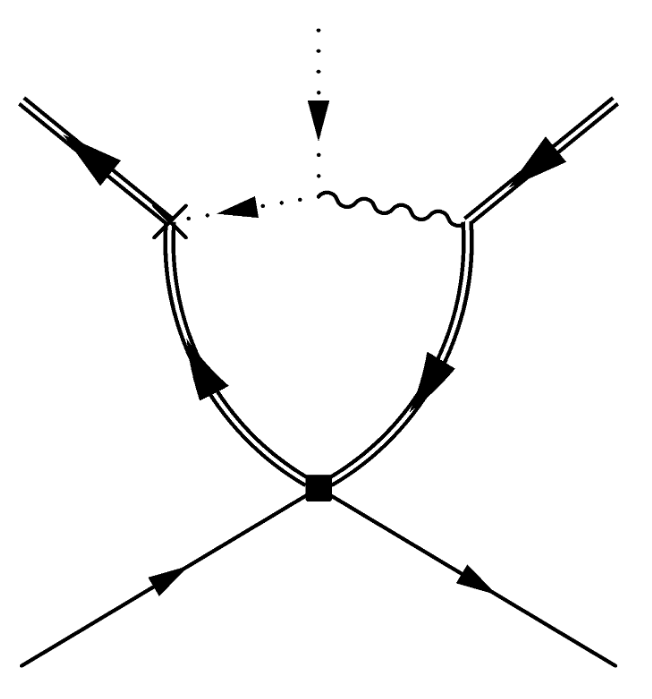}
    \caption[One-loop ``wine glass'' Feynman diagrams.]{One-loop ``wine glass'' Feynman diagrams. Insertions of the symmetry breaking operator are indicated by a cross (thus alluding to the chiral flip discussed in the text). Their divergent part is canceled between each other, so they give no contribution to the symmetry breaking. The single (double) lines represent physical right-handed (left-handed) fermions; analogous diagrams with opposite chiralities are also possible, as well as diagrams expressing violation of baryon and lepton numbers. The arrows also indicate the flux of the four-momenta introduced in the text.}
    \label{fig:Wine_Diagrams}
\end{figure}

Now, imagine that the previous Dirac structure appears in the calculation of the amplitude of a one-loop Feynman diagram with SDD=0. If \(k\) represents the integration variable, then the only contribution to the divergence comes from the term with only that momentum:

\begin{equation}
    \bar{\slashed{k}}\left[\hat{\slashed{k}}P_{R,L}-\hat{\slashed{k}}P_{L,R}\right]\hat{\slashed{k}}+\hat{\slashed{k}}\left[\hat{\slashed{k}}P_{R,L}-\hat{\slashed{k}}P_{L,R}\right]\bar{\slashed{k}}=\bar{\slashed{k}}\hat{\slashed{k}}\gamma_5\hat{\slashed{k}}+\hat{\slashed{k}}\hat{\slashed{k}}\gamma_5\bar{\slashed{k}} \, .
\end{equation}

\noindent Using the rules of the BMHV scheme it is easy to see that this is equal to \(\hat{k}^2\{\bar{\slashed{k}},\gamma_5\}=0\). Therefore, as already announced, \textit{there are no symmetry breaking contributions coming from diagrams with superficial degree of divergence zero, if the symmetry breaking operator is connected to an internal fermion}. Therefore, there are no contributions coming from diagrams with one insertion of a four-fermion operator and one insertion of the symmetry breaking operator in topologies consisting of a fermion loop with the following sets of external legs attached: one Higgs, one ghost and two gauge bosons; one ghost, two Higgses and one gauge bosons; one ghost and three gauge bosons; one ghost, and three Higgs fields.

The only diagrams with SDD=0 where the symmetry breaking operator appears attached to an external fermion leg are the ``wine glass'' diagrams, see Fig.~\ref{fig:Wine_Diagrams}. Nevertheless, they always appear in pairs, and a direct calculation
taking the gauge-fixing functional
$\partial_\mu A^\mu_\alpha$ (i.e., when the coupling of ghosts to gauge bosons involves only the anti-symmetric structure constants \cite{Weinberg2}) for any $\xi$-gauge shows that the divergent part of each of them is the same except for the structure constant \(C^{abc}\), that comes from the interaction vertex of two ghosts and one gauge boson. Due to the fact that the structure constant is antisymmetric, the contributions cancel between each other.
Obviously, such diagrams are possible only in the non-Abelian cases. The ${\rm SU}(3)$ case allows for diagrams where both left- and right-handed fermions interact with the internal ghost and gluon; with the same reasoning, in this case as well there is a cancellation among pairs of diagrams. The same holds for dimension-6 operators violating baryon and lepton numbers.
In conclusion, \textit{there is no symmetry breaking contribution coming from diagrams with SDD=0}.

The previous reasoning can also be used to simplify the computation of the diagrams with SDD=1. One can use the well-known relation \cite{Chetyrkin:1997fm}

\begin{equation}
    \label{Propagator_Decomposition}
    \frac{1}{\left(k+p\right)^2}=\frac{1}{k^2-m_R^2}-\frac{p^2+2kp+m_R^2}{k^2-m_R^2}\frac{1}{\left(k+p\right)^2},
\end{equation}

\noindent where $m_R$ is an arbitrary mass (that can be employed to regulate possible IR divergences), which allows the separation of the integrals in parts with different superficial degrees of divergence, because the first term behaves like \(k^{-2}\), while the second one behaves like \(k^{-3}\) in the UV. Then, for the divergent part of the amplitudes with SDD=1, one can find the contributions of the integral with all of the propagators substituted by \(1/(k^2-m^2_R)\) (SDD=1) and the case where one of them is substituted by the second term in the sum of Eq.~\eqref{Propagator_Decomposition} (SDD=0). Nonetheless, due to the fact that all the required Feynman diagrams with SDD=1 have the symmetry-breaking vertex attached to an internal fermion (as it will be shown), the previous reasoning shows that the second part of the decomposition (SDD=0) vanishes and only the case with all of the propagators written as \(1/(k^2-m^2_R)\) remains. Furthermore, the remaining terms that only contain the momentum \(k\) in the numerator (SDD=1) also vanish by the same argument used in the previous paragraph. Consequently, the only contribution comes from the case where all of the momenta of the amplitude are substituted by \(k\) except for the term in Eq.~\eqref{Dirac_Structure_SB} where one collects the terms with one \(k_g\):

\begin{equation}
    \label{Dirac_Structure_SB_1kg}
    \bar{\slashed{k}}_g\hat{\slashed{k}}\gamma_5\hat{\slashed{k}}-\bar{\slashed{k}}\hat{\slashed{k}}_gP_{L,R}\hat{\slashed{k}}+\hat{\slashed{k}}_g\hat{\slashed{k}}\gamma_5\bar{\slashed{k}}-\hat{\slashed{k}}\hat{\slashed{k}}_gP_{L,R}\bar{\slashed{k}}.
\end{equation}

An agreement has been found between these selection rules and the results displayed in Ref.~\cite{Belusca-Maito:2020ala}. 1PI diagrams with external BRST sources are convergent in our case, where our diagrams carry single insertions of dimension-6 four-fermion SMEFT operators.
(In other cases not discussed in this work, in general results depend on the specific choice of $\xi$-gauge, see e.g. Ref.~\cite{Belusca-Maito:2020ala}).

Now the results for the amplitudes that contribute to the finite obstructions to the Slavnov-Taylor identities are presented, as already mentioned, classified by the number of external Higgs fields.

\subsection{Amplitudes with no scalars} \label{sec:Amplitudes_No_Scalars}

Consider first the divergent Green's functions having no scalars, whose topologies are the ones shown in Fig.~\ref{fig:diagrams_no_scalars}. They can also be classified according to the number of external gauge bosons. Using the arguments presented before, it is clear that the cases with three or more external gauge bosons (\({\rm SSD}\leq0\)) do not contribute. The complete set of amplitudes is huge if one distinguishes between the different possible natures of the elements involved, i.e., the structure of the four-fermion operator and whether the external particles correspond to \({\rm SU}(3)\), \({\rm SU}(2)\) or \({\rm U}(1)\) ghost or gauge fields. However, all those expressions can be easily adapted from the ones of the most complicated case, which corresponds to the insertion of color-octet four-fermion operators $ Q_{q x}^{(8)} $, $x=u, d$, and $ Q_{u d}^{(8)} $ with external \({\rm SU}(3)\) ghost and gauge particles. For instance, the discussion of $\gamma_\mu \tau^I$ structures can be easily adapted from there, the only important difference being that the (fully) symmetric constant is now absent. Thus, to simplify the reading, only those amplitudes are presented here, given the fact that in the next section the full set of counterterms is displayed, which is the main purpose of this text.

In this subsection, $\xi$ denotes a fermion of left- or right-handed chirality, while the internal fermion carries left-handed chirality; cases proportional to the totally anti-symmetric symbol flip sign for an internal fermion carrying right-handed chirality. Consistently with this prescription, in this subsection no identity involving the contraction of the Levi-Civita symbol with the external fermion bilinear structure has been applied. Gauge couplings are omitted everywhere here. The constant $c$ displayed below is

\begin{equation}
    c = 2 i \frac{C_{\rm NP}}{3 \cdot 16 \pi^2},
\end{equation}

\noindent where $C_{\rm NP}$ is the Wilson coefficient of the dimension-6 SMEFT operator. The overall factor of $2$ is dropped when the internal degree of freedom in the loop is right-handed. With these considerations, the results of the amplitudes associated with the diagrams in Fig.~\ref{fig:diagrams_no_scalars} classified by the number of external gauge fields are the following:

\begin{figure}
    \centering
    \hspace{6mm} \includegraphics[width=0.20\linewidth]{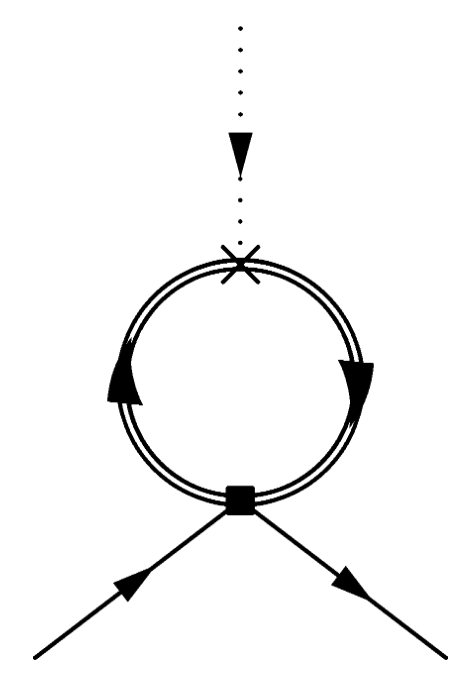} \hspace{11mm}
    \includegraphics[width=0.30\linewidth]{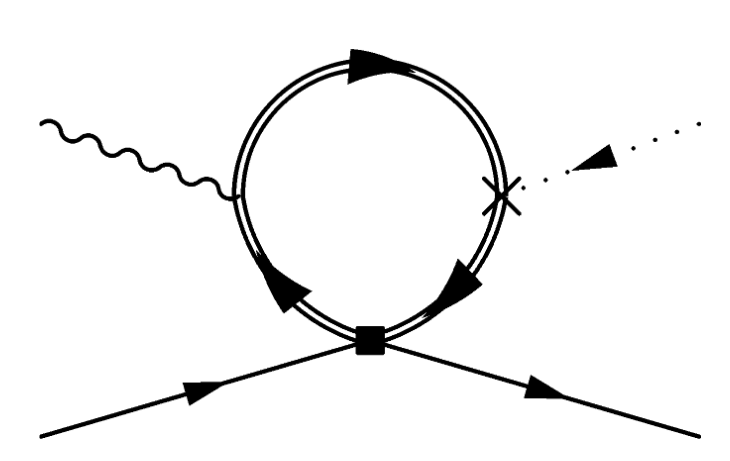} \hspace{5mm}
    \includegraphics[width=0.22\linewidth]{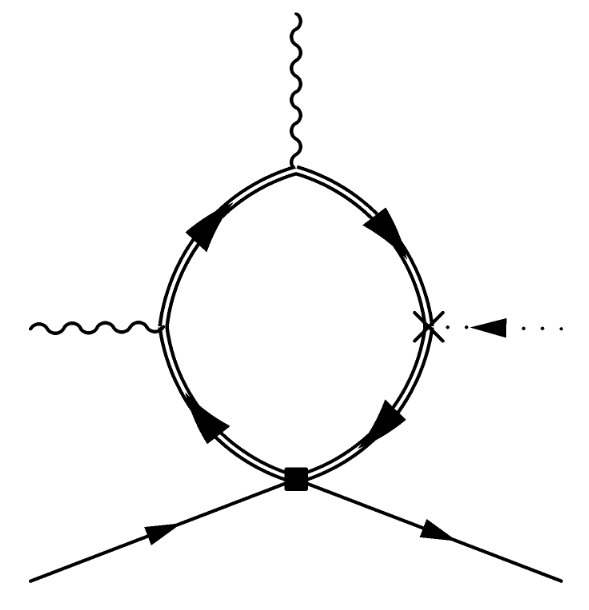}
    \caption[Examples of 1PI diagrams with single insertions of SMEFT four-fermion operators at one loop with no external scalars.]{Examples of 1PI diagrams with single insertions of SMEFT four-fermion operators at one loop with no external scalars. Insertions of the symmetry-breaking operator are indicated by a cross, four-fermion operators by filled squares and ghosts are represented by a dotted line. Gauge bosons are indicated by wiggly lines. The single (double) solid lines represent physical right-handed (left-handed) fermions; analogous diagrams with opposite chiralities are also possible.}
    \label{fig:diagrams_no_scalars}
\end{figure}

\noindent $ \bullet $ Amplitude with no gauge boson fields:

\begin{equation}\label{eq:ghostSU3}
    c \, {\rm tr} \{ T^A T^B \} \, [ \bar{\xi} \bar{\gamma}_\mu T^A \xi ] \, \bar{\partial}^2 \bar{\partial}^\mu \omega^B = c \frac{1}{2} \, [ \bar{\xi} \bar{\gamma}^\mu T^A \xi ] \, \bar{\partial}^2 \bar{\partial}_\mu \omega^A.
\end{equation}

\vspace{1mm}
\noindent
$ \bullet $ ``Anti-symmetric'' part of the amplitude with $ {\rm SU}(3) $ ghost, $ {\rm SU}(3) $ gauge boson:

\begin{equation}\label{eq:ghostSU3_GBSU3_asym}
\begin{split}
    & -i c \, {\rm tr} \{ T^A (T^B T^C - T^C T^B) \} \, [ \bar{\xi} \bar{\gamma}^\mu T^A \xi ] \left( \bar{\partial}^2 \omega^B G^C_\mu + 2 \bar{\partial}_\nu \omega^B \bar{\partial}^\nu G^C_\mu - 2 \bar{\partial}_\mu \omega^B \bar{\partial}^\nu G^C_\nu \right.\\&\hspace{6.5cm}\left. - 2 \bar{\partial}^\nu \bar{\partial}_\mu \omega^B G^C_\nu - 2 \bar{\partial}^\nu \omega^B \bar{\partial}_\mu G^C_\nu \right)\\
    & = \frac{1}{2} c \, C_{A B C} \, [ \bar{\xi} \bar{\gamma}^\mu T^A \xi ] \left( \bar{\partial}^2 \omega^B G^C_\mu + 2 \bar{\partial}_\nu \omega^B \bar{\partial}^\nu G^C_\mu - 2 \bar{\partial}_\mu \omega^B \bar{\partial}^\nu G^C_\nu\right.\\&\hspace{3.5cm}\left. - 2 \bar{\partial}_\mu \bar{\partial}^\nu \omega^B G^C_\nu - 2 \bar{\partial}^\nu \omega^B \bar{\partial}_\mu G^C_\nu \right). 
\end{split}
\end{equation}

\vspace{1mm}
\noindent
$ \bullet $ ``Symmetric'' part of the amplitude with $ SU(3) $ ghost, $ SU(3) $ gauge boson:

\begin{eqnarray}\label{eq:ghostSU3_GBSU3_sym}
    2 c \, {\rm Tr} \{ T^A ( T^B T^C + T^C T^B ) \} \, [\bar{\xi} \bar{\gamma}_\mu T^A \xi] \bar{\partial}_\rho G^B_\nu \bar{\partial}_\sigma \omega^C \epsilon^{\mu \nu \rho \sigma}.
\end{eqnarray}

\vspace{1mm}
\noindent
$ \bullet $ There are various contributions to the amplitude with one ghost and two gauge bosons. First of all, the symmetric structure:

\begin{equation}\label{eq:symmetric_structure}
\begin{split}
    - \frac{1}{2}&c \left( {\rm tr} \{ T^A T^B T^C T^D + T^A T^B T^D T^C \} \bar{\partial}_\alpha \omega^B G^C_\mu G^D_\beta \right. \\
    & + {\rm tr} \{ T^A T^C T^B T^D + T^A T^D T^B T^C \} G^C_\alpha \bar{\partial}_\mu \omega^B G^D_\beta \\
    &  + \left. {\rm tr} \{ T^A T^C T^D T^B + T^A T^D T^C T^B \} G^C_\alpha G^D_\mu \bar{\partial}_\beta \omega^B \right) [ \bar{\xi} \bar{\gamma}^\alpha \bar{\gamma}^\mu \bar{\gamma}^\beta T^A \xi ]. 
\end{split}
\end{equation}

\vspace{1mm}
\noindent
$ \bullet $ Also the following ``$\mathcal{A}$'' structure is present:

\begin{equation}\label{eq:A_totally_antisymmetric}
    i \frac{1}{2} c \, [\bar{\xi} \bar{\gamma}_\mu T^A \xi] G^B_\nu G^C_\rho \bar{\partial}_\sigma \omega^D \epsilon^{\mu \nu \rho \sigma} \mathcal{A}_{A}^{B C D} \,,
\end{equation}
where $\mathcal{A}_{A}^{B C D}$, which is totally antisymmetric
in $B, C$ and $D$, is defined by

\begin{equation}\label{eq:ASU3_definition}
\begin{split}
     \mathcal{A}_{A}^{B C D} &= {\rm tr} \{ T^A T^D ( T^B T^C - T^C T^B ) \}
     - {\rm tr} \{ T^A ( T^B T^D T^C - T^C T^D T^B ) \}\\& + {\rm tr} \{ T^A ( T^B T^C - T^C T^B ) T^D \} \\
    & = \frac{1}{4} i \left( C_{BCE} D_{ADE} - C_{DCE} D_{ABE} - C_{BDE} D_{ACE} \right).
\end{split}
\end{equation}

\noindent The property 

\begin{equation}\label{eq:identity_Haber}
\begin{split}
     {\rm tr} \{ T^A T^B T^C T^D \} &= \frac{1}{4 N} \left( \delta_{A B} \delta_{C D} - \delta_{A C} \delta_{B D} + \delta_{A D} \delta_{B C} \right) \\
    & + \frac{1}{8} \left( D_{A B E} D_{C D E} - D_{A C E} D_{B D E} + D_{A D E} D_{B C E} \right)  \\
    & + \frac{1}{8} i \left( D_{A B E} C_{C D E} + D_{A C E} C_{B D E} + D_{A D E} C_{B C E} \right), 
\end{split}
\end{equation}
with $N=N_c=3$ (and $N=2$ when adapting the discussion of this section to the operators $ Q_{\ell q}^{(3)}, Q_{q q}^{(3)} $, etc.), has been applied (see Ref.~\cite{Haber:2019sgz} for a systematic presentation).

\vspace{1mm}
\noindent
$ \bullet $ There is a ``$D \cdot C$'' structure accompanying a Levi-Civita symbol:

\begin{equation}\label{eq:d_f_structure}
\begin{split}
    & i c \, [\bar{\xi} \bar{\gamma}_\mu T^A \xi] G^B_\nu G^C_\rho \bar{\partial}_\sigma \omega^D \epsilon^{\mu \nu \rho \sigma} \, {\rm tr} \{ T^A ( T^B T^D T^C - T^C T^D T^B ) \} \\
    & = - c \, [\bar{\xi} \bar{\gamma}_\mu T^A \xi] G^B_\nu G^C_\rho \bar{\partial}_\sigma \omega^D \epsilon^{\mu \nu \rho \sigma} \, \frac{1}{4} \left( D_{A B E} C_{D C E} + D_{A D E} C_{B C E} + D_{A C E} C_{B D E} \right) \\
    & = - c \, [\bar{\xi} \bar{\gamma}_\mu T^A \xi] G^B_\nu G^C_\rho \bar{\partial}_\sigma \omega^D \epsilon^{\mu \nu \rho \sigma} \, \frac{1}{4} \left( 2 D_{A B E} C_{D C E} + D_{A D E} C_{B C E} \right). 
\end{split}
\end{equation}

\vspace{1mm}
\noindent
$ \bullet $ Finally, one finds the ``$C \cdot C$'' structure:

\begin{eqnarray}\label{eq:f_times_f}
    && c \, [\bar{\xi} \bar{\gamma}^\mu T^A \xi] G^B_\mu G^C_\nu \bar{\partial}^\nu \omega^D \, {\rm tr} \{ T^A T^D ( T^B T^C - T^C T^B ) \} \\
    && + c \, [\bar{\xi} \bar{\gamma}^\mu T^A \xi] G^B_\nu G^C_\mu \bar{\partial}^\nu \omega^D \, {\rm tr} \{ T^A ( T^B T^C - T^C T^B ) T^D \} \nonumber \\
    && = c \, [\bar{\xi} \bar{\gamma}^\mu T^A \xi] G^B_\mu G^C_\nu \bar{\partial}^\nu \omega^D \, {\rm tr} \{ ( T^A T^D - T^D T^A ) ( T^B T^C - T^C T^B ) \} \nonumber \\
    && = - c \, [\bar{\xi} \bar{\gamma}^\mu T^A \xi] G^B_\mu G^C_\nu \bar{\partial}^\nu \omega^D \frac{1}{2} C_{A D E} C_{B C E} \,. \nonumber
\end{eqnarray}

Sanity checks based on the Wess-Zumino consistency conditions (see Sec.~\ref{sec:Anomalies}) can be considered. For instance, it is straightforward to check explicitly that the coefficients that accompany the structures

\begin{eqnarray}
    && C_{A B C} C_{C E F} G^E_\mu \bar{\partial}^\nu g^F \bar{\partial}_\nu g^B [ \bar{\xi} \bar{\gamma}^\mu T^A \xi ] \,, \quad
    C_{A B C} C_{C E F} G^E_\nu \bar{\partial}_\mu g^F \bar{\partial}^\nu g^B [ \bar{\xi} \bar{\gamma}^\mu T^A \xi ] \,, \nonumber \\
    && C_{A B C} \bar{\partial}^\nu \bar{\partial}_\mu g^C \bar{\partial}_\nu g^B [ \bar{\xi} \bar{\gamma}^\mu T^A \xi ] \,, \hspace{1.5cm}
    C_{A B C} \bar{\partial}^2 g^C \bar{\partial}_\mu g^B [ \bar{\xi} \bar{\gamma}^\mu T^A \xi ] \,,
\end{eqnarray}
resulting from the combination of the amplitudes in Eqs.~\eqref{eq:ghostSU3} and \eqref{eq:ghostSU3_GBSU3_asym} (which involve different numbers of gauge bosons), vanish.

\subsection{Amplitudes with one scalar} \label{Amplitudes_One_Scalar}

\begin{figure}[H]
    \begin{minipage}{.15\textwidth}
        \centering
        \includegraphics[scale=0.15]{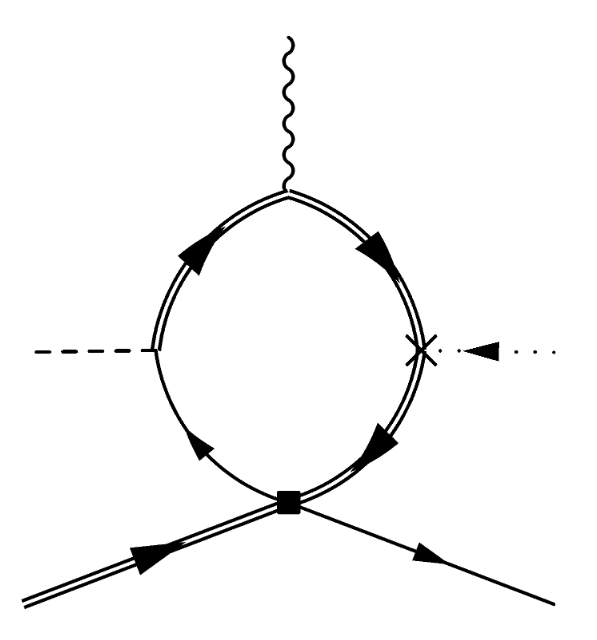} \\
        $ + \ldots $
    \end{minipage}
    \begin{minipage}{.85\textwidth}
        \centering
        \begin{eqnarray}\label{eq:1scalar_1GB}
            && = -\frac{1}{3 \cdot 16 \pi^2} C^{\mathcal{O},F}_{\psi \xi} \, Y^\dagger [ \bar{\psi} \{ (- C_L^g C_R^v - C_L^v C_R^g + 2 C_L^g C_L^v + 2 C_L^v C_L^g \nonumber\\
            && + 2 C_R^g C_R^v + 2 C_R^v C_R^g) C_6^{\mathcal{O}} \}_\mathcal{O} P_X \xi h ] \overline{k}_g^\mu A_\mu \omega \,,
        \end{eqnarray}
    \end{minipage}
\end{figure}

\begin{figure}[H]
    \begin{minipage}{.15\textwidth}
        \centering
        \includegraphics[scale=0.15]{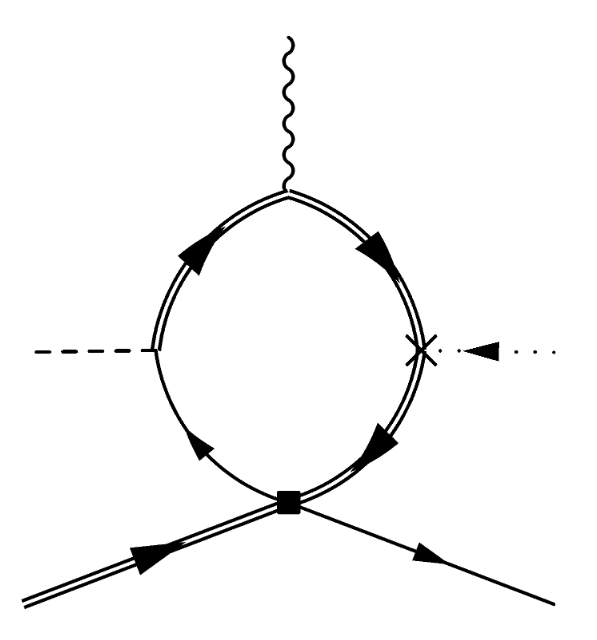} \\
        $ + \ldots $
    \end{minipage}
    \begin{minipage}{.84\textwidth}
        \centering
        \begin{eqnarray}\label{eq:1scalar_sigma_1GB}
            && = i \frac{4}{3 \cdot 16 \pi^2} C^{\mathcal{O},F}_{\psi \xi} \, Y^\dagger [ \bar{\psi} \{ (- C_L^g C_R^v + C_L^v C_R^g + 2 C_L^g C_L^v \nonumber\\
            && - 2 C_R^v C_R^g) C_6^{\mathcal{O}} \}_\mathcal{O} \bar{\sigma}_{\mu \nu} P_X \xi h ] \overline{k}_g^\mu A^\nu \omega \,.
        \end{eqnarray}
    \end{minipage}
\end{figure}

In the case of Green's functions with a single scalar, the amplitudes are presented in the following Eqs.~\eqref{eq:1scalar_0GB}-\eqref{eq:1scalar_sigma_1GB}. Examples of diagrams are shown, where single (double) fermion lines represent right-handed (respectively, left-handed) fields, which should make the identification of chiralities of the different diagrams easier for the reader; diagrams where the ghost and the gauge boson couple to different internal fermion lines are understood in the ellipses. One finds:

\begin{figure}[H]
    \begin{minipage}{.245\textwidth}
        \centering
        \includegraphics[scale=0.16]{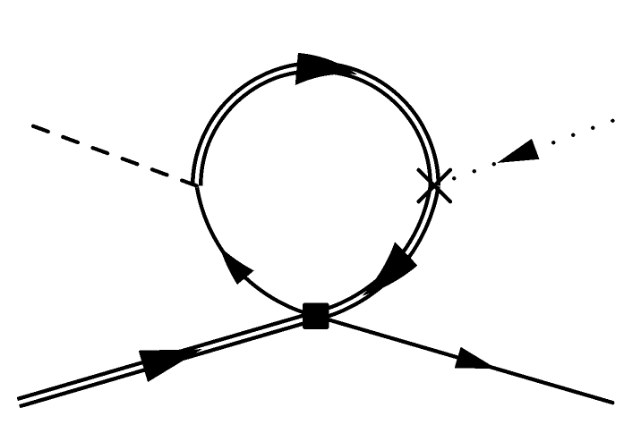} \\
        $ + \ldots $
    \end{minipage}
    \begin{minipage}{.745\textwidth}
        \centering
        \begin{equation}\label{eq:1scalar_0GB}
            = -\frac{1}{3 \cdot 16 \pi^2} C^{\mathcal{O},F}_{\psi \xi} \, Y^\dagger [ \bar{\psi} \{ (C^g_L - C^g_R) C_6^{\mathcal{O}} \}_\mathcal{O} P_X \xi h ] \overline{k}_g \cdot ( \overline{k}_g + 2 \overline{k}_h ) \omega \,,
        \end{equation}
    \end{minipage}
\end{figure}

\noindent where \(C^g_{L,R}\) are the couplings of the left- and right-handed fermions with the ghost, and \(h\) is employed to denote both \(\Phi\) and \(\Phi^c\) depending on the particular case. This amplitude vanishes when $ C^g_L = C^g_R $, as in the case of ${\rm SU}(3)$. The incoming four-momenta $ \overline{k}_g $ and $ \overline{k}_h $ are the ones of the ghost and the scalar, respectively.
The coefficients $C^{\mathcal{O},F}_{\psi \xi}$ collect the Wilson coefficients and pre-factors from the Fierz transformation used to put all relevant operators into a similar form (i.e., two fermion bilinears, each with a change of chirality), thus leading to a common calculation for the Feynman diagrams of all operator insertions. Then,

\noindent
where \(C_{L,R}^v\) are the couplings of the fermions with the gauge bosons. One finds $ C_L^v,\, C_L^g \to y_L,\, T^I_L,\, T^A $, and $ C_R^v,\, C_R^g \to y_R,\, 0,\, T^A $ for $ {\rm U}(1)_Y,\, {\rm SU}(2)_L,\, {\rm SU}(3) $, respectively. A trace $ \{ \cdot \}_\mathcal{O} \to {\rm tr} \{ \cdot \} $ over color indices applies, which depends on the operator $\mathcal{O}$ being considered, with $C_6^{\mathcal{O}} \to T^A$ for octet structures.

There are also structures in $\bar{\sigma}_{\mu \nu}$:

\begin{figure}[H]
    \begin{minipage}{.245\textwidth}
        \centering
        \includegraphics[scale=0.175]{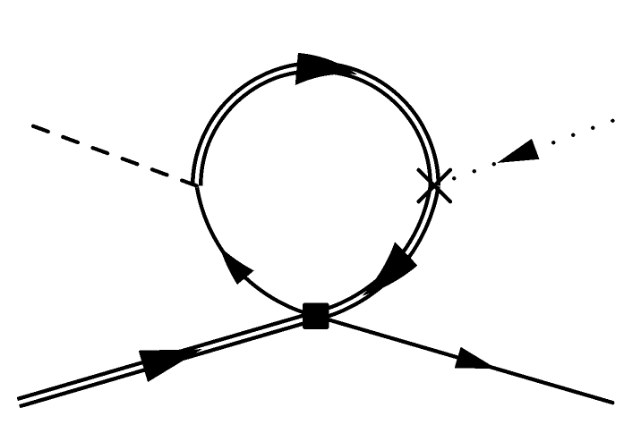} \\
        $ + \ldots $
    \end{minipage}
    \begin{minipage}{.745\textwidth}
        \centering
        \begin{equation}\label{eq:1scalar_sigma_0GB}
            = i \frac{8}{3 \cdot 16 \pi^2} C^{\mathcal{O},F}_{\psi \xi} \, Y^\dagger [ \bar{\psi} \{ (C^g_L + C^g_R) C_6^{\mathcal{O}} \}_\mathcal{O} \bar{\sigma}_{\mu \nu} P_X \xi h ] \overline{k}_g^\mu \overline{k}_h^\nu \omega \,,
        \end{equation}
    \end{minipage}
\end{figure}

\noindent
Terms with the antisymmetric Levi-Civita symbol contribute in an equal and constructive way to Eqs.~\eqref{eq:1scalar_sigma_0GB}-\eqref{eq:1scalar_sigma_1GB}.\footnote{
When considering cases carrying the anti-symmetric $\sigma_{\mu \nu} = i [\gamma_\mu, \gamma_\nu] / 2$ structure, an unexpected problem was found
while dealing with \texttt{Feyncalc} 9.2.0 (development version): for instance, the function \texttt{EpsChisholm} returns the same value for

\begin{center}
    $ \epsilon_{\mu \nu \alpha \beta} \bar{k}^\alpha_1 \bar{k}^\beta_2 \bar{\gamma}^{\mu} \bar{\gamma}^{\nu} P_R \;\; $ and $ \;\; \epsilon_{\mu \nu \alpha \beta} \bar{k}^\alpha_1 \bar{k}^\beta_2 \bar{\gamma}^{\nu} \bar{\gamma}^{\mu} P_R $,
\end{center}
which obviously should carry instead an opposite relative sign.
This issue has been corrected in subsequent versions, e.g., 10.0.0 (stable version).
Further discussion is found in Ref.~\cite{GitHub_FeynCalc}.
}

When considering the opposite chiralities (i.e., for insertions of the respective hermitian conjugated operators), Eqs.~\eqref{eq:1scalar_0GB} and \eqref{eq:1scalar_sigma_1GB} flip their overall signs.
This fact reflects in the need to introduce ghost number zero counterterm operators below with an $i$ in the case of the $\bar{\sigma}_{\mu \nu}$ structure contracted with two gauge bosons.
Also consistent with the change of signs of Eqs.~\eqref{eq:1scalar_0GB} and \eqref{eq:1scalar_sigma_1GB}, when taking the conjugate of the operators below, note that the BRST transformation of the scalar and its conjugate carry a relative sign
as per Eq.~\eqref{BRST_psi}.

\subsection{Amplitudes with two scalars} \label{Amplitudes_Two_Scalars}

Finally, shifting to Green's functions with two scalars,
an explicit calculation of this case gives

\begin{figure}[H]
    \begin{minipage}{.3\textwidth}
        \centering
        \includegraphics[scale=0.14]{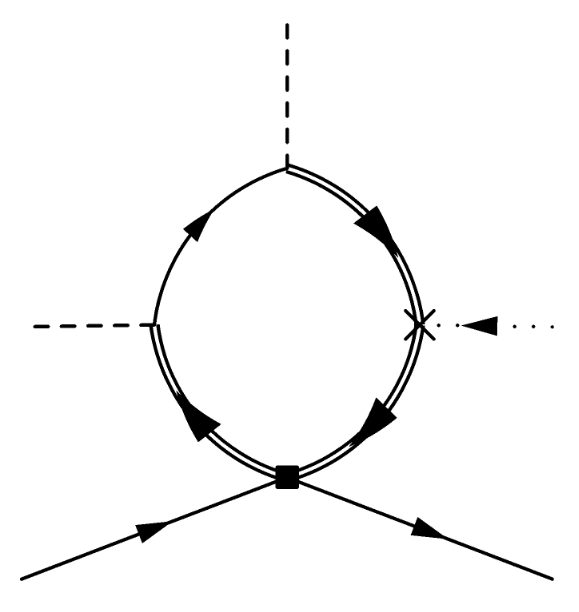} \\
        $ + \ldots $
    \end{minipage}
    \begin{minipage}{.69\textwidth}
        \centering
        \begin{equation}
            = C^{\mathcal{O}}_{\psi \xi} Y^\dagger Y \frac{1}{3 \cdot 16 \pi^2} \, [ \bar{\psi} \{ ( 2 C^g_{Y_1} - C^g_{Y_2} + 2 C^{g'}_{Y_3} ) C_6^\mathcal{O} \}_\mathcal{O} \overline{\slashed{k}}_g P_X \xi ] h^\dagger h \omega \,,
        \end{equation}
    \end{minipage}
\end{figure}

\noindent
where $Y_1, Y_2, Y_3 = L, R$, with $ Y_1 \neq Y_2 $ and $ Y_1 = Y_3 $; $C^g_{Y_1}$ and $C^{g'}_{Y_3}$ can be in principle different. The notation employed is the same as the one previously introduced in the case of Green's functions of a single scalar.

\section{Symmetry-Restoring Finite Counterterms} \label{Symmetry_FCTs}

In the previous section, the finite parts of the obstructions to the Slavnov-Taylor identities \(\left(\Gamma,\Gamma\right)=-2\Delta\cdot\Gamma\) have been obtained. Now, as already mentioned in Sec.~\ref{sec:Anomalies}, considering that no antifields are involved here, if the \(\Delta\cdot\Gamma\) terms can be expressed as the BRST transformation of some operator, then the obstructions are simply spurious anomalies that can be eliminated by adding the proper set of finite counterterms to the Lagrangian. These counterterms are the operators whose BRST transformation leads to the amplitudes in \(\Delta\cdot\Gamma\) up to the \(-2\) factor appearing in \(\left(\Gamma,\Gamma\right)=-2\Delta\cdot\Gamma\). Therefore, it is necessary to obtain the preimage of the BRST transformation of the amplitudes calculated in the previous section. Here the term preimage is used because the BRST operator has no inverse; therefore, \textit{the solution for the symmetry-restoring counterterms required in the finite renormalization program is not unique}: it is defined up to an arbitrary combination of operators that vanish under the BRST transformation; one can also employ the usual integration by parts to move around partial derivatives. Aside from requiring the ``simplicity'' of the counterterms, no strategy has been designed in this text to prescribe symmetry-restoring counterterms up to these BRST-invariant ambiguities; there are specific cases in which BRST-invariant ambiguities are absent, e.g., there are no gauge-invariant operators of dimension 6 coupling leptons to gluon fields. It is important to remark that gauge invariant structures, or BRST-exact ones, alone cannot provide the needed renormalization, since they both vanish under the BRST transformation. The results for the counterterms are given in this section in four dimensions: it is then made manifest that such counterterms are not evanescent.\footnote{Evanescent pieces to the following discussion could be added at will. The choice made in this work is not to include them.}

To compensate for the Green's functions of unit ghost number discussed in the previous section, one looks for BRST-exact polynomials of the fields, i.e., that are written as the BRST transformation of a monomial of matter and gauge fields. In the following, the fact that the theory in $D$ dimensions
is invariant under the global symmetry transformations that are part of any gauge symmetries is employed, and thus one can just consider candidates for counterterms that respect such global transformations. This is only possible due to the use of fictitious fields in the chosen scheme. In practice, the global symmetry implies that only the terms for which at least one derivative acts on the ghost fields survive, which arise, for instance, when the BRST transformation acts on a gauge field or on a derivative of a matter field. This result can be understood from the fact that the set of terms proportional to the ghost fields but not to their derivatives matches the effect of a global symmetry transformation (e.g., dropping the term $\partial^{\mu}\omega_{\alpha}\left(x\right)$ in the transformation of $A^{\mu}_{\alpha}\left(x\right)$ in Eq.~\eqref{BRST_A} deriving an expression analogous to the one obtained from the transformation of $A^{\mu}_{\alpha}\left(x\right)$ under a global symmetry). Implicitly, also ghost number conservation and four-dimensional Lorentz symmetry are employed. The potential counterterms sufficient to perform the finite renormalization have the following structures:

\noindent $ \bullet $ No scalars and two derivatives:

\begin{equation}
    ( \bar{\psi}_k \Gamma \xi_m )_b (\partial^2 A^b).
\end{equation}
When not indicated otherwise, indices $k$ and $m$ are contracted together, i.e., the Kronecker symbol $ \delta_{k m} $ is omitted.

\noindent $ \bullet $  No scalars and one derivative:

\begin{equation}
    ( \bar{\psi}_k \Gamma \xi_m )_{b c} (\partial A^b) A^c.
\end{equation}

\noindent $ \bullet $ No scalars and no derivatives:

\begin{equation}
    ( \bar{\psi}_k \Gamma \xi_m )_{a b c} A^a A^b A^c.
\end{equation}

\noindent $ \bullet $ One scalar and two derivatives:

\begin{equation}
    (\bar{\psi}_k \xi) (\partial^2 h_m).
\end{equation}

\noindent $ \bullet $ One scalar and one derivative:

\begin{equation}
    (\bar{\psi}_k \xi \partial h_m)_b A^b.
\end{equation}
\begin{equation}
    (\bar{\psi}_k \xi h_m)_b (\partial A^b).
\end{equation}

\noindent $ \bullet $ One scalar and no derivatives:

\begin{equation}
    (\bar{\psi}_k \xi h_m)_{a b} A^a A^b.
\end{equation}

\noindent $ \bullet $ Two scalars and no derivatives:

\begin{equation}
    ( \bar{\psi}_{k'} \Gamma \xi_{m'} h_{l}^\dagger h_{n} )_{a} A^a.
\end{equation}
Indices $k', m', l, n$ are contracted in the appropriate way.

From now on, the counterterms needed to perform the finite renormalization are shown, classified according to the number of scalar fields that they involve.

\subsection{Counterterms with no scalars} \label{subsec:Count_no_scalar}

The most difficult part of the whole calculation is to find the preimage of the BRST transformation in some specific cases, because the BRST transformation of one operator can contribute to different amplitudes. This will be illustrated obtaining the counterterms needed to eliminate the finite obstructions to the Slavnov-Taylor identities associated with amplitudes with insertions of octet operators and external \({\rm SU}(3)\) gauge and ghost fields. The Green's functions are listed in Eqs.~\eqref{eq:ghostSU3}-\eqref{eq:A_totally_antisymmetric} and Eqs.~\eqref{eq:d_f_structure} and \eqref{eq:f_times_f}. The notation followed in this calculation is the same as in Sec.~\ref{sec:Amplitudes_No_Scalars}, where those amplitudes were introduced.

For the candidate counterterm, one finds:

\begin{eqnarray}\label{eq:CT_f1_1GB}
    && s [ f_1^{(1)} \bar{\partial}^2 G^A_\mu ( \bar{\xi} \bar{\gamma}^\mu T^A \xi ) ] = f_1^{(1)} \bar{\partial}^2 ( \bar{\partial}_\mu \omega^A + C_{A B C} G^B_\mu \omega^C ) ( \bar{\xi} \bar{\gamma}^\mu T^A \xi ) \\
    && = f_1^{(1)} ( \bar{\partial}^2 \bar{\partial}_\mu \omega^A + C_{A B C} G^B_\mu \bar{\partial}^2 \omega^C + 2 C_{A B C} \bar{\partial}^\nu G^B_\mu \bar{\partial}_\nu \omega^C ) ( \bar{\xi} \bar{\gamma}^\mu T^A \xi ), \nonumber
\end{eqnarray}

\begin{eqnarray}\label{eq:CT_f2_1GB}
    && s [ f_2^{(1)} \bar{\partial}_\mu \bar{\partial}^\nu G^A_\nu ( \bar{\xi} \bar{\gamma}^\mu T^A \xi ) ] = f_2^{(1)} \bar{\partial}_\mu \bar{\partial}^\nu ( \bar{\partial}_\nu \omega^A + C_{A B C} G^B_\nu \omega^C ) ( \bar{\xi} \bar{\gamma}^\mu T^A \xi ) \\
    && = f_2^{(1)} ( \bar{\partial}^2 \bar{\partial}_\mu \omega^A + C_{A B C} G^B_\nu \bar{\partial}_\mu \bar{\partial}^\nu \omega^C + C_{A B C} \bar{\partial}_\mu G^B_\nu \bar{\partial}^\nu \omega^C + C_{A B C} \bar{\partial}^\nu G^B_\nu \bar{\partial}_\mu \omega^C ) ( \bar{\xi} \bar{\gamma}^\mu T^A \xi ), \nonumber
\end{eqnarray}
so that the amplitude in Eq.~\eqref{eq:ghostSU3} fixes the value of $ f_1^{(1)} + f_2^{(1)} = - c / 2 $.

Apart from these operators of coefficients $ f_1^{(1)} $ and $ f_2^{(1)} $, other candidate counterterms are:

\begin{eqnarray}\label{eq:CT_f1_2GB}
    && s [ f_1^{(2)} \bar{\partial}_\mu G^B_\nu G^{C \nu} ( \bar{\xi} \bar{\gamma}^\mu C_{A B C} T^A \xi ) ] \\
    && = f_1^{(2)} \left( G^C_\nu \bar{\partial}_\mu \bar{\partial}^\nu \omega^B + C_{B E F} G^{C \nu} G^E_\nu \bar{\partial}_\mu \omega^F + \bar{\partial}_\mu G^B_\nu \bar{\partial}^\nu \omega^C \right) ( \bar{\xi} \bar{\gamma}^\mu C_{A B C} T^A \xi ) \,, \nonumber
\end{eqnarray}

\begin{eqnarray}\label{eq:CT_f2_2GB}
    && s [ f_2^{(2)} \bar{\partial}^\nu G^B_\mu G^C_\nu ( \bar{\xi} \bar{\gamma}^\mu C_{A B C} T^A \xi ) ] \\
    && = f_2^{(2)} \left( G^C_\nu \bar{\partial}_\mu \bar{\partial}^\nu \omega^B + C_{B E F} G^C_\nu G^E_\mu \bar{\partial}^\nu \omega^F + \bar{\partial}^\nu G^B_\mu \bar{\partial}_\nu \omega^C \right) ( \bar{\xi} \bar{\gamma}^\mu C_{A B C} T^A \xi ), \nonumber
\end{eqnarray}

\begin{eqnarray}\label{eq:CT_f3_2GB}
    && s [ f_3^{(2)} \bar{\partial}^\nu G^B_\nu G^C_\mu ( \bar{\xi} \bar{\gamma}^\mu C_{A B C} T^A \xi ) ] \nonumber \\
    && = f_3^{(2)} \left( G^C_\mu \bar{\partial}^2 \omega^B + C_{B E F} G^E_\nu G^C_\mu \bar{\partial}^\nu \omega^F + \bar{\partial}^\nu G^B_\nu \bar{\partial}_\mu \omega^C \right) ( \bar{\xi} \bar{\gamma}^\mu C_{A B C} T^A \xi ) \nonumber \\
    && = f_3^{(2)} \left( G^C_\mu \bar{\partial}^2 \omega^B + \bar{\partial}^\nu G^B_\nu \bar{\partial}_\mu \omega^C \right) ( \bar{\xi} \bar{\gamma}^\mu C_{A B C} T^A \xi ) \\
    && + f_3^{(2)} \left( C_{B C E} C_{A D E} G^B_\mu G^C_\nu \bar{\partial}^\nu \omega^D + C_{B E F} C_{A B C} G^C_\nu G^E_\mu \bar{\partial}^\nu \omega^F \right) ( \bar{\xi} \bar{\gamma}^\mu T^A \xi ), \nonumber
\end{eqnarray}
where in the last equation the Jacobi identity

\begin{equation}
    C_{B E F} C_{A B C} = C_{E C B} C_{A B F} + C_{F C B} C_{A E B},
\end{equation}

\noindent has been employed. Looking at the structure $ C_{A B C} G^C_\mu \bar{\partial}^2 \omega^B ( \bar{\xi} \bar{\gamma}^\mu T^A \xi ) $, one finds that

\begin{equation}
    - f_1^{(1)} + f_3^{(2)} = - \frac{1}{2} c
\end{equation}
is fixed by the amplitude shown in Eq.~\eqref{eq:ghostSU3_GBSU3_asym}. Similarly,

\begin{equation}
    - f_2^{(1)} + f_1^{(2)} + f_2^{(2)} = c,
\end{equation}
is fixed after looking at the structure $ C_{A B C} G^C_\nu \bar{\partial}_\mu \bar{\partial}^\nu \omega^B ( \bar{\xi} \bar{\gamma}^\mu T^A \xi ) $;

\begin{equation}
    2 f_1^{(1)} + f_2^{(2)} = c,
\end{equation}
is obtained from the term $ C_{A B C} \bar{\partial}^\nu G^B_\mu \bar{\partial}_\nu \omega^C ( \bar{\xi} \bar{\gamma}^\mu T^A \xi ) $;

\begin{equation}
    f_2^{(1)} + f_1^{(2)} = - c,
\end{equation}
is derived from the structure $ C_{A B C} \bar{\partial}_\mu G^B_\nu \bar{\partial}^\nu \omega^C ( \bar{\xi} \bar{\gamma}^\mu T^A \xi ) $;

\begin{equation}
    f_2^{(1)} + f_3^{(2)} = - c,
\end{equation}
is fixed after looking at the  $ C_{A B C} \bar{\partial}^\nu G^B_\nu \bar{\partial}_\mu \omega^C ( \bar{\xi} \bar{\gamma}^\mu T^A \xi ) $ term. These relations among the coefficients are consistent, and imply

\begin{eqnarray}\label{eq:sols_0_1_2_GBs}
    && f_2^{(1)} = - \frac{c}{2} - f_1^{(1)} \,, \;\;\;\; f_1^{(2)} = f_1^{(1)} - \frac{c}{2} \,, \\
    && f_2^{(2)} = - 2 f_1^{(1)} + c \,, \;\;\;\; f_3^{(2)} = f_1^{(1)} - \frac{c}{2} = f_1^{(2)}. \nonumber
\end{eqnarray}
One particular case is $ f_1^{(1)} = c/2 $, but this is not the most general solution. This point is further elaborated below.

The symmetric structure of Eq.~\eqref{eq:symmetric_structure} requires the introduction of the following counterterm:

\begin{eqnarray}\label{eq:CT_f1_3GB}
    f^{(3)}_1 {\rm Tr} \{ T^A T^B T^C T^D \} G^B_\alpha G^C_\mu G^D_\beta ( \bar{\xi} \bar{\gamma}^\alpha \bar{\gamma}^\mu \bar{\gamma}^\beta T^A \xi ),
\end{eqnarray}
with $ f^{(3)}_1 = c $, in which case it acts as a counterterm for the amplitude in Eq.~\eqref{eq:symmetric_structure} and also the amplitude in Eq.~\eqref{eq:f_times_f}. However, it also introduces new ``$ D \cdot C $'' terms without a Levi-Civita symbol

\begin{eqnarray}
    && s [ f^{(3)}_1 {\rm Tr} \{ T^A T^B T^C T^D \} G^B_\alpha G^C_\mu G^D_\beta ( \bar{\xi} \bar{\gamma}^\alpha \bar{\gamma}^\mu \bar{\gamma}^\beta T^A \xi ) ] + \text{Eq.}~\eqref{eq:symmetric_structure} + \text{Eq.}~\eqref{eq:f_times_f} \nonumber \\
    && = i \frac{f_1^{(3)}}{8} ( \bar{\xi} \bar{\gamma}^\alpha \bar{\gamma}^\mu \bar{\gamma}^\beta T^A \xi ) \left[ C_{C D E} D_{A B E} \left( \bar{\partial}_\alpha \omega^B G^C_\mu G^D_\beta + G^C_\alpha G^D_\mu \bar{\partial}_\beta \omega^B \right) \right. \nonumber \\
    && \left. + \left( C_{C B E} D_{A D E} + C_{C D E} D_{A B E} + C_{B D E} D_{A C E} \right) G^C_\alpha \bar{\partial}_\mu \omega^B G^D_\beta \right].
\end{eqnarray}
These contributions are canceled by the addition of

\begin{eqnarray}\label{eq:CT_f1_3GB_fd}
    -i \frac{f_1^{(3)}}{8} ( \bar{\xi} \bar{\gamma}^\alpha \bar{\gamma}^\mu \bar{\gamma}^\beta T^A \xi ) \left( G^B_\alpha G^C_\mu G^D_\beta + G^C_\alpha G^B_\mu G^D_\beta + G^C_\alpha G^D_\mu G^B_\beta \right) C_{C D E} D_{A B E} .
\end{eqnarray}

Focusing on structures proportional to ``$ C \cdot C $'', introduced by the counterterms of coefficients $ f^{(2)}_{1,2,3} $, one also finds

\begin{eqnarray}\label{eq:CT_f4_3GB}
    && s [ f^{(3)}_4 ( \bar{\xi} \bar{\gamma}^\mu T^A \xi ) G^B_\mu G^C_\nu G^{D \nu} C_{A D E} C_{B C E} ] \\
    && = f^{(3)}_4 ( \bar{\xi} \bar{\gamma}^\mu T^A \xi ) ( \bar{\partial}_\mu \omega^B G^C_\nu G^{D \nu} + G^B_\mu \bar{\partial}_\nu \omega^C G^{D \nu} + G^B_\mu G^C_\nu \bar{\partial}^\nu \omega^D ) C_{A D E} C_{B C E} \nonumber \\
    && = f^{(3)}_4 ( \bar{\xi} \bar{\gamma}^\mu T^A \xi ) G^{C \nu} G^E_\nu \bar{\partial}_\mu \omega^F C_{A B C} C_{B E F} - f^{(3)}_4 ( \bar{\xi} \bar{\gamma}^\mu T^A \xi ) G^C_\nu G^E_\mu \bar{\partial}^\nu \omega^F C_{A B C} C_{B E F} \nonumber \\
    && + f^{(3)}_4 ( \bar{\xi} \bar{\gamma}^\mu T^A \xi ) G^B_\mu G^C_\nu \bar{\partial}^\nu \omega^D C_{A D E} C_{B C E}. \nonumber
\end{eqnarray}
Note that a similar structure identically vanishes

\begin{equation}
    ( \bar{\xi} \bar{\gamma}^\mu T^A \xi ) G^{E \nu} G^F_\nu G^C_\mu C_{A B C} C_{B E F} = 0.
\end{equation}
Therefore, the conditions $ f^{(3)}_4 + f^{(2)}_3 = 0 $,
$ f^{(3)}_4 + f^{(2)}_1 = 0 $, $ - f^{(3)}_4 + f^{(2)}_2 + f^{(2)}_3 = 0 $ are derived. Together with Eq.~\eqref{eq:sols_0_1_2_GBs}, this linear system admits the  solution:

\begin{eqnarray}
    f^{(3)}_4 = f^{(1)}_2 + c.
\end{eqnarray}
In the following, the choice $ f_1^{(1)} = c/2 $ is considered. Shifting this choice of $ f_1^{(1)} $ by a constant amounts to introducing a counterterm in the finite renormalization proportional to the gauge invariant structure:

\begin{equation}
    i \, \bar{\xi} \bar{\gamma}_\mu [\bar{D}_\nu, F^{\mu \nu}_A] T^A \xi = i \, \bar{\xi} \bar{\gamma}_\mu T^A \xi \, [\bar{D}_\nu, F^{\mu \nu}_A] \,,
\end{equation}
where $ \bar{D}_\mu $ is the covariant derivative in four dimensions. This shows what it has already been mentioned: \textit{the choice of the finite counterterms is not unique, since two solutions are equivalent if they differ by a set of BRST invariant operators}.

Turning to the remaining simplest structure,
Eq.~\eqref{eq:ghostSU3_GBSU3_sym} fixes the coefficient of:

\begin{eqnarray}\label{eq:CT_f4_2GB}
    && s [ f^{(2)}_4 {\rm Tr} \{ T^A ( T^B T^C + T^C T^B ) \} \, [\bar{\xi} \bar{\gamma}_\mu T^A \xi] \bar{\partial}_\rho G^B_\nu G^C_\sigma \epsilon^{\mu \nu \rho \sigma} ] \\
    && = f^{(2)}_4 {\rm Tr} \{ T^A ( T^B T^C + T^C T^B ) \} \, [\bar{\xi} \bar{\gamma}_\mu T^A \xi] \left( C_{B E F} G^C_\sigma G^E_\nu \bar{\partial}_\rho \omega^F + \bar{\partial}_\rho G^B_\nu \bar{\partial}_\sigma \omega^C \right) \epsilon^{\mu \nu \rho \sigma}, \nonumber
\end{eqnarray}
i.e., $ f^{(2)}_4 = - 2 c $.
The following counterterm is introduced to address the ``$ D \cdot C $'' structure of Eq.~\eqref{eq:d_f_structure}:

\begin{eqnarray}\label{eq:CT_f3_3GB}
    && s [ f^{(3)}_3 ( \bar{\xi} \bar{\gamma}_\mu T^A \xi ) G^B_\nu G^C_\rho G^D_\sigma \epsilon^{\mu \nu \rho \sigma} D_{A D E} C_{B C E} ] \\
    && = f^{(3)}_3 ( \bar{\xi} \bar{\gamma}_\mu T^A \xi ) \left( 2 \bar{\partial}_\nu \omega^B G^C_\rho G^D_\sigma + G^B_\nu G^C_\rho \bar{\partial}_\sigma \omega^D \right) \epsilon^{\mu \nu \rho \sigma} D_{A D E} C_{B C E} \nonumber \\
    && = f^{(3)}_3 ( \bar{\xi} \bar{\gamma}_\mu T^A \xi ) G^B_\nu G^C_\rho \bar{\partial}_\sigma \omega^D \epsilon^{\mu \nu \rho \sigma} D_{A D E} C_{B C E} \nonumber \\
    && - 2 f^{(3)}_3 ( \bar{\xi} \bar{\gamma}_\mu T^A \xi ) G^B_\nu G^C_\rho \bar{\partial}_\sigma \omega^D \epsilon^{\mu \nu \rho \sigma} D_{A B E} C_{D C E}, \nonumber
\end{eqnarray}
i.e., $ f^{(3)}_3 = c / 4 $, by direct comparison with Eq.~\eqref{eq:d_f_structure}. For the other piece of this amplitude, since

\begin{equation}
    {\rm Tr} \{ T^A ( T^B T^C + T^C T^B ) \} = \frac{1}{2} D_{A B C},
\end{equation}
and

\begin{eqnarray}
    && f^{(2)}_4 {\rm Tr} \{ T^A ( T^B T^C + T^C T^B ) \} \, [\bar{\xi} \bar{\gamma}_\mu T^A \xi] C_{B E F} G^C_\sigma G^E_\nu \bar{\partial}_\rho \omega^F \epsilon^{\mu \nu \rho \sigma} \\
    && = - f^{(2)}_4 \frac{1}{2} D_{A B E} \, [\bar{\xi} \bar{\gamma}_\mu T^A \xi] C_{D C E} G^B_\nu G^C_\rho \bar{\partial}_\sigma \omega^D \epsilon^{\mu \nu \rho \sigma}, \nonumber
\end{eqnarray}
one finds that $ - f^{(2)}_4 / 2 - 2 f^{(3)}_3 = c / 2 $, which holds true.

Finally, for the $ \mathcal{A} $-structure in Eq.~\eqref{eq:A_totally_antisymmetric} it is possible to introduce:

\begin{equation}\label{eq:CT_f2_3GB}
    s [ f^{(3)}_2 ( \bar{\xi} \bar{\gamma}_\mu T^A \xi ) G^B_\nu G^C_\rho G^D_\sigma \epsilon^{\mu \nu \rho \sigma} \mathcal{A}_{A}^{B C D} ] = 3 f^{(3)}_2 ( \bar{\xi} \bar{\gamma}_\mu T^A \xi ) G^B_\nu G^C_\rho \bar{\partial}_\sigma \omega^D \epsilon^{\mu \nu \rho \sigma} \mathcal{A}_{A}^{B C D} \,,
\end{equation}
where $ 3 f^{(3)}_2 = -i c / 2 $. To summarize, counterterms are found in Eqs.~\eqref{eq:CT_f1_1GB}, \eqref{eq:CT_f2_1GB}, \eqref{eq:CT_f1_3GB}, \eqref{eq:CT_f1_3GB_fd}, \eqref{eq:CT_f4_2GB}, \eqref{eq:CT_f3_3GB} and \eqref{eq:CT_f2_3GB}.

The counterterms for the amplitudes with other external fields and different four-fermion operator structures are, in general, much easier to obtain. Those cases (and the ones presented before) are listed  in the following part of the text. No derivation is presented, to simplify the reading, and because the techniques used are the same as in the previous calculation but in simplified cases.

\subsubsection{$\gamma_\mu$-structure counterterms}\label{sec:gamma_structure_CTs}

First of all, counterterm structures having a fermion bilinear of the form $ \bar{\xi} \bar{\gamma}_\mu \xi $ are considered, where $\xi$ is any left- or right-handed physical fermion, being a quark or a lepton of possibly different generations; the structure $ \bar{\xi} \bar{\gamma}_\alpha \bar{\gamma}_\beta \bar{\gamma}_\mu \xi $ can also appear (for example, as displayed below, it comes contracted with three gauge boson fields allowing further simplification), since it has been found more appropriate to keep the bilinear under such a Lorentz form in some instances. In order to obtain non-vanishing amplitudes, zero or at least two ${\rm SU(2)}_{\rm L}$ or ${\rm SU}(3)$ gauge bosons are needed. In the notation here, \(g_1\), \(g_2\) and \(g_3\) correspond to \(g'\), \(g_L\) and \(g_S\), respectively. A possible basis of counterterm operators includes

\begin{equation}
    \mathcal{Q}_1 = g_1 (\bar{\xi} \bar{\gamma}_\mu \xi) \bar{\partial}^2 B^\mu, \label{eq:T1} 
\end{equation}
which has a single gauge boson; then, the following cases having a term with two gauge bosons (which consist of the Chern-Simons form of QCD, and analogously for ${\rm U}(1)_{\rm Y}$ and ${\rm SU(2)}_{\rm L}$, contracted with a fermion bilinear) are also included\footnote{On the practical side, the BRST transformations of these combinations are terms with only a single gauge boson.}

\begin{eqnarray}
    && \mathcal{Q}_2 = g_1^2 (\bar{\xi} \bar{\gamma}_\mu \xi) \epsilon^{\mu \nu \rho \sigma} B_{\sigma} \bar{\partial}_\rho B_{\nu}, \label{eq:T1CS}\\ 
    && \mathcal{Q}_3 = (\bar{\xi} \bar{\gamma}_\mu \xi) \epsilon^{\mu \nu \rho \sigma} \left( g_2^2 W^b_\sigma \bar{\partial}_\rho W^b_\nu - g_2^3 \frac{1}{3} C^{SU(2)}_{b e f} W^b_\sigma W^e_\nu W^f_\rho \right), \label{eq:T2CS}\\
    && \mathcal{Q}_4 = (\bar{\xi} \bar{\gamma}_\mu \xi) \epsilon^{\mu \nu \rho \sigma} \left( g_3^2 G^B_\sigma \bar{\partial}_\rho G^B_\nu - g_3^3 \frac{1}{3} C_{B E F} G^B_\sigma G^E_\nu G^F_\rho \right); \label{eq:T3CS} 
\end{eqnarray}
finally, there are also various cases having three gauge bosons

\begin{eqnarray}
    \mathcal{Q}_5 = g_1^3 B_\alpha B_\mu B_\beta ( \bar{\xi} \bar{\gamma}^\alpha \bar{\gamma}^\mu \bar{\gamma}^\beta \xi ), \label{eq:T3A} 
\end{eqnarray}

\begin{eqnarray}
    && \mathcal{Q}_6 = g_1 g_2^2 W^b_\alpha W^b_\mu B_\beta \bar{\xi} ( \bar{\gamma}^\beta \bar{\gamma}^\alpha \bar{\gamma}^\mu + \bar{\gamma}^\alpha \bar{\gamma}^\beta \bar{\gamma}^\mu + \bar{\gamma}^\alpha \bar{\gamma}^\mu \bar{\gamma}^\beta ) \xi, \label{eq:T23A}\\ 
    && \mathcal{Q}_7 = g_1 g_3^2 G_\alpha^B G_\mu^B B_\beta \bar{\xi} ( \bar{\gamma}^\beta \bar{\gamma}^\alpha \bar{\gamma}^\mu + \bar{\gamma}^\alpha \bar{\gamma}^\beta \bar{\gamma}^\mu + \bar{\gamma}^\alpha \bar{\gamma}^\mu \bar{\gamma}^\beta ) \xi, \label{eq:T33A}
\end{eqnarray}

\begin{eqnarray}
    \mathcal{Q}_8 = g_3^3 D_{B C D} G^B_\alpha G^C_\mu G^D_\beta (\bar{\xi} \bar{\gamma}^\alpha \bar{\gamma}^\mu \bar{\gamma}^\beta \xi), \label{eq:T3d}
\end{eqnarray}

\begin{eqnarray}
    && \mathcal{Q}_9 = ( \bar{\xi} \bar{\gamma}_\mu \xi ) g_2^3 W^b_\nu W^c_\rho W^d_\sigma \epsilon^{\mu \nu \rho \sigma} C^{SU(2)}_{b c d}, \label{eq:T23f}\\
    && \mathcal{Q}_{10} = ( \bar{\xi} \bar{\gamma}_\mu \xi ) g_3^3 G^B_\nu G^C_\rho G^D_\sigma \epsilon^{\mu \nu \rho \sigma} C_{B C D}. \label{eq:T33f} 
\end{eqnarray}

The finite counterterm Lagrangian is

\begin{equation}
    \mathcal{L}_{\rm fct} \supset - \frac{C^{\mathcal{O}}_{\xi \xi}}{3 \cdot 16 \pi^2} \sum^{10}_{i=1} h^{\mathcal{O}}_{i} \, \mathcal{Q}_i + {\rm h.c.}
\end{equation}
Hermitian conjugation has to be included only for cases where the fermion bilinear involves two different flavors. The coefficients $h^{\mathcal{O}}$ are shown in Tab.~\ref{tab:gamma_gamma}.
This table is organized as follows: the first, second, third and fourth sectors of the first half correspond to $\ell$, $q$, $x=u, d$ and $y=e, \nu$ loops, respectively, when the Feynman diagrams are obtained by contracting the fermion fields that appear within the same bilinear of the four-fermion operators, following the convention of Tab.~\ref{tab:dimension_6}; the second half of the table corresponds to those other topologies obtained when contracting fermions in different bilinears of the four-fermion operators, again when following the Warsaw basis. These two halves can be related by the Fierz identities shown in Eqs.~\eqref{eq:Fierz_1}-\eqref{eq:Fierz_4}. As can be understood from Eq.~\eqref{eq:GellMann_identity_1}, ${{\rm SU}\rm(2)}_{\rm L}$ triplet and ${\rm SU}(3)$ octet structures can be obtained from ${\rm SU}(2)_{\rm L}$ and ${\rm SU}(3)$ singlet cases, and the converse is also true following Eq.~\eqref{eq:GellMann_identity_2}; the resulting counterterms are discussed in Secs.~\ref{sec:gamma_Pauli_structure_CTs}, \ref{sec:gamma_GellMann_structure_CTs}, \ref{sec:gamma_Pauli_GellMann_structure_CTs}.

It turns out that some of the different counterterm structures of Eqs.~\eqref{eq:T1}-\eqref{eq:T33f} appear in specific combinations

\begin{sidewaystable}
    \centering
    \begin{tabular}{|cc||c|c|c|c|c|c|c|c|c|c|}
        \hline
        $\mathcal{O}$ & & $ h^{\mathcal{O}}_{1} $ & $ h^{\mathcal{O}}_{2} $ & $ h^{\mathcal{O}}_{3} $ & $ h^{\mathcal{O}}_{4} $ & $ h^{\mathcal{O}}_{5} $ & $ h^{\mathcal{O}}_{6} $ & $ h^{\mathcal{O}}_{7} $ & $ h^{\mathcal{O}}_{8} $ & $ h^{\mathcal{O}}_{9} $ & $ h^{\mathcal{O}}_{10} $ \\
        \hline
        \hline
        $ Q_{\ell \ell} $ & $\delta_{fi}, \delta_{jk}$ & \multirow{4}{*}{$2 y^{\ell}_L$} & \multirow{4}{*}{$8 (y^{\ell}_L)^2$} & \multirow{4}{*}{$2$} & \multirow{4}{*}{$0$} & \multirow{4}{*}{$- 2 (y^{\ell}_L)^3$} & \multirow{4}{*}{$-\frac{1}{2} y^{\ell}_L$} & \multirow{4}{*}{$0$} & \multirow{4}{*}{$0$} & \multirow{4}{*}{$-\frac{1}{12}$} & \multirow{4}{*}{$0$} \\
        $ Q^{(1)}_{\ell q} $ & $\delta_{fi}$ & & & & & & & & & & \\
        $ Q_{\ell y} $ & $\delta_{fi}$ & & & & & & & & & & \\
        $ Q_{\ell x} $ & $\delta_{fi}$ & & & & & & & & & & \\
        \hline
        $ Q^{(1)}_{q q} $ & $\delta_{fi}, \delta_{jk}$ & \multirow{4}{*}{$2 N_c y^q_L$} & \multirow{4}{*}{$8 N_c (y^q_L)^2$} & \multirow{4}{*}{$2 N_c$} & \multirow{4}{*}{$4$} & \multirow{4}{*}{$- 2 N_c (y^q_L)^3$} & \multirow{4}{*}{$-\frac{N_c}{2} y^q_L$} & \multirow{4}{*}{$- y^q_L$} & \multirow{4}{*}{$-\frac{1}{2}$} & \multirow{4}{*}{$-\frac{N_c}{12}$} & \multirow{4}{*}{$-\frac{1}{6}$} \\
        $ Q^{(1)}_{\ell q} $ & $\delta_{jk}$ & & & & & & & & & & \\
        $ Q_{q y} $ & $\delta_{fi}$ & & & & & & & & & & \\
        $ Q^{(1)}_{q x} $ & $\delta_{fi}$ & & & & & & & & & & \\
        \hline
        $ Q_{x x} $ & $\delta_{fi}, \delta_{jk}$ & \multirow{5}{*}{$N_c y^x_R$} & \multirow{5}{*}{$- 4 N_c (y^x_R)^2$} & \multirow{5}{*}{$0$} & \multirow{5}{*}{$- 2$} & \multirow{5}{*}{$- N_c (y^x_R)^3$} & \multirow{5}{*}{$0$} & \multirow{5}{*}{$-\frac{1}{2} y^x_R$} & \multirow{5}{*}{$-\frac{1}{4}$} & \multirow{5}{*}{$0$} & \multirow{5}{*}{$\frac{1}{12}$} \\
        $ Q_{y x} $ & $\delta_{jk}$ & & & & & & & & & & \\
        $ Q^{(1)}_{u d} $ & $\delta_{fi}, \delta_{jk}$ & & & & & & & & & & \\
        $ Q_{\ell x} $ & $\delta_{jk}$ & & & & & & & & & & \\
        $ Q^{(1)}_{q x} $ & $\delta_{jk}$ & & & & & & & & & & \\
        \hline
        $ Q_{y y} $ & $\delta_{fi}, \delta_{jk}$ & \multirow{5}{*}{$y^y_R$} & \multirow{5}{*}{$- 4 (y^y_R)^2$} & \multirow{5}{*}{$0$} & \multirow{5}{*}{$0$} & \multirow{5}{*}{$- (y^y_R)^3$} & \multirow{5}{*}{$0$} & \multirow{5}{*}{$0$} & \multirow{5}{*}{$0$} & \multirow{5}{*}{$0$} & \multirow{5}{*}{$0$} \\
        $ Q_{y x} $ & $\delta_{fi}$ & & & & & & & & & & \\
        $ Q_{\nu e} $ & $\delta_{fi}, \delta_{jk}$ & & & & & & & & & & \\
        $ Q_{\ell y} $ & $\delta_{jk}$ & & & & & & & & & & \\
        $ Q_{q y} $ & $\delta_{jk}$ & & & & & & & & & & \\
        \hline
        \hline
        $ Q_{\ell \ell} $ & $\delta_{fk}, \delta_{ji}$ & $y^{\ell}_L$ & $4 (y^{\ell}_L)^2$ & $1$ & $0$ & $- (y^{\ell}_L)^3$ & $-\frac{1}{4} y^{\ell}_L$ & $0$ & $0$ & $-\frac{1}{24}$ & $0$ \\
        \hline
        $ Q^{(1)}_{q q} $ & $\delta_{fk}, \delta_{ji}$ & $y^q_L$ & $4 (y^q_L)^2$ & $1$ & $\frac{2}{N_c}$ & $- (y^q_L)^3$ & $-\frac{1}{4} y^q_L$ & $-\frac{1}{2 N_c} y^q_L$ & $-\frac{1}{4 N_c}$ & $-\frac{1}{24}$ & $-\frac{1}{12 N_c}$ \\
        \hline
        $ Q^{(3)}_{q q} $ & $\delta_{fk}, \delta_{ji}$ & $3 y^q_L$ & $12 (y^q_L)^2$ & $3$ & $\frac{6}{N_c}$ & $-3 (y^q_L)^3$ & $-\frac{3}{4} y^q_L$ & $-\frac{3}{2 N_c} y^q_L$ & $-\frac{3}{4 N_c}$ & $-\frac{1}{8}$ & $-\frac{1}{4 N_c}$ \\
        \hline
        $ Q_{x x} $ & $\delta_{fk}, \delta_{ji}$ & $y^x_R$ & $- 4 (y^x_R)^2$ & $0$ & $- \frac{2}{N_c}$ & $- (y^x_R)^3$ & $0$ & $-\frac{1}{2 N_c} y^x_R$ & $-\frac{1}{4 N_c}$ & $0$ & $\frac{1}{12 N_c}$ \\
        \hline
        $ Q_{y y} $ & $\delta_{fk}, \delta_{ji}$ & $y^y_R$ & $- 4 (y^y_R)^2$ & $0$ & $0$ & $- (y^y_R)^3$ & $0$ & $0$ & $0$ & $0$ & $0$ \\
        \hline
    \end{tabular}
    \caption[Coefficients in the finite counterterm Lagrangian which present a fermion bilinear of structure $ \bar{\xi} \bar{\gamma}_\mu \xi $.]{
    Coefficients in the finite counterterm Lagrangian which present a fermion bilinear of structure $ \bar{\xi} \bar{\gamma}_\mu \xi $.
    The fermion generation indices of the four-fermion operators of the Warsaw basis are $f i j k$, in this same ordering. The operators and the fermions contracted in the loop can be read from the first two columns.
    In the case of $ Q^{(1)}_{u d} $ ($ Q_{\nu e} $), since the indicated hypercharge $y^x_R$ ($y^y_R$, respectively) is the one of the fermion running inside the loop, $x = u$ for $\delta_{fi}$ and $x = d$ for $\delta_{jk}$ ($y = \nu$ for $\delta_{fi}$ and $y = e$ for $\delta_{jk}$, respectively).
    }
    \label{tab:gamma_gamma}
\end{sidewaystable}

\begin{eqnarray}
    && y_L^q \mathcal{Q}_1 - (y_L^q)^3 \mathcal{Q}_5 \;\; \text{or} \;\; y_L^\ell \mathcal{Q}_1 - (y_L^\ell)^3 \mathcal{Q}_5 \nonumber\\
    && \qquad \text{or} \;\; y_R^x \mathcal{Q}_1 - (y_R^x)^3 \mathcal{Q}_5 \;\; \text{or} \;\; y_R^y \mathcal{Q}_1 - (y_R^y)^3 \mathcal{Q}_5, \\
    && \mathcal{Q}_3 - \frac{1}{4} y_L^q \mathcal{Q}_6 - \frac{1}{24} \mathcal{Q}_9 \;\; \text{or} \;\; \mathcal{Q}_3 - \frac{1}{4} y_L^\ell \mathcal{Q}_6 - \frac{1}{24} \mathcal{Q}_9, \\
    && \mathcal{Q}_4 - \frac{1}{24} \mathcal{Q}_{10}, \\
    && \mathcal{Q}_8 + 2 y_L^q \mathcal{Q}_7 \;\; \text{or} \;\; \mathcal{Q}_8 + 2 y_L^\ell \mathcal{Q}_7.
\end{eqnarray}

To provide a concrete example of a particular structure in the finite counterterm Lagrangian, one finds:

\begin{eqnarray}
    && \mathcal{L}_{\rm fct} \supset \frac{1}{3 \cdot 16 \pi^2} \left( g_3^2 G^B_\nu \bar{\partial}_\rho G^B_\sigma + g_3^3 \frac{1}{3} C_{B E F} G^B_\nu G^E_\rho G^F_\sigma \right) \epsilon^{\mu \nu \rho \sigma} \\
    && \left[ C^{q q (1)}_{f i j k} 4 \delta_{fi} \bar{q}_j \bar{\gamma}_\mu q_k + C^{q q (1)}_{f i j k} 4 \delta_{jk} \bar{q}_f \bar{\gamma}_\mu q_i + C^{\ell q (1)}_{f i j k} 4 \delta_{jk} \bar{\ell}_f \bar{\gamma}_\mu \ell_i \right. \nonumber\\
    && + C^{q \nu}_{f i j k} 4 \delta_{fi} \bar{\nu}_j \bar{\gamma}_\mu \nu_k + C^{q e}_{f i j k} 4 \delta_{fi} \bar{e}_j \bar{\gamma}_\mu e_k \nonumber\\
    && + C^{q u (1)}_{f i j k} 4 \delta_{fi} \bar{u}_j \bar{\gamma}_\mu u_k + C^{q d (1)}_{f i j k} 4 \delta_{fi} \bar{d}_j \bar{\gamma}_\mu d_k \nonumber\\
    && - C^{u u}_{f i j k} 2 \delta_{fi} \bar{u}_j \bar{\gamma}_\mu u_k - C^{d d}_{f i j k} 2 \delta_{fi} \bar{d}_j \bar{\gamma}_\mu d_k \nonumber\\
    && - C^{u u}_{f i j k} 2 \delta_{jk} \bar{u}_f \bar{\gamma}_\mu u_i - C^{d d}_{f i j k} 2 \delta_{jk} \bar{d}_f \bar{\gamma}_\mu d_i \nonumber\\
    && - C^{\nu u}_{f i j k} 2 \delta_{jk} \bar{\nu}_f \bar{\gamma}_\mu \nu_i - C^{e u}_{f i j k} 2 \delta_{jk} \bar{e}_f \bar{\gamma}_\mu e_i \nonumber\\
    && - C^{\nu d}_{f i j k} 2 \delta_{jk} \bar{\nu}_f \bar{\gamma}_\mu \nu_i - C^{e d}_{f i j k} 2 \delta_{jk} \bar{e}_f \bar{\gamma}_\mu e_i \nonumber\\
    && - C^{u d (1)}_{f i j k} 2 \delta_{fi} \bar{d}_j \bar{\gamma}_\mu d_k - C^{u d (1)}_{f i j k} 2 \delta_{jk} \bar{u}_f \bar{\gamma}_\mu u_i \nonumber\\
    && - C^{\ell u}_{f i j k} 2 \delta_{jk} \bar{\ell}_f \bar{\gamma}_\mu \ell_i - C^{\ell d}_{f i j k} 2 \delta_{jk} \bar{\ell}_f \bar{\gamma}_\mu \ell_i \nonumber\\
    && - C^{q u (1)}_{f i j k} 2 \delta_{jk} \bar{q}_f \bar{\gamma}_\mu q_i - C^{q d (1)}_{f i j k} 2 \delta_{jk} \bar{q}_f \bar{\gamma}_\mu q_i \nonumber\\
    && + C^{q q (1)}_{f i j k} \frac{2}{N_c} \delta_{fk} \bar{q}_j \bar{\gamma}_\mu q_i + C^{q q (1)}_{f i j k} \frac{2}{N_c} \delta_{ji} \bar{q}_f \bar{\gamma}_\mu q_k \nonumber\\
    && + C^{q q (3)}_{f i j k} \frac{6}{N_c} \delta_{fk} \bar{q}_j \bar{\gamma}_\mu q_i + C^{q q (3)}_{f i j k} \frac{6}{N_c} \delta_{ji} \bar{q}_f \bar{\gamma}_\mu q_k \nonumber\\
    && - C^{u u}_{f i j k} \frac{2}{N_c} \delta_{fk} \bar{u}_j \bar{\gamma}_\mu u_i - C^{d d}_{f i j k} \frac{2}{N_c} \delta_{fk} \bar{d}_j \bar{\gamma}_\mu d_i \nonumber\\
    && \left. - C^{u u}_{f i j k} \frac{2}{N_c} \delta_{ji} \bar{u}_f \bar{\gamma}_\mu u_k - C^{d d}_{f i j k} \frac{2}{N_c} \delta_{ji} \bar{d}_f \bar{\gamma}_\mu d_k \right] + {\rm h.c.} \nonumber
\end{eqnarray}
No sum over generation indices $f, i, j, k$ is to be taken.
As this example illustrates, there are, for instance, lepton-gluonic finite counterterm structures, which are not possible when considering dimension-6 gauge-invariant operators.

Some generation assignments are identical: $ Q^{}_{e e; f i j k} = Q^{}_{e e; j k f i} $, to give an example. This symmetry is not employed to abbreviate the results,
i.e., Tab.~\ref{tab:gamma_gamma} includes $ \delta_{f i} $ and $ \delta_{j k} $ in such cases.
Eliminating redundancies leads to symmetry factors that can be absorbed into the Wilson coefficients.

\subsubsection{$\gamma_\mu \tau^I$-structure counterterms}\label{sec:gamma_Pauli_structure_CTs}

Consider now those structures that have the fermion bilinear $ \bar{\xi} \bar{\gamma}_\mu T_L^I \xi $, where $T_L^I$ is a generator of ${\rm SU(2)}_{\rm L}$; it is also convenient at times not to simplify structures with three $\gamma$-matrices (e.g., by employing the Chisholm identity). Non-vanishing obstructions are only possible when at least one ${\rm SU}(2)_{\rm L}$ gauge boson is radiated from the fermion loop. The following structure with a single gauge boson is found

\begin{table}[]
    \centering
    \begin{tabular}{|cc||c|c|c|c|}
    \hline
    $ \mathcal{O} $ & 
    & $h^{\mathcal{O}}_{11}$ 
    & $h^{\mathcal{O}}_{12}$ 
    & $h^{\mathcal{O}}_{13}$ 
    & $h^{\mathcal{O}}_{14}$ \\
    \hline
    \hline
    $ Q^{(3)}_{q q} $ 
    & $ \delta_{fi}, \delta_{jk} $ 
    & \multirow{2}{*}{$ 4 N_c $} 
    & \multirow{2}{*}{$ - 4 N_c $} 
    & \multirow{2}{*}{$ -2 N_c (y_L^q)^2 $} 
    & \multirow{2}{*}{$ -1 $} \\
    $ Q^{(3)}_{\ell q} $ 
    & $ \delta_{jk} $ 
    &  &  &  &  \\
    \hline
    $ Q^{(3)}_{\ell q} $ 
    & $ \delta_{fi} $ 
    & $ 4 $ 
    & $ -4 $ 
    & $ -2 (y_L^\ell)^2 $ 
    & $ 0 $ \\
    \hline
    \hline
    $ Q^{(1)}_{q q} $ 
    & $ \delta_{fk}, \delta_{ji} $ 
    & $ 2 $ 
    & $ -2 $ 
    & $ - (y_L^q)^2 $ 
    & $ - \frac{1}{2 N_c} $ \\
    \hline
    $ Q^{(3)}_{q q} $ 
    & $ \delta_{fk}, \delta_{ji} $ 
    & $ -2 $ 
    & $ 2 $ 
    & $ (y_L^q)^2 $ 
    & $ \frac{1}{2 N_c} $ \\
    \hline
    $ Q_{\ell \ell} $ 
    & $ \delta_{fk}, \delta_{ji} $ 
    & $ 2 $ 
    & $ -2 $ 
    & $ - (y_L^\ell)^2 $ 
    & $ 0 $ \\
    \hline
\end{tabular}
    \caption[Coefficients multiplying the counterterms in the finite counterterm Lagrangian which present a fermion bilinear of structure $ \bar{\xi} \bar{\gamma}_\mu T_L^I \xi $.]{
    Coefficients multiplying the counterterms in the finite counterterm Lagrangian which present a fermion bilinear of structure $ \bar{\xi} \bar{\gamma}_\mu T_L^I \xi $.
    The fermion generation indices of the four-fermion operators of the Warsaw basis are $f i j k$, in this same order. The operators and the fermions contracted in the loop can be read from the first two columns.
    }
    \label{tab:gammaTL_gammaTL}
\end{table}

\begin{eqnarray}
    \mathcal{Q}_{11} = - g_2 \frac{1}{2} \bar{\partial}^2 W^I_\mu ( \bar{\xi} \bar{\gamma}^\mu T^I_L \xi ) + g_2 \bar{\partial}_\mu \bar{\partial}^\nu W^I_\nu ( \bar{\xi} \bar{\gamma}^\mu T^I_L \xi ); \label{eq:Ftilde21}
\end{eqnarray}
There are also the following cases with three gauge bosons

\begin{eqnarray}
    \mathcal{Q}_{12} = g_2^3 {\rm Tr} \{ T^I_L T^J_L T^K_L T^L_L \} W^J_\alpha W^K_\mu W^L_\beta ( \bar{\xi} \bar{\gamma}^\alpha \bar{\gamma}^\mu \bar{\gamma}^\beta T^I_L \xi ), \label{eq:Ftilde13} 
\end{eqnarray}

\begin{eqnarray}
    && \mathcal{Q}_{13} = g_1^2 g_2 B_\alpha B_\mu W_\beta^I \bar{\xi} ( \bar{\gamma}^\beta \bar{\gamma}^\alpha \bar{\gamma}^\mu + \bar{\gamma}^\alpha \bar{\gamma}^\beta \bar{\gamma}^\mu + \bar{\gamma}^\alpha \bar{\gamma}^\mu \bar{\gamma}^\beta ) T^I_L \xi, \label{eq:Ftilde1AG}\\ 
    && \mathcal{Q}_{14} = g_2 g_3^2 G^B_\alpha G^B_\mu W_\beta^I \bar{\xi} ( \bar{\gamma}^\beta \bar{\gamma}^\alpha \bar{\gamma}^\mu + \bar{\gamma}^\alpha \bar{\gamma}^\beta \bar{\gamma}^\mu + \bar{\gamma}^\alpha \bar{\gamma}^\mu \bar{\gamma}^\beta ) T^I_L \xi. \label{eq:Ftilde2AG} 
\end{eqnarray}

In terms of this basis of structures, the finite counterterm Lagrangian is

\begin{equation}
    \mathcal{L}_{\rm fct} \supset - \frac{C^{\mathcal{O}}_{\xi \xi}}{3 \cdot 16 \pi^2} \sum^{14}_{i=11} h^{\mathcal{O}}_{i} \, \mathcal{Q}_i + {\rm h.c.}
\end{equation}
Coefficients are given in Tab.~\ref{tab:gammaTL_gammaTL}.
The structure of this table is as follows: the first and second sectors of the first half correspond to $q$ and $\ell$ loops, respectively, when contracting the fermion fields that appear within the same bilinear, following the convention of Tab.~\ref{tab:dimension_6}; the second half of the table corresponds to contracting fermions in different bilinears, again when following the Warsaw basis.
As seen from Tab.~\ref{tab:gammaTL_gammaTL}, operators other than $ Q^{(3)}_{q q}, Q^{(3)}_{\ell q} $ from the Warsaw basis also require finite renormalization by the structures shown above (${\rm SU}(2)_{\rm L}$-analogous identities of Eqs.~\eqref{eq:GellMann_identity_1} and \eqref{eq:GellMann_identity_2}).

Some of the different counterterm structures of Eqs.~\eqref{eq:Ftilde21}-\eqref{eq:Ftilde2AG} appear in specific combinations

\begin{equation}
    \mathcal{Q}_{11} - \mathcal{Q}_{12} - \frac{1}{2} (y_L^q)^2 \mathcal{Q}_{13} \;\; \text{or} \;\; \mathcal{Q}_{11} - \mathcal{Q}_{12} - \frac{1}{2} (y_L^\ell)^2 \mathcal{Q}_{13}.
\end{equation}

To provide a concrete example of a term present in the finite renormalization Lagrangian, one finds

\begin{table}[]
    \centering
    \begin{tabular}{|cc||c|c|c|c|c|c|c|c|}
    \hline
    $ \mathcal{O} $ & 
    & $h^{\mathcal{O}}_{15}$ 
    & $h^{\mathcal{O}}_{16}$ 
    & $h^{\mathcal{O}}_{17}$ 
    & $h^{\mathcal{O}}_{18}$ 
    & $h^{\mathcal{O}}_{19}$ 
    & $h^{\mathcal{O}}_{20}$ 
    & $h^{\mathcal{O}}_{21}$ 
    & $h^{\mathcal{O}}_{22}$ \\
    \hline
    \hline
    $Q_{qx}^{(8)}$ & $\delta_{fi}$ 
    & $2$ & $4$ & $-2$ & $\frac{1}{3}$ & $-2$ 
    & $-\frac{1}{2}$ & $- (y_L^q)^2$ & $-\frac{1}{4}$ \\
    \hline
    $Q_{qx}^{(8)}$ & $\delta_{jk}$ 
    & \multirow{2}{*}{$1$} 
    & \multirow{2}{*}{$-2$} 
    & \multirow{2}{*}{$-1$} 
    & \multirow{2}{*}{$-\frac{1}{6}$} 
    & \multirow{2}{*}{$-1$} 
    & \multirow{2}{*}{$\frac{1}{4}$} 
    & \multirow{2}{*}{$-\frac{1}{2} (y_R^x)^2$} 
    & \multirow{2}{*}{$0$} \\
    $Q_{ud}^{(8)}$ & $\delta_{fi}, \delta_{jk}$ 
    & & & & & & & & \\
    \hline
    \hline
    $Q_{qq}^{(1)}$ & $\delta_{fk}, \delta_{ji}$ 
    & $2$ & $4$ & $-2$ & $\frac{1}{3}$ & $-2$ 
    & $-\frac{1}{2}$ & $- (y_L^q)^2$ & $-\frac{1}{4}$ \\
    \hline
    $Q_{qq}^{(3)}$ & $\delta_{fk}, \delta_{ji}$ 
    & $6$ & $12$ & $-6$ & $1$ & $-6$ 
    & $-\frac{3}{2}$ & $-3 (y_L^q)^2$ & $-\frac{3}{4}$ \\
    \hline
    $Q_{xx}$ & $\delta_{fk}, \delta_{ji}$ 
    & $2$ & $-4$ & $-2$ & $-\frac{1}{3}$ & $-2$ 
    & $\frac{1}{2}$ & $- (y_R^x)^2$ & $0$ \\
    \hline
    \end{tabular}
    \caption[Coefficients of the finite counterterm Lagrangian for fermion bilinears of structure $ \bar{\xi} \bar{\gamma}_\mu T^A \xi $.]{
    Coefficients of the finite counterterm Lagrangian for fermion bilinears of structure $ \bar{\xi} \bar{\gamma}_\mu T^A \xi $.
    The fermion generation indices of the four-fermion operators of the Warsaw basis are $f i j k$, in this same order. The operators and the fermions contracted in the loop can be read from the first two columns.
    In the case of $ Q^{(8)}_{u d} $, $x = u$ for $\delta_{fi}$ and $x = d$ for $\delta_{jk}$.
    }
    \label{tab:gammaT_gammaT}
\end{table}

\begin{eqnarray}\label{eq:Lfct_gammamutauI}
    \mathcal{L}_{\rm fct} \supset && g_2 g_3^2 \frac{1}{6 \cdot 16 \pi^2} G^B_\alpha G^B_\mu W_\beta^I \\
    && \left[ C^{q q (3)}_{f i j k} \delta_{fi} \bar{q}_j S^{\alpha \mu \beta} T^I_L q_k + C^{q q (3)}_{f i j k} \delta_{jk} \bar{q}_f S^{\alpha \mu \beta} T^I_L q_i + C^{\ell q (3)}_{f i j k} \delta_{jk} \bar{\ell}_f S^{\alpha \mu \beta} T^I_L \ell_i \right. \nonumber\\
    && + C^{q q (1)}_{f i j k} \frac{1}{2 N_c} \delta_{fk} \bar{q}_j S^{\alpha \mu \beta} T^I_L q_i + C^{q q (1)}_{f i j k} \frac{1}{2 N_c} \delta_{ji} \bar{q}_f S^{\alpha \mu \beta} T^I_L q_k \nonumber\\
    && \left. - C^{q q (3)}_{f i j k} \frac{1}{2 N_c} \delta_{fk} \bar{q}_j S^{\alpha \mu \beta} T^I_L q_i - C^{q q (3)}_{f i j k} \frac{1}{2 N_c} \delta_{ji} \bar{q}_f S^{\alpha \mu \beta} T^I_L q_k \right] + {\rm h.c.} \,, \nonumber
\end{eqnarray}
where $ S^{\alpha \mu \beta} = \bar{\gamma}^\alpha (\bar{\gamma}^\mu \bar{\gamma}^\beta + \bar{\gamma}^\beta \bar{\gamma}^\mu) + (\bar{\gamma}^\mu \bar{\gamma}^\alpha \bar{\gamma}^\beta + \bar{\gamma}^\beta \bar{\gamma}^\alpha \bar{\gamma}^\mu) + (\bar{\gamma}^\mu \bar{\gamma}^\beta + \bar{\gamma}^\beta \bar{\gamma}^\mu) \bar{\gamma}^\alpha $.
In this example, also a lepton-gluonic counterterm appears.

\subsubsection{$\gamma_\mu T^A$-structure counterterms}\label{sec:gamma_GellMann_structure_CTs}

Analogously to Sec.~\ref{sec:gamma_Pauli_structure_CTs}, now fermion bilinears of structure $\bar{\xi} \bar{\gamma}^\mu T^A \xi$ are discussed (while cases with three $\gamma$-matrices are not simplified for convenience). Obstructions are only possible when at least one ${\rm SU}(3)$ gauge boson is radiated.
The following operator with a single gauge boson is defined
\begin{eqnarray}
    \mathcal{Q}_{15} = - g_3 \frac{1}{2} \bar{\partial}^2 G^A_\mu ( \bar{\xi} \bar{\gamma}^\mu T^A \xi )
    + g_3 \bar{\partial}_\mu \bar{\partial}^\nu G^A_\nu ( \bar{\xi} \bar{\gamma}^\mu T^A \xi ), \label{eq:F21}
\end{eqnarray}
followed by an operator with two gauge bosons

\begin{eqnarray}
    \mathcal{Q}_{16} = \frac{1}{2} g_3^2 D_{A B C} \, [\bar{\xi} \bar{\gamma}_\mu T^A \xi] G^C_\sigma \bar{\partial}_\rho G^B_\nu \epsilon^{\mu \nu \rho \sigma}; \label{eq:F42}
\end{eqnarray}
there are many more structures with three gauge bosons

\begin{eqnarray}
    \mathcal{Q}_{17} = g_3^3 {\rm Tr} \{ T^A T^B T^C T^D \} G^B_\alpha G^C_\mu G^D_\beta ( \bar{\xi} \bar{\gamma}^\alpha \bar{\gamma}^\mu \bar{\gamma}^\beta T^A \xi ), \label{eq:F13} 
\end{eqnarray}

\begin{equation}
    \mathcal{Q}_{18} = i g_3^3 ( \bar{\xi} \bar{\gamma}_\mu T^A \xi ) G^B_\nu G^C_\rho G^D_\sigma \epsilon^{\mu \nu \rho \sigma} \mathcal{A}_{A}^{B C D}, \label{eq:F23}
\end{equation}

\begin{eqnarray}
    \hspace{-0.2cm}\mathcal{Q}_{19} = - g_3^3 \frac{i}{8} ( \bar{\xi} \bar{\gamma}^\alpha \bar{\gamma}^\mu \bar{\gamma}^\beta T^A \xi ) \left( G^B_\alpha G^C_\mu G^D_\beta + G^C_\alpha G^B_\mu G^D_\beta + G^C_\alpha G^D_\mu G^B_\beta \right) C_{C D E} D_{A B E}, 
    \label{eq:F13DC}
\end{eqnarray}

\begin{eqnarray}
    \mathcal{Q}_{20} = g_3^3 ( \bar{\xi} \bar{\gamma}_\mu T^A \xi ) G^B_\nu G^C_\rho G^D_\sigma \epsilon^{\mu \nu \rho \sigma} D_{A D E} C_{B C E},\label{eq:F33}
\end{eqnarray}

\begin{eqnarray}
    && \mathcal{Q}_{21} = g_1^2 g_3 B_\alpha B_\mu G_\beta^C \bar{\xi} ( \bar{\gamma}^\beta \bar{\gamma}^\alpha \bar{\gamma}^\mu + \bar{\gamma}^\alpha \bar{\gamma}^\beta \bar{\gamma}^\mu + \bar{\gamma}^\alpha \bar{\gamma}^\mu \bar{\gamma}^\beta ) T^C \xi, \label{eq:F1AG}\\
    && \mathcal{Q}_{22} = g_2^3 g_3 W^b_\alpha W^b_\mu G_\beta^C \bar{\xi} ( \bar{\gamma}^\beta \bar{\gamma}^\alpha \bar{\gamma}^\mu + \bar{\gamma}^\alpha \bar{\gamma}^\beta \bar{\gamma}^\mu + \bar{\gamma}^\alpha \bar{\gamma}^\mu \bar{\gamma}^\beta ) T^C \xi. \label{eq:F2AG}
\end{eqnarray}
The operator $ \mathcal{Q}_{19} $ can be simplified using the Chisholm identity, and only the term proportional to $i$ times the Levi-Civita symbol remains, with an overall sign that depends on the chirality of $\xi$; in this way, it can be seen that this operator self-conjugates when the omitted generation indices of the fermion bilinear are the same. Using the same identity, the operator $\mathcal{Q}_{17}$ (and the analogous operator $\mathcal{Q}_{12}$ for ${\rm SU}(2)_{\rm L}$) can also be shown to self-conjugate when the two fermions carry the same flavor (from the identity in Eq.~\eqref{eq:identity_Haber}).

The finite counterterm Lagrangian is

\begin{equation}
    \mathcal{L}_{\rm fct} \supset - \frac{C^{\mathcal{O}}_{\xi \xi}}{3 \cdot 16 \pi^2} \sum^{22}_{i=15} h^{\mathcal{O}}_{i} \, \mathcal{Q}_i + {\rm h.c.}
\end{equation}

\noindent
Coefficients are given in Tab.~\ref{tab:gammaT_gammaT}.
In this table, the first and second sectors of the first half correspond to $q$ and $x = u, d$ loops, respectively, when contracting the fermion fields that appear within the same bilinear, following the convention of Tab.~\ref{tab:dimension_6}; the second half of the table corresponds to contracting fermions in different bilinears, again following the convention of Tab.~\ref{tab:dimension_6}.
Due to the identities Eqs.~\eqref{eq:GellMann_identity_1} and \eqref{eq:GellMann_identity_2},
operators other than $ Q^{(8)}_{q x}, Q^{(8)}_{u d} $ from the Warsaw basis also require finite renormalization by $ \bar{\gamma}_\mu T^A $ structures.
Some of the counterterms previously defined in Eqs.~\eqref{eq:F21}-\eqref{eq:F2AG} appear in specific combinations in the renormalization

\begin{eqnarray}
    && \mathcal{Q}_{16} + \frac{1}{12} \mathcal{Q}_{18} - \frac{1}{8} \mathcal{Q}_{20}, \\
    && \mathcal{Q}_{15} - \mathcal{Q}_{17} - \mathcal{Q}_{19} - \frac{1}{2} (y_L^q)^2 \mathcal{Q}_{21} \,, \; \text{or} \;\; \mathcal{Q}_{15} - \mathcal{Q}_{17} - \mathcal{Q}_{19} - \frac{1}{2} (y_R^x)^2 \mathcal{Q}_{21}. \nonumber\\
\end{eqnarray}

The following term provides a concrete example:

\begin{eqnarray}
    \mathcal{L}_{\rm fct} \supset && \frac{1}{3 \cdot 16 \pi^2} \left[ g_3^2 D_{A B C} G^B_\nu \bar{\partial}_\rho G^C_\sigma + g_3^3 G^B_\nu G^C_\rho G^D_\sigma \left( \frac{1}{4} D_{A D E} C_{B C E} - \frac{1}{6} i \mathcal{A}_{A}^{B C D} \right) \right] \nonumber\\
    && \epsilon^{\mu \nu \rho \sigma} \left[ C_{f i j k}^{q u (8)} \delta_{fi} \bar{u}_j \bar{\gamma}_\mu T^A u_k + C_{f i j k}^{q d (8)} \delta_{fi} \bar{d}_j \bar{\gamma}_\mu T^A d_k \right. \nonumber\\
    && - C_{f i j k}^{q u (8)} \delta_{jk} \bar{q}_f \bar{\gamma}_\mu T^A q_i - C_{f i j k}^{q d (8)} \delta_{jk} \bar{q}_f \bar{\gamma}_\mu T^A q_i \nonumber\\
    && - C_{f i j k}^{u d (8)} \delta_{fi} \bar{d}_j \bar{\gamma}_\mu T^A d_k - C_{f i j k}^{u d (8)} \delta_{jk} \bar{u}_f \bar{\gamma}_\mu T^A u_i \nonumber\\
    && + C_{f i j k}^{q q (1)} \delta_{fk} \bar{q}_j \bar{\gamma}_\mu T^A q_i + C_{f i j k}^{q q (1)} \delta_{ji} \bar{q}_f \bar{\gamma}_\mu T^A q_k \nonumber\\
    && + C_{f i j k}^{q q (3)} 3 \delta_{fk} \bar{q}_j \bar{\gamma}_\mu T^A q_i + C_{f i j k}^{q q (3)} 3 \delta_{ji} \bar{q}_f \bar{\gamma}_\mu T^A q_k \nonumber\\
    && - C^{uu}_{f i j k} 2 \delta_{fk} \bar{u}_j \bar{\gamma}_\mu T^A u_i - C^{dd}_{f i j k} 2 \delta_{fk} \bar{d}_j \bar{\gamma}_\mu T^A d_i \nonumber\\
    && \left. - C^{uu}_{f i j k} 2 \delta_{ji} \bar{u}_f \bar{\gamma}_\mu T^A u_k - C^{dd}_{f i j k} 2 \delta_{ji} \bar{d}_f \bar{\gamma}_\mu T^A d_k \right] + {\rm h.c.}
\end{eqnarray}

\begin{table}[]
    \centering
    \begin{tabular}{|cc||c|c|c|c|c|c|}
        \hline
        $ \mathcal{O} $ & & $h^{\mathcal{O}}_{23}$ & $h^{\mathcal{O}}_{24}$ & $h^{\mathcal{O}}_{25}$ & $h^{\mathcal{O}}_{26}$ & $h^{\mathcal{O}}_{27}$ & $h^{\mathcal{O}}_{28}$ \\
        \hline
        \hline
        $ Q^{(1)}_{qq} $ & $ \delta_{f k}, \delta_{j i} $ & $-y_L^q$ & $-1$ & $4$ & $\frac{3}{2}$ & $4$ & $\frac{3}{2}$ \\
        \hline
        $ Q^{(3)}_{qq} $ & $ \delta_{f k}, \delta_{j i} $ & $y_L^q$ & $1$ & $-4$ & $-\frac{3}{2}$ & $-4$ & $-\frac{3}{2}$ \\
        \hline
    \end{tabular}
    \caption[Coefficients of the finite counterterm Lagrangian for fermion bilinears of generic structure $ \bar{\xi} \bar{\gamma}_\mu T_L^I T^A \xi $.]{
    Coefficients of the finite counterterm Lagrangian for fermion bilinears of generic structure $ \bar{\xi} \bar{\gamma}_\mu T_L^I T^A \xi $.
    The fermion generation indices of the four-fermion operators of the Warsaw basis are $f i j k$, in this same order. The operators and the fermions contracted in the loop can be read from the first two columns.
    }
    \label{tab:gammaTTL_gammaTTL}
\end{table}

\subsubsection{$\gamma_\mu \tau^I T^A$-structure counterterms}\label{sec:gamma_Pauli_GellMann_structure_CTs}

Finally, there are special structures of the form $\bar{\gamma}_\mu \tau^I T^A$, which thus require at least one ${\rm SU}(2)_{\rm L}$, and at least one ${\rm SU}(3)$ external gauge boson. The required counterterms, having two or one gauge bosons, are

\begin{eqnarray}
    && \hspace{-0.5cm} \mathcal{Q}_{23} = g_1 g_2 g_3 W^a_\alpha G^B_\mu B_\beta \bar{\xi} S^{\beta \alpha \mu} T^B T^a_L \xi, \label{eq:TS111}\\
    && \hspace{-0.5cm}\mathcal{Q}_{24} = g_2 g_3^2 {\rm Tr} \{ T^A (T^B T^C + T^C T^B) \} {\rm Tr} \{ T^a_L T^b_L \} \, G^B_\nu \, W_\rho^b \, G^C_\sigma ( \bar{\xi} S^{\sigma \nu \rho} T^A T^a_L \xi ), \label{eq:TS12}\\
    && \hspace{-0.5cm} \mathcal{Q}_{25} = g_2 g_3 ( \bar{\xi} \bar{\gamma}_\mu T^A T^a_L \xi ) \bar{\partial}_\rho G^A_\nu W^a_\sigma \epsilon^{\mu \nu \rho \sigma}, \label{eq:T1eps11}\\ 
    && \hspace{-0.5cm} \mathcal{Q}_{26} = g_2 g_3^2 C_{A B C} ( \bar{\xi} \bar{\gamma}_\mu T^A T^a_L \xi ) W^a_\nu G^C_\rho G^B_\sigma \epsilon^{\mu \nu \rho \sigma}, \label{eq:T1eps12}\\ 
    && \hspace{-0.5cm} \mathcal{Q}_{27} = g_2 g_3 ( \bar{\xi} \bar{\gamma}_\mu T^A T^a_L \xi ) \bar{\partial}_\rho W^a_\nu G^A_\sigma \epsilon^{\mu \nu \rho \sigma}, \label{eq:T2eps11}\\ 
    && \hspace{-0.5cm} \mathcal{Q}_{28} = g_2^2 g_3 C^{SU(2)}_{a b c} ( \bar{\xi} \bar{\gamma}_\mu T^A T^a_L \xi ) G^A_\nu W^c_\rho W^b_\sigma \epsilon^{\mu \nu \rho \sigma}. \label{eq:T2eps21}
\end{eqnarray}
where $S$ is defined after Eq.~\eqref{eq:Lfct_gammamutauI}.

In terms of these operators, the finite counterterm Lagrangian is

\begin{equation}
    \mathcal{L}_{\rm fct} \supset - \frac{C^{\mathcal{O}}_{\xi \xi}}{3 \cdot 16 \pi^2} \sum^{28}_{i=23} h^{\mathcal{O}}_{i} \, \mathcal{Q}_i + {\rm h.c.},
\end{equation}
where the hermitian conjugation, as in all previous cases, applies only in cases where the fermion fields do not carry the same flavor. Coefficients are given in Tab.~\ref{tab:gammaTTL_gammaTTL}. It turns out that all the structures previously defined appear in the same combination

\begin{equation}
    y_L^q \mathcal{Q}_{23}
+ \mathcal{Q}_{24}
- 4 \left( \mathcal{Q}_{25} + \mathcal{Q}_{27} \right)
- \frac{3}{2} \left( \mathcal{Q}_{26} + \mathcal{Q}_{28} \right).
\end{equation}

To better explain the notation in use, consider the following example

\begin{eqnarray}
    \mathcal{L}_{\rm fct} \supset && y^q_L \frac{g_1 g_2 g_3}{3 \cdot 16 \pi^2} W^a_\alpha G^B_\mu B_\beta
    \left[ C^{qq (1)}_{f i j k} \delta_{f k} \bar{q}_j S^{\alpha \mu \beta} T^B T^a_L q_i + C^{qq (1)}_{f i j k} \delta_{j i} \bar{q}_f S^{\alpha \mu \beta} T^B T^a_L q_k \right. \nonumber\\
    && \left. - C^{qq (3)}_{f i j k} \delta_{f k} \bar{q}_j S^{\alpha \mu \beta} T^B T^a_L q_i - C^{qq (3)}_{f i j k} \delta_{j i} \bar{q}_f S^{\alpha \mu \beta} T^B T^a_L q_k \right] + {\rm h.c.},
\end{eqnarray}
which has three different gauge bosons.

\subsection{Counterterms with one scalar}\label{sec:one_scalar_results}

Consider now the finite counterterms with one scalar needed to restore the gauge symmetry. The structures are the following:

\begin{equation}\label{eq:CT_Th}
    T_h = Y^\dagger \bar{\psi} P_X \xi \bar{\partial}^2 h,
\end{equation}

\begin{eqnarray}
    && T_{h1} = g_1 Y^\dagger \bar{\psi} \bar\sigma^{\mu \nu} P_X \xi B_\mu \bar{\partial}_\nu h, \label{eq:Th1}\\
    && T_{h2} = g_2 Y^\dagger \bar{\psi} T^a_L \bar\sigma^{\mu \nu} P_X \xi W^a_\mu \bar{\partial}_\nu h, \label{eq:Th2}\\
    && T_{h3} = g_3 Y^\dagger \bar{\psi} T^\alpha \bar\sigma^{\mu \nu} P_X \xi G^\alpha_\mu \bar{\partial}_\nu h, \label{eq:Th3}
\end{eqnarray}

\begin{eqnarray}
    && T^{11}_g = g_1^2 Y^\dagger (\bar{\psi} P_X \xi h) \bar{\eta}^{\mu \nu} B_\mu B_\nu, \label{eq:Tg11}\\
    && T^{22}_g = g_2^2 Y^\dagger (\bar{\psi} \mathcal{G}^{a b} P_X \xi h) \bar{\eta}^{\mu \nu} W^a_\mu W^b_\nu, \label{eq:Tg22}\\
    && T^{33}_g = g_3^2 Y^\dagger (\bar{\psi} \delta^{\alpha \beta} P_X \xi h) \bar{\eta}^{\mu \nu} G^\alpha_\mu G^\beta_\nu, \label{eq:Tg33}\\
    && \tilde{T}^{33}_g = g_3^2 Y^\dagger (\bar{\psi} \mathcal{G}^{\alpha \beta} P_X \xi h) \bar{\eta}^{\mu \nu} G^\alpha_\mu G^\beta_\nu, \label{eq:Ttildeg33}\\
    && T^{12}_g = g_1 g_2 Y^\dagger (\bar{\psi} T^a_L P_X \xi h) \bar{\eta}^{\mu \nu} B_\mu W^a_\nu, \label{eq:Tg12}\\
    && T^{13}_g = g_1 g_3 Y^\dagger (\bar{\psi} T^\alpha P_X \xi h) \bar{\eta}^{\mu \nu} B_\mu G^\alpha_\nu, \label{eq:Tg13}\\
    && T^{23}_g = g_2 g_3 Y^\dagger (\bar{\psi} T^a_L T^\alpha P_X \xi h) \bar{\eta}^{\mu \nu} W^a_\mu G^\alpha_\nu, \label{eq:Tg23}
\end{eqnarray}

\begin{eqnarray}
    && T^{22}_\sigma = i g_2^2 Y^\dagger (\bar{\psi} T^a_L T^b_L \bar\sigma^{\mu \nu} P_X \xi h) W^a_\mu W^b_\nu, \label{eq:T22sigma}\\
    && \tilde{T}^{33}_\sigma = i g_3^2 Y^\dagger (\bar{\psi} T^\alpha T^\beta \bar\sigma^{\mu \nu} P_X \xi h) G^\alpha_\mu G^\beta_\nu, \label{eq:Ttilde33sigma}\\
    && T^{12}_\sigma = i g_1 g_2 Y^\dagger (\bar{\psi} T^a_L \bar\sigma^{\mu \nu} P_X \xi h) B_\mu W^a_\nu, \label{eq:T12sigma}\\
    && T^{13}_\sigma = i g_1 g_3 Y^\dagger (\bar{\psi} T^\alpha \bar\sigma^{\mu \nu} P_X \xi h) B_\mu G^\alpha_\nu, \label{eq:T13sigma}\\
    && T^{23}_\sigma = i g_2 g_3 Y^\dagger (\bar{\psi} T^a_L T^\alpha \bar\sigma^{\mu \nu} P_X \xi h) W^a_\mu G^\alpha_\nu, \label{eq:T23sigma}
\end{eqnarray}

\noindent
where $ \mathcal{G}^{a b} = \frac{1}{2} ( T^a_L T^b_L + T^b_L T^a_L ) = \frac{1}{4} \delta^{a b} \mathbb{I}_2 $, with $T^a_L = \tau^a/2$ and where $\tau^a$ are the Pauli matrices; and
$ \mathcal{G}^{\alpha \beta} = \frac{1}{2} ( T^\alpha T^\beta + T^\beta T^\alpha ) $, which is not proportional to the identity due to the contribution from the (totally) symmetric symbol $ D^{\alpha \beta \gamma} $.
The normalization used is $ {\rm tr} \{ T^\alpha T^\beta \} = \frac{1}{2} \delta^{\alpha \beta} $.
The BRST transformation of each one of the operators listed in Eqs.~\eqref{eq:CT_Th}-\eqref{eq:T23sigma} consists of a fermion bilinear, a ghost on which one or two derivatives act, one scalar field and up to one gauge boson.

A linear combination of these counterterms is considered to perform the finite renormalization in order to restore the Slavnov-Taylor identities at one-loop order:

\begin{eqnarray}\label{eq:Lfct_one_scalar}
    \mathcal{L}_{\rm fct} \supset && C^{\mathcal{O}}_{\psi \xi} \frac{1}{3 \cdot 16 \pi^2} ( \alpha_h^\mathcal{O} \, T_h + \beta_{ij}^\mathcal{O} \, T^{ij}_g + \tilde{\beta}_{33}^\mathcal{O} \, \tilde{T}^{33}_g ) \nonumber \\
    && \quad + C^{\mathcal{O}}_{\psi \xi} \frac{8}{3 \cdot 16 \pi^2} ( \alpha_{hi}^\mathcal{O} \, T_{hi} + \zeta_{mn}^\mathcal{O} \, T^{mn}_\sigma + \tilde{\zeta}_{33}^\mathcal{O} \, \tilde{T}^{33}_\sigma ) + {\rm h.c.} \,,
\end{eqnarray}
where $i=1,2,3$, and $ij=11, 12, 13, 22, 23, 33$, $mn = 12, 13, 22, 23$; and $C^{\mathcal{O}}_{\psi \xi}$ is the Wilson coefficient.
A sum over SMEFT dimension-6 operators and different possible contractions of their fermion fields via the scalar coupling is implicit.
Eq.~\eqref{eq:Lfct_one_scalar} is schematic in the sense that it does not explicitly show the identities of the fermions $\psi, \xi$ (whether left- or right-handed quarks or leptons), nor the dependence on the scalar field $h$ (whether $\Phi$ or $\Phi^c$, or their complex conjugates) and the Yukawa matrix (whether $Y_d, Y_u, Y_e, Y_\nu$, or their complex conjugates). This point will also be studied here.
The values of the coefficients are given in Tabs.~\ref{tab:CTs_one_scalar_1} and \ref{tab:CTs_one_scalar_2}.
It is convenient to define $ y^{i j} = 2 (y^i_L)^2 - y^i_L \, y^j_R + 2 (y^j_R)^2 $.
In deriving these results, it has been used that $ y_h = y_L - y_R $, where $ y_h = 1/2 $ for $ \Phi $, and $ y_h = -1/2 $ for $ \Phi^c $.

It turns out that
some of the counterterm operators previously defined always appear in specific combinations:

\begin{eqnarray}
    && T_h + y^{q x} T^{11}_g \;\; \text{or} \;\;
    T_h + y^{\ell y} T^{11}_g, \\
    && (y_L^q + y_R^x) T^{13}_g + \tilde{T}^{33}_g, \\
    && T_{h2} - \frac{3}{2} y_R^x T^{12}_\sigma - T^{22}_\sigma \;\; \text{or} \;\; T_{h2} - \frac{3}{2} y_R^y T^{12}_\sigma - T^{22}_\sigma, \\
    && T_{h3} + \frac{3}{4} (y_L^q - y_R^x) T^{13}_\sigma.
\end{eqnarray}

The fermion bilinear structures and the generation indices for the counterterms can be established straightforwardly from the first two columns in Tabs.~\ref{tab:CTs_one_scalar_1} and \ref{tab:CTs_one_scalar_2}, while the scalar field ($\Phi^{(\dagger)}$ or $\Phi^{c(\dagger)}$) is the one needed to ensure global symmetry transformations. To clarify the schematic notation in use, the finite counterterm Lagrangian for the structures $T_h$ is presented

\begin{eqnarray}
    \mathcal{L}_{\rm fct} \supset && \frac{1}{3 \cdot 16 \pi^2} \Big[ C^{\ell e d q}_{f i j k} N_c (Y^\dagger_d)_{kj} \bar{\ell}_m^f e^i \bar{\partial}^2 \Phi^m + (C^{\ell e d q}_{f i j k})^\ast (Y^\dagger_e)_{fi} \bar{q}_m^k d^j \bar{\partial}^2 \Phi^m \\
    && + C^{\ell \nu u q}_{f i j k} N_c (Y^\dagger_u)_{kj} \bar{\ell}_m^f \nu^i \bar{\partial}^2 \Phi^{cm} + (C^{\ell \nu u q}_{f i j k})^\ast (Y^\dagger_\nu)_{fi} \bar{q}_m^k u^j \bar{\partial}^2 \Phi^{cm} \nonumber\\
    && - (C^{\ell e q u (1)}_{f i j k})^\ast N_c (Y^\dagger_u)_{jk} \bar{e}^i \ell_m^f \bar{\partial}^2 \Phi^{\dagger m} - (C^{\ell e q u (1)}_{f i j k})^\ast (Y^\dagger_e)_{fi} \bar{u}^k q_m^j \bar{\partial}^2 \Phi^{c\dagger m} \nonumber\\
    && + (C^{\ell \nu q d (1)}_{f i j k})^\ast N_c (Y^\dagger_d)_{jk} \bar{\nu}^i \ell_m^f \bar{\partial}^2 \Phi^{c\dagger m} + (C^{\ell \nu q d (1)}_{f i j k})^\ast (Y^\dagger_\nu)_{fi} \bar{d}^k q_m^j \bar{\partial}^2 \Phi^{\dagger m} \nonumber\\
    && + (C^{\ell \nu \ell e}_{f i j k})^\ast (Y^\dagger_e)_{jk} \bar{\nu}^i \ell_m^f \bar{\partial}^2 \Phi^{c\dagger m} + (C^{\ell \nu \ell e}_{f i j k})^\ast (Y^\dagger_\nu)_{fi} \bar{e}^k \ell_m^j \bar{\partial}^2 \Phi^{\dagger m} \nonumber\\
    && + (C^{\ell \nu \ell e}_{f i j k})^\ast \frac{1}{2} (Y^\dagger_e)_{fk} \bar{\nu}^i \ell_m^j \bar{\partial}^2 \Phi^{c\dagger m} + (C^{\ell \nu \ell e}_{f i j k})^\ast \frac{1}{2} (Y^\dagger_\nu)_{ji} \bar{e}^k \ell_m^f \bar{\partial}^2 \Phi^{\dagger m} \nonumber\\
    && + (C^{q u q d (1)}_{f i j k})^\ast N_c (Y^\dagger_d)_{jk} \bar{u}^i q_m^f \bar{\partial}^2 \Phi^{c\dagger m} + (C^{q u q d (1)}_{f i j k})^\ast N_c (Y^\dagger_u)_{fi} \bar{d}^k q_m^j \bar{\partial}^2 \Phi^{\dagger m} \nonumber\\
    && + (C^{q u q d (1)}_{f i j k})^\ast \frac{1}{2} (Y^\dagger_d)_{fk} \bar{u}^i q_m^j \bar{\partial}^2 \Phi^{c\dagger m} + (C^{q u q d (1)}_{f i j k})^\ast \frac{1}{2} (Y^\dagger_u)_{ji} \bar{d}^k q_m^f \bar{\partial}^2 \Phi^{\dagger m} \nonumber\\
    && + (C^{q u q d (8)}_{f i j k})^\ast \frac{C_A}{2} (Y^\dagger_d)_{fk} \bar{u}^i q_m^j \bar{\partial}^2 \Phi^{c\dagger m} + (C^{q u q d (8)}_{f i j k})^\ast \frac{C_A}{2} (Y^\dagger_u)_{ji} \bar{d}^k q_m^f \bar{\partial}^2 \Phi^{\dagger m} \nonumber\\
    && - C^{\ell \nu}_{f i j k} 2 (Y^\dagger_\nu)_{ij} \bar{\ell}_m^f \nu^k \bar{\partial}^2 \Phi^{cm} - (C^{\ell \nu}_{f i j k})^\ast 2 (Y^\dagger_\nu)_{kf} \bar{\ell}_m^i \nu^j \bar{\partial}^2 \Phi^{cm} \nonumber\\
    && - C^{\ell e}_{f i j k} 2 (Y^\dagger_e)_{ij} \bar{\ell}_m^f e^k \bar{\partial}^2 \Phi^m - (C^{\ell e}_{f i j k})^\ast 2 (Y^\dagger_e)_{kf} \bar{\ell}_m^i e^j \bar{\partial}^2 \Phi^m \nonumber\\
    && - C^{q d (1)}_{f i j k} 2 (Y^\dagger_d)_{ij} \bar{q}_m^f d^k \bar{\partial}^2 \Phi^m - (C^{q d (1)}_{f i j k})^\ast 2 (Y^\dagger_d)_{kf} \bar{q}_m^i d^j \bar{\partial}^2 \Phi^m \nonumber\\
    && - C^{q u (1)}_{f i j k} 2 (Y^\dagger_u)_{ij} \bar{q}_m^f u^k \bar{\partial}^2 \Phi^{cm} - (C^{q u (1)}_{f i j k})^\ast 2 (Y^\dagger_u)_{kf} \bar{q}_m^i u^j \bar{\partial}^2 \Phi^{cm} \nonumber\\
    && - C^{q d (8)}_{f i j k} 2 C_A (Y^\dagger_d)_{ij} \bar{q}_m^f d^k \bar{\partial}^2 \Phi^m - (C^{q d (8)}_{f i j k})^\ast 2 C_A (Y^\dagger_d)_{kf} \bar{q}_m^i d^j \bar{\partial}^2 \Phi^m \nonumber\\
    && - C^{q u (8)}_{f i j k} 2 C_A (Y^\dagger_u)_{ij} \bar{q}_m^f u^k \bar{\partial}^2 \Phi^{cm} - (C^{q u (8)}_{f i j k})^\ast 2 C_A (Y^\dagger_u)_{kf} \bar{q}_m^i u^j \bar{\partial}^2 \Phi^{cm} \Big] \nonumber\\
    && + {\rm h.c.} \nonumber
\end{eqnarray}

\begin{sidewaystable}
    \centering
    \renewcommand{\arraystretch}{1.0}
    \begin{tabular}{|cc||c|c|c|c|c|c|c|c|}
        \hline
        $\mathcal{O}$ & $Y^{\dagger}$ & $\alpha_h^\mathcal{O}$ & $\beta_{11}^\mathcal{O}$ & $\beta_{12}^\mathcal{O}$ & $\beta_{13}^\mathcal{O}$ & $\beta_{22}^\mathcal{O}$ & $\beta_{23}^\mathcal{O}$ & $\beta_{33}^\mathcal{O}$ & $\tilde{\beta}_{33}^\mathcal{O}$ \\
        \hline
        \hline
        $Q_{\ell e d q}$ & $(Y^{\dagger}_d)_{kj}$ & \multirow{2}{*}{$N_c$} & \multirow{2}{*}{$N_c y^{q x}$} & \multirow{2}{*}{$N_c (4 y^q_L - y^x_R)$} & \multirow{2}{*}{$0$} & \multirow{2}{*}{$2 N_c$} & \multirow{2}{*}{$0$} & \multirow{2}{*}{$\frac{3}{2}$} & \multirow{2}{*}{$0$} \\
        $Q_{\ell \nu u q}$ & $(Y^{\dagger}_u)_{kj}$ & & & & & & & & \\
        \hline
        $Q_{\ell e d q}^\dagger$ & $(Y^{\dagger}_e)_{fi}$ & \multirow{2}{*}{$1$} & \multirow{2}{*}{$y^{\ell y}$} & \multirow{2}{*}{$4 y^\ell_L - y^y_R$} & \multirow{2}{*}{$0$} & \multirow{2}{*}{$2$} & \multirow{2}{*}{$0$} & \multirow{2}{*}{$0$} & \multirow{2}{*}{$0$} \\
        $Q_{\ell \nu u q}^\dagger$ & $(Y^{\dagger}_\nu)_{fi}$ & & & & & & & & \\
        \hline
        $Q_{\ell e q u}^{(1) \dagger}$ & $(Y^{\dagger}_u)_{jk}$ & \multirow{2}{*}{$\mp N_c$} & \multirow{2}{*}{$\mp N_c y^{q x}$} & \multirow{2}{*}{$\pm N_c (4 y^q_L - y^x_R)$} & \multirow{2}{*}{$0$} & \multirow{2}{*}{$\mp 2 N_c$} & \multirow{2}{*}{$0$} & \multirow{2}{*}{$\mp \frac{3}{2}$} & \multirow{2}{*}{$0$} \\
        $Q_{\ell \nu q d}^{(1) \dagger}$ & $(Y^{\dagger}_d)_{jk}$ & & & & & & & & \\
        \hline
        $Q_{\ell e q u}^{(1) \dagger}$ & $(Y^{\dagger}_e)_{fi}$ & \multirow{2}{*}{$\mp 1$} & \multirow{2}{*}{$\mp y^{\ell y}$} & \multirow{2}{*}{$- (4 y^\ell_L - y^y_R)$} & \multirow{2}{*}{$0$} & \multirow{2}{*}{$\mp 2$} & \multirow{2}{*}{$0$} & \multirow{2}{*}{$0$} & \multirow{2}{*}{$0$} \\
        $Q_{\ell \nu q d}^{(1) \dagger}$ & $(Y^{\dagger}_\nu)_{fi}$ & & & & & & & & \\
        \hline
        \multirow{2}{*}{$Q_{\ell \nu \ell e}^\dagger$} & $(Y^{\dagger}_e)_{jk}$ & \multirow{2}{*}{$1$} & \multirow{2}{*}{$y^{\ell y}$} & \multirow{2}{*}{$- (4 y^\ell_L - y^y_R)$} & \multirow{2}{*}{$0$} & \multirow{2}{*}{$2$} & \multirow{2}{*}{$0$} & \multirow{2}{*}{$0$} & \multirow{2}{*}{$0$} \\
         & $(Y^{\dagger}_\nu)_{fi}$ & & & & & & & & \\
        \hline
        \multirow{2}{*}{$Q_{q u q d}^{(1) \dagger}$} & $(Y^{\dagger}_d)_{jk}$ & \multirow{2}{*}{$N_c$} & \multirow{2}{*}{$N_c y^{q x}$} & \multirow{2}{*}{$- N_c (4 y^q_L - y^x_R)$} & \multirow{2}{*}{$0$} & \multirow{2}{*}{$2 N_c$} & \multirow{2}{*}{$0$} & \multirow{2}{*}{$\frac{3}{2}$} & \multirow{2}{*}{$0$} \\
         & $(Y^{\dagger}_u)_{fi}$ & & & & & & & & \\
        \hline
        \multirow{2}{*}{$Q_{q u q d}^{(8) \dagger}$} & $(Y^{\dagger}_d)_{jk}$ & \multirow{2}{*}{$0$} & \multirow{2}{*}{$0$} & \multirow{2}{*}{$0$} & \multirow{2}{*}{$\frac{3}{2} (y^q_L + y^x_R)$} & \multirow{2}{*}{$0$} & \multirow{2}{*}{$- \frac{3}{2}$} & \multirow{2}{*}{$- \frac{3}{4 N_c}$} & \multirow{2}{*}{$\frac{3}{2}$} \\
         & $(Y^{\dagger}_u)_{fi}$ & & & & & & & & \\
        \hline
        \hline
        $Q_{\ell y}$ & $(Y^{\dagger}_y)_{ij}$ & \multirow{2}{*}{$-2$} & \multirow{2}{*}{$-2 y^{\ell y}$} & \multirow{2}{*}{$-2 (4 y^\ell_L - y^y_R)$} & \multirow{2}{*}{$0$} & \multirow{2}{*}{$-4$} & \multirow{2}{*}{$0$} & \multirow{2}{*}{$0$} & \multirow{2}{*}{$0$} \\
        $Q_{\ell y}^\dagger$ & $(Y^{\dagger}_y)_{kf}$ & & & & & & & & \\
        \hline
        $Q_{q x}^{(1)}$ & $(Y^{\dagger}_x)_{ij}$ & \multirow{2}{*}{$-2$} & \multirow{2}{*}{$-2 y^{q x}$} & \multirow{2}{*}{$-2 (4 y^q_L - y^x_R)$} & \multirow{2}{*}{$-6 (y^q_L + y^x_R)$} & \multirow{2}{*}{$-4$} & \multirow{2}{*}{$-6$} & \multirow{2}{*}{$0$} & \multirow{2}{*}{$-6$} \\
        $Q_{q x}^{(1) \dagger}$ & $(Y^{\dagger}_x)_{kf}$ & & & & & & & & \\
        \hline
        $Q_{q x}^{(8)}$ & $(Y^{\dagger}_x)_{ij}$ & \multirow{2}{*}{$-2 C_A$} & \multirow{2}{*}{$-2 C_A y^{q x}$} & \multirow{2}{*}{$-2 C_A (4 y^q_L - y^x_R)$} & \multirow{2}{*}{$\frac{3}{N_c} (y^q_L + y^x_R)$} & \multirow{2}{*}{$-4 C_A$} & \multirow{2}{*}{$\frac{3}{N_c}$} & \multirow{2}{*}{$- \frac{3}{2}$} & \multirow{2}{*}{$\frac{3}{N_c}$} \\
        $Q_{q x}^{(8) \dagger}$ & $(Y^{\dagger}_x)_{kf}$ & & & & & & & & \\
        \hline
        \multirow{2}{*}{$Q_{\ell \nu \ell e}^\dagger$} & $(Y^{\dagger}_e)_{fk}$ & \multirow{2}{*}{$\frac{1}{2}$} & \multirow{2}{*}{$\frac{1}{2} y^{\ell y}$} & \multirow{2}{*}{$\pm \frac{1}{2} (4 y^\ell_L - y^y_R)$} & \multirow{2}{*}{$0$} & \multirow{2}{*}{$1$} & \multirow{2}{*}{$0$} & \multirow{2}{*}{$0$} & \multirow{2}{*}{$0$} \\
         & $(Y^{\dagger}_\nu)_{ji}$ & & & & & & & & \\
        \hline
        \multirow{2}{*}{$ Q_{q u q d}^{(1) \dagger} $} & $(Y^{\dagger}_d)_{fk}$ & \multirow{2}{*}{$\frac{1}{2}$} & \multirow{2}{*}{$\frac{1}{2} y^{q x}$} & \multirow{2}{*}{$\pm \frac{1}{2} (4 y^q_L - y^x_R)$} & \multirow{2}{*}{$\frac{3}{2} (y^q_L + y^x_R)$} & \multirow{2}{*}{$1$} & \multirow{2}{*}{$\pm \frac{3}{2}$} & \multirow{2}{*}{$0$} & \multirow{2}{*}{$\frac{3}{2}$} \\
         & $(Y^{\dagger}_u)_{ji}$ & & & & & & & & \\
        \hline
        \multirow{2}{*}{$ Q_{q u q d}^{(8) \dagger} $} & $(Y^{\dagger}_d)_{fk}$ & \multirow{2}{*}{$\frac{C_A}{2}$} & \multirow{2}{*}{$\frac{C_A}{2} y^{q x}$} & \multirow{2}{*}{$\pm \frac{C_A}{2} (4 y^q_L - y^x_R)$} & \multirow{2}{*}{$- \frac{3}{4 N_c} (y^q_L + y^x_R)$} & \multirow{2}{*}{$C_A$} & \multirow{2}{*}{$\mp \frac{3}{4 N_c}$} & \multirow{2}{*}{$\frac{3}{8}$} & \multirow{2}{*}{$- \frac{3}{4 N_c}$} \\
         & $(Y^{\dagger}_u)_{ji}$ & & & & & & & & \\
        \hline
    \end{tabular}
    \caption[Coefficients of the set of counterterms needed in the finite renormalization of Green's functions carrying a single scalar field.]{
    Coefficients of the set of counterterms needed in the finite renormalization of Green's functions carrying a single scalar field.
    The notation is $x=u, d$, and $y=\nu, e$.
    The generation indices of the fields composing the operators are $fijk$.
    When two signs are given, the upper (lower) sign corresponds to the upper (lower, respectively) case in the same row.
    }
    \label{tab:CTs_one_scalar_1}
\end{sidewaystable}

\begin{sidewaystable}
    \centering
    \renewcommand{\arraystretch}{1.0}
    \begin{tabular}{|cc||c|c|c|c|c|c|c|c|}
        \hline
        $ \mathcal{O} $ & $Y^{\dagger}$ & $ \alpha^\mathcal{O}_{h1} $ & $ \alpha^\mathcal{O}_{h2} $ & $ \alpha^\mathcal{O}_{h3} $ & $ \zeta^\mathcal{O}_{12} $ & $ \zeta^\mathcal{O}_{13} $ & $ \zeta^\mathcal{O}_{22} $ & $ \zeta^\mathcal{O}_{23} $ & $ \tilde{\zeta}^\mathcal{O}_{33} $ \\
        \hline
        \hline
        $ Q_{\ell e q u}^{(3) \dagger} $ & $(Y^{\dagger}_u)_{jk}$ & \multirow{2}{*}{$\mp N_c(y^q_L + y^x_R)$} & \multirow{2}{*}{$\pm N_c$} & \multirow{2}{*}{$0$} & \multirow{2}{*}{$\mp \frac{3 N_c}{2} y^x_R$} & \multirow{2}{*}{$0$} & \multirow{2}{*}{$\mp N_c$} & \multirow{2}{*}{$0$} & \multirow{2}{*}{$0$} \\ 
        $ Q_{\ell \nu q d}^{(3) \dagger} $ & $(Y^{\dagger}_d)_{jk}$ & & & & & & & & \\
        \hline
        $ Q_{\ell e q u}^{(3) \dagger} $ & $(Y^{\dagger}_e)_{fi}$ & \multirow{2}{*}{$\mp (y^\ell_L + y^y_R)$} & \multirow{2}{*}{$- 1$} & \multirow{2}{*}{$0$} & \multirow{2}{*}{$\frac{3}{2} y^y_R$} & \multirow{2}{*}{$0$} & \multirow{2}{*}{$1$} & \multirow{2}{*}{$0$} & \multirow{2}{*}{$0$} \\
        $ Q_{\ell \nu q d}^{(3) \dagger} $ & $(Y^{\dagger}_\nu)_{fi}$ & & & & & & & & \\
        \hline
        \hline
        \multirow{2}{*}{$Q_{\ell \nu \ell e}^\dagger$} & $(Y^{\dagger}_e)_{fk}$ & \multirow{2}{*}{$\frac{1}{8} (y^\ell_L + y^y_R)$} & \multirow{2}{*}{$\pm \frac{1}{8}$} & \multirow{2}{*}{$0$} & \multirow{2}{*}{$\mp \frac{3}{16} y^y_R$} & \multirow{2}{*}{$0$} & \multirow{2}{*}{$\mp \frac{1}{8}$} & \multirow{2}{*}{$0$} & \multirow{2}{*}{$0$} \\
         & $(Y^{\dagger}_\nu)_{ji}$ & & & & & & & & \\
        \hline
        \multirow{2}{*}{$ Q_{q u q d}^{(1) \dagger} $} & $(Y^{\dagger}_d)_{fk}$ & \multirow{2}{*}{$\frac{1}{8} (y^q_L + y^x_R)$} & \multirow{2}{*}{$\pm \frac{1}{8}$} & \multirow{2}{*}{$\frac{1}{4}$} & \multirow{2}{*}{$\mp \frac{3}{16} y^x_R$} & \multirow{2}{*}{$\frac{3}{16} (y^q_L - y^x_R)$} & \multirow{2}{*}{$\mp \frac{1}{8}$} & \multirow{2}{*}{$\pm \frac{3}{16}$} & \multirow{2}{*}{$- \frac{1}{16}$} \\ 
         & $(Y^{\dagger}_u)_{ji}$ & & & & & & & & \\
        \hline
        \multirow{2}{*}{$ Q_{q u q d}^{(8) \dagger} $} & $(Y^{\dagger}_d)_{fk}$ & \multirow{2}{*}{$\frac{C_A}{8} ( y^q_L + y^x_R )$} & \multirow{2}{*}{$\pm \frac{C_A}{8}$} & \multirow{2}{*}{$- \frac{1}{8 N_c}$} & \multirow{2}{*}{$\mp \frac{3 C_A}{16} y^x_R$} & \multirow{2}{*}{$- \frac{3}{32 N_c} (y^q_L - y^x_R)$} & \multirow{2}{*}{$\mp \frac{C_A}{8}$} & \multirow{2}{*}{$\mp \frac{3}{32 N_c}$} & \multirow{2}{*}{$\frac{1}{32 N_c}$} \\ 
         & $(Y^{\dagger}_u)_{ji}$ & & & & & & & & \\
        \hline
    \end{tabular}
    \caption{Continuation of Tab.~\ref{tab:CTs_one_scalar_1}.}
    \label{tab:CTs_one_scalar_2}
\end{sidewaystable}

\newpage

\subsection{Counterterms with two scalars}\label{sec:two_scalars_results}

Finally, the counterterms with two scalar fields needed for the finite renormalization are

\begin{eqnarray}\label{eq:Thh1}
    && T^{(1)}_{h^\dagger h} = g_1 [ \bar{\psi} \bar{\gamma}^\mu P_X \xi ] h^\dagger h B_\mu, \\
    && T^{(2)}_{h^\dagger h} = g_2 [ \bar{\psi} T_L^a \bar{\gamma}^\mu P_X \xi ] h^\dagger h W_\mu^a, \label{eq:Thh2}\\
    && T^{(3)}_{h^\dagger h} = g_3 [ \bar{\psi} T^\alpha \bar{\gamma}^\mu P_X \xi ] h^\dagger h G_\mu^\alpha, \label{eq:Thh3}
\end{eqnarray}

\begin{eqnarray}
    && \tilde{T}^{(1)}_{h^\dagger h} = g_1 [ \bar{\psi} T_L^a \bar{\gamma}^\mu P_X \xi ] h^\dagger T_L^a h B_\mu, \label{eq:Ttildehh1}\\
    && \tilde{T}^{(2)}_{h^\dagger h} = g_2 [ \bar{\psi} \bar{\gamma}^\mu P_X \xi ] h^\dagger T_L^a h W_\mu^a, \label{eq:Ttildehh2}\\
    && \tilde{T}^{(3)}_{h^\dagger h} = g_3 [ \bar{\psi} T_L^a T^\alpha \bar{\gamma}^\mu P_X \xi ] h^\dagger T_L^a h G_\mu^\alpha. \label{eq:Ttildehh3}
\end{eqnarray}
The BRST transformation of the operators in Eqs.~\eqref{eq:Thh1}-\eqref{eq:Ttildehh3} all lead to structures carrying a fermion bilinear, a ghost on which a single derivative acts and two scalar fields.

The counterterm Lagrangian is

\begin{eqnarray}
    \mathcal{L}_{\rm fct} \supset && - C^{\mathcal{O}}_{\psi \xi} \frac{1}{3 \cdot 16 \pi^2} \left( \gamma^{\mathcal{O}}_{1} T^{(1)}_{h^\dagger h} + \gamma^{\mathcal{O}}_{2} T^{(2)}_{h^\dagger h} + \gamma^{\mathcal{O}}_{3} T^{(3)}_{h^\dagger h} \right. \\
    && \left. \quad \tilde\gamma^{\mathcal{O}}_{1} \tilde{T}^{(1)}_{h^\dagger h} + \tilde\gamma^{\mathcal{O}}_{2} \tilde{T}^{(2)}_{h^\dagger h} + \tilde\gamma^{\mathcal{O}}_{3} \tilde{T}^{(3)}_{h^\dagger h} \right) + {\rm h.c.}, \nonumber
\end{eqnarray}
where the coefficients are found in Tabs.~\ref{tab:CTs_two_scalars_1} (when fermions within the same bilinear are contracted, when referring to the Warsaw basis) and \ref{tab:CTs_two_scalars_2} (when fermions in different bilinears are contracted, with respect to the Warsaw basis).
Hermitian conjugation has to be included only for non--self-conjugate cases (such as when two different flavors are present). As before, the explicit counterterm operator structures can be built from the information contained in the first two columns of Tabs.~\ref{tab:CTs_two_scalars_1} and \ref{tab:CTs_two_scalars_2}.
To better illustrate the notation in use, consider the structures $T^{(3)}_{h^\dagger h}$

\begin{eqnarray}\label{eq:example_2_Higgs}
    \mathcal{L}_{\rm fct} \supset && - \frac{g_3}{16 \pi^2} \Big\{ \frac{1}{2} C^{qu(8)}_{fijk} [(Y^\dagger_u Y_u)_{if} + (Y^\dagger_d Y_d)_{if}] ( \bar{u}^j T^\alpha \bar{\gamma}^\mu u^k ) \Phi^\dagger \Phi G_\mu^\alpha \\
    && + \frac{1}{2} C^{qd(8)}_{fijk} [(Y^\dagger_u Y_u)_{if} + (Y^\dagger_d Y_d)_{if}] ( \bar{d}^j T^\alpha \bar{\gamma}^\mu d^k ) \Phi^\dagger \Phi G_\mu^\alpha \nonumber\\
    && + \frac{1}{2} C^{qu(8)}_{fijk} (Y_u Y^\dagger_u)_{kj} ( \bar{q}^f T^\alpha \bar{\gamma}^\mu q^i ) \Phi^\dagger \Phi G_\mu^\alpha
       + \frac{1}{2} C^{qd(8)}_{fijk} (Y_d Y^\dagger_d)_{kj} ( \bar{q}^f T^\alpha \bar{\gamma}^\mu q^i ) \Phi^\dagger \Phi G_\mu^\alpha \nonumber\\
    && + \frac{1}{2} C^{ud(8)}_{fijk} (Y_d Y^\dagger_d)_{kj} ( \bar{u}^f T^\alpha \bar{\gamma}^\mu u^i ) \Phi^\dagger \Phi G_\mu^\alpha
       + \frac{1}{2} C^{ud(8)}_{fijk} (Y_u Y^\dagger_u)_{if} ( \bar{d}^j T^\alpha \bar{\gamma}^\mu d^k ) \Phi^\dagger \Phi G_\mu^\alpha \nonumber\\
    && + \frac{1}{2} C^{qq(1)}_{fijk} [ (Y^\dagger_u Y_u)_{kf} + (Y^\dagger_d Y_d)_{kf} ] ( \bar{q}^j T^\alpha \bar{\gamma}^\mu q^i ) \Phi^\dagger \Phi G_\mu^\alpha \nonumber\\
    && + \frac{1}{2} C^{qq(1)}_{fijk} [ (Y^\dagger_u Y_u)_{ij} + (Y^\dagger_d Y_d)_{ij} ] ( \bar{q}^f T^\alpha \bar{\gamma}^\mu q^k ) \Phi^\dagger \Phi G_\mu^\alpha \nonumber\\
    && + \frac{3}{2} C^{qq(3)}_{fijk} [ (Y^\dagger_u Y_u)_{kf} + (Y^\dagger_d Y_d)_{kf} ] ( \bar{q}^j T^\alpha \bar{\gamma}^\mu q^i ) \Phi^\dagger \Phi G_\mu^\alpha \nonumber\\
    && + \frac{3}{2} C^{qq(3)}_{fijk} [ (Y^\dagger_u Y_u)_{ij} + (Y^\dagger_d Y_d)_{ij} ] ( \bar{q}^f T^\alpha \bar{\gamma}^\mu q^k ) \Phi^\dagger \Phi G_\mu^\alpha \nonumber\\
    && + C^{uu}_{fijk} (Y_u Y_u^\dagger)_{kf} ( \bar{u}^j T^\alpha \bar{\gamma}^\mu u^i ) \Phi^\dagger \Phi G_\mu^\alpha
       + C^{uu}_{fijk} (Y_u Y_u^\dagger)_{ij} ( \bar{u}^f T^\alpha \bar{\gamma}^\mu u^k ) \Phi^\dagger \Phi G_\mu^\alpha \nonumber\\
    && + C^{dd}_{fijk} (Y_d Y_d^\dagger)_{kf} ( \bar{d}^j T^\alpha \bar{\gamma}^\mu d^i ) \Phi^\dagger \Phi G_\mu^\alpha
       + C^{dd}_{fijk} (Y_d Y_d^\dagger)_{ij} ( \bar{d}^f T^\alpha \bar{\gamma}^\mu d^k ) \Phi^\dagger \Phi G_\mu^\alpha \Big\} \nonumber\\
    && + {\rm h.c.} \nonumber
\end{eqnarray}

\begin{table}[]
    \centering
    \begin{tabular}{|cc||c|c|c|c|c|c|}
        \hline
        $\mathcal{O}$ & & $\gamma^\mathcal{O}_1$ & $\gamma^\mathcal{O}_2$ & $\gamma^\mathcal{O}_3$ & $\tilde{\gamma}^\mathcal{O}_1$ & $\tilde{\gamma}^\mathcal{O}_2$ & $\tilde{\gamma}^\mathcal{O}_3$ \\
        \hline
        \hline
        \multirow{2}{*}{$ Q_{q q}^{(1)} $} & $ (Y^\dagger_x Y_x)_{kj} $, & \multirow{5}{*}{$ N_c (4 y_L^q - y_R^x) $} & \multirow{5}{*}{$0$} & \multirow{5}{*}{$0$} & \multirow{5}{*}{$0$} & \multirow{5}{*}{$\pm 4 N_c$} & \multirow{5}{*}{$0$} \\
         & $ (Y^\dagger_x Y_x)_{if} $ & & & & & & \\
        $ Q_{\ell q}^{(1)} $ & $ (Y^\dagger_x Y_x)_{kj} $ & & & & & & \\
        $ Q_{q u}^{(1)} $, $ Q_{q d}^{(1)} $ & $ (Y^\dagger_x Y_x)_{if} $ & & & & & & \\
        $ Q_{q y} $ & $ (Y^\dagger_x Y_x)_{if} $ & & & & & & \\
        \hline
        \multirow{2}{*}{$ Q_{\ell \ell} $} & $ (Y^\dagger_y Y_y)_{kj} $, & \multirow{5}{*}{$ 4 y_L^\ell - y_R^y $} & \multirow{5}{*}{$0$} & \multirow{5}{*}{$0$} & \multirow{5}{*}{$0$} & \multirow{5}{*}{$\pm 4$} & \multirow{5}{*}{$0$} \\
         & $ (Y^\dagger_y Y_y)_{if} $ & & & & & & \\
        $ Q_{\ell q}^{(1)} $ & $ (Y^\dagger_y Y_y)_{if} $ & & & & & & \\
        $ Q_{\ell \nu} $, $ Q_{\ell e} $ & $ (Y^\dagger_y Y_y)_{if} $ & & & & & & \\
        $ Q_{\ell x} $ & $ (Y^\dagger_y Y_y)_{if} $ & & & & & & \\
        \hline
        \multirow{2}{*}{$ Q_{x x} $} & $ (Y_x Y^\dagger_x)_{kj} $, & \multirow{7}{*}{$ N_c (4 y_R^x - y_L^q) $} & \multirow{7}{*}{$0$} & \multirow{7}{*}{$0$} & \multirow{7}{*}{$0$} & \multirow{7}{*}{$\mp N_c$} & \multirow{7}{*}{$0$} \\
         & $ (Y_x Y^\dagger_x)_{if} $ & & & & & & \\
        $ Q_{y x} $ & $ (Y_x Y^\dagger_x)_{kj} $ & & & & & & \\
        \multirow{2}{*}{$ Q_{u d}^{(1)} $} & $ (Y_d Y^\dagger_d)_{kj} $, & & & & & & \\
         & $ (Y_u Y^\dagger_u)_{if} $ & & & & & & \\
        $ Q_{\ell x} $ & $ (Y_x Y^\dagger_x)_{kj} $ & & & & & & \\
        $ Q_{q x}^{(1)} $ & $ (Y_x Y^\dagger_x)_{kj} $ & & & & & & \\
        \hline
        \multirow{2}{*}{$ Q_{y y} $} & $ (Y_y Y^\dagger_y)_{kj} $, & \multirow{7}{*}{$4 y_R^y - y_L^\ell$} & \multirow{7}{*}{$0$} & \multirow{7}{*}{$0$} & \multirow{7}{*}{$0$} & \multirow{7}{*}{$\mp1$} & \multirow{7}{*}{$0$} \\
         & $ (Y_y Y^\dagger_y)_{if} $ & & & & & & \\
        $ Q_{y x} $ & $ (Y_y Y^\dagger_y)_{if} $ & & & & & & \\
        \multirow{2}{*}{$ Q_{\nu e} $} & $ (Y_\nu Y^\dagger_\nu)_{if} $, & & & & & & \\
         & $ (Y_e Y^\dagger_e)_{kj} $ & & & & & & \\
        $ Q_{\ell y} $ & $ (Y_y Y^\dagger_y)_{kj} $ & & & & & & \\
        $ Q_{q y} $ & $ (Y_y Y^\dagger_y)_{kj} $ & & & & & & \\
        \hline
        $ Q_{d u \nu e} $ & $ (Y_u Y^\dagger_d)_{if} $ & $0$ & $0$ & $0$ & $0$ & $-N_c$ & $0$ \\ 
        \hline
        $ Q_{d u \nu e}^\dagger $ & $ (Y_\nu Y^\dagger_e)_{jk} $ & $0$ & $0$ & $0$ & $0$ & $-1$ & $0$ \\ 
        \hline
        \multirow{2}{*}{$ Q_{q q}^{(3)} $} & $ (Y^\dagger_x Y_x)_{kj} $, & \multirow{3}{*}{$0$} & \multirow{3}{*}{$4 N_c$} & \multirow{3}{*}{$0$} & \multirow{3}{*}{$\pm 4 N_c ( 4 y_L^q - y_R^x )$} & \multirow{3}{*}{$0$} & \multirow{3}{*}{$0$} \\
         & $ (Y^\dagger_x Y_x)_{if} $ & & & & & & \\
        $ Q_{\ell q}^{(3)} $ & $ (Y^\dagger_x Y_x)_{kj} $ & & & & & & \\
        \hline
        $ Q_{\ell q}^{(3)} $ & $ (Y^\dagger_y Y_y)_{if} $ & $0$ & $4$ & $0$ & $ \pm 4 ( 4 y_L^\ell - y_R^y )$ & $0$ & $0$ \\
        \hline
        $ Q_{q u}^{(8)} $, $ Q_{q d}^{(8)} $ & $ (Y^\dagger_x Y_x)_{if} $ & $0$ & $0$ & $\frac{3}{2}$ & $0$ & $0$ & $0$ \\
        \hline
        $ Q_{q x}^{(8)} $ & $ (Y_x Y^\dagger_x)_{kj} $ & \multirow{3}{*}{$0$} & \multirow{3}{*}{$0$} & \multirow{3}{*}{$\frac{3}{2}$} & \multirow{3}{*}{$0$} & \multirow{3}{*}{$0$} & \multirow{3}{*}{$0$} \\
        \multirow{2}{*}{$ Q_{u d}^{(8)} $} & $ (Y_d Y^\dagger_d)_{kj} $, & & & & & & \\
         & $ (Y_u Y^\dagger_u)_{if} $ & & & & & & \\
        \hline
    \end{tabular}
    \caption[ Coefficients of the set of counterterms needed in the finite renormalization of Green's functions carrying two scalar fields.]{
    Coefficients of the set of counterterms needed in the finite renormalization of Green's functions carrying two scalar fields.
    The notation is $x=u, d$, and $y=\nu, e$.
    The generation indices of the fields composing the operators are $fijk$.
    In the cases of the coefficients $\tilde{\gamma}^\mathcal{O}_1$, $\tilde{\gamma}^\mathcal{O}_2$ and $\tilde{\gamma}^\mathcal{O}_3$, the upper (lower) sign corresponds to $x=d$ or $y=e$ ($x=u$ or $y=\nu$, respectively).
    }
    \label{tab:CTs_two_scalars_1}
\end{table}

\begin{table}[]
    \centering\begin{tabular}{|cc||c|c|c|c|c|c|}
        \hline
        $\mathcal{O}$ & & $\gamma^\mathcal{O}_1$ & $\gamma^\mathcal{O}_2$ & $\gamma^\mathcal{O}_3$ & $\tilde{\gamma}^\mathcal{O}_1$ & $\tilde{\gamma}^\mathcal{O}_2$ & $\tilde{\gamma}^\mathcal{O}_3$ \\
        \hline
        \hline
        \multirow{2}{*}{$ Q_{\ell \ell} $} & $ (Y^\dagger_y Y_y)_{kf} $, & \multirow{2}{*}{$ \frac{1}{2} (4 y_L^\ell - y_R^y) $} & \multirow{2}{*}{$ 2 $} & \multirow{2}{*}{$ 0 $} & \multirow{2}{*}{$ \pm 2 (4 y_L^\ell - y_R^y) $} & \multirow{2}{*}{$ \pm 2 $} & \multirow{2}{*}{$ 0 $} \\
         & $ (Y^\dagger_y Y_y)_{ij} $ & & & & & & \\
        \hline
        \multirow{2}{*}{$ Q_{q q}^{(1)} $} & $ (Y^\dagger_x Y_x)_{kf} $, & \multirow{2}{*}{$ \frac{1}{2} (4 y_L^q - y_R^x) $} & \multirow{2}{*}{$ 2 $} & \multirow{2}{*}{$ \frac{3}{2} $} & \multirow{2}{*}{$ \pm 2 (4 y_L^q - y_R^x) $} & \multirow{2}{*}{$ \pm 2 $} & \multirow{2}{*}{$ \pm 6 $} \\
         & $ (Y^\dagger_x Y_x)_{ij} $ & & & & & & \\
        \hline
        \multirow{2}{*}{$ Q_{q q}^{(3)} $} & $ (Y^\dagger_x Y_x)_{kf} $, & \multirow{2}{*}{$ \frac{3}{2} (4 y_L^q - y_R^x) $} & \multirow{2}{*}{$- 2$} & \multirow{2}{*}{$\frac{9}{2}$} & \multirow{2}{*}{$ \mp 2 (4 y_L^q - y_R^x) $} & \multirow{2}{*}{$\pm 6$} & \multirow{2}{*}{$\mp 6$} \\
         & $ (Y^\dagger_x Y_x)_{ij} $ & & & & & & \\
        \hline
        \multirow{2}{*}{$ Q_{y y} $} & $ (Y_y Y^\dagger_y)_{kf} $, & \multirow{2}{*}{$4 y_R^y - y_L^\ell$} & \multirow{2}{*}{$0$} & \multirow{2}{*}{$0$} & \multirow{2}{*}{$0$} & \multirow{2}{*}{$\mp 1$} & \multirow{2}{*}{$0$} \\
         & $ (Y_y Y^\dagger_y)_{ij} $ & & & & & & \\
        \hline
        \multirow{2}{*}{$ Q_{x x} $} & $ (Y_x Y^\dagger_x)_{kf} $, & \multirow{2}{*}{$4 y_R^x - y_L^q$} & \multirow{2}{*}{$0$} & \multirow{2}{*}{$3$} & \multirow{2}{*}{$0$} & \multirow{2}{*}{$\mp 1$} & \multirow{2}{*}{$0$} \\
         & $ (Y_x Y^\dagger_x)_{ij} $ & & & & & & \\
        \hline
        $ Q_{\nu e} $ & $ (Y_\nu Y^\dagger_e)_{ij} $ & \multirow{2}{*}{$0$} & \multirow{2}{*}{$0$} & \multirow{2}{*}{$0$} & \multirow{2}{*}{$0$} & \multirow{2}{*}{$-1$} & \multirow{2}{*}{$0$} \\ 
        $ Q_{\nu e}^\dagger $ & $ (Y_\nu Y^\dagger_e)_{fk} $ & & & & & & \\
        \hline
        $ Q_{u d}^{(1)} $ & $ (Y_u Y^\dagger_d)_{ij} $ & \multirow{2}{*}{$0$} & \multirow{2}{*}{$0$} & \multirow{2}{*}{$0$} & \multirow{2}{*}{$0$} & \multirow{2}{*}{$- 1$} & \multirow{2}{*}{$0$} \\ 
        $ Q_{u d}^{(1) \dagger} $ & $ (Y_u Y^\dagger_d)_{fk} $ & & & & & & \\
        \hline
        $ Q_{u d}^{(8)} $ & $ (Y_u Y^\dagger_d)_{ij} $ & \multirow{2}{*}{$0$} & \multirow{2}{*}{$0$} & \multirow{2}{*}{$0$} & \multirow{2}{*}{$0$} & \multirow{2}{*}{$- C_A$} & \multirow{2}{*}{$0$} \\ 
        $ Q_{u d}^{(8) \dagger} $ & $ (Y_u Y^\dagger_d)_{fk} $ & & & & & & \\
        \hline
    \end{tabular}
    \caption{Continuation of Tab.~\ref{tab:CTs_two_scalars_1}.}
    \label{tab:CTs_two_scalars_2}
\end{table}

Some operator counterterms appear in the following combinations

\begin{equation}
    T^{(2)}_{h^\dagger h} \pm (4 y_L^q - y_R^x) \tilde{T}^{(1)}_{h^\dagger h} \;\; \text{or} \;\; T^{(2)}_{h^\dagger h} \pm (4 y_L^\ell - y_R^y) \tilde{T}^{(1)}_{h^\dagger h},
\end{equation}
where the upper (lower) sign corresponds to a right-handed down-type (up-type, respectively) quark or lepton running inside the internal fermion line.

At this point, now that all the counterterms have been presented, it is important to remark that full Lagrangian needs to be hermitian. This condition is obviously fulfilled in the cases where the hermitian conjugate needs to be explicitly added, but not so evident in all the other situations. The operators that might have an imaginary part are the ones in Eqs.~\eqref{eq:Ftilde13}, \eqref{eq:F13}, \eqref{eq:F23} and  \eqref{eq:F13DC}. However, as mentioned before, it has been explicitly checked that when evaluating the traces of the generators and considering the symmetry properties of the \(C\) and \(D\) constants, the imaginary parts vanish, and the Lagrangian is indeed hermitian. This calculation can be interpreted as a sanity check of the results.

\subsection{Discussion and comparison with the spurion approach} \label{sec:discussion_spurion}

Having found finite counterterms that cure the obstructions to the Slavnov-Taylor identities, it is explicitly shown that the theory is anomaly free at one loop-order. This is expected as mentioned in Sec.~\ref{sec:Anomalies}, since the fermion content of the SM is such that no anomalies in its local symmetries are generated (of course, adding a sterile right-handed neutrino as done here does not change the conclusion). The total number of needed operators, having in mind that some of them appear in specific combinations, is 28, when omitting the nature of the fermions (whether quarks or leptons), their chiralities, and generation indices, in linear combinations that may depend on the various hypercharges. They consist of a single fermion bilinear contracted with a varying number of gauge boson fields, divided into three categories according to the number of scalar fields. The zeros in the tables of the counterterms correspond to the absence of a particular structure, due to various reasons, such as some fields being singlets under ${\rm SU}(2)_{\rm L}$ and/or ${\rm SU}(3)$, the tracelessness of the ${\rm SU(2)}_{\rm L}$ and ${\rm SU}(3)$ generators, or the impossibility of writing a particular gauge-variant structure of dimension-6 that could possibly act as a counterterm (for instance, in the case of four-fermion operators, which must respect global symmetries in the scheme employed where fictitious fields are introduced).

As already mentioned, there is also a calculation of the same counterterms (which indeed covers all dimension-6 operators of SMEFT) performed using the spurion method. Both sets of results depend on diagrams calculated with single-insertions of the four-fermion operators, insertions of the the fermion-gauge boson and Yukawa vertices and the evanescent symmetry-breaking vertex, that in the case with the algebraic approach consists of a fermion-ghost vertex, while in the case of Ref.~\cite{Fuentes-Martin:2025meq} introduces a spurion field $\Omega$.
The background field gauge of Ref.~\cite{Fuentes-Martin:2025meq} leads to Feynman rules that are very similar to the ones shown in Eq.~\eqref{eq:SM_nuR_Lagrangian}, with gauge bosons and the Higgs replaced by background fields for the purpose of this calculation; the propagators in both setups are the same. One would expect then to obtain the same set of finite symmetry-restoring counterterms at one loop as in Ref.~\cite{Fuentes-Martin:2025meq} up to BRST-invariant operators. The reason for this is that their regularized classical Lagrangian is equivalent to the one presented in this text in the limit $\Omega \to \mathbb{I}$, and their resulting finite amplitudes
are polynomial in $\Omega - \mathbb{I}$, such that interchanging the limit on the spurion field and the integration over loop momentum should commute.

However, a small number of differences with respect to the published version of Ref.~\cite{Fuentes-Martin:2025meq} have been identified. These differences cannot all be explained by the use of two different but equivalent bases of counterterms, i.e., a change of renormalization scheme; for instance, both results for $Q_{\ell e d q}$ are equivalent, differing only by a gauge invariant combination that consists of $ D^\mu D_\mu H^{(\dagger)} $ multiplying the lepton or the quark bilinear.
On the other hand, for example, in the case of the $Q_{ee}$ operator, the difference between both results (Eq.~(2.37) in their supplemental material) in the finite counterterm Lagrangian is ($ y^e_R = -1 $):
\begin{equation}
\label{Difference_Granada}
    \frac{C^{ee}_{fijj}}{16 \pi^2} \left[ \left( - \frac{4}{3} y^e_R -\frac{8}{3} \right) g_1 (\bar{e}_f \bar{\gamma}_\mu e_i) \bar{\partial}^2 B^\mu + \left( \frac{16}{3} (y^e_R)^2 -\frac{8}{3} \right) g_1^2 (\bar{e}_f \bar{\gamma}_\mu e_i) \epsilon^{\mu \nu \rho \sigma} B_{\sigma} \bar{\partial}_\rho B_{\nu} \right]  + {\rm h.c.},
\end{equation}
which is not invariant under a BRST transformation.\footnote{The linearized Slavnov-Taylor operator is not introduced in this text, see e.g. Ref.~\cite{Belusca-Maito:2020ala}, which acts non-trivially only on external fields with respect to the BRST transformation. From the analysis done here of four-fermion operators, there are no finite counterterms depending on these external sources to the BRST transformations at one loop.} In this latter expression, the two shown operators are the $\mathcal{Q}_1$ and $\mathcal{Q}_2$ operators, respectively, with right-handed charged leptons, and $ f, i, j $ are generation indices (when $f=i$ hermitian conjugation in the last equation is not taken).

\thispagestyle{empty}

\chapter{\boldmath Left-Right Models. General Bounds on the Masses of the \(W_R\) and \(Z_R\) Gauge Bosons}
\markboth{5. LR MODELS. GENERAL BOUNDS ON THE LR SCALE}{}

\section*{Introduction}

In the previous chapters the SMEFT approach for finding new physics effects has been discussed. It relies on the fact that the high-energy theory that describes the fundamental interactions of nature is still unknown, so it is convenient to study the low-energy effects of all the possible new interactions through the associated effective operators. However, the main goal of theoretical physics is to discover the principles that govern the laws of the universe. This means that proposals for UV-complete models are also needed. In the context of particle physics there are two usual ways to extend the Standard Model: add new particles to the theory or introduce a new symmetry that contains the one of the SM. The second scenario is the one that will be studied here. In particular, in this text the LR models will be considered. They have been one of the most popular extensions of the SM since their introduction in the mid-to-late 1970s by the works of Pati, Mohapatra, Salam and Senjanovic \cite{Pati:1974yy,Mohapatra:1974hk,Senjanovic:1975rk,Senjanovic:1978ev}.

The original idea behind LR models was that \textit{the breaking of the parity symmetry by the electroweak interactions might be generated from a left-right-symmetric theory with an SSB mechanism}. In these models, \textit{a right-handed counterpart for the weak interaction} is added, leading to the gauge group \({\rm SU}(3)_{\rm QCD}\otimes{\rm SU}(2)_{\rm L}\otimes{\rm SU}(2)_{\rm R}\otimes{\rm U}(1)_{\rm X}\) with the usual normalization \({\rm X}=(B-L)/2\). It is important to remark that \textit{the addition of right-handed neutrinos is mandatory} in this scenario. The parity symmetry is restored by imposing the condition \(g_L=g_R\) in the coupling constants associated with the \({\rm SU}(2)\) parts of the group. Nonetheless, as already mentioned in the first chapter, it is nowadays known that \textit{these models can be embedded into larger gauge grand-unification groups} like \({\rm SO}(10)\) \cite{Fritzsch:1974nn,PhysRevLett.52.1072,PhysRevD.30.1052,PhysRevD.31.1718}. This means that a parity-symmetry can be restored at even higher energy scales and the running of the gauge couplings can produce different values \(g_L\neq g_R\) in the TeV range. Consequently, \textit{it will not be generally assumed during this text that the gauge couplings are equal}, so the general scenario will be considered.

In the first section of this chapter the gauge and fermion content in the minimal scenario of these theories will be presented, paying special attention to the cancellation of anomalies. In the second section the two most usual SSB mechanisms will be studied. One of them needs the addition of one bidoublet and two doublets of scalars, while the second one uses the same bidoublet but two scalar triplets. The main difference between them (apart from different interactions in the Higgs sector) is that in the former neutrinos are Dirac particles while in the latter they are Majorana fermions. The third section contains the presentation of a new way of breaking the parity symmetry based on an effective model with only two doublets. The fourth, fifth and sixth sections contain a derivation of the lower bounds on the masses of the new gauge bosons of the theory and on the  LR symmetry-breaking scale given the current experimental data. The fourth section is devoted to flavor-physics bounds while the other two study the results from collider physics. With the collider searches \textit{it is possible to derive bounds on the LR scale and on the masses of the new gauge bosons that do not depend on the rest of the parameters of the theory} as long as they stay in the perturbative regime.

It is important to remark that the last four sections are part of the original work developed by the author of this text and the results were already published in \cite{Solera:2023kwt,Solera:2024wjz,Solera:2024jrm}. 

\section{Matter Content in LR Models. Cancellation of Anomalies} \label{sec:Matter_Gauge_LR}

As already mentioned, the main idea of the LR models is to introduce a new interaction analogous to the weak interaction of the left-handed fermions but for their right-handed counterparts. As a consequence, the right-handed fields must be included into doublets of a new gauge group \({\rm SU}(2)_{\rm R}\), but this is only possible if the theory contains right-handed neutrinos. Therefore, the matter content of the models in the minimal case (with no extra fermions apart from the ones of the SM and the new neutrinos) with its representation under the group \({\rm SU}(3)_{\rm QCD}\otimes{\rm SU}(2)_{\rm L}\otimes{\rm SU}(2)_{\rm R}\otimes{\rm U}(1)_{\rm X}\) is given by

\begin{equation}
    \label{Matter_Content_LR}
    \begin{split}
        &q_L^{i}=\left(\begin{array}{c}
        u_L^{i}\\d_L^{i}
        \end{array}\right)\sim(\mathbf{3},\mathbf{2},\mathbf{1})_{x_{q_L}},\hspace{0.8cm}q_R^{i}=\left(\begin{array}{c}
        u_R^{i}\\d_R^{i}
        \end{array}\right)\sim(\mathbf{3},\mathbf{1},\mathbf{2})_{x_{q_R}}, \\&\ell_L^i=\left(\begin{array}{c}
        \nu_L^{i}\\e_L^i\end{array}\right)\sim(\mathbf{1},\mathbf{2},\mathbf{1})_{x_{\ell_L}},\hspace{0.8cm} \ell_R^i=\left(\begin{array}{c}
        \nu_R^{i}\\e_R^i\end{array}\right)\sim(\mathbf{1},\mathbf{1},\mathbf{2})_{x_{\ell_R}},
    \end{split}
\end{equation}

\noindent where the superscript \(i=1,2,3\) represents the family number.

The values of the generalized \({\rm U}(1)_{\rm X}\) charges can be determined by the conditions of the cancellation of anomalies, using Eqs.~\eqref{Chiral_Generator} and~\eqref{Vanishing-Symmetric} one finds:

\begin{itemize}
    \item \({\rm SU}(3)_{\rm QCD}-{\rm SU}(3)_{\rm QCD}-{\rm U}(1)_{\rm X}\):
    \begin{equation}
        \label{Anomaly33X}
        2x_{q_L}-2x_{q_R}=0.
    \end{equation}
    \item \({\rm SU}(2)_{\rm L}-{\rm SU}(2)_{\rm L}-{\rm U}(1)_{\rm X}\):
    \begin{equation}
        \label{AnomalyLLX}
        3x_{q_L}+x_{\ell_L}=0.
    \end{equation}
    \item \({\rm SU}(2)_{\rm R}-{\rm SU}(2)_{\rm R}-{\rm U}(1)_{\rm X}\):
    \begin{equation}
        \label{AnomalyRRX}
        -3x_{q_R}-x_{\ell_R}=0.
    \end{equation}
\end{itemize}

\noindent The combination of these equations leads to the constraint \(x_{\ell_L}=x_{\ell_R}=-x_{q_L}/3=-x_{q_R}/3\). This condition automatically cancels the \({\rm U}{(1)}^3\) and the gravitational anomalies:

\begin{itemize}
    \item \({\rm U}(1)_{\rm X}-{\rm U}(1)_{\rm X}-{\rm U}(1)_{\rm X}\):
    \begin{equation}
        \label{AnomalyXXX}
        6x^3_{q_L}-6x^3_{q_R}+2x^3_{\ell_L}-2x^3_{\ell_R}=0.
    \end{equation}
    \item \({\rm Graviton}-{\rm Graviton}-{\rm U}(1)_{\rm X}\):
    \begin{equation}
        \label{AnomalyGGX}
        6x_{q_L}-6x_{q_R}+2x_{\ell_L}-2x_{\ell_R}=0.
    \end{equation}
\end{itemize}

\noindent The canonical normalization is to take \(x_\ell=-1/2\) (other options will result in different values for the conventions of the electric charges up to an overall normalization) in such a way that

\begin{equation}
    \label{Matter_Content_Transformation_LR}
    q_L^i\sim(\mathbf{3},\mathbf{2},\mathbf{1})_{1/6}, \hspace{0.5cm} q_R^i\sim(\mathbf{3},\mathbf{1},\mathbf{2})_{1/6},  \hspace{0.5cm}
    \ell_L^i\sim(\mathbf{1},\mathbf{2},\mathbf{1})_{-1/2}, \hspace{0.5cm}
    \ell_R^i\sim(\mathbf{1},\mathbf{1},\mathbf{2})_{-1/2}.
\end{equation}

\noindent Given the transformations of the fermions under the gauge group, it is not difficult to show the following relation between the electric charge operator and the generators of the Lie algebra of the LR models: \(Q=T^3_L+T^3_R+{\rm X}\).

In the LR models the generalized charge is always proportional to \(B-L\), particularly, with this normalization it can be expressed as \({\rm X}=(B-L)/2\). This is highly remarkable for two reasons. On the one hand, \textit{the consistency condition of cancellation of anomalies completely determines (apart from normalization) the values of the charges of the matter fields}. As shown in Sec.~\ref{sec:Strong_EW}, in the SM there are two possible solutions and the correct one needs to be chosen by looking at the experimental evidence. After the SSB this leads to a prediction for the values of the electric charges. On the other hand, \(B-L\) is the only non-anomalous quantity associated with a global transformation that is conserved in the SM with the addition of right-handed neutrinos\footnote{As already mentioned in Sec.~\ref{sec:CKM}, in the absence of right-handed neutrinos the differences between lepton numbers per families are also conserved.} once non-perturbative effects are considered \cite{Weinberg2}. But, as pointed out in Sec.~\ref{sec:SMEFT}, the accidental global symmetries of a theory are usually broken once the higher order operators are considered. Furthermore, there are hints coming from the theoretical results of quantum gravity which point to the fact that only charges that are gauged can be conserved due to the elimination of information caused by black holes \cite{Hawking:1976ra}. Thus, taking the conservation of the \(B-L\) charge as a principle of a theory by gauging it, instead of being an accidental result, is natural given its conservation in the SM. This is automatically done in LR models, where \textit{all the conserved charges are associated with gauge symmetries}.

The gauge and fermion part of the Lagrangian of the LR models is simply given by:

\begin{equation}
    \label{Lagrangian_LR_Fermion_Gauge}
    \begin{split}
    \mathcal{L}&=i\sum_f\bar{\psi}_f\slashed{D}\psi_f-\frac{1}{4}\sum_{\alpha=1}^8G^{\alpha\mu\mu}G^{\alpha}_{\mu\nu}-\frac{1}{4}\sum_{i=1}^3W_L^{i\mu\nu}W_{L\mu\nu}^i\\&-\frac{1}{4}\sum_{i=1}^3W_R^{i\mu\nu}W_{R\mu\nu}^i-\frac{1}{4}W_{\rm X}^{\mu\nu}W_{\rm X\mu\nu},
    \end{split}
\end{equation}

\noindent where the covariant derivatives are now

\begin{equation}
    \label{Covariant_Derivative_LR}
    D_{\mu}\psi=\left(\partial_{\mu}-ig_{\rm X}{\rm X}W_{\rm X\mu}-ig_LT_L^iW^i_{L\mu}-ig_RT_R^iW^i_{R\mu}-ig_ST_S^{\alpha}G^\alpha_{\mu}\right)\psi
\end{equation}

\noindent and \(W_{\rm X}^{\mu\nu}\), \(W_L^{i\mu\nu}\), \(W_R^{i\mu\nu}\) and \(G^{\alpha\mu\nu}\) are the associated field-strength tensors. It is important to remark that right-handed counterparts for the left-handed gauge bosons have been added. Again, it is convenient to define the fields \(W_{1,2}^{\pm\mu}\coloneqq\left(W_{L,R}^{1\mu}\mp iW_{L,R}^{2\mu}\right)/\sqrt{2}\) which, as it will be shown in the following section, are in good approximation the mass eigenstates of the charged gauge bosons of the theory. The reason for this is that, if the scale of the breaking of the LR symmetry into the SM one \(v_R\) is much larger than the electroweak scale (\(v_R\gg v_{EW}\)) as the experimental evidence shows, then \textit{the breaking of  \({\rm SU}(2)_{\rm L}\otimes{\rm SU}(2)_{\rm R}\otimes{\rm U}(1)_{\rm X}\) into \({\rm U}(1)_{\rm QED}\) is done in two approximately independent steps}: first \({\rm SU}(2)_{\rm R}\otimes{\rm U}(1)_{\rm X}\to{\rm U}(1)_{\rm Y}\) and then \({\rm SU}(2)_{\rm L}\otimes{\rm U}(1)_{\rm Y}\to{\rm U}(1)_{\rm QED}\). Then, by analogy with the SM, the diagonalization of the mass matrices of the neutral gauge bosons will be done with the following transformation with a new mixing parameter \(\gamma\):

\begin{equation}
    \label{Approx_Diag_Neutral}
    \begin{split}
    \left(
    \begin{array}{c}
    Z_L^{\mu} \\
    Z_R^{\mu} \\
    A^{\mu}
    \end{array}\right)&\approx\left(
    \begin{array}{ccc}
    \cos\theta_W & 0 & -\sin\theta_W \\
    0 & 1 & 0 \\
    \sin\theta_W & 0 & \cos\theta_W
    \end{array}\right)\left(
    \begin{array}{c}
    W_L^{3\mu} \\
    Z_R^{\mu} \\
    B^{\mu}
    \end{array}\right)\\&\approx\left(
    \begin{array}{ccc}
    \cos\theta_W & 0 & -\sin\theta_W \\
    0 & 1 & 0 \\
    \sin\theta_W & 0 & \cos\theta_W
    \end{array}\right)\left(
    \begin{array}{ccc}
    1 & 0 & 0 \\
    0 & \cos\gamma & -\sin\gamma  \\
    0 & \sin\gamma & \cos\gamma
    \end{array}\right)\left(
    \begin{array}{c}
    W_L^{3\mu} \\
    W_R^{3\mu} \\
    W_{X}^{\mu}
    \end{array}\right).
    \end{split}
\end{equation}

Indeed, there is an exact relation at tree level analogous to the one of the SM between the coupling constants of the theory. It comes from the combination of \(e=g_L\sin\theta_W=g_Y\cos\theta_W\) and \(g_Y=g_R\sin\gamma=g_X\cos\gamma\):

\begin{equation}
    \label{Couplings_Relations}
    e=g_L\sin\theta_W=g_R\sin\gamma\cos\theta_W=g_{X}\cos\gamma\cos\theta_W.
\end{equation}

\noindent This means that the determination of the mixing parameter \(\gamma\) leads to the values of the coupling constants \(g_R\) and \(g_X\) since \(e\) and \(\theta_W\) are experimentally known. In fact, imposing the parity symmetry \(g_L=g_R\) gives the relation \(\sin\gamma=\tan\theta_W\), from which the value \(\gamma\approx33^{\circ}\) is obtained. Moreover, if one wants to stay in the perturbative regime \(g_R,g_X<1\), then it is necessary for \(\gamma\) to satisfy

\begin{equation}
    \label{gamma_perturbative}
    \sqrt{1-\frac{e^2}{\cos^2\theta_W}}>\sin\gamma>\frac{e}{\cos\theta_W},
\end{equation}

\noindent which leads to the range \(70^{\circ}\gtrsim\gamma\gtrsim20^{\circ}\).

\section{Spontaneous Symmetry Breaking in LR Models} \label{sec:SSB_LR}

In the literature there are mainly two specific realizations of the SSB mechanism of the LR models:  one of them requires the presence of one bidoublet and two scalar doublets, and the other one contains the same bidoublet but two triplets. The main difference between the two scenarios from the phenomenological point of view is that in the former all the fermions are Dirac particles, while in the latter neutrinos are Majorana fermions. However, they share some common features because they both contain the same scalar bidoublet, which is responsible in both cases for the generation of the masses of the quarks through Yukawa interactions.  Thus, it is convenient to first study these common properties.

\subsection{The bidoublet field and quark Yukawa interactions} \label{subsec:bidoublet}

To use a field \(\Phi\) to generate Yukawa terms of the form \(\bar{\psi}_L\Phi\psi_R\) with \(\psi=q,\ell\), it is clear that, given the transformation rules of the fermions in Eq.~\eqref{Matter_Content_Transformation_LR}, it must get transformed under the representation \(\Phi\sim\left(\mathbf{1},\mathbf{2},\mathbf{2}\right)_{0}\) of the gauge group.\footnote{In the literature it is sometimes written \(\Phi\sim\left(\mathbf{1},\mathbf{2},\bar{\mathbf{2}}\right)_{0}\), but the representations \(\mathbf{2}\) and \(\bar{\mathbf{2}}\) of \({\rm SU}(2)\) are equivalent.} Therefore, it is possible to write \(\Phi\) as

\begin{equation}
    \label{Bidoublet_Definition}
    \Phi=\left(
    \begin{array}{cc}
    \phi_1^0 & \phi_1^+ \\
    \phi_2^- &  \phi_2^0
    \end{array}\right),
\end{equation}

\noindent where the electric charges of the components are determined using that they are eigenstates of the operator \(Q=T^3_L+T^3_R+{\rm X}\). Consequently, the most general form of the VEV \(\left<\Phi\right>\) of the bidoublet that leaves \(Q\) as an unbroken generator is

\begin{equation}
    \label{VEV_Bidoublet}
    \left<\Phi\right>=\frac{1}{\sqrt{2}}\left(
    \begin{array}{cc}
    \kappa_1 & 0\\
    0 & \kappa_2
    \end{array}\right),
\end{equation}

\noindent where \(\kappa_1\) and \(\kappa_2\) can carry in principle complex phases \cite{Zhang:2007da}.

Introducing the charge conjugated field \(\tilde{\Phi}\coloneqq\sigma^2\Phi^{\ast}\sigma^2\sim\left(\mathbf{1},\mathbf{2},\mathbf{2}\right)_{0}\) of the bidoublet, it is possible to write the most general Yukawa Lagrangian for the quarks as

\begin{equation}
    \label{Yukawa_Quars_LR}
    \mathcal{L}_Y^q=-\sum_{i,j}\bar{q}_{L}^{i}\left(r_{ij}\Phi+s_{ij}\tilde{\Phi}\right)q_{R}^{j}+\mathrm{h.c.},
\end{equation}

\noindent where \(r\) and \(s\) are two completely general complex matrices. If one imposes the parity symmetry, since  \(\mathcal{P}:\Phi\rightarrow\Phi^{\dagger}\), the matrices \(r\) and \(s\) would become hermitian. Imposing charge conjugation symmetry, since \(\mathcal{C}:\Phi\rightarrow\Phi^{\ast}\),\footnote{Other definitions of these operators can be found at \cite{Ecker:1980at,Branco:1985ng}.} one obtains that they are real. Nonetheless, \textit{these discrete symmetries will not be assumed in this text}. Taking the VEV of the bidoublet, one gets the following mass matrices for the up- and down-type quarks:

\begin{equation}
    \label{u_d_quark_matrices_LR}
    \left(M_u\right)_{ij}\coloneqq\frac{1}{\sqrt{2}}\left(r_{ij}\kappa_1+s_{ij}\kappa_2^{\ast}\right),\hspace{0.8cm} \left(M_d\right)_{ij}\coloneqq\frac{1}{\sqrt{2}}\left(r_{ij}\kappa_2+s_{ij}\kappa_1^{\ast}\right).
\end{equation}

\noindent Since the mass matrices are proportional to \(\kappa_1\) and \(\kappa_2\), this means that \textit{the VEVs of the bidoublet must be of the order of the electroweak scale}. Now it is necessary to diagonalize those mass matrices, and it will be shown that the situation is different from the one of the SM. The kinetic term of the Lagrangian in the quark sector has a \({\rm U}(3)_f^2\) flavor symmetry, so it remains invariant under the transformation

\begin{equation}
    \label{Transformation_Flavor_Quarks}
    q_{L}^{i}\rightarrow\left(A_{L}^{u}\right)^{i}_{j}\,q_{L}^{j}, \hspace{0.8cm} q_{R}^{i}\rightarrow\left(A_{R}^{u}\right)^{i}_{j}\,q_{R}^{j},
\end{equation}

\noindent where \(A_{L}^{u}\) and \(A_{R}^{u}\) are unitary matrices. Using the singular value decomposition it is possible to show that these transformations can be used to diagonalize, for instance, \(M_u\). It is important to remark that if \(M_u\) is diagonalized by the transformation \(M_u\rightarrow \left(A^{u}_{L}\right)^{\dagger}M_{u}A_{R}^{u}\), then it is also diagonalized by \(M_u\rightarrow\left(K^{u}\right)^{\dagger}\left(A^{u}_{L}\right)^{\dagger}M_{u}A_{R}^{u}K^{u}\), where \(K^{u}\) is a diagonal matrix of phases. However, in order to diagonalize \(M_d\) an additional flavor transformation of the down-type quarks is needed. Then, the most general transformation that diagonalizes the two mass matrices is

\begin{equation}
    \label{Transformation_Flavor_Quarks_2}
    u_{L,\,R}^{i}\rightarrow(A_{L,\,R}^{u}K^{u})_{j}^{i}\,u_{L,\,R}^{j}, \hspace{0.8cm} d_{L,\,R}^{i}\rightarrow(A_{L,\,R}^{u}A_{L,\,R}^{d}K^{d})_{j}^{i}\,d_{L,\,R}^{j}, 
\end{equation}

\noindent where the \(A_{L,\,R}^{d}\) are unitary matrices and \(K^{d}\) is a diagonal matrix of phases.

Defining the left and right CKM matrices \(V_{L,\,R}^{\textrm{CKM}}\coloneqq \left(K^{u}\right)^{\dagger}A_{L,\,R}^{d}K^{d}\) and using the transformations in Eq.~\eqref{Transformation_Flavor_Quarks_2} it is possible to write the Lagrangian associated with charged currents in the quark sector mediated by gauge bosons as

\begin{equation}
    \label{Charged_Currents_Quarks_Gauge_LR}
    \mathcal{L}_{CC}^q=\frac{1}{\sqrt{2}}\bar{u}\,\gamma_{\mu}\left(g_LW_{1}^{+\mu}V_{L}^{\textrm{CKM}}P_{L}+g_RW^{+\mu}_{2}V_{R}^{\textrm{CKM}}P_{R}\right)d+\mathrm{h.c.}
\end{equation}

\noindent The main difference with respect to the SM is that now \textit{two different CKM matrices are needed}. It is possible to choose the freedom in defining \(K^u\) and \(K^d\) in a such a way that \(V_L^{\rm CKM}\) takes the same form as in the SM, with three real and one complex parameter; but then \(V_R^{\rm CKM}\) will remain as a completely general unitary matrix,\footnote{It is possible to reduce the number of parameters imposing additional symmetries. For instance, if one imposes the CP symmetry on the Lagrangian and considers that its breaking comes from the complex phases in \(\kappa_1\) and \(\kappa_2\) (SSB of the CP symmetry), then it is easy to show that \(|V_{L}^{\rm CKM}|=|V_{R}^{\rm CKM}|\) \cite{Langacker:2017uah}.} with \textit{three real and six complex parameters}. Even though this has the inconvenience to add a huge number of free parameters to the quark flavor sector which makes the theory less predictive, \textit{it has the advantage of having more sources of CP violation than the SM, which could be used in principle to explain the observed matter-antimatter asymmetry in the universe}.

Since the matrices \(V_{L,R}^{\rm CKM}\) are unitary and the interactions between fermions and gauge bosons do not mix the left- and right-handed parts of the spinors, \textit{there are no FCNCs at tree level mediated by gauge fields}. Furthermore, the \textit{GIM mechanism operates in both left- and right-handed sectors}. Taking the transformation in Eq.~\eqref{Approx_Diag_Neutral} for the neutral gauge bosons (which neglects the mixing between the \(Z_L\) and the \(Z_R\)) it is easy to show that the Lagrangian of the neutral currents mediated by the photon and the \(Z_L\) has the same form as in the SM, while the interactions with the \(Z_R\) are described by

\begin{equation}
    \label{Z_R_quarks_lagrangian}
    \begin{split}
    \mathcal{L}^q_{RNC}&=-\frac{1}{12}\frac{e}{\cos\theta_W}\tan\gamma\,Z_R^{\mu}\,\bar{u}\gamma_{\mu}\left\{\left(2-3\cot^2\gamma\right)-\left(3\cot^2\gamma\right)\gamma_5\right\}u
        \\& -\frac{1}{12}\frac{e}{\cos\theta_W}\tan\gamma\,Z_R^{\mu}\,\bar{d}\gamma_{\mu}\left\{\left(2+3\cot^2\gamma\right)+\left(3\cot^2\gamma\right)\gamma_5\right\}d.
    \end{split}
\end{equation}

\noindent However, the situation with the neutral currents is very different when looking at the scalar sector. Considering that, if \(|\kappa_1|\neq|\kappa_2|\) the relations in Eq.~\eqref{u_d_quark_matrices_LR} can be inverted to write

 \begin{equation}
     \label{Inverse_Mass_Matrix_Relation}
     r=\frac{\sqrt{2}}{\left|\kappa_1\right|^2-\left|\kappa_2\right|^2}\left(\kappa_1^{\ast}M_{u}-\kappa_2^{\ast}M_d\right), \hspace{0.8cm} s=\frac{\sqrt{2}}{\left|\kappa_1\right|^2-\left|\kappa_2\right|^2}\left(\kappa_1 M_{d}-\kappa_2M_{u}\right)
 \end{equation}

\noindent and using the transformations in Eq.~\eqref{Transformation_Flavor_Quarks_2} that diagonalize the mass matrices, it is possible to write the Lagrangian of interaction between the scalars and the mass eigenstates of the quarks as

\begin{equation}
    \label{Scalar_Quark_Charged_Interactions}
    \begin{split}
        \mathcal{L}^{I}_{QF,\,C}&=-\frac{\sqrt{2}}{\left|\kappa_1\right|^2-\left|\kappa_2\right|^2}\left\{\bar{u}_{L}\left[\left(\kappa_1^{\ast}\phi_{1}^{+}+\kappa_2\phi_{2}^{+}\right)M_{u}^{\textrm{diag}}V_{R}^{\textrm{CKM}}\right.\right.\\&\left.\left.-\left(\kappa_2^{\ast}\phi_{1}^{+}+\kappa_1\phi_{2}^{+}\right)V_{L}^{\textrm{CKM}}M_{d}^{\textrm{diag}}\right]d_{R}+\bar{d}_{L}\left[\left(\kappa_1^{\ast}\phi_{2}^{-}+\kappa_2\phi_{1}^{-}\right)\left(V_{L}^{\textrm{CKM}}\right)^{\dagger}M_{u}^{\textrm{diag}}\right.\right.\\&\left.\left.-\left(\kappa_2^{\ast}\phi_{2}^{-}+\kappa_1\phi_{1}^{-}\right)M_{d}^{\textrm{diag}}\left(V_{R}^{\textrm{CKM}}\right)^{\dagger}\right]u_{R}\right\}+\textrm{h.c.},
    \end{split}
\end{equation}

\begin{equation}
    \label{Scalar_Quark_Neutral_Interactions}
    \begin{split}
        \mathcal{L}^{I}_{QF,\,N}=-\frac{\sqrt{2}}{\left|\kappa_1\right|^2-\left|\kappa_2\right|^2}&\left\{\left(\kappa_1^{\ast}\phi_{1}^{0}-\kappa_2\phi_{2}^{0\,\dagger}\right)\bar{u}_{L}M_{u}^{\textrm{diag}}u_{R}+\left(\kappa_1\phi_{1}^{0\,\dagger}-\kappa_2^{\ast}\phi_{2}^{0}\right)\bar{d}_{L}M_{d}^{\textrm{diag}}d_{R}\right.\\&\left.+\left(\kappa_1\phi_{2}^{0\,\dagger}-\kappa_2^{\ast}\phi_{1}^{0}\right)\bar{u}_{L}V_{L}^{\textrm{CKM}}M_{d}^{\textrm{diag}}\left(V_{R}^{\textrm{CKM}}\right)^{\dagger}u_{R}\right.\\&\left.+\left(\kappa_1^{\ast}\phi_{2}^{0}-\kappa_2\phi_{1}^{0\,\dagger}\right)\bar{d}_{L}\left(V_{L}^{\textrm{CKM}}\right)^{\dagger}M_{u}^{\textrm{diag}}V_{R}^{\textrm{CKM}}d_{R}\right\}+\textrm{h.c.},
    \end{split}
\end{equation}

\noindent where Eq.~\eqref{Scalar_Quark_Charged_Interactions} contains the charged currents, Eq.~\eqref{Scalar_Quark_Neutral_Interactions} the neutral ones and \(M_{u,d}^{\textrm{diag}}\) are the diagonalized mass matrices. So \textit{there are FCNCs in the scalar sector}, but it can be shown that the ones mediated by the SM-like Higgs are suppressed. Imposing discrete symmetries on the theory this property has been the traditional way to put strong lower bounds on the masses of the scalar mediators using meson mixing. This leads to typical values of around \(10\,{\rm TeV}\) depending on the mass of the \(W_R\) \cite{Maiezza:2010ic}.

Given the VEV in Eq.~\eqref{VEV_Bidoublet} it is easy to see that, if \(\kappa_1\neq\kappa_2\), only the generators \(T_L^3+T_R^3\) and \((B-L)/2\) of the generalized electroweak sector (or any linear combination of them) remain unbroken. This means that group \({\rm SU(2)}_{\rm L}\otimes{\rm SU}(2)_{\rm R}\otimes{\rm U}(1)_{\rm X}\) is spontaneously broken by the bidoublet into a \({\rm U}(1)_{\rm X}\otimes{\rm U}(1)'\), not into \({\rm U}(1)_{\rm QED}\). This means that additional scalar degrees of freedom are required and, as it will be shown, this will lead to a \textit{big number of new Higgs bosons}. So a natural question is whether it is possible to reduce the number of independent degrees of freedom in the bidoublet. It will be shown that if the bidoublet is the only responsible for the masses of the quarks this is not the case. To the best of the authors' knowledge, this is the first time that this proof appears in the literature.

The only way to reduce the number of degrees of freedom of the bidoublet is to impose linear constraints between its components, since this is the only option that might be preserved after a gauge transformation. From the transformation rule \(\Phi\rightarrow \exp\{-i\,\beta_L^j\,\sigma^j/2\}\Phi\exp\{i\,\beta_R^k\,\sigma^k/2\}\) of the bidoublet, it is easy to obtain how its components get transformed in the infinitesimal case:

\begin{eqnarray}
\label{Infinitesimal_Transformation_Components}
\left\{
\begin{array}{l}
\phi_1^0\rightarrow\phi^0_1-\frac{i}{2}\left(\beta_L^3-\beta_R^3\right)\phi_1^0-\frac{i}{2}\left(\beta_L^1-i\beta_L^2\right)\phi_2^-+\frac{i}{2}\left(\beta_R^1+i\beta_R^2\right)\phi_1^+ \\
\\
\phi_2^0\rightarrow\phi^0_2+\frac{i}{2}\left(\beta_L^3-\beta_R^3\right)\phi_2^0-\frac{i}{2}\left(\beta_L^1+i\beta_L^2\right)\phi_1^++\frac{i}{2}\left(\beta_R^1-i\beta_R^2\right)\phi_2^- \\
\\
\phi_1^+\rightarrow \phi_1^+-\frac{i}{2}\left(\beta_L^3+\beta_R^3\right)\phi_1^+-\frac{i}{2}\left(\beta_L^1-i\beta_L^2\right)\phi_2^0+\frac{i}{2}\left(\beta_R^1-i\beta_R^2\right)\phi_1^0 \\
\\
\phi_2^-\rightarrow \phi_2^-+\frac{i}{2}\left(\beta_L^3+\beta_R^3\right)\phi_2^--\frac{i}{2}\left(\beta_L^1+i\beta_L^2\right)\phi_1^0+\frac{i}{2}\left(\beta_R^1+i\beta_R^2\right)\phi_2^0
\end{array}\right. .
\end{eqnarray}

\noindent Any relation between the components must preserve the previous equations and can only be imposed between fields of the same electric charge. For instance, the constraint \(\phi_2^0=a\phi_1^0\), with \(a\) an arbitrary complex number, is clearly incompatible with the first two equations in \eqref{Infinitesimal_Transformation_Components}, due to the different sign in the \(i(\beta_L^3-\beta_R^3)/2\) factor. As a consequence, the only possibility is \(\phi_2^0=a\phi_1^{0\dagger}\),\footnote{The most general relation between the neutral components is \(\phi_2^0=a_1\phi_1^{0}+a_2\phi_1^{0^\dagger}\), but having a non-zero value for \(a_1\) is incompatible with the gauge transformations.} but it is also incompatible if no relation is imposed between the charged components, so it is mandatory to simultaneously take \(\phi_2^-=b\left(\phi_1^+\right)^{\dagger}=b\phi_1^-\), with \(b\) an arbitrary complex number. Consider the transformation in Eq.~\eqref{Infinitesimal_Transformation_Components} with these constraints imposed:

\begin{eqnarray}
    \label{Infinitesimal_Transformation_Components_2}
\left\{
\begin{array}{l}
\phi_1^0\rightarrow\phi^0_1-\frac{i}{2}\left(\beta_L^3-\beta_R^3\right)\phi_1^0-\frac{i}{2}\left(\beta_L^1-i\beta_L^2\right)b\phi_1^-+\frac{i}{2}\left(\beta_R^1+i\beta_R^2\right)\phi_1^+ \\
\\
a\phi_1^{0\dagger}\rightarrow a\phi^{0\dagger}_1+\frac{i}{2}\left(\beta_L^3-\beta_R^3\right)a\phi_1^{0\dagger}-\frac{i}{2}\left(\beta_L^1+i\beta_L^2\right)\phi_1^++\frac{i}{2}\left(\beta_R^1-i\beta_R^2\right)b\phi_1^- \\
\\
\phi_1^+\rightarrow \phi_1^+-\frac{i}{2}\left(\beta_L^3+\beta_R^3\right)\phi_1^+-\frac{i}{2}\left(\beta_L^1-i\beta_L^2\right)a\phi_1^{0\dagger}+\frac{i}{2}\left(\beta_R^1-i\beta_R^2\right)\phi_1^0 \\
\\
b\phi_1^-\rightarrow b\phi_1^-+\frac{i}{2}\left(\beta_L^3+\beta_R^3\right)b\phi_1^--\frac{i}{2}\left(\beta_L^1+i\beta_L^2\right)\phi_1^0+\frac{i}{2}\left(\beta_R^1+i\beta_R^2\right)a\phi_1^{0\dagger}
\end{array}\right. .
\end{eqnarray}

\noindent Therefore, it is easy to see that to preserve these relations one must impose \(ba^{\ast}=-1\) and \(b=-a\), which leads to \(a=-b=e^{i\theta}\) with \(\theta\in\mathbb{R}\). This means that the VEVs are related by \(\kappa_2=e^{i\theta}\kappa_1^{\ast}\). Now consider the following gauge transformation of the VEV:

\begin{equation}
    \label{Gauge_transformation_VEV_bidoublet}
    \left<\Phi\right>\rightarrow\exp\left\{-i\beta_L\,\frac{\sigma^3}{2}\right\}\left<\Phi\right>\exp\left\{i\,\beta_R\,\frac{\sigma^3}{2}\right\}=\frac{1}{\sqrt{2}}\left(\begin{array}{cc}
    e^{\frac{i}{2}\left(\beta_R-\beta_L\right)}\kappa_1 & 0 \\
    0 & e^{i\theta}e^{\frac{i}{2}\left(\beta_L-\beta_R\right)}\kappa_1^{\ast}
    \end{array}\right).
\end{equation}

\noindent It is straightforward to see that it is always possible to choose a gauge convention in which \(\kappa_1\) is real. In that situation the mass matrices are \(M_u=\kappa_1(r+e^{-i\theta}s)/\sqrt{2}\) and \(M_d=\kappa_1(e^{i\theta}r+s)/\sqrt{2}\), so \(M_u=e^{-i\theta}M_d\). This implies that the modulus of the eigenvalues of those matrices will be the same, so the up- and down-type quarks will have the same mass, a situation that does not correspond to what is observed in nature. This means that \textit{it is not possible to reduce the number of degrees of freedom of the bidoublet and obtain the correct phenomenological effects if this field is the only responsible for quark masses}.

\subsection{LR models with the bidoublet and two scalar doublets} \label{subsec:doublets}

To spontaneously break the \({\rm SU}(2)_{\rm R}\otimes{\rm U}(1)_{\rm X}\) symmetry into \({\rm U}(1)_{\rm Y}\) one can add a scalar doublet \(\chi_R\sim(\mathbf{1},\mathbf{1},\mathbf{2})_{1/2}\). Using the charge operator \(Q\) it is easy to see that it can be decomposed as

\begin{equation}
\label{Right_Handed_Doublet}
\chi_{R}=\left(
\begin{array}{c}
\chi_{R}^+ \\
\chi_{R}^0
\end{array}\right).
\end{equation}

\noindent Then, in order for \(Q\) not to be a broken generator, the VEV must have the form

\begin{equation}
    \label{VEV_Right_Handed_Doublet}
    \left<\chi_{R}\right>=\frac{1}{\sqrt{2}}\left(
    \begin{array}{c}
    0 \\
    v_{R}
    \end{array}\right).
\end{equation}

\noindent Thus, the only generators that are not broken by the VEV of the \(\chi_R\) doublet are linear combinations of the \(T_L^i\) with \(i=1,2,3\) and \(Y=T_R^3+{\rm X}\), apart from the generators of \({\rm SU}(3)_{\rm QCD}\). However, as already mentioned, the only unbroken generators when considering the VEV of the bidoublet are linear combinations of \(T_L^3+T_R^3\) and X. As a consequence, with the doublet and the bidoublet the only unbroken generator of the generalized electroweak sector is \(Q=T_L^3+T_R^3+{\rm X}\). Therefore, in principle it is not necessary to introduce more degrees of freedom, but another doublet \(\chi_{L}\sim(\mathbf{1},\mathbf{2},\mathbf{1})_{1/2}\) is usually added. The reason for this is that it is necessary if one wants to explicitly preserve the left-right symmetry. Since no additional discrete symmetries are considered in this text, it is not mandatory to introduce the left-handed doublet. Nonetheless, in order to be as general as possible it will be considered. Its properties are analogous to the ones of the right-handed doublet:

\begin{equation}
\label{Left_Handed_Doublet}
\chi_{L}=\left(
\begin{array}{c}
\chi_{L}^+ \\
\chi_{L}^0
\end{array}\right), \hspace{0.8cm}
    \left<\chi_{L}\right>=\frac{1}{\sqrt{2}}\left(
    \begin{array}{c}
    0 \\
    v_{L}
    \end{array}\right).
\end{equation}

To the best of the authors' knowledge, the most general potential including \(\chi_L\), \(\chi_R\) and \(\Phi\) with no additional discrete symmetries was first introduced in \cite{Solera:2023kwt}. It is given by

\begin{equation}
\label{Doublet_Potential}
    \begin{split}
    V_D&=-\mu_1^2\,\mathrm{tr}\left\{\Phi\Phi^{\dagger}\right\}+\left(-\mu_2^2\,\mathrm{tr}\left\{\tilde{\Phi}\Phi^{\dagger}\right\}+\mathrm{h.c.}\right)-\mu^2_L\,\chi_L^{\dagger}\chi_L-\mu^2_R\,\chi_R^{\dagger}\chi_R\\&
    +\lambda_1\,\mathrm{tr}\left\{\Phi\Phi^{\dagger}\right\}^2+\lambda_2\,\mathrm{tr}\left\{\tilde{\Phi}\Phi^{\dagger}\right\}\mathrm{tr}\left\{\Phi\tilde{\Phi}^{\dagger}\right\}
    \\&+\left(\lambda_3\,\mathrm{tr}\left\{\tilde{\Phi}\Phi^{\dagger}\right\}^2+\lambda_4\,\mathrm{tr}\left\{\Phi\Phi^{\dagger}\right\}\mathrm{tr}\left\{\tilde{\Phi}\Phi^{\dagger}\right\}+\mathrm{h.c.}\right)\\&+
    \lambda_L\left(\chi_L^{\dagger}\chi_L\right)^2+\lambda_R\left(\chi_R^{\dagger}\chi_R\right)^2+\lambda_{LR}\left(\chi_L^{\dagger}\chi_L\right)\left(\chi_R^{\dagger}\chi_R\right) \\& +\left(\mu_1'\chi_L^{\dagger}\Phi\chi_R+\mu_2'\chi_L^{\dagger}\tilde{\Phi}\chi_R+\mathrm{h.c.}\right)\\& +\alpha_{1,L}\,\mathrm{tr}\left\{\Phi\Phi^{\dagger}\right\}\chi_L^{\dagger} \chi_L+\left(\alpha_{2,L}\,\mathrm{tr}\left\{\tilde{\Phi}\Phi^{\dagger}\right\}\chi_L^{\dagger}\chi_L+\mathrm{h.c.}\right)
    \\&+\alpha_{3,L}\,\chi_L^{\dagger}\Phi\Phi^{\dagger}\chi_L+\alpha_{4,L}\,\chi_L^{\dagger}\tilde{\Phi}\tilde{\Phi}^{\dagger}\chi_L
    \\&+\alpha_{1,R}\,\mathrm{tr}\left\{\Phi\Phi^{\dagger}\right\}\chi_R^{\dagger}\chi_R+\left(\alpha_{2,R}\,\mathrm{tr}\left\{\tilde{\Phi}\Phi^{\dagger}\right\}\chi_R^{\dagger}\chi_R+\mathrm{h.c.}\right)
    \\&+\alpha_{3,R}\,\chi_R^{\dagger}\Phi^{\dagger}\Phi\chi_R+\alpha_{4,R}\,\chi_R^{\dagger}\tilde{\Phi}^{\dagger}\tilde{\Phi}\chi_R
    ,
    \end{split}
\end{equation}

\noindent where the parameters are real when no hermitian conjugate is explicitly added. Already in the original texts of the LR models \cite{Senjanovic:1975rk} it was shown that, even with simplified versions of this potential, there are configurations of the parameters that can lead to very different energy scales\footnote{This will be explicitly shown in the following section.} \(\left|v_R\right|\gg\left|\kappa_1\right|,\left|\kappa_2\right|,\left|v_L\right|\); something that is required for phenomenological reasons and that would spontaneously break the LR symmetry if imposed. It is important to remark that one can always eliminate some phases of the VEVs in order to simplify the results. For instance, in the previous subsection it was mentioned that the transformation \(\left<\Phi\right>\rightarrow\exp\{-i\beta_L\,\sigma^3/2\}\left<\Phi\right>\exp\{i\,\beta_R\,\sigma^3/2\}\) induces the phase-shifts \(\kappa_1\to\kappa_1\exp\{i(\beta_R-\beta_L)/2\}\) and \(\kappa_2\to\kappa_2\exp\{-i(\beta_R-\beta_L)/2\}\), so it is possible to make either \(\kappa_1\) or \(\kappa_2\) real. Moreover, the VEVs of the doublets get transformed under the \({\rm U}(1)_{\rm X}\) group as \(v_{L,R}\to v_{L,R}\exp\{-i\beta_{\rm X}/2\}\), so it is also possible to cancel one of their complex phases. However, \textit{in this text all of the VEVs will be kept as complex numbers since it does not add difficulties to the analysis done}.

There is no possible way to construct a renormalizable interaction between the fermions and these two doublets, so the analysis done in the previous subsection for the quark sector is valid and the situation with the leptons will be completely analogous. This means that, \textit{in this model all of the fermions are Dirac particles, so there is no natural way to explain why the masses of neutrinos are so small}. The expressions will be the same as the ones for the quarks but changing \(u\to\nu\), \(d\to e\) and \(V_{L,R}^{\rm CKM}\to V_{L,R}^{\rm PMNS}\), with the sole exception of the interactions with the \(Z_R\) gauge boson, which now become

\begin{equation}
    \label{ecu.2.15}    
    \begin{split}
    \mathcal{L}^\ell_{RNC}&=\frac{1}{4}\frac{e}{\cos\theta_W}\tan\gamma\,Z_R^{\mu}\,\bar{e}\gamma_{\mu}\left\{\left(2-\cot^2\gamma\right)-\left(\cot^2\gamma\right)\gamma_5\right\}e
    \\& +\frac{1}{2}\frac{e}{\cos\theta_W}\tan\gamma\, Z_R^{\mu}\,\bar{\nu}_L\gamma_{\mu}\nu_L+\frac{1}{2}\frac{e}{\cos\theta_W}\tan\gamma\left(1+\cot^2\gamma\right)\,Z_R^{\mu}\,\bar{\nu}_R\gamma_{\mu}\nu_R.
    \end{split}
\end{equation}

\noindent There is no interaction at tree level between the \(Z_L\) and the right-handed neutrino if one neglects the \(Z_L-Z_R\) mixing. The \(V_{L,R}^{\rm PMNS}\) are the so-called Pontecorvo-Maki-Nakagawa-Sakata matrices for leptons, which are completely analogous to the \(V_{L,R}^{\rm CKM}\) matrices for quarks. It is always possible to take field redefinitions that make the left-handed matrix a unitary matrix with three real parameters and one complex parameter, but then the right-handed counterpart remains as a general unitary matrix with three real and six complex parameters. The current experimental constraints on the mixing parameters contained in \(V_L^{\rm PMNS}\) coming from neutrino experiments can be found in \cite{ParticleDataGroup:2024cfk}. It is important to remark that \textit{no diagonal structure as the one in the CKM case is found for that matrix}.

The mass matrices for the gauge bosons are obtained from the kinetic terms of the scalars:

\begin{equation}
    \label{Scalar_kinetic_Term_Doublet}
    \mathcal{L}^K_{\textrm{scalar}}=\textrm{tr}\left\{\left(D_{\mu}\Phi\right)^{\dagger}\left(D^{\mu}\Phi\right)\right\}+\left(D_{\mu}\chi_L\right)^{\dagger}\left(D^{\mu}\chi_L\right)+\left(D_{\mu}\chi_R\right)^{\dagger}\left(D^{\mu}\chi_R\right),
\end{equation}

\noindent where the covariant derivatives are given by

\begin{equation}
    \label{Covariant_Derivative_Biboublet}
    D_{\mu}\Phi=\partial_{\mu}\Phi-ig_LW_{L\mu}^k\frac{\sigma^k}{2}\Phi+ig_RW_{R\mu}^k\Phi\frac{\sigma^k}{2}
\end{equation}

 \noindent and

\begin{equation}
    \label{Covariant_Derivative_Doublets}
    D_{\mu}\chi_{L,R}=\left(\partial_{\mu}-ig_{L,R}W_{L,R\mu}^k\frac{\sigma^k}{2}-\frac{i}{2}g_{\rm X}W_{\rm X\mu}\right)\chi_{L,R}.
\end{equation}

\noindent When evaluated in the VEVs, the Lagrangian in Eq.~\eqref{Scalar_kinetic_Term_Doublet} leads to the following mass terms 

\begin{equation}
     \label{Mass_Terms_Gauge_Bosons_Doublets}
     \mathcal{L}^{\textrm{M}}_{\textrm{gauge}}=\left(W_{1\mu}^{\dagger},W_{2\mu}^{\dagger}\right)M_{C}^2\left(
     \begin{array}{c}
     W^{\mu}_1 \\
     W^{\mu}_2
     \end{array}\right)+\frac{1}{2}\left(W_{L\mu}^3,W_{R\mu}^3,W_{\rm X\mu}\right)M_{N}^2\left(
     \begin{array}{c}
     W_{L}^{3\mu} \\
     W_{R}^{3\mu} \\
      W_{\rm X}^{\mu}
      \end{array}\right),
 \end{equation}

\noindent where the mass matrices for the charged and neutral gauge bosons \(M_C^2\) and \(M_N^2\) are given by

\begin{eqnarray}
    \label{Mass_Matrix_Doublets_Gauge_Charged}
    M_{C}^2=\frac{1}{4}\,\left(
    \begin{array}{cc}
    g_L^2(\left|\kappa_1\right|^2+\left|\kappa_2\right|^2+\left|v_L\right|^2) & -2g_Lg_R\kappa_1^{\ast}\kappa_2 \\
    -2g_Lg_R\kappa_1\kappa_2^{\ast} & g_R^2(\left|\kappa_1\right|^2+\left|\kappa_2\right|^2+\left|v_R\right|^2)
    \end{array}\right),
\end{eqnarray} 

\begin{equation}
    \label{Mass_Matrix_Doublets_Gauge_Charged}
    M_{N}^2=\frac{1}{4}\left(
    \begin{array}{ccc}
    g^2_L(\left|\kappa_1\right|^2+\left|\kappa_2\right|^2+\left|v_L\right|^2) & -g_Lg_R(\left|\kappa_1\right|^2+\left|\kappa_2\right|^2) & -g_Lg_{\rm X}\left|v_L\right|^2 \\
    -g_Lg_R(\left|\kappa_1\right|^2+\left|\kappa_2\right|^2) & g_R^2(\left|\kappa_1\right|^2+\left|\kappa_2\right|^2+\left|v_R\right|^2) & -g_Rg_{\rm X}\left|v_R\right|^2  \\ -g_Lg_{\rm X}\left|v_L\right|^2 &
    -g_Rg_{\rm X}\left|v_R\right|^2 & g_{\rm X}^2(\left|v_L\right|^2+\left|v_R\right|^2)
    \end{array}\right).
\end{equation}

 The exact mass eigenstates \(W_{L,R}^{\pm}\) at tree level for the charged gauge bosons are obtained using the transformations

\begin{equation}
    \label{Diagonalization_Charged_Gauge}
    \left(
    \begin{array}{c}
    W_L^{\mu}  \\
    W_R^{\mu} 
    \end{array}\right)=\left(
    \begin{array}{cc}
    \cos\xi_W & -\sin\xi_W\,e^{-i\lambda}  \\
    \sin\xi_W\,e^{i\lambda} & \cos\xi_W
    \end{array}\right)\left(
    \begin{array}{c}
    W_1^{\mu}   \\
    W_2^{\mu}  
    \end{array}\right),
\end{equation}

\noindent where, defining \(\kappa_1\kappa_2^{\ast}\coloneqq\left|\kappa_1\right|\left|\kappa_2\right|e^{i\beta}\), it is easy to see that

\begin{equation}
    \label{ecu.2.15}
    \lambda=\beta,\hspace{0.8cm}\tan{2\xi_W}=\frac{4g_Lg_R|\kappa_1||\kappa_2|}{g_R^2|v_R|^2-g_L^2|v_L|^2+\left(g_R^2-g_L^2\right)\left(|\kappa_1|^2+|\kappa_2|^{2}\right)}.
\end{equation}

\noindent The mass matrix of the neutral gauge bosons is diagonalized using the transformation

\begin{equation}
    \label{Transformation_Nuetral_Gauge}
    \left(\!\!
    \begin{array}{c}
    Z_L^{\mu} \\
    Z_R^{\mu} \\
    A^{\mu}
    \end{array}\!\!\right)=\left(
    \begin{array}{ccc}
    c_\alpha & -s_\alpha & 0 \\
    s_\alpha & c_\alpha & 0 \\
    0 & 0 & 1
    \end{array}\right)\left(
    \begin{array}{ccc}
    c_{\theta_W} & 0 & -s_{\theta_W} \\
    0 & 1 & 0 \\
    s_{\theta_W} & 0 & c_{\theta_W}
    \end{array}\right)\left(
    \begin{array}{ccc}
    1 & 0 & 0 \\
    0 & c_\gamma & -s_\gamma  \\
    0 & s_\gamma & c_\gamma
    \end{array}\right)\left(\!\!
    \begin{array}{c}
    W_L^{3\mu} \\
    W_R^{3\mu} \\
    W_{X}^{\mu}
    \end{array}\!\!\right),
\end{equation}

\noindent where \(c_\alpha\coloneqq\cos \alpha\), etc. The \(\alpha\) parameter satisfies the relation

{\small\begin{equation}
    \label{ZL_ZR_Mixing_Doublets}
    \tan(2\alpha)=\frac{2(g_{\rm X}^2|v_L|^2-g_R^2(|\kappa_1|^2+|\kappa_2|^2))\dfrac{g_Lg_Rg_{\rm X}}{e}}{(g_R^2+g_X^2)^2|v_R|^2+g_{\rm X}^4|v_L|^2+g_R^4(|\kappa_1|^2+|\kappa_2|^2)-\left(\dfrac{g_Lg_Rg_{\rm X}}{e}\right)^2(|\kappa_1|^2+|\kappa_2|^2+|v_L|^2)},
\end{equation}}

\noindent where the formula 

\begin{equation}
    \label{Quadratic_Relation_Couplings}
    \left(\frac{g_Lg_Rg_{\rm X}}{e}\right)^2=(g_Lg_R)^2+(g_Lg_{\rm X})^2+(g_{R}g_{\rm X})^2,
\end{equation}

\noindent which can be easily derived from Eq.~\eqref{Couplings_Relations}, has been used to simplify the expressions. 

It is clear that \textit{both the \(W_L-W_R\) and \(Z_L-Z_R\) mixings are highly suppressed if \(\left|v_R\right|\gg\left|\kappa_1\right|,\left|\kappa_2\right|,\left|v_L\right|\)}. Indeed, in this scenario, they behave as \(|v_{EW}/v_R|^2\). Furthermore, the mass eigenstates are 

\begin{equation}
    \label{Mass_Eigenstates_Gauge_Boson_Doublet}
    \begin{array}{ll}
    M_{W_L}^2\approx\dfrac{1}{4}g_L^2\left\{\left|\kappa_1\right|^2+\left|\kappa_2\right|^2+\left|v_L\right|^2\right\}, &
    M^2_{W_R}\approx\dfrac{1}{4}g_R^2\left|v_R\right|^2, \\
    M_{Z_L}^2\approx\dfrac{g_L^2g_R^2+g_L^2g_{\rm X}^2+g_R^2g_{\rm X}^2}{4\left(g_{\rm X}^2+g_R^2\right)}\left\{\left|\kappa_1\right|^2+\left|\kappa_2\right|^2+\left|v_L\right|^2\right\}, &
    M_{Z_R}^2\approx\dfrac{1}{4}\left(g_R^2+g_{\rm X}^2\right)\left|v_R\right|^2, \\
    \end{array}
\end{equation}

\noindent plus corrections of order \(\mathcal{O}(|v_R|^{-2})\) for the \(W_L\) and \(Z_L\) masses; and corrections of order \(\mathcal{O}(|v_{EW}|^{2})\) for the masses of the \(W_R\) and \(Z_R\). This means that the electroweak scale can be identified with \(v_{EW}^2=\left|\kappa_1\right|^2+\left|\kappa_2\right|^2+\left|v_L\right|^2\) and, at leading order in \(|v_{EW}/v_R|^2\), one can find

\begin{equation}
    \label{ZWMass_Relations_Doublets}
    M_{W_{L}}\approx M_{Z_L}\cos\theta_W,\hspace{0.8cm} M_{W_R}\approx M_{Z_{R}}\cos\gamma,
\end{equation}

\noindent meaning that the \(\rho\) parameter defined as \(\rho \coloneqq M_{W_L}^2/(M_{Z_L}^2\cos^2\theta_W)\) is equal to one at tree level with the approximations done. In the following sections a recast of the collider bounds in the general scenario on the masses of the heavy \(W_R\) and \(Z_R\) gauge bosons is presented. Furthermore, it is easy to see that \textit{this model contains six neutral and two charged scalars}. The only way to reduce the number of degrees of freedom is to eliminate the \(\chi_L\) doublet, leading to a scalar content with one charged and four neutral scalars. It is important to remark that \textit{only one of those scalars has a mass proportional to the electroweak scale}, so it is identified with the Higgs field of the SM. The rest of them have a mass proportional to \(|v_R|\).

\subsection{LR models with the bidoublet and two scalar triplets} \label{subsec:triplets}

The main problem with the SSB mechanism with two doublets is that it cannot explain why neutrino masses are small. Instead of using the \(\chi_R\) scalar to produce the \({\rm SU}(2)_{\rm R}\otimes{\rm U}(1)_{\rm X}\to{\rm U}(1)_{\rm Y}\) SSB pattern, it is possible to introduce a right-handed scalar triplet \(\Delta_R\sim(\mathbf{1},\mathbf{1},\mathbf{3})_{1}\). Again, adding the left-handed counterpart \(\Delta_L\sim(\mathbf{1},\mathbf{3},\mathbf{1})_{1}\) is not mandatory, but in order to be as general as possible, it will be done in this text. Both triplets can be written as

\begin{equation}
    \label{Right_Triplet}
    \Delta_{L,R}=\left(
    \begin{array}{cc}
    \delta_{L,R}^+/\sqrt{2}     &  \delta^{++}_{L,R} \\
    \delta_{L,R}^0     &  -\delta_{L,R}^+/\sqrt{2}
    \end{array}
    \right).
\end{equation}

\noindent In order not to break the \({\rm U}(1)_{\rm QED}\) generator \(Q\), the VEVs must have the form

\begin{equation}
    \label{Triplet_VEV}
    \left<\Delta_{L,R}\right>=\frac{1}{\sqrt{2}}\left(
    \begin{array}{cc}
    0 & 0 \\
    v_{L,R} & 0 
    \end{array}\right).
\end{equation}

\noindent In this scenario, the author of this text found that the most general potential is 

\begin{equation}
\label{Scalar_Potential_Triplets}
    \begin{split}
    V_T&=-\mu_1^2\,\mathrm{tr}\left\{\Phi\Phi^{\dagger}\right\}-\left(\mu_2^2\,\mathrm{tr}\left\{\tilde{\Phi}\Phi^{\dagger}\right\}+\mathrm{h.c.}\right)-\mu^2_L\,\mathrm{tr}\left\{\Delta_L\Delta_L^{\dagger}\right\}-\mu^2_R\,\mathrm{tr}\left\{\Delta_R\Delta_R^{\dagger}\right\}\\&
    +\lambda_1\,\mathrm{tr}\left\{\Phi\Phi^{\dagger}\right\}^2+\lambda_2\,\mathrm{tr}\left\{\tilde{\Phi}\Phi^{\dagger}\right\}\mathrm{tr}\left\{\Phi\tilde{\Phi}^{\dagger}\right\}
    \\&+\left(\lambda_3\,\mathrm{tr}\left\{\tilde{\Phi}\Phi^{\dagger}\right\}^2+\lambda_4\,\mathrm{tr}\left\{\Phi\Phi^{\dagger}\right\}\mathrm{tr}\left\{\tilde{\Phi}\Phi^{\dagger}\right\}+\mathrm{h.c.}\right)\\&
    +\rho_{1,L}\,\mathrm{tr}\left\{\Delta_L\Delta_L^{\dagger}\right\}^2+\rho_{2,L}\,\mathrm{tr}\left\{\Delta_L\Delta_L\right\}\mathrm{tr}\left\{\Delta_L^{\dagger}\Delta_L^{\dagger}\right\}\\&+\rho_{1,R}\,\mathrm{tr}\left\{\Delta_R\Delta_R^{\dagger}\right\}^2+\rho_{2,R}\,\mathrm{tr}\left\{\Delta_R\Delta_R\right\}\mathrm{tr}\left\{\Delta_R^{\dagger}\Delta_R^{\dagger}\right\}\\& +\rho_3\,\mathrm{tr}\left\{\Delta_L\Delta_L^{\dagger}\right\}\left\{\Delta_R\Delta_R^{\dagger}\right\}+\left(\rho_4\,\mathrm{tr}\left\{\Delta_L\Delta_L\right\}\left\{\Delta_R^{\dagger}\Delta_R^{\dagger}\right\}+\mathrm{h.c.}\right) \\&+\alpha_{1,L}\,\mathrm{tr}\left\{\Phi\Phi^{\dagger}\right\}\mathrm{tr}\left\{\Delta_L\Delta_L^{\dagger}\right\}+\left(\alpha_{2,L}\,\mathrm{tr}\left\{\tilde{\Phi}\Phi^{\dagger}\right\}\mathrm{tr}\left\{\Delta_L\Delta_L^{\dagger}\right\}+\mathrm{h.c.}\right)
    \\&+\alpha_{1,R}\,\mathrm{tr}\left\{\Phi\Phi^{\dagger}\right\}\mathrm{tr}\left\{\Delta_R\Delta_R^{\dagger}\right\}+\left(\alpha_{2,R}\,\mathrm{tr}\left\{\tilde{\Phi}\Phi^{\dagger}\right\}\mathrm{tr}\left\{\Delta_R\Delta_R^{\dagger}\right\}+\mathrm{h.c.}\right)
    \\&+\alpha_{3,L}\,\mathrm{tr}\left\{\Phi\Phi^{\dagger}\Delta_L\Delta_L^{\dagger}\right\}
    +\alpha_{3,R}\,\mathrm{tr}\left\{\Phi^{\dagger}\Phi\Delta_R\Delta_R^{\dagger}\right\}\\& + \left(\beta_1\mathrm{tr}\left\{\Phi\Delta_R\Phi^{\dagger}\Delta_L^{\dagger}\right\}+\beta_2\mathrm{tr}\left\{\tilde{\Phi}\Delta_R\Phi^{\dagger}\Delta_L^{\dagger}\right\}+\beta_3\mathrm{tr}\left\{\Phi\Delta_R\tilde{\Phi}^{\dagger}\Delta_L^{\dagger}\right\}+\mathrm{h.c.}\right),
    \end{split}
\end{equation}

\noindent  where the coefficients are real when the hermitian conjugated is not explicitly added.\footnote{The properties of the scalar spectrum in relation to the imposed discrete symmetries were studied in \cite{Branco:1999fs,Barenboim:2001vu}.}

The kinetic term of the triplets is given by

\begin{equation}
\label{e}
    \mathcal{L}_{K,\,\Delta}=\mathrm{tr}\left\{\left(D_{\mu}\Delta_{L}\right)^{\dagger}\left(D^{\mu}\Delta_{L}\right)\right\}+\mathrm{tr}\left\{\left(D_{\mu}\Delta_{R}\right)^{\dagger}\left(D^{\mu}\Delta_{R}\right)\right\},
\end{equation}

\noindent where the covariant derivatives can be written as

\begin{equation}
    \label{Covariant_Derivative_Triplets}
    D^{\mu}\Delta_{L,R}=\partial^{\mu}\Delta_{L,R}-ig_{L,R}\left[W_{L,R}^{k\mu}T^k_{L,R},\Delta_{L,\,R}\right]-ig_{\rm X}W_{\rm X}^{\mu}\Delta_{L,R}.
\end{equation}

\noindent Therefore, the mass matrices for charged and neutral gauge bosons are, respectively:

\begin{eqnarray}
    \label{Mass_Matrix_Triplets_Gauge_Charged}
    M_{C}^2=\frac{1}{4}\,\left(
    \begin{array}{cc}
    g_L^2(\left|\kappa_1\right|^2+\left|\kappa_2\right|^2+2\left|v_L\right|^2) & -2g_Lg_R\kappa_1^{\ast}\kappa_2 \\
    -2g_Lg_R\kappa_1\kappa_2^{\ast} & g_R^2(\left|\kappa_1\right|^2+\left|\kappa_2\right|^2+2\left|v_R\right|^2)
    \end{array}\right),
\end{eqnarray}

\begin{equation}
    \label{Mass_Matrix_Triplets_Gauge_Charged}
    M_{N}^2=\frac{1}{4}\left(
    \begin{array}{ccc}
    g^2_L(\left|\kappa_1\right|^2+\left|\kappa_2\right|^2+4\left|v_L\right|^2) & -g_Lg_R(\left|\kappa_1\right|^2+\left|\kappa_2\right|^2) & -4g_Lg_{\rm X}\left|v_L\right|^2 \\
    -g_Lg_R(\left|\kappa_1\right|^2+\left|\kappa_2\right|^2) & g_R^2(\left|\kappa_1\right|^2+\left|\kappa_2\right|^2+4\left|v_R\right|^2) & -4g_Rg_{\rm X}\left|v_R\right|^2  \\ -4g_Lg_{\rm X}\left|v_L\right|^2 &
    -4g_Rg_{\rm X}\left|v_R\right|^2 & 4g_{\rm X}^2(\left|v_L\right|^2+\left|v_R\right|^2)
    \end{array}\right).
\end{equation}

\noindent It is easy to see that they are equivalent to the ones in the model with doublets but with the exchange \(|v_{L,R}|^2\to2|v_{L,R}|^2\) for the charged bosons and \(|v_{L,R}|^2\to4|v_{L,R}|^2\) for the neutral ones. Then, the mass eigenstates can be found making those transformations in the expressions for the doublets model. In particular, the masses of the bosons are, at leading order in \(|v_{EW}/v_R|^2\):

\begin{equation}
    \label{Mass_Eigenstates_Gauge_Boson_Triplet}
    \begin{array}{ll}
    M_{W_L}^2\approx\dfrac{1}{4}g_L^2\left\{\left|\kappa_1\right|^2+\left|\kappa_2\right|^2+2\left|v_L\right|^2\right\}, &
    M^2_{W_R}\approx\dfrac{1}{2} g_R^2\left|v_R\right|^2, \\
    M_{Z_L}^2\approx\dfrac{g_L^2g_R^2+g_L^2g_{\rm X}^2+g_R^2g_{\rm X}^2}{4\left(g_{\rm X}^2+g_R^2\right)}\left\{\left|\kappa_1\right|^2+\left|\kappa_2\right|^2+4\left|v_L\right|^2\right\}, &
    M_{Z_R}^2\approx\left(g_R^2+g_{\rm X}^2\right)\left|v_R\right|^2. \\
    \end{array}
\end{equation}

\noindent This means that, in order to have \(\rho \coloneqq M_{W_L}^2/(M_{Z_L}^2\cos^2\theta_W)\approx 1\) in this scenario, \textit{it is necessary to impose the condition \(|v_R|^2\gg|\kappa_1|^2+|\kappa_2|^2=v_{EW}^2\gg|v_L|^2\)}, which is radiatively stable \cite{Mohapatra:1986uf,Mohapatra:1998rq}. It is possible to prove that there are configurations of the parameters in Eq.~\eqref{Scalar_Potential_Triplets} that can fulfill this requirement \cite{Barenboim:1996rn,Barenboim:2001vu}. Now the masses of the \(Z\) and \(W\) gauge bosons with this approximation are related by:

\begin{equation}
    \label{ZWMass_Relations_Triplets}
    M_{W_{L}}\approx M_{Z_L}\cos\theta_W,\hspace{0.8cm} M_{W_R}\approx\frac{1}{\sqrt{2}}M_{Z_{R}}\cos\gamma.
\end{equation}

\noindent Finally, it is straightforward to show that, in this scenario, \textit{there are six neutral, two singly-charged and two doubly-charged physical Higgs bosons}.

The Yukawa sector for quarks in the models with doublets and triplets are equal. However, it is possible to add the following operators that combine the scalar triplets and leptons:

\begin{equation}
    \label{Triplet_Lepton_Yukawa}
    \mathcal{L}_{\rm Majorana}=-\sum_{m,n}\left\{\bar{\ell}_R^{\,c\,m}\left(h_R\right)_{mn}\,i\sigma^2\Delta_R\,\ell_R^{\,n}+\bar{\ell}_L^{\,c\,m}\left(h_L\right)_{mn}\,i\sigma^2\Delta_L\,\ell_L^{\,n}\right\}+\textrm{h.c.}
\end{equation}

\noindent As a consequence, \textit{while in the model with doublets both baryon and lepton number where conserved, in the model with triplets the operators in Eq.~\eqref{Triplet_Lepton_Yukawa} violate lepton number in two units, but the baryon number is still preserved}. When considering the VEVs of the scalars, the combination of the Yukawa term with the bidoublet and with the triplets leads to the following Majorana mass term for neutrinos:

\begin{eqnarray}
    \label{Majora_Neutrinos}
    \mathcal{L}_{m,\,\nu}=-\frac{1}{2}\left(\bar{\nu}_L,\bar{\nu}_R^{\,c}\right)\,M_{\nu}\left(
    \begin{array}{c}
    \nu_L^{\,c} \\
    \nu_{R}
    \end{array}
    \right) +\textrm{h.c.}, \hspace{0.8cm}
    M_{\nu}\coloneqq\left(
    \begin{array}{cc}
    M_L & M_D \\
    M^T_D & M_R
    \end{array}
    \right),
\end{eqnarray}

\noindent with \(M_D\coloneqq\left(t\kappa_1+u\kappa_2^{\ast}\right)/\sqrt{2}\), \(M_L\coloneqq\sqrt{2}v_Lh_L\), \(M_R\coloneqq\sqrt{2}v_Rh_R\) and \(t\) and \(u\) are the analogous of the \(r\) and \(s\) Yukawa matrices for quarks. The mass terms for charged leptons are the same as in the doublets model. Moreover, since \(\bar{\psi}^{c\,m}_{L,R}\psi_{L,R}^n=\bar{\psi}_{L,R}^{c\,n}\psi_{L,R}^m\), it is always possible to take symmetric \(h_{L,R}\) matrices, so the whole \(M_{\nu}\) mass matrix is symmetric. 

Now it is necessary to diagonalize the mass matrices of leptons. Again, since the kinetic terms have a \({\rm U(3)}^2_f\) flavor symmetry in the leptonic sector, it is possible to take a transformation \(\ell_{L,R}^{i}\rightarrow(A_{L,R}^{e}K^e)^{i}_{j}\,\ell_{L,R}^{j}\), where the \(A_{L,R}^{e}\) are unitary matrices and \(K^e\) is a diagonal matrix of phases, that diagonalizes the mass matrix of the charged leptons and renders it positive. This induces the transformations \(\tilde{M}_D=K^{e\dagger}(A_L^e)^{\dagger}M_DA_R^eK^e\) and \(\tilde{M}_{L,R}=K^e(A_{L,R}^e)^{\textrm{T}}M_{L,R}A_{L,R}^eK^e\) in the neutrino mass matrices, but \(\tilde{M}_{\nu}\) is still symmetric. Generally speaking, the diagonalization of \(\tilde{M}_{\nu}\) requires a \(6\times6\) transformation. Nevertheless, in the case where the eigenvalues of \(\tilde{M}_R\) are much bigger than the ones of \(\tilde{M}_{D}\) and \(\tilde{M}_L\), the transformation

\begin{eqnarray}
    \label{Neutrino_Seesaw_Transformation}
    \left(
    \begin{array}{c}
    \nu_l \\
    \nu_h
    \end{array}
    \right)=\left(
    \begin{array}{cc}
    \mathbb{I} & -\tilde{M}_D^{\ast}\left(\tilde{M}_R^{-1}\right)^{\ast} \\
    \tilde{M}_R^{-1} \tilde{M}_D^{\textrm{T}} & \mathbb{I}
    \end{array}\right)\left(
    \begin{array}{c}
    \nu_L^{\,c} \\
    \nu_R
    \end{array}\right)
\end{eqnarray}

\noindent approximately leads to the following mass terms

\begin{equation}
    \label{Seesaw_Neutrino_Mass_Lagrangian}
    \mathcal{L}_{m,\,\nu}\approx-\frac{1}{2}\left(\bar{\nu}_l^{\,c},\,\bar{\nu}_h^{\,c}\right)\left(
    \begin{array}{cc}
    \tilde{M}_L-\tilde{M}_D\tilde{M}_R^{-1}\tilde{M}_D^{\textrm{T}} & 0 \\
    0 & \tilde{M}_R
    \end{array}\right)\left(
    \begin{array}{c}
    \nu_l \\
    \nu_h
    \end{array}\right)+{\rm h.c.}
\end{equation}

\noindent This approximation holds because \(|v_R|\gg|\kappa_1|,|\kappa_2|\gg|v_L|\). Furthermore, it is clear that the heavy neutrinos \(\nu_h\) have masses proportional to \(|v_R|\), while the ones of the light neutrinos \(\nu_l\) behave as \(v_L-v_{EW}^2/v_R\). Neglecting the value of \(|v_L|\) for simplicity of the discussion, one can find that the heavier the \(\nu_h\) neutrinos are, the lighter the \(\nu_l\) are. Moreover, from Eq.~\eqref{Neutrino_Seesaw_Transformation} it is easy to see that the \(\nu_l\) can approximately be identified with the \(\nu_L^c\) and \(\nu_h\) with the right-handed \(\nu_R\). This is the \textit{specific realization of the  seesaw mechanism in the LR models, which explains why the mass of the observed neutrinos are so small}. It is a type I seesaw mechanism due to the presence of the heavy right-handed neutrino, but the non-vanishing VEV of \(\Delta_L\) leads to a direct type II seesaw Majorana mass term \cite{Langacker:2017uah}.

Finally, one still needs to diagonalize the mass matrices of the light and heavy neutrinos. Since they are both symmetric, this is done by introducing the transformations  \(\nu_l\rightarrow A_l^{\nu}\,\nu_l\) and \(\nu_h\rightarrow A_h^{\nu}\,\nu_h\), where \(A_l^{\nu}\) and \(A_h^{\nu}\) are unitary matrices. Here, no additional diagonal matrix of phases can be added because they do not leave invariant the mass term. Taking the approximations \(\nu_l\approx\nu_L^c\) and \(\nu_h\approx\nu_R\) one can show that the gauge charged currents can be written as

\begin{equation}
    \label{ecu.1.96}
    \mathcal{L}^{I}_{CC,\,\ell}\approx\frac{1}{\sqrt{2}}\left\{g_L\bar{\nu}_l^{c}\gamma_{\mu}W_{1}^{\mu\dagger}V_l^{\textrm{PMNS}}e_L+g_R\bar{\nu}_h\gamma_{\mu}W_{2}^{\mu\dagger}V_h^{\textrm{PMNS}}e_R\right\}+\textrm{h.c.},
\end{equation}

\noindent where \(V_l^{\textrm{PMNS}}\coloneqq\left(A_l^{\nu}\right)^{\textrm{T}}K^e\) and \(V_h^{\textrm{PMNS}}\coloneqq\left(A_h^{\nu}\right)^{\dagger}K^e\). Now, it is possible to use the freedom in choosing \(K^e\) and the fact that there is no non-physical phase associated with the conservation of lepton number, to eliminate three complex parameters in \(V_l^{\rm PMNS}\).  A usual parametrization for this matrix is the Chau-Keung parametrization:

 \begin{equation}
 \label{Majorana_PMNS_Matrix}
 \begin{split}
     V_{l}^{\textrm{PMNS}}&=\left(
     \begin{array}{ccc}
         1 & 0 & 0 \\
         0 & c_{23} & s_{23} \\
         0 & -s_{23} & c_{23}
     \end{array}
     \right)\left(
     \begin{array}{ccc}
         c_{13} & 0 & s_{13}\,e^{-i\delta} \\
         0 & 1 & 0 \\
         -s_{13}\,e^{i\delta} & 0 & c_{13}
     \end{array}
     \right)\left(
     \begin{array}{ccc}
         c_{12} & s_{12} & 0 \\
         -s_{12} & c_{12} & 0\\
         0 & 0 & 1
     \end{array}
     \right)\\&\times\left(
     \begin{array}{ccc}
     e^{i\alpha_1} & 0 & 0 \\
     0 & e^{i\alpha_2} & 0 \\
     0 & 0 & 1
     \end{array}\right),
     \end{split}
\end{equation}

\noindent where \(\alpha_1\) and \(\alpha_2\) are the \textit{Majorana phases} that do not appear if neutrinos are Dirac fermions. The right-handed counterpart will remain as a general unitary matrix.

\section{\boldmath The \(\chi_L+\chi_R\) Effective LR Model} \label{sec:Effective_LR}

As already mentioned, the usual realizations of the SSB of the gauge symmetry of the LR models, which have been presented in the previous section, require a big number of new scalar degrees of freedom: six neutral and two singly-charged Higgs fields in the doublets case, plus two additional doubly-charged bosons in the scenario with triplets. Furthermore, if no additional discrete symmetries are imposed in the potentials in Eqs.~\eqref{Doublet_Potential} and~\eqref{Scalar_Potential_Triplets}, \textit{the diagonalization of the mass matrices (particularly the one of the neutral scalars) is highly non-trivial}. These are the reasons why a model with only the minimal scalar sector required to spontaneously break the symmetry was introduced in \cite{Solera:2023kwt} by the author of this text: \textit{the \(\chi_L+\chi_R\) effective LR model}. In this theory, the \(\chi_R\) doublet is used to produce the \({\rm SU}(2)_{\rm R}\otimes{\rm U}(1)_{\rm X}\to{\rm U}(1)_{\rm Y}\) SSB pattern and only the \(\chi_L\) is added to complete the \({\rm SU}(2)_{\rm L}\otimes{\rm U}(1)_{\rm Y}\to{\rm U}(1)_{\rm QED}\) breaking. The scalar sector is so simple that one can take a gauge \({\rm SU}(2)_{\rm L}\otimes{\rm SU}(2)_{\rm R}\) transformation to write the doublets as

\begin{equation}
    \label{Unitary_Gauge_Effective_LR}
    \chi_{L,R}=\frac{1}{\sqrt{2}}\left(
    \begin{array}{c}
    0 \\
    v_{L,R}+\chi_{L,R}^{0r}
    \end{array}
    \right),
\end{equation}

\noindent where both \(v_{L,R}\) and \(\chi_{L,R}^{0r}\)  are real, so \textit{this model contains only two physical neutral scalars}. This gauge fixing condition is analogous to the one of the unitary gauge of the SM. 

The mass matrices of the gauge bosons can be obtained considering the ones of the model with two doublets and the bidoublet taking the limit \(\kappa_{1,2}\to0\). Then, it is clear that the \(W_{1,2}^{\pm}\) are the exact \(W_{L,R}^{\pm}\) charged mass eigenstates at tree level and their masses are given by

\begin{equation}
    \label{W_Bosons_Mass_Effective_LR}
    M_{W_L}=\frac{1}{2}g_Lv_L,\hspace{0.8cm} M_{W_R}=\frac{1}{2}g_Rv_R,
\end{equation}

\noindent with no further approximations. As a consequence, \textit{there is no \(W_L-W_R\) mixing at tree-level}. Therefore, \(v_L\) is identified with the electroweak scale and, for phenomenological reasons, the condition \(v_R\gg v_L\) is assumed. The mass matrix for the neutral gauge bosons is 

\begin{equation}
    \label{Mass_Matrix_Neutral_Gauge_Effective_LR}
    M_N^2=\frac{1}{4}\left(
    \begin{array}{ccc}
    g_L^2v_L^2 & 0 & -g_Lg_{\rm X}v_L^2 \\
    0 & g_R^2v_R^2 & -g_{R}g_{\rm X}v_R^2 \\
    -g_Lg_{\rm X}v_L^2 & -g_{R}g_{\rm X}v_R^2 & g_{\rm X}^2\left(v_L^2+v_R^2\right)
    \end{array}
    \right),
\end{equation}

\noindent and it is diagonalized by the transformation in Eq.~\eqref{Transformation_Nuetral_Gauge} with the \(\alpha\) parameter that mediates the \(Z_L-Z_R\) mixing satisfying the condition:

\begin{equation}
    \label{ZL_ZR_Mixing_Effective_LR}
    \tan\left(2\alpha\right)=\frac{2g_{\rm X}^2\sqrt{g_L^2g_{\rm X}^2+g_{R}^2g_{\rm X}^2+g_R^2g_L^2}\,v_L^2}{\left(g_R^2+g_{\rm X}^2\right)^2v_R^2-\left(g_L^2g_{\rm X}^2+g_{R}^2g_{\rm X}^2+g_R^2g_L^2-g_{\rm X}^4\right)v_L^2}.
\end{equation}

\noindent Assuming that \(v_R\gg v_L\) the mass eigenstates are approximately:

\begin{equation}
    \label{Z_Boson_Masses_Effective_LR}
    M_{Z_R}^2\approx\frac{1}{4}\left(g_R^2+g_{\rm X}^2\right)v_R^2,\hspace{0.8cm}M^2_{Z_L}\approx\frac{1}{4}\frac{g_L^2g_{\rm X}^2+g_{R}^2g_{\rm X}^2+g_R^2g_L^2}{g_R^2+g_{\rm X}^2}\,v_L^2.
\end{equation}

\noindent Obviously the relations \(M_{W_L}\approx M_{Z_L}\cos\theta_W\) and \(M_{W_{R}}\approx M_{Z_R}\cos\gamma\) are also satisfied in this model.

The most general renormalizable scalar potential with two doublets is

\begin{equation}
    \label{Scalar_Potential_Doublets}
    V=-\mu_L^2\chi_L^{\dagger}\chi_L-\mu_R^2\chi_R^{\dagger}\chi_R+\lambda_L(\chi_L^{\dagger}\chi_L)^2+\lambda_R(\chi_R^{\dagger}\chi_R)^2+\lambda_{LR}(\chi_L^{\dagger}\chi_L)(\chi_R^{\dagger}\chi_R).
\end{equation}

\noindent Hermiticity implies that all the parameters must be real. Furthermore, imposing that it must be bounded from below, one gets the following conditions: \(\lambda_L>0\), \(\lambda_R>0\) and \(4\lambda_L\lambda_R-\lambda_{LR}^2>0\). These requirements also lead to a positive definite mass matrix for the scalars. The minima of the potential are obtained from solving

\begin{equation}
    \label{Minima_Potential_Effective_LR}
    \left\{
    \begin{array}{c}
    v_L\left(-\mu_L^2+\lambda_Lv_L^2+\lambda_{LR}v_R^2/2\right)=0 \\
    \\
    v_R\left(-\mu_R^2+\lambda_Rv_R^2+\lambda_{LR}v_L^2/2\right)=0
    \end{array}\right. .
\end{equation}

\noindent Having vanishing values for \(v_L\) or \(v_{R}\) will imply, for instance, that the charged gauge bosons are massless. Thus, the physical solutions are

 \begin{equation}
     \label{Minima_Solutions_Effective_LR}
     v_L^2=\frac{4\lambda_R\mu_L^2-2\lambda_{LR}\mu_R^2}{4\lambda_L\lambda_R-\lambda_{LR}^2},\hspace{0.8cm} v_R^2=\frac{4\lambda_L\mu_R^2-2\lambda_{LR}\mu_L^2}{4\lambda_L\lambda_R-\lambda_{LR}^2}.
 \end{equation}

\noindent As pointed out in the previous section, it is not difficult to find configurations of the parameter space that lead to the condition \(v_R\gg v_L\). For instance, one can take \(\lambda_{LR}\ll \lambda_L,\,\lambda_R\) and \(\lambda_L\mu_R^2\gg\lambda_R\mu_L^2\). Nonetheless, \textit{it is not possible to satisfy this requirement having a LR symmetric potential}, while for the models with the bidoublet this is perfectly possible. In the basis \(\left\{\chi_R^{0r},\,\chi_L^{0r}\right\}\) the mass matrix of the scalars is

\begin{equation}
    \label{Mass_Matrix_Scalars_Effective_LR}
    M_{NS}^2=\left(
    \begin{array}{cc}
    2\lambda_Rv_R^2 & v_Lv_R\lambda_{LR} \\
    v_Lv_R\lambda_{LR} & 2\lambda_L v_L^2
    \end{array}\right).
\end{equation}

\noindent It can be easily diagonalized just by taking the transformation

\begin{equation}
    \label{NS_Mass_Eigenstates_Effective_LR}
    \left(
    \begin{array}{c}
    H \\
    h
    \end{array}\right)=\left(\begin{array}{cc}
    \cos\theta & -\sin\theta \\
    \sin\theta & \cos\theta
    \end{array}\right)\left(\begin{array}{c}
    \chi_R^{0r} \\
    \chi_L^{0r}
    \end{array}\right)
\end{equation}

\noindent with the mixing parameter \(\theta\) satisfying

\begin{equation}
    \label{Mixing_NS_Effective_LR}
    \tan\left(2\theta\right)=\frac{v_Lv_R\lambda_{LR}}{\lambda_Lv_L^2-\lambda_Rv_R^2}.
\end{equation}

\noindent Therefore, the scalar mixing behaves as \(v_L/v_R\). In the limit \(v_R\gg v_L\) \(H\) can be identified with a heavy Higgs boson while \(h\) will be a SM-like Higgs.\footnote{The hierarchy problem also appears in LR models. The mass of the light Higgs in these models is raised up to the \(v_R\) scale due to quantum loop corrections that include heavy states like \(W_R\), \(Z_R\) and the heavy scalars. If no additional symmetries, like supersymmetry, are imposed on the theory and the couplings of the light Higgs with the heavy states are of order one, this points to a \(v_R\) scale of a few TeV. This can be in contradiction with the explanation of the masses of the light neutrinos.} In that situation, their masses are approximately given by

\begin{equation}
    \label{NS_Masses_Effective_LR}
    M^2_{H}\approx 2\lambda_R v_R^2,\hspace{0.8cm}M^2_{h}\approx\frac{4\lambda_L\lambda_R-\lambda_{LR}^2}{2\lambda_R}\,v_L^2.
\end{equation}

The reason why this model is called effective is that \textit{there is no way to generate the masses of the fermions through renormalizable operators}. This is not in contradiction with the philosophy of this work, where it is assumed that the LR models are the low energy limits of other theories with typical energy scale \(\Lambda\gg v_R\gg v_{EW}\). Thus, defining \(\tilde{\chi}_{L,R}\coloneqq i\sigma^2\chi_{L,R}^{\ast}\), it is possible to introduce the following \textit{Yukawa terms of dimension five} for quarks and leptons:

\begin{equation}
    \label{Yukawa_Quark_Effective_LR}
    \mathcal{L}^q_Y=-\frac{1}{\Lambda}\sum_{ij}C_d^{ij}\bar{q}_L^{\,i}\chi_L\chi_R^{\dagger}q_R^{\,j}-\frac{1}{\Lambda}\sum_{ij}C_u^{ij}\bar{q}_L^{\,i}\tilde{\chi}_L\tilde{\chi}_R^{\dagger}q_R^{\,j}+\textrm{h.c.},
\end{equation}

\begin{equation}
    \label{Yukawa_Lepton_Effective_LR}
    \begin{split}
        \mathcal{L}^\ell_Y=&-\frac{1}{\Lambda}\sum_{ij}C_e^{ij}\bar{\ell}_L^{\,i}\chi_L\chi_R^{\dagger}\ell_R^{\,j}-\frac{1}{\Lambda}\sum_{ij}C_{\nu_{D}}^{ij}\bar{\ell}_L^{\,i}\tilde{\chi}_L\tilde{\chi}_R^{\dagger}\ell_R^{\,j}\\
        &-\frac{1}{\Lambda}\sum_{ij}C_{\nu_{L,M}}^{ij}\bar{\ell}_L^{\,i}\tilde{\chi}_L\tilde{\chi}_L^{\mathrm{T}}l_L^{j\,c}-\frac{1}{\Lambda}\sum_{ij}C_{\nu_{R,M}}^{ij}\bar{\ell}_R^{\,c\,i}\tilde{\chi}_R^{\ast}\tilde{\chi}_R^{\dagger}\ell_R^{\,j}+\textrm{h.c.}
    \end{split}
\end{equation}

\noindent This is the complete set of dimension-five operators in this model, and the ones in the second line are analogous to the Weinberg operator in SMEFT. Notice that the second line of Eq.~\eqref{Yukawa_Lepton_Effective_LR} contains \textit{Majorana terms that break lepton number by two units}, so in this effective model neutrinos are Majorana particles. The mass terms are

\begin{equation}
    \label{Fermion_Mass_Term_Effective_LR}
    \mathcal{L}_m=-\sum_{f=u,\,d,\,e}\bar{f}_L\,M_f\,f_R-\frac{1}{2}\left(\bar{\nu}_L\,\bar{\nu}_R^{\,c}\right)\left(\begin{array}{cc}
     M_{\nu_{L,M}} & M_{\nu_{D}} \\
    M^{\mathrm{T}}_{\nu_{D}} & M_{\nu_{R,M}}
    \end{array}\right)\left(\begin{array}{c}
    \nu_L^{c} \\
    \nu_R
    \end{array}\right)+\textrm{h.c.},
\end{equation}

\noindent where the mass matrices are given by

 \begin{equation}
    \label{Mass_Matrices_Effective_LR}
    M_f\coloneqq \frac{1}{2}\frac{v_Lv_R}{\Lambda}C_{f},\hspace{0.4cm}M_{\nu_{D}}\coloneqq\frac{1}{2}\frac{v_Lv_R}{\Lambda}C_{\nu_{D}}, \hspace{0.4cm}M_{\nu_{L,M}}\coloneqq\frac{v_L^2}{\Lambda}C_{\nu_{L,M}},\hspace{0.4cm}M_{\nu_{R,M}}\coloneqq\frac{v_R^2}{\Lambda}C_{\nu_{R,M}}.
\end{equation}

\noindent The diagonalization of these matrices and the interactions with the gauge bosons are completely analogous to the ones of the model with one bidoublet and two triplets. A \textit{seesaw mechanism also appears here}: the masses of the heavy neutrinos are proportional to \(v_R^2/\Lambda\) while the masses of the light ones to \(v_L^2/\Lambda\). The masses of the other fermions are proportional to \(v_Lv_R/\Lambda\).\footnote{At tree level, the top quark mass has the form $ v_LC^{(5)} v_R / \Lambda $, where $ C^{(5)} $ is the Wilson coefficient of the dimension-five operator (with flavor indices omitted). Since \(\Lambda\gg v_R\) and \(v_L\) is identified with the electroweak scale, the Wilson coefficient falls outside the perturbative regime since \(C^{(5)}\) has to be substantially larger than one. Apart from this feature, the effective model has the advantage of explaining why the masses of the other fermions are small compared to those of the gauge bosons.} Finally, the interactions between scalars and charged fermions are given by

\begin{equation}
    \label{Scalars_Charged_Fermions_Effective_LR}
    \mathcal{L}^{Y}_{u,d,e}=-\left(1+\frac{\chi_L^{0r}}{v_L}\right)\left(1+\frac{\chi_R^{0r}}{v_R}\right)\sum_{f=u,\,d,\,e}\bar{f}\mathcal{M}_ff,
\end{equation}

\noindent where \(\mathcal{M}_f\) is the diagonalized mass matrix. Then, it is clear that in this model \textit{there are no FCNCs mediated by scalars} \cite{Glashow:1970gm}.

At this point, it is important to remark that possible UV-completions that lead to this mechanism of generation of fermion masses involve the introduction of vector-like fermions, whose masses are not generated by the scalar sector and are not related with the EW scale \cite{Davidson:1987mh,Rajpoot:1986nv,Rajpoot:1987fca,Chang:1986bp,Mohapatra:1987nx,Babu:1988yq,Babu:1988mw,Babu:1989rb,Balakrishna:1987qd,Balakrishna:1988ks,Balakrishna:1988bn}. The implications of this approach have been studied in the recent literature, e.g., Refs.~\cite{Babu:2018vrl,Babu:2022ikf,Dey:2022tbp,Dcruz:2022rjg,Dcruz:2023mvf,Babu:2023srr}.

\section{Flavor Physics Bounds on the LR Scale}

From this point onward, the purpose of this text is to obtain general lower bounds on the masses of the  new \(W_R^{\pm}\) and \(Z_R\) gauge bosons, and consequently, on the LR scale \(v_R\), with no further assumption like imposing an additional discrete symmetry on the theory. Flavor observables in LR models with discrete symmetries have been widely studied in the literature, e.g., Refs~\cite{Chang:1984hr,Basecq:1985cr,Kenmoku:1987fm,Hou:1985ur,Bernard:2015boz,ValeSilva:2016dcg,Zhang:2007da,Blanke:2011ry,Dekens:2021bro,Bertolini:2014sua}. In this section, the analysis of the flavor observables done by the author of this text in \cite{Solera:2024jrm} will be extended, adding new theoretical developments for the \(\chi_L+\chi_R\) effective model. Nonetheless, it is important to remark that the conclusion will be the same: \textit{the derivation of lower bounds on the masses of the new particles in the general scenario using flavor observables is non-trivial and collider bounds are preferred for this purpose.}

Consider first the effective model with doublets. It might seem that there is an inconsistency in this scenario when one calculates the usual one-loop box diagrams that appear in the computation of some flavor observables. The diagrams with two \(W_R\) gauge bosons are completely analogous to the ones of the SM where two \(W_L\) bosons are present. Consequently, the calculation of both contributions leads to perfectly consistent and finite gauge invariant results. However, there are additional contributions coming from diagrams with a \(W_L\) and a \(W_R\) gauge bosons. In the doublets and triplets models, the gauge invariant calculation also requires the addition of diagrams that contain FCNCs mediated by scalars \cite{Chang:1984hr,Basecq:1985cr}. Nonetheless, it was shown that, in general, the calculations of the box diagrams for meson mixing observables in LR theories will be gauge invariant if FCNCs mediated by scalars are not present \cite{Kenmoku:1987fm}, which is the case in the effective model. It will be shown that, \textit{when considering the contributions from dimension five operators, one also gets a gauge invariant result in that scenario.}

\begin{figure}
\centering
\includegraphics[scale=0.57]{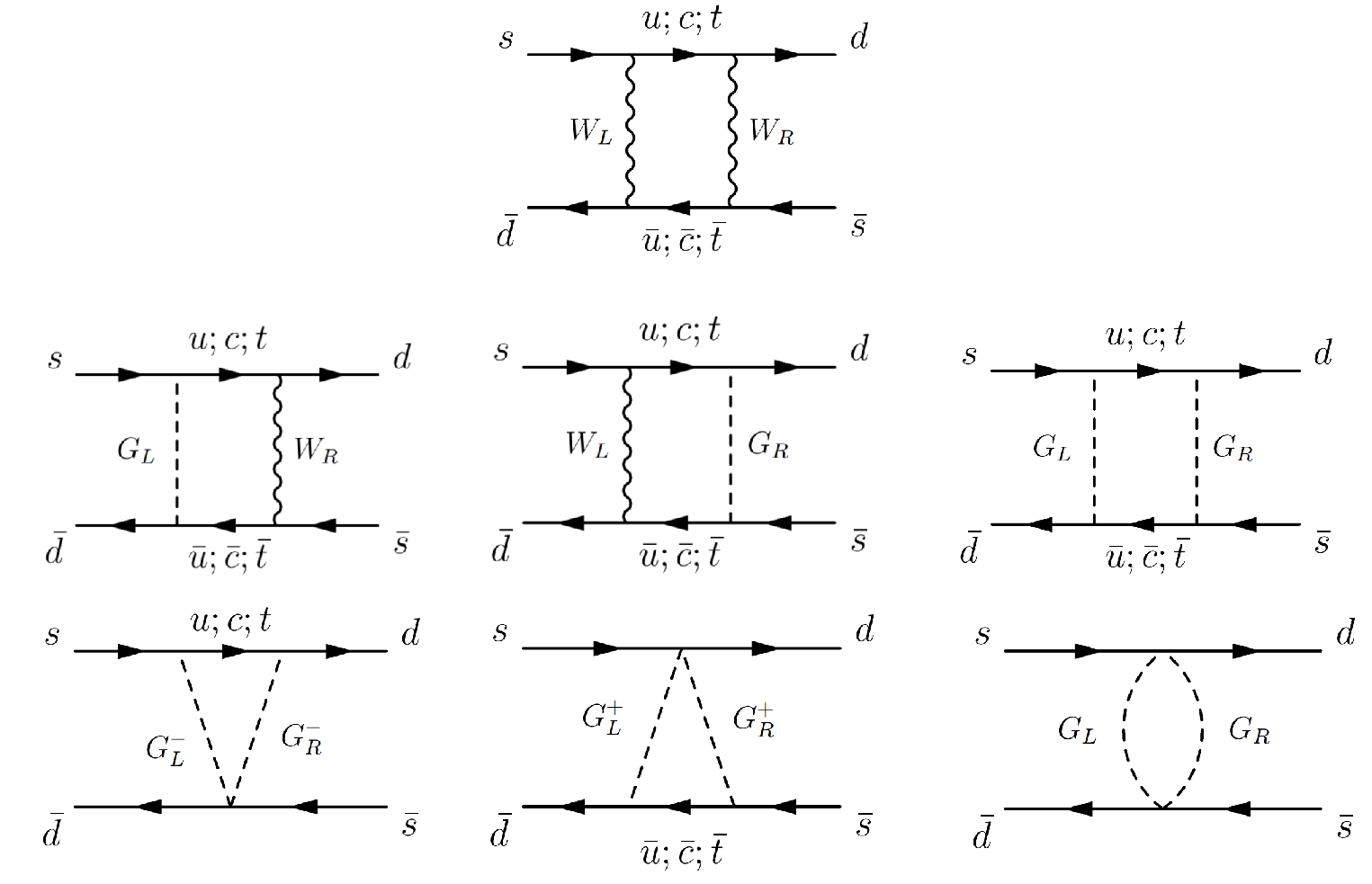}
\caption{Feynman diagrams that contain the left and right gauge or Goldstone states and contribute to the Kaon mixing in the \(\chi_L+\chi_R\) effective LR model at one loop in a general \(R_{\xi}\) gauge with operators of dimension five or less.}
\label{ds_Diagrams}
\end{figure}

To take a particular example, consider the \(K_L-K_S\) mixing for which the computation of the amplitude \(\bar{d}s\rightarrow\bar{s}d\) is required. To show that the result is gauge invariant, it is necessary to do the calculation in the general \(R_{\xi}\) gauges presented in Sec.~\ref{sec:SSB} and prove that there is no final \(\xi\) dependence. It is easy to identify the components \(\chi_{L,R}^{\pm}\) with the Goldstone bosons \(G_{L,R}^{\pm}\) from which the \(W_{L,R}^{\pm}\) get their masses. The dimension-five Yukawa couplings in Eq.~\eqref{Yukawa_Quark_Effective_LR} generate the following interaction between these Goldstone modes and the quarks:

\begin{equation}
    \label{Goldstone_Quarks_Dim5}
    \begin{split}
        \mathcal{L}_{ffGG}=-\frac{1}{2}\frac{g_Lg_R}{M_{W_L}M_{W_R}}&\left\{\bar{d}_L(V_L^{\rm CKM})^{\dagger}M_uV_R^{\rm CKM}d_RG_R^+G_L^-\right.\\&\left.+\bar{u}_LV_L^{\rm CKM}M_d(V_R^{\rm CKM})^{\dagger}u_RG_L^+G_R^-\right\}+\textrm{h.c.},
    \end{split}
\end{equation}

\noindent where \(M_u\) and \(M_d\) are the diagonalized mass matrices for the up- and down-type quarks, respectively. With this interaction Lagrangian, it is clear that the complete set of one-loop diagrams that contribute to \(\bar{d}s\rightarrow\bar{s}d\) and contain left and right gauge or Goldstone states are the ones shown in Fig.~\ref{ds_Diagrams}.

The calculation of the first four diagrams in Fig.~\ref{ds_Diagrams} was first performed in \cite{Chang:1984hr}. Eliminating the \(W_{L}-W_{R}\) mixing (which is vanishing in this model) and neglecting the external momenta, the integrand \(M_1\) of the four one-loop integrals can be decomposed as \(M_1=M-N_1-N_2-N_2\), where \(M\) is the result in the unitary gauge, and thus free from the \(\xi\) parameter, and the \(N_i\) factors are

\begin{figure}
\centering
\includegraphics[scale=0.4]{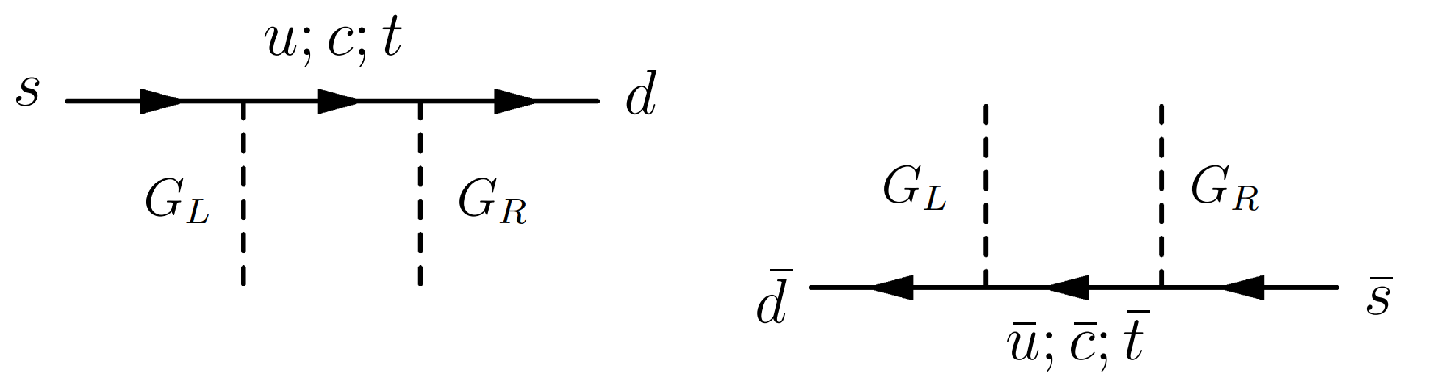}
\caption{Feynman diagrams of the processes associated to the amplitudes \(A_1\left(k\right)\) and \(A_{2}\left(k\right)\).}
\label{Amplitudes_A1_A2}
\end{figure}

\begin{equation}
    \label{N1_Factor}
        N_1=\frac{i}{2}\frac{g_L\,g_R}{M_{W_L}M_{W_R}}\left[(V_L^{\rm CKM})^{\dagger}M_uV_R^{\rm CKM}\right]_{ds}(\bar{d}_Ls_R)\frac{1}{k^2-\xi\,M_{W_L}^2}\frac{1}{k^2-\xi M_{W_R}^2}A_1\left(k\right),
\end{equation}

\begin{equation}
    \label{N2_Factor}
        N_2=\frac{i}{2}\frac{g_L\,g_R}{M_{W_L}M_{W_R}}\left[(V_R^{\rm CKM})^{\dagger}M_uV_L^{\rm CKM}\right]_{ds}(\bar{d}_Rs_L)\frac{1}{k^2-\xi\,M_{W_L}^2}\frac{1}{k^2-\xi M_{W_R}^2}A_2\left(k\right),
\end{equation}

\begin{equation}
    \label{N3_Factor}
    N_3=\frac{1}{4}\frac{g_L^2\,g_R^2}{M_{W_L}^2M_{W_R}^2}\frac{[(V_L^{\rm CKM})^{\dagger}M_uV_R^{\rm CKM}]_{ds}}{k^2-\xi\,M_{W_L}^2}\frac{[(V_R^{\rm CKM})^{\dagger}M_uV_L^{\rm CKM}]_{ds}}{k^2-\xi M_{W_R}^2}\left(\bar{d}_Rs_L\right)(\bar{d}_Ls_R).
\end{equation}

\noindent The internal momentum is represented by \(k\) and \(A_1\left(k\right)\) and \(A_{2}\left(k\right)\) are the amplitudes associated with the diagrams shown in Fig.~\ref{Amplitudes_A1_A2} with momentum \(k\) for the up-type quarks. Nevertheless, it is easy to see from the Lagrangian in Eq.~\eqref{Goldstone_Quarks_Dim5} that \(N_1\), \(N_2\) and \(N_3\) are precisely the amplitudes associated with the last three diagrams in Fig.~\ref{ds_Diagrams}. This means that \textit{the addition of the seven amplitudes is exactly equal to the result in the unitary gauge}, free from the \(\xi\) parameter and hence gauge invariant.

Even though it has been shown that the result satisfies gauge invariance, there is still one problem from the phenomenological point of view.  \textit{The divergences of the contributions with only \(W_L\) or \(W_R\) bosons vanish due to the unitarity of the CKM matrices, a generalized GIM mechanism.} However, the calculation with a \(W_L\) and a \(W_R\) can be easily computed in the unitary gauge, and it leads to the following amplitude:

\begin{equation}
    \label{Complete_Amplitude_Unitary_Gauge}
    \begin{split}
    A_{LR}\coloneqq&i\frac{g_L^2g_R^2}{4}(\bar{d}_Ls_R)(\bar{d}_Rs_L)\sum_{i,j=u,c,t}m_im_j\lambda_{i}^{LR}\lambda_j^{RL}\\&\times\int\frac{{\rm d}^4k}{\left(2\pi\right)^4}\frac{1}{D_LD_RD_iD_j}\left\{4-k^2\left(\frac{1}{M_{W_L}^2}+\frac{1}{M_{W_R}^2}\right)+\frac{\left(k^2\right)^2}{M_{W_L}^2M_{W_R}^2}\right\},
    \end{split}
\end{equation}

\noindent where \(\lambda_{i}^{LR}\coloneqq(V_L^{\rm CKM})_{id}^{\ast}(V_R^{\rm CKM})_{is}\), \(\lambda_{j}^{RL}\coloneqq(V_R^{\rm CKM})_{jd}^{\ast}(V_L^{\rm CKM})_{js}\), \(D_{L,\,R}\coloneqq k^2-M_{W_{L,\,R}}^2\), \(D_{i}\coloneqq k^2-m_i^2\) and \(m_i\) are the masses of the up-type quarks. There is no analogous GIM mechanism when both left- and right-handed states are considered, so this amplitude is divergent. If FCNCs mediated by scalars are present, there is a cancellation of this contribution leading to the expected finite result, but they are absent in the effective model. As a consequence, \textit{the complete result not only requires the addition of dimension-five operators, but also the addition of operators with dimension six}. This is expected from the fact that the masses of the quarks are proportional to \(1/\Lambda\), so the amplitude of the box diagrams behaves as \(1/\Lambda^2\).

In this situation, there are two possible options: one can try to do a calculation considering the operators of dimension six, or simply study flavor processes that do not include box diagrams. \textit{The complete basis of dimension-six fermionic operators was also obtained by the author of this text} and it is presented in Tab.~\ref{Tab_Dimension6_LR}, although it has not been shown that it is minimal.\footnote{As double-insertions of operators of dimension five require, in general, the introduction of counterterms of dimension six, the basis of required operators of dimension eight needed in the analogous renormalization of four-fermion operators of dimension six can be derived from the one in \cite{ValeSilva:2022tph}; meson-mixing is then used to constrain the Wilson coefficients for the later operators.} It can be easily derived from the fact that the LR gauge symmetry is more restrictive than the SM one, so it is only necessary to take the SMEFT+\(\nu\) \cite{Liao:2016qyd} basis of fermionic dimension six operators and consider only the linear combinations of them that are invariant under the \({\rm SU}(3)_{\rm QCD}\otimes{\rm SU}(2)_{\rm L}\otimes{\rm SU}(2)_{\rm R}\otimes{\rm U}(1)_{\rm X}\) group.

The dimension-6 operator that contributes to the aforementioned process can be obtained imposing flavor-symmetry constraints. The Lagrangian in Eq.~\eqref{Yukawa_Quark_Effective_LR} is invariant under the global flavor \({\rm U}(3)_{\rm L}\otimes{\rm U}(3)_{\rm R}\)  symmetry, with \(q_{L,R}\to U_{L,R}q_{L,R}\), provided that the coefficients transform as \(C_q\to U_LC_qU_R^{\dagger}\). The only dimension-6 operators that are invariant under both the LR and flavor symmetries, and contain the \(LRRL\) structure needed to cancel the divergence in Eq.~\eqref{Complete_Amplitude_Unitary_Gauge}, have the form \(Q_{q_1q_2}^{LRRL}\coloneqq(\bar{q}_L^aC_{q_1}q_R^b)(\bar{q}_R^bC_{q_2}^{\dagger}q_L^a)\), where \(a\) and \(b\) are \({\rm SU}(2)\) indices. Considering the relations in Eq.~\eqref{Mass_Matrices_Effective_LR} it is clear that, when written in terms of the mass eigenstates, the \(Q_{uu}^{LRRL}\) operator contains the desired structure

\begin{equation}
    \label{Flavor_Divergence_Structure}
    \frac{1}{\Lambda^2}Q_{uu}^{LRRL}\to\frac{4}{v_L^2v_R^2}[\bar{d}_L(V_L^{\rm CKM})^{\dagger}\mathcal{M}_uV_R^{\rm CKM}d_R][\bar{d}_R(V_R^{\rm CKM})^{\dagger}\mathcal{M}_uV_L^{\rm CKM}d_L].
\end{equation}

The problem with this procedure is that it introduces a new free parameter to the theory: the Wilson coefficient of the dimension six operator. It is not trivial to derive constraints on the masses of the new particles in this scenario, so \textit{this research line will not be pursued further in this work}. The other option is to use dipole observables like \(b\to s\gamma\), which only receive contributions from penguin diagrams at one loop, not from boxes. However, as it will be shown, \textit{the main contributions to these processes in LR theories come from the \(W_L-W_R\) mixing}. Since this mixing is not present at tree-level in the effective model, the bounds on the \(v_R\) scale that one will derive are not as strong as the ones coming from collider physics. As a consequence, \textit{flavor observables are not the preferred ones to obtain bounds on the masses of the new heavy states in the effective LR model.}

\begin{table}[t!]
\renewcommand{\arraystretch}{1.7}
    \begin{tabular}{|m{2.5em}|m{11em}||m{3em}|m{12em}|}
    \hline
    \multicolumn{2}{|c||}{{\normalsize\(\left(\bar{L}L\right)\left(\bar{L}L\right)\)}}&\multicolumn{2}{c|}{{\normalsize\(\left(\bar{R}R\right)\left(\bar{R}R\right)\)}}\\ [0.5ex] \hline
    \(Q^q_L\) & \(\left(\bar{q}_L\gamma_{\mu}q_L\right)\left(\bar{q}_L\gamma^{\mu}q_L\right)\) & \(Q^q_R\) & \(\left(\bar{q}_R\gamma_{\mu}q_R\right)\left(\bar{q}_R\gamma^{\mu}q_R\right)\)  \\ [0.7ex]
    \(Q^{q\,(3)}_L\) & \(\left(\bar{q}_L\gamma_{\mu}\sigma_L^Iq_L\right)\left(\bar{q}_L\gamma^{\mu}\sigma_L^Iq_L\right)\) & \(Q_R^{q\,(3)}\)  & \(\left(\bar{q}_R\gamma_{\mu}\sigma_R^Iq_R\right)\left(\bar{q}_R\gamma^{\mu}\sigma_R^Iq_R\right)\)  \\ [0.7ex]
    \(Q_L^\ell\) & \(\left(\bar{\ell}_L\gamma_{\mu}\ell_L\right)\left(\bar{\ell}_L\gamma^{\mu}\ell_L\right)\) & \(Q_R^\ell\) & \(\left(\bar{\ell}_R\gamma_{\mu}\ell_R\right)\left(\bar{\ell}_R\gamma^{\mu}\ell_R\right)\) \\[0.5ex] 
    \(Q_L^{\ell\,(3)}\) & \(\left(\bar{\ell}_L\gamma_{\mu}\sigma_L^I\ell_L\right)\left(\bar{\ell}_L\gamma^{\mu}\sigma_L^I\ell_L\right)\) & \(Q_R^{\ell\,(3)}\) & \(\left(\bar{\ell}_R\gamma_{\mu}\sigma_R^I\ell_R\right)\left(\bar{\ell}_R\gamma^{\mu}\sigma_R^I\ell_R\right)\)\\ [0.7ex]
    \(Q_L^{q\ell}\) & \(\left(\bar{q}_L\gamma_{\mu}q_L\right)\left(\bar{\ell}_L\gamma^{\mu}\ell_L\right)\) & \(Q_R^{q\ell}\) & \(\left(\bar{q}_R\gamma_{\mu}q_R\right)\left(\bar{\ell}_R\gamma^{\mu}\ell_R\right)\) \\[0.5ex]
    \(Q_L^{q\ell\,(3)}\) & \(\left(\bar{q}_L\gamma_{\mu}\sigma_L^Iq_L\right)\left(\bar{\ell}_L\gamma^{\mu}\sigma_L^I\ell_L\right)\) & \(Q_R^{q\ell\,(3)}\) & \(\left(\bar{q}_R\gamma_{\mu}\sigma_R^Iq_R\right)\left(\bar{\ell}_R\gamma^{\mu}\sigma_R^I\ell_R\right)\)\\[0.7ex]
    \hline
    \end{tabular}
    \begin{tabular}{|m{2.5em}|m{11em}||m{3em}|m{12em}|}
    \hline
    \multicolumn{2}{|c||}{{\normalsize \(\left(\bar{L}L\right)\left(\bar{R}R\right)\)}}&\multicolumn{2}{c|}{{\normalsize \(\left(\bar{L}R\right)\left(\bar{R}L\right)\) and \(\left(\bar{L}R\right)\left(\bar{L}R\right)\)}} \\ [0.7ex] \hline
    \(Q_{LR}^q\) & \(\left(\bar{q}_L\gamma_{\mu}q_L\right)\left(\bar{q}_R\gamma^{\mu}q_R\right)\) & \(Q^q_{LRLR}\) & \(\varepsilon_{ij}\varepsilon_{kl}(\bar{q}_L^{\,i}q_R^k)(\bar{q}_L^{\,j}q_R^l)\) \\ [0.7ex]
    \(Q^{q\,(8)}_{LR}\) & \(\left(\bar{q}_L\gamma_{\mu}T^Aq_L\right)\left(\bar{q}_R\gamma^{\mu}T^Aq_R\right)\)  & \(Q_{LRLR}^{q\,(8)}\) & \(\varepsilon_{ij}\varepsilon_{kl}(\bar{q}_L^{\,i}T^Aq_R^k)(\bar{q}_L^{\,j}T^Aq_R^l)\) \\ [0.7ex]
    \(Q^{\ell}_{LR}\) & \(\left(\bar{\ell}_L\gamma_{\mu}\ell_L\right)\left(\bar{\ell}_R\gamma^{\mu}\ell_R\right)\) & \(Q^{\ell}_{LRLR}\) & \(\varepsilon_{ij}\varepsilon_{kl}(\bar{\ell}_L^{\,i}\ell_R^k)(\bar{\ell}_L^{\,j}\ell_R^l)\) \\ [0.7ex]
    \(Q^{q\ell}_{LR}\) &  \(\left(\bar{q}_L\gamma_{\mu}q_L\right)\left(\bar{\ell}_R\gamma^{\mu}\ell_R\right)\) & \(Q^{\ell q}_{LRRL}\) & \((\bar{\ell}_L^{\,i}\ell_R^{\,j})(\bar{q}_R^{\,j}q_L^{\,i})\) \\ [0.7ex]
    \(Q^{\ell q}_{LR}\) & \(\left(\bar{\ell}_L\gamma_{\mu}\ell_L\right)\left(\bar{q}_R\gamma^{\mu}q_R\right)\) & \(Q^{\ell q}_{LRLR}\) & \(\varepsilon_{ij}\varepsilon_{kl}(\bar{\ell}_L^{\,i}\ell_R^{\,k})(\bar{q}_L^{\,j}q_R^{\,l})\)\\[0.7ex]
    &  & \(Q^{\ell q\,(T)}_{LRLR}\) & \(\varepsilon_{ij}\varepsilon_{kl}(\bar{\ell}_L^{\,i}\sigma_{\mu\nu}\ell_R^{\,k})(\bar{q}_L^{\,j}\sigma^{\mu\nu}q_R^{\,l})\) \\[0.7ex]
    \hline
    \end{tabular}
    \begin{tabular}{|m{2.5em}|c|}
    \hline
    \multicolumn{2}{|c|}{{\normalsize \(B\) and \(L\) Violating}} \\[0.7ex]
    \hline
    \(Q^{BL}_L\) & \(\varepsilon^{\alpha\beta\gamma}\varepsilon_{ij}\varepsilon_{kl}\left[(q_L^{\,\alpha\,i})^{\mathrm{T}}\mathcal{C}q_L^{\,\beta\,k}\right]\left[(q_L^{\,\gamma\,l})^{\mathrm{T}}\mathcal{C}\ell_L^{\,j}\right]\) \\ [0.7ex]
    \(Q^{BL}_R\) & \(\varepsilon^{\alpha\beta\gamma}\varepsilon_{ij}\varepsilon_{kl}\left[(q_R^{\,\alpha\,i})^{\mathrm{T}}\mathcal{C}q_R^{\,\beta\,k}\right]\left[(q_R^{\,\gamma\,l})^{\mathrm{T}}\mathcal{C}\ell_R^{\,j}\right]\) \\ [0.7ex]
    \(Q^{BL}_{LR}\) & \(\varepsilon^{\alpha\beta\gamma}\varepsilon_{ij}\varepsilon_{kl}\left[(q_L^{\,\alpha\,i})^{\mathrm{T}}\mathcal{C}q_L^{\,\beta\,j}\right]\left[(q_R^{\,\gamma\,k})^{\mathrm{T}}\mathcal{C}\ell_R^{\,l}\right]\) \\ [0.7ex]
    \(Q^{BL}_{RL}\) & \(\varepsilon^{\alpha\beta\gamma}\varepsilon_{ij}\varepsilon_{kl}\left[(q_R^{\,\alpha\,i})^{\mathrm{T}}\mathcal{C}q_R^{\,\beta\,j}\right]\left[(q_L^{\,\gamma\,k})^{\mathrm{T}}\mathcal{C}\ell_L^{\,l}\right]\) \\[0.7ex]
    \hline
    \end{tabular}
    \caption[Complete set of dimension six effective fermionic operators of a LR theory.]{Complete set of dimension six effective fermionic operators of a LR theory. This basis is, however, given in a non-minimal form. Flavor indices have been omitted.}
    \label{Tab_Dimension6_LR}
\end{table}

For the other two scenarios with the bidoublet, it is possible to identify the order of magnitude of the flavor-physics bounds on \(v_R\) using the observables based in the transitions \(b\to s\gamma\) and \(b\to s\ell\ell\). As already mentioned, the diagonalization of the scalar potential is highly non-trivial. However, since the bounds on the scalar masses derived from their contribution to FCNCs are greater than the ones of the heavy gauge bosons \cite{Maiezza:2010ic}, it is reasonable to neglect their effects if only a rough estimation of the bound on \(v_R\) is desired. Furthermore, for the sake of illustration, only the model with the bidoublet and the two doublets will be considered. The main advantage of this model from the phenomenological point of view is that neutrinos are Dirac particles, so both the left- and right-handed chiralities have the same extremely light mass. As done in \cite{Solera:2024jrm}, using the normalization

\begin{equation}
    \label{Dipole_Lagrangian}
    \mathcal{L}_{\mathrm{Eff}}=\frac{1}{16\pi^2}\left\{C_7\left(\bar{s}\sigma^{\mu\nu}P_R b\right)F_{\mu\nu}+C_7^{\prime}\left(\bar{s}\sigma^{\mu\nu}P_L b\right)F_{\mu\nu}+C_9^{\prime \ell}\left(\bar{s}\gamma_{\mu}P_Rb\right)\left(\bar{\ell}\gamma^{\mu}\ell\right)\right\}+{\rm h.c.},
\end{equation}

\noindent the new physics part of the Wilson coefficients that contribute to \(b\to s\gamma\) and \(b\to s\ell\ell\) can be computed using the results in \cite{Bishara:2021buy} to obtain

\begin{equation}
    \label{LR_C7}
    \left.C_7\right|_{\mathrm{NP}}\approx\frac{1}{\sqrt{2}}4em_bG_F\sum_{i=u,c,t}\frac{g_R}{g_L}\frac{m_i}{m_b}\sin\xi_W e^{-i\lambda}\left(V_L^{\mathrm{CKM}}\right)_{is}^{\ast}\left(V_R^{\mathrm{CKM}}\right)_{ib}\tilde{F}(x^{f_i}_{W_L}),
\end{equation}

\begin{equation}
    \label{LR_C7_prime}
    \left.C_7^{\prime}\right|_{\mathrm{NP}}\approx\frac{1}{\sqrt{2}}4em_bG_F\sum_{i=u,c,t}\frac{g_R}{g_L}\frac{m_i}{m_b}\sin\xi_W e^{i\lambda}\left(V_R^{\mathrm{CKM}}\right)_{is}^{\ast}\left(V_L^{\mathrm{CKM}}\right)_{ib}\tilde{F}(x^{f_i}_{W_L}),
\end{equation}

\begin{equation}
    \label{C9prime}
     C_9^{\prime \ell}\approx\frac{4}{9}\left(\frac{eg_R}{M_{W_R}}\right)^2\sum_{i=u,c,t}\left(V_R^{\mathrm{CKM}}\right)^{\ast}_{is}\left(V_R^{\mathrm{CKM}}\right)_{ib}\ln\left(\frac{m_i}{m_0}\right),
\end{equation}

\noindent where \(\tilde{F}\left(x\right)\coloneqq\left(-20+31x-5x^2\right)/12\left(x-1\right)^2-x\left(2-3x\right)\ln x/2\left(x-1\right)^3\), the mass ratios are \(x_{W_L}^{f_i}\coloneqq m_{f_i}^2/M^2_{W_L}\), \(m_0\) is the mass of a particular up-type quark, \(G_F\) is the Fermi constant and \(\xi_W\) and \(\lambda\) are the parameters of the \(W_L-W_R\) mixing. The corrections of order \(\mathcal{O}(v_R^{-4})\) have been neglected, so Wilson coefficients of this order, like \( C_9^{\ell}\), \(C_{10}^{\ell}\) and \(C_{10}^{\prime \ell}\), do not appear.

Even though the contributions of the scalars have been neglected, there is still a large number of free parameters due to the presence of the \(V_R^{\rm CKM}\) matrix. Consequently, if no additional discrete symmetry that relates \(V_R^{\rm CKM}\) with \(V_L^{\rm CKM}\) is imposed, \textit{using flavor observables to put a bound on \(v_R\) will require a global fit}, it cannot be done studying a single process.\footnote{The situation will be the same if one uses electroweak precision observables.} Furthermore, the bound that one might expect is not as strong as the general one coming from collider physics, which will be derived in the following sections. To see this, one can estimate a bound taking the simplified case with \(g_L=g_R\), the diagonal structure \(V_R^{\rm CKM}={\rm diag}(1,1,1)\), \(\lambda=0\) and a mixing given by \(\sin\xi_W=\left(v_{\rm EW}/v_R\right)^2\). In this scenario, the main constraint comes from \(\left.C_7\right|_{\mathrm{NP}}\), and using the global fit presented in \cite{LHCb:2021byf} one obtains \(v_R\sim7\,{\rm TeV}\), \textit{which is close to but not as strong as the bound that will be derived using collider processes.}

\section{Collider Bounds on the LR Scale. Theoretical Predictions} \label{sec:Collider_Theory}
\markboth{5. LR MODELS. GENERAL BOUNDS ON THE LR SCALE}{5.5. COLLIDER BOUNDS ON THE LR SCALE. THEORY}
As mentioned in the previous section, collider observables are the preferred option to put lower bounds on the masses of the new gauge bosons and, consequently, on the LR scale. It will be shown that \textit{it is possible to constrain those parameters of the theory in the perturbative regime with no further assumptions about the symmetries of the model, the scalar content or the values of the different couplings}. The observables that will be studied are cross sections of processes like \(Z_R\to \bar{f}f\) and \(W_R\to f\bar{f'}\), from which the lowest possible direct lower bounds on \(M_{Z_R}\) and \(M_{W_R}\) are obtained, using techniques like the equivalence theorem.\footnote{In this context, the lowest lower bound means that it is associated with a particular value of the parameters of the theory in such a way that any other option leads to  a higher lower bound.} Furthermore, due to the relations between the masses of the gauge bosons and the mixing parameter \(\gamma\), it is also possible to derive indirect bounds on both \(M_{Z_R}\) and \(M_{W_R}\). Combining the results one obtains the general lowest lower bounds on the masses and on the LR scale. This section is devoted to the presentation of the theoretical analysis.

\subsection{Production and decay of the \(Z_R\)} \label{subsec:prod_dec_ZR}

In this part of the text, direct searches involving \(Z_R\) decaying into a pair of leptons \(\ell^+\ell^-\) will be studied. 
This process provides a straightforward way of putting bounds on the LR scale, since at tree level there is no dependence on the leptonic mixing matrices. It will be shown that \textit{the only unknown parameters are the angle \(\gamma\) and the mass of the \(Z_R\) boson}. Consider the case where the \(Z_R\) is produced on-shell after the collision of two protons. The main goal is to calculate the cross section of the process \(pp\rightarrow Z_RX\rightarrow \ell^+\ell^-X\), where \(X\) represents an arbitrary state. All the couplings needed are summarized in App.~\ref{sec:summary_couplings}. 

The interaction Lagrangian of the \(Z_R\) and two fermions can be written in the following two equivalent ways

\begin{equation}
    \label{ecu.3.1}
    \mathcal{L}_{Z_R}^{ff}=\frac{1}{2}Z_R^{\mu}\bar{f}\gamma_{\mu}\left(g_V^f-g_A^f\gamma_5\right)f=Z_R^{\mu}\bar{f}\gamma_{\mu}\left(g_L^fP_L+g_R^fP_R\right)f.
\end{equation}

\noindent In the context of \(Z_R\) searches, it is common to define the functions

\begin{equation}
    \label{ecu.3.2}
    c_q^f\coloneqq \frac{1}{2}\left( g_V^{q\,2}+g_A^{q\,2} \right)\mathrm{Br}\left(Z_R\rightarrow f\bar{f}\right)=\left(g_L^{q\,2}+g_R^{q\,2}\right)\mathrm{Br}\left(Z_R\rightarrow f\bar{f}\right),
\end{equation}

\noindent because for narrow resonances one obtains\footnote{More details will be provided in the next section when discussing $W_R^\pm$ production.}
 \cite{Ellis:1996mzs,ParticleDataGroup:2024cfk}
\begin{equation}
    \label{ecu.3.3}
    \sigma\left(pp\rightarrow Z_RX\rightarrow f\bar{f}X\right)\approx\frac{\pi}{6s}\sum_qc_q^{f}\,\omega_q\!\left(s,M_{Z_R}^2\right),
\end{equation}

\begin{table}[!t]
    \centering
    \renewcommand{\arraystretch}{1.1}
    \begin{tabular}{|lcc|c|}
        \hline
        Mode & Scenario & Eqs. & Comments \\
        \hline
        \hline
        $ Z_R \to f \bar{f} $ & All & \eqref{ecu.3.4} & \\ 
        $ Z_R \to W_L^+ W_L^- $ & All & \eqref{ecu.3.5} to \eqref{ecu.3.7} & Equivalence Theorem \\
        $ Z_R \to Z_L h^0 $ & All & \eqref{ecu.3.5} to \eqref{ecu.3.7} & Equivalence Theorem \\
        \hline
        \hline
        $ Z_R \to Z_L H^0 $ & All & \eqref{ecu.3.5} to \eqref{ecu.3.7} & ``Light new scalars'', Equiv. Theo. \\
        $ Z_R \to W_L^\pm H^\mp $ & D, T & \eqref{ecu.3.5} and \eqref{ecu.3.6} & ``Light new scalars'', Equiv. Theo. \\
        $ Z_R \to h^0 H^0 $ & All & \eqref{ecu.3.5} to \eqref{ecu.3.7} & ``Light new scalars'' \\
        $ Z_R \to W_L^\pm W_R^\mp $ & D, T & -- & Negligible \\
        \hline
        \hline
        $ Z_R \to H_1 H_2 $ & D, T & \eqref{ecu.3.5} and \eqref{ecu.3.6} & ``Light new scalars'' \\
        $ Z_R \to W_R^\pm H^\mp $ & D, T & -- & Negligible \\ 
        $ Z_R \to W_R^+ W_R^- $ & All & \eqref{ecu.3.10} and \eqref{ecu.3.11} & \\
        \hline
        \hline
        $ W_R^\pm \to f \bar{f}' $ & All & \eqref{ecu.3.15} & \\
        $ W_R^\pm \to W_L^\pm Z_L $ & D, T & \eqref{ecu.3.16} and \eqref{ecu.3.17} & Equivalence Theorem \\
        $ W_R^\pm \to W_L^\pm h^0 $ & D, T & \eqref{ecu.3.16} and \eqref{ecu.3.17} & Equivalence Theorem \\
        \hline
        \hline
        $ W_R^\pm \to W_L^\pm H^0 $ & D, T & \eqref{ecu.3.16} and \eqref{ecu.3.17} & ``Light new scalars'', Equiv. Theo. \\
        $ W_R^\pm \to H^\pm Z_L $ & D, T & \eqref{ecu.3.16} and \eqref{ecu.3.17} & ``Light new scalars'', Equiv. Theo. \\
        $ W_R^\pm \to H^\pm h^0 $ & D, T & \eqref{ecu.3.16} and \eqref{ecu.3.17} & ``Light new scalars'' \\
        $ W_R^\pm \to W_L^\mp H^{\pm\pm} $ & T & -- & Negligible \\
        \hline
        \hline
        $W_R^{\pm}\to H^{\pm} A$ & -- & -- & No contribution \\
        \hline
        \hline
        $ W_R^\pm \to H_1 H_2 $ & D, T & \eqref{ecu.3.16} and \eqref{ecu.3.17} & ``Light new scalars'' \\
        \hline
    \end{tabular}
    \caption[Set of $Z_R$ and $W_R$ two-body decays.]{Set of $Z_R$ and $W_R$ two-body decays. The masses of the $Z_R$ and $W_R$ are assumed to lay much above the EW scale (which is corroborated by, e.g., the present analysis), and $ M_{W_R} \leq M_{Z_R} $ in the models analysed here: (D) two doublets and one bi-doublet, (T) two triplets and one bi-doublet, (Eff) two doublets and no bi-doublet. The masses of the RH neutrinos are discussed in the text. The notation is as follows: $ h^0, f, Z_L, W_L^\pm $ are the SM-like Higgs, fermions, neutral and charged weakly-coupled massive gauge bosons, respectively, while $A$ stands for the photon; $ Z_R, W_R^\pm $ are the new neutral and charged weakly-coupled gauge bosons, respectively; $ H $ represents the new physical scalar sector (neutral, singly or doubly charged), where their CP nature in the case of neutral scalars has not been explicitly written (also an indexing when multiple scalars of the same charge are possible has been generally omitted in this table). The last column gives the ingredients used to set upper bounds on the partial widths of the $Z_R$ and $W_R$ gauge bosons, that translate into a more model-independent bound on their masses, where considering the limit of light new scalars is used only to determine the maximum of the heavy gauge-boson widths.}
    \label{tab:ZR_WR_two_body}
\end{table}

\noindent where \(s\) is the square of the center-of-mass energy and the \(\omega_q\!\left(s,M_{Z_R}^2\right)\) contain all the information about the Parton Distribution Functions (PDFs) of the quarks inside the proton (see the analogous Eq.~\eqref{ecu.3.14} below for the case of charged currents). The largest functions are  the ones for the up and down quarks, $q=u, d$. Since the couplings of the \(Z_R\) are generation-independent one only needs to calculate \(c_u^\ell\) and \(c_d^\ell\). Note that, although $g_{V, A}^q$ or $g_{L, R}^q$ do not depend on the specific realization of the scalar sector, $c^{\ell}_{u,d}$ do.

Consider the computation of \(\mathrm{Br}\left(Z_R\rightarrow  \ell^+\ell^-\right)\), for which the total width of the \(Z_R\) boson \(\Gamma_{Z_R}\) is needed.
All calculations in this part of the text are performed at the tree level. The possible two-body contributions are displayed in Tab.~\ref{tab:ZR_WR_two_body}.\footnote{Higher multiplicity decays are comparatively suppressed by the allowed phase space, and by powers of the weak couplings.
They cannot be enhanced
by factors of $ M_{heavy}^{n+1}/M_{light}^n \sim v_R^{n+1} / v_{\rm EW}^n $, $ n \geq 1 $, where $ M_{light} $ is the mass of an SM-like boson, since the massless limit $ M_{light} \to 0^+ $ should be continuous in a theory with SSB (for an example in which the gauge coupling of the spontaneously broken symmetry is vanishingly small, see Ref.~\cite{Weinberg2}).}
The explicit calculation of the processes that involve scalars requires diagonalizing their mass matrices.
The spectrum depends on the scalar potential, which involves a large number of parameters even when discrete symmetries are enforced.
In order to avoid this inconvenience, \textit{only an upper bound on those partial widths is derived, which ultimately results in a model-independent bound on the mass of the \(Z_R\) boson}.

Start by looking at the partial width into fermion pairs. It is easy to show that  if \(M_{Z_R}\gg m_f\), then

\begin{equation}
    \label{ecu.3.4}
    \Gamma\left(Z_R\rightarrow f\bar{f}\right)\approx N_C^f\frac{M_{Z_R}}{48\pi}\left(g_V^{f\,2}+g_A^{f\,2}\right)=N_C^f\frac{M_{Z_R}}{24\pi}\left(g_L^{f\,2}+g_R^{f\,2}\right),
\end{equation}

\noindent where \(N_C^f\) represents the number of colors of the fermion \(f\). Considering the mass of the fermions leads to corrections of order \((m_f/M_{Z_R})^2\), e.g., for \(M_{Z_R}\sim1\,\rm TeV\) one has for the top quark \((m_t/M_{Z_R})^2\sim 0.03\), and thus these effects are totally negligible for the SM-like fermions. However, such kinematic correction might be important for RH neutrinos because their mass can be proportional to the LR scale. To avoid introducing new unknown parameters, one can consider the following two limiting cases: 1) very heavy RH neutrinos, where the process \(Z_R\rightarrow\nu_R\bar{\nu}_R\) is forbidden; 2) light RH neutrinos, that would give the maximum possible value of the total width \(\Gamma_{Z_R}\) when their masses are neglected. This corresponds to the minimum value of \(\mathrm{Br}\left(Z_R\rightarrow \ell^+\ell^-\right)\), leading to less constraining bounds on the \(Z_R\) mass. Consequently, only the second case will be studied in this text.

Consider now processes that involve at least one light gauge boson in the final state.
For \(Z_R\rightarrow W_L^+W_L^-\) and \(Z_R\rightarrow Z_L h^0\) the masses of the particles in the final states are completely negligible compared to their energy, because the $Z_R$ is much heavier. In the decays \(Z_R\rightarrow  H^{\pm}W_L^{\mp}\) and \(H^0Z_L\), either the mass difference between the heavy states is small and the partial widths are suppressed by the phase space, or the mass difference is enough to allow the light particle to have a high energy compared to its own mass. To set an upper bound on the partial width of these processes, one only considers the second case, in which there are only light particles in the final state carrying a large momentum. Consequently, the dominant contribution to the partial widths will come from situations in which the gauge bosons in the final states have longitudinal polarization. Thus, \textit{one can calculate the amplitudes substituting the spin-1 fields by their associated Goldstone bosons using the equivalence theorem} \cite{Cornwall:1974km,Lee:1977eg}, and these processes will be collectively called \(Z_R\rightarrow 2\,\mathrm{Scalars}\). In applying this theorem, the $ v_{EW}^2/v_R^2 $ corrections coming from the transverse degrees of freedom are neglected.

The masses of the scalar sector also determine whether the channels $Z_R \to h^0 H^0, H_1 H_2$ are allowed.
In App.~\ref{sec:Non-Fermionic_Decays} it is shown that \textit{one can set an upper bound on the total width of $Z_R$ using the limit where the mass of all scalars is set to zero}. Once more, this translates into the least constraining lower bounds on the mass of the \(Z_R\). It is worth noting that when considering ``light'' scalars one avoids detailing the specific realization of the scalar sector further, i.e., whether a discrete symmetry is enforced in the scalar potential. Finally, one gets for doublets and triplets

\begin{equation}
    \label{ecu.3.5}
    \Gamma_D\left(Z_R\rightarrow 2\,\mathrm{Scalars}\right)<\frac{M_{Z_R}}{96\pi}\frac{e^2}{\cos^2\theta_W}\left\{2\tan^2\gamma+3\cot^2\gamma\right\},
\end{equation}

\begin{equation}
    \label{ecu.3.6}
    \Gamma_T\left(Z_R\rightarrow 2\,\mathrm{Scalars}\right)<\frac{M_{Z_R}}{16\pi}\frac{e^2}{\cos^2\theta_W}\left\{\cot^2\gamma+2\tan^2\gamma\right\}.
\end{equation}

\begin{figure}[!t]
    \centering
    \includegraphics[scale=0.7]{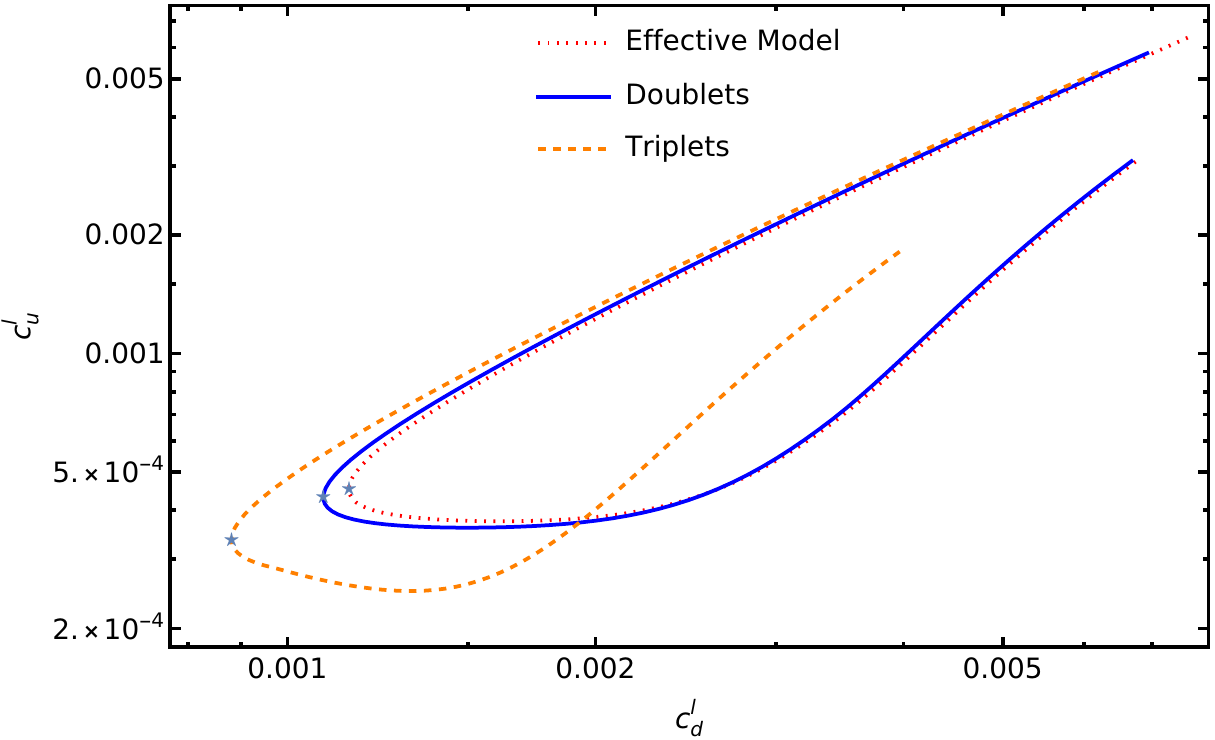}
    \caption[Parametric plot of the functions \(c^\ell_d\) and \(c^\ell_u\) depending on the angle \(\gamma\).]{Parametric plot of the functions \(c^\ell_d\) and \(c^\ell_u\) depending on the angle \(\gamma\). Varying $\gamma$ in the range $ [20^{\circ}, 70^{\circ}] $, the plotted curves are transited counterclockwise. The points used for setting the bounds on the mass of $Z_R$ are the ones marked with a star, see Fig.~\ref{fig:Gamma_doublets_triplets}. They correspond to values of \(\gamma\) around \(40^{\circ}\) and \(41^{\circ}\).}
    \label{fig:cu_cd_plane}
\end{figure}

Among the previous processes, the only ones that have to be considered in the Effective LR Model are \(Z_R\rightarrow  Z_Lh^0,Z_LH^0,h^0H^0\) and \(W_L^+W_L^-\). The first three decays can be taken into account by putting a bound on \(\Gamma_{\mathrm{Eff}}\left(Z_R\rightarrow2\,\mathrm{Neutral\,Scalars}\right)\) as it was done for the other two models. One can compute exactly the partial width of \(Z_R\rightarrow W_L^+W_L^-\) using the Feynman rule in Eq.~\eqref{ecu.A.9}:

\begin{equation}\label{ecu.3.7}
    \begin{split}
    & \Gamma_{\mathrm{Eff}}\left(Z_R\rightarrow 2\,\mathrm{Neutral\,Scalars}\right)+\Gamma_{\mathrm{Eff}} \left(Z_R\rightarrow W_L^+W_L^-\right)
    \\& \;\;\;\;\;\;\; <\frac{M_{Z_R}}{192\pi}\frac{e^2}{\cos^2\theta_W}\left\{2+\cot^2\gamma+3\tan^2\gamma\right\}.
    \end{split}
\end{equation}
 
 It is still necessary to study the decays \(Z_R\rightarrow W_R^{\pm}W_L^{\mp},\,H^{\pm}W_R^{\mp}\) and \(W_R^+W_R^-\). In App.~\ref{sec:Non-Fermionic_Decays} it is shown that the contribution of the first two is negligible since the partial width is, at most, proportional to \(v_{EW}^2/v_R\). Finally, the exact value of the partial width for the process \(Z_R\rightarrow W_R^+W_R^-\) can be written as

\begin{equation}
    \label{ecu.3.10}
    \Gamma_D\left(Z_R\rightarrow W_R^+W_R^-\right)=\frac{M_{Z_R}}{192\pi}\left(\frac{e}{\cos\theta_W}\right)^2\!\! \; \frac{\left(1-4\cos^2\gamma\right)^{3/2}}{\sin^2\gamma\cos^2\gamma}\!\left\{1+20\cos^2\gamma+12\cos^4\gamma\right\},
\end{equation}

\begin{equation}
    \label{ecu.3.11}
    \Gamma_T\left(Z_R\rightarrow W_R^+W_R^-\right)=\frac{M_{Z_R}}{48\pi}\left(\frac{e}{\cos\theta_W}\right)^2\!\! \; \frac{\left(1-2\cos^2\gamma\right)^{3/2}}{\sin^2\gamma\cos^2\gamma}\!\left\{1+10\cos^2\gamma+3\cos^4\gamma\right\}.
\end{equation}

\noindent As evident, Eq.~(\ref{ecu.3.10}) also applies to the effective LR model with only doublet scalars.
This process gives a result only if \(\gamma\in\left[60^{\circ},70^{\circ}\right]\) or \(\gamma\in\left[45^{\circ},70^{\circ}\right]\) in the models with doublets or triplets, respectively, given the requirement $ M_{Z_R} > 2 \, M_{W_R} $, and that one stays in the perturbative regime (\(70^{\circ}\gtrsim\gamma\gtrsim20^{\circ}\)). 

\begin{figure}
    \centering
    \includegraphics[scale=0.67]{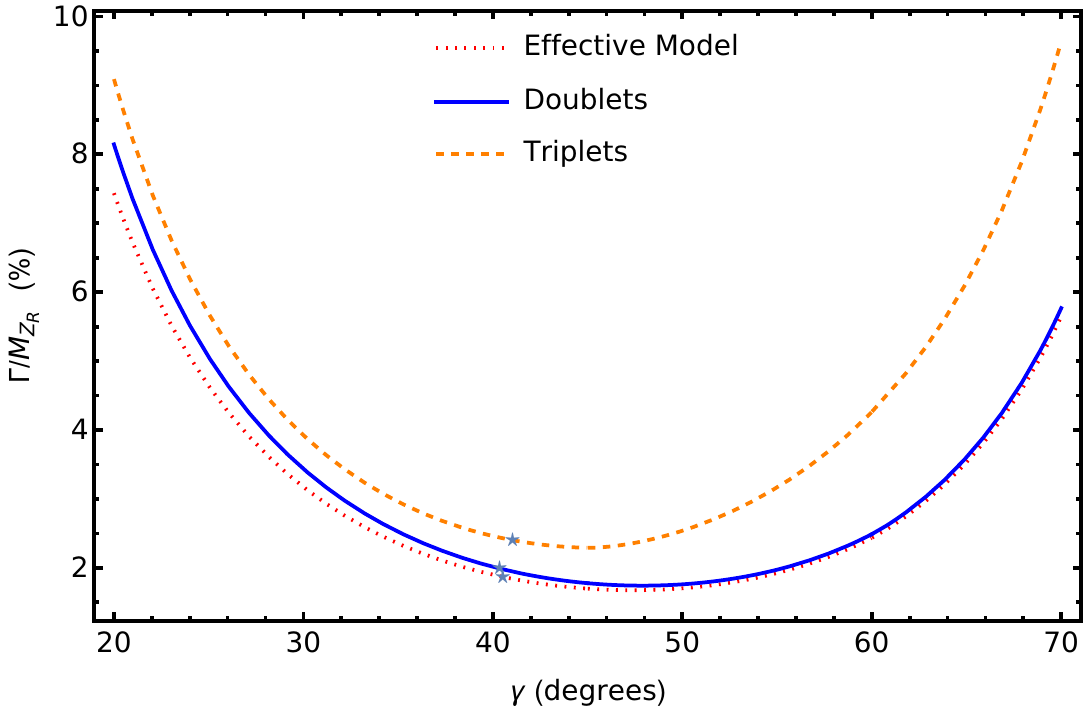}
    \caption[Total width of the $Z_R$ as a function of the angle $\gamma$.]{Total width of the $Z_R$ as a function of the angle $\gamma$. The points marked with a star are the ones used for setting the bounds on the mass $M_{Z_R}$, see Fig.~\ref{fig:cu_cd_plane}.}
    \label{fig:Gamma_doublets_triplets}
\end{figure}

All the partial widths are proportional to \(M_{Z_R}\), with no further explicit dependence on this quantity. Thus, both \(c_u^\ell\) and \(c_d^\ell\) will only depend on \(\gamma\). This dependence is displayed in Fig.~\ref{fig:cu_cd_plane} for the three LR models considered. The lowest cross sections are obtained when $c_d^\ell$ reaches its minimum possible value allowed by this model; the results are very close to the minimum of $c_u^\ell$, which shows a flatter dependence on $\gamma$ compared to $c_d^\ell$. Thus, when comparing to the experimental data, the three points marked with a star in Fig.~\ref{fig:cu_cd_plane} will give the less constraining lower bound on the $Z_R$ mass, which in
turn are close to the LR limit $g_L = g_R$. For the values of \(\gamma\) used to set bounds on the mass of $Z_R$, the values of the total width over the mass of the \(Z_R\) gauge boson are below \(3\%\), as shown in Fig.~\ref{fig:Gamma_doublets_triplets}. Using the \textit{narrow-resonance approximation} is thus justified.

\subsection{ Production and decay of the \(W_R\)} \label{subsec:prod_dec_WR}

Consider now the searches involving the production and decay to a pair of fermions of the \(W_R\) boson in colliders. In this case, there are more free parameters because the elements of the mixing matrix \(V_R^{\mathrm{CKM}}\) appear explicitly. The main objective is to calculate the cross section of the process \(pp\rightarrow W_R^\pm X \rightarrow f\bar{f'} X\).
The general expression
in the s-channel and at leading order in \(\alpha_s\)
is the following \cite{Ellis:1996mzs}:

\begin{eqnarray}\label{eq:beyond_NWA_1}
    \lefteqn{\sigma\left(pp\rightarrow W_R^\pm X \rightarrow f\bar{f'} X\right) = \frac{2\pi}{3} \frac{\Gamma_{W_R}}{M_{W_R}} \alpha_R\sum_{ij}|\!\left(V_R^{\mathrm{CKM}}\right)_{ij}\!|^2} && \\
    & \times & \int_{0}^1 {\rm d}x_1 {\rm d}x_2 \left[u_i\left(x_1,\,\mu\right)\bar{d}_j\left(x_2,\,\mu\right)+\bar{u}_i\left(x_2,\,\mu\right)d_j\left(x_1,\,\mu\right)\right] \frac{\hat{s} \, \mathrm{Br}\left(W_R^\pm \rightarrow f\bar{f}'\right)}{(\hat{s} - M^2_{W_R})^2 + M^2_{W_R} \, \Gamma^2_{W_R}}\, , \nonumber
\end{eqnarray}

\noindent where $ \hat{s} \coloneqq x_1 x_2 s $. For narrow resonances, one can substitute\footnote{Equivalently, for numerical purposes one can take $ \hat{s} $ in the integration range $ [(M_{W_R} - \Delta)^2, (M_{W_R} + \Delta)^2] $ (together with requiring $ \hat{s} \leq x s $, $ x $ being the other integration variable), where $ \Gamma_{W_R} \ll \Delta \ll M_{W_R} $.}

\begin{equation}
    \frac{1}{(\hat{s} - M^2_{W_R})^2 + M^2_{W_R} \, \Gamma^2_{W_R}} \;\to\; \frac{\pi}{M_{W_R} \, \Gamma_{W_R}} \,\delta (\hat{s} - M^2_{W_R})\, ,
\end{equation}

\noindent leading to:

\begin{equation}
    \label{ecu.3.12}
    \sigma\left(pp\rightarrow W_R^\pm X \rightarrow f\bar{f}' X\right)\approx\sigma\left(pp\rightarrow W_R^\pm X \right)\mathrm{Br}\left(W_R^\pm \rightarrow f\bar{f}'\right),
\end{equation}

\noindent after integration over $ \hat{s} $. The $W_R^\pm$ production cross section is \cite{ParticleDataGroup:2024cfk}

\begin{equation}
    \label{ecu.3.13}
    \sigma\left(pp\rightarrow W_R^\pm X \right)\approx\frac{2\pi^2}{3s}\alpha_R\sum_{ij}|\!\left(V_R^{\mathrm{CKM}}\right)_{ij}\!|^2\,\omega_{ij}\!\left(M^2_{W_R}/s,\,M_{W_R}\right),
\end{equation}

\noindent where \(s\) is the square of the center-of-mass energy, \(\alpha_R\coloneqq g_R^2/(4\pi)\) and the functions \(\omega_{ij}\left(z,\mu\right)\) are given by

\begin{equation}
    \label{ecu.3.14}
    \omega_{ij}\left(z,\,\mu\right)=\int_{z}^1\frac{{\rm d}x}{x} \left[u_i\left(x,\,\mu\right)\bar{d}_j\left(z/x,\,\mu\right)+\bar{u}_i\left(x,\,\mu\right)d_j\left(z/x,\,\mu\right)\right],
\end{equation}

\noindent with \(q_i\left(x,\,\mu\right)\) the PDF of the quark \(q_i\) inside the proton at the factorization scale \(\mu\) and parton momentum fraction \(x\) (sub-leading effects in \(\alpha_s\) will be discussed ahead).
Note that there is an explicit dependence on the elements of the matrix \(V_R^{\mathrm{CKM}}\) in Eq.~(\ref{ecu.3.13}).
Feynman rules for the $W_R$ couplings to fermion pairs are found in Eqs.~(\ref{ecu.A.11}) and (\ref{ecu.A.12}).

As derived from Eq.~\eqref{eq:beyond_NWA_1}, not restricting the $W_R$ from being produced with an invariant mass close to its resonance peak largely enhances the total cross section for large values of $ M_{W_R} $, compared to the narrow-width approximation, when $ M_{W_R} $ approaches $ \sqrt{s} $: in this case $\omega_{ij}$ in Eq.~\eqref{ecu.3.14} approaches $0$ since $z$ approaches $1$, while non-vanishing values are achieved from the original expression Eq.~\eqref{eq:beyond_NWA_1} when not restricting the value of $\hat{s}$. At the same time, the SM background increases when allowing for small values of $\hat{s}$, so one loses out on sensitivity towards signal events.

Now the calculation of the decay width of the \(W_R\) boson \(\Gamma_{W_R}\) at leading order in \(g_R\) is presented. At tree level, the relevant processes are listed in Tab.~\ref{tab:ZR_WR_two_body}. Considering that the fermions in the final state have a mass much smaller than the one of the \(W_R\) boson it is easy to obtain

\begin{equation}
    \label{ecu.3.15}
    \Gamma\left(W_R\rightarrow f_i\bar{f}_j\right)\approx\frac{\alpha_R}{12}|\!\left(V_R\right)_{ij}\!|^2N^f_CM_{W_R},
\end{equation}

\noindent where \(N^f_C\) represents the number of colors of the fermion $f$ and \(V_R=V_R^{\mathrm{CKM}}\) or \(V_R=V_R^{\mathrm{PMNS}}\) for quarks and leptons, respectively. In order to avoid the introduction of new free parameters, one can consider the two limiting cases for the masses of the RH neutrinos. If they are extremely heavy only the quark decay modes are kinematically open and
one obtains \(\Gamma\left(W_R\rightarrow f\bar{f}'\right) \coloneqq \sum_{ij} \Gamma\left(W_R\rightarrow f_i\bar{f}_j\right)
\approx (3/4)\alpha_RM_{W_R}\), while for light RH neutrinos the expression is \(\Gamma\left(W_R\rightarrow f\bar{f}'\right)\approx\alpha_RM_{W_R}\). The less stringent bound on the mass of the \(W_R\) results from the second case.

When using the equivalence theorem \cite{Cornwall:1974km,Lee:1977eg}, it is easy to see that one can set a bound on the sum of partial widths of the decays \(W_R^{\pm}\rightarrow W_L^{\pm}Z_L\), $ W_L^{\pm}h^0 $, $ W_L^{\pm}H^0 $, $H^{\pm}Z_L$, $H^{\pm}h^0$, $H^{\pm}H^0$ and \(H^{\pm\pm}H^{\mp}\), by calculating a bound on \(\Gamma\left(W_R\rightarrow 2 \, \mathrm{Scalars}\right)\). The contribution of the process \(W_R^{\pm}\rightarrow W_L^{\mp}H^{\pm\pm}\) for the models with triplets is negligible. The decay \(W_R^{\pm}\to H^{\pm}A\) is forbidden at the tree level. The details are given in App.~\ref{sec:Non-Fermionic_Decays}.
By adding all fermionic and bosonic final states, one gets the following bounds on the total width of the \(W_R\) boson for the doublet and triplet models:

\begin{figure}[t]
\centering
\includegraphics[scale=0.4]{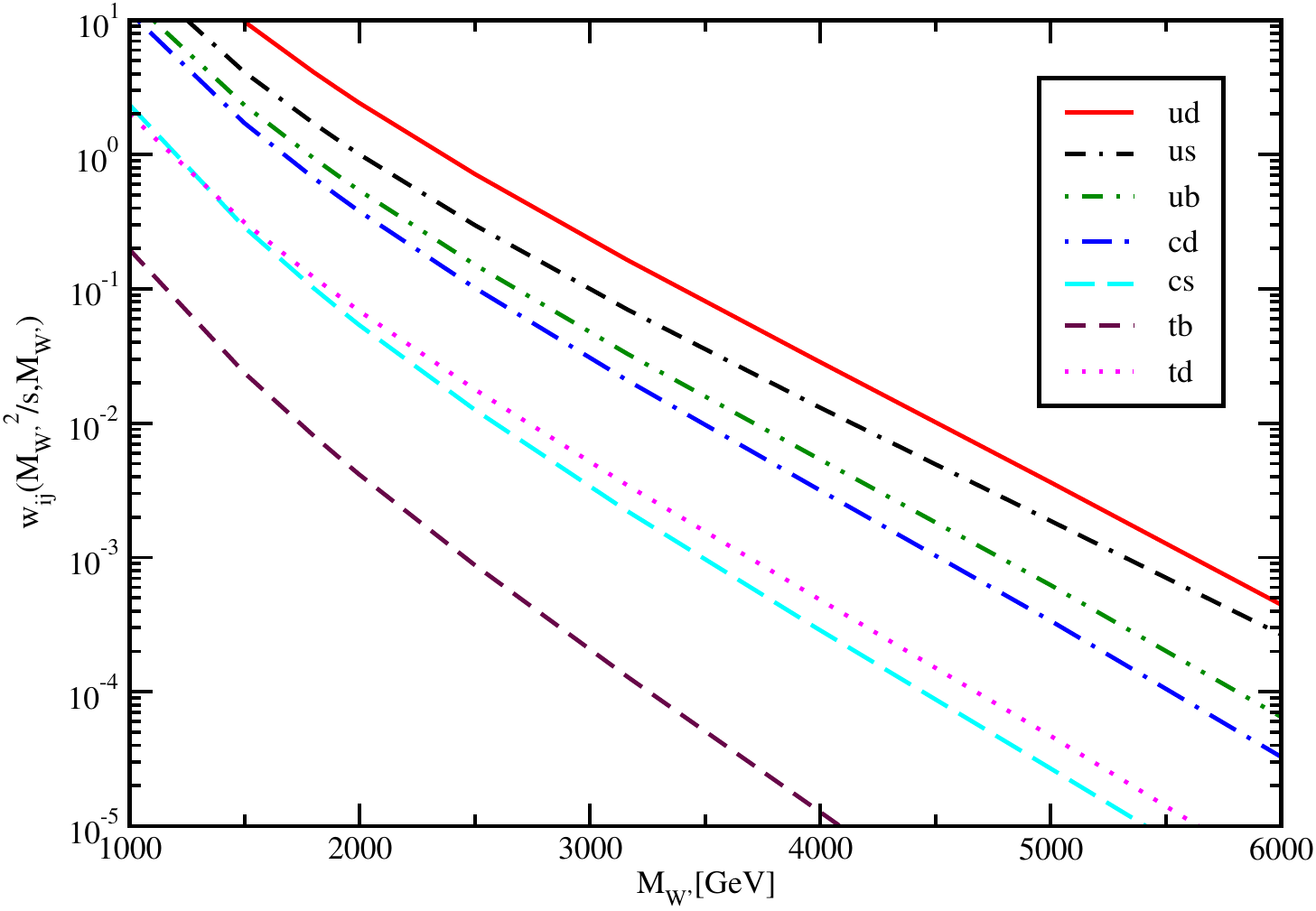}
\caption[The quantity $w_{ij}(M_{W_R}^2/s,M_{W_R})$ for up and down type quarks of various generations as a function of the $W' \coloneqq W_R$ mass.]{The quantity $w_{ij}(M_{W_R}^2/s,M_{W_R})$ for up and down type quarks of various generations as a function of the $W' \coloneqq W_R$ mass. Figure extracted from Ref.~\cite{Bernard:2020cyi}.}
\label{fig:pdfatlas}
\end{figure}

\begin{equation}
    \label{ecu.3.16}
    \Gamma_{W_R}^D< \frac{9}{8}\alpha_R M_{W_R},
\end{equation}

\begin{equation}
    \label{ecu.3.17}
    \Gamma_{W_R}^T< \frac{5}{4}\alpha_R M_{W_R}.
\end{equation}

The non-fermionic processes \(W_R^{\pm}\rightarrow W_L^{\pm} Z_L,W_L^{\pm}h^0\) and \(W_L^{\pm}H^0\)
are not allowed in the Effective LR Model because there is no mixing between the \(W_L\) and the \(W_R\) gauge bosons.
Thus, the following bound by considering light RH neutrinos is obtained:

\begin{equation}
    \label{ecu.18}
     \Gamma_{W_R}^{\mathrm{Eff}}< \alpha_R M_{W_R}.
\end{equation}

The bound on the mass of the \(W_R\) boson obtained from direct searches in colliders depends on the specific structure of the \(V_R^{\mathrm{CKM}}\) matrix. Therefore, following the spirit of this text, one must find the values of the matrix elements  \(|\left(V_R^{\mathrm{CKM}}\right)_{ij}|\) that minimize the cross section in Eq.~(\ref{ecu.3.13}).
The functions \(\omega_{ij}\) are displayed in Fig.~\ref{fig:pdfatlas}, and have been obtained using the NNPDF23LO PDF set \cite{Ball:2012cx} (at a factorization scale of $ 6 $~TeV, and for $ \sqrt{s} = 13 $~TeV). The figure clearly shows that the main contributions come from \(\omega_{ud}\) and \(\omega_{us}\); then, one has to take \(\left(V_R^{\mathrm{CKM}}\right)_{ud}=\left(V_R^{\mathrm{CKM}}\right)_{us}=0\). After the previous two terms, the one proportional to \(\omega_{ub}\) is dominant; in this case, it is not possible to take \(\left(V_R^{\mathrm{CKM}}\right)_{ub}=0\), as the matrix $V_R^{\mathrm{CKM}}$ must be unitary. The following leading term in the $W_R$ production is the one proportional to \(\omega_{cd}\), for which one then can set \(\left(V_R^{\mathrm{CKM}}\right)_{cd}=0\). Taking the above into account, the following structure is derived:

\begin{equation}
    \label{ecu.3.19}
    V_R^{\mathrm{CKM}}=\left(
    \begin{array}{ccc}
    0 & 0 & 1 \\
    0 & 1 & 0 \\
    1 & 0 & 0 
    \end{array}\right)
\end{equation}

\noindent where complex phases are irrelevant in the present context. \textit{This anti-diagonal matrix is associated with the minimum possible value of the cross section in the s-channel, and one gets from it the less stringent bounds on \(M_{W_R}\) from di-jet and di-lepton searches} that are discussed in the following section (see Figs.~\ref{fig:WR_Dijet} and \ref{fig:WR_Lepton} and Tab.~\ref{tab:summary_bounds}).
These bounds can be somewhat improved by considering NLO corrections in QCD, namely, associated production of the $W_R$ with a bottom quark. This is phase-space suppressed, and moreover suppressed by the small value of the strong coupling at the high perturbative scales relevant for the process, but depends on the PDF of the gluon.\footnote{
The $(V^{\rm CKM}_R)_{cb}$ coupling has been studied in the context of $B$-meson flavor-changing charged-current anomalies,
and it was noticed that with a future LHC luminosity of 300~fb$^{-1}$ one can probe $ \mathcal{O}(1) $ couplings of a $W_R$ of mass around $ 2.4 $~TeV in the tau decay channel, see Ref.~\cite{Altmannshofer:2017poe}; see also Refs.~\cite{Abdullah:2018ets,Blanke:2022pjy}.} Just for comparison, in Figs.~\ref{fig:WR_Dijet} and \ref{fig:WR_Lepton}, and Tab.~\ref{tab:summary_bounds}
the stronger limits that would be obtained with a diagonal structure 
of the RH quark mixing matrix are provided, for which the latter associated $W_R$ production, like the t-channel production, is negligible compared to the dominant s-channel production \cite{ParticleDataGroup:2024cfk}.

Now the detailed discussion of the di-jet decay mode of the \(W_R\) boson is presented.
This channel has the advantage of not depending on the characteristics of the leptonic sector, namely, neutrino masses and RH lepton mixing matrix.
In order to obtain the cross section of the process \(pp\rightarrow W_R\rightarrow j_1j_2\), one needs to calculate \(\Gamma\left(W_R\rightarrow q_1\bar{q}_2\right)\). Looking at Eq.~(\ref{ecu.3.15}), it is easy to get

\begin{equation}
    \label{ecu.3.20}
    \Gamma\left(W_R\rightarrow q_1\bar{q}_2\right)\approx\frac{\alpha_R}{4}M_{W_R}\sum_{i=u,c,t}\sum_{j=d,s,b}\left|\left(V_R^{\mathrm{CKM}}\right)_{ij}\right|^2= \frac{3}{4}\alpha_RM_{W_R}.
\end{equation}

\noindent Consequently, the following lower bounds on the branching fractions are obtained:

\begin{equation}
    \label{ecu.3.21}
    \mathrm{Br}\left(W_R\rightarrow j_1j_2\right)_{\mathrm{Eff}}> \frac{3}{4},
\end{equation}

\begin{equation}
    \label{ecu.3.22}
    \mathrm{Br}\left(W_R\rightarrow j_1j_2\right)_D>\frac{2}{3},
\end{equation}

\begin{equation}
    \label{ecu.3.23}
    \mathrm{Br}\left(W_R\rightarrow j_1j_2\right)_T> \frac{3}{5}.
\end{equation}

\noindent In the hadronic $W_R^\pm$ decay width the QCD correction factor $\delta_{\mathrm{QCD}}\approx 1 + \frac{\alpha_s}{\pi}\ge 1$ has been neglected. It is small at the high scale $\mu=M_{W_R}$, and since it would increase the branching fractions, these lower bounds are even more conservative. For a given mixing matrix $V_R^{\mathrm{CKM}}$, the minimum value of the production cross section corresponds to the minimum value of \(\alpha_R = \alpha/(\sin^2{\gamma} \cos^2{\theta_W}) \), 
for which one finds \(\Gamma_{W_R}^D/M_{W_R}< 1.2\%\), \(\Gamma_{W_R}^T/M_{W_R}< 1.3\%\) and \(\Gamma_{W_R}^{\mathrm{Eff}}/M_{W_R}< 1.1\%\), clearly in the narrow resonance regime.

\textit{If the masses of the right-handed neutrinos are negligible compared to the mass of the \(W_R\),}\footnote{This is precisely the case in the model with two doublets and one bidoublet.} \textit{one can also consider the leptonic decay modes} \(W_R\rightarrow \ell_i \bar{\nu}_R\), $ \ell_i = e, \mu, \tau $, where the RH neutrinos are stable (their flavors are omitted). In this scenario, adding the contributions from the different neutrinos, one gets \(\Gamma\left(W_R\rightarrow \ell_i \bar{\nu}_R\right)\approx\alpha_R M_{W_R}/12\). Therefore, the bounds on the branching fractions are

\begin{equation}
    \label{ecu.3.24}
    \mathrm{Br}\left(W_R\rightarrow \ell_i\bar{\nu}_R\right)_{\mathrm{Eff}}\gtrsim\frac{1}{12},
\end{equation}

\begin{equation}
    \label{ecu.3.25}
    \mathrm{Br}\left(W_R\rightarrow \ell_i\bar{\nu}_R\right)_D>\frac{2}{27},
\end{equation}

\begin{equation}
    \label{ecu.3.26}
    \mathrm{Br}\left(W_R\rightarrow \ell_i\bar{\nu}_R\right)_T>\frac{1}{15}.
\end{equation}

\section{Collider Bounds on the LR Scale. Experimental Data} \label{sec:Collider_Data}

This section is devoted to the presentation of the collider bounds on the masses of the \(Z_R\) and \(W_R\) gauge bosons, and on the LR scale \(v_R\) that are obtained once the cross sections of the previous part of the text are compared with the experimental evidence. \textit{The data that have been used come from Run 2 of the LHC.}

\subsection{Direct bounds on the \(Z_R\) mass} \label{subsec:Z_R_Experimental_Bound}

The current experimental information needed to put a bound on the \(Z_R\) mass using the cross section of the process  \(pp\rightarrow Z_RX\rightarrow \ell^+\ell^-X\) can be extracted from the following LHC searches for two charged leptons in the final state:

\begin{itemize}
    \item CMS ($ \sqrt{s} = 13 $~TeV) \cite{CMS:2021ctt}, in the di-electron (137~fb$^{-1}$) and di-muon (140~fb$^{-1}$) channels;
    \item ATLAS ($ \sqrt{s} = 13 $~TeV, 139~fb$^{-1}$) \cite{ATLAS:2019erb}, in the di-electron and di-muon channels;
    \item ATLAS ($ \sqrt{s} = 13 $~TeV, 36.1~fb$^{-1}$) \cite{ATLAS:2017eiz}, in the di-tau channel.
\end{itemize}

\noindent They achieve very similar bounds on the $Z_R$ mass in the former two cases (i.e., di-electron and di-muon searches), which are better than in the third case (i.e., di-tau searches, of lower statistics). In this text there will be no attempt at combining results from ATLAS and CMS. The impact of the total width of the $Z_R$ on the present bounds on its mass is modest in the experimental searches of Refs.~\cite{CMS:2021ctt,ATLAS:2019erb}, for values of the mass $ M_{Z_R} \gtrsim 4 $~TeV, and when $ \Gamma_{Z_R} \lesssim 10 \% $ of the gauge boson mass.
Theoretical uncertainties result from PDF uncertainties and higher-orders, and grow with the invariant mass, reaching about 20\% at scales of many TeV. Based on Ref.~\cite{CMS:2021ctt} for the combination of di-electron and di-muon modes, the achieved bounds by the author of this text are shown in Tab.~\ref{tab:summary_bounds} \cite{Solera:2023kwt}. The very conservative lower bounds on the \(Z_R\) mass presented here remain valid even if the condition \(g_R,\,g_X\leq 1\) is relaxed, because for values of \(\gamma\) smaller than \(20^{\circ}\) or greater than \(70^0\), one would get larger values for both \(c_d^\ell\) and \(c_u^\ell\), as shown in Fig.~\ref{fig:cu_cd_plane}.

There are also available
di-jet searches from CMS ($ \sqrt{s} = 13 $~TeV, 137~fb$^{-1}$) \cite{CMS:2019gwf} and ATLAS ($ \sqrt{s} = 13 $~TeV, 139~fb$^{-1}$) \cite{ATLAS:2019fgd}, and searches based on the di-boson decay modes $ Z_R \to W_L^+ W_L^- $ and $ Z_R \to Z_L h^0 $, Refs.~\cite{ATLAS:2019nat,ATLAS:2020fry,CMS:2021fyk,CMS:2021klu}. Better bounds, however, are achieved from the previous di-lepton searches, see also Refs.~\cite{Osland:2020onj,Osland:2022ryb}.

\subsection{Direct bounds on the \(W_R\) mass} \label{subsec:W_R_Experimental_Bound}

The less conservative direct bound on the mass of the \(W_R\) in the effective LR model and in the model with triplets is obtained using the dijet decay mode \(pp\rightarrow W_R\rightarrow j_1j_2\), because the derived smallest cross section does not depend on the masses of the right-handed neutrinos, which can be comparable to the one of the \(W_R\) if they are Majorana particles. For this mode, there are data from:

\begin{itemize}
    \item CMS ($ \sqrt{s} = 13 $~TeV, 139~fb$^{-1}$) \cite{CMS:2019gwf,CMS_supplementary_material_dijet} and
    \item ATLAS ($ \sqrt{s} = 13 $~TeV, 139~fb$^{-1}$) \cite{ATLAS:2019fgd,ATLAS_supplementary_material_dijet}.
\end{itemize}

\begin{figure}[t!]
\centering
\includegraphics[scale=0.6]{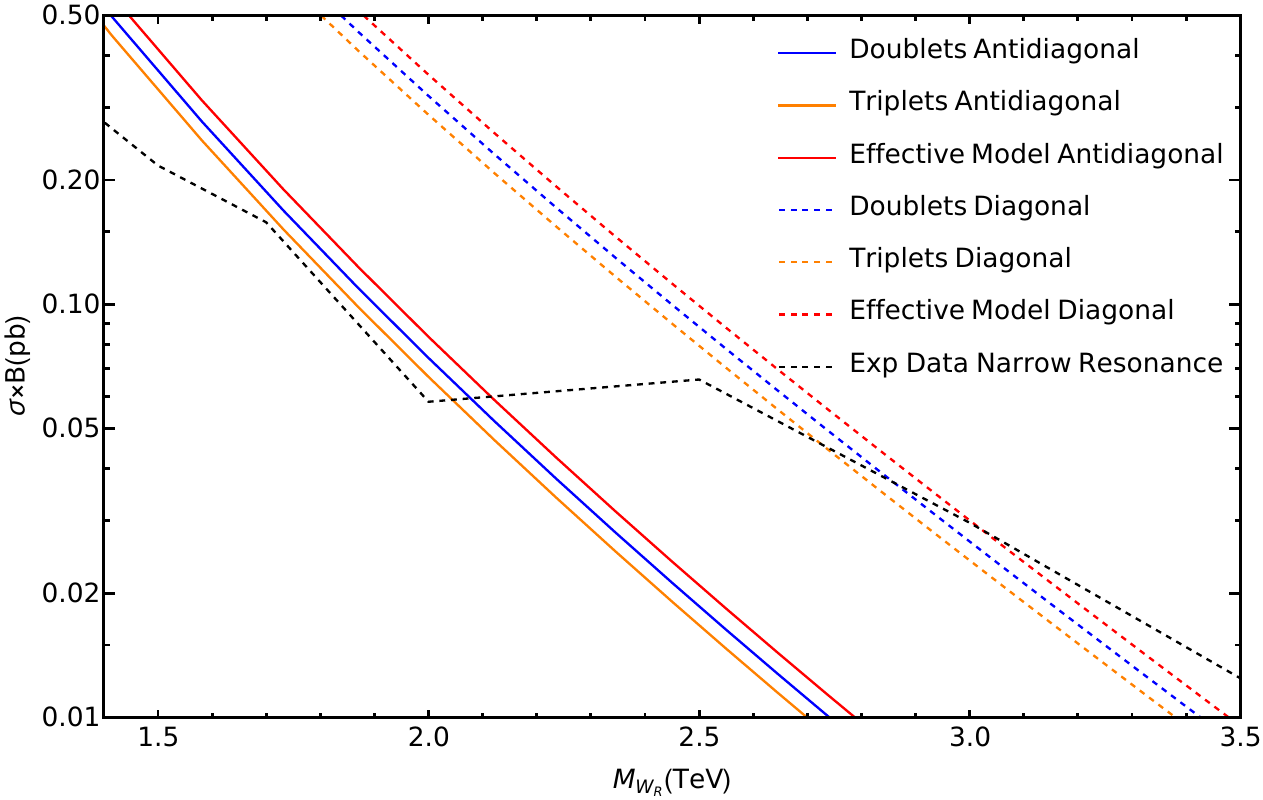}
\caption[Comparison between the theoretical prediction of the cross section times the branching fraction of the process \(W_R\rightarrow jj\) for different LR symmetric models and an anti-diagonal or diagonal structure of $V_R^{\mathrm{CKM}}$.]{Comparison between the theoretical prediction of the cross section times the branching fraction of the process \(W_R\rightarrow jj\) for different LR symmetric models and an anti-diagonal or diagonal structure of $V_R^{\mathrm{CKM}}$. The experimental bound obtained by ATLAS \cite{ATLAS:2019fgd} in the narrow resonance regime is also shown.}
\label{fig:WR_Dijet}
\end{figure}

They achieve very similar bounds on the total cross section up to $ \lesssim 4$~TeV; as previously stated, there will be no attempt at combining results from ATLAS and CMS in this text. Note that in Eq.~\eqref{ecu.3.20} the top flavor has been included, which is highly boosted and explicitly considered in the simulation of signal events in Ref.~\cite{ATLAS:2019fgd}, but it is also present to some extent in the final di-jet mode of Ref.~\cite{CMS:2019gwf,andreas-hinzmann}. The results are shown in Fig.~\ref{fig:WR_Dijet} and Tab.~\ref{tab:summary_bounds} for Ref.~\cite{ATLAS:2019fgd}, where the acceptance from Ref.~\cite{ATLAS_supplementary_material_dijet} for the $W'$ model has been taken into account. In the comparison to data, a $K$ factor of 1.3 \cite{CMS:2019gwf,Sullivan:2002jt} to take into account the higher order corrections of QCD is employed.

\begin{figure}[t]
\centering
\includegraphics[scale=0.77]{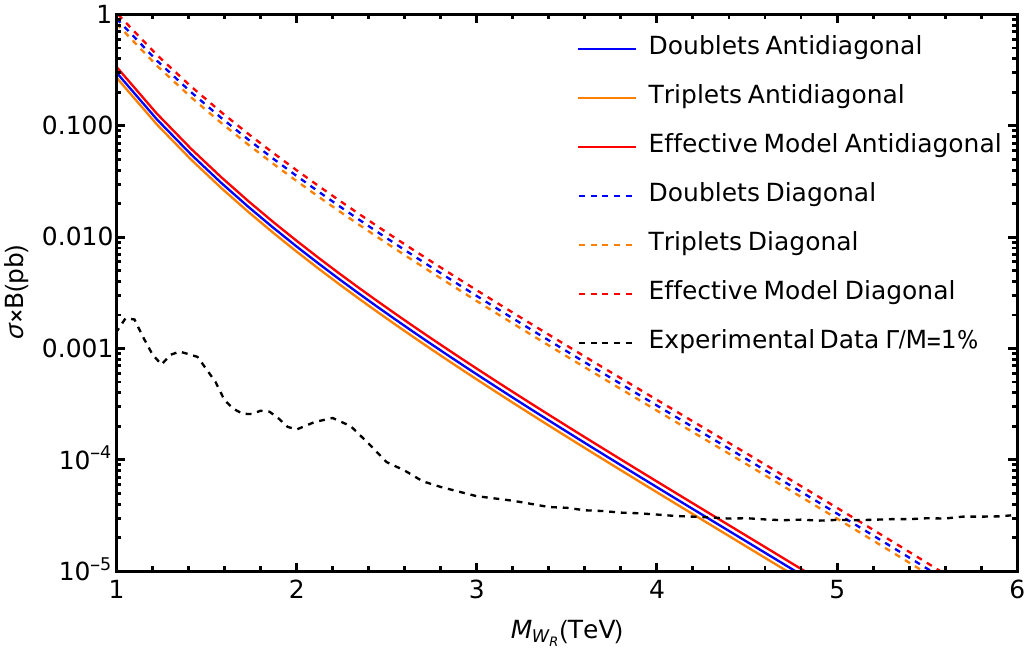}
\caption[Comparison between the theoretical prediction of the cross section times the branching fraction of the process \(W_R\rightarrow \ell_i \bar{\nu}_R\) for different LR symmetric models and an anti-diagonal or diagonal structure of $V_R^{\mathrm{CKM}}$.]{Comparison between the theoretical prediction of the cross section times the branching fraction of the process \(W_R\rightarrow \ell_i \bar{\nu}_R\) for different LR symmetric models and an anti-diagonal or diagonal structure of $V_R^{\mathrm{CKM}}$. Also shown is the experimental bound obtained by ATLAS \cite{ATLAS:2019lsy}, combining the data from the decays to \(e \bar{\nu}_R\) and \(\mu \bar{\nu}_R\) in the narrow resonance regime. The data corresponds to the case $ m_{\ell \nu} > 0.3 \, M_{W_R}$.}
\label{fig:WR_Lepton}
\end{figure}

As already mentioned, in the model with a bidoublet and two doublets neutrinos are Dirac particles, so they have a small mass, and hence the leptonic decay mode  \(W_R\rightarrow \ell \bar{\nu}_R\) is available. The associated experimental data are:

\begin{itemize}
    \item ATLAS ($ \sqrt{s} = 13 $~TeV, 139~fb$^{-1}$) \cite{ATLAS:2019lsy,ATLAS_supplementary_material_ellnu} and
    \item CMS ($ \sqrt{s} = 13 $~TeV, 138~fb$^{-1}$) \cite{CMS:2022yjm,CMS_supplementary_material_ellnu},
\end{itemize}

\noindent and the tauonic mode:

\begin{itemize}
    \item ATLAS ($ \sqrt{s} = 13 $~TeV, 139~fb$^{-1}$) \cite{ATLAS:2021bjk} and
    \item CMS ($ \sqrt{s} = 13 $~TeV, 35.9~fb$^{-1}$) \cite{CMS:2018fza}.
\end{itemize}

\noindent They achieve very similar bounds on the $W_R$ mass in the combined light lepton case, which are better than those bounds in the latter tauonic mode.
The combined data of \(W_R\rightarrow e \bar{\nu}_R\) and \(W_R\rightarrow \mu \bar{\nu}_R\) from Ref.~\cite{ATLAS:2019lsy} are taken for the present study.
Theoretical uncertainties from PDFs and higher-orders are small. A constant $K$ factor of 1.3 is employed, valid for $ M_{W_R} \simeq 3.0 $~TeV \cite{CMS:2022yjm}.
The results of the bound on \(M_{W_R}\) are shown in Fig.~\ref{fig:WR_Lepton} and Tab.~\ref{tab:summary_bounds}. The experimental data of the fiducial case $ m_{\ell \nu} > 0.3 \, M_{W_R}$ are employed, which is very close to the on-shell limit.
The less constraining bound is achieved for the case with the minimum value of \(\alpha_R\), which is shown therein. These bounds are better than the ones produced by searches of the $ W_R^\pm \to W_L^\pm h^0 $ (which depend on the realization of the scalar sector) and $ W_R^\pm \to W_L^\pm Z_L $ decay modes, see Ref.~\cite{Workman:2022ynf} and references therein; note that both channels are not present in the Effective LR Model.

Relaxing the condition \(g_R,\, g_X\leq 1\) in the direct searches of the \(W_R\) gauge boson has an impact in the bound on its mass. This is due to the fact that, if one allows \(g_X\) to take greater values, then \(g_R\) could be smaller and consequently we would get a smaller cross section.

\subsection{Combined bounds on \(M_{W_R}\) and \(M_{Z_R}\)} \label{subsec:Combined_Bounds}

\begin{figure}
    \centering
    \includegraphics[scale=0.463]{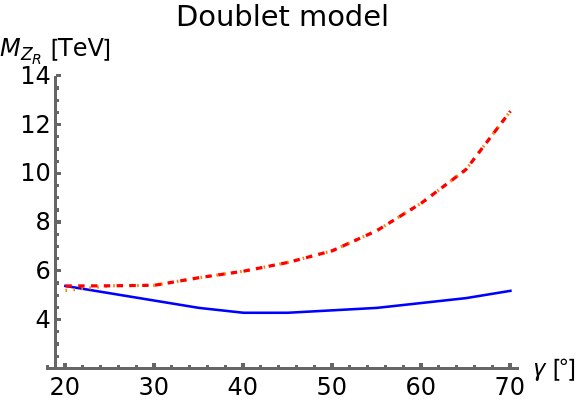} \hspace{5mm}
    \includegraphics[scale=0.463]{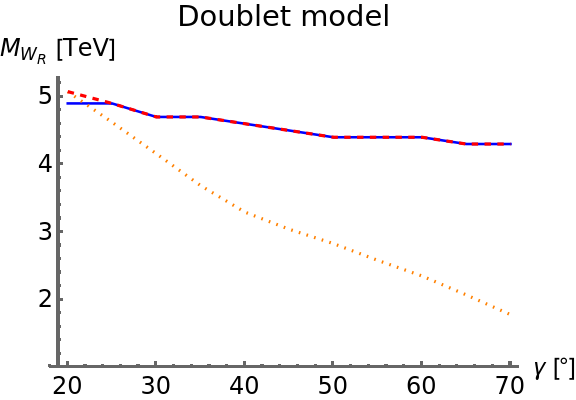} \\
    \vspace{5mm}
    \includegraphics[scale=0.37]{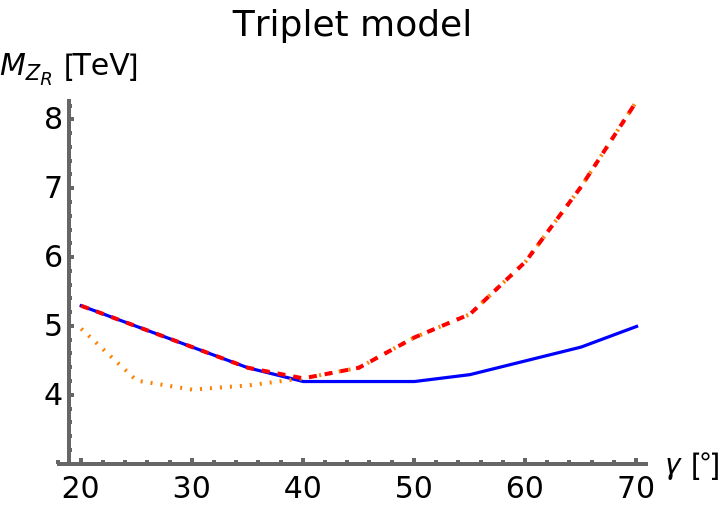} \hspace{5mm}
    \includegraphics[scale=0.37]{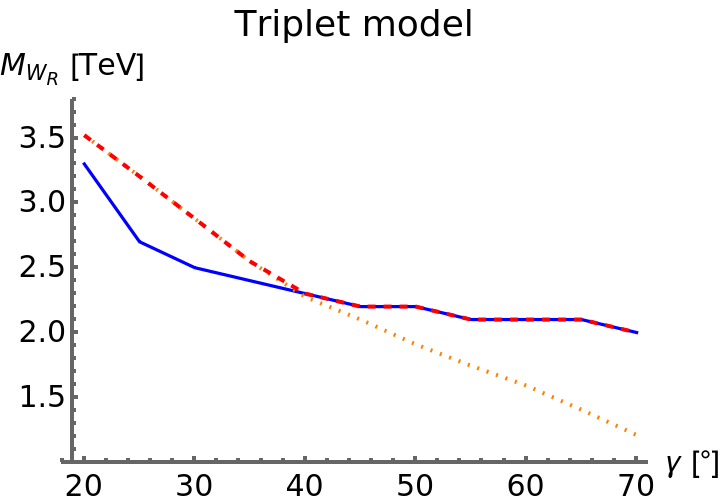} \\
    \vspace{5mm}
    \includegraphics[scale=0.463]{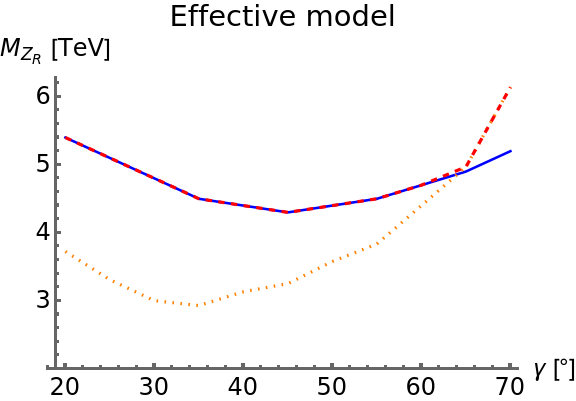} \hspace{5mm}
    \includegraphics[scale=0.463]{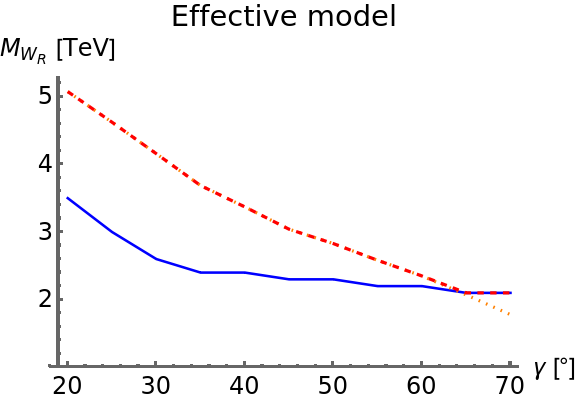}
    \caption[Bounds on 
    $M_{Z_R}$ (left panels) and $M_{W_R}$ (right panels) in
    the doublet (upper panels), triplet (middle panels) and effective models (lower panels).]{Bounds on 
    $M_{Z_R}$ (left panels) and $M_{W_R}$ (right panels) in
    the doublet (upper panels), triplet (middle panels) and effective models (lower panels). The solid blue curves correspond to the bounds on $M_{Z_R}$ ($M_{W_R}$) derived from direct searches of the $Z_R$ (respectively, $W_R$) gauge boson; the dotted orange lines show the bounds on the $Z_R$ ($W_R$) mass derived from direct searches of the $W_R$ (respectively, $Z_R$) gauge boson, based on the relation between the two heavy gauge boson masses in the LR models realizations under discussion, see Eqs.~\eqref{eq:relation_masses_doublets} and \eqref{eq:relation_masses_triplets}.
    The strongest limits are displayed in dashed red.}
    \label{fig:direct_indirect_mass_bounds}
\end{figure}

Hereby the bounds achieved in the previous sections are summarized.
They are displayed with solid blue curves in Fig.~\ref{fig:direct_indirect_mass_bounds} as function of the mixing angle $\gamma$ for the three scalar sectors analyzed: (D), (T) and (Eff). The left and right panels show, respectively, the  \begin{table}[h!]
    \centering
    \renewcommand{\arraystretch}{1.2}
    \begin{tabular}{|lcc|c|c|c|}
        \hline
        Channel & Refs. & & $\Phi$ + $\chi_{L,R}$ (D) & $\Phi$ + $\Delta_{L,R}$ (T) & $\chi_{L,R}$ (Eff) \\
        \hline
        \hline
        $Z_R \to \ell_i \bar{\ell}_i$ & \cite{CMS:2021ctt} & $M_{Z_R}$ & 4.3 & 4.2 & 4.3 \\
        \hline
        \hline
        $W_R \to j j$, anti-diag. & \cite{ATLAS:2019fgd} & \multirow{2}{*}{$M_{W_R}$} & 2.1 & 2.0 & 2.1 \\
        \cline{4-6}
        $W_R \to \ell_i \bar{\nu}_R$, anti-diag. & \cite{ATLAS:2019lsy} & & 4.3 & 4.2 & 4.3 \\
        \hline
        \hline
        $W_R \to j j$, diag. & \cite{ATLAS:2019fgd} & \multirow{2}{*}{$M_{W_R}$} & 2.9 & 2.7 & 3.0 \\
        \cline{4-6}
        $W_R \to \ell_i \bar{\nu}_R$, diag. & \cite{ATLAS:2019lsy} & & 5.1 & 5.0 & 5.1 \\
        \hline
    \end{tabular}
    \caption[Summary of the direct bounds on the new gauge boson masses $ M_{Z_R} $ or $ M_{W_R} $ (given in TeV), according to the case indicated in the first column.]{Summary of the direct bounds on the new gauge boson masses $ M_{Z_R} $ or $ M_{W_R} $ (given in TeV), according to the case indicated in the first column. 
    The more conservative bounds on $M_{W_R}$ are obtained with an anti-diagonal RH quark-mixing matrix. For comparison, also the stronger limits emerging from a diagonal (i.e., CKM-like) texture are shown. The case $W_R \to \ell_i \bar{\nu}_R$ concerns light RH neutrinos. Bounds combine information from electronic and muonic modes.}
    \label{tab:summary_bounds}
\end{table} lower limits on $M_{Z_R}$ and $M_{W_R}$. 
The $W_R$ figures correspond to an anti-diagonal RH quark-mixing matrix that imposes less restrictive constraints.
In the triplet and effective cases, where Majorana neutrinos are possible, they are based on data from the $ W_R \to jj $ mode, which is independent of the neutrino sector. In the doublet case, where neutrinos are purely Dirac fermions, the blue curve shows the more constraining bounds on $ M_{W_R} $ obtained from the leptonic decay mode $ W_R \to \ell \bar{\nu}_R $ with light right-handed neutrinos.

The values in Tab.~\ref{tab:summary_bounds} result from the less stringent constraints in scenarios (D), (T) and (Eff), i.e., they correspond to the value of $\gamma$ giving the weakest lower bound in the given scenario, assuming a perturbative theory. Therefore,
masses are excluded below these values (with a high significance, of around 95\% CL).
In all three cases, the limits on $M_{W_R}$ obtained from the hadronic and leptonic decay modes of the $W_R$ are displayed, and with diagonal and anti-diagonal structures of the RH quark mixing matrix. The masses are constrained to lie above a few TeV in both cases.
The most important bound on the $ W_R $ mass comes from the leptonic decay mode, which depends on the assumption that RH neutrinos are light enough. The bounds on the $ W_R $ mass from di-jet final states are independent of this assumption. Note that the consideration of different structures of the mixing matrix for right-handed quarks leads to differences of $ \lesssim 1 $~TeV in the bounds. The bounds are very similar across the three LRMs that are discussed in this work, which differ in their scalar content, implying that collider searches based on fermions in the final state are mostly independent of the impact of the scalar sector on the heavy gauge-boson total width.

Now a detailed discussion of indirect bounds derived from
the use of Eqs.~\eqref{Mass_Eigenstates_Gauge_Boson_Doublet}, \eqref{Mass_Eigenstates_Gauge_Boson_Triplet}, \eqref{W_Bosons_Mass_Effective_LR} and \eqref{Z_Boson_Masses_Effective_LR}
as a function of the couplings of the theory is performed.  Namely, in the two cases having doublet fields:

\begin{equation}\label{eq:relation_masses_doublets}
    M_{W_R} = \cos \gamma \, M_{Z_R} \,,
\end{equation}
while in the model with triplets

\begin{equation}\label{eq:relation_masses_triplets}
    \sqrt{2} \, M_{W_R} = \cos \gamma \, M_{Z_R} \,.
\end{equation}

\noindent
As seen from these expressions, the masses of the \(W_R\) and \(Z_R\) gauge bosons are related by \(\cos\gamma\), where $\gamma$ is defined in Eq.~\eqref{Couplings_Relations}. Consequently, \textit{by imposing constraints on one of these masses, one can establish limits on the mass of the other gauge boson}, taking into account the permissible range of \(\gamma\) as determined by perturbativity considerations. In scenarios (D), (T) and (Eff), it is clear that $ M_{W_R} \leq M_{Z_R} $.\footnote{As such, Eqs.~\eqref{eq:relation_masses_doublets} or \eqref{eq:relation_masses_triplets} would not hold, for instance, under the breaking pattern $ SU(2)_R \times U(1)_X \to U(1)_{3R} \times U(1)_X \to U(1)_Y $ \cite{Mohapatra:1998rq}, not considered in this work. In the latter case, one could still exploit that $ M_{W_R} > M_{Z_R} $, see also Ref.~\cite{Mohapatra:1986uf}.}

\begin{figure}
    \centering
    \includegraphics[scale=0.29]{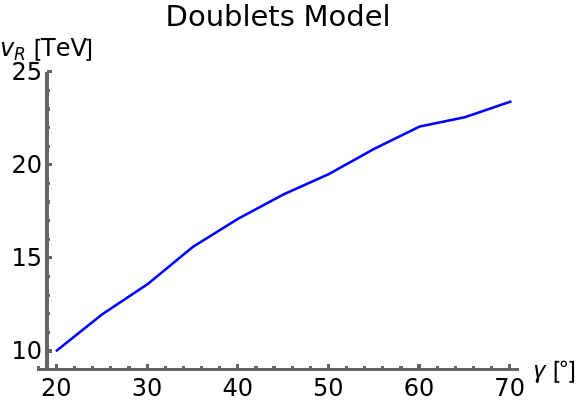} \hspace{5mm}
    \includegraphics[scale=0.29]{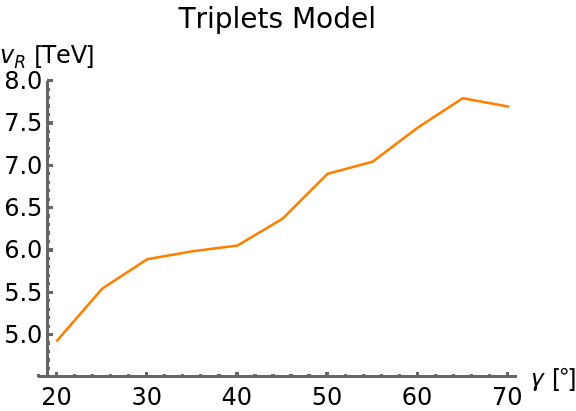} \hspace{5mm}
    \includegraphics[scale=0.29]{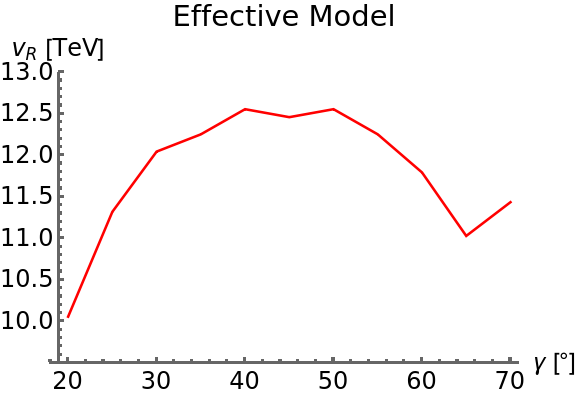}
    \caption{Bounds on the LR scale 
    in the doublet (left panel, in blue), triplet (middle panel, in orange) and effective (right panel, in red) models, as functions of the mixing angle \(\gamma\).
    }
    \label{fig:vR:bounds}
\end{figure}

The use of these relations is illustrated with dotted orange curves in
Fig.~\ref{fig:direct_indirect_mass_bounds}.
Note that the expression relating the masses $ M_{Z_R} $ and $ M_{W_R} $ depends on the specific scenario analyzed, as shown in Eqs.~\eqref{eq:relation_masses_doublets} and \eqref{eq:relation_masses_triplets}. Consequently, the indirect constraints vary according to whether the doublet or triplet realizations of the scalar sector are considered.
Depending on the value of $\gamma$ inside the perturbative region, the ratio of direct vs. indirect bounds on the gauge boson masses from searches in colliders can vary by as much as a factor of 2. In particular, this is the case for the LR symmetric scenario (\(\gamma\approx33^{\circ}\)) for the mass of the \(W_R\) boson in the effective model.
Fig.~\ref{fig:direct_indirect_mass_bounds} also demonstrates that for lower (higher) values of $\gamma$, one can achieve up to about a 50\% improvement in the bounds on the $W_R$ (respectively, $Z_R$) gauge boson mass. This highlights the complementarity of the two strategies in the context of the Doublet, Triplet and Effective LR models.

Only the data corresponding to the narrow-width approximation for all the values of \(\gamma\) has been used. As stated previously, considering different values of the \(Z_R\) width in the range that corresponds to the present case (\(\Gamma_{Z_R}/M_{Z_R}\) between 2\% and 10\%) has little impact on the bounds on its mass. On the other hand,
\(\Gamma_{W_R}/M_{W_R}\lesssim5\%\) when \(\gamma>30^{\circ}\); small discrepancies with respect to the narrow width approximation could then only be found for smaller values of the mixing angle, i.e., \(\gamma<30^{\circ}\). However, the bounds for small values of $\gamma$ are dominated by the indirect bounds derived from $Z_R$ production.

\begin{table}
    \centering
    \begin{tabular}{|c|c|c|c|}
    \hline
         & \(\Phi+\chi_{L,R}\) (D) & \(\Phi+\Delta_{L,R}\) (T) & \(\chi_{L,R}\) (Eff) \\ \hline\hline
      \(M_{W_R}\)    & 4.3 & 2.0 & 2.1 \\ \hline
      \(M_{Z_R}\)   & 5.4 & 4.2 & 4.3 \\ \hline
      \(v_{R}\)   & 10 & 4.9 & 10 \\ 
    \hline
    \end{tabular}
    \caption{Summary of the bounds on the masses of the heavy gauge bosons and on the LR scale (given in TeV) for the three different SSB mechanisms considering the combination of both direct and indirect methods.}
    \label{Final_Summary_Tab}
\end{table}

As seen from the dashed red curves in Fig.~\ref{fig:direct_indirect_mass_bounds}, combining direct and indirect bounds, masses are constrained to lie above: (D) $M_{W_R} > 4.3$~TeV and $M_{Z_R} > 5.4$~TeV in the case of the ``bidoublet + two doublets'' model; (T) $M_{W_R} > 2.0$~TeV and $M_{Z_R} > 4.2$~TeV in the case of the ``bidoublet + two triplets'' model; (Eff) $M_{W_R} > 2.1$~TeV and $M_{Z_R} > 4.3$~TeV in the case of the effective ``two doublets'' model. In the first (D) case, neutrinos are necessarily of Dirac kind, and thus of negligible masses, ultimately implying the larger values for the bounds. These bounds are summarized in Tab.~\ref{Final_Summary_Tab}.

\textit{Given the values of the lower bounds on the gauge boson masses as a function of $\gamma$, and the perturbativity requirement on the gauge couplings, one can extract 
the lower bounds on the LR energy scale shown }in Fig.~\ref{fig:vR:bounds} from Eqs.~\eqref{Mass_Eigenstates_Gauge_Boson_Doublet}, \eqref{Mass_Eigenstates_Gauge_Boson_Triplet}, \eqref{W_Bosons_Mass_Effective_LR} and \eqref{Z_Boson_Masses_Effective_LR}. The less constraining limits are obtained for the lowest allowed value of the mixing angle 
\(\gamma=20^{\circ}\), which results in the absolute bounds summarized in Tab.~\ref{Final_Summary_Tab}.

\newpage

\thispagestyle{empty}

\chapter*{Results, Conclusions and Future Lines of Research}
\addcontentsline{toc}{chapter}{Results, Conclusions and Future Lines of Research}
\markboth{RESULTS, CONCLUSIONS AND FUTURE LINES OF RESEARCH}{RESULTS, CONCLUSIONS AND FUTURE LINES OF RESEARCH}

Chiral gauge theories are currently a major focus of research in theoretical physics. Due to the breaking of parity in the electroweak sector of the SM, they play a fundamental role in the understanding of the interactions between elementary particles. As a consequence, any possible new physics model needs to take into account the chiral nature of the fermions. For instance, the dimension-6 four-fermion operators of the SMEFT are classified according to the chiralities of those particles. Furthermore, obtaining a deeper explanation of the observed breaking of parity in the current experiments can be used as a guide to propose extensions of the SM. This was precisely the original motivation for the LR models, where parity-breaking can be derived using an SSB mechanism.

However, a mathematically rigorous calculation of the radiative corrections in chiral gauge theories is highly non-trivial due to the presence of the \(\gamma_5\) matrix. Nowadays, the preferred regularization prescription to take into account the divergences of the amplitudes in such calculations is dimensional regularization. Nevertheless, although the four-dimensional version of the Lorentz symmetry is preserved in this scheme, chiral gauge symmetries are violated already at the classical level, when the BMHV prescription for \(\gamma_5\) is employed, which is the only known mathematically consistent method. Using the tools of algebraic renormalization, it can be shown that obstructions to the Slavnov-Taylor identities, generated by single-insertions of a symmetry breaking operator (which is simply the BRST transformation of the \(D\)-dimensional Lagrangian) in 1PI diagrams, are present in this situation. Depending on their nature, these obstructions can be classified as evanescent (\(\varepsilon\)-terms), divergent (\(1/\varepsilon\)-terms) or finite (\(\varepsilon/\varepsilon\)-terms). The latter can be computed from the divergent part of a certain set of Feynman diagrams using the Bonneau method. In particular, in a theory which is free of physical anomalies, these finite obstructions to the Slavnov-Taylor identities can be eliminated adding the appropriate set of finite counterterms.

In this PhD thesis, a pedagogical introduction to the topics of chiral gauge theories and dimensional regularization has been presented. Special attention has been payed to the BRST and Batalin-Vilkovisky formalisms to quantize gauge theories, from which it is easy to prove (even in the presence of effective operators) that they are renormalizable; and that the SM, the SMEFT and the LR models are anomaly-free. The proof of renormalizability is completed using the BPH prescription, the forest formula and Weinberg's theorem. Also proofs of the most important identities of algebraic renormalization, particularized for dimensional regularization, are displayed; namely: the QAP and Bonneau identities. Using them, it is shown how to identify and eliminate the finite obstructions to the Zinn-Justin equation generated when using the BMHV scheme for \(\gamma_5\).

One of the main results of this text (Tabs.~\ref{tab:gamma_gamma} to~\ref{tab:CTs_two_scalars_2}) is the calculation of the finite symmetry-restoring counterterms needed for the one-loop renormalization of the SMEFT\(+\nu_R\) four-fermion operators, using the methods of algebraic renormalization. The scheme employed is minimal subtraction with dimensional regularization, using the BMHV prescription for \(\gamma_5\). Since the calculation has been done in the unbroken phase, fictitious fermions with opposite chiralities are added to the kinetic terms for each of the SM\(+\nu_R\) fermion fields, preserving the global symmetries of the theory, like the one associated with hypercharge conservation. The finite obstructions to the Slavnov-Taylor identities are obtained using the Bonneau method; the major difficulty of this procedure is to find the preimage under the BRST transformation of those Green's functions, needed to get the appropriate set of finite counterterms. The total number of structures needed are 28, and the set of operators is compared with the ones obtained using the spurion approach. Up to the point of completion of this text, no complete agreement between the two results has been found, there are some minor differences that cannot be explained by the addition of BRST invariant operators.

The situation with LR models has changed since their introduction in the 1970s. It is known that they can be embedded into grand unified theories, which can address the problem of the breaking of parity. Due to the running of the couplings, one can find \(g_L\neq g_R\) at the \(v_R\) scale, so it is interesting to study these models without additional discrete symmetries. The problem is that this would increase the number of free parameters, specially in the flavor sector, due to having a new \(V_{R}^{\rm CKM}\) matrix unrelated with its left-handed counterpart. In this scenario, the bounds on the masses of the new \(W_R\) and \(Z_R\) gauge bosons and on the LR scale are relaxed.

The other main result of the thesis is the derivation of lower bounds on the masses of the \(W_R\) and \(Z_R\) gauge bosons, and on the \(v_R\) scale, without dependence on the rest of the parameters of the theory. These bounds are obtained for the usual models where the SSB is realized using a bidoublet, and two doublets or two triplets; and for an effective model proposed by the author of this text, where only two doublets are required, and the masses of the fermions are generated through dimension-5 effective operators. It has been shown that the easiest way to get those bounds is using data from collider experiments, since flavor observables will require a global fit due to the presence of a huge number of free parameters in the quark-mixing matrix \(V_R^{\rm CKM}\).

To obtain a bound on the masses and on the energy scale that does not depend on the rest of the parameters of the theory, a lowest lower bound has been derived, in the sense that any other configuration of the parameter space will lead to a higher constraint. The tools employed for the theoretical calculation are the equivalence theorem, and the fact that the branching fraction of the decay of a gauge boson into a final state with one or two scalars is maximum in the limit where their masses are zero. Then, one only needs to find the value of the mixing angle \(\gamma\) that minimizes the cross section of production of the \(Z_R\), and the value of \(\gamma\) in the perturbative regime and the texture of the \(V_R^{\rm CKM}\) matrix that minimizes the production of the \(W_R\). The data employed come from the ATLAS and CMS experiments during the LHC Run 2, and are associated with the decays \(Z_R\to \ell\ell\) and \(W_R\to j_1j_2\), and \(W_R\to \ell\nu_R\) for the model with the bidoublet and two doublets. Then, from the direct bound on the mass of one state and the relation between the masses of the two gauge bosons and the angle \(\gamma\), it is possible to obtain an indirect bound on the mass of the other state. Combining the direct and indirect constraints, one gets the final bounds on the masses and on the \(v_R\) scale, which are displayed in Tab. \ref{Final_Summary_Tab}.

The following conclusions can be highlighted from the analysis done in this text:

\begin{itemize}
    \item Gauge symmetry can be employed, using the BRST and Batalin-Vilkovisky formalism, to constraint the divergences of the Green's functions in perturbative calculations. It can be shown that the infinities can be eliminated by redefinitions of the original parameters and fields of the Lagrangian. This statement is also true for EFTs.
    \item In a chiral gauge theory, it is possible to find obstructions to the Slavnov-Taylor identities, known as gauge anomalies, which can be classified into physical and spurious. In the SM, the SMEFT and LR models, all the gauge anomalies are spurious, so they can be eliminated adding the appropriate set of counterterms to the Lagrangian.
    \item The only known mathematically-consistent prescription for \(\gamma_5\) in dimensional regularization is the BMHV scheme, which introduces gauge symmetry-breaking already at the classical level, due to the necessity of regulating fermion propagators.
    \item Using the QAP and the Bonneau method, it can be shown that the finite obstructions to the Slavnov-Taylor identities can be calculated from divergences of 1PI diagrams with single insertions of a symmetry-breaking operator, the Dvo\v{r}\'ak operator \(\check{\Delta}\).
    \item From the calculation of the finite symmetry-restoring counterterms needed for the one-loop renormalization of the SMEFT\(+\nu_R\) four-fermion operators, it has been explicitly shown that the theory is anomaly-free at that order in the perturbative expansion. 
    \item The major difficulty in using the algebraic renormalization techniques in this calculation, was to find the preimage of the BRST transformation of the finite counterterms.
    \item Since the LR models can be embedded into other BSM theories, it is natural to study those models in the case where no additional discrete symmetries are assumed, and to propose a model where the masses of the fermions are generated through effective operators. 
    \item Model independent bounds using flavor observables on the masses of the \(W_R\) and \(Z_R\) gauge bosons, and on the \(v_R\) scale, will require a global fit to address the ensemble of free parameters.
    \item It is possible to find such global bounds, assuming only that the theory is perturbative, using collider-physics data from the Run 2 of the LHC experiment.
\end{itemize}

Taking into account the results of this thesis, some future lines of research might include:

\begin{itemize}
    \item The calculation using the methods of algebraic renormalization of the full set of finite symmetry-restoring counterterms needed to perform the one-loop renormalization of the SMEFT\(+\nu_R\), beyond the four-fermion operators already analyzed here.
    \item The application of these results to compute physical observables at one-loop order, and make comparisons with other renormalization schemes.
    \item The design of an appropriate software to automatize the calculation of these counterterms within the algebraic renormalization formalism.
    \item The calculation (probably using the software mentioned before) of the symmetry-restoring counterterms in SMEFT (or in other physical theories of interest) at higher loop-order and/or considering operators of higher dimension.
    \item An update of the collider bounds obtained for the masses of the \(W_R\) and \(Z_R\) gauge bosons, and on the LR scale \(v_R\), using the data from the Run 3 of the LHC.
    \item A calculation of the bounds on the masses of the new gauge bosons in other BSM theories using the techniques developed here, to get results without dependence on the other physical parameters.
    \item A derivation of a minimal basis of dimension-6 operators in the effective LR model, and applications of it to produce phenomenological results.
    \item A global fit including quark and lepton flavor and electroweak precision observables to constrain the parameters of the LR models in the case where no discrete symmetries are assumed.
\end{itemize}

\thispagestyle{empty}

\newpage

\bibliography{mybib}{}\addcontentsline{toc}{chapter}{Bibliography}
\bibliographystyle{unsrturl}
\newpage
\thispagestyle{empty}

\appendix
\counterwithin{equation}{chapter}
\chapter{Summary of Couplings in LR Models}\label{sec:summary_couplings}
\markboth{APPENDIX A. SUMMARY OF COUPLINGS IN LR MODELS}{APPENDIX A. SUMMARY OF COUPLINGS IN LR MODELS}

Here all the terms of the Lagrangian that have been used to calculate the widths and cross sections of the different $1 \to 2$ processes are displayed. The fields have been decomposed as \(\chi_{L,R}^0\coloneqq(v_{L,R}+\chi_{L,R}^{0r}+i\,\chi_{L,R}^{0i})/\sqrt{2}\), \(\delta_{L,R}^0\coloneqq(v_{L,R}+\delta_{L,R}^{0r}+i\,\delta_{L,R}^{0i})/\sqrt{2}\) and \(\phi_{1,2}^0\coloneqq(\kappa_{1,2}+\phi_{1,2}^{0r}\pm\,i\,\phi_{1,2}^{0\,i})/\sqrt{2}\).
All the couplings are given at leading order in powers of \(v_{EW}/v_R\).

\begin{itemize}
    \item \(Z_R\) with two fermions:
    \begin{equation}
    \label{ecu.A.1}
        \begin{split}
        &-\frac{1}{12}\frac{e}{\cos\theta_W}\tan\gamma\,Z_R^{\mu}\,\bar{u}\gamma_{\mu}\left\{\left(2-3\cot^2\gamma\right)-\left(3\cot^2\gamma\right)\gamma_5\right\}u
        \\& -\frac{1}{12}\frac{e}{\cos\theta_W}\tan\gamma\,Z_R^{\mu}\,\bar{d}\gamma_{\mu}\left\{\left(2+3\cot^2\gamma\right)+\left(3\cot^2\gamma\right)\gamma_5\right\}d
        \\& +\frac{1}{4}\frac{e}{\cos\theta_W}\tan\gamma\,Z_R^{\mu}\,\bar{e}\gamma_{\mu}\left\{\left(2-\cot^2\gamma\right)-\left(\cot^2\gamma\right)\gamma_5\right\}e
        \\& +\frac{1}{2}\frac{e}{\cos\theta_W}\tan\gamma\, Z_R^{\mu}\,\bar{\nu}_L\gamma_{\mu}\nu_L+\frac{1}{2}\frac{e}{\cos\theta_W}\tan\gamma\left(1+\cot^2\gamma\right)\,Z_R^{\mu}\,\bar{\nu}_R\gamma_{\mu}\nu_R.
        \end{split}
    \end{equation}
    
    \item \(Z_R\) with two neutral scalars in the models with doublets:
    \begin{equation}
    \label{ecu.A.2}
        \begin{split}
        Z_R^{\mu}&\left\{\frac{1}{2}\,g_R\cos\gamma\left[\phi_1^{0r}\partial_{\mu}\phi_1^{0i}-\phi_1^{0i}\partial_{\mu}\phi_1^{0r}+\phi_2^{0r}\partial_{\mu}\phi_2^{0i}-\phi_2^{0i}\partial_{\mu}\phi_2^{0r}\right]\right.\\&+\frac{1}{2}\,g_{\rm X}\sin\gamma\left[\chi_L^{0r}\partial_{\mu}\chi_L^{0i}-\chi_L^{0i}\partial_{\mu}\chi_L^{0r}\right]\\&\left.+\frac{1}{2}\,\left(g_R\cos\gamma+g_{\rm X}\sin\gamma\right)\left[\chi_R^{0r}\partial_{\mu}\chi_R^{0i}-\chi_R^{0i}\partial_{\mu}\chi_R^{0r}\right]\right\}.
        \end{split}
    \end{equation}
    
    \item \(Z_R\) with two neutral scalars in the models with triplets:
    \begin{equation}
    \label{ecu.A.3}
        \begin{split}
        Z_R^{\mu}&\left\{\frac{1}{2}\,g_R\cos\gamma\left[\phi_1^{0r}\partial_{\mu}\phi_1^{0i}-\phi_1^{0i}\partial_{\mu}\phi_1^{0r}+\phi_2^{0r}\partial_{\mu}\phi_2^{0i}-\phi_2^{0i}\partial_{\mu}\phi_2^{0r}\right]\right.\\&+\,g_{\rm X}\sin\gamma\left[\delta_L^{0r}\partial_{\mu}\delta_L^{0i}-\delta_L^{0i}\partial_{\mu}\delta_L^{0r}\right]\\&\left.+\left(g_R\cos\gamma+g_{\rm X}\sin\gamma\right)\left[\delta_R^{0r}\partial_{\mu}\delta_R^{0i}-\delta_R^{0i}\partial_{\mu}\delta_R^{0r}\right]\right\}.
        \end{split}
    \end{equation}

    \item \(Z_R\) with two singly charged scalars in the models with doublets:
    \begin{equation}
    \label{ecu.A.4}
        \begin{split}
        i\,Z_R^{\mu}&\left\{\frac{1}{2}\,g_R\cos\gamma\left[\phi_1^{-}\partial_{\mu}\phi_1^{+}-\phi_1^{+}\partial_{\mu}\phi_1^{-}+\phi_2^{-}\partial_{\mu}\phi_2^{+}-\phi_2^{+}\partial_{\mu}\phi_2^{-}\right]\right.\\&+\frac{1}{2}\,g_{\rm X}\sin\gamma\left[\chi_L^{+}\partial_{\mu}\chi_L^{-}-\chi_L^{-}\partial_{\mu}\chi_L^{+}\right]\\&\left.+\frac{1}{2}\left(g_R\cos\gamma-g_{\rm X}\sin\gamma\right)\left[\chi_R^{-}\partial_{\mu}\chi_R^{+}-\chi_R^{+}\partial_{\mu}\chi_R^{-}\right]\right\}.
        \end{split}
    \end{equation}

    \item \(Z_R\) with two singly charged scalars in the models with triplets:
    \begin{equation}
    \label{ecu.A.5}
        \begin{split}
        i\,Z_R^{\mu}&\left\{\frac{1}{2}\,g_R\cos\gamma\left[\phi_1^{-}\partial_{\mu}\phi_1^{+}-\phi_1^{+}\partial_{\mu}\phi_1^{-}+\phi_2^{-}\partial_{\mu}\phi_2^{+}-\phi_2^{+}\partial_{\mu}\phi_2^{-}\right]\right.\\&+g_{\rm X}\sin\gamma\left[\delta_L^{+}\partial_{\mu}\delta_L^{-}-\delta_L^{-}\partial_{\mu}\delta_L^{+}+\delta_R^{+}\partial_{\mu}\delta_R^{-}-\delta_R^{-}\partial_{\mu}\delta_R^{+}\right]\bigg\}.
        \end{split}
    \end{equation}

    \item \(Z_R\) with two doubly charged scalars:
    \begin{equation}
    \label{ecu.A.6}
    \begin{split}
    i\,Z^{\mu}_R&\left\{\left(g_R\cos\gamma-g_{\rm X}\sin\gamma\right)\left[\delta_R^{--}\partial_{\mu}\delta_R^{++}-\delta_R^{++}\partial_{\mu}\delta_R^{--}\right]\right.\\&\left.-g_{\rm X}\sin\gamma\left[\delta_L^{--}\partial_{\mu}\delta_L^{++}-\delta_L^{++}\partial_{\mu}\delta_L^{--}\right]\right\}.
    \end{split}
    \end{equation}

    \item \(Z_R\) with \(W_R\) and a singly charged scalar in the models with doublets:
    \begin{equation}
    \label{ecu.A.7}
        -\frac{e}{\cos\theta_W}\sin\gamma M_{Z_R}\chi_R^{-}W_{R,\mu}^+Z_R^{\mu}+\mathrm{h.c.}
    \end{equation}
     
    \item \(Z_R\) with \(W_R\) and a singly charged scalar in the models with triplets: 
    \begin{equation}
        \label{ecu.A.8}
        -\frac{1}{\sqrt{2}}\frac{e}{\cos\theta_W}\sin\gamma\left(2+\cot^2\gamma\right)M_{Z_R}\delta_R^{-}W_{R,\mu}^+Z_R^{\mu}+\mathrm{h.c.}
    \end{equation}

    \item \(Z_R\) with two \(W_L\) bosons:
    \begin{equation}
        \label{ecu.A.9}
        \begin{split}
        i e \cot \theta_W \sin \alpha &\left\{(\partial^{\mu}W_L^{-\nu})(W_{L,\mu}^+Z_{R,\nu}-W_{L,\nu}^+Z_{R,\mu})+(\partial^{\mu}W_L^{+\nu})(W_{L,\nu}^-Z_{R,\mu}\right.\\&\left.-W_{L,\mu}^-Z_{R,\nu})+(\partial^{\mu}Z_R^{\nu})(W_{L,\mu}^-W_{L,\nu}^+-W_{L,\nu}^-W_{L,\mu}^+)\right\},
    \end{split}
    \end{equation}
    where $ \sin \alpha = \dfrac{\tan\gamma \tan \theta_W}{\cos \theta_W}\dfrac{M_{W_L}^2}{M_{Z_R}^2} $ results from the mixing of the massive neutral gauge bosons 
    in the \(\chi_L+\chi_R\) Effective LR Model.

    \item \(Z_R\) with two \(W_R\) bosons:
    \begin{equation}
        \label{ecu.A.10}
        \begin{split}
        i\frac{e}{\cos\theta_W}\cot\gamma&\left\{(\partial^{\mu}W_R^{-\nu})(W_{R,\mu}^+Z_{R,\nu}-W_{R,\nu}^+Z_{R,\mu})+(\partial^{\mu}W_R^{+\nu})(W_{R,\nu}^-Z_{R,\mu}\right.\\&\left.-W_{R,\mu}^-Z_{R,\nu})+(\partial^{\mu}Z_R^{\nu})(W_{R,\mu}^-W_{R,\nu}^+-W_{R,\nu}^-W_{R,\mu}^+)\right\}.
    \end{split}
    \end{equation}

    \item \(W_R\) with two fermions (Dirac neutrinos):
    \begin{equation}
    \label{ecu.A.11}
    \frac{g_R}{\sqrt{2}}W_R^{+\,\mu}\left\{\bar{u}\gamma_{\mu}V_R^{\mathrm{CKM}}\mathcal{P}_Rd+\bar{\nu}\gamma_{\mu}V_R^{\mathrm{PMNS}}\mathcal{P}_Re\right\}+\mathrm{h.c.}
    \end{equation}

    \item \(W_R\) with two fermions (Majorana neutrinos):
    \begin{equation}
    \label{ecu.A.12}
    \frac{g_R}{\sqrt{2}}W_R^{+\,\mu}\left\{\bar{u}\gamma_{\mu}V_R^{\mathrm{CKM}}\mathcal{P}_Rd+\bar{\nu}_h\gamma_{\mu}V_h^{\textrm{PMNS}}\mathcal{P}_R e
    \right\}+\mathrm{h.c.}
    \end{equation}

    \item \(W_R\) with a singly charged and a neutral scalar in the models with doublets:
    \begin{equation}
    \label{ecu.A.13}
    \begin{split}
    &\frac{g_R}{2}W_R^{+\mu}\left\{\phi_1^-\partial_{\mu}\phi_1^{0i}-\phi_1^{0i}\partial_{\mu}\phi_1^-+\phi_2^{0i}\partial_{\mu}\phi_2^--\phi_2^-\partial_{\mu}\phi_2^{0i}+\chi_R^{0i}\partial_{\mu}\chi_R^--\chi_R^-\partial_{\mu}\chi_R^{0i}\right.\\ &\left.-i\left[\phi_1^-\partial_{\mu}\phi_1^{0r}-\phi_1^{0r}\partial_{\mu}\phi_1^-+\phi_2^{0r}\partial_{\mu}\phi_2^--\phi_2^-\partial_{\mu}\phi_2^{0r}+\chi_R^{0r}\partial_{\mu}\chi_R^--\chi_R^-\partial_{\mu}\chi_R^{0r}\right]\right\}+\mathrm{h.c.}
    \end{split}
    \end{equation}

    \item \(W_R\) with a singly charged and a neutral scalar in the models with triplets:
    \begin{equation}
    \label{ecu.A.14}
    \begin{split}
    &\frac{g_R}{2}W_R^{+\mu}\left\{\phi_1^-\partial_{\mu}\phi_1^{0i}-\phi_1^{0i}\partial_{\mu}\phi_1^-+\phi_2^{0i}\partial_{\mu}\phi_2^--\phi_2^-\partial_{\mu}\phi_2^{0i}+\sqrt{2}\left(\delta_R^{0i}\partial_{\mu}\delta_R^--\delta_R^-\partial_{\mu}\delta_R^{0i}\right)\right.\\ &\left.-i\left[\phi_1^-\partial_{\mu}\phi_1^{0r}-\phi_1^{0r}\partial_{\mu}\phi_1^-+\phi_2^{0r}\partial_{\mu}\phi_2^--\phi_2^-\partial_{\mu}\phi_2^{0r}+\sqrt{2}\left(\delta_R^{0r}\partial_{\mu}\delta_R^--\delta_R^-\partial_{\mu}\delta_R^{0r}\right)\right]\right\}\\&+\mathrm{h.c.}
    \end{split}
    \end{equation}

     \item \(W_R\) with a doubly and a singly charged scalars:
    \begin{equation}
        \label{ecu.A.15}
       - ig_RW_R^{+\mu}\left\{\delta_R^{--}\partial_{\mu}\delta_R^+-\delta_R^+\partial_{\mu}\delta_R^{--}\right\}+\mathrm{h.c.}
    \end{equation}

    \item \(W_R\) with \(W_L\) and a doubly charged scalar:
    \begin{equation}
        \label{ecu.A.16}
        \sqrt{2}g_R^2v_R e^{i\lambda}\sin\xi\cos\xi W_R^{-\mu}W_{L\mu}^-\delta_R^{++}+\mathrm{h.c.}
    \end{equation}

    \item \(W_R\) with a singly charged scalar and a photon in the model with a bidoublet and two doublets (exact result up to \(W_L-W_R\) mixing coefficients):
    \begin{equation}
        \label{ecu.A.17}
        \frac{1}{2}eg_RW_R^{+\mu}A_{\mu}\left\{v_R\chi_R^{-}+\kappa_2^{\ast}\phi_2^--\kappa_1\phi_1^-\right\}+\mathrm{h.c.}
    \end{equation}
    \item \(W_R\) with a singly charged scalar and a photon in the model with a bidoublet and two triplets (exact result up to \(W_L-W_R\) mixing coefficients):
    \begin{equation}
        \label{ecu.A.18}
         \frac{1}{2}eg_RW_R^{+\mu}A_{\mu}\left\{\sqrt{2}v_R\delta_R^{-}+\kappa_2^{\ast}\phi_2^--\kappa_1\phi_1^-\right\}+\mathrm{h.c.}
    \end{equation}
    \item \(W_R\) with scalars in the model with a bidoublet and two doublets (exact result up to \(W_L-W_R\) mixing coefficients):
    \begin{equation}
        \label{ecu.A.19}
        -\frac{i}{2}g_RW_R^{+\mu}\partial_{\mu}\left\{v_R\chi_R^{-}+\kappa_2^{\ast}\phi_2^--\kappa_1\phi_1^-\right\}+\mathrm{h.c.}
    \end{equation}
    \item \(W_R\) with scalars in the model with a bidoublet and two triplets (exact result up to \(W_L-W_R\) mixing coefficients):
    \begin{equation}
        \label{ecu.A.20}
        -\frac{i}{2}g_RW_R^{+\mu}\partial_{\mu}\left\{\sqrt{2}v_R\delta_R^{-}+\kappa_2^{\ast}\phi_2^--\kappa_1\phi_1^-\right\}+\mathrm{h.c.}
    \end{equation}
\end{itemize}

Eqs.~\eqref{ecu.A.19} and \eqref{ecu.A.20} are used to identify the combinations of scalar degrees of freedom that result being the Goldstone bosons, see App.~\ref{sec:Non-Fermionic_Decays}. There is no interaction vertex between \(W_R\), \(W_L\) and \(A\) at the tree level; there are couplings of the form \(W_1W_1A\) and \(W_2W_2A\) (while \(W_1 W_2 A\) is not possible) but when the transformation in Eq.~\eqref{Diagonalization_Charged_Gauge} is applied, the terms that contain the vertex \(W_LW_RA\) cancel each other due to the unitarity of the mixing matrix. Neither cubic nor quartic interaction vertices among only neutral gauge bosons can be present. Indeed, the gauge self-interactions are generated by terms proportional to \(\mathrm{tr}\left\{\partial^{\mu}W^{\nu}\left[W_{\mu},W_{\nu}\right]\right\}\) and \(\mathrm{tr}\left\{\left[W^{\mu},W^{\nu}\right]\left[W_{\mu},W_{\nu}\right]\right\}\), but the commutator \(\left[W_{\mu},W_{\nu}\right]\) does not contain products of the form \(W_{\mu}^3W_{\nu}^3\) since \(\left[\sigma^3,\sigma^3\right]=0\).
\newpage

\thispagestyle{empty}

\chapter{\boldmath Non-Fermionic Decays of the \(Z_R\) and \(W_R\) Gauge Bosons}\label{sec:Non-Fermionic_Decays}
\markboth{APPENDIX B. NON-FERMIONIC DECAYS OF THE \(Z_R\) AND \(W_R\)}{APPENDIX B. NON-FERMIONIC DECAYS OF THE \(Z_R\) AND \(W_R\)}

Here the details of the calculation of the bounds on the partial widths of the non-fermionic decays of the \(W_R\) and \(Z_R\) gauge bosons are shown. Consider first the latter. To calculate the partial width of the process \(Z_R\) decaying to two particular neutral scalars, the Lagrangian of the interaction is needed. It can be written as

\begin{equation}
    \label{ecu.C.1}
    \mathcal{L}=-Z_R^{\mu}\sum_{a,\,b}C_{ab}\left\{\psi_a\partial_{\mu}\psi_b-\psi_b\partial_{\mu}\psi_a\right\},
\end{equation}

\noindent where the \(C_{ab}\) are the coupling constants and \(\left\{\psi_a\right\}_{a=1}^{8}\) is just a basis of the neutral fields. Looking at Eqs.~(\ref{ecu.A.2}) and (\ref{ecu.A.3}) it is easy to see that Eq.~(\ref{ecu.C.1}) can be used to describe both the cases of doublets and triplets with the bi-doublet. In the interaction basis the non-vanishing couplings are \(C_{12}^D=C_{12}^T=C_{34}^D=C_{34}^T=\frac{1}{2}g_R\cos\gamma\), \(C_{56}^D=\frac{1}{2}\,C_{56}^T=\frac{1}{2}g_X\sin\gamma\) and \(C_{78}^D=\frac{1}{2}\,C_{78}^T=\frac{1}{2}\,\left(g_R\cos\gamma+g_X\sin\gamma\right)\). Now, it is necessary to make a rotation into the basis of the physical fields \(\left\{\psi'_a\right\}_{a=1}^{8}\). The transformation can be expressed as \(\psi_a=\sum_i\mathbb{R}_{ai}\psi_i'\). Then, defining the matrix \(C'\coloneqq\mathbb{R}^{\mathrm{T}}\,C\,\mathbb{R}\) one can write

\begin{equation}
    \label{ecu.C.2}
    \mathcal{L}=-Z_R^{\mu}\sum_{i,\,j}C_{ij}'\left\{\psi'_i\partial_{\mu}\psi'_j-\psi'_j\partial_{\mu}\psi'_i\right\}.
\end{equation}

\noindent Therefore, the partial width of the process \(Z_R\rightarrow\psi_i'\psi_j'\) is just

\begin{equation}
    \label{ecu.C.3}
    \Gamma\left(Z_R\rightarrow\psi_i'\psi_j'\right)=\frac{M_{Z_R}}{48\pi}\left(C'_{ij}-C'_{ji}\right)^2\left[1-2\left(x_i+x_j\right)+\left(x_i-x_j\right)^2\right]^{3/2},
\end{equation}

\noindent where \(x_i\coloneqq m_i^2/M_{Z_R}^2\). It is not difficult to show that the maximum possible value of the partial width is achieved when \(x_i=x_j=0\), i.e., in the massless case. Thus, one can derive the following bound:

\begin{equation}
    \label{ecu.C.4}
    \begin{split}
    \Gamma\left(Z_R\rightarrow2\,\mathrm{Neutral\,Scalars}\right)&<\frac{M_{Z_R}}{48\pi}\sum_{j>i}\left(C'_{ij}-C'_{ji}\right)^2\\&=\frac{M_{Z_R}}{48\pi}\frac{1}{2}\mathrm{tr}\left\{\left(C'-C'^{\mathrm{T}}\right)\left(C'^{\mathrm{T}}-C'\right)\right\}.
    \end{split}
\end{equation}

\noindent Using the properties of the trace and the definition of \(C'\) it is also not difficult to show that 

\begin{equation}
    \label{ecu.C.5}
    \Gamma\left(Z_R\rightarrow2\,\mathrm{Neutral\,Scalars}\right)<\frac{M_{Z_R}}{48\pi}\sum_{a,\,b}C_{ab}^2.
\end{equation}

It is possible to do exactly the same for the decay to charged scalars. From Eqs.~(\ref{ecu.A.4}) and (\ref{ecu.A.5}) it is easy to see that the interaction Lagrangian of the \(Z_R\) and 2 singly charged scalars can be written as

\begin{equation}
    \label{ecu.C.6}
    \mathcal{L}=-i\,Z_R^{\mu}\sum_aC_a\left\{\psi^+_a\partial_{\mu}\psi^-_a-\psi^-_a\partial_{\mu}\psi_a^+\right\}.
\end{equation}

\noindent
In this case, in the interaction basis one obtains \(C_1^D=C_1^T=C_2^D=C_2^T=\frac{1}{2}\,g_R\cos\gamma\), \(C_3^D=\frac{1}{2}\,C_3^T=\frac{1}{2}\,C_4^T=-\frac{1}{2}g_X\sin\gamma\) and \(C_4^D=\frac{1}{2}\,\left(g_R\cos\gamma-g_X\sin\gamma\right)\). Now, it is necessary to make the transformation into the physical basis \(\psi_a^{-}=\sum_i U_{ai}\psi'^{-}_i\), where \(U\) is a unitary matrix. Using the same arguments as in the previous case one can find that

\begin{equation}
    \label{ecu.C.7}
    \Gamma\left(Z_R\rightarrow2\,\mathrm{Singly\,Charged\,Scalars}\right)<\frac{M_{Z_R}}{48\pi}\sum_{i,\,j}\left|C_{ij}\right|^2,
\end{equation}

\noindent where \(C_{ij}\coloneqq\sum_a C_a U_{ai}^{\ast}U_{aj}\). Therefore, it is not difficult to prove that

\begin{equation}
    \label{ecu.C.8}
    \Gamma\left(Z_R\rightarrow2\,\mathrm{Singly\,Charged\,Scalars}\right)<\frac{M_{Z_R}}{48\pi}\sum_{a}C_a^2.
\end{equation}

If one wants to calculate the bound on the partial width of the decay of the \(Z_R\) to two bosons, it is only necessary to add the contribution of \(Z_R\rightarrow\mathrm{Doubly\,Charged\,Scalars}\) in the case of the model with triplets. From Eq.~(\ref{ecu.A.6}), it is easy to see that this calculation is completely analogous to the one for singly charged scalars.

Now the fact that the partial width of the decay \(Z_R\rightarrow W_R^{\pm}W_L^{\mp}\) is negligible will be shown. The Feynman rule of the interaction of three gauge bosons is given by a function of the form \(\eta_{\mu\nu}\left(q-p\right)_{\lambda}+\eta_{\lambda\mu}\left(p-r\right)_{\nu}+\eta_{\nu\lambda}\left(r-q\right)_{\mu}\) with \(p\), \(q\) and \(r\) the incoming momenta of the particles. Then, the dependence on the mass of the partial width of the decay of one gauge boson with mass \(M\) into other two with masses \(m_1\) and \(m_2\) is given by

\begin{equation}
    \label{ecu.C.9}
    \begin{split}
    f\left(M,m_1,m_2\right)=\frac{M^5}{m_1^2m_2^2}&\left[1-2\frac{m_1^2+m_2^2}{M^2}+\left(\frac{m_1^2-m_2^2}{M^2}\right)^2\right]^{3/2}\\&\times\left\{1+10\frac{m_1^2+m_2^2}{M^2}+\frac{m_1^4+m_2^4+10m_1^2m_2^2}{M^4}\right\}.
    \end{split}
\end{equation}

\noindent In the present case, one can neglect the mass of the \(W_L\) boson and write

\begin{equation}
    \label{ecu.C.10}
    f\left(M_{Z_R},M_{W_R},M_{W_L}\right)\approx\frac{M^5_{Z_R}}{M_{W_L}^2M_{W_R}^2}\left(1-\frac{M^2_{W_R}}{M^2_{Z_R}}\right)^3\left[1+10\left(\frac{M_{W_R}}{M_{Z_R}}\right)^2+\left(\frac{M_{W_R}}{M_{Z_R}}\right)^4\right].
\end{equation}

\noindent
The process \(Z_R\rightarrow W_R^{\pm}W_L^{\mp}\) can only occur at the tree level if there is a mixing between the \(W_L\) and the \(W_R\) bosons. The amplitude is proportional to \(\sin\xi\), so one gets

\begin{equation}
    \label{ecu.C.11}
    \Gamma\left(Z_R\rightarrow W_R^{\pm}W_L^{\mp}\right)\sim\frac{M^5_{Z_R}}{M_{W_L}^2M_{W_R}^2}\sin^2\xi\sim\frac{v_{EW}^2}{v_R}.
\end{equation}

\noindent
Consequently, this term can be neglected in the calculation of the width of the \(Z_R\).

It is also possible to prove that the partial width of the decay \(Z_R\to H^{\pm}W_R^{\mp}\) is negligible. The process is absent in the \(\chi_L+\chi_R\) Effective LR Model since there are no physical charged scalars. On the other hand, as one can see in Eqs. \eqref{ecu.A.7} and \eqref{ecu.A.8}, the scalars that mediate the decay in the interaction eigenstates basis in the models with a bidoublet and two doublets or two triplets are \(\chi_R^{\pm}\) and \(\delta_R^{\pm}\), respectively. Looking at Eqs. \eqref{ecu.A.19} and \eqref{ecu.A.20} one can see that the linear combinations \(v_R\chi_R^{-}+\kappa_2^{\ast}\phi_2^--\kappa_1\phi_1^-\), \(\sqrt{2}v_R\delta_R^{-}+\kappa_2^{\ast}\phi_2^--\kappa_1\phi_1^-\) and their hermitian conjugates are proportional to a Goldstone boson, i.e., this interacting term can be canceled by an appropriate choice of gauge-fixing Lagrangian. Consequently, the mixing of the fields \(\chi_R^{\pm}\) and \(\delta_R^{\pm}\) with physical scalars is \(\mathcal{O}\left(v_{EW}/v_R\right)\). Thus, the partial width is, at most, proportional to \(v_{EW}^2/v_R\), so it can be neglected.

Finally, it is necessary to calculate the partial width of the decay \(Z_R\rightarrow W_R^+W_R^-\). Using the coupling shown in Eq.~(\ref{ecu.A.10}) one can see that 

\begin{equation}
    \label{ecu.C.15}
    \Gamma\left(Z_R\rightarrow W_R^+W_R^-\right)=\frac{M_{Z_R}}{192\pi}\left(\frac{e}{\cos\theta_W}\right)^2\frac{\cot^2\gamma}{x_w^2}\left(1-4x_w\right)^{3/2}\left\{1+20x_w+12x_w^2\right\}.
\end{equation}

Consider now the non-fermionic decays of the \(W_R\) boson. As indicated in the text, one can set a bound on the partial widths of these processes by calculating the upper bound on  \(\Gamma\left(W_R\rightarrow 2\,\mathrm{Scalars}\right)\). Start with the contributions with a neutral and a singly charged scalar, whose interaction Lagrangians are shown in Eqs.~(\ref{ecu.A.13}) and (\ref{ecu.A.14}). One can write them as

\begin{equation}
    \label{ecu.C.16}
    \mathcal{L}=-W_R^+\sum_{ij}C_{ij}\left\{\psi_i^0\partial_{\mu}\psi_j^--\psi_j^-\partial_{\mu}\psi_i^0\right\}+\mathrm{h.c.}
\end{equation}

\noindent It is necessary to make the transformations into the mass eigenstates \(\psi_i^0=\sum_k\mathbb{R}_{ik}\psi'^{0}_k\) and \(\psi_j^-=\sum_l\mathrm{U}_{jl}\psi'^{-}_l\). Then, one obtains:

\begin{equation}
    \label{ecu.C.17}
    \mathcal{L}=-W_R^+\sum_{ij}C'_{ij}\left\{\psi'^0_i\partial_{\mu}\psi'^-_j-\psi'^-_j\partial_{\mu}\psi'^0_i\right\}+\mathrm{h.c.},
\end{equation}

\noindent where \(C':=\mathbb{R}^{\mathrm{T}}C\mathrm{U}\). Again, the upper bound on the partial width corresponds to the case in which the particles in the final state have a negligible mass compared to the one of the \(W_R\) boson:

\begin{equation}
    \label{ecu.C.18}
    \Gamma\left(W^{\pm}_R\rightarrow\psi'^0\psi'^{\pm}\right)<\frac{M_{W_R}}{48\pi}\sum_{ij}\left|C'_{ij}\right|^2=\frac{M_{W_R}}{48\pi}\mathrm{tr}\left\{\mathbb{R}^{\mathrm{T}}C\mathrm{U}\mathrm{U}^{\dagger}C^{\dagger}\mathbb{R}\right\}=\frac{M_{W_R}}{48\pi}\sum_{ij}C_{ij}^2.
\end{equation}

The calculation for the case with a singly and a doubly charged scalars Eq.~\eqref{ecu.A.15} is completely analogous. One can also show that the contribution of \(W_R^{\pm}\rightarrow W_L^{\mp}H^{\pm\pm}\) in the models with triplets is negligible. At tree level, this process is mediated by the coupling term in Eq.~\eqref{ecu.A.16}. It can be written as

\begin{equation}
    \label{ecu.C.19}
    \mathcal{L}=CW_R^{-\mu}W_{L\mu}^-\delta_R^{++}+\mathrm{h.c.}.
\end{equation}

\noindent
Since \(\sin\xi\sim v_{EW}^2/v_R^2\), then \(|C|\sim v_{EW}^2/v_R\). Now, make a transformation into the mass eigenstates \(\delta_R^{++}=\sum_i \mathrm{U}_{2i}\psi^{++}_i\), where \(\mathrm{U}\) is a unitary transformation. Thus, one finds that

\begin{equation}
    \label{ecu.C.20}
    \mathcal{L}=CW_R^{-\mu}W_{L\mu}^-\sum_i \mathrm{U}_{2i}\psi^{++}_i+\mathrm{h.c.}.
\end{equation}

\noindent Then, it can be shown that the partial width of the process \(W_R^{\pm}\rightarrow W_L^{\mp}\psi_i^{\pm\pm}\) is 

\begin{equation}
    \label{ecu.C.21}
    \Gamma\left(W_R^{\pm}\rightarrow W_L^{\mp}\psi^{\pm\pm}_i\right)\approx\frac{1}{24\pi}\frac{|C|^2}{M_{W_R}^3}\left|\mathrm{U}_{2i}\right|^2\left(M_{W_R}^2-M_i^2\right)\left\{1+\frac{\left(M_{W_R}^2-M_i^2\right)^2}{8M_{W_L}^2M_{W_R}^2}\right\},
\end{equation}

\noindent where \(M_i\) is the mass of \(\psi_i^{\pm\pm}\) and the approximation \(M_{W_R}\gg M_{W_L}\) has been taken. Since \(\left|\mathrm{U}_{2i}\right|^2\sim\mathcal{O}(1)\) at most, then \(\Gamma\left(W_R^{\pm}\rightarrow W_L^{\mp}\psi^{\pm\pm}_i\right)\sim v_{EW}^2/v_R\). Consequently, the partial width \(\Gamma\left(W_R^{\pm}\rightarrow W_L^{\mp}H^{\pm\pm}\right)\) is negligible.

Finally, one can show that the process \(W_R^{\pm}\to H^{\pm}A\) is forbidden at tree level. Since there are no physical charged scalars in the \(\chi_L+\chi_R\) Effective LR model the result is straightforward in this case. The important interaction vertices in the models with a bidoublet and two doublets or two triplets are presented in Eqs.~\eqref{ecu.A.17} and \eqref{ecu.A.18}, respectively. Nonetheless, looking at Eqs.~\eqref{ecu.A.19} and \eqref{ecu.A.20}, it is easy to see that those combinations of scalars correspond to Goldstone bosons. This is so because of gauge invariance of the electromagnetic interaction, i.e., a vertex of the form \(W_R^{\pm\mu}A_{\mu}H^{\mp}\) with physical fields would break gauge invariance.
Consequently, there are no Feynman rules for this process with physical Higgses at tree level.

\end{document}